# Construction of Molecular Dynamics Like Cellular Automata Models for Simulation of Compressible Fluid Dynamic Systems

**Himanshu Agrawal**

**Department of Aerospace Engineering**
INDIAN INSTITUTE OF TECHNOLOGY, KANPUR
February 1998

# Construction of Molecular Dynamics Like Cellular Automata Models for Simulation of Compressible Fluid Dynamic Systems

A Dissertation Submitted
in Partial Fulfillment of the Requirements
for the Degree of
**Doctor of Philosophy**

*by*

**Himanshu Agrawal**

*to the*

**Department of Aerospace Engineering**
INDIAN INSTITUTE OF TECHNOLOGY, KANPUR—208 016, INDIA

February 1998

**Revision: 2. (by Himanshu Agrawal; October 2, 1998)**

# Certificate

It is certified that the work contained in this thesis entitled *"Construction of Molecular Dynamics Like Cellular Automata Models for Simulation of Compressible Fluid Dynamic Systems"* is an original research work carried out by *"Himanshu Agrawal"* under my supervision, and that this work has not been submitted elsewhere for a degree.

(Prof. E. Rathakrishnan)
*Department of Aerospace Engineering*
*Indian Institute of Technology, Kanpur—208 016, INDIA*

February 6, 1998

*To*

**Mom, Dad, Swati, Udit, and Madhuri**

# Acknowledgments

*. . . chukle chukle chukle . . .*
Verbalize.
*Who comes first?*
. . . ?!
*The one who plants or the one who nurtures.*
. . . !
*Ah! Naturally. . . .*

The support required for completion of this work was provided by Dr. E. Rathakrishnan, my thesis supervisor. Had he not given me an unusual freedom—to choose my path and my tools as if I knew the destination and the terrain to cross, to roam around as if I were lost among tall mountains, to drink innumerable cups of tea as if there were no other edibles, to embrace fire as if it were the essence of life, and to sleep for unusually long hours or not to sleep at all as if I had serious work to do in both the worlds—this work would never even have been undertaken, what to say of its conclusion. He has been the driving force behind this work—supporting at all strange and critical bends, inspiring, renewing confidence, and invigorating me with energy again and again without as much as a single question. He is an extremely demanding yet freedom loving man who knows its meaning and knows that others need it too, a man of faith who knows that it works wonders and can be held in place only with threads of unquestioning honesty reaching from end to end. Dr. E. Rathakrishnan has been a most observant, extremely patient, and careful thesis supervisor.

I was introduced to cellular automata in the context of image processing in a course in electrical engineering taught by Dr. E. G. Rajan. And a peculiar turn that my thought took also occurred in that course when he asked me to model gas dynamics using cellular automata, as a full semester assignment problem. It was a tough job, but I did not know that and took it up. Knowing fully well the difficulties and the implications of success, he was constantly vigilant and helped me at every point in all possible ways that he could find. He left the institute after the course was over, giving me an "A" grade. He did not seem to think that I had achieved much and also did not know that he had changed my entire thought process. As a consequence of the course and the assignment problem, I had developed a way of looking at things in which the entire world was discrete. In the next semester, during the literature survey, I was convinced that nothing but microscopic approach will give me what I want; only if molecular dynamics can be made faster. I *knew* it can be done using cellular automata, but "How?" was the problem. Dr. E. G. Rajan is the one who implanted the seeds for this work in my mind which fructified after fifteen thousand cups of tea, twenty five thousand cigarettes and thousands of sleepless nights.

During this work I had to expand my field of knowledge from gas dynamics to computer science, discrete mathematics, and physics (particularly theories of space-time, complex systems, interparticle interactions, and many body problems). During this process I feel fortunate to have held enlightening discussions with Deepak Murthy and Jitendra Das on issues related to many aspects of computing, computer science, formal systems, and computer architecture. Discussions with Venu Gopal Rao, a man of great experience and expertise, on computational methods, particularly finite element method, have found their place in this work in the form of certain abstractions.

Lengthy and indepth discussions with P. V. K. Reddy and Shreesh Jadhav on various aspects of human life, particularly society, history, philosophy, religion, and psychology were the major factors in my unlocking of ideas. Without them, perhaps, I would never express myself in the way I do now.

I am convinced that nobody, other than the lab mates and coworkers, can understand the problems and frustrations of a research scholar and give appropriate support when needed. The company of T. J. Ignatius, S. Elangovan, K. Srinivasan, Rahul Konnur, Bhadri Srinivas, Sujay Datta, Shanti Lal Das, Alok Sharan, Alok Srivastava, Shashi Bhushan Verma, Kumar Neeraj Sachdev, Manish Shreekhande, G. K. Singh, Sirshendu De, Subir Bhattacharjee, Swapan Chakraborty, Raghunath Behera, Dipak Sathpati, Rajesh Khanna, Brijesh Pandey, and Sudipto Ghosh had been very pleasant and supportive through out this work.

It is hard to accept that ideas can be generated while teaching even in a brain that is almost always clogged with them. I experienced it, and had to accept, while explaining the Kac master kinetic equation, workings and philosophy of the direct simulation Monte-Carlo method, and their interrelationship to V. Manoj Kumar while he was working on his Masters' dissertation with my supervisor. The insight gained from monologues delivered to him turned out to be very useful in the investigations presented herein.

The help given by Suresh Mishra, Sharad Chauhan, and S. S. Chauhan simplified some administrative jobs and saved many days of precious time. The computer center staff and system engineers, particularly Aftab Alam and Nirmal Roberts, have been very helpful continuously. Through A. K. P. Rajpal, a very practical man with great knowledge and experience in construction and working of electronic devices, I came to know of a number of shortcomings in the attitude of typical scientists and novices aiming to be scientists and pitfalls in their scientific career.

*(Himanshu Agrawal)*
February 1998

# Synopsis

This study aims at finding a method for constructing molecular dynamics like models using the formalism of cellular automata for fast simulation of fluid dynamic systems (including compressible phenomena). In as much as the results indicate, the attempt is successful. A systematic method for constructing cellular automata models of fluid dynamic systems is discovered and proposed following a review of the existing developments. The considerations required for constructing such models for fluid dynamic systems consisting of particles with arbitrary interaction potentials and existing over arbitrary spatial lattices are outlined. The method is illustrated by constructing a model for simulation of systems of particles moving with unit speed along the links of an underlying square spatial lattice. Using this model, two two-dimensional systems are simulated and studied for a number of model and system parameters. For almost all the model and system parameters, the results are found to be in complete agreement with the available theoretical predictions. For some model parameters, results show (expected) departure from theoretical predictions which is explained. The layout, contents, and major findings of various chapters are as described below.

**Chapter 1:** Broad overview of the scope of fluid dynamics, general aspects of modeling, and modeling of fluid dynamic systems using conventional calculus based approach is presented. Distinction is established between *computational models* and *computational methods*. Difficulties in modeling and computer simulation of fluid dynamic systems using existing calculus based models are discussed. Use of a formalism other than calculus for modeling of fluid dynamic systems is found to be the only remedy for difficulties encountered in study of fluid dynamic systems using calculus based models. Cellular automata are proposed as plausible alternate formalism for developing microscopic models of fluid dynamic systems which, besides being insusceptible to the problems encountered in using conventional calculus based models, are computer friendly and computationally fast. The broad and the exact objectives of the present investigation are outlined. The exact objective being *"to find a method for constructing molecular dynamics or molecular dynamics like models using cellular automata for simulation of compressible fluid dynamic systems"*.

**Chapter 2:** Broad overview of the formalism of cellular automata is presented. Wolfram's conception of cellular automata as fully discrete abstract mathematical systems [1,2] and the basic elements of cellular automata, *viz.*, lattice of sites, evolution axis, lattice sites values, evolution neighborhood, and evolution rules are described. It is found that basic elements of cellular automata, owing to their abstraction, can be interpreted in many different ways for modeling of physical systems and thus, many different types of models can be developed. The existing cellular automata models of physical systems, known as *lattice gases*, are renamed as *"multiparticle lattice gases"* because in them lattice sites values are interpreted as representing the state of more than one particle. Growth of multiparticle lattice gases starting from origin of the concept in 1963 [3] till most recent (*i.e.*, till 1997) and multifarious developments in modeling, simulation, and analysis (*e.g.*, [4]) are reviewed. Different representations of multiparticle lattice gases, *viz.*, *partitioned spatial lattice representation* and *multiple particle representation*, are outlined. Some multiparticle lattice gases of interest in the present investigation, *viz.*, HPP gas [5,6], BBMCA model [7,8], TM gas [8], and FHP gas [9–11], are described. Generalized definition of multiparticle lattice gases and major steps involved in their mathematical analysis which leads to various kinetic and hydrodynamic equations are outlined. Some problems with multiparticle lattice gases, *viz.*, violation of Galilean invariance and incorrect simulation of compressible systems and phenomena, and their causes explained in the literature are critically reviewed. Arguments contesting the validity and completeness of these explanations are presented. It is found that if these problems can be overcome, multiparticle lattice gases would make good molecular dynamics like cellular automata models of physical systems. Finally, it is concluded that problems encountered with multiparticle lattice gases need to be analyzed afresh.

**Chapter 3:** The conceptual representations of multiparticle lattice gases are analyzed. The partitioned spatial lattice representation, although computationally efficient, is found to be physically inconsistent. The multiple particle representation is found to be correct and physically consistent representation for multiparticle lattice gases. The twin problems with multiparticle lattice gases are rigorously analyzed afresh. It is observed that nonlinear terms of the momentum (Navier-Stokes) equation encode the mechanism of spatial momentum redistribution in physical systems. In physical systems two different microscopic processes, *viz.*, translation of particles and interparticle interactions, cause redistribution of momentum in space [12]. Thus, nonlinear terms of the momentum equation encode, and originate from, the combined effect of these two microscopic mechanisms of spatial momentum redistribution. In multiparticle lattice gases interactions occur among particles occupying the same lattice site (by definition). As a result, in them particles behave like *rigid point particles* and spatial momentum redistribution does not occur during interparticle interactions. In view of this, it is concluded that in multiparticle lattice gases violation of Galilean invariance and incorrect simulation of compressible phenomena occur because of lack of spatial momentum redistribution during interparticle interactions. Since interparticle interactions are encoded into collision rules of multiparticle lattice gases, it is also concluded that the structure of collision rules plays key role in the macroscopic dynamics of multiparticle lattice gases.

**Chapter 4:** Requirements and methods of overcoming the problems with multiparticle lattice gases are discussed and whether or not these lattice gases can be improved is analyzed. Restoring spatial momentum redistribution during interparticle interactions is found to be necessary for overcoming the problems. This can be done by incorporating a mechanism into lattice gases so that interactions occur among particles occupying different lattice sites. This condition, although necessary, is insufficient by itself. Analysis shows that the problems can be overcome by allowing at most one or zero particle to occupy a lattice site at any time step (necessary and sufficient condition). This constraint is termed as *"single particle exclusion principle"*. It is found that this exclusion principle cannot be incorporated into multiparticle lattice gases without altering their basic definition and structure; which makes the resulting lattice gases very different from multiparticle lattice gases. In view of this it is concluded that the twin problems cannot be removed from multiparticle lattice gases. The lattice gases based on this exclusion principle are termed as *"single particle lattice gases"*. Consequences of the single particle exclusion principle, *e.g.*, being able to associate desired interaction potentials with particles, are discussed. It is found and concluded that single particle lattice gases are the desired fully discrete analogs of molecular dynamics.

**Chapter 5:** The considerations which are necessary for constructing single particle lattice gases, *viz.*, considerations related to length scales, time scales, interaction potential, range of interactions, types of interparticle interactions, interaction neighborhoods, discrete velocity sets, and determination of time step $\Delta t$ and lattice parameters of single particle lattice gases in terms of corresponding physical parameters, are outlined. It is found that for simulation of fluid dynamic systems single particle lattice gases can be constructed under the following assumptions: (i) at any time step particles undergo zero or one collision with one or many neighbors, and (ii) $\Delta t_{\mathrm{C}} \ll \Delta t$, where $\Delta t_{\mathrm{C}}$ is collision time. These assumptions are equivalent to the first basic assumption in Boltzmann's analysis of gases, *viz.*, $\Delta t_{\mathrm{C}} \ll t_{\mathrm{mft}}$, where $t_{\mathrm{mft}}$ is the mean free time or the relaxation time [12,13]. Generalized definition of single particle lattice gases, including generalized description of evolution strategy wherein evolution during each time step is decomposed into two sequential sub-steps, *viz.*, (i) processing of interparticle interactions, and then (ii) processing of particle translation, are presented. This decomposition is equivalent to decomposing the evolution operator $\mathcal{E}$ of the system into interaction operator $\mathcal{C}$ and translation operator $\mathcal{T}$ as $\mathcal{E} = \mathcal{T}\mathcal{C}$. The operators $\mathcal{T}$ and $\mathcal{C}$ are, in general, non-commutative. This is in contrast with multiparticle lattice gases.

It is found that due to single particle exclusion principle a new type of global coupling arises among particles in single particle lattice gases despite the fact that all interparticle interactions are local. This global coupling is termed as *"exclusion global coupling"*. This global coupling cannot be eliminated and necessitates that the interaction rules be developed by considering the entire spatial lattice as interaction neighborhood of particles. This being infeasible, in turn, necessitates that overall interaction operator $\mathcal{C}$ be expanded recursively in terms of fractional interaction operator $\mathcal{C}'$ as

$$\mathcal{C} = \underbrace{\mathcal{C}'\mathcal{C}' \cdots \mathcal{C}'}_{\mathcal{N}_{\mathrm{iter}}}$$

where $\mathcal{N}_{\mathrm{iter}}$ is total number of terms in the expansion and $\mathcal{C}'$ is constructed over a small (compared to the spatial lattice) interaction neighborhood. This recursive decomposition makes the processing of interparticle interaction an iterative process. The operators $\mathcal{C}$ and $\mathcal{C}'$, in general, contain probabilistic elements; which inhibits *a priori* determination of $\mathcal{N}_{\mathrm{iter}}$. Finally, the generalized method of construction of fractional interaction rules is described.

**Chapter 6:** A two-dimensional single speed single species single particle lattice gas existing over square spatial lattice is constructed for demonstrating the method outlined in chapter 5. Some aspects of mathematical analysis of this lattice gas are presented. This lattice gas is termed as *"SPLG-1"* model or *"SPLGA-1"* model for identification in the following chapters. The acronym SPLG(A) stands for **S**ingle **P**article **L**attice **G**as (**A**utomata).

**Chapter 7:** Results of simulation of two-dimensional system of particles enclosed in a box at equilibrium carried out using the SPLG-1 model for five different model parameters and two different densities are presented. Mean displacement and mean square displacement (second integral of velocity autocorrelation function) of tagged particles and time correlation function for velocity at lattice sites are sampled. Correlation function is found to decay as $t^{-1}$ at long times. Other results agree with predictions of the Langevin equation. Departure, where ever seen, is explained and found to be controllable feature of the model.

**Chapter 8:** Results of simulation of relaxation of strong density perturbation in finite length two-dimensional tube carried out using the SPLG-1 model are presented. Mean density and mean $x$- and $y$-momentum are sampled along the centerline of the tube. Observed dynamics of formation, propagation, and reflection (from a solid wall) of density and momentum waves in the tube is analyzed. The results are first of their kind in the literature and no rigorous comparison is feasible. The profile of the leading edge of density wave, however, is compared with existing theoretical results for other systems and both are found to be similar.

**Chapter 9:** Achievements, failures, and difficulties encountered during the study are summarized. The major conclusions are: (i) it is indeed possible to develop molecular dynamics like models for simulation of fluid dynamic systems using cellular automata, and (ii) a systematic and algorithmic method has been discovered for constructing such models.

**Chapter 10:** Some plausible offshoots and extensions of this study are pointed out and directions for further work along similar lines are given.

# Contents

# List of Abbreviations

**BBGKY:** Short form made from initials of N. N. **B**ogoliubov, M. **B**orn, H. S. **G**reen, G. **K**irkwood, and J. **Y**von. Used in phrases like "BBGKY equations" or "BBGKY hierarchy" to refer to a set of kinetic equations derived by them.

**HPP :** Short form made from initials of J. **H**ardy, O. de **P**azzis, and Y. **P**omeau. Used to refer to their lattice gas automata model as in the phrase "HPP gas".

**BBMCA:** Short form made from **B**illiards **B**all **M**odel **C**ellular **A**utomata. A cellular automata model introduced by Margolus for the billiards ball model of computation introduced by Fredkin. Used to refer to this cellular automata model as in the phrase "BBMCA model".

**FHP :** Short form made from initials of U. **F**risch, B. **H**asslacher, and Y. **P**omeau. Used to refer to their lattice gas automata model as in the phrase "FHP gas".

**PSLR:** Abbreviation for **P**artitioned **S**patial **L**attice **R**epresentation.

**MPR:** Abbreviation for **M**ultiple **P**article **R**epresentation.

**BCC :** Abbreviation for **B**asic **C**ollision **C**onfiguration.

**STP :** Abbreviation for **S**tate **T**ransition **P**robability.

**SPLG(A):** Abbreviation for **S**ingle **P**article **L**attice **G**as (**A**utomata).

**NIN :** Abbreviation of **N**earest **I**nteraction **N**eighbors.

**NNIN:** Abbreviation of **N**ext **N**earest **I**nteraction **N**eighbors.

# List of Tables

# List of Figures

# Chapter 1

# Introduction

. . . and in the past, great men have done wonderful things. You must start by learning them.
*You mean history, sire! It is so boring. It makes me feel sleepy.*
You will have to fight sleep, then.
*But sire, why must I learn history? It is all over, the present is its consequence.*
Yes.
. . .

In this chapter broad overview of fluid dynamics, modeling, modeling of fluid dynamic systems, and difficulties encountered in simulation of fluid dynamic systems has been presented. It has been argued that an alternate formalism for modeling and simulation of fluid dynamic systems is required for overcoming these difficulties. The formalism of cellular automata has been proposed as an appropriate alternate formalism. On the whole, the aim of this chapter is to systematically bring out the motivations and objectives of this study.

## 1.1 The Scope of Fluid Dynamics

Old as the study of fluid flow is, its scope has been increasing rapidly since the last 200 years of documented scientific study. It is the result of innumerable in-depth investigations carried out by equally innumerable number of researchers in the past that the field of fluid dynamics proudly boasts of possessing amazingly wide horizons at present. In Newton's times, it perhaps would have been correct to say that fluid dynamics concerned itself only with the study of fluids (liquids and gases) in motion, whether reacting or non-reacting, and so did the fluid dynamicists. However, numerous scientific and engineering applications of the cumulative knowledge of fluid flow behavior gained over the years have continuously encouraged researchers to discover, resolve, and incorporate new and challenging problems in the field; thus adding many branches to it and widening its horizons considerably. As a result, fluid dynamics no longer restricts itself merely to the study of *fluids in motion* as the term would have stood in the past. At present, the study of almost any aspect of any physical system that is, or can in some way be, related to or used for understanding the dynamical behavior of arbitrary fluid systems can, broadly and legitimately, be categorized to be a part of fluid dynamics; and a *fluid dynamicist* can be found engaged in pursuing it.

Looking from the prevalent classical pedagogic perspective, the presently evident scope of fluid dynamics spans the study of various phenomena occurring in systems ranging from those involving highly rarefied flow conditions to those involving onset of solid-liquid phase transitions and multi-phase chemically reacting flows with embedded solid particles, to point out a few outstanding examples. A deeper look into fluid dynamics from the most sophisticated feasible scientific and technological application's point of view shows that bits and pieces of quantum-mechanical and relativistic thought are also scattered into it. In view of the strong influence that the existing and emergent technological trends at any time have on fluid dynamics, it is not very surprising that, except for its classical components, not much thought has been spared for its quantum-mechanical and relativistic bits. Consequently, ideas and developments related to these have largely remained unknown, neglected, and ignored even though they found their way-in many decades ago.

Being an inherently applied subject, the path that fluid dynamic thought has taken over the years can only be attributed to be a consequence of the demands imposed as well as tools provided by the associated technological development during the same period. Thus, in order to understand, even intuitively, how the currently emerging technological trends will affect the fluid dynamic thought in future, *i.e.*, what will the future fluid dynamic thought be like, what path will it take, what are and will be the major factors influencing it, it is necessary to isolate the factors that have influenced it in the past and understand the exact nature of their influence. This exercise should only be carried out in the hope that it will allow isolation of factors that are influencing it now and will most likely affect it in future.

All fluid dynamic investigations, whether they happen to be experimental, theoretical, or computational in nature, are centered around an extremely vital human act known as *modeling*. It is this act through which the emergent technological trends enter into fluid dynamics and affect the fluid dynamic thought. This is because modeling, in order to be successfully carried out, requires tools which may be mathematical or technological in nature. Thus, any investigation that involves modeling is bound to be affected by the emergent technological trends. Experimental investigations are directly affected by the emergence of technology for better technology implies better precision which allows the design and execution of more sophisticated experiments that have not been feasible in the past. Similarly, computational investigations are also directly affected by the emergence of more sophisticated computational devices for better computational devices permit one to carry out computations faster and at those levels of complexity and accuracy that were not attainable in the past. It might appear that theoretical investigations sail clear off emerging technological trends for in these investigations one only needs mathematical tool—theorems and the like. More careful thought, however, reveals that this is at best an illusion, for all theory needs to be validated and substantiated with hard experimental evidence.

The above does much towards substantiating the observation that at any point in time and at any stage of its development the direction that fluid dynamic thought has taken, takes, and will take depends upon the tools for modeling that are available. Naturally, the direction that fluid dynamic thought must take at present and will take in future depends upon the most powerful tool, *viz.*, *computers*, that is and will be available for modeling and simulation of physical systems, particularly fluid dynamical systems. This minor, yet important observation, lies at the core of this thesis.

## 1.2 Modeling: Philosophical Aspects

**Definition:** *Modeling* is one of the most important and indispensable steps in the study of physical systems. It is inherent to all the analytical sciences, irrespective of whether they are experimental, theoretical, or computational in nature. As a result, it has been defined from many different perspectives by researchers working in different fields. Among these, the effort made by the dictionaries of English language in defining this word is particularly notable. The dictionaries try to bring out the meaning of modeling as *an act, the act of making models, with some artistic element in it*, *e.g.*,

> **Modeling**[1)] *(verbal substantive)*: 3. The action or art of making model; the act of constructing representations of things in clay, wax, plaster, or the like; ...
>
> — The Oxford English Dictionary,
> Vol. VI, L–M,
> Clarendon Press, 1933.

Despite vast differences in the contexts of their origin, all the definitions of modeling share the fact that it involves at least two different entities, namely: (i) *the model*, and (ii) *the system*. In terms of these, *modeling* can essentially be viewed as an act of development of a mapping of the desired aspects of *the system*. In view of this, the following definition of *modeling* seems appropriate in the context of this thesis.

> **Modeling:** The act of development of alternate representation of given system that has desired features in common with the system.

**Types of Modeling:** Depending upon the context and objectives involved in model development, modeling can be classified into physical modeling (construction of prototypes, scaled-up and scaled-down physical replicas of systems), mathematical modeling (construction of mathematical description of systems), geometric modeling (construction of geometric description of systems), conceptual modeling (construction of conceptual description of systems), symbolic modeling (construction of symbolic description of systems), *etc.* In general, solution of real-life (physical) problems requires that more than one kind of modeling be carried out simultaneously, *e.g.*, physical and mathematical modeling. Consequently, an actual modeling exercise is an extremely complex process. As a result, to simplify the model building process, the modelers follow certain systematic steps as described below.

**Steps Involved in Model Development:** Modeling, being an act, must always be driven by and towards an objective, *i.e.*, an answer to the question *"Why does one want to model a system?"* must exist. The answer of this question happens to be the key to the steps that must be followed during the model building process. The exact nature of steps involved in any modeling exercise varies strongly depending upon the context and the objectives of the exercise. As a result, an attempt at generalized description of the steps involved in an arbitrary modeling exercise will turn out to be futile here. Not only this, such an attempt will also go beyond the scope of this study, which is restricted to modeling and simulation of fluid dynamical systems. The major models of fluid dynamical systems, their basis, and difficulties encountered in using these models are described in the following sections.

---

1) The dictionary uses British spelling.

## 1.3 Modeling of Fluid Dynamic Systems

Solution of fluid dynamic problems has emerged as one of the most challenging task of this century. It has grabbed much attention of the scientific community, causing rapid inflow of ideas into fluid dynamics. As a result, many experimental, theoretical, and computational models and methods have been developed for studying fluid dynamic systems of varied complexity. Out of these, computational methods have steadily and rapidly captured wider attention compared to the other two. This is because they offer better flexibility and wider scope in terms of problem complexity that can be approached, at the same time being more economical and less time consuming.

Grossly, all the theoretical and computational models used for studying fluid dynamic systems are based on the length and time scales at which it is desired to investigate the system. Alternately, the models of fluid dynamic systems are developed based on whether the target of the study is to determine the macroscopic, microscopic, or mesoscopic (finer than macroscopic but coarser than microscopic) behavior of the system. If the target is to determine the macroscopic behavior of the system, it is advantageous to employ the Eulerian description of the system. On the other hand, if the target is to determine the microscopic behavior of the system, the Lagrangian description is useful. For determining the mesoscopic behavior, it is advantageous to employ appropriate statistical description.

### 1.3.1 Computational Models versus Computational Methods

To be scrupulous, all the computational tools that are available at present for simulation of physical systems on digital computers can be broadly classified into two different categories represented by the terms *computational methods* and *computational models*. These terms have inherently different meanings and are used in the context of fundamentally different types of computational tools. Thus, these terms and computational tools represented by them should be differentiated clearly. The difference between these two terms arises because the computational tools which they represent are based on radically different underlying concepts, and thus, have fundamentally different nature.

#### 1.3.1.1 Computational Methods

The term *computational methods* represents those computational tools which are used for obtaining numerical solutions of theoretical models, *e.g.*, various differential equations, using computer for investigating the phenomena represented by the theoretical models. At present large number of computational methods are available. Some examples being finite difference methods, finite volume methods, finite element methods, panel methods, *etc*. In fact, most of the computational tools that are available at present fall under this category, *i.e.*, they are *computational methods* and not *computational models*.

Since theoretical models represent various phenomena, it can be said that computational methods are used for studying the phenomena (represented by the theoretical models) numerically. An essential aspect of usage of computational methods for studying a phenomena is that such a usage presupposes the existence of an appropriate theoretical model of the desired phenomena. In the absence of an appropriate theoretical model of the desired phenomena, all computational methods are rendered useless as far as computational investigation of the phenomena is concerned.

The usage of a computational method for studying the phenomena represented by a theoretical model gives an *alternate representation of the theoretical model* that is suitable for implementation on a computer for obtaining numerical solutions of the theoretical model. This alternate representation is obtained using a formal rigorous procedure called *discretization*. Hence, it can be called as *discretized form of the theoretical model* or simply as *discretized model*. At times discretized models are referred to as *computational models*. This usage, however, should be avoided because the term *computational models* stands for a very specific type of computational tools whose existence and usability for investigating a phenomena is independent of the existence of theoretical models.

#### 1.3.1.2 Computational Models

The term *computational models* represents those computational tools which are used for studying various phenomena by directly mimicking the processes underlying them on a computer. Computational models are developed by understanding the process underlying the phenomena of interest and then formalizing this understanding directly in terms of algorithms—the basic elements of computation. This formalization allows one to mimic the processes directly on a computer and leads to a *computational model*. Thus, the method of development of computational models is nearly identical to the method of development of theoretical models. The sole difference between the two being that while theoretical models are developed by formalizing the understanding of processes in terms of basic elements of an appropriate theoretical framework, *e.g.*, calculus, the computational models are developed by formalizing the understanding in terms of basic elements of computation.

An essential aspect of computational models is that at no stage in their development does one need to use theoretical models, *i.e.*, understanding of the fundamental laws and processes giving rise to the phenomena is sufficient for developing computational models. As a result, the usage of computational models for investigating various phenomena does neither need nor presuppose the existence of theoretical models of the phenomena. Another important aspect of computational models is that it may be possible to establish *equivalence* of computational models with (appropriate) theoretical models. The equivalence, however, can be established (if at all possible)[2)] only *after* computational models have been developed. Furthermore, while attempting to establish such an equivalence, one does not have *a-priory* knowledge of the theoretical model. Instead, the theoretical model is the outcome of establishing the desired equivalence. Alternatively, the equivalence of a computational model with some appropriate theoretical model is established through a systematic analysis of the computational model; the theoretical model being obtained as the end result of this analysis. As a result, while attempting to establish such an equivalence, one may, at times, discover new theoretical models.

The establishment of equivalence between a computational model and a theoretical model does in no way alter the status of the computational model from *computational model* to *computational method* (for obtaining numerical solutions of the theoretical model). Furthermore, it is not necessary that such an equivalence *should* be established

[2)] Here I am not saying that *"if a theoretical model that is equivalent to the computational model exists"*, instead, I am saying that *"if at all it is possible to establish the desired equivalence"*. I have pointed it out because whether or not the desired equivalence can be established depends upon the existence and availability of appropriate methods and tools that are to be used. One can, very correctly, argue that every computational model is necessarily equivalent to some existing or yet to be discovered theoretical model.

to show the validity of computational models. This is because the validity of a model, whether theoretical or computational, depends upon the correctness of the steps that have been followed and the arguments that have been used for developing it. In case of doubt about these, validity of the model can be ascertained only by comparing the results obtained using it with those from actual *experiments*. The establishment of such an equivalence, however, is very useful because if such an equivalence has been established, then the computational model can be used freely, instead of the theoretical model, for studying all the phenomena represented by the theoretical model. This is a very important and useful point because most of the theoretical models, being nonlinear, are not amenable to theoretical or computational methods of analysis without severe approximations. But, if an appropriate computational model exists, one can bypass all the problems by employing the computational model instead.

At times *computational models* are loosely referred to as *computational methods*. This usage possibly originates from the fact that computational models, just like computational methods, provide a way or a method of investigating desired phenomena using computers. Although this utilitarian point of similarity between the two exists, the usage is incorrect and should be avoided because the computational tools represented by the terms *computational models* and *computational methods* have different nature as elaborated above. At present very few *computational models* of physical systems and phenomena, especially fluid dynamic systems and phenomena, are available. Some examples being the well known direct simulation Monte-Carlo method [1] and lattice gases (the subject of the present study).

The direct simulation Monte-Carlo method, though looked upon as a method, is truly a computational model. It has not been developed as a computational method for obtaining numerical solutions of any theoretical model using computer; although earlier it was thought that the direct simulation Monte-Carlo method has been derived from the Boltzmann equation as a statistical numerical method for obtaining numerical solutions of the equation, but later findings revealed that it is not so. The origin of this method lies in the understanding of the stochastic nature of processes which occur in particle dynamical systems, specifically in gases, in a way in which the processes themselves can be mimicked on a computer using appropriate algorithms. The *equivalence* of solution obtained through this methods with the solutions of the Kac master kinetic equation was established only after the method was developed and used for a long time for solving a number of problems. Once the equivalence was established, it was accepted that the direct simulation Monte-Carlo method can be *viewed* as a method for solving the Kac master kinetic equation. At this point, it should be noted that the direct simulation Monte-Carlo method is actually *not* a *method* for solving the Kac master kinetic equation. It is only *viewed* to be so because the solutions obtained through it are *equivalent* to the solutions of Kac master kinetic equation.

Similarly, lattice gases are also truly computational models and not merely computational methods for obtaining numerical solutions of specific equations or theoretical models. Lattice gases, as they exist at present, are fully discrete microscopic or mesoscopic computational models of physical systems. They are developed by understanding the processes that go on in physical systems at microscopic or mesoscopic levels of description and then recasting or formalizing this understanding in terms of the fundamental elements of cellular automata. At no stage in the development of lattice gases does one encounter or use the existing theoretical models in any way. Once a lattice gas has been developed, it can be analyzed using available tools in a desired framework and its *equivalence* with existing (or, at times new) models of physical systems can be established. For example, there exists a lattice gas, called the FHP gas (described later), whose dynamics has been found to be equivalent to the incompressible Navier-Stokes equation [2–4]. This equivalence, however, does not alter the status of the FHP gas from a truly computational model to a computational method for obtaining numerical solutions of the incompressible Navier-Stokes equation. This equivalence only guarantees that the FHP gas can be used freely for investigating all those phenomena for which the incompressible Navier-Stokes equation is used.

Finding a systematic method for developing lattice gases happens to be the primary objective of the present study. As a result, more details about them will start appearing soon. But, first it is necessary to know— *"why are lattice gases needed?"*

### 1.3.2 Roots of Fluid Dynamic Models: Eulerian, Lagrangian, and Kinetic Descriptions

All the models of fluid dynamic systems, irrespective of whether they are computational or theoretical, employ either Eulerian, Lagrangian, or kinetic (statistical) description of fluids.

In the Eulerian description, fluid is treated as a continuum and governing dynamical equations are derived and solved to understand its behavior. This description is macroscopic in nature. Hence, the governing equations employ only macroscopic variable, *e.g.*, pressure ($p$), temperature ($T$), density ($\rho$), *etc.* The validity of these equations breaks down with the break down of the underlying "continuum-hypothesis". The Navier-Stokes and Euler equations are examples of this description.

In the Lagrangian description, fluid is treated as a conglomeration of discrete particles (atoms, molecules, *etc.*) and its behavior is described in terms of the collective behavior of individual particles. The dynamical behavior of each of the particles is described through appropriate equations of motion, *e.g.*, the Newton's laws, which take care of its interactions with all the other particles comprising the system. This description is microscopic in nature and being inherently discrete (inherently discrete in that each particle is treated individually) it is applicable to all fluid dynamic systems irrespective of the validity of continuum-hypothesis, *i.e.*, irrespective of the Knudsen number. Molecular dynamics, the direct simulation Monte-Carlo method, and some lattice gases (including the ones being proposed in this study) are examples of this description.

In the kinetic or statistical description, fluid is treated as a conglomeration of discrete particles (atoms, molecules, *etc.*) and its behavior is described in terms of the collective behavior of the particles. The particles themselves are not treated individually, which is contrary to what is done in the Lagrangian descriptions. Rather, they are treated collectively in a statistical manner using probability distribution functions. As a result, the kinetic equations employ probability distribution functions. The macroscopic properties and correlation functions are computed by evaluating various moments of the distribution functions. This description is mesoscopic description and applicable as long as the length and time scales for which the description has been developed are not compromised. The Boltzmann equation and the BBGKY (Bogoliubov, Born, Green, Kirkwood, and Yvon) hierarchy of equations are examples of this description [5].

### 1.3.3 Motivation of This Study: Difficulties in Simulation of Fluid Dynamic Systems

To understand the motivation of this study clearly, the following points of difficulty regarding simulation of fluid dynamic systems using theoretical models, particularly calculus based models, *e.g.*, differential equations, should be noted.

#### 1.3.3.1 Break-Down of Continuum-Hypothesis

The continuum-hypothesis breaks down under rarefied conditions, *i.e.*, when the Knudsen number (Kn) is not negligible compared to unity. This break down, rather than being sharp, is a continuous phenomena. It starts at $\mathrm{Kn} \approx 0.01$ and becomes fully visible when $\mathrm{Kn} \approx 1$. As a result, it is difficult to analyze flow fields for which $\mathrm{Kn} \in [0.01, 1]$. This is because in this range of Knudsen number the Eulerian description is not applicable due to break down of the continuum-hypothesis and models based on the Lagrangian and kinetic descriptions are computationally too expensive to apply. This range of Kn characterizes the well known slip ($\mathrm{Kn} \in [0.01, 0.1]$) and transition ($\mathrm{Kn} \in [0.1, 1]$) flow regimes.

#### 1.3.3.2 Flow Speed and Nature of Equations in Eulerian Description

In the Eulerian description of fluid dynamic systems, the nature of the governing equations changes depending upon Mach number ($M$). The governing equations are elliptic for $M < 1$, parabolic for $M = 1$, and hyperbolic for $M > 1$. For most of nontrivial problems, the governing equations are not amenable to analytical methods and can only be solved numerically using appropriate computational methods. The appropriateness of a computational method for obtaining numerical solutions of an equation, however, is dictated by the nature of the equation, among other factors. This poses severe problems in the analysis of flow fields having mixed subsonic ($M < 1$), transonic ($M = 1$), and supersonic ($M > 1$) flow. The analysis becomes further complicated and time consuming if unsteady flow fields with this type of mixed flow need to be investigated.

#### 1.3.3.3 Transport Coefficients and the Eulerian Description

The use of Eulerian equations, *e.g.*, the Navier-Stokes equations, for flow analysis assumes *a-priori* knowledge of transport coefficients. Consequently, transport coefficients have to be determined separately prior to the commencement of flow analysis. This is a resource consuming exercise and requires the existence of appropriate experimental or computational tools. At times, this exercise may require more resources than the flow analysis itself. In other situations, *e.g.*, when the fluid density is high enough so that many ($> 2$) particle collisions become important, appropriate method for determining the transport coefficients may not be available—thus, hampering the flow analysis completely.

#### 1.3.3.4 Uncomputability in Continuum

An aspect common to all the theoretical calculus based models of fluid dynamic systems[3)] is that modeling exercise leading to them is carried out in *continuum*, *i.e.*, both the range

[3)] In fact, here and at most of the places in this thesis one can use the more general term *"physical systems"* instead of the highly restricted term *"fluid dynamic systems"* without incurring any error.

and domain of modeling functions/variables, *e.g.*, velocity $\boldsymbol{V}$, pressure $p$, temperature $T$, density $\rho$, probability-density function $f$, number density $n$, *etc.*, belong to *continuum*. Mathematically, for a function $\xi$ in these models,

$$\begin{aligned} \mathsf{Domain}(\xi) &\subset \mathcal{R} \\ \mathsf{Range}(\xi) &\subset \mathcal{R} \end{aligned} \tag{1.1}$$

where $\mathcal{R}$ is the set of all real numbers (or, the *continuum*). Henceforth, all the models that satisfy Eq. (1.1) will be referred to as *"continuum models"*.

The study of physical systems using continuum models requires that the models be amenable to theoretical methods of analysis and that appropriate methods of analysis be available. In the event of unavailability of appropriate theoretical methods of analysis, which is generally the case with models for most of nontrivial systems, it becomes necessary to revert to computational methods. The computational methods for continuum models use *discretization* to convert the domain of functions from $\mathcal{R}$ to $\mathcal{I}$, where the discrete domain $\mathcal{I}$ is the set of all integers. The discretization of a continuum model yields an alternate representation of the model, to be referred to as *"discretized model"*, which is suitable for implementation on computers for obtaining numerical solution of the continuum model. Mathematically, the range and domain of functions in discretized models are

$$\begin{aligned} \mathsf{Domain}(\xi_i) &\subset \mathcal{I} \\ \mathsf{Range}(\xi_i) &\subset \mathcal{R} \end{aligned} \tag{1.2}$$

where $\xi_i$ is some function, $i \in \mathcal{I}$ and $\xi_i \equiv \xi(i)$. In the above expressions $\xi_i$ has been used instead of $\xi$ because in discretized models $\xi$ exists only at discrete points $i$ in the domain.

Let, the domain discretization parameter that is used to obtain a discretized model be denoted by $\Delta$, the numerical solution of the modeling function obtained using the discretized model be denoted by $\{\xi_i\} \equiv \{\xi(i)\}$, and the analytical solution of the modeling function obtained using the continuum model be denoted by $\xi$. It is readily evident that $\{\xi_i\} \to \xi$ only in the limit $\Delta \to 0$, *i.e.*, the numerical solution of the modeling function becomes identical with its analytical solution only when the discrete domain becomes identical with the continuum domain. This implies that an *exact* study of the system using discretized models requires that computations should be carried out in the limit $\Delta \to 0$.

Now, let us assume that the discretized model can be implemented on a computer using an algorithm whose computational time complexity is $\mathcal{O}(n)$, where $n$ is the number of points in the discrete domain.[4] Since, $n$ varies like $1/\Delta$ for a given continuum domain, the computational time also follows the same trend. As a result, the computation time (or, the number of logical operations) required for computing in the limit $\Delta \to 0$ goes to $\infty$. This implies that it is impossible to know *exactly* what is happening in any arbitrary (space-time) domain in finite amount of time using discretized models. This fact was also observed by Feynman as quoted below

> *"It always bothers me that, according to the laws as we understand them today, it takes a computing machine an infinite number of logical operations to figure out what goes on in no matter how tiny a region of space and no matter how tiny a region of*

[4] Note that it is not possible to have still faster algorithms for studying the dynamics of physical systems, especially fluid dynamic systems. This is simply because if there are $n$ points in the domain, then at least one logical operation is necessarily needed for computing the value at each one of them, *i.e.*, at least $n$ logical operations are required for computing the values at all the points in the domain.

> *time. How can all that be going on in that tiny space? Why should it take an infinite amount of logic to figure out what one tiny piece of space-time is going to do?"*
>
> — Richard Feynman [6]

#### 1.3.3.5 Nature of Computation on Digital Computers

Present day digital computers are *finite precision* machines. By finite precision it is meant that only finitely many bits are (can be) devoted for representing any number (or, symbol in general) inside the machine. This limitation comes because of finiteness of memory that can be installed and finiteness of speed with which logical operations can be carried out.

An important consequence of finite precision is that the range and domain of functions computed on digital computers are necessarily discrete. This implies that in all computational investigations carried out using discretized models, the range of functions automatically becomes discrete as soon the models are implemented on finite precision computers. Mathematically, the range and domain of functions computed on digital computers are

$$\begin{aligned} \mathsf{Domain}(\xi_i^c) &\subset \mathcal{I} \\ \mathsf{Range}(\xi_i^c) &\subset \mathcal{I} \end{aligned} \tag{1.3}$$

where $\xi_i^c$ is some function, $i \in \mathcal{I}$ and the superscript $c$ denotes that it has been computed using digital computers.

Automatic conversion of range of functions from continuum to discrete during computations implies that in any computational investigation $\xi_i^c \neq \xi_i$, in general. Thus, even after reverting to computational methods and employing discretization, it is not possible to know the exact value of functions, $\xi_i$, at grid points. Numerical computations only give a value $\xi_i^c$ as a representative of the exact value $\xi_i$. As far as the relationship of $\xi_i^c$ and $\xi_i$ is concerned one can at best *expect* that $\xi_i^c = \xi_i + \epsilon$, where $\epsilon$ is a small number compared to $\xi_i$. This expectation is fulfilled only if the range of parameters for which computations have been performed does not fall into or cross the chaotic domain, if any, for the system. In case the parameters happen to cross or lie within the chaotic domain, no meaningful statement about the relationship of $\xi_i^c$ and $\xi_i$ can be made, other than that $\xi_i^c$ could be absurdly far away from $\xi_i$. This is because changes in values during computations affected by round-off errors amount to changes in initial conditions; and it is known that small changes in initial conditions, in the chaotic domain, lead to drastic changes in the final state.

### 1.3.4 Overcoming the Difficulties: An Alternate Way of Modeling and Simulation of Fluid Dynamic Systems

Problems associated with the use of Eulerian description for studying fluid dynamic systems, particularly the problems related to break-down of continuum hypothesis, change in the nature of governing equations, and *a-priori* knowledge of transport coefficients, can be overcome by employing Lagrangian or kinetic descriptions in the study. Lagrangian description leads to dynamic models, *e.g.*, molecular dynamics, whereas kinetic description leads to stochastic models, *e.g.*, direct simulation Monte-Carlo method. Dynamic models give more detailed information compared to stochastic models. In fact, in a comparative study of a system at some length and time scales of interest using both the dynamical and stochastic models, all the information obtained from the stochastic model can be derived from the information obtained from the dynamical model but not vice versa. As a result, in an *exact* and detailed investigation it is preferable to employ Lagrangian description rather than kinetic description for arriving at the models to be used.

The major problem associated with the use of Lagrangian models, *e.g.*, molecular dynamics, for studying fluid dynamic systems is that the number of variables becomes very large and humanly intractable if the Knudsen number is small enough compared to unity or, alternatively, if the system is sufficiently dense and sufficiently large. This problem can be easily overcome if computations are carried out on a computer having sufficiently large memory. Molecular dynamics[5)] computations, however, have their own problems. The first one is that the time required for carrying out the computations being very large makes the study very expensive. The second one is that computations carried out on digital computers never give exact values because both the range and domain of functions are discrete (*c.f.*, Sec. 1.3.3.4 and 1.3.3.5); the main cause of concern because of this, besides that theoretical expectations about exactness of results are never fulfilled, is that one might obtain absurd results if the parameters happen to drift into a (known or possibly unknown) chaotic domain due to round-off errors.

#### 1.3.4.1 Need of an Alternate Formalism

From the point of view of computation, the main cause of large computation time required for carrying out molecular dynamics computations is that these computations involve very large number of logical operations per particle at each time step. Since computation time is (almost) directly proportional to the number of logical operations, it can be reduced only by reducing the number of logical operations. The number of logical operations per particle per time step cannot be reduced unless the formalism which is used for carrying out molecular dynamics is changed. This because, in a given formalism, irrespective of what is done, one has to go through the same steps and use the same set of relations (thus, same number of logical operations) for computing new positions of particles. Thus, in order to reduce the computation time required for molecular dynamics simulations, one needs to change the very formalism which is used for making the models and carrying out computations.

Regarding the accuracy of computed values nothing much can be done because this problem originates from the very nature of computation on digital computers and not from molecular dynamics itself. The doubts which arise on the accuracy of computed values due to round-off errors—the doubts like *"How far away are the computed values from the exact values?"*—can be eliminated by eliminating these errors. This can be done by switching over to a formalism that is closer to the nature of computation on digital computers compared to differential and integral calculus for developing the models. This implies that the new formalism should be fully discrete because computations carried out on digital computers are fully discrete (*c.f.*, Eq. (1.3)).

The above brings out that an essential element of the formalism sought for developing fast molecular dynamics or molecular dynamics like models and eliminating round-off errors due to finite precision computations on digital computers is that the formalism should be *fully discrete*. This implies that the models developed using such a formalism will also be fully discrete. With this, however, the following basic questions come up: (i) *Does such a formalism exist?"* (ii) *What constraints should such a formalism satisfy*

[5)] To be more general, the term *"Lagrangian"* should be used instead of the term *"molecular dynamics"*. In the present investigation, however, the term *"molecular dynamics"* will be used and should be understood as a representative of all possible Lagrangian models in continuous (as opposed to discrete) space and time.

*in order that it can be used for developing molecular dynamics or molecular dynamics like models?* (iii) *How can such a formalism be used for actually developing models and carrying out computations?*

The questions (i), (ii), and (iii) raised above can be answered rigorously. The answer of question (iii) occupies this entire work and follows starting from Sec. 1.4 onwards after a clearer rephrasal. The questions (i) and (ii) are answered in the next section. These questions form a natural sequence of query in that it is imperative that question (ii) be answered before question (i). Thus, the next section begins with answer of question (ii).

#### 1.3.4.2 The Alternate Formalism

Modeling, as defined in Sec. 1.2, is the development of an alternate representation of a given system. While developing an alternate representation of a system, one chooses a formalism and then translates various elements of the system in terms of basic components of the formalism. During translation, the translator associates very specific interpretations with various components of the formalism.[6] This association is necessary for the alternate representation to be a valid model of the system in the chosen formalism and also for the results obtained through it to be translated back sensibly in the context of the original system. Thus, in order that a formalism can be used for modeling of a system, it must be possible to unambiguously represent all the elements of the system in terms of the basic components of the formalism, *i.e.*, the formalism should have enough number and variety (or, type) of basic components to facilitate such a representation. To determine the constraints that a formalism should satisfy in order that it can be used for modeling of physical systems, it is necessary and sufficient to determine the number and types of basic components that are required for representing physical systems (and their dynamics).

In general, modeling of any physical system is carried out in terms of some parameters of interest and is said to be completed when the model or the law governing the behavior of the system has been defined in terms of these parameters. All physical systems (and thus, the parameters of interest in terms of which they are modeled) can be represented in terms of symbols which exist in position space. Thus, *if a formalism provides enough number and variety of components to represent the position space and the symbol space and to define the evolution rule or the governing law in terms of these, it can be used for modeling of physical systems*. In general, position space can also be represented in terms of symbols and the governing law can be viewed as a set of interrelationship among the symbols. Thus, it can be said that *if a formalism allows one to construct arbitrary number of symbols and define interrelationships among them, it can be used for modeling of physical systems*.

Now that the constraints which should be satisfied by the sought after formalism are known, the question (i) raised in Sec. 1.3.4.1—*"Does such a formalism exist?"*—can be

---

[6] Some like to assert that

"During translation, very specific interpretations *get associated* ..." (A1)

instead of asserting that

"During translation, *the translator associates* very specific interpretations ..." (A2)

and give endless arguments to support their viewpoint which I do not find quite acceptable. This is because the assertion (A1) has the connotation that formation of associations is an automatic and involuntary process beyond the control of the translator, whereas the assertion (A2) has the connotation that formation of associations is an intentional and voluntary act of the translator and thus fully controlled by the translator. Since modeling is an intentional and voluntary act, the assertion (A2) is correct.

answered. A fully discrete formalism which allows one to construct arbitrary number of symbols and define interrelationships among them does indeed exist. It is known as *cellular automata* [7]. The questions which arise now, *e.g.*, *"what is cellular automata?"*, *etc.*, have been answered in detail starting from Sec. 2.1 onwards.

## 1.4 Objective of This Investigation

An important conclusion of previous section (Sec. 1.3.4.2) is that the formalism of cellular automata can be used for modeling and simulation of physical systems. With this, the question (iii) raised in Sec. 1.3.4.1 needs to be rephrased as *"How can cellular automata be used for modeling and simulation of fluid dynamic systems?"* Conclusion of Sec. 1.3.4 is that one needs to develop molecular dynamics or molecular dynamics like models for overcoming the problems pointed out in Sec. 1.3.3. In view of this, the above question can be rephrased in a more precise form as *"How can cellular automata be used for developing molecular dynamics or molecular dynamics like models for simulation of fluid dynamic systems?"* Looking broadly, the objective of this investigation is to answer this question.

Molecular dynamics models of fluid dynamic systems can be developed at many different length and time scales. Since, different phenomena occur at different length and time scales, it is quite possible that there might be different ways (and possibly many different ways) of developing molecular dynamics or molecular dynamics like models using cellular automata. In view of this, to answer the above question rigorously means to find all possible ways of developing the required models using cellular automata. This, however, is not feasible in a single investigation. As a result, *the exact objective of this investigation is to find a method for constructing molecular dynamics or molecular dynamics like models using cellular automata for simulation of (compressible) fluid dynamic systems*.

## 1.5 Layout of the Remaining Chapters

In the last few decades many such models (called lattice gases) have been developed and studied. These models, however, have many problems and do not contain all the elements of molecular dynamics models. Hence, they cannot be called as "molecular dynamics" or "molecular dynamics like" models, even though they appear to be similar. As a result, the objective of this investigation has been fulfilled as follows: In chapter 2 cellular automata, lattice gases, problems encountered in using the lattice gases for simulation of fluid dynamic systems, and the cause(s) and origin of these problems (explained in the literature) have been reviewed; counter arguments suggesting invalidity of explanations given in the literature and implications of the problems have been discussed. In chapter 3 the problems have been rigorously analyzed to determine their actual cause and origin. In chapter 4 a method of overcoming the problems and its consequence have been discussed. Other methods of overcoming the problems and possibility and impossibility of existence of other simpler methods have been discussed. In chapter 5 considerations about length scale, time scale, interaction potential, and method of determining interaction neighborhood for developing lattice gases for arbitrary systems along the lines suggested in chapter 4 are presented. The method of construction of evolution rules and computer algorithms for these lattice gases is outlined. In chapter 6 a lattice gas model (called SPLG-1 model) for particles moving with unit speed along the links of an underlying square spatial lattice is constructed as an example for demonstrating the application of the method described in chapter 5. In chapters 7 and 8 simulation results for two problems obtained using the SPLG-1 model have been presented. Although the conclusions of each chapter have been given at the end of the chapter, the overall conclusions of this investigation have been reconsolidated and outlined in chapter 9. The scope and directions for further work along the lines proposed and followed in this investigation have been outlined and discussed in chapter 10.

## 1.6 Conclusions

The major conclusions of this chapter are the following:

**1)** The formalism of cellular automata satisfies all the requirements of being an appropriate formalism for modeling of physical systems.

**2)** Cellular automata models of fluid dynamic systems, being fully discrete, will most probably not be subject to the problems that have been encountered in simulation of physical systems on digital computers using integral and differential calculus based models existing in continuum space-time.

**3)** It might be possible to develop molecular dynamics like models of fluid dynamic systems using the formalism of cellular automata.

In view of the suitability of formalism of cellular automata for modeling of fluid dynamic systems broadly seen and outlined above, and the possibility of these models having many advantages over the conventional integral and differential calculus based models, it is desirable to carry out further investigations along this direction.

# Chapter 2

# Cellular Automata and Lattice Gases

Sailors must know the sea well, before they can sail. It has two faces one bright, one dark. The bright is blinding, but it lacks depth.
*Yes, sire.*
What is the depth of darkness?
*Sire?*
How deep is the sea?
*I do not know, sire.*
Shouldn't you?
...

In view of the conclusions of chapter 1, a review of cellular automata and lattice gases has been presented here. The objective of this review is to introduce cellular automata and lattice gases in their fullest generality and to bring out the problems that have been encountered with lattice gases during simulation of compressible fluid dynamic systems.

## 2.1 Cellular Automata

*Cellular automata* were introduced by Ulam and von Neumann [1] in connection with evolutionary biological systems.[1)] Cellular automata have been conceived, described, and defined in many different ways depending upon the context of their application. Among these, one of the most general definition and description has been conceived and put forth by Wolfram. This conception has been described in the following section along with examples.

### 2.1.1 Wolfram's Conception of Cellular Automata

**Generalized Description:** In Wolfram's conception, cellular automata are fully discrete abstract mathematical systems which can be described as follows: In the most general case, a cellular automaton has a $\mathcal{D}$-dimensional spatial lattice of sites. The spatial lattice is regular and can have any regular structure possible in $\mathcal{D}$-dimensional space. All the lattice sites are identical. Lattice sites can have finitely many possible values. The values and value set can only be discrete and the value set necessarily contains finitely many elements. Entire spatial lattice evolves in discrete time steps according to an evolution

[1)] An additional fact is that cellular automata have been invented independently many times in different contexts and in many cases independent developments have taken place [2].

rules. Evolution is synchronous, *i.e.*, all the lattice sites evolve simultaneously. An evolution is essentially change in the values of lattice sites according to the evolution rule. During each evolution, new value of a lattice site depends upon the old values of lattice sites in its complete spatio-temporal neighborhood. Both the spatial as well as temporal neighborhood of each lattice site are finite. Thus, evolution rule is a function of the values of all the lattice sites in the spatio-temporal neighborhood of a lattice site. Since lattice sites values are discrete, the evolution rule is necessarily a discrete function.

**Basic Defining Characteristics:** Cellular automata have many basic defining characteristics which, as put forth by Wolfram [2,3], are:

1) **Discrete in space:** They consist of discrete grid of spatial cells or sites.

2) **Discrete in time:** The value of each cell is updated in a sequence of discrete time steps.

3) **Discrete states:** Each cell has a finite number of possible values.

4) **Homogeneous:** All cells are identical, and are arranged in a regular array.

5) **Synchronous updating:** All cell values are updated in synchrony, each depending on the *previous values of neighboring cells* (this means values of all the cells in the complete spatio-temporal neighborhood).

6) **Deterministic rule:** Each cell values is updated according to a fixed, deterministic, rule. As a further generalization, the rule can be stochastic as well.

7) **Spatially local rule:** The rule at each site depends only on the values of a local neighborhood of sites around it.

8) **Temporally local rule:** The rule for the new values of a site depends only on values of a fixed number of preceding steps (usually just one).

**Generalized Mathematical Representation:** The evolution during one time step of a lattice site (or, the value at a lattice site) in *any* cellular automaton can be written as

$$A_{\{i\}}^{(t+1)} = \Phi\left(\left\{A_{\{i\}}^{(t)}\right\}, \left\{A_{\{i\}}^{(t-1)}\right\}, \ldots, \left\{A_{\{i\}}^{(t-\tau+1)}\right\}\right) \tag{2.1}$$

$$= \Phi\left(\left\{A_{\{i\}}^{(t)}\right\}^{(\tau)}\right) \tag{2.2}$$

and one time step evolution of an entire system can be written as

$$\Gamma^{(t+1)} = \Phi\left(\Gamma^{(t)}, \Gamma^{(t-1)}, \ldots, \Gamma^{(t-\tau+1)}\right) \tag{2.3}$$

$$= \Phi\left(\left\{\Gamma^{(t)}\right\}^{(\tau)}\right) \tag{2.4}$$

where the notation is as explained in table 2.1.

| Symbol | Meaning/representation |
|---|---|
| $\mathcal{D}$ | Dimensionality of the spatial lattice of sites or cells. |
| $\{i\}$ | Coordinate of a lattice site in $\mathcal{D}$-dimensional space. |
| $t$ | Time step. |
| $\mathcal{A}$ | Discrete value (or, symbol) set. It, necessarily, contains finitely many elements (values or symbols). |
| $A_{\{i\}}^{(t)}$ | Value of lattice site $\{i\}$ at time step $t$. $A_{\{i\}}^{(t)} \in \mathcal{A}$. |
| $\{A_{\{i\}}\}$ | Values of lattice sites in the spatial neighborhood of lattice site $\{i\}$. Spatial neighborhood necessarily contains finitely many lattice sites. |
| $\left\{A_{\{i\}}^{(t)}\right\}$ | Values of lattice sites in the spatial neighborhood of lattice site $\{i\}$ at time step $t$. Spatial neighborhood necessarily contains finitely many lattice sites. |
| $\tau$ | Number of preceding time steps on which the new value (*i.e.*, value at time step $t+1$) of lattice site $\{i\}$ depends. $\tau$ is necessarily finite and $\tau = 1, 2, 3, \ldots$. |
| $\left\{A_{\{i\}}^{(t)}\right\}^{(\tau)}$ | Values of lattice sites in the spatio-temporal neighborhood of lattice site $\{i\}$. The temporal neighborhood extends to $\tau$ preceding time steps, *i.e.*, from time step $\tau$ to time step $t-\tau+1$. Spatial neighborhood at each time step necessarily contains finitely many lattice sites and different preceding time steps can, in general, be different. If $\tau = 1$, the superscript $(\tau)$ is omitted. Thus, $$\left\{A_{\{i\}}^{(t)}\right\}^{(1)} \equiv \left\{A_{\{i\}}^{(t)}\right\}$$ |
| $\Gamma^{(t)}$ | State of the entire system, *i.e.*, all the lattice sites comprising the system, at time step $t$. |
| $\left\{\Gamma^{(t)}\right\}^{(\tau)}$ | State of the entire system, *i.e.*, all the lattice sites comprising the system, at all time steps in the temporal neighborhood, *i.e.*, from time step $t$ to time step $t-\tau+1$. If $\tau = 1$, the superscript $(\tau)$ is omitted. Thus, $$\left\{\Gamma^{(t)}\right\}^{(1)} \equiv \Gamma^{(t)}$$ |
| $\Phi$ | Evolution rule governing one time step evolution of each lattice site as well as the entire system. |

**Table 2.1:** Explanation of the notation used in Eqs. (2.1), (2.2), (2.3), and (2.4).

### 2.1.2 Basic Elements of Cellular Automata

In view of the description given in Sec. 2.1.1, cellular automata consist of five basic elements. These are: i) lattice of sites, ii) evolution axis, iii) lattice site values, iv) evolution neighborhood (spatial as well as temporal), and v) evolution rule. In terms of these five basic elements, a cellular automaton can be said to be unambiguously known only if all of these basic elements have been unambiguously defined, *i.e.*, unambiguous specification of a cellular automaton consists of unambiguous specification of its five basic elements.

In order to arrive at the five basic elements listed above, Wolfram's description of cellular automata has been further generalized. The generalization is as follows: The "time axis" has been generalized to "evolution axis", and the "spatial lattice of sites" has been generalized to "lattice of sites". The details of this generalization and of the five basic elements listed above are given in the following sections.

#### 2.1.2.1 Lattice of Sites

The lattice of sites is a regular $\mathcal{D}$-dimensional lattice of identical sites. Its boundaries can be finite, infinite, periodic, or of any other conceivable type. The lattice sites are necessarily identical in all possible respects. The links connecting the lattice sites may or may not be orthogonal. As a further generalization, irregular lattices can also be included.

Note that the term "lattice of sites" is a slight generalization of the term "spatial lattice of sites" used by Wolfram. This generalization has been carried out because of the following: Usually the term "space", unless prefixed or suffixed otherwise, means the "physical position space". It is not necessary that "lattice sites" should always be interpreted as coordinate points in discrete "physical position space" and the lattice sites values (or, symbols occupying the lattice sites) should always be viewed to be located in the "physical position space". Indeed, whichever space they happen to be located in, is necessarily their "position space". But this position space need not be the "physical position space". In general, position space could be "phase space", "velocity space", "momentum space", "energy-momentum space", *etc.* The exact interpretation of lattice of sites (or, position space) in a cellular automaton depends upon the context in which the cellular automaton is being used.

Examples of the topology of some one- and two-dimensional regular lattices of sites along with the evolution axis have been illustrated in Fig. 2.1.

#### 2.1.2.2 Evolution Axis

The evolution axis is a discrete axis just like the lattice of sites with the exception that it is necessarily orthogonal to the lattice of sites. The entire lattice of sites is replicated at every point along the evolution axis (see Fig. 2.1). In each replica, the values at corresponding lattice sites may or may not be different (generally they will be different).

Note that the term "evolution axis" is a slight generalization of the term "time axis" implied by the term "time steps" used by Wolfram. This generalization has been carried out because of the following: It not necessary that (model) systems should always evolve in time or along the "time axis". In principle, systems and their models can evolve along an axis other than the "time axis", *e.g.*, the "spatial axis", or the "energy axis", or the "entropy axis", *etc.* The exact interpretation of the nature of evolution axis in a cellular automaton depends upon the context in which the cellular automaton is being used.

From Wolfram's definition, it is not clear whether the duration of each evolution (or, the time step) should be equal or not. In the literature each evolution has been taken to be of equal duration; no counter citations could be found. Possibly, this unstated and unexplained consistency in the interpretation of duration of each evolution has got established in the literature because during numerical analysis[2)] of (calculus based) unsteady state dynamical models the time step is (almost) invariably taken to be a constant. In the case of cellular automata, an appropriate physical or mathematical explanation of "Why should each evolution be (or, not be) of equal duration?" is not available yet. In as much as I understand, the duration of each evolution can, in principle, be allowed to be different. This does not give rise to any mathematical absurdities, except that it will make the dynamical analysis of the system very difficult. As a result, *in this investigation each evolution (of cellular automata) has been taken to be of equal duration.*

---

[2)]Here, numerical analysis means development of discretized models (*c.f.*, Sec. 1.3.3.4).

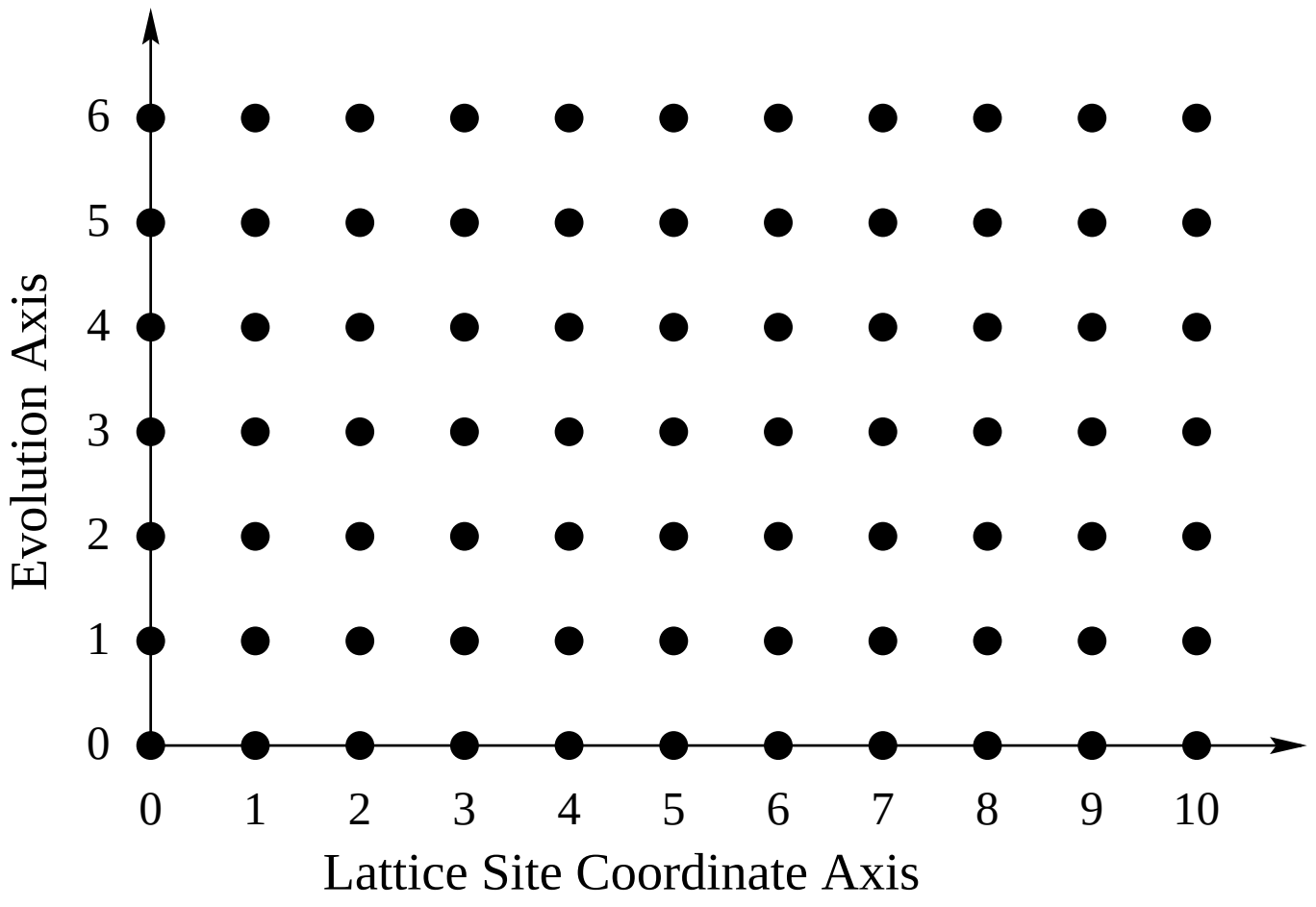

(a) One-dimensional regular lattice of sites with evolution axis.

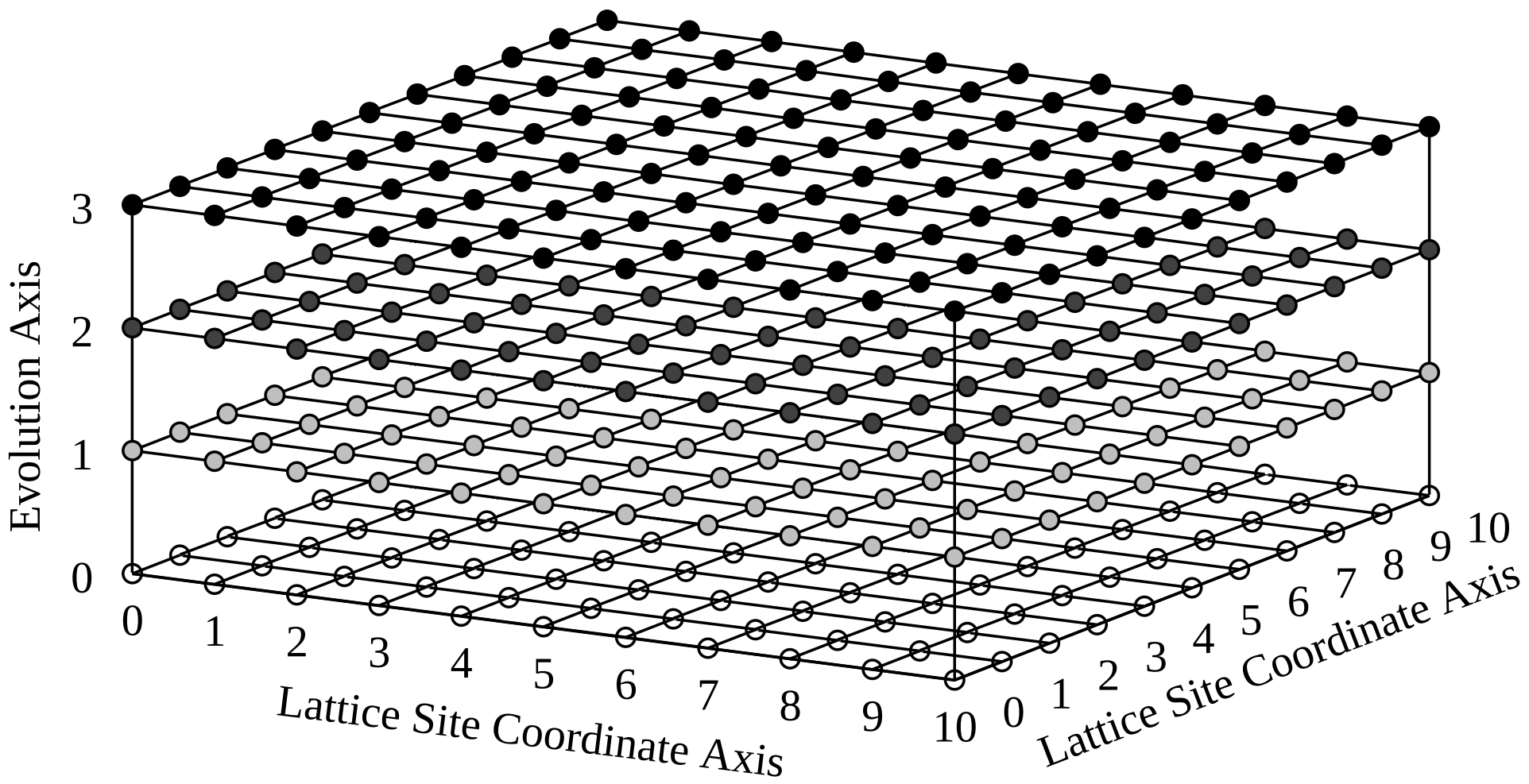

(b) Two-dimensional regular lattice of sites with evolution axis.

**Figure 2.1:** Example of one- and two-dimensional regular lattice of sites along with the evolution axis in cellular automata. The evolution axis is orthogonal to the lattice of sites. Lattice sites are marked with shaded circles. In two-dimensional case, different shades indicate different time steps. The two-dimensional lattice shown is a square lattice.

#### 2.1.2.3 Lattice Site Values

The lattice site values are the states that lattice sites can have in a cellular automaton. In cellular automata, lattice sites can only have finitely many states. As a result, lattice sites values are necessarily discrete. The value set (or, the state space) of lattice sites is the set of all possible values (or, states) that the lattice sites can have. Since lattice sites can have only finitely many values and the lattice site values are discrete, the value set is also discrete and contains only finitely many elements (or, values).

A cellular automaton in which lattice sites can have $\mathcal{N}$ different states is called as an $\mathcal{N}$-state cellular automaton. In an $\mathcal{N}$-state cellular automaton, *any* $\mathcal{N}$ different symbols can be chosen to represent the states of lattice sites. Mathematically, the choice of symbols is immaterial and does not alter any characteristic of the cellular automaton. Since lattice site values have to be represented using symbols, the value set or the state space of lattice sites in a cellular automaton can also be termed as the symbol set of the cellular automaton. With this, the lattice site values can also be termed as symbols occupying (or, posited on) the lattice sites. The exact interpretation of lattice site values in a cellular automaton depends upon the context in which the cellular automaton is being used.

As an example, in a 2-state cellular automaton, the states of lattice sites can be represented using any one of the following pair of symbols: (i) “0” and “1”, or (ii) “2” and “5”, or (iii) “173” and “10”, or (iv) “$\uparrow$” and “$\downarrow$”, or (v) “$\rightarrow$” and “$\leftarrow$”, or (vi) “$\circ$” and “$\bullet$”, or (vii) “ON” and “OFF”, or (viii) “EXISTENCE” and “NON-EXISTENCE”, or (ix) “SPACE” and “PARTICLE”, or (x) “DhI” and “ADN”, or (xi) “$(1, 1)$” and “$(1, -1)$”, or using any other conceivable pair of symbols.

#### 2.1.2.4 Evolution Neighborhood

In cellular automata, new state of a lattice site depends upon the states of finitely many lattice sites lying in its neighborhood at finitely many previous evolutions. The complete domain around each lattice site on which its new state depends is called the evolution neighborhood (of lattice sites). In a cellular automaton, the evolution neighborhood of all the lattice sites is identical because all the lattice sites are necessarily identical (by definition) in all possible respects. Furthermore, total number of lattice sites contained in the evolution neighborhood in any cellular automaton is necessarily finite.

Properties of information transfer mechanism in cellular automata are completely dictated by the topology of the evolution neighborhood. In general, the evolution neighborhood can have any conceivable topology. Topology of some possible evolution neighborhoods in one-dimensional cellular automata has been illustrated in Fig. 2.2.

A cellular automaton in which the evolution neighborhood contains $\mathcal{R}$ lattice sites is called as $\mathcal{R}$-neighborhood cellular automaton. In general, cellular automata with large $\mathcal{R}$ show more complex behavior compared to cellular automata with small $\mathcal{R}$. The complexity, however, also depends strongly upon detailed topology of evolution neighborhood, total number of possible states of lattice sites, and evolution rule.

#### 2.1.2.5 Evolution Rule

In cellular automata the function used for determining new states of lattice sites during evolutions is called as the evolution rule. Since the states of lattice sites are constrained to be discrete, the evolution rule is necessarily a discrete function. In a $\mathcal{R}$-neighborhood cellular automaton, the evolution rule is a function of $\mathcal{R}$ variables; variables being the states of lattice sites comprising the evolution neighborhood.

The evolution neighborhood of a $\mathcal{N}$-state $\mathcal{R}$-neighborhood cellular automaton can have total of $\mathcal{N}^{\mathcal{R}}$ different state configurations. Thus, the evolution rule can have a total of $\mathcal{N}^{\mathcal{R}}$ different possible inputs and a valid output for each one of the inputs. Since output of the evolution rule is state of lattice sites, there can only be $\mathcal{N}$ different outputs. Thus, all cellular automata rules are many-to-one discrete mappings (see Fig. 2.3). Using the notation given in table 2.1, the input-output mapping provided by any evolution rule can be written as a two column table with elements of the form

$$\left\{A_{\{i\}}^{(t)}\right\}^{(\tau)} \mapsto A_{\{i\}}^{(t+1)} \tag{2.5}$$

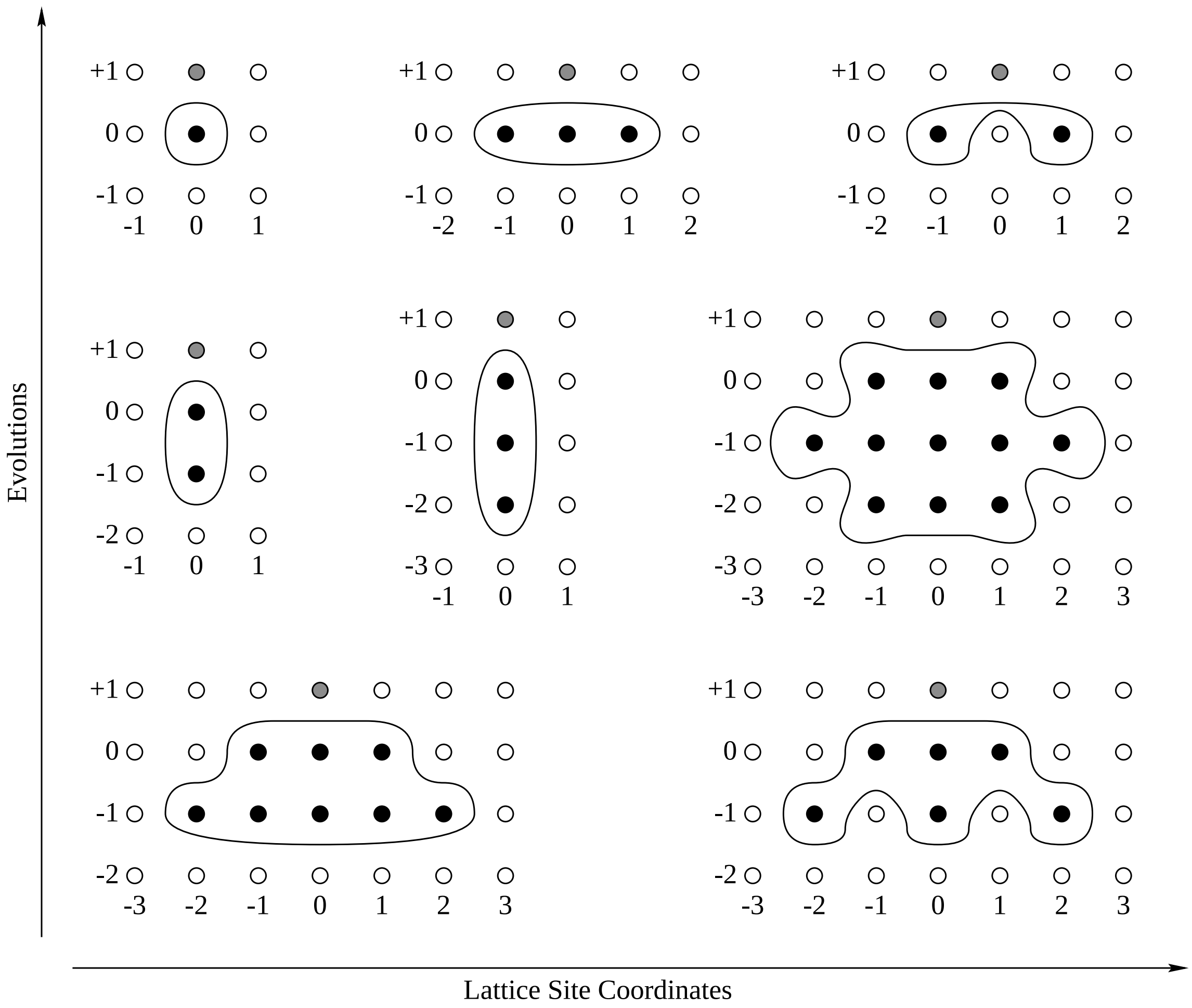

**Figure 2.2:** Some examples of evolution neighborhood in one-dimensional cellular automata. Lattice sites are marked with hollow, dotted, and black circles. Reference points for evolutions and lattice site coordinates are arbitrary. The evolution to be computed is $+1$; previous evolutions are $0$, $-1$, $-2$, *etc.* Lattice site coordinates are relative to the lattice site to be computed (marked with dotted circle at evolution $+1$). Black circles indicate lattice sites comprising the evolution neighborhood of the lattice site to be computed. Evolution neighborhoods are enclosed in boxes.

The table, naturally, contains $\mathcal{N}^{\mathcal{R}}$ rows. Each row of the table corresponds to one input-output pair; the input (or, configuration of the evolution neighborhood) being written in the first column and the output (or, new state of lattice site) in the second column. This table is called the *evolution rule table* or simply the *rule table* of cellular automata.

In $\mathcal{N}$-state $\mathcal{R}$-neighborhood cellular automata there can be total of $\mathcal{N}$ different valid outputs for each one of the inputs. This is because the only constraint on the output of evolution rule for any input is that it should be a valid state of lattice sites. As a result, in $\mathcal{N}$-state $\mathcal{R}$-neighborhood cellular automata there can be a total of $\mathcal{N}^{\mathcal{N}^{\mathcal{R}}}$ different valid input-output mappings or evolution rules. By devising an appropriate convention, each one of the input states of $\mathcal{N}$-state $\mathcal{R}$-neighborhood cellular automata can be numbered from 0 to $\mathcal{N}^{\mathcal{R}}-1$ and each mapping or evolution rules can be numbered from 0 to $\mathcal{N}^{\mathcal{N}^{\mathcal{R}}}-1$. One such numbering convention can be found in [4]. Number of possible states of evolution neighborhood and total number of rules in $\mathcal{N}$-state $\mathcal{R}$-neighborhood cellular automata for some values of $\mathcal{N}$ and $\mathcal{R}$ have been given in table 2.2.

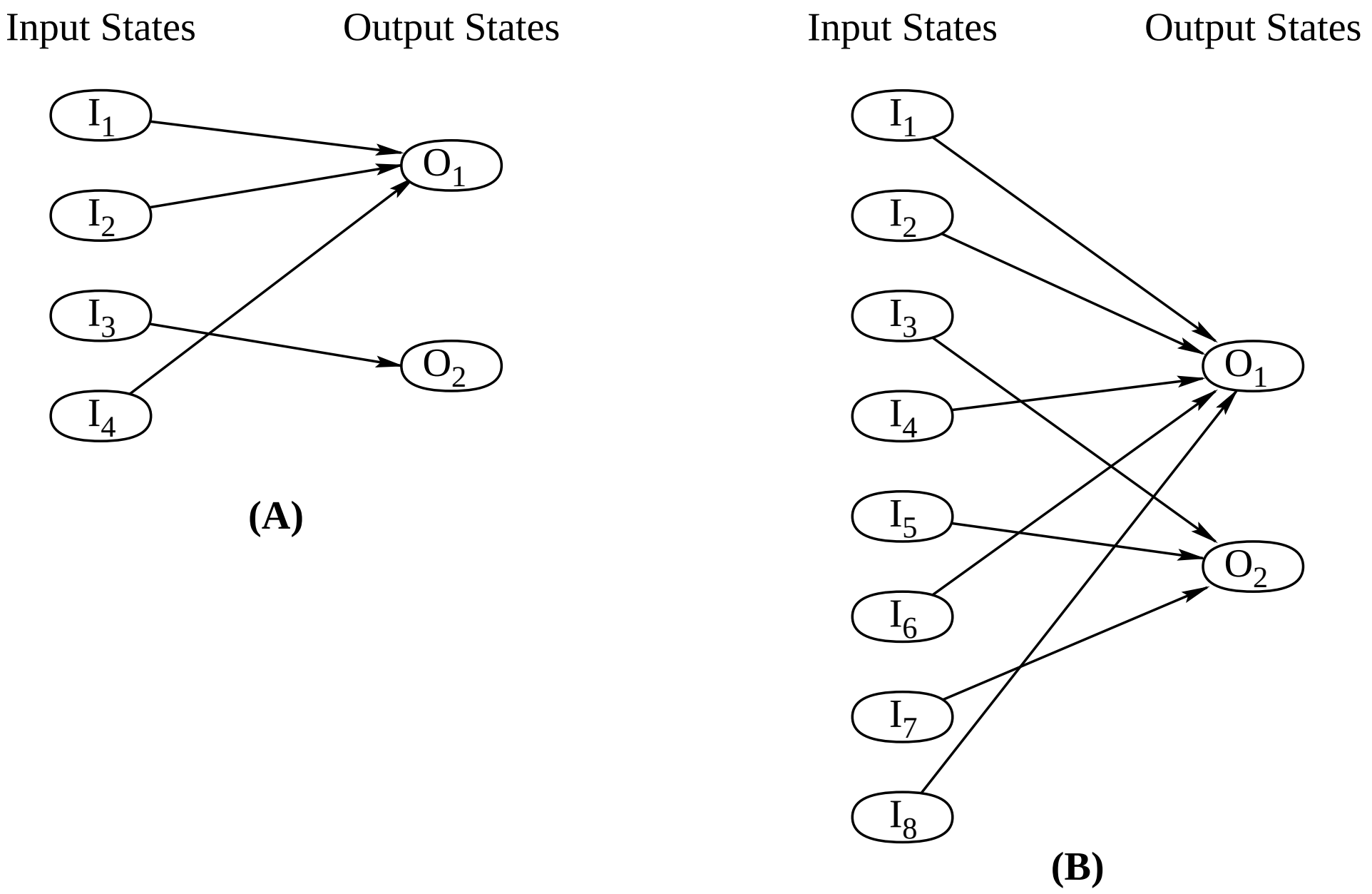

**Figure 2.3:** Examples of many-to-one mappings. Mapping (A) is for four inputs and two outputs. Mapping (B) is for eight inputs and two outputs.

| $\mathcal{N}$ | $\mathcal{R}$ | $\mathcal{N}^{\mathcal{R}}$ | Total number of evolution rules ($=\mathcal{N}^{\mathcal{N}^{\mathcal{R}}}$) | Order |
|---|---|---|---|---|
| 2 | 1 | 2 | 4 | $[\mathcal{O}(10^0)]$ |
| 2 | 2 | 4 | 16 | $[\mathcal{O}(10^1)]$ |
| 3 | 1 | 3 | 27 | $[\mathcal{O}(10^1)]$ |
| 4 | 1 | 4 | 256 | $[\mathcal{O}(10^2)]$ |
| 2 | 3 | 8 | 256 | $[\mathcal{O}(10^2)]$ |
| 5 | 1 | 5 | 3125 | $[\mathcal{O}(10^3)]$ |
| 3 | 2 | 9 | 19683 | $[\mathcal{O}(10^4)]$ |
| 4 | 2 | 16 | 4294967296 | $[\mathcal{O}(10^9)]$ |
| 3 | 3 | 27 | 7625597484987 | $[\mathcal{O}(10^{12})]$ |
| 5 | 2 | 25 | 298023223876953125 | $[\mathcal{O}(10^{17})]$ |
| 4 | 3 | 64 | 340282366920938463463374607431768211456 | $[\mathcal{O}(10^{38})]$ |
| 5 | 3 | 125 | $2.35098870164457501593747307444449135 56 \times 10^{87}$ | $[\mathcal{O}(10^{87})]$ |

**Table 2.2:** Number of possible states of evolution neighborhood ($\mathcal{N}^{\mathcal{R}}$) and total number of evolution rules in $\mathcal{N}$-state $\mathcal{R}$-neighborhood cellular automata for some small values of $\mathcal{N}$ and $\mathcal{R}$. Exponential growth in the states of evolution neighborhood with $\mathcal{R}$ and Super-exponential growth in the number of evolution rules with $\mathcal{N}$ and $\mathcal{R}$ is notable.

The simplest example of cellular automata are 2-state 1-neighborhood cellular automata. In these cellular automata only 2 different input states and 2 different output states are possible. Each one of the input states maps to one of the possible output states. Thus, there can be 4 different evolution rules or input-output mappings in these cellular automata. All these evolution rules have been outlined in table 2.3.

| Input state number | Input states | Output states for evolution rule number | | | |
|---|---|---|---|---|---|
| | | 0 | 1 | 2 | 3 |
| 0 | ○ | ○ | ● | ○ | ● |
| 1 | ● | ○ | ○ | ● | ● |

**Table 2.3:** All the evolution rules (or, input-output mappings) in 2-state 1-neighborhood cellular automata. The symbols ○ and ● represent the states of lattice sites.

If total number of rules in $\mathcal{N}$-state $\mathcal{R}$-neighborhood cellular automata is taken to be an *intuitive indicator* of their complexity,[3)] the examples of slightly more complex cellular automata are the 2-state 2-neighborhood cellular automata and then 3-state 1-neighborhood cellular automata (see table 2.2). All the possible evolution rules for these cellular automata have been outlined in tables 2.4 and 2.5. The next examples, in this order of complexity, are the 4-state 1-neighborhood cellular automata and 2-state 3-neighborhood cellular automata. Both of these have equal complexity. The 2-state 3-neighborhood cellular automata have been outlined and studied extensively by Wolfram on one dimensional lattice of sites with nearest neighbor configuration of evolution neighborhood defined over the immediately preceding evolution [4]. Such generalized studies involving more complex cellular automata have never been attempted in the literature.[4)] In fact, such generalized studies (involving computer simulation) for more complex cellular automata do not appear to be feasible because total number of evolution rules that need to be studied becomes very large; neither can the evolution rules for more complex cellular automata be tabulated on paper.

### 2.1.3 Classification of Cellular Automata

It has been pointed out in Sec. 2.1.2.4 that information transfer mechanism in cellular automata is completely dictated by the topology of evolution neighborhood. Since information is the raw material as well as the end product of any computation, the topology of evolution neighborhood provides a broad basis for classification of cellular automata. Depending upon the topology of evolution neighborhood cellular automata can be broadly classified into two types: (i) trivial cellular automata, and (ii) nontrivial cellular automata. A finer classification of cellular automata on the basis of their gross evolutionary behavior has been proposed by Wolfram [6]. This classification, however, is of no immediate relevance here.

[3)] Note that the measure being used here is merely an *intuitive indicator* of the complexity of very broad *class of cellular automata*, *e.g.*, the class of $\mathcal{N}$-state $\mathcal{R}$-neighborhood cellular automata for some $\mathcal{N}$ and $\mathcal{R}$. The class of $\mathcal{N}$-state $\mathcal{R}$-neighborhood cellular automata consists of all the $\mathcal{N}$-state $\mathcal{R}$-neighborhood cellular automaton taken together, *i.e.*, it has exactly $\mathcal{N}^{\mathcal{N}^{\mathcal{R}}}$ members. Also note that there are many other parameters, *e.g.*, topology of the evolution neighborhood, which are not being considered here.

The measure of complexity being used here does not help in measuring complexity of a cellular automaton. The concept of complexity of a cellular automaton is very different from the complexity of an entire class of cellular automata that has been proposed and used here. Extensive discussion on complexity of a cellular automaton and its measure can be found in [5,6].

[4)] No such attempt shall be made in this investigation either.

| Input state number | Input states | Output states for evolution rule number | | | | | | | | | | | | | | | |
|---|---|---|---|---|---|---|---|---|---|---|---|---|---|---|---|---|---|
| | | 0 | 1 | 2 | 3 | 4 | 5 | 6 | 7 | 8 | 9 | 10 | 11 | 12 | 13 | 14 | 15 |
| 0 | $\circ$ $\circ$ | $\circ$ | $\bullet$ | $\circ$ | $\bullet$ | $\circ$ | $\bullet$ | $\circ$ | $\bullet$ | $\circ$ | $\bullet$ | $\circ$ | $\bullet$ | $\circ$ | $\bullet$ | $\circ$ | $\bullet$ |
| 1 | $\circ$ $\bullet$ | $\circ$ | $\circ$ | $\bullet$ | $\bullet$ | $\circ$ | $\circ$ | $\bullet$ | $\bullet$ | $\circ$ | $\circ$ | $\bullet$ | $\bullet$ | $\circ$ | $\circ$ | $\bullet$ | $\bullet$ |
| 2 | $\bullet$ $\circ$ | $\circ$ | $\circ$ | $\circ$ | $\circ$ | $\bullet$ | $\bullet$ | $\bullet$ | $\bullet$ | $\circ$ | $\circ$ | $\circ$ | $\circ$ | $\bullet$ | $\bullet$ | $\bullet$ | $\bullet$ |
| 3 | $\bullet$ $\bullet$ | $\circ$ | $\circ$ | $\circ$ | $\circ$ | $\circ$ | $\circ$ | $\circ$ | $\circ$ | $\bullet$ | $\bullet$ | $\bullet$ | $\bullet$ | $\bullet$ | $\bullet$ | $\bullet$ | $\bullet$ |

**Table 2.4:** All the evolution rules (or, input-output mappings) in 2-state 2-neighborhood cellular automata. The symbols $\circ$ and $\bullet$ represent the states of lattice sites.

| Input state number | Input states | Output states for evolution rule number | | | | | | | | | | | | | |
|---|---|---|---|---|---|---|---|---|---|---|---|---|---|---|---|
| | | 0 | 1 | 2 | 3 | 4 | 5 | 6 | 7 | 8 | 9 | 10 | 11 | 12 | 13 |
| 0 | $\circ$ | $\circ$ | $\bullet$ | $\star$ | $\circ$ | $\bullet$ | $\star$ | $\circ$ | $\bullet$ | $\star$ | $\circ$ | $\bullet$ | $\star$ | $\circ$ | $\bullet$ |
| 1 | $\bullet$ | $\circ$ | $\circ$ | $\circ$ | $\bullet$ | $\bullet$ | $\bullet$ | $\star$ | $\star$ | $\star$ | $\circ$ | $\circ$ | $\circ$ | $\bullet$ | $\bullet$ |
| 2 | $\star$ | $\circ$ | $\circ$ | $\circ$ | $\circ$ | $\circ$ | $\circ$ | $\circ$ | $\circ$ | $\circ$ | $\bullet$ | $\bullet$ | $\bullet$ | $\bullet$ | $\bullet$ |

| Input state number | Input states | Output states for evolution rule number | | | | | | | | | | | | |
|---|---|---|---|---|---|---|---|---|---|---|---|---|---|---|
| | | 14 | 15 | 16 | 17 | 18 | 19 | 20 | 21 | 22 | 23 | 24 | 25 | 26 |
| 0 | $\circ$ | $\star$ | $\circ$ | $\bullet$ | $\star$ | $\circ$ | $\bullet$ | $\star$ | $\circ$ | $\bullet$ | $\star$ | $\circ$ | $\bullet$ | $\star$ |
| 1 | $\bullet$ | $\bullet$ | $\star$ | $\star$ | $\star$ | $\circ$ | $\circ$ | $\circ$ | $\bullet$ | $\bullet$ | $\bullet$ | $\star$ | $\star$ | $\star$ |
| 2 | $\star$ | $\bullet$ | $\bullet$ | $\bullet$ | $\bullet$ | $\star$ | $\star$ | $\star$ | $\star$ | $\star$ | $\star$ | $\star$ | $\star$ | $\star$ |

**Table 2.5:** All the evolution rules (or, input-output mappings) in 3-state 1-neighborhood cellular automata. The symbols $\circ$, $\bullet$, and $\star$ represent the states of lattice sites.

#### 2.1.3.1 Trivial Cellular Automata

Trivial cellular automata are those cellular automata in which all the lattice sites comprising the evolution neighborhood lie on a straight line[5)] which passes through the lattice site whose new state is to be computed. Because of this restriction, information propagates only along parallel straight lines in $(\mathcal{D}+1)$-dimensional space without dispersing or spreading to neighboring lattice sites. Since there is no dispersion, information propagating along each straight line remains (or, gets) locked at it and no *meaningful computation*[6)] occurs (information contained long each line never changes). As a result, the dynamical behavior produced by these cellular automata is trivial, *i.e.*, lacking complexity.

In these cellular automata the evolution axis can be made parallel to the straight lines along which information propagates using simple coordinate transformation. Such a transformation explicitly brings out an additional fact that in these cellular automata, information, essentially, remains confined at the lattice sites. As a result, in these cellular automata each lattice site behaves as an isolated system. Since, isolated systems do not interact, it is evident that these cellular automata cannot be used for simulation of physical systems.

---

[5)] In cellular automata having $\mathcal{D}$-dimensional lattice of sites, the straight line lies in $(\mathcal{D}+1)$-dimensional space which consists of the lattice of sites and evolution axis taken together.

[6)] *Meaningful computation* involves generation and annihilation (in essence, *change*) of information, which cannot occur unless there is dispersion and mixing of information from various sources.

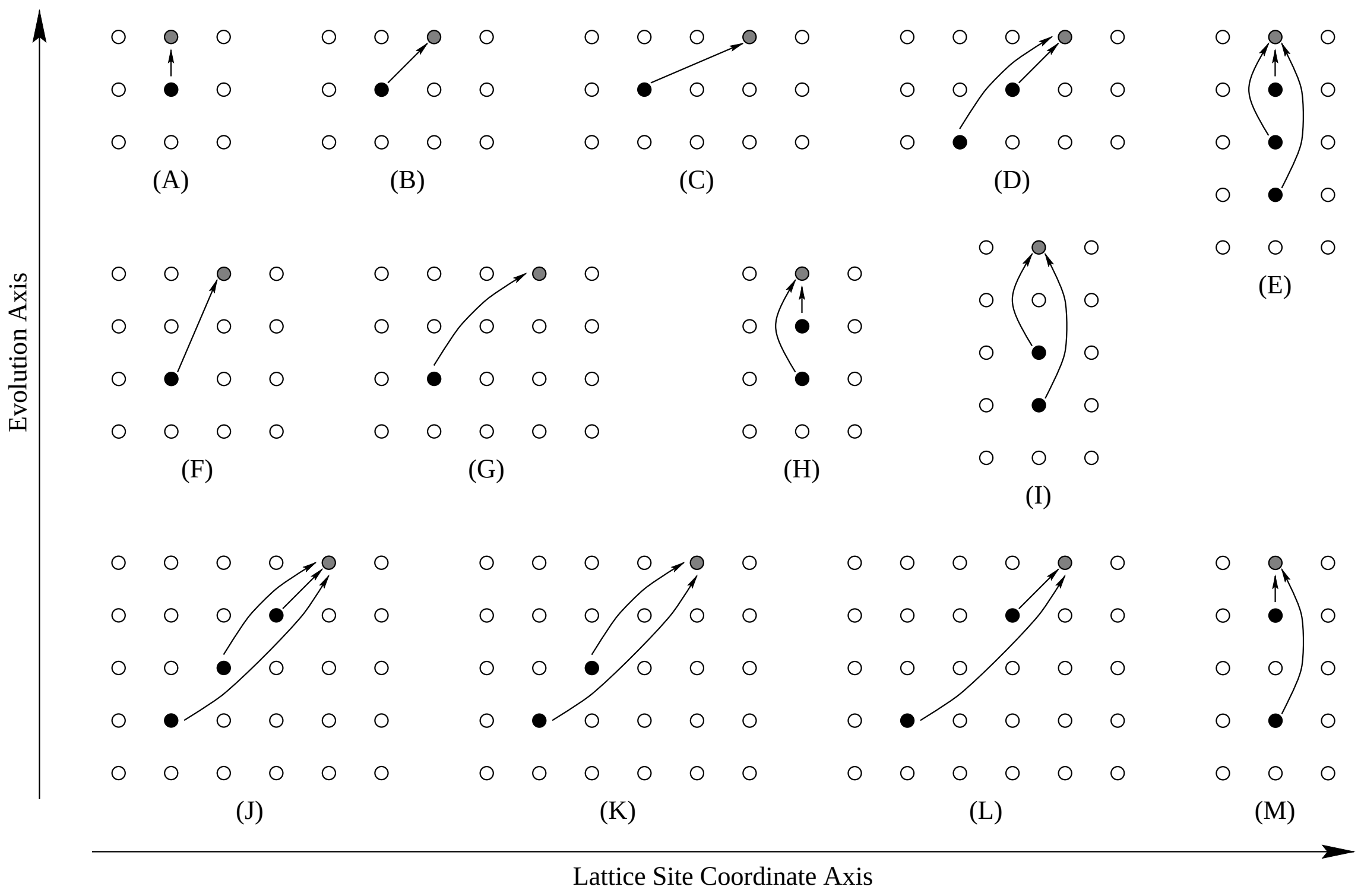

**Figure 2.4:** Examples of evolution neighborhoods over one-dimensional lattice of sites which give rise to trivial cellular automata. Lattice sites to be computed are marked with dotted circle. Black circles indicate lattice sites comprising the evolution neighborhood of the lattice site to be computed. Hollow circles represent other lattice sites. Arrows pointing from back circles to dotted circles indicate the direction of flow of information.

Note that the above definition of trivial cellular automata is considerable generalization over that used and proposed in [7,8] wherein cellular automata whose evolution neighborhood consists of just one lattice site have been defined as trivial cellular automata.

Some examples of evolution neighborhood over one-dimensional lattice of sites which give rise to trivial cellular automata have been illustrated in Fig. 2.4.

#### 2.1.3.2 Non-trivial Cellular Automata

The simplest definition of nontrivial cellular automata is negation of that of trivial cellular automata. To be rigorous, nontrivial cellular automata are those cellular automata in which all the lattice sites comprising the interaction neighborhood do not lie on one straight line passing through the lattice site whose new state is to be computed. Because of this restriction, each lattice site falls in the evolution neighborhood of at least two lattice sites at each new evolution. As a result, information from each lattice site comprising the evolution neighborhood of any lattice site propagates (disperses) in at least two different directions at each evolution. This causes mixing of information coming from at least two directions. This dispersion and mixing of information causes generation of new information at each evolution and leads to meaningful computation. As a result, the dynamical behavior produced by these cellular automata is nontrivial, *i.e.*, shows complexity. In these cellular automata, lattice sites behave as interacting systems and information at each lattice site changes at each evolution. As a result, these cellular automata can be used for simulation of (appropriate) physical systems. Some examples of evolution neighborhood over one-dimensional lattice of sites which give rise to nontrivial cellular automata are illustrated in Fig. 2.5.

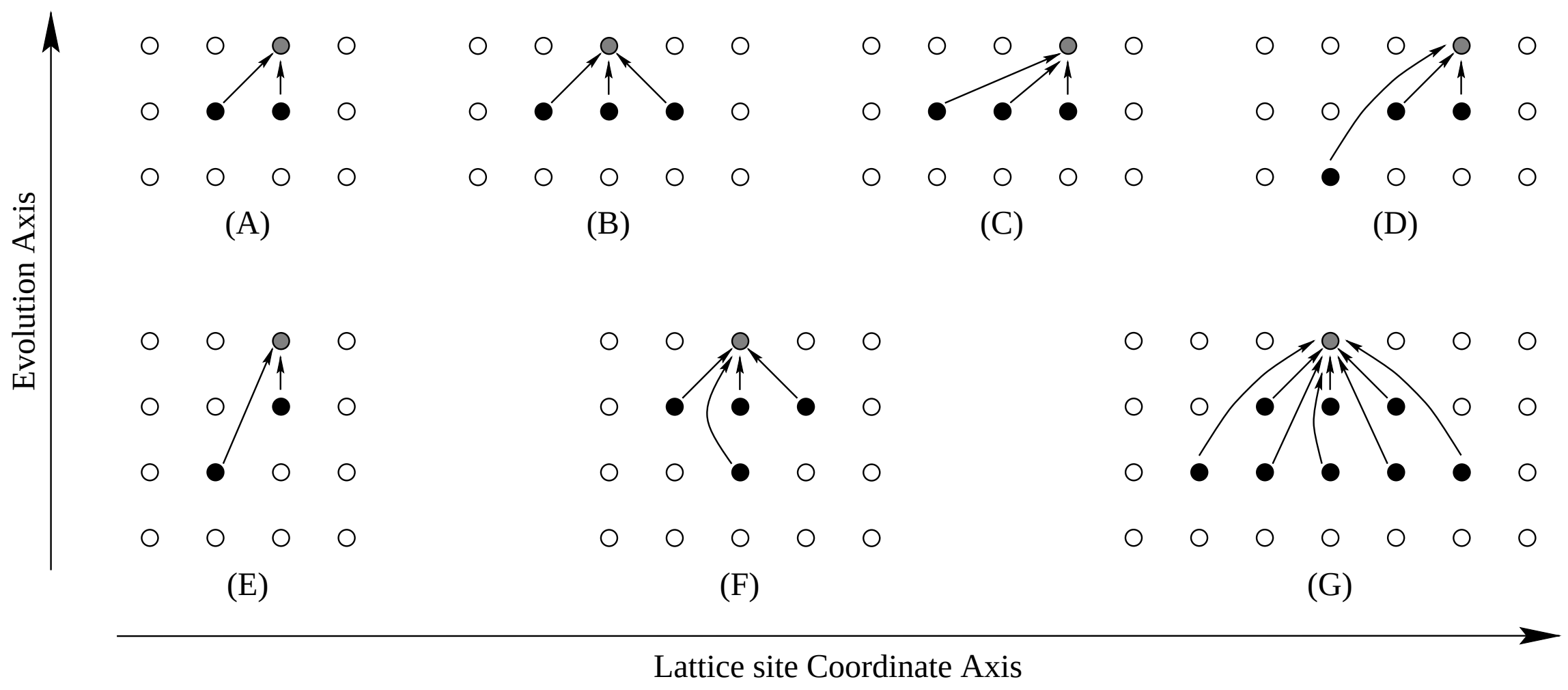

**Figure 2.5:** Examples of evolution neighborhoods over one-dimensional lattice of sites which give rise to nontrivial cellular automata. Lattice sites to be computed are marked with dotted circle. Black circles indicate lattice sites comprising the evolution neighborhood of the lattice site to be computed. Hollow circles represent other lattice sites. Arrows pointing from back circles to dotted circles indicate the direction of flow of information.

### 2.1.4 Cellular Automata Modeling of Physics and Physical Systems

Although cellular automata were introduced in connection with evolutionary biological systems, being abstract mathematical formalism, they find many applications in non-biological fields as well [4]. Among these, modeling of physics and physical systems has emerged as one of the primary areas of application of cellular automata. Many independent investigations have confirmed the feasibility of such application [4,9–11].[7)] As a result, cellular automata have also been envisioned as alternative to the usual calculus based approach for modeling of physics and physical systems [12]. In fact, the discussion outlined in chapter 1 and earlier sections of this chapter also shows that such an application is not only possible but also needed at present. Presently, cellular automata models of physical systems are known as *lattice gas automata* or simply as *lattice gases* in the literature.

Cellular automata, by themselves, are mathematical abstractions. As a result, while applying cellular automata for modeling of any system, their basic elements (*c.f.*, Sec. 2.1.2) need to be interpreted in an appropriate manner (*c.f.*, 1.3.4.2) to establish unambiguous correspondence between the model and the system and behavior of the two. In order to model physical systems using cellular automata, the basic elements of cellular automata can be interpreted in many different ways. One of the possible interpretations

[7)] An enhanced view point, with some additional arguments and features, regarding desirability and feasibility of such application of cellular automata is as outlined in chapter 1 (Secs. 1.3.3 and 1.3.4).

| Correspondence Between Elements of | |
|---|---|
| Cellular Automata | Physical Systems |
| Lattice of sites | Physical position space. |
| Evolution axis | Time axis. |
| Lattice site values | Particles and their velocities. Necessarily many (at least two) particles. |
| Evolution neighborhood | Interaction neighborhood of particles. |
| Evolution rule | Dynamical law. |

**Table 2.6:** Interpretation of basic elements of cellular automata in existing lattice gases for establishing correspondence between the model and the physical system.

that has been employed extensively in the literature for cellular automata modeling of physical systems and that makes the basis of existing lattice gases is outlined in table 2.6.

## 2.2 Lattice Gases

### 2.2.1 The Concept: Early Developments and Opinions

As a concept, lattice gases have been known for more than four decades now. Their appearance in text books of statistical mechanics goes as far back as 1963 in [13] wherein they have been described as follows

> *A lattice gas is a collection of atoms whose positions can take on only discrete values. These discrete values form a lattice of given geometry with $\gamma$ nearest neighbors to each lattice site. Each lattice site can be occupied by at most one atom. ...*

The earliest theoretical studies on the lattice gases of the type mentioned above have been carried out under the assumption that only nearest neighbors interact and that kinetic energy of an atom is negligible. After imposing very simple two-body interaction potential

$$\Phi(r) = \begin{cases} \infty & (r = 0) \\ -\epsilon & (r = \text{nearest-neighbor distance}) \\ 0 & (\text{otherwise}) \end{cases} \tag{2.6}$$

the total energy, partition function, grand partition function, and equation of state of these lattice gases have been determined. The equivalence of lattice gases with Ising models has been established by adopting a method known as "correct Boltzmann counting" for counting the number of configurations. Theoretical studies on the behavior of lattice gases near critical point, *e.g.*, determination of specific heats, have also been carried out [14].

Despite much theoretical developments, early opinions on the usability of lattice gases and their correspondence with physical systems were not quite hopeful. The interest shown in them was purely from academic and mathematical points of views. The following quote from Huang [13] does much to clarify these apprehensions.

> *The lattice gas does not directly correspond to any real system in nature. If we allow the lattice constant to approach zero, however, and then add to the resulting equation of state the pressure of an ideal gas, the model corresponds to a real gas of atoms interacting with one another through a zero-range potential. Thus it may be interesting to study the phase transition of a lattice gas.*

> *The lattice gas has also been used as a model for the melting of a crystal lattice. When it is so used, however, the lattice constant must be kept finite. The kinetic energy of the atoms in the crystal lattice is appended in some ad hoc fashion. Such a model would only have a mathematical interest, because it is not clear that it described melting.*

### 2.2.2 Recognition of Lattice Gases as Cellular Automata

The origin of the concept of lattice gases precedes that of the concept of cellular automata in that the concept of lattice gases appeared in text books as early as 1963 [13] whereas cellular automata were introduced by von Neumann in 1966 [1] (according to Wolfram [4] in 1963). In their early days, lattice gases were simulated using the Ising model. They were not looked upon as cellular automata even after the concept of cellular automata had been introduced. The equivalence of the two, in fact, was not established until a series of papers by Hardy, de Pazzis, and Pomeau [15–17] appeared a decade later.

In their (now classical) papers Hardy, de Pazzis, and Pomeau developed and used a highly simplified model for studying the time evolution and transport properties of a two-dimensional gas (henceforth HPP gas). Their studies, in as much as was their aim, involved systematic theoretical analysis of the model and derivation of coarse grained hydrodynamical equations, transport coefficients, and correlation functions. The studies showed that even a very simple model of particle dynamical systems is capable of reproducing very complex dynamical phenomena. The dynamics of the model studied, however, was very far away from that of any real system. As a result, the study was perceived as a primarily academic investigation and was not followed up until much later. Furthermore, Hardy, de Pazzis, and Pomeau had not perceived their model as a cellular automaton and no statements in this regard were made in their papers. As a result, even after their papers, lattice gases were not recognized as cellular automata explicitly in the literature and interrelationship between them remained unknown until much later.

Explicit recognition of lattice gases as cellular automata came after one more decade of silence in the literature. This recognition was brought about through an extensive review of cellular automata by Wolfram [4] and investigations presented by Margolus [9] and Vichniac [10] in an interdisciplinary workshop on cellular automata [18] in 1983. In the workshop, two different *billiards ball model cellular automaton* (henceforth BBMCA) were proposed by Margolus [9]. One of these models is similar to the HPP gas in many ways. In studies on simulating physics with cellular automata conducted by Vichniac [10], which include simulation of the Ising model with cellular automata, the HPP gas was cited as an example of cellular automaton capable of meaningful physical computation.

### 2.2.3 Recent Development of Lattice Gases

#### 2.2.3.1 Underlying Causes and Driving Factors

After it was shown that lattice gases are cellular automata, lattice gases have seen extremely rapid progress during the last decade and a half. This process is reflected in the form of growing number of publications and specialized conferences on this topic (see [19–23] and other references). This progress can be attributed largely to the simplicity of lattice gases and the ease with which they can be developed and implemented on computers. Their simplicity makes them widely accessible to scientists working in many different and unrelated fields, causing rapid inflow and hybridization of new ideas which naturally leads to diverse developments. Another major factor that has attracted attention of scientists from various disciplines to lattice gases is the advantage offered by lattice gases in terms of simple, accurate (round-off error free), and very fast simulation of extremely complex dynamical systems (*c.f.*, Sec. 1.3.4 and 1.4). To elaborate, lattice gases can be used with equal ease for simulation of simple systems, *e.g.*, flow of a simple hard sphere gas in an ideal friction-less pipe, to exceedingly complex systems, *e.g.*, flow of a mixture of many particle species having complex interactions in systems with geometrically complex boundary conditions.

Yet another powerful driving factor in the development of lattice gases has been construction of extremely fast massively parallel specialized computers known as *cellular automata machines* [24–26]. These computers provide naturally adopted base for lattice gas simulations and are almost as inexpensive as a typical personal computer. The prospects of being able to use such powerful yet inexpensive computers for studying the dynamics of physical systems have also proven to be powerful driving factors in the development of lattice gases.

#### 2.2.3.2 Turning Point in the Development of Lattice Gases

Although the formal introduction of lattice gases as cellular automata is attributed to Hardy, de Pazzis, and Pomeau, the turning point which brought them out as acceptable realistic models of physical systems, rather than as objects of purely academic or mathematical curiosity, and considerably accelerated their development lies elsewhere. This turning point came after the HPP gas and BBMCA models, with the introduction of a new lattice gas by Frisch, Hasslacher, and Pomeau in 1986 [27] (henceforth FHP gas). Rigorous derivation of the coarse grained hydrodynamic equations for the FHP gas by Wolfram [28] and Frisch *et. al.* in the followup [29] of their letter [27] showed that the macrodynamics of the FHP gas is equivalent to that represented by the incompressible Navier-Stokes equations.

The recovery of incompressible Navier-Stokes equations for the FHP gas brought about the revelation that as simple and fully discrete models as the FHP gas are indeed capable of simulating dynamical processes which are as complicated as the Navier-Stokes equations themselves. This revelation, combined with the simplicity of lattice gases (they are easy to understand, construct, and implement on computer), caught the attention of scientists from many different fields and opened the path for variety of developments.

Although the coarse grained hydrodynamic equations obtained for the HPP gas are also nonlinear, its introduction did not turn out be as impacting as that of the FHP gas. This is because the HPP gas, owing to its unacceptably unphysical dynamical behavior, could not be used for studying realistic physical systems.

#### 2.2.3.3 Systems Modeled using Lattice Gases

Following the FHP gas, lattice gases have developed very rapidly. Presently, large number of LGA models are available for simulation of a wide variety of systems and phenomena. For example, lattice gas models have been developed for simulation of two- and three-dimensional hydrodynamics [28–31], three-dimensional external flows [32], free boundary flows [33], heat conduction [34], phase-transitions [35–38], reaction-diffusion systems (including porous media) and reaction transport process [39–44], Ising systems [45,46], immiscible two-phase flow [47], critical phenomena in immiscible systems [48], soliton turbulence [49], polymer fluids [50], high Reynolds number incompressible flows [51], one-dimensional diffusion [52], segregation of two-species granular flow [53], diffusion limited aggregation [54], systems with attractive and repulsive interactions [55], cage effects [56], coarsening in two species driven diffusive systems [57], density waves of granular flow [58], formation of Liesegang patterns [59], domain growth [60], collective biological motion [61], sound propagation in one-dimension [62], diffusive and dissipative systems [63], phase separation of crystal surface [64], long-range correlations in nonequilibrium systems [65], dendritic growth [66], and continuous phase transition from active to inactive phase [67]. Some of the recent studies on lattice gases are concerned with development of quantum lattice gases [68–74]. Quantum lattice gases, however, are of no interest in the present investigation.

This list of lattice gases is not exhaustive. Many more lattice gases have been developed and are available in the literature. These can be traced through the references cited above and elsewhere in the present investigation.

#### 2.2.3.4 Simulation Studies

Lattice gases pointed out in the previous section have been the subject of many simulation studies. Some among these (besides the ones mentioned above) are: simulation of propagating fronts [75], hydrodynamics in two-dimensions [76], porous media flow [77–80], diffusion [81], reaction-diffusion systems [82,83], stable and unstable interfaces [84], velocity autocorrelation function in four-dimensions [85], three-dimensional hydrodynamics [86], viscous fingering in porous media [87], frequency-dependent permeability of porous media [88], circular Couette flow and chaotic mixing [89], chemical wave fronts [90], one-dimensional phase transitions [91], anisotropic polydomain structures in systems with repulsive interactions [92], density profile in two-phase systems [93], rupture and coalescence [94], interface roughening [95], dynamical behavior of immiscible fluids and microemulsions [96], $\mathcal{P}$-completeness [46], immiscible fluids and microemulsions [97], self-organization in driven lattice gas [98], interface dynamics [99], and hydrodynamic behavior of gas-solid fluidized beds [100]. This list of investigations is merely indicative of the depth and breadth of simulation studies carried out using lattice gases. It is not exhaustive. Many more simulation studies have been conducted using lattice gases. These can be traced through the references cited above and elsewhere in the present investigation.

It is worthwhile to note that most of the simulation studies pointed out above have been carried out either for understanding or for demonstrating the capabilities and limitations of the subject lattice gases in simulating (usually highly abstracted) physical systems. Through these investigation much insight has been gained into macrodynamics of lattice gases and many limitations of lattice gases have been found out.

It is not with the scope of the present investigation to enlist and elaborate on *all* the capabilities and limitations of lattice gases. Some of these limitations, however, have been addressed in this investigation. These have been pointed out in Sec. 2.5.

#### 2.2.3.5 Theoretical Studies

Alongside the development of lattice gases and simulation studies using them, a large number of theoretical studies have also been carried out in the literature. To cite some examples from a small cross section of the available lattice gas literature, these studies include investigations on hydrodynamics [28,29], Reynolds number scaling of hydrodynamics [101], heat transfer (Fourier's law and Green-Kubo formula) [102,103], Knudsen layer theory [104], self-diffusion [105], staggered transport coefficients [106], velocity autocorrelation function [107,108], density-functional theory and ordering transitions [109,110], relaxation dynamics (mode-coupling theory of ideal glass transition) [111], generalized hydrodynamics and dispersion relations [112], statistical hydrodynamics [113], pair-correlation function [114], scaling of fluctuations [115], instabilities and pattern formation [116], surface tension and interface fluctuations in immiscible fluids [117], phase separation [118], correlations and renormalization [119], long-time tails [120], pattern formation [121], long-range correlations in nonequilibrium systems [65], wetting [122], driven diffusion [123], cooperative diffusion [124], dynamics of lattice gases with arbitrary number of particles moving in each direction [125], and computational complexity of Lorentz lattice gas [126]. This list is merely indicative of the depth and breadth of theoretical investigation carried out on lattice gases. It is not exhaustive. Many more theoretical studies have been conducted on lattice gases. These can be traced through the references cited above and elsewhere in the present investigation.

Like the simulation studies, the objective of theoretical studies has also been to analyze and understand the dynamics of lattice gases, primarily their macrodynamics. From these studies immense amount of information has been obtained on the nature of coarse grained dynamics of lattice gases. The quantitative information furnished by these studies includes "exact" expressions for coarse grained dynamical equations and transport coefficients among many other dynamical parameters of interest. Besides this, these studies have shown that a number of methods that are used for analysis of continuum models are equally well applicable to lattice gases with appropriate limiting considerations.

### 2.2.4 Some Basic Lattice Gases of Interest in the Present Investigation

If viewed with the wide perspective of a fluid dynamicist, all the lattice gases that have been used for simulation of fluid dynamic systems are of interest in this investigation. Thus, within this perspective, almost all the lattice gases pointed out in Sec. 2.2.3 are of interest in this investigation. Analysis of advantages and disadvantages of all these lattice gases and separate elaboration on each one of them, however, is not feasible here (simply because of their number) and thus will not be attempted. Instead, all these lattice gases will be addressed collectively in the following chapter after generalizing them as members of a wide class containing infinitely many lattice gases. A few selected lattice gases, however, are of special interest here either because they occupy special place in the development that has occurred till now or because results from simulations carried out using them can be used for comparison later on. These lattice gases, even though they belong to the wide class to be defined and addressed later, will be addressed individually also. These lattice gases and the reasons of special interest in them are as follows:

Three basic lattice gases that are of direct interest at present are the HPP gas, the FHP gas, and the BBMCA model. The HPP and FHP gases are of interest because the entire development of lattice gases outlined in Sec. 2.2.3 is pivoted, either directly or indirectly, at these two basic lattice gases. The BBMCA model is of interest because it is one of the earliest models and chronologically their development falls between the HPP and FHP gases. As a result, only these lattice gases have been pointed out explicitly in Sec. 2.2.3.

The second reason of explicitly pointing out only these three lattice gases in Sec. 2.2.3 is that they are fluid lattice gases and scope of this investigation is restricted to fluid lattice gases only. Because of this constraint analysis of quantum lattice gases and lattice gases which show phase transitions, *e.g.*, the two-phase lattice gas due to Appert and Zaleski [36] which shows liquid-gas phase transition, goes beyond the scope of this investigation. The third reason is that the example lattice gas that will be developed and studies in this investigation is a *single speed* lattice gas,[8)] *i.e.*, in this lattice gas the speed of all the particles is always equal. Hence its structure and dynamical behavior can be compared with that of other single speed lattice gases only. The three lattice gases mentioned above are also single speed lattice gases and thus will be useful for a comparative study.

Besides the three single speed lattice gases mentioned above there exists one more single speed lattice gas due to Toffoli and Margolus [24] known as the TM gas. This lattice gas is very similar to the HPP gas and the BBMCA model. The dynamical behavior of this lattice gas has been studied, though not in much depth, in the literature and has been found to be quite different from that of the HPP gas. This lattice gas is of interest in this investigation because its dynamical exponents (asymptotic exponents) [26] are quite close to those of truly two-dimensional gases [127,128] and will be used for comparison with those of the lattice gas to be developed and studied in the following chapters.

### 2.2.5 Conceptual Representations of Existing Lattice Gases

There are two different conceptual representations of the existing lattice gases. In the first one, the spatial lattice is viewed as partitioned into blocks consisting of more than one cell. Each one of the cells can either contain exactly one particle or be empty. No cell can contain more than one particle at any time. Henceforth this conceptual representation will be referred to as *"partitioned spatial lattice representation"*. In the second one, spatial lattice is viewed as non-partitioned and consisting of cells which can hold more than one (but, finitely many) particles at the same time. Henceforth this conceptual representation will be referred to as *"multiple particle representation"*.

Both these representations, except for some peculiarities (*c.f.*, chapter 3), are are essentially equivalent. If a lattice gas can be represented using one of them, then it can be represented using the other also. *All the existing lattice gases* can be represented using both of them. These representations differ in the way spatial lattice is viewed in them, and in the way evolution rules are presented (in descriptive as well as graphical ways), and (as will be seen later) also in the nature of certain conclusions which can be *directly* drawn from them. Details on these conceptual representations are as follows:

#### 2.2.5.1 Partitioned Spatial Lattice Representation

In this representation of lattice gases, the spatial lattice is partitioned into non-overlapping blocks of cells using some appropriate partitioning scheme. Since each block consists of more than one cell, the spatial lattice (irrespective of its topological details) can be partitioned in at least two different ways. Partitioning is done in all possible ways and each partition is assigned an identification tag (a "number" or a "name" or some other identifier), *e.g.*, "even" and "odd" partitions, "solid" and "dashed" partitions, or "0" and "1" partitions. Since all blocks are identical, all the partitions are also identical. All the partitions are unique in that the blocks of no two partitions fully overlap.

[8)] In the lattice gas literature *"single speed"* lattice gases are also known as *"athermal"* lattice gases. This is because in these lattice gases temperature can not be defined and the energy equation (or, the law of conservation of energy) becomes identical with the continuity equation (or, the law of conservation of mass).

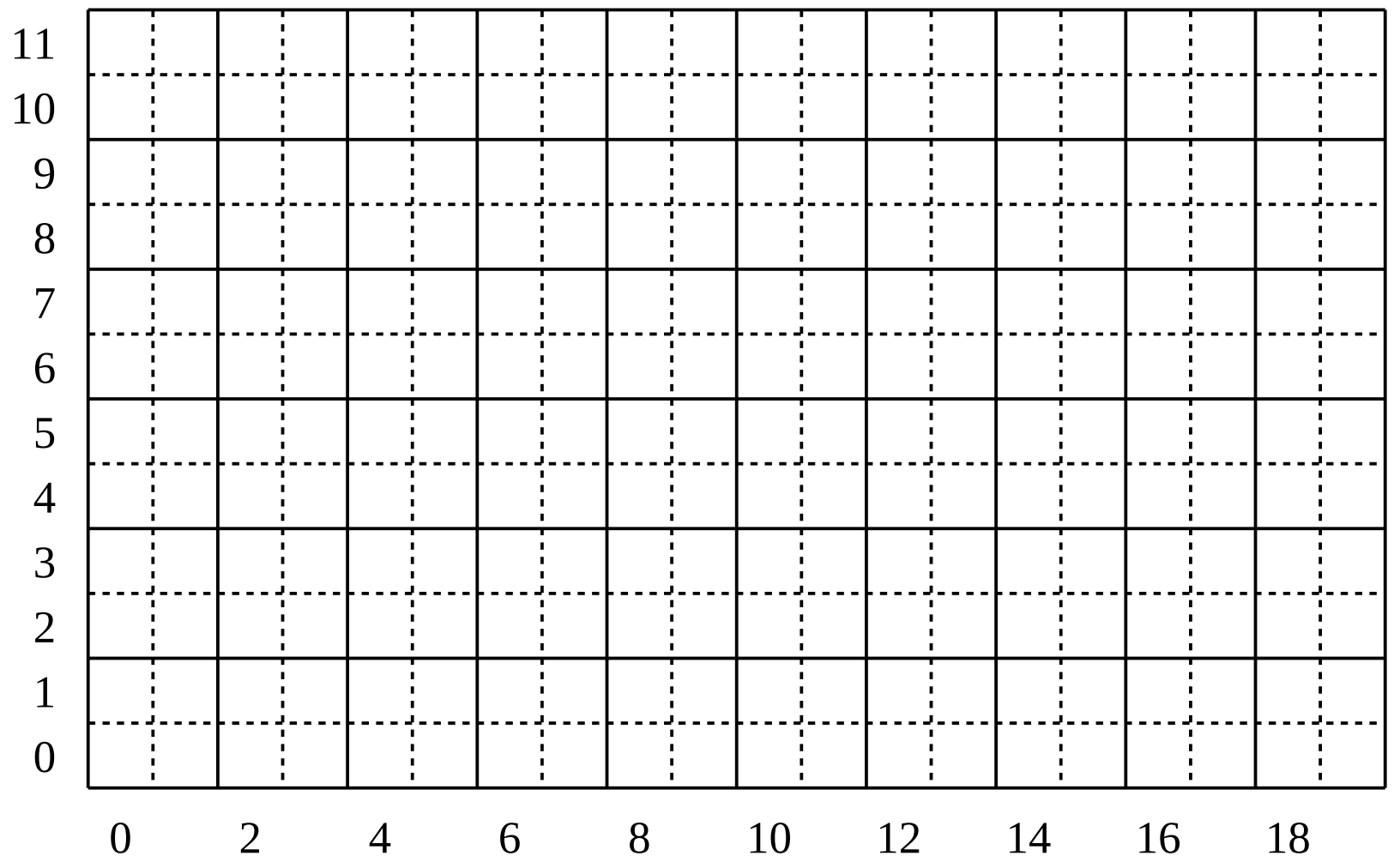

**Figure 2.6:** Partitioning of the square spatial lattice into blocks of $2 \times 2$ cells using the Margolus neighborhood. Partition which starts at cells with even $x$- and even $y$-coordinates is represented by solid lines (called "even" or "solid" partition) and the one which starts at cells with odd $x$- and odd $y$-coordinates is represented by dashed lines (called "odd" or "dashed" partition).

Example of partitioning of square spatial lattice using the *Margolus neighborhood*[9] into blocks of $2 \times 2$ cells is shown in Fig. 2.6. In this figure two partitions, one starting at cells with even $x$- and even $y$-coordinates and the other starting at cells with odd $x$- and odd $y$-coordinates, have been shown. Henceforth these partitions will be referred to as "even" partition and "odd" partition, respectively. Variedly, they will also be referred to as "solid" partition and "dashed" partition, respectively. Two more partitions, one starting at cells with even $x$- and odd $y$-coordinates and the other starting at cells with odd $x$- and even $y$-coordinates are also possible. These have not been shown in the figure.

After partitioning, the partitions to be used during simulations are identified and an appropriate conceptual scheme specifying the sequence in which they will be used during simulations is developed. All the partitions need not be considered. After this, each block is treated as a "super-cell" and evolution rules are constructed for updating the contents of the super-cells (or, blocks) of each partition. Since evolution rules are defined for updating the blocks, the motion of particles at any time step is restricted to be from one lattice site to another within each block. Similarly, at any time step collisions also occur only among particles present within the same block.

If each block consists of $\mathcal{R}$ cells and each cell can have $\mathcal{N}_\mathrm{s}$ different states (one state for representing space and $\mathcal{N}_\mathrm{p} = \mathcal{N}_\mathrm{s} - 1$ states for representing particle species), then each block can have a total of $\mathcal{N}_\mathrm{s}^{\mathcal{R}}$ different initial states or configurations. As a result, construction of evolution rules for each partition consists of specifying the final states for each one of the $\mathcal{N}_\mathrm{s}^{\mathcal{R}}$ possible initial states of the blocks. If total of $\mathcal{P}$ partitions are used during simulations, then evolution rules contain specification of the final states for at least of $\mathcal{N}_\mathrm{s}^{\mathcal{R}}$ and at most of $\mathcal{P}\mathcal{N}_\mathrm{s}^{\mathcal{R}}$ initial states of the blocks. The minimum is when all the partitions use the same evolution rule and the maximum is when each partition uses a different evolution rule.

[9] *Margolus neighborhood* is square block of $2 \times 2$ cells. In general, it need not be square [24].

During simulations, partitions are switched in accordance with the predecided conceptual scheme. At each time step one partition is selected and all the blocks in this partitions are updated according to the evolution rules for this partition. Then, a new partition is selected for the next time step and evolved in an identical way. The number of partitions is necessarily finite because each block consists of finitely many cells. Furthermore, in the partition selection scheme no segment is repeated, *i.e.*, no partition occurs more than once. As a result, the scheme for selecting partitions is used cyclically, *i.e.*, once all the partitions have been selected the selection process starts again with the first partition in the scheme.

Construction of lattice gases using this representation consists of specifying the species of particles, structure of spatial lattice, geometry of blocks in terms of which the spatial lattice is to be partitioned, the partitions, the scheme to be used for switching partitions during simulations, and evolution rules for each partition. The velocity with which particles move over the spatial lattice is not specified. Instead, it is deduced from the evolution rules. Since each block evolves independently of others at every time step, the evolution rules are defined such that all the laws governing dynamics of the system, *e.g.*, the laws of conservation of mass, momentum, and energy, are satisfied within each block during its evolution.

This conceptual representation of lattice gases was developed by Toffoli and Margolus [24]. Using this representation, lattice gases can be implemented on *cellular automata machines* in less memory compared to that required when multiple particle representation is used. This representation is adequate for all the existing lattice gases.[10)] It, however, has some disadvantages in that it hides certain aspects of lattice gases which readily point out their inadequacy as idealized models of physical systems.[11)]

#### 2.2.5.2 Multiple Particle Representation

In this representation of lattice gases, the spatial lattice consists of cells which can hold more than one particle at any time step subject to a maximum. The spatial lattice is not partitioned. Instead, evolution of the system during one time step is decomposed into two sequential sub-steps, namely: (i) the collision step, and (ii) the translation step. Evolution rules are developed separately for each of these steps and used in a fixed sequence, *i.e.*, collision step followed by translation step or vice versa, during simulations. The sequence is fixed in that it remains same at all time steps during a simulation.

In this conceptual representation, in the early stages of development of lattice gases, particles occupying the same lattice site were subjected to an *exclusion principle*: *"No two particles of the same species occupying the same lattice site can have the same velocity"*. This exclusion principle, however, being externally imposed (*i.e.*, not being an element of the representation itself), is not a limitation of this conceptual representation. As a result, many particle of the same species having the same velocity can also be permitted to occupy the same lattice site simultaneously as done in some recent developments [125].

Construction of lattice gases in this representation requires specification of particle species, structure of spatial lattice, velocity with which particles of each species move over the spatial lattice, total number of particles of the same species having same velocity that can occupy the same lattice site at any time step, evolution rules for the translation

[10)] No classical (as opposed to quantum) lattice gas has been found in the literature for which this representation is inadequate. This excludes the lattice gases which will be proposed in this manuscript.

[11)] This statement is based on a specific analysis of lattice gases which will be presented later.

step (henceforth, translation rules), and evolution rules for the collision step (henceforth, collision rules). In this representation, states of lattice sites encode the number of particles, velocity vector of each particle, and species of each particle occupying the lattice sites. As a result, in a lattice gas for $\mathcal{N}_{\mathrm{p}}$ types of particle species with $\mathcal{N}_{\mathrm{v}}^{(i)}$ velocity vectors for particles of species $i$, $i = 1, \ldots, \mathcal{N}_{\mathrm{p}}$, total of $\mathcal{N}_{\mathrm{sym}}$

$$\mathcal{N}_{\mathrm{sym}} = 1 + \sum_{i=1}^{\mathcal{N}_{\mathrm{p}}} \mathcal{N}_{\mathrm{v}}^{(i)}$$

symbols are required for uniquely representing the velocity and species of particles including one symbol for space or absence of particles. If $\mathcal{N}_{\mathrm{m}}^{(i,j)}$ particles of species $i$ moving with velocity $\boldsymbol{v}_j^{(i)}$, $j = 1, \ldots, \mathcal{N}_{\mathrm{v}}^{(i)}$, are permitted to occupy the same lattice site simultaneously, then maximum of $\mathcal{N}_{\mathrm{t}}$

$$\mathcal{N}_{\mathrm{t}} = \sum_{i=1}^{\mathcal{N}_{\mathrm{p}}} \sum_{j=1}^{\mathcal{N}_{\mathrm{v}}^{(i)}} \mathcal{N}_{\mathrm{m}}^{(i,j)} \tag{2.7}$$

particles can occupy the same lattice site simultaneously and total of $\mathcal{N}_{\mathrm{s}}$

$$\mathcal{N}_{\mathrm{s}} = \prod_{i,j=1,1}^{\mathcal{N}_{\mathrm{p}}, \mathcal{N}_{\mathrm{v}}^{(i)}} \left(1 + \mathcal{N}_{\mathrm{m}}^{(i,j)}\right) \tag{2.8}$$

different states are possible for each lattice site at any time step.

Collision rules encode the dynamics of collisions among particles. Collisions occur among particles occupying the same lattice site. As a result, collision rules specify final state of individual lattice sites for all possible ($\mathcal{N}_{\mathrm{s}}$) initial states. During the collision step, the initial state of a lattice site is its state at the end of the previous translation step, and the final state of a lattice site is its state after it (or, the entire system) has been evolved through the collision rules. Since each lattice site evolves independently of others during the collision step at every time step, the collision rules are defined such that all the laws governing dynamics of the system, *e.g.*, the laws of conservation of mass, momentum, and energy, are satisfied at each lattice site during its evolution in the collision step.

Translation rules simply reposition the particles on new lattice sites as pointed by their velocity vectors. Since these rules change only the position of particles, all the laws governing the dynamics remain unchanged during the translation step. That no change occurs due to generation or annihilation of particles during repositioning is guaranteed by the fact that evolution of the system is synchronous (*i.e.*, all the lattice sites evolve simultaneously) and all the lattice sites are identical in all possible respects (*i.e.*, the maximum number of particles of each species having the same velocity that can occupy a lattice site simultaneously is identical for all the lattice sites).

This conceptual representation of lattice gases was first conceived and used by Hardy, de Pazzis, and Pomeau for describing the HPP gas [16,17]. Later, it was adopted by Frisch *et. al.* [27,29] and Wolfram [28] for describing the FHP gas. Since then it has been used widely, in preference to the partitioned spatial lattice representation, for describing lattice gases. It is preferred because of two reasons: (i) it is more elegant compared to the partitioned spatial lattice representation and (ii) construction of evolution rules (collision rules and translation rules) in this representation is simpler compared to that in the partitioned spatial lattice representation. This representation, like the partitioned spatial lattice representation, is adequate for describing all the existing lattice gases of interest in this investigation.[12)] It, however, has some disadvantages in that it hides certain aspects of lattice gases which readily point out their inadequacy as idealized models of physical systems.[13)]

### 2.2.6 Structure of Some Basic Lattice Gases

Structure of the basic lattice gases pointed out in Sec. 2.2.4 will be described in this section. These lattice gases, as pointed out in Sec. 2.2.5, can be described using both the partitioned spatial lattice representation as well as the multiple particle representation. Since both these representations are essentially identical, only one of them will be used in the following descriptions. The BBMCA model and the TM gas will be described using the partitioned spatial lattice representation and the FHP gas will be described using the multiple particle representation. The HPP gas, however, will be described using both these representations to illustrate the similarities and differences between them.

Description of lattice gases using partitioned spatial lattice representation requires description of partitioning and partition selection scheme among other things. The BBMCA model and the HPP and TM gases have identical partitioning and partition selection scheme besides some other features. As a result, it will be advantageous to describe all the common features of these lattice gases together under a separate title. The advantage being that such a description will clearly show the correlations among these lattice gases. This has been done below, before describing the individual lattice gases.

#### 2.2.6.1 Common Features of HPP, BBMCA, and TM gases

The HPP, BBMCA, and TM gases exist over the square spatial lattice. In partitioned spatial lattice representation of these lattice gases, the spatial lattice is partitioned into blocks of $2 \times 2$ cells using the Margolus neighborhood. Identical partitioning and identical partition selection scheme is used in all these three basic lattice gases.

As pointed out in Sec. 2.2.5.1, square spatial lattice can be partitioned in four different ways using the Margolus neighborhood. Out of these four partitions only two are used in the lattice gases mentioned above. Selection of partitions is constrained in that the partitions to be used should be *diagonally* (not horizontally or vertically) displaced relative to each other. Thus, there are two possibilities of selecting the partitions. The first possibility is to use the group of partitions in which one partition starts at cells (or, lattice sites) with even $x$- and even $y$-coordinates and the other starts at cells with odd $x$- and odd $y$-coordinates. The second possibility is to use the group of partitions in which one partition starts at cells with even $x$- and odd $y$-coordinates and the other starts at cells with odd $x$- and even $y$-coordinates. Both these groups are related to each other through linear transformation of coordinates along one axis (*i.e.*, either $x$-axis or $y$-axis) by one cell unit in either positive or negative direction in any one of them. As a result, any one of these two groups of partitions can be used to describe these basic lattice gases without the loss of generality.

In the following sections, the three basic lattice gases will be described using the first group of partitions outlined above. The structure of the spatial lattice with this

[12)] No classical (as opposed to quantum) lattice gas has been found in the literature for which this representation is inadequate. This excludes the lattice gases which will be proposed in this manuscript.

[13)] This statement is based on a specific analysis of lattice gases which will be presented later.

partitioning has been illustrated in Fig. 2.6. The partition which starts at cells with even $x$- and even $y$-coordinates has been drawn using solid lines in the figure. Henceforth, this partition will be referred to as the "solid" or "even" partition. The partition which starts at cells with odd $x$- and odd $y$-coordinates has been drawn using dashed lines in the figure. Henceforth, this partition will be referred to as the "dashed" or "odd" partition.

In these lattice gases partitions can be switched in only one way, *i.e.*, to switch over to a different partition at every time step (or, to use the same partition at even (or, odd) time steps), during simulations. This is because only two partitions are employed for developing these lattice gases. The choice of partition that should be used at even (or, odd) time steps lies free. It does not alter the *macrodynamics* (*i.e.*, macroscopic dynamics or coarse grained dynamics) of these lattice gases.

In these three basic lattice gases all the particles are of same species (*i.e.*, $\mathcal{N}_\mathrm{p} = 1$ and $\mathcal{N}_\mathrm{s} = \mathcal{N}_\mathrm{p} + 1 = 2$) and each block consists of 4 cells. As a result, total of $2^4 = 16$ different states are possible for each block at each time step in these lattice gases. All these states have been shown in Fig. 2.7. During simulations, any one of these states can be the initial (as well as the final) state of a block at any time step. As a result, if the evolution rules are written in a tabular format, the rule table for each partition must contain total of 16 entries. Many of these states, however, are isometrically equivalent.[14)] This, combined with the fact that the fundamental laws of physics remain invariant under isometric transformations, implies that specification of the final state for any one of the isometrically equivalent initial states determines the final states of all the others also.[15)] Thus, it is sufficient to specify the final states for only 6 initial states, instead of all the 16 initial states, in these three basic lattice gases. The 6 basic initial states whose final states shall be specified in the descriptions given in following sections have been shown in Fig. 2.8. All the 16 states shown in Fig. 2.7 can be easily obtained from those shown in Fig. 2.8 through isometric transformations; more specifically through rotations of $\pm n\pi/2$ radians, where $n$ is an integer.

#### 2.2.6.2 The HPP Gas

In this section, structure of the HPP gas is described in both the partitioned spatial lattice representation [24] as well as the multiple particle representation [16,17].

**Multiple Particle Representation:** The HPP gas exists over square spatial lattice. The particles reside on the vertices of the lattice. The particles are indistinguishable and have unit mass. They move with one of the four velocity vectors belonging to the discrete

[14)] Two states or configurations A and B are said to be isometrically equivalent if they are related through isometric transformations (*i.e.*, rotations, reflections through plain mirror, and translations). For example, the states (2), (3), (4), and (5) shown in Fig. 2.7 are isometrically equivalent because they can be obtained from each other simply by rotations of $\pm n\pi/2$ radians, where $n$ is an integer.

[15)] The fundamental laws of physics remain invariant under isometric transformations. This implies that in any model of physical systems the final states for each one of the isometrically equivalent initial states should also be isometrically equivalent. Moreover, the final states should be related to each other through the same isometric transformations through which their respective initial states are related. That is, if final states for the initial states (2) and (3) shown in Fig. 2.7 are denoted $(2^\star)$ and $(3^\star)$ and the states (2) and (3) are related through an isometric transformation $\mathcal{T}$, *i.e.*, $\mathcal{T}[(2)] = (3)$, then the final states $(2^\star)$ and $(3^\star)$ should also be related through the isometric transformation $\mathcal{T}$ only, *i.e.*, $\mathcal{T}[(2^\star)] = (3^\star)$ should hold. Thus, in models of physical systems, it is sufficient to specify the final state for any one of isometrically equivalent initial states. The final states for the remaining initial states can be obtained from the specified state through isometric transformations.

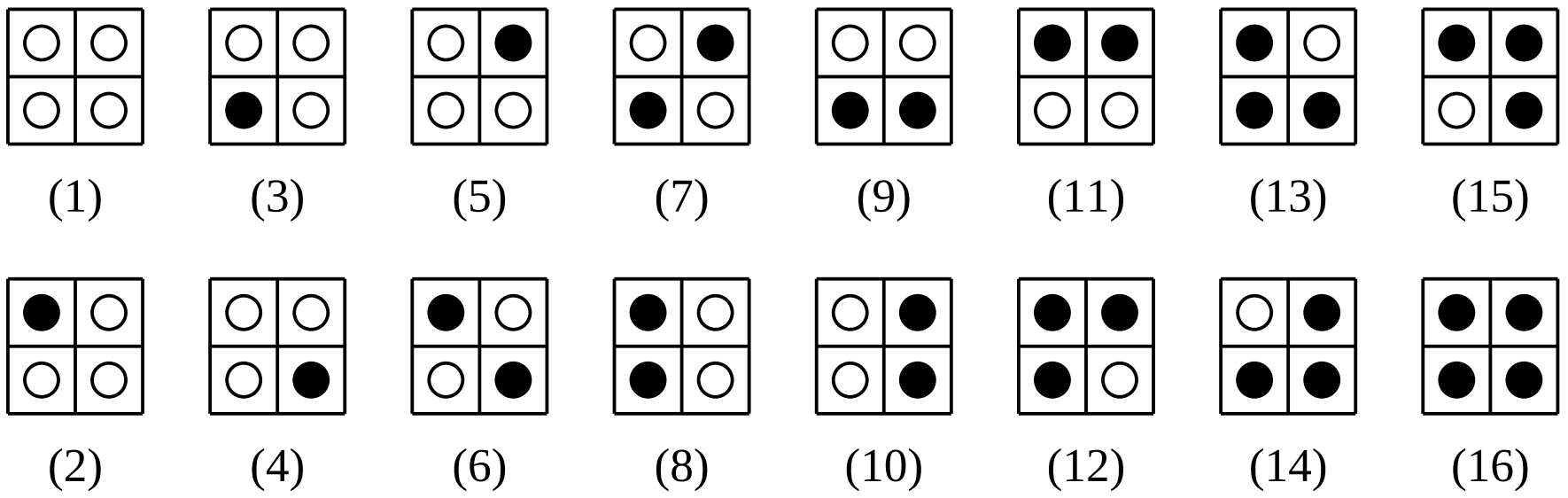

**Figure 2.7:** Possible states of 4 ($= 2 \times 2$) cell blocks for single species systems existing over square spatial lattice. Hollow circles represent space (or, empty cell) and solid circles represent particles.

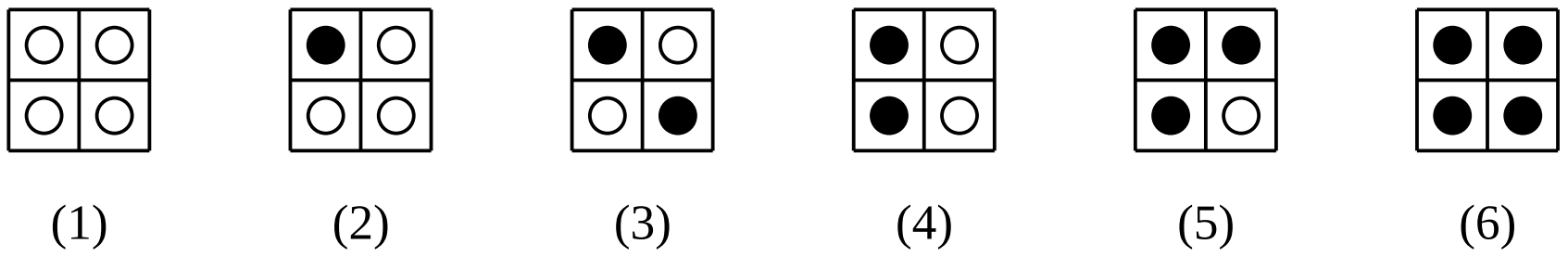

**Figure 2.8:** Six basic states of 4 ($= 2 \times 2$) cell blocks for single species systems existing over square spatial lattice. Hollow circles represent space (or, empty cell) and solid circles represent particles.

velocity set $\mathcal{V}_{\text{HPP}} \equiv \{\boldsymbol{v}_i : \boldsymbol{v}_i = (\cos(i\pi/2), \sin(i\pi/2)); i \in [0, 3]\} = \{(\pm 1, 0), (0, \pm 1)\}$ along the links of the lattice. At the most four particles can occupy a lattice site (or, vertex) simultaneously at any time step subject to the exclusion principle that no two particles occupying the same lattice site can have the same velocity.

Evolution of the system during one time step is decomposed into two sub steps, namely, (i) translation and (ii) collision. During translation step, particles move to neighboring lattice sites as indicated by their velocity vectors; no collisions occur. During collision step, collisions occur among particles occupying the same lattice site; the particles do not move. Collisions occur only on those lattice sites which are occupied by exactly two particles with velocity vectors pointing in opposite directions. Thus, there are only two possible configurations, the configurations (a) and (b) shown in Fig. 2.9, in which collisions occur. All other configurations remain unchanged. The collisions rotate the velocity vectors of colliding particles by $\pi/2$ radians (either clockwise or counter-clockwise). Thus, the configurations (a) and (b) shown in Fig. 2.9 replace each other during collision step, *i.e.*, if configuration (a) is found on a lattice site it is replaced by configuration (b) and vice versa.

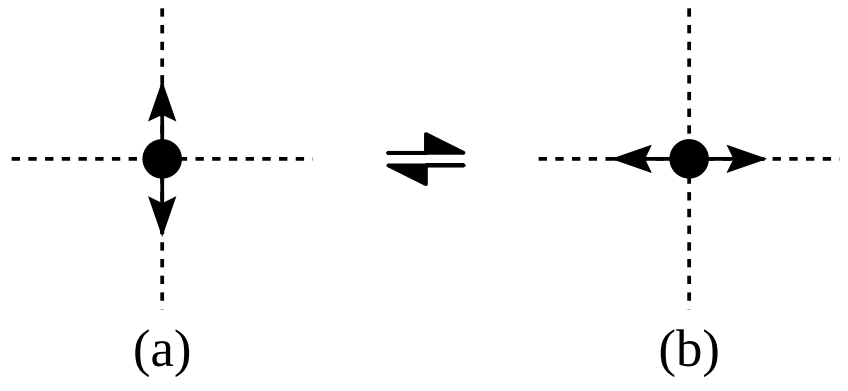

**Figure 2.9:** Collision rules of the HPP gas in multiple particle representation. Only the configurations which are changed by collisions have been shown along with their final state.

| State # | $(\alpha_0, \alpha_1, \alpha_2, \alpha_3)$ for Initial State $\mapsto$ Final State | State # | $(\alpha_0, \alpha_1, \alpha_2, \alpha_3)$ for Initial State $\mapsto$ Final State |
|---|---|---|---|
| 0 | (0,0,0,0) $\mapsto$ (0,0,0,0) | 8 | (1,0,0,0) $\mapsto$ (1,0,0,0) |
| 1 | (0,0,0,1) $\mapsto$ (0,0,0,1) | 9 | (1,0,0,1) $\mapsto$ (1,0,0,1) |
| 2 | (0,0,1,0) $\mapsto$ (0,0,1,0) | **10** | (1,0,1,0) $\mapsto$ (0,1,0,1) |
| 3 | (0,0,1,1) $\mapsto$ (0,0,1,1) | 11 | (1,0,1,1) $\mapsto$ (1,0,1,1) |
| 4 | (0,1,0,0) $\mapsto$ (0,1,0,0) | 12 | (1,1,0,0) $\mapsto$ (1,1,0,0) |
| **5** | (0,1,0,1) $\mapsto$ (1,0,1,0) | 13 | (1,1,0,1) $\mapsto$ (1,1,0,1) |
| 6 | (0,1,1,0) $\mapsto$ (0,1,1,0) | 14 | (1,1,1,0) $\mapsto$ (1,1,1,0) |
| 7 | (0,1,1,1) $\mapsto$ (0,1,1,1) | 15 | (1,1,1,1) $\mapsto$ (1,1,1,1) |

**Table 2.7:** Collision rule table of the HPP gas in the multiple particle representation. The value of $\alpha_i$ denotes presence (1) or absence (0) of particle with velocity $(\cos(i\pi/2), \sin(i\pi/2))$, $i = 0, \ldots, 3$, on the lattice site. Initial states with collisions have been numbered in bold. Note that collisions change the initial states 5 and 10 into each other.

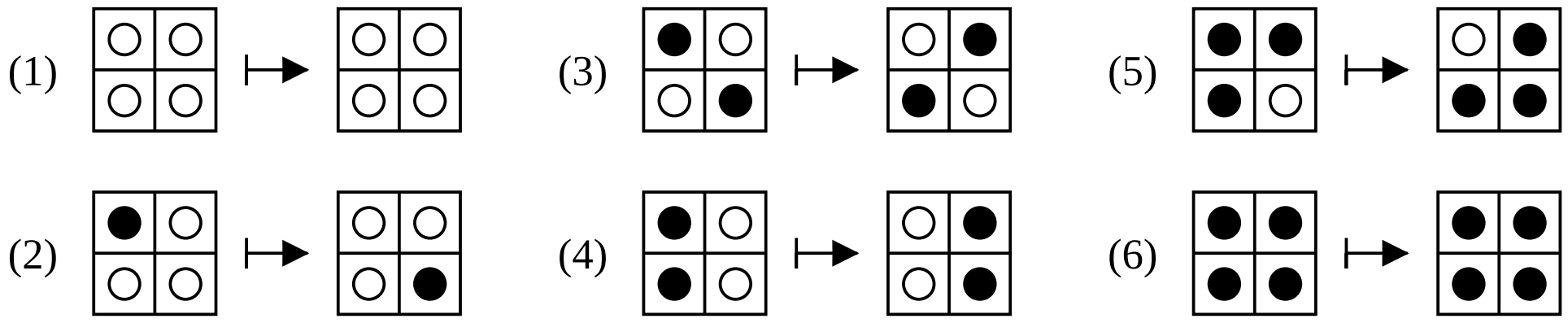

**Figure 2.10:** Evolution rules of the HPP gas for both "even" and "odd" partitions in the partitioned spatial lattice representation. Hollow circles represent space or empty cells and solid circles represent particles.

Since at most four particles with different velocities can occupy a lattice site, each lattice site can have any one of 16 different possible states during the collision and/or translation step at any time step. The state of a lattice site can be uniquely specified using a 4-tuple $(\alpha_0, \alpha_1, \alpha_2, \alpha_3)$, where the value of $\alpha_i$ denotes the presence (1) or absence (0) of particle having velocity $\boldsymbol{v}_i$, $i = 0, \ldots, 3$, on the lattice site. The elements of this 4-tuple can be interpreted as bits giving a sequence number to the state in binary notation. Using this convention, collision rules of the HPP gas for all the initial states are outlined in table 2.7.

**Partitioned Spatial Lattice Representation:** Partitioning of the spatial lattice and partition selection scheme employed in the HPP gas in this representation is as described in Sec. 2.2.6.1. All the particles are of equal mass and indistinguishable. The evolution rules for both "even" and "odd" partitions are identical and have been shown in Fig. 2.10.

Simulation of simple systems, *e.g.*, those shown in Fig. 2.11, using these evolution rules shows that particles move diagonally with velocity vectors $\{(\pm\sqrt{2}, \pm\sqrt{2})\}$ over the spatial lattice instead of horizontally and vertically with velocity vectors $\{(0, \pm 1), (\pm 1, 0)\}$ as in the multiple particle representation. Because of this it appears that the dynamics of the HPP gas in both these representations might differ. This, however, is not so. The evolution rules of the HPP gas in both the multiple particle representation and the partitioned spatial lattice representation are equivalent and can be easily transformed into each other. The difference in the velocity vectors and trajectory of particles in these two representations is because of difference in the meaning of lattice site values in both the cases. This difference comes in during transformation of one representation into the other.

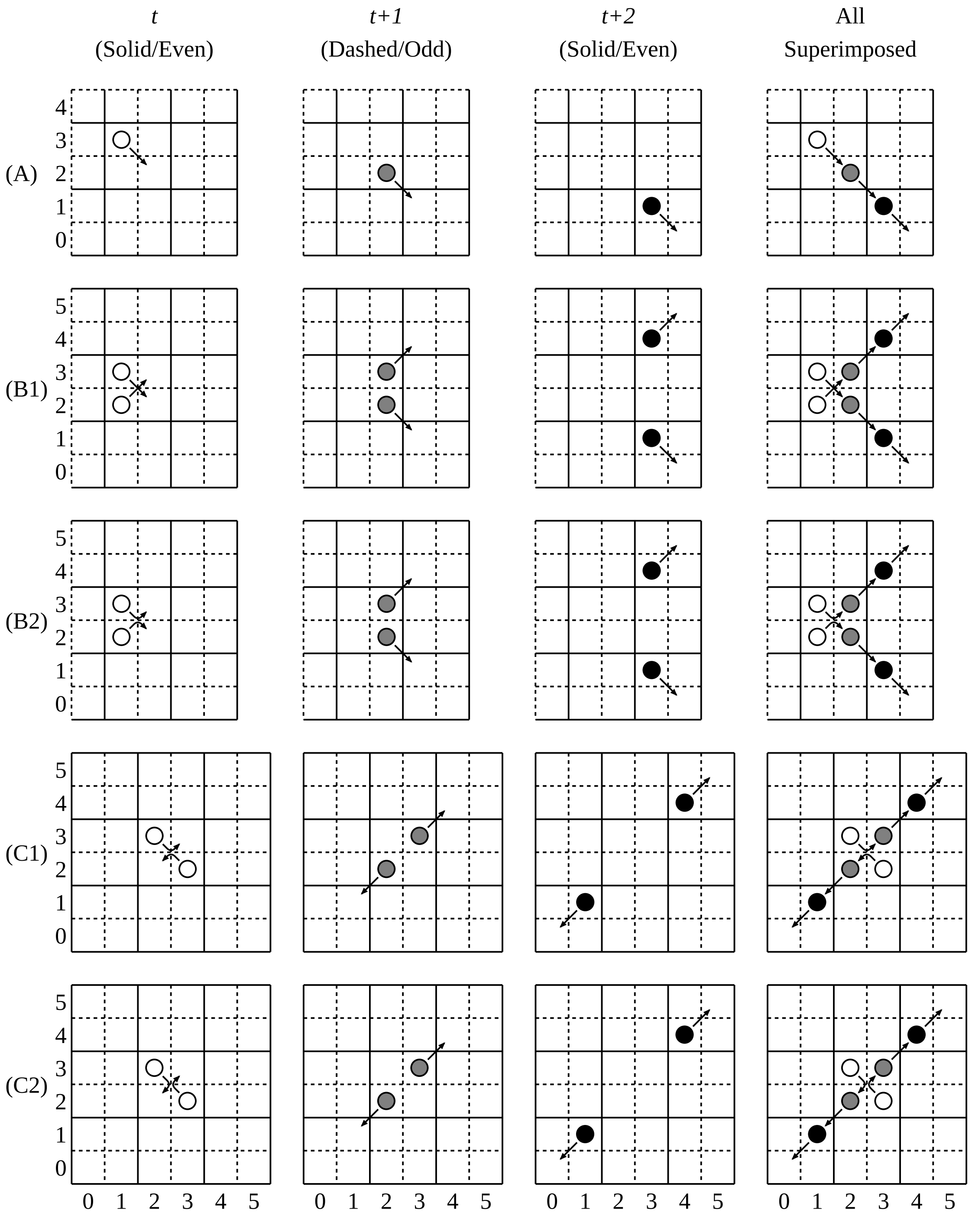

**Figure 2.11:** Evolution of some simple configurations over square spatial lattice in the HPP gas for three consecutive time steps ($t$, $t+1$, and $t+2$). At time step $t$, "solid" (or "even") partition is used for evolving the system. Hollow, dotted, and solid circles inside cells represent particles at time steps $t$, $t+1$, and $t+2$, respectively. All particles are indistinguishable; different shades have been used for distinguishing time steps in the superimposed picture. Other (unmarked) cells are empty. Arrows show the path taken by the particles. For enhancing clarity, the last column shows superimposition of configurations at all the three time steps.

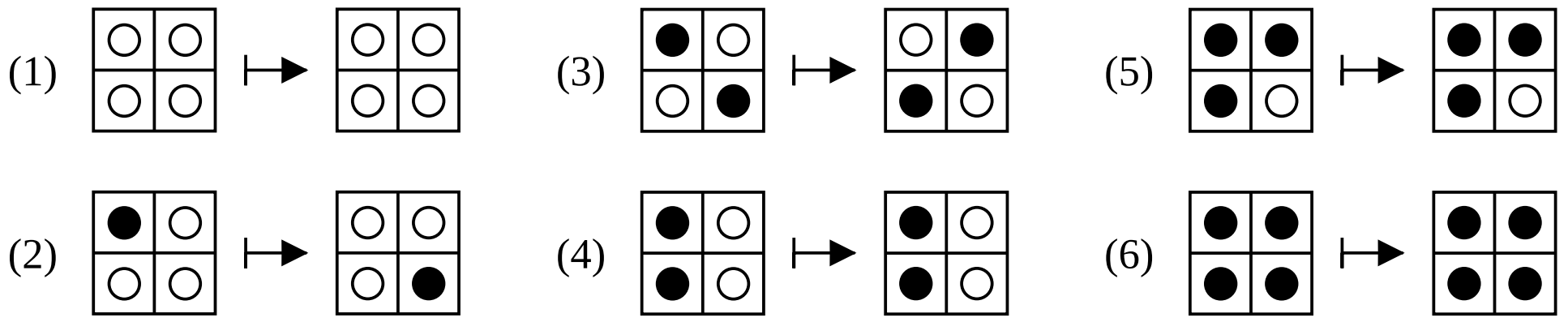

**Figure 2.12:** Evolution rules of the BBMCA model for both "even" and "odd" partitions in the partitioned spatial lattice representation. Hollow circles represent space or empty cells and solid circles represent particles.

Note that the configurations (B1) and (B2) shown in Fig. 2.11 are identical at all time steps. From the overall dynamics of particle shown in these configurations, two different trajectories of particles, as marked in the configurations at time step $t$, can be deduced. These trajectories arise and imply the presence of two different underlying dynamical processes. The trajectory shown in (B1) implies that particles move to new lattice sites without collision, whereas that shown in (B2) implies that particles collide while moving to new lattice sites. The final configuration at time step $t+1$, however, is identical in both these cases. As a result, to keep the evolution rules (and the dynamics encoded in them) in conformity with the original definition given by HPP [16,17], the trajectory shown in (B1) is taken to be correct. Similarly, two trajectories are possible in the collision configurations also. An example is shown in Fig. 2.11 through configurations (C1) and (C2). Both these configurations are identical at all time steps. Still, two different trajectories, as marked at time step $t$, are possible in them. In this case, however, both the trajectories are taken to be correct. This is because both of these trajectories are equally likely and the dynamics implied by both of them conforms with the original definition given by HPP.

Both the cases mentioned above arise because the particles are indistinguishable and hence it cannot be said which particle went where. Since the trajectories of the particles have to be deduced only from their initial and final positions, one solution in the configurations (B1) and (B2) and two solutions in the configurations (C1) and (C2), which conform with the original definition of the evolution rules given by HPP, are possible. As far as dynamics of the system is concerned, it does not matter which of these trajectories are correct because the initial and final velocities of particle in both of them are identical. If fact, it cannot be deduced (without coloring the particles differently) which one of the two trajectories shown in the configurations (C1) and (C2) is correct.

#### 2.2.6.3 The BBMCA Model

In this section, structure of the BBMCA model is described in the partitioned spatial lattice representation [9,24].

**Partitioned Spatial Lattice Representation:** Partitioning of the spatial lattice and partition selection scheme employed in the BBMCA model in this representation is as described in Sec. 2.2.6.1. All the particles are of equal mass and indistinguishable. The evolution rules for both "even" and "odd" partitions are identical and have been shown in Fig. 2.12.

Similarity between the evolution rules of the BBMCA model and the HPP gas is notable. The evolution rules of the two models are different only by the final states that have been assigned to the initial states (4) and (5) in them. This difference, however, makes the dynamical behavior of the two models completely different. One aspect of this difference is that in the BBMCA model stationary particles, in addition to the particles moving with four velocities as in the HPP gas, are also permitted. Another difference is that particle velocities go to zero for one time step during collisions and then the particles retrace their old trajectory with reversed velocities in the BBMCA model (*i.e.*, collisions take finite time and violate momentum conservation), whereas in the HPP gas the collisions are instantaneous and momentum conserving and the particles continue their motion on a new trajectory. Dynamical behavior of the BBMCA model differs from that of the TM gas also. These differences have been explained in details in [9,24] and thus will not be reproduced here.

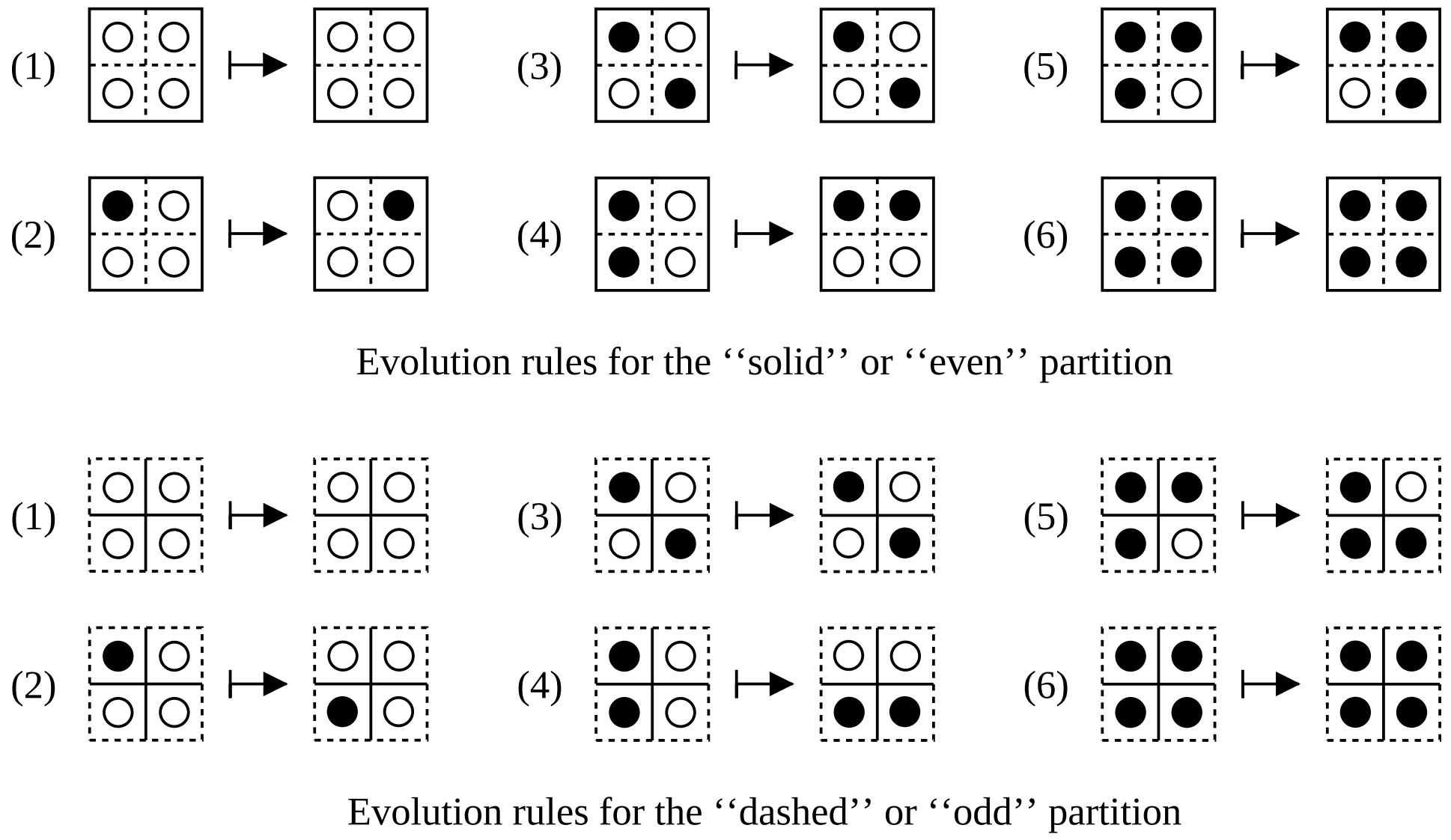

**Figure 2.13:** Evolution rules of the TM gas for both "even" and "odd" partitions in the partitioned spatial lattice representation. Hollow circles represent space or empty cells and solid circles represent particles.

#### 2.2.6.4 The TM Gas

In this section, structure of the TM gas is described in the partitioned spatial lattice representation [24].

**Partitioned Spatial Lattice Representation:** Partitioning of the spatial lattice and partition selection scheme employed in the TM gas in this representation is as described in Sec. 2.2.6.1. All the particles are of equal mass and indistinguishable. Evolution rules for both the "even" and "odd" partitions are different and have been illustrated in Fig. 2.13.

Simulation of simple systems similar to those shown in Fig. 2.11 through few time steps using these evolution rules shows that in the TM gas particles move horizontally and vertically with velocity vectors $\{(\pm 1, 0), (0, \pm 1)\}$ over the spatial lattice instead of diagonally with velocity vectors $\{(\pm\sqrt{2}, \pm\sqrt{2})\}$ as in the HPP gas in the partitioned spatial lattice representation. This difference arises because the evolution rules of TM gas, as can be seen from Fig. 2.13, for the "even" ("odd") partition are obtained by rotating all the initial configurations by $\pi/2$ radians clockwise (counter clockwise) except for the configuration (3) which is left unchanged, whereas the evolution rules of the HPP gas are obtained by rotating all the initial configurations by $\pi$ radians clockwise ($\equiv$ counter clockwise $\equiv$ swapping on the diagonal) except for the configuration (3) which is rotated only by $\pi/2$ radians clockwise ($\equiv$ counter clockwise since the particles are indistinguishable).

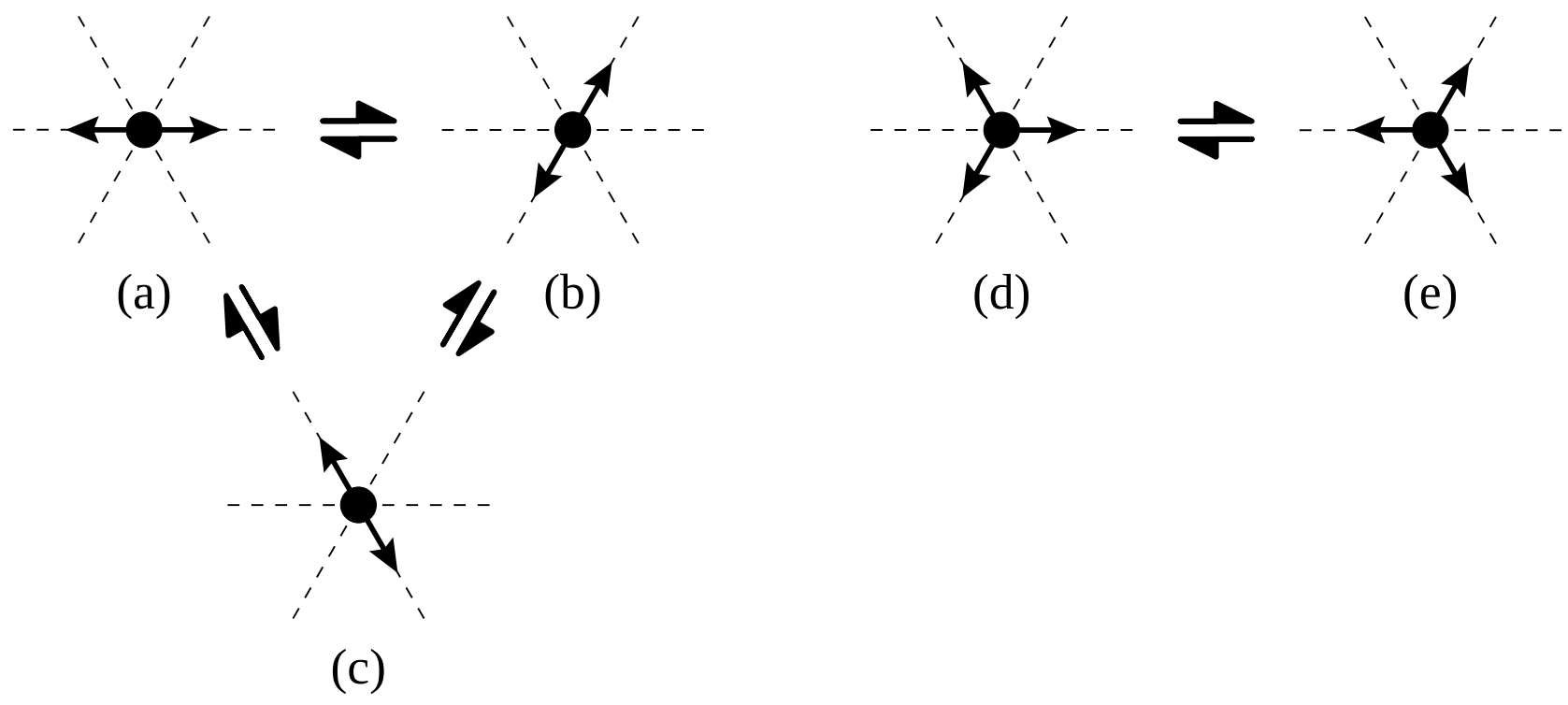

**Figure 2.14:** Collision rules of the FHP gas in multiple particle representation. Only the configurations which are changed by collisions have been shown along with their final states.

An important consequence of the above mentioned difference in the evolution rules of the HPP and TM gases in the partitioned spatial lattice representation is that in this representation the impact parameter is always zero in the HPP gas whereas it is non-zero in the TM gas. This results in different behavior of the long time tail of the *time autocorrelation function for velocity at lattice sites* $\nu(t)$ for both the gases [26,24] in the partitioned spatial lattice representation. The asymptotic exponent of $\nu(t)$ is $-2/3$ for the HPP gas and $-1$ for the TM gas. These values of the exponent imply that that the dynamical behavior of the HPP gas is essentially equivalent to that of one-dimensional gases (even though, geometrically it appears to be two-dimensional gas) whereas the behavior of the TM gas is equivalent to that of truly two-dimensional gases [127,128].

#### 2.2.6.5 The FHP Gas

In this section, structure of the FHP gas is described in the multiple particle representation [27–29].

**Multiple Particle Representation:** The FHP gas exists over triangular spatial lattice. The particles reside on the vertices of the lattice. The particles are indistinguishable and have unit mass. They move with one of the six velocity vectors belonging to the discrete velocity set $\mathcal{V}_{\mathrm{FHP}} \equiv \{\boldsymbol{v}_i : \boldsymbol{v}_i = (\cos(i\pi/3), \sin(i\pi/3); i \in [0,5]\} \equiv \{(\pm 1, 0), (\pm 1/2, \pm\sqrt{3}/2)\}$ along the links of the lattice. At the most six particles can occupy a lattice site (or, vertex) simultaneously at any time step subject to the exclusion principle that no two particles occupying the same lattice site can have the same velocity.

Evolution of the system during one time step is decomposed into two sub steps, namely, (i) translation and (ii) collision. During translation step, particles move to neighboring lattice sites as indicated by their velocity vectors; no collisions occur. During collision step, collisions occur among particles occupying the same lattice site; the particles do not move. Collisions occur only on those lattice sites which are occupied by either exactly two particles with velocity vectors pointing in opposite directions or exactly three particles with velocity vectors rotated by $2\pi/3$ radians relative to each other.[16)] Thus, as shown in Fig. 2.14, three binary collision configurations and two triple collision configurations are possible. Collisions alter the velocity of particles in these configurations. All other configurations remain unchanged. In binary as well as triple collisions, the velocity vectors of colliding particles rotate by $\pi/3$ radians either clockwise or counter-clockwise (same direction for all the colliding particles). As a result, as shown in Fig. 2.14, two final states are possible for each binary collision configuration and one final state is possible for each triple collision configuration. In binary collision configurations, both final states are allowed by assigning (usually equal) *ad-hoc* probability of occurrence to them.

Since at most six particles with different velocities can occupy a lattice site, each lattice site can have any one of 64 different possible states during the collision and/or translation step at any time step. The state of a lattice site can be uniquely specified using a 6-tuple $(\alpha_0, \alpha_1, \alpha_2, \alpha_3, \alpha_4, \alpha_5)$, where the value of $\alpha_i$ denotes presence (1) or absence (0) of particle having velocity $\boldsymbol{v}_i$, $i = 0, \ldots, 5$, on the lattice site. The elements of this 6-tuple can be interpreted as bits giving a sequence number to the state in binary notation. With this convention, collision rules of the FHP gas for all the 64 initial states are given in table 2.8.

### 2.2.7 Offshoots of Lattice Gases

In the entire development of lattice gases outlined in Sec. 2.2.3 many factors have remained common among almost all the lattice gases despite the diversity exhibited by phenomena for which they have been developed.[17)] The most important of these common factors is the philosophy that lies at the core of these lattice gases and projects them as cellular automata models of physical systems. This philosophy, in essence, consists of physical interpretations that have been assigned to the basic elements of cellular automata for establishing correspondence between the resulting models (*i.e.*, lattice gases) and physical systems (*c.f.*, Sec. 1.3.4.2 and 2.1.4). These interpretations are outlined in table 2.6. This underlying philosophy originates from the HPP gas (*c.f.*, Sec. 2.2.5.2). It has remained unaltered throughout the development of lattice gases because all the lattice gases find their roots, either directly or indirectly, in the HPP gas; with the FHP gas itself being rooted at the HPP gas.

Because of the common underlying philosophy, many similar problems, *e.g.*, breaking of Galilean invariance and incorrect behavior in the compressible flow regime, are encountered during simulation of physical systems with almost all the existing lattice gases.[18)]

---

[16)] In fact, in this lattice gas collisions can be defined to occur among two, three, as well as four particles [28,29]. Three particles collisions can be symmetric as well as asymmetric (a two particle collision in presence of another particle). Four particle collisions are always asymmetric. Thus, it is possible to choose the type of collisions (or, collision rules) in a model. This choice allows development of different versions of the FHP gas. These versions differ only in parameters like viscosity and presence or absence of spurious conservation laws. The overall form of the course grained hydrodynamic equations in all of them remains same (*i.e.*, unaffected by the collision rules). The version of FHP gas with only two particle collisions has spurious conservation laws [28,29]. The version of FHP gas with two particle and symmetric three particle collisions, as defined above, has no spurious conservation laws. This version, through termed FHP gas in this manuscript, is referred to as FHP-1 gas in the literature [29]. Details can be obtained from [28,29].

[17)] Here, the quantum lattice gases, being outside the scope of present study, have not been accounted for.

[18)] The cause and effect relationship pointed out in this statement is subject to (and also, is consequence of) argumentation which follows in Sec. 2.5 and chapter 3. It differs strongly from that explained in the literature. In the literature it is thought that discreteness of the spatial lattice is the cause of problems encountered in simulation of physical systems using lattice gases. This explanation, however, has been

| State # | $(\alpha_0,\alpha_1,\alpha_2,\alpha_3,\alpha_4,\alpha_5)$ for Initial State $\mapsto$ Final State | State # | $(\alpha_0,\alpha_1,\alpha_2,\alpha_3,\alpha_4,\alpha_5)$ for Initial State $\mapsto$ Final State |
|---|---|---|---|
| 0 | (0,0,0,0,0,0) $\mapsto$ (0,0,0,0,0,0) | 32 | (1,0,0,0,0,0) $\mapsto$ (1,0,0,0,0,0) |
| 1 | (0,0,0,0,0,1) $\mapsto$ (0,0,0,0,0,1) | 33 | (1,0,0,0,0,1) $\mapsto$ (1,0,0,0,0,1) |
| 2 | (0,0,0,0,1,0) $\mapsto$ (0,0,0,0,1,0) | 34 | (1,0,0,0,1,0) $\mapsto$ (1,0,0,0,1,0) |
| 3 | (0,0,0,0,1,1) $\mapsto$ (0,0,0,0,1,1) | 35 | (1,0,0,0,1,1) $\mapsto$ (1,0,0,0,1,1) |
| 4 | (0,0,0,1,0,0) $\mapsto$ (0,0,0,1,0,0) | **36** | (1,0,0,1,0,0) $\mapsto$ (0,0,1,0,0,1) |
| 5 | (0,0,0,1,0,1) $\mapsto$ (0,0,0,1,0,1) | | or (0,1,0,0,1,0) |
| 6 | (0,0,0,1,1,0) $\mapsto$ (0,0,0,1,1,0) | 37 | (1,0,0,1,0,1) $\mapsto$ (1,0,0,1,0,1) |
| 7 | (0,0,0,1,1,1) $\mapsto$ (0,0,0,1,1,1) | 38 | (1,0,0,1,1,0) $\mapsto$ (1,0,0,1,1,0) |
| 8 | (0,0,1,0,0,0) $\mapsto$ (0,0,1,0,0,0) | 39 | (1,0,0,1,1,1) $\mapsto$ (1,0,0,1,1,1) |
| **9** | (0,0,1,0,0,1) $\mapsto$ (0,1,0,0,1,0) | 40 | (1,0,1,0,0,0) $\mapsto$ (1,0,1,0,0,0) |
| | or (1,0,0,1,0,0) | 41 | (1,0,1,0,0,1) $\mapsto$ (1,0,1,0,0,1) |
| 10 | (0,0,1,0,1,0) $\mapsto$ (0,0,1,0,1,0) | **42** | (1,0,1,0,1,0) $\mapsto$ (0,1,0,1,0,1) |
| 11 | (0,0,1,0,1,1) $\mapsto$ (0,0,1,0,1,1) | 43 | (1,0,1,0,1,1) $\mapsto$ (1,0,1,0,1,1) |
| 12 | (0,0,1,1,0,0) $\mapsto$ (0,0,1,1,0,0) | 44 | (1,0,1,1,0,0) $\mapsto$ (1,0,1,1,0,0) |
| 13 | (0,0,1,1,0,1) $\mapsto$ (0,0,1,1,0,1) | 45 | (1,0,1,1,0,1) $\mapsto$ (1,0,1,1,0,1) |
| 14 | (0,0,1,1,1,0) $\mapsto$ (0,0,1,1,1,0) | 46 | (1,0,1,1,1,0) $\mapsto$ (1,0,1,1,1,0) |
| 15 | (0,0,1,1,1,1) $\mapsto$ (0,0,1,1,1,1) | 47 | (1,0,1,1,1,1) $\mapsto$ (1,0,1,1,1,1) |
| 16 | (0,1,0,0,0,0) $\mapsto$ (0,1,0,0,0,0) | 48 | (1,1,0,0,0,0) $\mapsto$ (1,1,0,0,0,0) |
| 17 | (0,1,0,0,0,1) $\mapsto$ (0,1,0,0,0,1) | 49 | (1,1,0,0,0,1) $\mapsto$ (1,1,0,0,0,1) |
| **18** | (0,1,0,0,1,0) $\mapsto$ (1,0,0,1,0,0) | 50 | (1,1,0,0,1,0) $\mapsto$ (1,1,0,0,1,0) |
| | or (0,0,1,0,0,1) | 51 | (1,1,0,0,1,1) $\mapsto$ (1,1,0,0,1,1) |
| 19 | (0,1,0,0,1,1) $\mapsto$ (0,1,0,0,1,1) | 52 | (1,1,0,1,0,0) $\mapsto$ (1,1,0,1,0,0) |
| 20 | (0,1,0,1,0,0) $\mapsto$ (0,1,0,1,0,0) | 53 | (1,1,0,1,0,1) $\mapsto$ (1,1,0,1,0,1) |
| **21** | (0,1,0,1,0,1) $\mapsto$ (1,0,1,0,1,0) | 54 | (1,1,0,1,1,0) $\mapsto$ (1,1,0,1,1,0) |
| 22 | (0,1,0,1,1,0) $\mapsto$ (0,1,0,1,1,0) | 55 | (1,1,0,1,1,1) $\mapsto$ (1,1,0,1,1,1) |
| 23 | (0,1,0,1,1,1) $\mapsto$ (0,1,0,1,1,1) | 56 | (1,1,1,0,0,0) $\mapsto$ (1,1,1,0,0,0) |
| 24 | (0,1,1,0,0,0) $\mapsto$ (0,1,1,0,0,0) | 57 | (1,1,1,0,0,1) $\mapsto$ (1,1,1,0,0,1) |
| 25 | (0,1,1,0,0,1) $\mapsto$ (0,1,1,0,0,1) | 58 | (1,1,1,0,1,0) $\mapsto$ (1,1,1,0,1,0) |
| 26 | (0,1,1,0,1,0) $\mapsto$ (0,1,1,0,1,0) | 59 | (1,1,1,0,1,1) $\mapsto$ (1,1,1,0,1,1) |
| 27 | (0,1,1,0,1,1) $\mapsto$ (0,1,1,0,1,1) | 60 | (1,1,1,1,0,0) $\mapsto$ (1,1,1,1,0,0) |
| 28 | (0,1,1,1,0,0) $\mapsto$ (0,1,1,1,0,0) | 61 | (1,1,1,1,0,1) $\mapsto$ (1,1,1,1,0,1) |
| 29 | (0,1,1,1,0,1) $\mapsto$ (0,1,1,1,0,1) | 62 | (1,1,1,1,1,0) $\mapsto$ (1,1,1,1,1,0) |
| 30 | (0,1,1,1,1,0) $\mapsto$ (0,1,1,1,1,0) | 63 | (1,1,1,1,1,1) $\mapsto$ (1,1,1,1,1,1) |
| 31 | (0,1,1,1,1,1) $\mapsto$ (0,1,1,1,1,1) | | |

**Table 2.8:** Collision rule table of the FHP gas in the multiple particle representation. The value of $\alpha_i$ denotes presence (1) or absence (0) of particle with velocity $(\cos(i\pi/3), \sin(i\pi/3)$, $i = 0, \ldots, 5$, on the lattice site. Initial states with collisions have been numbered in bold. Note that collisions change the initial states 9, 18, and 36 into one another and the states 21 and 42 into each other.

---

argued to be doubtful in Sec. 2.5 and is not quite acceptable any more. The problems have been critically analyzed in chapter 3 and it has been found that they arise because of some physical inconsistency (to be outlined in chapter 3) in the structure of these lattice gases. This physical inconsistency originates from the philosophy that is embedded into these lattice gases; or, more specifically, from the interpretation assigned to the lattice sites values in these lattice gases.

It is noteworthy that the above mentioned finding is the primary point of departure from literature in this investigation. This departure, in fact, is so radical that all that follows from chapter 3 onwards has no direct or indirect relationship with the existing development of lattice gases outlined in Sec. 2.2. The consequence of this departure is a method of overcoming the problems which requires changes in the very philosophy that lies at the core of existing lattice gases. This change leads to totally new type of lattice gases and new method of constructing them. The structure and evolution rules of these new lattice gases

Although some solutions to these problems have been suggested in the literature, the problems still persist. As a result, in attempts to bypass these problems, two important offshoots of lattice gases have developed. These are known as the lattice Boltzmann models [129–157] and the lattice Bhatanager-Gross-Krook models [158–163]. These models are very similar to lattice gases and their dynamics is also analyzed in almost identical fashion as that of the lattice gases. These models, however, will not be discussed further in this investigation because according to strict classification they are not lattice gases.

## 2.3 Generalization of Existing Lattice Gases: Multiparticle Lattice Gases

The existing lattice gases, owing to common underlying philosophy, can be generalized and looked upon as members of a class containing infinitely many lattice gases. All the members of this class can be described using either one of the conceptual representations outlined in Sec. 2.2.5. In the following, however, the structure of a generalized member of this class (and hence, the class) will be described using the multiple particle representation. As a result, this class will, henceforth, be referred to as the *class of multiple particle (or, multiparticle) lattice gases* and its members will be referred to as *multiparticle lattice gases*. The reason of preferring multiple particle representation over partitioned spatial lattice representation for the description, despite their equivalence (*c.f.*, Sec. 2.2.5), is that the partitioned spatial lattice representation has certain unacceptable peculiarities (explained and analyzed in chapter 3) which must be bypassed to avoid ambiguities.

Details on multiparticle lattice gases are as follows:

**Definition:** Multiparticle lattice gases exist over an underlying $\mathcal{D}$-dimensional spatial lattice[19)] and operate in discrete time steps $\tau$. In a multiparticle lattice gas particle of $\mathcal{N}_\mathrm{p}$ different species exist over the spatial lattice. Particles of species $i$, $i = 1, \ldots, \mathcal{N}_\mathrm{p}$, move with the discrete velocities, $\boldsymbol{v}_j^{(i)}$, $j = 1, \ldots, \mathcal{N}_\mathrm{v}^{(i)}$, over the spatial lattice. The velocities for particles of species $i$ form a finite discrete velocity set $\mathcal{V}^{(i)} = \left\{\boldsymbol{v}_j^{(i)} : j = 1, \ldots, \mathcal{N}_\mathrm{v}^{(i)}\right\}$ containing $\mathcal{N}_\mathrm{v}^{(i)}$ velocity vectors. At any time step, $\mathcal{N}_\mathrm{m}^{(i,j)}$ particles of species $i$ moving with velocity $\boldsymbol{v}_j^{(i)}$, $j = 1, \ldots, \mathcal{N}_\mathrm{v}^{(i)}$, can occupy a lattice site simultaneously.[20)] The maximum number of particles $\mathcal{N}_\mathrm{t}$ that can occupy a lattice site at any time step simultaneously and the total number of symbols $\mathcal{N}_\mathrm{s}$ that are required for uniquely representing all possible states of the lattice sites are as given by Eqs. (2.7) and (2.8).

**Evolution Strategy:** Dynamics of particles over the spatial lattice is computed by decomposing the evolution of the system during one time step into two sequential sub-steps, namely: (i) *translation step*, and (ii) *collision step*. Evolution rules are developed separately for each of these sub-steps and used in a fixed sequence during simulations. The are radically different from those of the existing lattice gases. All the chapters, starting from chapter 4 are devoted to these developments. It is noteworthy that these new lattice gases are the kind of cellular automata models that are being sought in this investigation (*c.f.*, Sec. 1.4).

19) Usually, structure of the spatial lattice is regular. In general, however, it may be irregular. If the structure is irregular, some additional constraints on the velocity of particles at various lattice sites also come up. These are logically deducible from the structure and will not be elaborated here.

20) In early developments of multiparticle lattice gases, $\mathcal{N}_\mathrm{m}^{(i,j)}$ was constrained to be unity, *i.e.*, at most one (either one or zero) particle of species $i$ moving with velocity $\boldsymbol{v}_j^{(i)}$ was allowed to occupy a lattice site at any time step. This condition was called as exclusion principle based on velocity of particles. This condition, however, is not necessary and can be relaxed (*c.f.*, Sec. 2.2.5.2).

sequence is fixed in that it remains same at all time steps during a simulation. The sequence in which the evolution rules are used is same as the sequence of sub-steps in which the dynamics of the system during an evolution (or, one time step) is decomposed. The order of decomposition, *i.e.*, whether evolution during one time step is decomposed into collision step followed by translation step or vice versa, is a matter of free choice. This choice does not alter the overall dynamics (or, macrodynamics), although the microdynamics is different in both cases.

The evolution rules for the translation and collision sub-steps (henceforth referred to as the translation rules and collision rules) are developed in such a way that all the conserved quantities, *e.g.*, mass, momentum, and energy, remain conserved during each sub-step. This guarantees that all the required conservation laws hold during each time step and hence during the complete time evolution of the system also.

**Translation Step:** During the translation step particles move to the lattice sites indicated by their velocity vectors. The motion is completed in time $(\tau - \delta\tau)$ in the limit $\delta\tau \to 0$. Since translation rules simply reposition the particles on new lattice sites, all the laws governing the dynamics of the system and all the conserved quantities remain unchanged during the translation step. That no change occurs due to generation or annihilation of particles during repositioning, is guaranteed by the fact that evolution of the system is synchronous (*i.e.*, all the lattice sites evolve simultaneously) and all the lattice sites are identical in all possible respects (*i.e.*, the maximum number of particles of each species having the same velocity that can occupy a lattice site simultaneously is identical for all the lattice sites).

**Collision Step:** During the collision step various effects of interparticle interactions (or, collisions) are computed.[21)] Since more than one particle can occupy the same lattice site simultaneously at any time step, *collisions occur among particles occupying the same lattice site*. As a result, interparticle interactions are computed only among particle occupying the same lattice site. The computation is carried out by evolving each lattice site according to predefined *collision rules*. Henceforth, the process of computing interparticle interactions will be called as *collision resolution* process. This process is completed in time $\delta\tau$ in the limit $\delta\tau \to 0$. Usually it is assumed that collisions are instantaneous, *i.e.*, $\delta\tau = 0$.

During collision step, the initial state of a lattice site is its state at the end of the previous translation step, and the final state of a lattice site is its state after it (or, the entire system) has been evolved through the collision rules. Since the velocities and species of particles occupying a lattice site can change during the collision step, the laws governing the dynamics of the system are not (and, cannot be) satisfied automatically as in the translation step. Instead, the collision rules need to be defined explicitly in such a manner that all the laws governing the dynamics of the system remain satisfied during the collision step.

**Anatomy of Collision Rules:** Collision rules of multiparticle lattice gases are symbolic statements representing the dynamics of *collisions which occur among particles occupying the same lattice site*. Since at the beginning of collision step a lattice site can have any one of $\mathcal{N}_\mathrm{s}$ possible states, collision rules must specify the final states of lattice sites for all possible initial states.

Let the possible states of lattice sites be numbered sequentially from 0 to $\mathcal{N}_\mathrm{s} - 1$. Let $\mathcal{S}_k$ be the $k^{\mathrm{th}}$ state in this sequence. Let $\mathcal{S}_i$ be an initial state of the lattice sites and $\mathcal{S}_{i,f}$,

[21)] This includes chemical reactions also.

| Initial States | | Final States $\mathcal{S}_{i,f}$ for $f =$ | | | | | | |
|---|---|---|---|---|---|---|---|---|
| $i$ | $\mathcal{S}_i$ | 0 | 1 | 2 | ... | $f_k$ | ... | $f_{\max}-1$ |
| 0 | $\mathcal{S}_0$ | $\mathcal{S}_{0,0}$ | — | — | ... | — | ... | — |
| 1 | $\mathcal{S}_1$ | $\mathcal{S}_{1,0}$ | $\mathcal{S}_{1,1}$ | — | ... | — | ... | — |
| $\vdots$ | $\vdots$ | $\vdots$ | $\vdots$ | $\vdots$ | | $\vdots$ | | $\vdots$ |
| $k$ | $\mathcal{S}_k$ | $\mathcal{S}_{k,0}$ | $\mathcal{S}_{k,1}$ | $\mathcal{S}_{k,2}$ | ... | $\mathcal{S}_{k,f_k}$ | ... | — |
| $\vdots$ | $\vdots$ | $\vdots$ | $\vdots$ | $\vdots$ | | $\vdots$ | | $\vdots$ |
| $n$ | $\mathcal{S}_n$ | $\mathcal{S}_{n,0}$ | $\mathcal{S}_{n,1}$ | $\mathcal{S}_{n,2}$ | ... | $\mathcal{S}_{n,f_k}$ | ... | $\mathcal{S}_{n,f_n-1}$ |
| $\vdots$ | $\vdots$ | $\vdots$ | $\vdots$ | $\vdots$ | | $\vdots$ | | $\vdots$ |
| $\mathcal{N}_s-1$ | $\mathcal{S}_{\mathcal{N}_s-1}$ | $\mathcal{S}_{\mathcal{N}_s-1,0}$ | — | — | ... | — | ... | — |
| 0 | 1 | 2 | 3 | 4 | ... | $f_k+2$ | ... | $f_{\max}+1$ |

Column Numbers

**Table 2.9:** Schematic representation of the collision rule table of an arbitrary multiparticle lattice gas. Note that $f_i$ is in general different for different values of $i$ and in this table $f_{\max} = f_n$. The column numbers have been shown explicitly at the foot of the table.

$f = 0, \ldots, f_i-1$, be the corresponding final states, where $f_i$ is the number of final states for the $i^{\text{th}}$ initial state.[22] Let $f_{\max} = \max[f_0, f_1, \ldots, f_{\mathcal{N}_s-1}]$. Then, in the simplest approach, the collision rules can be written in the form of a table. This table can be visualized as a multicolumn listing with $\mathcal{N}_s$ rows and $f_{\max}+2$ columns, the columns being numbered from 0 to $f_{\max}+1$, of the initial states and the corresponding final states of lattice sites. In this table, the $i^{\text{th}}$ initial state appears in column 1 of row $i$ and the corresponding final states appear in columns $2, \ldots, f_i+1$ of the same row. The column 0 of the table contains the sequence number of the initial states.[23] A generalized schematic representation of the collision rule table has been depicted in table 2.9. More concrete examples are the collision rules of HPP and FHP gases depicted in tables 2.7 and 2.8.

Let $\mathcal{S} = \{\mathcal{S}_k : k = 0, \ldots, \mathcal{N}_s - 1\}$ be the set of all possible states of lattice sites. Then, $\mathcal{S}_i \in \mathcal{S}$ and $\mathcal{S}_{i,f} \in \mathcal{S}$ because the initial states $\mathcal{S}_i$ and the corresponding final states $\mathcal{S}_{i,f}$ are valid states of lattice sites. It is also worthwhile to observe that if a lattice site is unoccupied at the beginning of the collision step than it remains unoccupied at the end also. This example illustrates that the initial and final states of lattice sites can be the same. This, in general, is true for all possible initial states. As a result, if $\mathcal{S}^{(i)} = \{\mathcal{S}_{i,f} : f = 0, \ldots, f_i\}$ is the set of all possible final states for the $i^{\text{th}}$ initial state of lattice sites, $i = 0, \ldots, \mathcal{N}_s - 1$, then $\mathcal{S}_i \in \mathcal{S}^{(i)}$. This implies that $f_i \geq 1$, always. This ensures that each initial state necessarily has a final state, even though both may be identical. Usually, the final states for any given initial state are numbered in such a manner that $\mathcal{S}_{i,0} = \mathcal{S}_i$; this, however, is just a convention and can be changed. In collision rules of the HPP and FHP gases depicted in tables 2.7 and 2.8, the final states for the initial states with collisions[24] have been originally defined to be different from the initial states. This constraint, however, can be relaxed to obtain more general collision rules in which dynamics can be controlled probabilistically.

[22] In multiparticle lattice gases, in general, more than one final state is possible for any possible initial state of the lattice sites. As an example, see the collision rules of the FHP gas given in table 2.8.

[23] This is just for the sake of giving a conventional appearance to the table. If all the symbols are represented through numbers, as is done during actual simulations, columns 0 and 1 become identical and either one of them can be eliminated without the loss of generality.

[24] States 5 and 10 in HPP gas and states 9, 18, 21, 36, and 42 in the FHP gas are the states with collisions.

**Collision Resolution Process:** The process of resolving collisions in multiparticle lattice gases is essentially a symbolic substitution process. In this process, the symbol $\mathcal{S}_i$ existing at a lattice site denoting the current state (or, initial state) of the lattice site (or, particles occupying the lattice site) is replaced by a new symbol $\mathcal{S}_{i,f}$ denoting the new state of the lattice site after collisions. The replacement is done according to the collision rules.

By the definition of collision rules, every initial state of lattice sites has at least one final state. If there is only one final state for an initial state, the replacement is unambiguous and additional considerations are not needed. On the other hand, if there can be more than one final states for an initial state, then an appropriate method is needed for selecting one of these states as the actual final state of a lattice site. Usually the method is statistical wherein a random number is generated and one final state is selected depending upon its value. This requires that various $\mathcal{S}_i \mapsto \mathcal{S}_{i,f}$ state transition pairs be assigned probabilities (to be called as *"state transition probabilities"*). Let $p_{i,f}$ be the state transition probability for $\mathcal{S}_i \mapsto \mathcal{S}_{i,f}$ state transition. Then, the state transition probabilities of all the possible state transitions for each initial state must satisfy the relation

$$\sum_{f=0}^{f_i-1} p_{i,f} = 1 \qquad i = 1, \ldots, \mathcal{N}_\mathrm{s} - 1$$

This is because during the collision step a transition must occur for each initial state, *i.e.*, each initial state must have at least one final state.

Use of a statistical algorithm for selecting one final state out of many makes the evolution of the lattice gas irreversible. If required, reversibility can be restored by devising and using an appropriate non-statistical algorithm. A very simple example of such an algorithm is to substitute the initial state $\mathcal{S}_i$ by the final state $\mathcal{S}_{i,n}$ at time step $t$, where $n = t \bmod f_i$. In this algorithm, selection of the final state depends only on the time step in a very simple and cyclic manner. Algorithms in which selection of the final state depends upon the time step in more complicated (*e.g.*, pseudo-random) manner or depends upon the time step as well as lattice site coordinates can also be devised, if needed.

The collision resolution process described above is inherently massively parallel process because collisions are resolved separately at each lattice site. Consequently, in multiparticle lattice gases all the lattice sites can be processed simultaneously on an appropriate massively parallel computer or sequentially on a single processor computer with identical results. The particle translation step can also be processed in the same way. As a result, the evolution algorithms of multiparticle lattice gases are inherently massively parallel algorithms.

**Construction of Collision Rules:** For any multiparticle lattice gas, *i.e.*, for given particle species and discrete velocity sets for particles of each species, all the $\mathcal{N}_\mathrm{s}$ possible states of lattice sites can be determined and tabulated as described earlier. During the collision step only these states can be the initial and final state of lattice sites. As a result, construction of collision rules essentially reduces to determining which of these $\mathcal{N}_\mathrm{s}$ states can be used for replacing a given initial state. This can be done in the following way:

First of all the conserved quantities for all the states are determined. Let the conserved quantities be represented through a vector $\boldsymbol{Q}$ defined as

$$\boldsymbol{Q} = \left\{ \begin{array}{c} q_0 \\ q_1 \\ \vdots \\ q_{\mathcal{N}_\mathrm{q}-1} \end{array} \right\}$$

where $q_j$ is the $j^{\mathrm{th}}$ conserved quantity and $\mathcal{N}_\mathrm{q}$ is the total number of conserved quantities for the system.[25)] Let $\boldsymbol{Q}_i$ be the vector representing the value of conserved quantities for the $i^{\mathrm{th}}$ state $\mathcal{S}_i$, $i = 0, \ldots, \mathcal{N}_\mathrm{s} - 1$, of lattice sites. Then, determination of conserved quantities for all the possible states of lattice sites means that all $\boldsymbol{Q}_i$, $i = 0, \ldots, \mathcal{N}_\mathrm{s} - 1$, are known. Since the species and velocities of all the particles corresponding to each $\mathcal{S}_i$ are known, the $\boldsymbol{Q}_i$ can be determined easily. For example, the component of $\boldsymbol{Q}_i$ which represents total mass of particles can be determined by summing up the mass of all the particles as encoded in $\mathcal{S}_i$. The other components of $\boldsymbol{Q}_i$, *e.g.*, total kinetic energy of particles and various components of momentum, can also be determined in a similar way.

Once the $\boldsymbol{Q}_i$ for all the states have been determined, the states with equal $\boldsymbol{Q}_i$ are grouped together. All the states within a group will be called *equivalent* to each other[26)] in terms of the conserved quantities. Replacement of equivalent states by each other does not lead to violation of the conservation laws of the system. As a result, if there are no other constraints on the dynamics of the system, all the equivalent states can be the final states of each other during the collision step. This gives the collision rules for systems which are non-reacting or in which all the chemical reactions are reversible.

Dynamics of systems in which at least one irreversible chemical reaction occurs has an additional constraint which needs to be addressed separately while defining the collision rules. The constraint is that though the state representing the reactants $\mathcal{S}_\mathrm{R}$ can be replaced by the state representing the products $\mathcal{S}_\mathrm{P}$, the converse is not permissible. This is despite the fact that the states $\mathcal{S}_\mathrm{R}$ and $\mathcal{S}_\mathrm{P}$ are equivalent in terms of the conservation laws, *i.e.*, $\boldsymbol{Q}_\mathrm{R} = \boldsymbol{Q}_\mathrm{P}$. This constraint appears from the chemical kinetics of the system. In this case also $\mathcal{S}_\mathrm{R}$ and $\mathcal{S}_\mathrm{P}$ can be final states of themselves depending upon the rate of reaction.

## 2.4 Mathematical Analysis of Multiparticle Lattice Gases

In this section a generalized mathematical description of the dynamics of multiparticle lattice gases has been outlined. This description is possible because of the common philosophy which lies at the core of these lattice gases and permits their description as members of a highly generalized class as in Sec. 2.3. In this description the overall form of the dynamical equations, procedure of their derivation, and expressions for ensemble averaged quantities will be outlined. Description of the procedure of arriving at the overall dynamical equations, however, will be restricted only to the most general steps which are common for all multiparticle lattice gases. This is because the exact sequence of steps and the exact form of the dynamical equations depends strongly upon the finer details, *i.e.*, structure of

[25)] For systems in which the laws of conservation of mass, momentum, and energy hold, a conserved quantity $q_j$ could be either mass, or some component of momentum, or energy. If such a system exists in $\mathcal{D}$-dimensional physical position space, then it will have a total of $\mathcal{N}_\mathrm{q} = \mathcal{D} + 2$ conserved quantities namely, mass, $\mathcal{D}$ components of momentum, and energy.

[26)] Two states $\mathcal{S}_\alpha$ and $\mathcal{S}_\beta$ are called *equivalent* to each other if $\boldsymbol{Q}_\alpha = \boldsymbol{Q}_\beta$.

spatial lattice, particle species, velocity set for particles of each species, and the collision rules of the lattice gas. Details on finer aspects of mathematical analysis of multiparticle lattice gases can be obtained from the references (*e.g.*, [16,17,28,29,118,119,125]).

### 2.4.1 Description of the Lattice Gas

The following definitions and analysis are for a multiparticle lattice gas existing over $\mathcal{D}$-dimensional spatial lattice. The lattice gas comprises of particles of $\mathcal{N}_{\rm p}$ different species numbered sequentially from 1 to $\mathcal{N}_{\rm p}$. Particles of species $a$, $a = 1, \ldots, \mathcal{N}_{\rm p}$, move with the velocities $\boldsymbol{v}_b^{(a)}$, $b = 1, \ldots, \mathcal{N}_{\rm v}^{(a)}$, over the spatial lattice. At any time step maximum of $\mathcal{N}_{\rm m}^{(a,b)}$ particles of species $a$ having velocity $\boldsymbol{v}_b^{(a)}$ can occupy a lattice site simultaneously. The system evolves in discrete time steps of equal duration $\Delta t$. The mass of particles of species $a$ is $m^{(a)}$. Particles do not posses other physical properties, *e.g.*, charge, spin, *etc.*[27)]

For the multiparticle lattice gas described above, maximum number of particles $\mathcal{N}_{\rm t}$ that can occupy a lattice site simultaneously at any time step is given by

$$\mathcal{N}_{\rm t} = \sum_{a=1}^{\mathcal{N}_{\rm p}} \sum_{b=1}^{\mathcal{N}_{\rm v}^{(a)}} \mathcal{N}_{\rm m}^{(a,b)} \tag{2.9}$$

and total number of states of lattice sites $\mathcal{N}_{\rm s}$ and thus the total number of symbols required to represent these states is given by

$$\mathcal{N}_{\rm s} = \prod_{a,b=1,1}^{\mathcal{N}_{\rm p},\mathcal{N}_{\rm v}^{(a)}} \left(1 + \mathcal{N}_{\rm m}^{(a,b)}\right) \tag{2.10}$$

### 2.4.2 Conventions, Definitions and Assumptions

An exact analysis of the multiparticle lattice gas described in Sec. 2.4.1 can be carried out only in terms of the joint $N$-particle probability distribution function. Such an analysis, however, is neither feasible nor needed for studying macroscopic phenomena. Most of phenomena can be studied by analyzing the system using reduced distribution functions in which most degrees of freedom have been averaged out. In particular, for studying macroscopic phenomena it is sufficient to analyze the system in terms of one-particle distribution functions [164–168]. This is the approach that has been usually followed in the literature and will also be following herein. The symbols that will be used for representing various quantities, *e.g.*, mass, momentum, energy, distribution function, *etc.*, and the convention that will be used of constructing symbols in various contexts are as explained in table 2.10.

Note that the symbols with and without parenthesis are not identical, *i.e.*, $\boldsymbol{Q} \not\equiv \boldsymbol{Q}(\boldsymbol{x}, t)$, $\boldsymbol{Q}^{(a)} \not\equiv \boldsymbol{Q}^{(a)}(\boldsymbol{x}, t)$, $\boldsymbol{Q}_b^{(a)} \not\equiv \boldsymbol{Q}_b^{(a)}(\boldsymbol{x}, t)$. Presence of parenthesis indicates functional dependence of quantity on position and time, *i.e.*, the quantity may change with space and time.

[27)] If required, other physical properties can also be included and treated along similar lines as mass and velocity. To maintain simplicity, however, they will not be addressed in this investigation.

| Symbol | Meaning |
|---|---|
| $\boldsymbol{Q}$ | A vector quantity, *e.g.*, momentum. |
| $Q$ | A scalar quantity, *e.g.*, mass, energy. |
| $n$ | Number of particles. |
| $m$ | Mass. |
| $\rho$ | Density. |
| $\boldsymbol{P}$ | Momentum. |
| $\varepsilon$ | Energy (kinetic energy). |
| $F$ | Unnormalized one particle probability distribution function. |
| $f$ | Normalized one particle probability distribution function. |
| $\boldsymbol{Q}^{(a)}$ | Value of quantity $\boldsymbol{Q}$ for particles of species $a$. |
| $\boldsymbol{Q}_b^{(a)}$ | Value of quantity $\boldsymbol{Q}$ for particles of species $a$ having velocity $\boldsymbol{v}_b^{(a)}$. |
| $\boldsymbol{Q}(\boldsymbol{x},t)$ | Ensemble averaged value of quantity $\boldsymbol{Q}$ at position $\boldsymbol{x}$ at time $t$. |
| $\boldsymbol{Q}^{(a)}(\boldsymbol{x},t)$ | Ensemble averaged value of quantity $\boldsymbol{Q}$ at position $\boldsymbol{x}$ at time $t$ for particles of species $a$. |
| $\boldsymbol{Q}_b^{(a)}(\boldsymbol{x},t)$ | Ensemble averaged value of quantity $\boldsymbol{Q}$ at position $\boldsymbol{x}$ at time $t$ for particles of species $a$ having velocity $\boldsymbol{v}_b^{(a)}$. |

**Table 2.10:** Explanation of the mathematical notation used in Sec. 2.4.

With this, some remarks on one particle distribution function are in order: Let $F_b^{(a)}(\boldsymbol{x},t)$ be the ensemble averaged value of the number of particles of species $a$ having velocity $\boldsymbol{v}_b^{(a)}$ at position $\boldsymbol{x}$ at time $t$. Naturally, $F_b^{(a)}(\boldsymbol{x},t)$ takes values in the real domain $[0,\mathcal{N}_{\mathrm{m}}^{(a,b)}]$. $F_b^{(a)}(\boldsymbol{x},t)$ can be normalized to obtain the one particle distribution function $f_b^{(a)}(\boldsymbol{x},t)$ as

$$f_b^{(a)}(\boldsymbol{x},t) = \frac{F_b^{(a)}(\boldsymbol{x},t)}{\mathcal{N}_{\mathrm{m}}^{(a,b)}} \tag{2.11}$$

As a result, $F_b^{(a)}(\boldsymbol{x},t)$ can be viewed as unnormalized probability distribution function.

### 2.4.3 Basic Quantities

In terms of the one particle distribution functions the ensemble averaged values of number, mass, momentum, and energy for particles of species $a$ at position $\boldsymbol{x}$ at time $t$ are given by

$$n^{(a)}(\boldsymbol{x},t) = \sum_{b=1}^{\mathcal{N}_{\mathrm{v}}^{(a)}} F_b^{(a)}(\boldsymbol{x},t) \tag{2.12}$$

$$\rho^{(a)}(\boldsymbol{x},t) = \sum_{b=1}^{\mathcal{N}_{\mathrm{v}}^{(a)}} m^{(a)} F_b^{(a)}(\boldsymbol{x},t) = m^{(a)} n^{(a)}(\boldsymbol{x},t) \tag{2.13}$$

$$\boldsymbol{P}^{(a)}(\boldsymbol{x},t) = \sum_{b=1}^{\mathcal{N}_{\mathrm{v}}^{(a)}} \boldsymbol{P}_b^{(a)} F_b^{(a)}(\boldsymbol{x},t) = \rho^{(a)}(\boldsymbol{x},t)\boldsymbol{v}^{(a)}(\boldsymbol{x},t) \tag{2.14}$$

$$\varepsilon^{(a)}(\boldsymbol{x},t) = \sum_{b=1}^{\mathcal{N}_{\mathrm{v}}^{(a)}} \varepsilon_b^{(a)} F_b^{(a)}(\boldsymbol{x},t) \tag{2.15}$$

where $\boldsymbol{P}_b^{(a)} = m^a \boldsymbol{v}_b^{(a)}$ is the momentum and $\varepsilon_b^{(a)} = \frac{1}{2} m^{(a)} \boldsymbol{v}_b^{(a)} \cdot \boldsymbol{v}_b^{(a)}$ is the energy of particles of species $a$ having velocity $\boldsymbol{v}_b^{(a)}$. Similarly, the ensemble averaged values of mass, momentum, and energy at position $\boldsymbol{x}$ at time $t$ are given by

$$\rho(\boldsymbol{x},t) = \sum_{a=1}^{\mathcal{N}_\mathrm{p}} \rho^{(a)}(\boldsymbol{x},t) = \sum_{a=1}^{\mathcal{N}_\mathrm{p}} \sum_{b=1}^{\mathcal{N}_\mathrm{v}^{(a)}} m^{(a)} F_b^{(a)}(\boldsymbol{x},t) \tag{2.16}$$

$$\boldsymbol{P}(\boldsymbol{x},t) = \sum_{a=1}^{\mathcal{N}_\mathrm{p}} \boldsymbol{P}^{(a)}(\boldsymbol{x},t) = \sum_{a=1}^{\mathcal{N}_\mathrm{p}} \sum_{b=1}^{\mathcal{N}_\mathrm{v}^{(a)}} \boldsymbol{P}_b^{(a)} F_b^{(a)}(\boldsymbol{x},t) = \rho(\boldsymbol{x},t)\boldsymbol{v}(\boldsymbol{x},t) \tag{2.17}$$

$$\varepsilon(\boldsymbol{x},t) = \sum_{a=1}^{\mathcal{N}_\mathrm{p}} \varepsilon^{(a)}(\boldsymbol{x},t) = \sum_{a=1}^{\mathcal{N}_\mathrm{p}} \sum_{b=1}^{\mathcal{N}_\mathrm{v}^{(a)}} \varepsilon_b^{(a)} F_b^{(a)}(\boldsymbol{x},t) \tag{2.18}$$

### 2.4.4 Kinetic Equations

#### 2.4.4.1 Boltzmann Equation

In multiparticle lattice gases two processes, *viz.*, motion of particles and collisions among them, lead to change in $F_b^{(a)}(\boldsymbol{x},t)$. These can be described through the master equation

$$F_b^{(a)}(\boldsymbol{x} + \boldsymbol{v}_b^{(a)}\Delta t, t + \Delta t) - F_b^{(a)}(\boldsymbol{x},t) = \Delta t \Omega_b^{(a)}(\boldsymbol{x},t) \tag{2.19}$$

where the collision term $\Omega_b^{(a)}(\boldsymbol{x},t)$ gives change in $F_b^{(a)}(\boldsymbol{x},t)$ per unit time due to collisions among particles (of all types) and $\Delta t \ll 1$. All quantities are in consistent units. Taylor series expansion of $F_b^{(a)}(\boldsymbol{x} + \boldsymbol{v}_b^{(a)}\Delta t, t + \Delta t)$ around $(\boldsymbol{x},t)$ in Eq. (2.19) gives

$$\begin{aligned}\Delta t \left[\frac{\partial F_b^{(a)}(\boldsymbol{x},t)}{\partial t} + \boldsymbol{v}_b^{(a)} \cdot \boldsymbol{\nabla} F_b^{(a)}(\boldsymbol{x},t)\right] + (\Delta t)^2 \left[\frac{1}{2}\frac{\partial^2 F_b^{(a)}(\boldsymbol{x},t)}{\partial t^2} + \right.\\ \left. \boldsymbol{v}_b^{(a)} \cdot \nabla \frac{\partial F_b^{(a)}(\boldsymbol{x},t)}{\partial t} + \frac{1}{2}(\boldsymbol{v}_b^{(a)} \cdot \nabla)^2 F_b^{(a)}(\boldsymbol{x},t)\right] + \mathcal{O}(\Delta t)^3 = \Delta t \Omega_b^{(a)}(\boldsymbol{x},t)\end{aligned} \tag{2.20}$$

If variations in $F_b^{(a)}(\boldsymbol{x},t)$ are small, then terms of order higher than $\mathcal{O}(\Delta t)$ can be neglected from Eq. (2.20). With this, Eq. (2.20) becomes

$$\frac{\partial F_b^{(a)}(\boldsymbol{x},t)}{\partial t} + \boldsymbol{v}_b^{(a)} \cdot \boldsymbol{\nabla} F_b^{(a)}(\boldsymbol{x},t) = \Omega_b^{(a)}(\boldsymbol{x},t) \tag{2.21}$$

The collision term $\Omega_b^{(a)}(\boldsymbol{x},t)$ in Eqs. (2.19)–(2.21), in general, depends upon two-particle and higher order distribution functions. Two-particle distribution functions in turn satisfy an equation involving three-particle and higher order distribution functions, and so on. This results in an exact BBGKY hierarchy of equations [165] of which Eq. (2.21) is the first one.

The Boltzmann transport equation is an approximation of Eq. (2.21) with the assumption that the collision term depends only on one-particle distribution functions. In particular, recovery of the Boltzmann transport equation from Eq. (2.21) involves invocation of the assumption of molecular chaos according to which all particles are taken to be statistically uncorrelated before each collision. As a result, multiparticle distribution functions can be written as products of one-particle distribution functions. Thus, with appropriate expression for $\Omega_b^{(a)}(\boldsymbol{x},t)$ obtained by invoking the assumption of molecular chaos, Eq. (2.21) is same as the Boltzmann transport equation for $F_b^{(a)}(\boldsymbol{x},t)$ accurate to $\mathcal{O}(\Delta t)$. In the analysis that follows, Eq. (2.21) will be used in the sense of Boltzmann transport equation.

#### 2.4.4.2 Equation of Change

Let $Q$ be a quantity associated with particles and let $Q$ be either a constant or a function of velocity of particles. Then, multiplying Eq. (2.21) with $Q$ and summing up the resulting equation over all possible particle velocities gives the equation of change for $Q$ as

$$\sum_{b=0}^{\mathcal{N}_\mathrm{v}^{(a)}} Q\frac{\partial F_b^{(a)}(\boldsymbol{x},t)}{\partial t} + \sum_{b=0}^{\mathcal{N}_\mathrm{v}^{(a)}} Q\boldsymbol{v}_b^{(a)}\cdot\boldsymbol{\nabla}F_b^{(a)}(\boldsymbol{x},t) = \sum_{b=0}^{\mathcal{N}_\mathrm{v}^{(a)}} Q\Omega_b^{(a)}(\boldsymbol{x},t) \tag{2.22}$$

After slight rearrangement, this equation becomes

$$\frac{\partial}{\partial t}\left(\sum_{b=0}^{\mathcal{N}_\mathrm{v}^{(a)}} QF_b^{(a)}(\boldsymbol{x},t)\right) + \boldsymbol{\nabla}\cdot\left(\sum_{b=0}^{\mathcal{N}_\mathrm{v}^{(a)}} Q\boldsymbol{v}_b^{(a)}F_b^{(a)}(\boldsymbol{x},t)\right) = \Delta^{(a)}[Q] \tag{2.23}$$

where

$$\Delta^{(a)}[Q] = \sum_{b=0}^{\mathcal{N}_\mathrm{v}^{(a)}} Q\Omega_b^{(a)}(\boldsymbol{x},t) \tag{2.24}$$

represents change in $Q$ at position $\boldsymbol{x}$ at time $t$ due to collisions among particles.

If $Q$ is a collisional invariant (*e.g.*, $m$, $m\boldsymbol{v}$, $\frac{1}{2}m\boldsymbol{v}.\boldsymbol{v}$) or a summational invariant (*e.g.*, $A_0m + A_1\cdot m\boldsymbol{v} + \frac{1}{2}A_2m\boldsymbol{v}.\boldsymbol{v}$, $A_i$'s are arbitrary constants), then

$$\Delta^{(a)}[Q] = 0$$

Thus, for collisional and summational invariants the equation of change becomes

$$\frac{\partial}{\partial t}\left(\sum_{b=0}^{\mathcal{N}_\mathrm{v}^{(a)}} QF_b^{(a)}(\boldsymbol{x},t)\right) + \boldsymbol{\nabla}\cdot\left(\sum_{b=0}^{\mathcal{N}_\mathrm{v}^{(a)}} Q\boldsymbol{v}_b^{(a)}F_b^{(a)}(\boldsymbol{x},t)\right) = 0 \tag{2.25}$$

### 2.4.5 Hydrodynamics

#### 2.4.5.1 Continuity Equation

The equation of change Eq. (2.25), after substituting $Q = m^{(a)}$, gives

$$\frac{\partial}{\partial t}\left(\sum_{b=0}^{\mathcal{N}_\mathrm{v}^{(a)}} m^{(a)}F_b^{(a)}(\boldsymbol{x},t)\right) + \boldsymbol{\nabla}\cdot\left(\sum_{b=0}^{\mathcal{N}_\mathrm{v}^{(a)}} m^{(a)}\boldsymbol{v}_b^{(a)}F_b^{(a)}(\boldsymbol{x},t)\right) = 0 \tag{2.26}$$

This equation, in view of Eqs. (2.13) and (2.14), gives

$$\frac{\partial\rho^{(a)}(\boldsymbol{x},t)}{\partial t} + \boldsymbol{\nabla}\cdot\left(\rho^{(a)}(\boldsymbol{x},t)\boldsymbol{v}^{(a)}(\boldsymbol{x},t)\right) = 0 \tag{2.27}$$

which is the usual continuity equation of hydrodynamics.

#### 2.4.5.2 Momentum Equation

The equation of change Eq. (2.25), after substituting $Q = m^{(a)}\boldsymbol{v}_b^{(a)}$, gives

$$\frac{\partial}{\partial t}\left(\sum_{b=0}^{\mathcal{N}_\mathrm{v}^{(a)}} m^{(a)}\boldsymbol{v}_b^{(a)}F_b^{(a)}(\boldsymbol{x},t)\right) + \boldsymbol{\nabla}\cdot\left(\sum_{b=0}^{\mathcal{N}_\mathrm{v}^{(a)}} m^{(a)}\boldsymbol{v}_b^{(a)}\boldsymbol{v}_b^{(a)}F_b^{(a)}(\boldsymbol{x},t)\right) = 0 \tag{2.28}$$

This equation, in view of Eqs. (2.13) and (2.14), gives

$$\frac{\partial}{\partial t}\left(\rho^{(a)}(\boldsymbol{x},t)\boldsymbol{v}^{(a)}(\boldsymbol{x},t)\right) + \boldsymbol{\nabla}\cdot\boldsymbol{\Pi}^{(a)}(\boldsymbol{x},t) = 0 \tag{2.29}$$

where $\boldsymbol{\Pi}^{(a)}(\boldsymbol{x},t)$ is the momentum flux density tensor defined as

$$\boldsymbol{\Pi}^{(a)}(\boldsymbol{x},t) = \sum_{b=0}^{\mathcal{N}_\mathrm{v}^{(a)}} m^{(a)}\boldsymbol{v}_b^{(a)}\boldsymbol{v}_b^{(a)}F_b^{(a)}(\boldsymbol{x},t)$$

Simple macroscopic expressions for $\boldsymbol{\Pi}^{(a)}(\boldsymbol{x},t)$ cannot be obtained directly from Eqs. (2.13) and (2.14) alone. Determination of the exact form of $\boldsymbol{\Pi}^{(a)}(\boldsymbol{x},t)$ requires Chapman-Enskog expansion of $F_b^{(a)}(\boldsymbol{x},t)$ in terms of macroscopic parameters $\boldsymbol{v}^{(a)}(\boldsymbol{x},t)$ and $\rho^{(a)}(\boldsymbol{x},t)$ and solution of the kinetic equation Eq. (2.21) including collision term with appropriate constraints on the symmetry properties of tensors of rank 3 and 4 which arise after the Chapman-Enskog expansion. The Chapman-Enskog expansion of $F_b^{(a)}(\boldsymbol{x},t)$ is carried out under the assumptions that local equilibrium exists and $\boldsymbol{v}^{(a)}(\boldsymbol{x},t)$ and $\rho^{(a)}(\boldsymbol{x},t)$ vary slowly with $\boldsymbol{x}$ and $t$. Solution of the kinetic equation Eq. (2.21) requires determination of the collision term. Since the collision term involves two-particle and higher order distribution functions, determination of its exact form is a nontrivial task. As a result, usually the form of the collision term is determined approximately in the Boltzmann limit, *i.e.*, under the assumption of molecular chaos. For obtaining hydrodynamic behavior, the symmetry conditions imposed on tensors of rank 3 and 4 are that these tensors should be isotropic [28].

The above procedure can be carried out only if all the details of the lattice gas are known. This is because determination of the form of collision term requires knowledge of the exact collision rules. As a result, a generalized analysis leading to determination of the exact form of $\boldsymbol{\Pi}^{(a)}(\boldsymbol{x},t)$ for arbitrary multiparticle lattice gases is not possible. The overall form of the macroscopic equation, however, can be determined without the knowledge of exact values of these parameters through the procedure outlined by Wolfram [28]. If required symmetry conditions are satisfied by the lattice gas, a momentum equation of the form[28)]

$$\frac{\partial\rho\boldsymbol{v}}{\partial t} + \boldsymbol{\nabla}\cdot(g\rho\boldsymbol{v}\boldsymbol{v}) = -\boldsymbol{\nabla}p + \eta\nabla^2\boldsymbol{v} + \left(\zeta + \frac{1}{\mathcal{D}}\eta\right)\boldsymbol{\nabla}(\boldsymbol{\nabla}\cdot\boldsymbol{v}) \tag{2.30}$$

[28)] According to Wolfram the standard Navier-Stokes equation for a continuum fluid in $\mathcal{D}$-dimensions can be written in the form

$$\underbrace{\frac{\partial\rho\boldsymbol{v}}{\partial t}}_{(\mathrm{A})} + \underbrace{g\rho(\boldsymbol{v}\cdot\boldsymbol{\nabla})\boldsymbol{v}}_{(\mathrm{B})} = \underbrace{\boldsymbol{\nabla}p}_{(\mathrm{C})} + \underbrace{\eta\nabla^2\boldsymbol{v}}_{(\mathrm{D})} + \underbrace{\left(\zeta + \frac{1}{\mathcal{D}}\eta\right)\boldsymbol{\nabla}(\boldsymbol{\nabla}\cdot\boldsymbol{v})}_{(\mathrm{E})} \tag{2.6.1}$$

The number of this equation is the same as that given in Sec. 2.6 of [28] (different symbols have been used for some variables). This form, however, is incorrect because of the following: Presence of term (E) implies that incompressibility has not been assumed. With this, irrespective of the place where $g$ has to is obtained for particles of each species in $\mathcal{D}$-dimensional space [28]. In this equation $p$ is pressure, $\eta$ is shear viscosity, $\zeta$ is bulk viscosity, and the coefficient $g$ of the convective term is a function of $\rho$. In physical fluid dynamic systems the coefficient $g$ is always constrained to be 1 by Galilean invariance. The coefficient of the last term in this equation is determined by the requirement that the term in momentum flux density tensor proportional to $\eta$ be traceless [28]. Note that in this equation $\rho$, $\boldsymbol{v}$, and $p$ depend on $\boldsymbol{x}$ and $t$. This dependence, however, has not been shown explicitly in the equation for maintaining clarity. Furthermore, the superscript $(a)$ denoting particle species has been from the parameters in this equation. In general, the viscosities $\zeta$ and $\eta$ vary considerably with temperature and pressure and hence they also depend on $\boldsymbol{x}$ and $t$ [169,170]. In Eq. (2.30), however, the viscosities have been assumed to be constant just to obtain a compact form of the equation.

If the required symmetry conditions are not satisfied by a multiparticle lattice gas, as is the case with the HPP gas, the macroscopic equations for momentum transport obtained by following the above procedure cannot be put in the standard Navier-Stokes form [28,29].

### 2.4.6 Second Order Corrections

Eqs. (2.27) and (2.29) have been derived from Eq. (2.21) or equivalently from Eq. (2.25). These equations are accurate only to $\mathcal{O}(\Delta t)$ because Eq. (2.21) itself is accurate only to $\mathcal{O}(\Delta t)$. This approximation is expected to be accurate only when the density is low and the velocity is also small compared to the velocity of sound [28]. Higher order accurate forms of Eqs. (2.27) and (2.29) can be derived by keeping higher order terms in the Taylor series expansion Eq. (2.20) of $F_b^{(a)}(\boldsymbol{x}, t)$. In some cases correction terms might be obtained which may be important in high density and high speed (*e.g.*, supersonic) flows. Whether corrections terms will appear in the continuity and momentum equations by keeping terms of $\mathcal{O}(\Delta t)^2$ in Eq. (2.20) has been analyzed below. Exact form of these and other higher order correction terms can be determined by following the procedure outlined in [28].

The kinetic equation obtained by keeping $\mathcal{O}(\Delta t)^2$ terms in Eq. (2.20) is

$$\left\{\left[\frac{\partial}{\partial t} + \boldsymbol{v}_b^{(a)} \cdot \boldsymbol{\nabla}\right] + \frac{\Delta t}{2}\left[\frac{\partial}{\partial t} + \boldsymbol{v}_b^{(a)} \cdot \boldsymbol{\nabla}\right]^2\right\} F_b^{(a)}(\boldsymbol{x}, t) = \Omega_b^{(a)}(\boldsymbol{x}, t) \tag{2.31}$$

This equation gives the equation of change for collisional invariant $Q$ as

$$\sum_{b=1}^{\mathcal{N}_{\mathrm{v}}^{(a)}} \left\{\left[\frac{\partial}{\partial t} + \boldsymbol{v}_b^{(a)} \cdot \boldsymbol{\nabla}\right] + \frac{\Delta t}{2}\left[\frac{\partial}{\partial t} + \boldsymbol{v}_b^{(a)} \cdot \boldsymbol{\nabla}\right]^2\right\} Q F_b^{(a)}(\boldsymbol{x}, t) = 0 \tag{2.32}$$

---

appear, the form of the term (B) is incorrect. The correct form should have been $g\boldsymbol{\nabla} \cdot (\rho \boldsymbol{v}\boldsymbol{v})$ or $\boldsymbol{\nabla} \cdot (g\rho \boldsymbol{v}\boldsymbol{v})$ depending upon the source and mechanism of appearance of $g$.

Had incompressibility been assumed, term (E) would not be present in Eq. (2.6.1) because $\boldsymbol{\nabla} \cdot \boldsymbol{v} = 0$ by continuity equation. In addition to this, the first term, term (A), should have been $\rho \frac{\partial \boldsymbol{v}}{\partial t}$ instead of $\frac{\partial \rho \boldsymbol{v}}{\partial t}$.

Analysis of the context connecting Secs. 2.6 and 2.5 of [28] reveals that the incorrect form Eq. (2.6.1) of the Navier-Stokes equation has come about because the objective had been to establish equivalence between Eq. (2.5.11) and the Navier-Stokes equation and the form of Eq. (2.5.11) appears similar to Eq. (2.6.1). Further rearrangement of Eq. (2.5.11) with the piece given in Eq. (2.5.12) after judicious removal of some terms as argued in Sec. 2.6 of [28], however, leads to an equation of the form Eq. (2.30) as in the text on page 55 herein and not to an equation of the form Eq. (2.6.1) in [28] (see Appendix A for details).

**Correction in Continuity Equation:** Substituting $Q = m^{(a)}$ in Eq. (2.32) and then rearranging the terms gives

$$\begin{aligned}
&\left[\frac{\partial \rho^{(a)}(\boldsymbol{x},t)}{\partial t} + \boldsymbol{\nabla}\cdot\Big(\rho^{(a)}(\boldsymbol{x},t)\boldsymbol{v}^{(a)}(\boldsymbol{x},t)\Big)\right] \quad + \\
&\frac{\partial}{\partial t}\left[\frac{\partial \rho^{(a)}(\boldsymbol{x},t)}{\partial t} + \boldsymbol{\nabla}\cdot\Big(\rho^{(a)}(\boldsymbol{x},t)\boldsymbol{v}^{(a)}(\boldsymbol{x},t)\Big)\right] \quad + \\
&\boldsymbol{\nabla}\cdot\left[\frac{\partial}{\partial t}\Big(\rho^{(a)}(\boldsymbol{x},t)\boldsymbol{v}^{(a)}(\boldsymbol{x},t)\Big) + \boldsymbol{\nabla}\cdot\boldsymbol{\Pi}^{(a)}(\boldsymbol{x},t)\right] \quad = \quad 0
\end{aligned} \tag{2.33}$$

This equation vanishes if $F_b^{(a)}(\boldsymbol{x},t)$ is such that Eqs. (2.27) and (2.29) are satisfied. Thus, no corrections appear in the continuity equation at $\mathcal{O}(\Delta t)^2$.

**Correction in Momentum Equation:** Substituting $Q = m^{(a)}\boldsymbol{v}_b^{(a)}$ in Eq. (2.32) and then rearranging the terms gives

$$\begin{aligned}
&\underbrace{\left[\frac{\partial}{\partial t}\Big(\rho^{(a)}(\boldsymbol{x},t)\boldsymbol{v}^{(a)}(\boldsymbol{x},t)\Big) + \boldsymbol{\nabla}\cdot\boldsymbol{\Pi}^{(a)}(\boldsymbol{x},t)\right]}_{\text{Term-1}} \quad + \\
&\underbrace{\frac{\Delta t}{2}\frac{\partial}{\partial t}\left[\frac{\partial}{\partial t}\Big(\rho^{(a)}(\boldsymbol{x},t)\boldsymbol{v}^{(a)}(\boldsymbol{x},t)\Big) + \boldsymbol{\nabla}\cdot\boldsymbol{\Pi}^{(a)}(\boldsymbol{x},t)\right]}_{\text{Term-2}} \quad + \\
&\underbrace{\sum_{b=1}^{\mathcal{N}_v^{(a)}}\frac{\Delta t}{2}\left[\boldsymbol{v}_b^{(a)}\cdot\boldsymbol{\nabla}\frac{\partial}{\partial t} + \Big(\boldsymbol{v}_b^{(a)}\cdot\boldsymbol{\nabla}\Big)^2\right] m^{(a)}\boldsymbol{v}_b^{(a)}F_b^{(a)}(\boldsymbol{x},t)}_{\text{Term-3}} \quad = \quad 0
\end{aligned} \tag{2.34}$$

The first and second terms of this equation vanish if $F_b^{(a)}(\boldsymbol{x},t)$ is such that Eq. (2.29) is satisfied. The third term, however, does not vanish because of a piece trilinear in $\boldsymbol{v}_b^{(a)}$ [28]. As a result, corrections appear in the momentum equation at $\mathcal{O}(\Delta t)^2$.

## 2.5 Problems with Multiparticle Lattice Gases

In the course of development, analysis, and usage of multiparticle lattice gases for simulation of physical systems (*c.f.*, Sec. 2.2, 2.3, and 2.4), many complex problems related to closeness of dynamical behavior of multiparticle lattice gases with that of physical fluid dynamic systems have surfaced. Two of these problems, being well known and of direct interest in this investigation, are: (i) breaking of Galilean invariance by multiparticle lattice gases (or, non-Galilean invariance of multiparticle lattice gases), and (ii) the inability of simulating compressible systems and phenomena correctly using multiparticle lattice gases. Although these problems have been addressed by many authors and some methods of overcoming them have been suggested [28,29,39,47,118,119,125], the problems still persist. The suggested methods are largely kludges rather than being satisfactory solutions for eliminating the problems completely.[29)] These problems are directly related to usability of multiparticle lattice gases for studying physical systems and phenomena and also,

[29)] This statement has been made to highlight the severity of the problems and the nature of the methods that have been suggested for overcoming them. From this statement it should not be inferred or assumed that in this investigation I am going to suggest a permanent or a magical remedy for these problems in

in a wider context, to usability of the formalism of cellular automata itself for modeling and simulation of physical systems. As a result, they will be reviewed and analyzed in the following sections.

Another problem that one might like to address is isotropy/anisotropy of multiparticle lattice gases (or, of various tensors that appear during coarse graining of multiparticle lattice gases). This issue, however, is not of interest in this investigation because it has been addressed in great details elsewhere [28,29,118,119]. Presently it is known that isotropy is a purely geometrical property that depends upon the structure of the underlying spatial lattice. Thus, its presence or absence in lattice gases can be controlled as desired (or, dictated by the simulation requirements) by developing lattice gases over appropriate underlying spatial lattices using appropriate discrete velocity sets for particles.

## 2.5.1 Violation of Galilean Invariance

### 2.5.1.1 Evidences of Violation of Galilean Invariance

The dynamics of (*most of*) multiparticle lattice gases is not invariant under Galilean transformations. This fact reflects in the form of appearance of a density dependent multiplicative factor, commonly denoted as $g(\rho)$, in the nonlinear terms of coarse grained hydrodynamic equations of these models (*c.f.*, Sec. 2.4). This factor appears during Chapman-Enskog analysis of lattice gas dynamics and when not equal to unity (which is usually the case) indicates violation of Galilean invariance by the models. As a result, this factor is called as the *Galilean invariance breaking parameter* or the *non-Galilean invariance factor*.

The effect of violation of Galilean invariance in multiparticle lattice gases is also reflected during simulations in the form of need of renormalizing either the flow velocity or the viscosity in order to be able to compare the simulation results with experimental observations. This renormalization is needed even at very low particle densities and very low flow speeds (*i.e.*, when compressibility effects are not important) [39,76,141]. At high flow speeds (*i.e.*, when compressibility effects become important), high particle densities, for mixtures of particle species, and for multiphase systems even renormalization of lattice gas parameters cannot make the simulation results conform with experimental observations. That it has to be so is clearly evident from Wolfram's [28] and Frisch *et. al.*'s [29] analysis of the dynamics of multiparticle lattice gases.[30)] As a result, under these conditions simulations using (most of) multiparticle lattice gases become useless and are avoided.

---

multiparticle lattice gases. This, simply, is not the objective of this investigation. In this investigation, to fulfill the objective outlined in Sec. 1.4, I have only attempted to analyze the problems and the cause of their origin in multiparticle lattice gases and, in view of the findings thereof, attempted to find a new method of developing lattice gases so that (if possible) these problems do not appear at all in the new models. The new lattice gases that come out from the method are not multiparticle lattice gases. I have not attempted to find a remedy for these problems in multiparticle lattice gases.

"Why the methods suggested in the literature for overcoming the problems are no satisfactory?" and "Why have they been termed as *kludges* here?" will become clear from the following sections.

[30)] Although Wolfram's analysis as presented in [28] revolves primarily around single species and (mostly) single speed multiparticle lattice gases, it is widely applicable, as pointed out by him, to other multiparticle lattice gases also. Same is the case with the analysis of Frisch *et. al.* [29] also.

#### 2.5.1.2 Explanations and Cause of Violation of Galilean Invariance

With the introduction of multiparticle lattice gases and simultaneous discovery that most of them violate Galilean invariance, many attempts have been made to explain and find the cause of this violation. Among these, the earliest attempts are due to Wolfram [28] (two explanations) and Frisch *et. al.* [29] (one explanation). Their explanations about the violation and its cause are as follows:

**(A) Wolfram's First Explanation:** Wolfram has attempted to explain the appearance of the Galilean invariance breaking parameter (denoted by $\mu$) in [28] on page 484 as

> *The convective term in Eq. (2.5.11) has the same structure as in the Navier-Stokes equation (2.6.1), but includes a coefficient*
>
> $$\mu = \frac{1}{4}c^{(2)} \tag{2.6.4}$$
>
> *which is not in general equal to 1. In continuum fluids, the covariant derivative*[31)] *usually has the form* $D_t = \partial_t + \boldsymbol{u}\cdot\boldsymbol{\nabla}$ *which is implied by Galilean invariance. The cellular automaton fluid acts, however, as a mixture of components, each with velocities* $\boldsymbol{e}_a$*, and these components can contribute with different weights to covariant derivatives of different quantities, leading to convective terms with different coefficients.*

This explanation is based on change in the form of covariant derivative because cellular automaton fluid acts as a mixture of components of particles moving with different velocities.[32)] This explanation, however, appears to be incorrect in view of the following:

The derivation of the form of covariant derivative [169][33)] suggests that (in continuum) the convective term represents instantaneous time rate of change brought about by fluid motion in some quantity $\xi$ due the existence of spatial gradient of $\xi$. Because of this the convective term will be zero if either the flow velocity is zero or $\xi$ is uniform in space. The continuum in fluids, however, is an approximation in that even an infinitesimal fluid element is necessarily taken to be large enough to contain many fluid particles (atoms or molecules). As a result, the overall form of the convective term must appear from the microscopic world only. The following analysis illustrates the mechanism of appearance of the convective term from the microscopic world:

31) The *covariant derivative* is variedly known as *substantial/total/material derivative*, also.

32) It is noteworthy that in this mixture the components, instead of consisting of particles of different species, consist of particles of the same species moving with different velocities.

33) The following derivation is reproduced from [169] with some generalizations and other minor changes.

The derivative $\frac{D\xi}{Dt}$ denotes rate of change of $\xi$ of a given fluid particle as it moves about in space. This derivative has to be expressed in terms of quantities referring to points fixed in space. To do so, we notice that change $D\xi$ in $\xi$ of the given fluid particle during the time $Dt$ is composed of two parts, namely (i) the change during $Dt$ in $\xi$ at a point fixed in space, and (ii) the difference between $\xi$'s (at the same instant) at two points $D\boldsymbol{r}$ apart, where $D\boldsymbol{r} = \boldsymbol{v}Dt$ is the distance moved by the given fluid particle during the time $Dt$. The first part is $\frac{\partial\xi}{\partial t}Dt$, where the derivative $\frac{\partial\xi}{\partial t}$ is taken for constant $x$, $y$, $z$, *i.e.*, at the given point in space. The second part is

$$Dx\frac{\partial\xi}{\partial x} + Dy\frac{\partial\xi}{\partial y} + Dz\frac{\partial\xi}{\partial z} = (D\boldsymbol{r}\cdot\boldsymbol{\nabla})\xi$$

Thus

$$D\xi = \frac{\partial\xi}{\partial t}Dt + (D\boldsymbol{r}\cdot\boldsymbol{\nabla})\xi$$

or, dividing both sides by $Dt$,

$$\frac{D\xi}{Dt} = \frac{\partial\xi}{\partial t} + (\boldsymbol{v}\cdot\boldsymbol{\nabla})\xi$$

Let a fluid element contain large number of identical particles so that continuum is valid. The number of particles $N$ contained in the fluid element is necessarily finite. Particles have been assumed to be identical only for the sake of simplicity. Let a fraction $f_i = N_i/N$ of these particles be moving with velocity $\boldsymbol{v}_i$. Then, the fluid element moves with the velocity

$$\boldsymbol{v} = \sum_i f_i \boldsymbol{v}_i \tag{2.35}$$

Contribution of a particle moving with velocity $\boldsymbol{v}_i$ to the convective change of $\xi$ is $\frac{1}{N}(\boldsymbol{v}_i \cdot \boldsymbol{\nabla})\xi$, and the contribution due to all the particles moving with velocity $\boldsymbol{v}_i$ is $f_i(\boldsymbol{v}_i \cdot \boldsymbol{\nabla})\xi$. Thus, total convective change in $\xi$ due to all the particles comprising the fluid element is

$$\sum_i f_i(\boldsymbol{v}_i \cdot \boldsymbol{\nabla})\xi \tag{2.36}$$

or, alternatively

$$\left(\left[\sum_i f_i \boldsymbol{v}_i\right] \cdot \boldsymbol{\nabla}\right)\xi \tag{2.37}$$

which, in view of Eq. (2.35), becomes

$$(\boldsymbol{v} \cdot \boldsymbol{\nabla})\xi \tag{2.38}$$

Thus, although particles moving with different velocities contribute with different weights to the convective derivative, the overall form of the convective term (and thus the covariant derivative) does not depend on these weights. The weights get subsumed into mean velocity of the fluid element and do not appear separately. Thus, Wolfram's explanation about the appearance of Galilean invariance breaking parameter quoted above is incorrect.

**(B) Wolfram's Second Explanation:** Although the explanation about appearance of Galilean invariance breaking parameter put forth by Wolfram on page 484 in [28] has been argued to be incorrect above, another unrelated explanation is also available in Sec. 4.7 of [28]. This is as follows:

Wolfram's analysis outlined in Sec. 4.7 of [28] shows that the Galilean invariance breaking parameter takes its form due to Fermi-Dirac nature of the equilibrium distribution function of particles. The Fermi-Dirac nature of the one-particle distribution function, in turn, comes about because of exclusion principle based on velocity of particles that has been imposed in the multiparticle lattice gases analyzed in [28]. Further analysis by Wolfram for the case in which the exclusion principle is relaxed and arbitrarily large number of particles having same velocity are allowed to occupy the same lattice site shows that though the Galilean invariance breaking parameter loses its density dependence, it still remains different from unity. With this mathematical finding, Wolfram has concluded that the violation of Galilean invariance is associated with fixed speed of particles.

Although the above explanation seems to be acceptable for the specific type of single speed multiparticle lattice gases analyzed in [28], it can neither be generalized to all possible types of single speed lattice gases nor to all possible types of lattice gases. In fact, whether or not this explanation gives the correct cause of appearance of $g(\rho)$ even for single speed multiparticle lattice gases analyzed in [28] is also doubtful. This is because Wolfram's conclusion seems to be a straight jump from his mathematical analysis and findings about the form of $g(\rho)$ presented in Sec. 4.7 of [28] without acceptable intermediate arguments connecting the two (*i.e.*, the conclusion and the mathematical findings). In other words, it is not clear "how does change in the form of $g(\rho)$ on relaxation of the exclusion principle imply that the violation of Galilean invariance is associated with fixed speed of particles?"; arguments connecting the two are unclear/missing.

Further arguments and counterarguments on this explanation about violation of Galilean invariance by multiparticle lattice gases are being postponed till the next section.

**(C) Frisch *et. al.*'s Explanation:** Frisch *et. al.*'s analysis [29] of $\mathcal{D}$-dimensional single speed multiparticle lattice gases with exclusion principle based on velocity of particles also shows the appearance of the Galilean invariance breaking parameter in the coarse grained hydrodynamic equations of the models. Like Wolfram [28], they also find that the nature of equilibrium distribution function is Fermi-Dirac and that this nature comes about because of the exclusion principle. They observe that in the macrodynamical equations for these lattice gases $g(\rho)$ appears (from the distribution function) as a multiplicative factor in advection term of the momentum flux tensor [29, Sec. 7]. Furthermore, in the incompressible limit $g(\rho)$ comes out as the multiplicative factor of the advection term $\boldsymbol{v} \cdot \boldsymbol{\nabla} \boldsymbol{v}$. Following this they assert that the equations obtained in the incompressible limits are not Galilean invariant and conclude that this obviously reflects lack of Galilean invariance at the lattice level (see paragraph beginning on page 678 in [29]).

Frisch *et. al.*'s explanation has been accepted and used widely in a number of studies related to overcoming the problem of Galilean invariance in multiparticle lattice gases [39,47,119]. In their studies on Galilean invariance in mixtures d'Humières *et. al.* [39] assert (about the original FHP gas [27]) that

> *The absence of Galilean invariance of the original model is due to the use of a finite set of directions for the velocity of particles and to the exclusion principle that leads to boolean character of particles.*[34)]

Similarly, Gunstensen and Rothman [47] assert (about the basic FHP model [27]) that[35)]

> *... the second is an expression for the conservation of momentum similar to the usual Navier-Stokes equation except for the $g(\rho)$ factor preceding the inertial term. This extra factor is the result of the discretization of the particle velocities, all of which are of unit magnitude, and of the lattice, which has only six possible directions. The presence of this factor causes the FHP model to lack Galilean invariance.*[36)]
>
> ---
>
> *... Physically, it (the $g(\rho)$ factor) is a manifestation of the discreteness of particle velocities and of the lattice.*
>
> ---
>
> *Essentially, the FHP gas is not Galilean invariant because (in it) all the particles move with unit speed. ...*

Similarly, Boghosian and Taylor [119] assert that

---

[34)] It is noteworthy that boolean character of particles is the cause of the equilibrium distribution function being Fermi-Dirac.

[35)] Note that these statements appear in a scattered manner in [47].

[36)] Here one should note that the cause and effect relationship pointed out in the footmarked statement is incorrect. The $g(\rho)$ factor appears in the equations because of lack of Galilean invariance. It is not the cause of lack of Galilean invariance instead it is the effect (or consequence) of lack of Galilean invariance.

> *.... Thus the presence of the $g(\rho_0)$ factor in the inertial term is reflective of a breakdown of Galilean invariance. As has been pointed out by numerous authors [1,2,21],*[37] *this is not surprising since the lattice itself constitutes a preferred Galilean frame of reference.*

The above shows that both Frisch *et. al.*'s explanation and Wolfram's second explanation are arrived at by proceeding along identical lines. As a result, although both of them read differently (or are phrased differently), they are equivalent as far as their essential content is concerned. Their equivalence is clearly evident from d'Humières *et. al.*'s and, Gunstensen and Rothman's assertions both of which, although derived from Frisch *et. al.*'s analysis [29], are phrased along the lines of Wolfram's second explanation.

An important consequence of equivalence of Wolfram's second explanation and Frisch *et. al.*'s explanation is that the later also becomes subject to the same doubts as the former.

**(D) Summary of Explanations:** Various explanations about violation of Galilean invariance in multiparticle lattice gases outlined above, although phrased differently, originate from identical mathematical analysis (namely, Boltzmann and Chapman-Enskog analysis) of the dynamics and are in essence identical. To see the interconnections and various similarities among them, these explanations can be summarized together as follows:

During mathematical analysis of multiparticle lattice gases the Galilean invariance breaking parameter appears as coefficient of the second order term in the Chapman-Enskog expansion of one-particle distribution function from the Fermi-Dirac nature of the equilibrium distribution function[38] [28,29,39,47,118,119]. This implies that at the microscopic level the non-Galilean invariance arises from the structure and discreteness of the underlying spatial lattice. Alternatively, it can also be said that non-Galilean invariance arises from discreteness of spatial lattice and velocity vectors and finiteness of the magnitude and number of velocity vectors (*i.e.*, total number of velocity vectors is finite as well as the magnitude of each velocity vector is also finite). In yet another alternate fashion, it can be said that the non-Galilean invariance of lattice gases arises from breaking of translation invariance which occurs because the underlying discrete spatial lattice provides a preferred Galilean reference frame and constrains the motion of particles to be only along the links of the lattice.

Note that the above summary does not include the explanation proposed by Wolfram [28] and Gunstensen and Rothman [47] that non-Galilean invariance arises because of fixed speed of particles. This is because this explanation is specific to the FHP gas (or, more generally to single speed multiparticle lattice gases) and becomes invalid as soon as one takes into consideration multispeed multiparticle lattice gases, *e.g.*, extension of the FHP with one or more rest particles. Furthermore, it is evident from Wolfram's [28] and Boghosian *et. al.*'s [125] analysis that most of multiparticle lattice gases, irrespective of whether they are single speed or multispeed, violate Galilean invariance (except under certain specific conditions, *e.g.*, when particles with different speed are present in specific ratios).

[37] The citations [1,2,21] in [119] are the citations [27,29,28] in this manuscript.

[38] In view of Wolfram's analysis [28, Sec. 4.7] of single speed multiparticle lattice gases without exclusion principle (*i.e.*, when arbitrarily large number of particles with same velocity are allowed to occupy the same lattice site simultaneously) one might conclude that the footmarked statement is incorrect. This conclusion, however, is incorrect because of the following: Boghosian *et. al.* [125] have carried out a very general analysis of single speed multiparticle lattice gases in which a maximum of $2^N - 1$ particles having the same velocity are allowed to occupy the same lattice site simultaneously. Their analysis shows that distribution function of each bit is Fermi-Dirac and distribution function for each direction is weighted sum of the distribution functions for individual bits for that direction. Thus, the distribution function for each direction obtained in the limit $N \to \infty$, in essence, consists of Fermi-Dirac distributions only.

#### 2.5.1.3 Counter Arguments to Explanations

Although the explanations about the cause of appearance of $g(\rho)$ outlined above seem plausible, they raise an important question—"Is it possible to carry out truly Galilean invariant simulations of particle dynamical systems using lattice gases?" This question stems from the argument that if non-Galilean invariance actually arises from the discreteness and structure of the underlying spatial lattice and/or from discreteness and finiteness (in number and magnitude) of velocity vectors, then it can never be removed and will be present in all the lattice gases. This is because by definition lattice gases exist in fully discrete space-time and the number and magnitude of velocity vectors with which particles of each species move over the spatial lattice is finite. This argument is further supported by the observation that athermal (*i.e.*, single speed) and thermal (*i.e.*, multispeed) lattice gases have been developed over almost all one-, two-, and three-dimensional regular lattices including some four-dimensional lattices (*c.f.*, Sec. 2.2.3.3) and all these lattice gases, with some exceptions under certain specific conditions, violate Galilean invariance.

The objection raised above on the acceptability of explanations and cause of violation of Galilean invariance in multiparticle lattice gases furnished in the literature and outlined in Sec. 2.5.1.2 is valid. At this stage of inquiry, this objection, however, does not imply that truly Galilean invariant simulations of particle dynamical systems cannot be carried out using lattice gases. It only implies that the explanations furnished in the literature are not unobjectionable and that the violation of Galilean invariance in multiparticle lattice gases needs to be investigated further to arrive at concrete and unobjectionable conclusions about its actual physical cause. The first step towards this investigation is to examine the exceptions that have been pointed out. These exceptions and their limitations are as follows:

Since the very introduction of multiparticle lattice gases beginning with the FHP gas it has been opined that non-Galilean invariance can be overcome. In the case of single species athermal multiparticle lattice gases it is possible to overcome non-Galilean invariance at very low flow speeds and particle densities by simply rescaling the parameters as outlined in [27–29]. Although this rescaling restores Galilean invariance at macroscopic level, it is important to note that the lattice gases themselves still remain non-Galilean invariant. Furthermore, this rescaling procedure becomes ineffective if the flow speed and particle density are not small or if the system contains particles of more than one species. Thus, as far as development of truly Galilean invariant lattice gases is concerned and as far as carrying out truly Galilean invariant simulations of all types of particle dynamical systems is concerned, rescaling of parameters is not an effective solution.

Another method that has been suggested for overcoming non-Galilean invariance is modification of the basic (athermal) lattice gases by addition of rest particles. Using this method new multiparticle lattice gases can be developed which are Galilean invariant under the right combination of rest particles and particle density [39]. Problem with this method, however, is that it is too restrictive since it places limits on the range of variation of density and *requires* explicit inclusion of rest particles in the model. Furthermore, this method does not lead to lattice gases for which $g(\rho) = 1$ uniformly for all values of $\rho$. Rather, the resulting lattice gases are Galilean invariant only at very specific values of density (generally two values) for specific number of rest particles and collision rules. In other words, this method leads to lattice gases for which $1 - \epsilon < g(\rho) < 1 + \epsilon$ for $\rho_{\min} < \rho < \rho_{\max}$, where $\epsilon$ is small compared to unity and $\rho$ is normalized particle density, *i.e.*, $\rho = n/N_{\max}$, where $n$ is the number of particles on a lattice site and $N_{\max}$ is the maximum number of particles (including the rest particles) that can occupy a lattice site at any time step. In fact, it can be seen from [125] that this method does not lead to lattice gases for which $g(\rho) = 1$ uniformly for all values of $\rho$ even if the number of particles having the same (non-zero) velocity is made variable. Thus, this method is also not an effective solution as far as development of truly Galilean invariant lattice gases and carrying out truly Galilean invariant simulations of all types of particle dynamical systems is concerned.

Perhaps non-Galilean invariance can also be overcome by developing lattice gases using sufficiently large number of discrete velocity vectors (possibly over irregular spatial lattices) and by allowing multiple particle having the same velocity to occupy the same lattice site simultaneously. This leads to generalized multiparticle lattice gas described in Sec. 2.4.1. The development of such lattice gases beyond the level of their general description, however, seems to be a very complex task and has not been accomplished yet. The primary problem in this task is the need of a simple enough (and humanly comprehensible) method of defining collision rules and also the need of a way for analysis of the resulting lattice gas. In case it is possible to develop Galilean invariant lattice gases in this way, there still remains another doubt to be answered—"Can Galilean invariant athermal lattice gases be developed?" and ignoring the issue of isotropy one might also like to add "on a square spatial lattice" to it, which further complicates the problem.

#### 2.5.1.4 Need of Further Exploration

In view of the present developments, the answer to the questions raised in the previous section seems to be a flat "no". This implies that one of the following two conclusions should be true: (i) truly Galilean invariant simulations of particle dynamical systems cannot be carried out using lattice gases, or (ii) the explanations about violation of Galilean invariance furnished in the literature are not rigorous enough.

To accept the first conclusion, *i.e.*, truly Galilean invariant simulations of particle dynamical systems cannot be carried out using lattice gases, one needs a formal proof which is lacking. To arrive at such a proof one cannot take the help of mathematical analysis of the kind put forth by Wolfram [28] and Frisch *et. al.* [29] for determination of the $g(\rho)$ factor. This is because this analysis requires knowledge of the structure of lattice gases and thus will have to be carried out once for each and every multiparticle lattice gas. This is not feasible because there are infinitely many multiparticle lattice gases. Thus, to arrive at a formal proof (or, otherwise) a path which bypasses Wolfram's and Frisch *et. al.*'s mathematical analysis will have to be found out.

To accept the second conclusion, *i.e.*, the explanations about violation of Galilean invariance furnished in the literature are not rigorous enough, again a formal proof is needed. For this proof the cause of violation of Galilean invariance in multiparticle lattice gases will have to be investigated afresh using a method which bypasses Wolfram's and Frisch *et. al.*'s mathematical analysis for determination of $g(\rho)$. This is also because of the reason described above. This proof can be accomplished by showing that violation of Galilean invariance does not occur because of discreteness of spatial lattice and discreteness and finiteness (in number and magnitude) of velocity vectors.

It is evident that formal proof of either one of the conclusions outlined above requires determination of the actual cause of violation of Galilean invariance in multiparticle lattice gases. Thus, further investigation on the cause of violation of Galilean invariance in multiparticle lattice gases are required. These have been outlined in chapter 3.

It is worthwhile to note that an affirmative proof of the first conclusion above does not imply that lattice gases cannot be viewed as true alternative to calculus based modeling approach as has been envisioned by cellular automata theorists [12]. It only implies that multiparticle lattice gases are not good enough alternatives to the calculus based modeling approach. It is possible to develop many other kinds of cellular automata models of particle dynamical systems also (as will be seen later). These models are also lattice gases. As a result, the vision of cellular automata theorists stands unharmed.

### 2.5.2 Incompressibility

The second major problem that has been encountered with multiparticle lattice gases is their inability in simulating compressible systems and phenomena correctly. In the following text this problem will be referred to as *"incompressibility of multiparticle lattice gases"*.[39] The explanation and cause of this problem which are accepted in the literature and the arguments which originate from further analysis in the present investigation are as follows:

It is evident from elaborations given in Sec. 2.5.1 that in the literature all investigations on Galilean invariance and its violation in multiparticle lattice gases are mathematically oriented. The deductive process followed in these investigations is one of back calculation rather than being one of direct deduction in the sense that all inferences about whether or not Galilean invariance is being violated by a lattice gas are made only after the exact form of the $g(\rho)$ factor appearing in coarse grained equations has been determined mathematically. At present there is no rigorous method, other than mathematical analysis leading to the exact form of the $g(\rho)$ factor, using which one can deduce whether or not Galilean invariance will be violated by a multiparticle lattice gas. The only other alternative to mathematical analysis is computer simulation which is not a very rigorous method.

Because of this peculiar back calculative nature of investigations, the inability of lattice gases in correctly simulating compressible systems and phenomena is also attributed to the appearance and form of the $g(\rho)$ factor. Alternatively, in the literature it is accepted that the incompressibility of multiparticle lattice gases is because of (or, a consequence of) their non-Galilean invariance.[40] This view can be accepted as largely correct because unless coarse grained equations of multiparticle lattice gases are identical with the Navier-Stokes equations their macroscopic dynamics will be different from that of the Navier-Stokes equations. As a result, multiparticle lattice gases with $g(\rho) \neq 1$ will not be able to reproduce compressible phenomena that are consistent with those reproduced by the Navier-Stokes equations.

---

[39] It is likely that some confusion might arise from the phrase *"incompressibility of multiparticle lattice gases"*. The following elaboration would clarify the confusion: This phrase does not mean that the lattice fluid in these lattice gases is not compressible or that the multiparticle particle lattice gases cannot be run in the compressible domain. The lattice fluid is compressible and the multiparticle lattice gases can be run in the compressible domain. The results obtained from these simulations (runs), however, cannot be mapped to experimental observations. The results obtained from multiparticle lattice gas simulations can be mapped to experimental observations only in incompressible domain. As a result, the practical utility of multiparticle lattice gases is restricted to simulation of incompressible systems and phenomena only. The phrase under debate refers only to this particular limitation of multiparticle lattice gases.

[40] The major implication of this statement is that the explanations and cause of inability of multiparticle lattice gases in correctly simulating compressible systems and phenomena are the same as those for violation of Galilean invariance detailed in Sec. 2.5.1.

The interrelationship of non-Galilean invariance and incompressibility of multiparticle lattice gases outlined above is logically sound and acceptable. Problems surface when this interrelationship is stretched to infer, which is usually the case, that recovery of Galilean invariance in multiparticle lattice gases will allow correct simulation of compressible systems and phenomena using them. This inference, however, seems to be incorrect because of the following: In Galilean invariant lattice Boltzmann models problems related to simulation of compressible systems which are identical with those in multiparticle lattice gases have been encountered [141,171]. These observations, since lattice Boltzmann models are minor extensions of multiparticle lattice gases and are treated in an identical fashion as the multiparticle lattice gases, imply that the incompressibility of lattice gases is not only because of their non-Galilean invariance. It is, however, certain that Galilean invariance is a necessary condition that must be met before truly compressible simulations can be correctly carried out using multiparticle lattice gases. Boghosian also accepts the finding that recovery of Galilean invariance might not allow correct simulation of compressible systems and phenomena [172]. It, however, remains to be proved that Galilean invariance is only a necessary but not sufficient conditions for compressibility. Thus, presently it seems hard to comment on the exact role played by non-Galilean invariance in the incompressibility of lattice gases beyond the conclusions that have been already derived above.

Implications of the incompressibility of multiparticle lattice gases are almost identical to those outlined in the previous section; perhaps more far reaching. This is because, unless truly compressible simulations can be carried out using multiparticle lattice gases, they will remain restricted to the Mach numbers $\leq 0.3$ (incompressible flow range) which excludes the complete open domain of Mach numbers $> 0.3$ where one finds some of the most complex and most interesting physical phenomena.

Counter indications and insufficiency of explanations about the exact cause of incompressibility of multiparticle lattice gases in the literature necessitate further investigation of this problem. These have been outlined in chapter 3.

## 2.6 Conclusions

The review of the formalism of cellular automata and lattice gases presented in this chapter corroborates the findings and expectations of chapter 1 and thus, eliminating the possibility of baselessness, puts them on strong grounds. Besides this, the major conclusions from chapter are as follows:

**1)** Cellular automata are abstract mathematical systems having all the elements required for modeling and simulation of physical systems. Their application for modeling and simulation of physical systems, however, requires appropriate interpretation of their basic elements because of their abstraction.

**2)** Interpretation of the basic elements of cellular automata, being not internal to the formalism itself, has a certain context dependent flexibility in the sense that many different interpretations of the basic elements are possible depending upon the context. As a result, the nature and properties of cellular automata models of physical systems are bound to be dependent largely upon the interpretations.

**3)** The formalism of cellular automata can indeed be used for developing molecular dynamics like models of physical systems by interpreting their basic elements as outlined in table 2.6. These models are termed as "multiparticle lattice gases".[41)]

**4)** Multiparticle lattice gases have seen vast and rapid development in the last decade. Specifically, many multiparticle lattice gases has been developed, analyzed, and successfully used for simulation of large number of physical systems including fluid dynamic systems. These lattice gases, however, have problems which limit their utility to simulation of incompressible fluid dynamic systems.

**5)** Two major problems (among others) with multiparticle lattice gases are violation of Galilean invariance and incompressibility. These problems have been addressed by many authors and attempts have been made to explain the cause of their appearance and to eliminate them. It has been found that non-Galilean invariance can be overcome in specific systems if conditions of incompressible flow are satisfied.

It is thought that non-Galilean invariance is because of (i) Fermi-Dirac nature of equilibrium distribution function, (ii) structure and discreteness of the underlying spatial lattice and discreteness and finiteness of magnitude and number of velocity vectors of particles, and (iii) breaking of translation invariance because the underlying discrete spatial lattice provides a preferred Galilean reference frame and constrains the motion of particles to be only along the links of the lattice. All these explanations are, in essence, identical and subject to strong counter arguments. As a result, these explanations appear to be incomplete and are not acceptable without further investigations.

**6)** In the literature, incompressibility of multiparticle lattice gases is taken to be a consequence of their non-Galilean invariance. From this interlink, usually, it is inferred that recovery of Galilean invariance would allow simulation of compressible systems. This inference, however, being subject to strong counter evidence arising from studies conducted on lattice Boltzmann automata, is incorrect. This implies that non-Galilean invariance is not the *only* cause of incompressibility of multiparticle lattice gases, thus necessitating further investigations in this direction.

Multiparticle lattice gases are cellular automata models of physical systems and appear similar to molecular dynamics models. These lattice gases, however, have several problems which constrain their utility to simulation of incompressible fluid dynamic systems.

[41)] Earlier, in the literature, these models were termed as "lattice gases". In this investigations, however, they have been renamed as "multiparticle lattice gases". This is because of the following: Traditionally, all cellular automata models of physical systems are termed as "lattice gases". This term does not involve distinctions based on the interpretation of basic elements of cellular automata (possibly because it was never needed before). It is possible that many different interpretations of basic elements of cellular automata lead to valid models of physical systems. Traditionally all these models are "lattice gases". In this investigation, however, lattice gases based on different interpretations need to be differentiated and thus named differently. One way of naming them differently is to prefix the term "lattice gases" with some key aspects of the interpretation employed for developing them. Thus, the term "multiparticle lattice gases" for the earlier cellular automata models. Naturally, all "multiparticle lattice gases" are "lattice gases", but not vice versa.

Many attempts have been made in the literature to find the cause of these problems and to eliminate them. These attempts, however, have met with only partial success. The explanations about the cause of non-Galilean invariance and incompressibility of multiparticle lattice gases furnished in the literature, being subject to counter arguments and evidences against them, are insufficient and unacceptable without further investigation. If these problems can be overcome, multiparticle lattice gases might turn out be the tools that are being sought. In that case the objective of this investigation (*c.f.*, Sec. 1.4) might get fulfilled easily because the method of development of these lattice gases is well developed. It is possible that for overcoming the problems some changes might be needed in the method of development of these lattice gases. Such changes, if any, would of course have to be incorporated.

In view of the above, further investigations on the cause of non-Galilean invariance and incompressibility of multiparticle lattice gases are required.

# Chapter 3

# Cause of non-Galilean Invariance and Incompressibility of Lattice Gases

---

What did you find?
*Darkness, sire.*
Describe it.
*I can't, sire.*
Did you not take the light?
*I did, sire. It didn't help.*
The source?
. . .

---

In this chapter the problems of non-Galilean invariance and incompressibility of multiparticle lattice gases have been analyzed following the conclusions of chapter 2. The objective of these investigations is to find the actual cause of these problems. In view of the discussion presented in Secs. 2.5.1.4 and 2.5.2, mathematical methods of analysis like those outlined by Wolfram [1] and Frisch *et. al.* [2] have not been used in these investigations. Instead, arguments pivoted at the findings of classical literature are used to arrive at concrete and widely applicable conclusions. Before proceeding further it is necessary to analyze and resolve certain critical issues arising from multiplicity of conceptual representation of multiparticle lattice gases (*c.f.*, Sec. 2.2.5) and also from misinterpretation of some findings about existing lattice gas literature. These are addressed in Secs. 3.1 and 3.2 and problems of non-Galilean invariance and incompressibility are analyzed from Sec. 3.3 onwards.

## 3.1 Conceptual Representation of Multiparticle Lattice Gases

In Sec. 2.2.5 two different conceptual representations of multiparticle lattice gases, namely, (i) partitioned spatial lattice representation, and (ii) multiple particle representation, were described. It was pointed out that, except for some peculiarities in them, both these representations are essentially equivalent in that all multiparticle lattice gases can be described using both of them. In Sec. 2.3 the multiple particle representation was chosen in preference to the partitioned spatial lattice representation for giving generalized definition of multiparticle lattice gases. This choice was made with a remark that the partitioned spatial lattice representation has certain unacceptable peculiarities which must be bypassed to avoid ambiguities. The objective of this section is to describe these peculiarities and to point out and clarify the ambiguities that might arise from them. The simplest way to begin in this direction is to see the equivalence of the two representations and then to look at their differences and peculiarities and ambiguities which might arise from them.

### 3.1.1 Equivalence of the Two Representations

The equivalence of partitioned spatial lattice representation (PSLR) and multiple particle representation (MPR) can be established by constructing a general procedure of transforming multiparticle lattice gases from one representation to the other and demonstrating it with an example. Any such transformation procedure will necessarily involve at least two steps, namely, (i) transformation of the structure of spatial lattice, and (ii) transformation of the evolution rules. This is because descriptions of the structure of the spatial lattice as well as the evolution rules differ in both the representations. The general procedure of transformation of multiparticle lattice gases from PSLR to MPR is described below. The procedure of transformation from MPR to PSLR can be constructed in a similar manner.

**Procedure for Transformation of Multiparticle Lattice Gases from PSLR to MPR:** The method of describing multiparticle lattice gases using partitioned spatial lattice representation and multiple particle representation has been described in detail in Sec. 2.2.5. With this description, the general procedure for transformation of multiparticle lattice gases from PSLR to MPR can be constructed easily as done and described below. Note that this procedure involves three major steps, namely, (i) transformation of the structure of spatial lattice, (ii) construction of collision rules in MPR (from evolution rules in PSLR), and (iii) construction of translation rules in MPR (from evolution rules in PSLR). The steps (ii) and (iii) together complete the transformation of evolution rules from PSLR to MPR. This is because in PSLR the motion of particles from one cell to another within a block and interactions among particles are computed in a single step by the evolution rules, whereas in MPR the motion and interactions are decoupled from each other and computed in two different sequential steps. As a result, the transformation of evolution rules from PSLR to MPR involves construction of the collision as well as translation rules in MPR. All these three steps have been developed with maximum possible generalization and described in detail in the following paragraphs.

**(I) Transformation of the Structure of Spatial Lattice:** In PSLR (*c.f.*, Sec. 2.2.5.1) each block can be treated as a single unit because states of particles occupying the same block are interrelated and evolution rules are applied separately to each block. In MPR (*c.f.*, Sec. 2.2.5.2) each lattice site can contain multiple particles and interactions occur only among particles occupying the same lattice site. Also note that in PSLR all blocks are identical and in MPR all lattice sites are identical. This implies that the blocks in PSLR correspond to lattice sites in MPR. Furthermore, in PSLR no two blocks in the same partition overlap, blocks in different partitions are slightly offset relative to each other, and no two blocks in different partitions overlap completely. Thus, transformation of the structure of spatial lattice from PSLR to MPR can be accomplished by treating each block in each partition as a single unit and collapsing it as a lattice site located at its centroid.

An important aspect of the above transformation procedure is that the individuality of cells comprising the blocks (in PSLR) is lost after the transformation. This is because in PSLR each cell has a unique coordinate, whereas in MPR each lattice site has a unique coordinate. As a result, as soon as a block is collapsed into a lattice site the coordinates of all the cells comprising the block become identical to the coordinates of the lattice site and the cells no longer remain distinguishable on the basis of their coordinates.

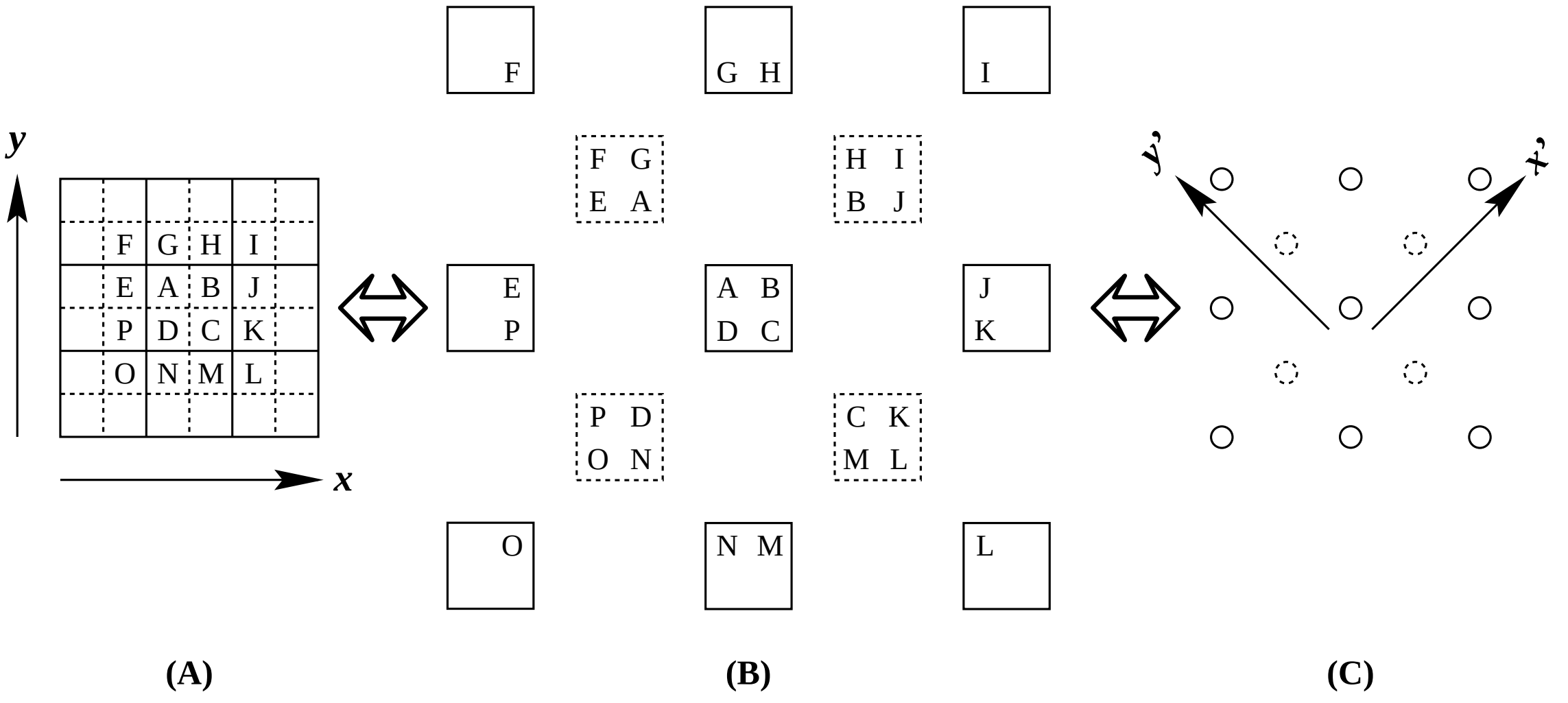

**Figure 3.1:** Geometrical description of transformation of the spatial lattice of HPP and TM gases from partitioned spatial lattice representation to multiple particle representation. The structure of the spatial lattice in partitioned spatial lattice representation is labeled (A), in multiple particle representation (C), and in an intermediate stage (B). Different line styles show blocks in different partitions and lattice sites formed from them.

Transformation of the spatial lattice of the HPP and TM gases from PSLR to MPR using the above procedure is illustrated in Fig. 3.1. In this figure, the stages (A) and (C) show the structure of the spatial lattice in PSLR and MPR, respectively. The stage (B) is an intermediate stage in which each block in each partition has been shown in a separated out fashion for the sake of additional clarity in visualization. Another example involving square spatial lattice partitioned into three partitions using blocks of $3 \times 3$ cells is shown in Fig. 3.2. The notation used in this figure is the same as that used in Fig. 3.1.

During transformation of the structure of spatial lattice from PSLR to MPR, blocks in PSLR are collapsed to lattice sites in MPR. In this process, the cells comprising each block on the spatial lattice in PSLR get replicated on as many lattice sites on the spatial lattice in MPR as the number of partitions in PSLR.[1)] This is because in PSLR each cell is necessarily contained in one block in each partition. As an example see (A) and (B) in Fig. 3.1. In this figure the cell marked A is contained once in solid partition and once in dashed partition and, as a result, gets replicated on lattice sites coming from these partitions. Another example of square spatial lattice partitioned into three partitions using blocks of $3 \times 3$ cells is shown in Fig. 3.2. In this example each cell comprising the blocks on spatial lattice in PSLR gets replicated on three lattice sites on spatial lattice in MPR when the structure of spatial lattice is transformed from PSLR to MPR, *e.g.*, the

[1)]The replication of each cell comprising the spatial lattice in PSLR on many lattice sites comprising the spatial lattice in MPR implies that system in MPR consists of multiple replicas of the basic system in PSLR. This means that simulation of a system using MPR requires more memory compared to that using PSLR. If fact, if the spatial lattice in PSLR consists of $N$ partitions, then each cell gets replicated on exactly $N$ lattice sites on the spatial lattice in MPR and simulation of the system using MPR requires exactly $N$ times more memory compared to that required when PSLR is used.

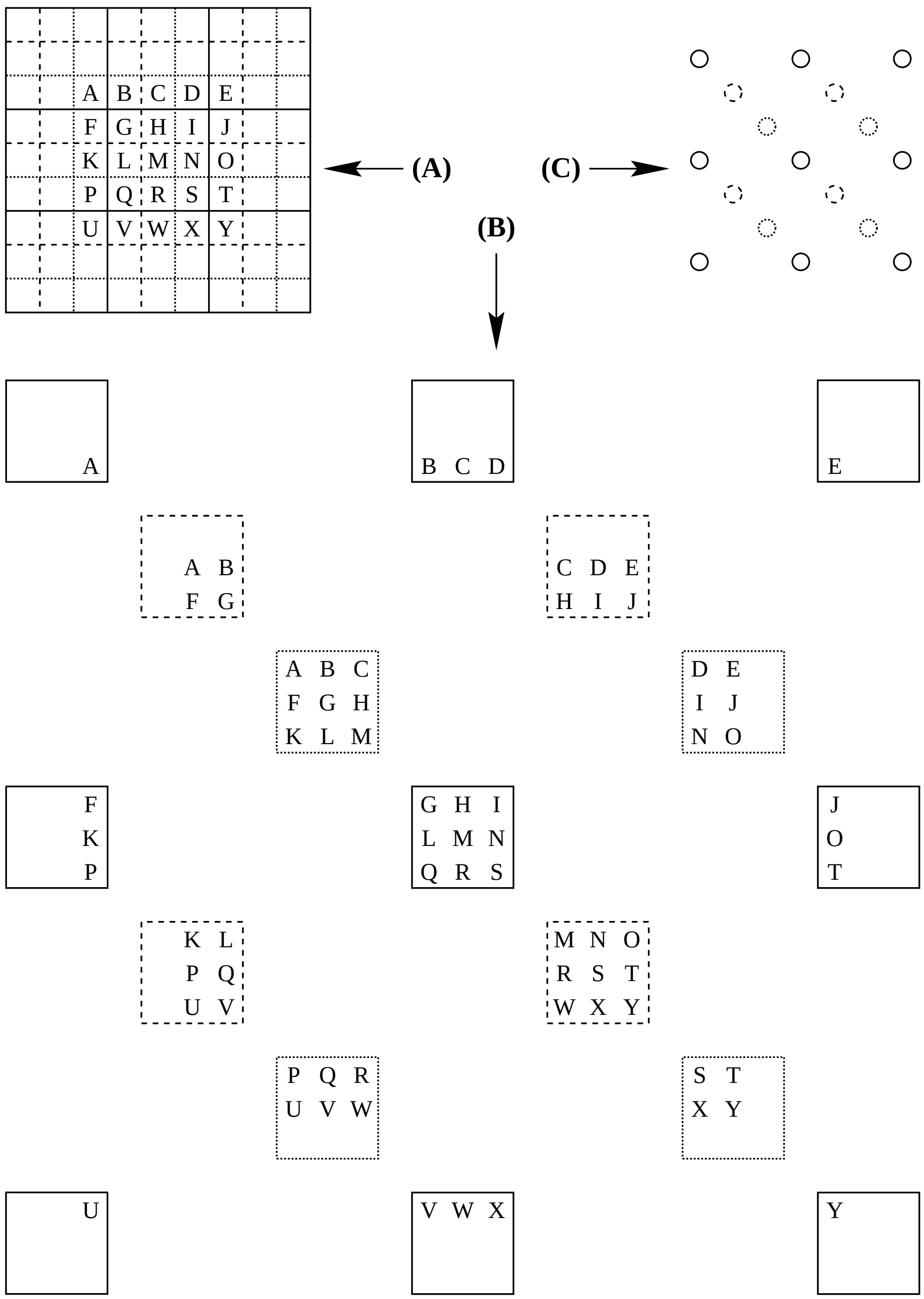

**Figure 3.2:** Geometrical description of transformation of square spatial lattice, partitioned into three partitions using blocks of $3\times3$ cells, from partitioned spatial lattice representation to multiple particle representation. The structure of the spatial lattice in partitioned spatial lattice representation is labeled (A), in multiple particle representation (C), and in an intermediate stage (B). Different line styles show blocks in different partitions and lattice sites formed from them.

cell marked A is contained in exactly one block in each one of solid, dotted, and dashed partitions, and gets replicated in the lattice sites coming from blocks in these partitions.

**(II) Construction of Collision Rules in MPR:** In MPR evolution during each time step is carried out by sequentially applying the rules of collision and translation steps to all the lattice sites (*c.f.*, Sec. 2.2.5.2), and then repeating the procedure for the next time step. In PSLR evolution during each time step is carried out by selecting a partition and applying evolution rules of that partition to all the blocks in that partition (*c.f.*, Sec. 2.2.5.1), and then repeating the procedure for the next time step. The important point to be noted about these two evolution procedures is that in MPR both collision and translation rules change the states of lattice sites, whereas in PSLR the states of cells change only when evolution rules are applied to the blocks and not during partition switching. Furthermore, in PSLR evolution rules are local to each block, whereas in MPR only collision rules are local to each lattice site and locality of the translation rules is spread out in a small spatial neighborhood around each lattice site. Also, while transforming the structure of spatial lattice from PSLR to MPR each block is collapsed into a lattice site. All these, when viewed together, imply that block rules in PSLR directly correspond to collision rules in MPR. As a result, evolution rules in PSLR can be mapped/transformed directly to collision rules in MPR.

A key point in the above transformation is that while constructing the collision rules[2)] in MPR from the evolution rules of PSLR the particles should be distinguished only by their state parameters[3)] and not by the coordinates of the cells occupied by them within a block in the evolution rules of PSLR. This is because coordinates of all the cells comprising a block in PSLR become identical when the blocks are collapsed into lattice sites during transformation of the structure of spatial lattice from PSLR to MPR. As a result, before the transformation can be carried out it is necessary to determine the state parameters of particles occupying various cells in the initial and final states of blocks in PSLR. The general procedure of determining the initial and final states of particles encoded into the evolution rules of PSLR is identical for all the state parameters. This procedure is demonstrated below with reference to determination of velocity of particles in arbitrary evolution rules.

The velocity of particles occupying various cells in the initial and final states of blocks in the evolution rules for a partition in PSLR can be deduced easily from those block rules (of that partition) in which there is only one particle inside the blocks. This is because these evolution rules encode free motion of particles and leave their state parameters unchanged.[4)] In such an evolution rule, let $\boldsymbol{x}_\alpha^{(\mathrm{I})}$ be the coordinate of a particle (or the cell occupied by it) in the initial state of a block and $\boldsymbol{x}_\beta^{(\mathrm{F})}$ be its coordinate in the final state of the block.[5)] Then, in natural units of the lattice system velocity $\boldsymbol{v}$ of the particle is

---

[2)] This involves identification of the initial and final states of lattice sites and construction of the symbols representing these states.

[3)] State parameters for particles are species, velocity, charge, spin, *etc.* Specification of species involves specification of mass and pseudo parameters if any, *e.g.*, color. At times, as in many well established theories, energy and momentum are considered as state parameters instead of mass and velocity. Specifically, the coordinates of particles in physical position space are not part of their state parameters.

[4)] This statement is true only in the absence of external fields. Multiparticle lattice gases incorporating the effect of external fields, although plausible, are not available in the literature. Such cases will not be discussed in this investigation either.

[5)] Here, the superscripts (I) and (F) have been used to explicitly indicate that the coordinates are with reference to the initial and final states of a block. This is because the unsuperscripted coordinates $\boldsymbol{x}_\alpha$ and $\boldsymbol{x}_\beta$ are valid coordinates in both the initial as well as the final state and as a result can cause confusion.

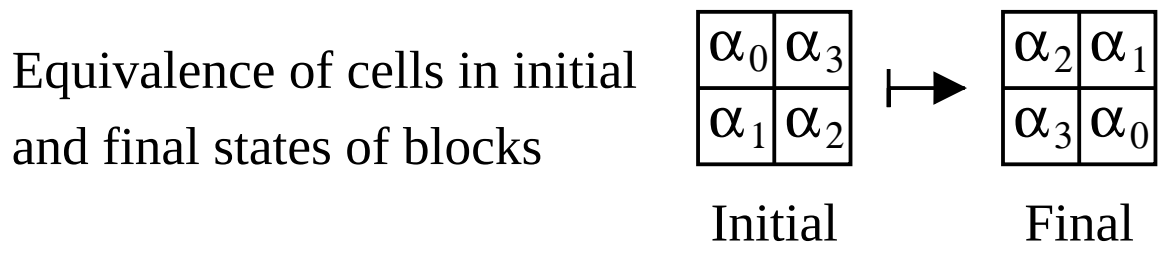

**Figure 3.3:** Equivalence of cells in the initial and final states of blocks in partitioned spatial lattice representation of the HPP gas on the basis of equivalence of velocity of particles occupying them. Cells marked with the same symbol in the initial and final states of the block are equivalent in that the state (here, mass and velocity) of the particle occupying them is the same.

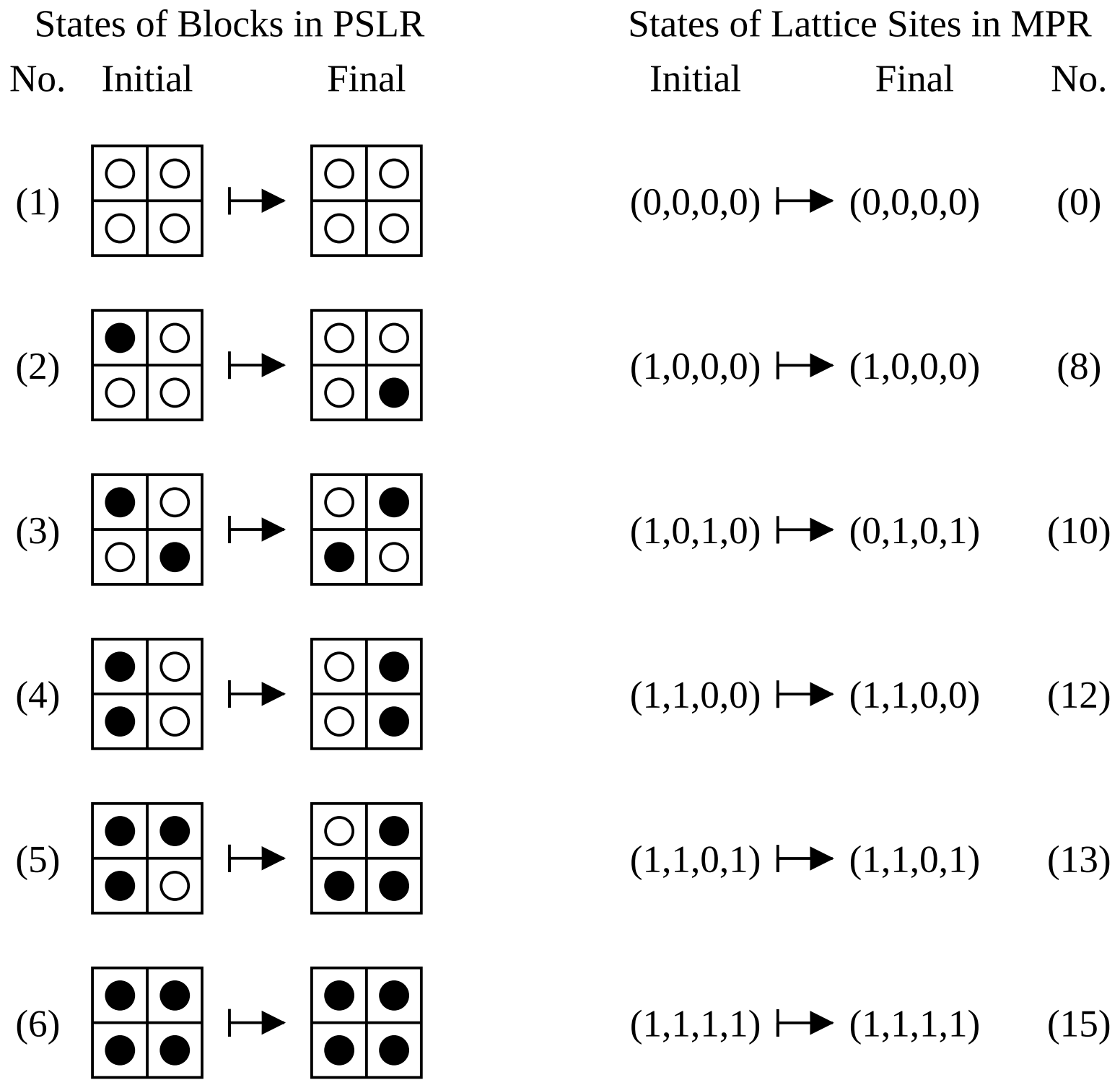

**Figure 3.4:** Transformation of block rules in partitioned spatial lattice representation to collision rules in multiple particle representation for the HPP gas. The evolution rules in partitioned spatial lattice representation are reproduced from Fig. 2.10. The collision rules in multiple particle representation are written using the symbolism used in table 2.7.

given by $\boldsymbol{v} = \boldsymbol{x}_\beta^{(\mathrm{F})} - \boldsymbol{x}_\alpha^{(\mathrm{I})}$. In fact, if any particle occupies these cells (*i.e.*, the cell located at $\boldsymbol{x}_\alpha^{(\mathrm{I})}$ in the initial state and/or the cell located at $\boldsymbol{x}_\beta^{(\mathrm{F})}$ in the final state of a block) in any evolution rule for that partition in PSLR its velocity is necessarily $\boldsymbol{v}$. Thus, in terms of velocity of particles, the cell located at $\boldsymbol{x}_\alpha^{(\mathrm{I})}$ in the initial state and at $\boldsymbol{x}_\beta^{(\mathrm{F})}$ in the final state of a block in the evolution rules are equivalent. Note that this equivalence is valid only for the partition(s) on which the evolution rules are applied. This is because in PSLR different partitions can have different evolution rules as in the TM gas (*c.f.*, Sec. 2.2.6.4). For the HPP gas this equivalence is shown in Fig. 3.3. Transformation of the evolution rules in PSLR to collision rules in MPR for the HPP gas using the symbolism defined in Sec. 2.2.6.2 is shown in Fig. 3.4.

**(III) Construction of Translation Rules in MPR:** In MPR, in the translation step particles are simply repositioned on new lattice sites as pointed out by their velocity vectors. Development of rules for this step (*i.e.*, the translation rules) is straight forward if velocities of particles are known (*c.f.*, Secs. 2.2.5.2 and 2.3). As a result, it might appear that further analysis will not be needed for construction of translation rules in MPR if evolution rules in PSLR have already been transformed to collision rules in MPR because velocities of particles[6)] would already have been determined for carrying out this transformation. This, however, is not the case because while transforming the evolution rules in PSLR to collision rules in MPR the velocities of particles are determined (i) only over the spatial lattice in PSLR, and (ii) only for identifying the states of particles occupying various cells in the initial and final states of blocks in evolution rules of PSLR.[7)] Furthermore, since the structure of spatial lattice in PSLR and MPR is in general different (see for example Fig. 3.1 and for a more explicit example Fig. 3.2), the velocities of particles over the spatial lattice in PSLR are expected to be different from those in MPR. As a result, velocities of particles over spatial lattice in MPR have to be determined afresh. This is done as follows:

In PSLR particles move when evolution rules are applied to blocks in currently selected partition. In this process, particles move from one cell to another within each block but remain confined within the blocks. The evolution during one time step in PSLR, however, is said to be completed when new partition has been selected for the next evolution. As a result, when evolution for a time step is over the particles have actually moved to blocks in the next partition. Since the blocks in different partitions in PSLR give rise to different lattice sites in MPR, partition switching in PSLR is equivalent to motion of particles from one lattice site to another in MPR. This has been illustrated in intermediate states (B) shown in Figs. 3.1 and 3.2. This implies that velocity of particles in MPR can be computed easily from partition switching scheme and block evolution rules in PSLR. This is done as follows: Let coordinates of blocks in different partitions in PSLR be identified by the coordinates of their centroids. Let a particle occupy the block located at $\boldsymbol{x}_\alpha$ in the currently selected partition and the block located at $\boldsymbol{x}_\beta$ in the next partition selected according to partition selection scheme. Then, the velocity $\boldsymbol{v}$ of particle in MPR is given by $\boldsymbol{v} = \boldsymbol{x}_\beta - \boldsymbol{x}_\alpha$.

The velocity vectors of particles in the HPP gas in MPR, computed using the procedure outlined above from the partition selection scheme outlined in Sec. 2.2.6.2 and the block evolution rules in PSLR shown in Fig. 2.10, are $\{(\pm\sqrt{2}, \pm\sqrt{2})\}$. Note that these velocity vectors are in the coordinate system used for identifying the coordinates of cells comprising the spatial lattice in PSLR as shown in (A) in Fig. 3.1. If the coordinate system in MPR is changed, these velocity vectors will also have to be changed accordingly. For the HPP gas the coordinate system used in MPR is rotated by 45° (either clockwise or counter clockwise) relative to that used in PSLR as shown (B) in Fig. 3.1. In this coordinate system the velocity vectors of particles in the HPP gas transform to $\{(\pm 1, 0), (0, \pm 1)\}$.

### 3.1.2 Peculiarities of Partitioned Spatial Lattice Representation

As far as description of multiparticle lattice gases is concerned PSLR and MPR are equivalent as has been shown in Sec. 3.1.1. Despite this equivalence, PSLR has certain inherent

[6)] In fact, all the state parameters of particles.

[7)] This done so that unique symbols can be assigned to lattice sites in different states for constructing the collision rules in MPR from the evolution rules in PSLR (see the two preceding paragraphs).

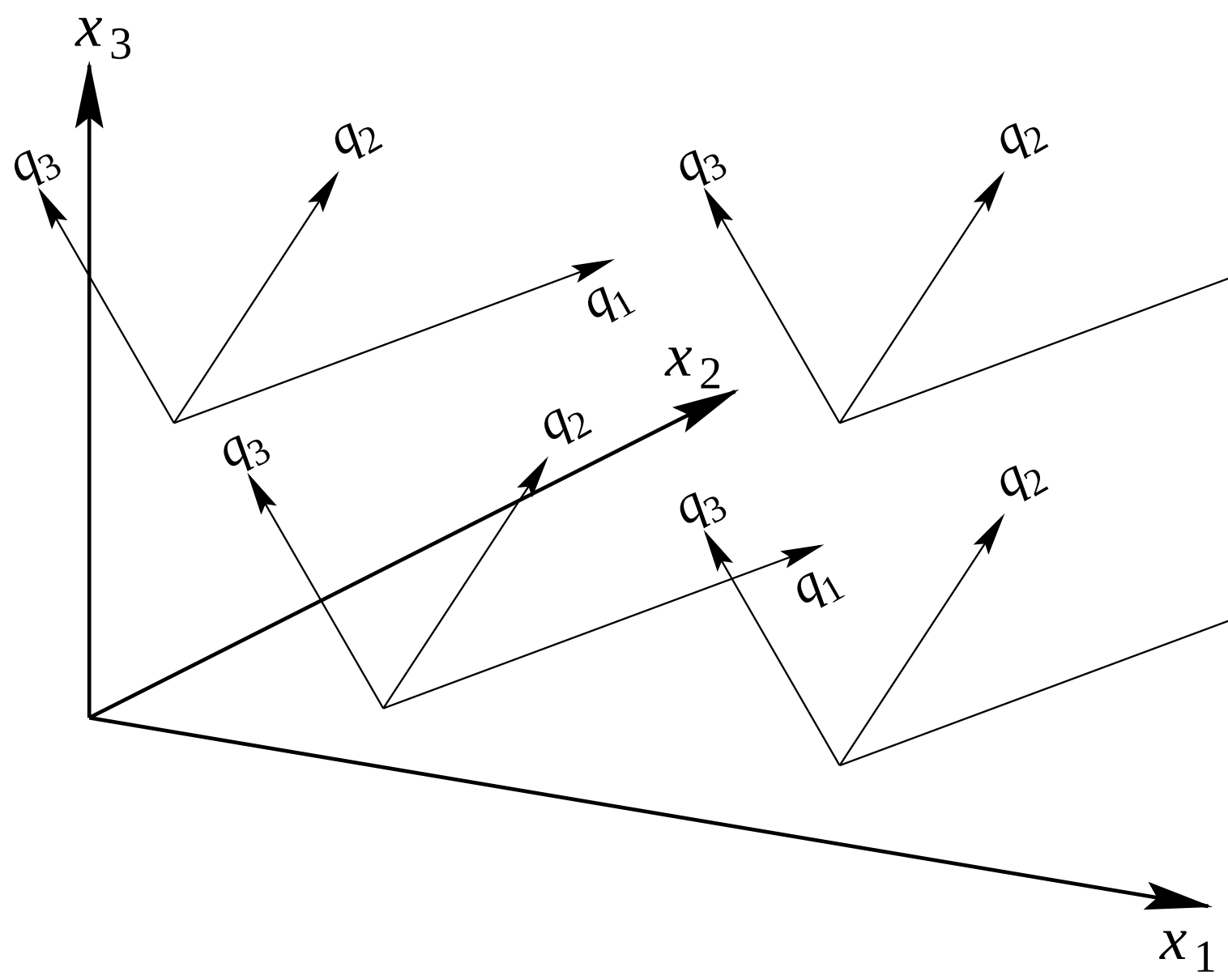

**Figure 3.5:** Structure of phase space of single particle in physical systems in terms of the coordinate systems describing its location in the physical position space and the momentum space.

peculiarities which make it different from MPR in subtle manner. These peculiarities arise from differences in the structure of phase space and value of impact parameter in the two representations. The details on these are as follows:

#### 3.1.2.1 Structure of Phase Space

In Lagrangian description of a system of $N$ particles the phase space is a $6N$-dimensional structure. In this phase space the entire system is described by a single coordinate point consisting of the position and momentum coordinates of all the particles taken together. This phase space cannot be visualized easily because of its dimensionality. The aspect that is of importance at present, however, can be visualized in a simple manner by visualizing the phase space of a single particle as follows:

Consider a physical position space (simplest structure being the orthogonal Cartesian coordinate system) in which particles move from one location to another. Let, for simplicity, all particles be identical hard spheres. Then the phase space of a single particle in this system can be viewed as its momentum space superimposed over its physical position space. In terms of coordinate systems used for describing the location on the particles in the physical position space and momentum space, one can visualize the structure of the phase space of the particle as a coordinate system for the position space in which coordinate system for the entire momentum space is fixed at every point as shown in Fig. 3.5. An essential property that readily comes out from this phase space is that a particle can occupy any point in the physical position space and have any momentum. Alternatively, all values of momentum are permissible at all points in the physical position space and there is no correlation between the position and momentum coordinates of particles.

In multiparticle lattice gases both the physical position space and the momentum space are discrete and the momentum space has finitely many coordinate points. Even in this case, observations about the structure of phase space outlined above should hold with appropriate modifications arising out of the discreteness and finiteness. Thus, in two-dimensional systems in which particles of unit mass move with any one of $\mathcal{N}_{\mathrm{v}}$ different velocity vectors over square spatial lattice the momentum space[8)] should have $\mathcal{N}_{\mathrm{v}}$ coordinate points at every coordinate point in the physical position space for every particle. For one particle, this phase space can be visualized schematically as in Fig. 3.5. Such a visualization, however, does not give much insight because total number of dimensions in the phase space exceed three and as a result it cannot be visualized well on two-dimensional surfaces.

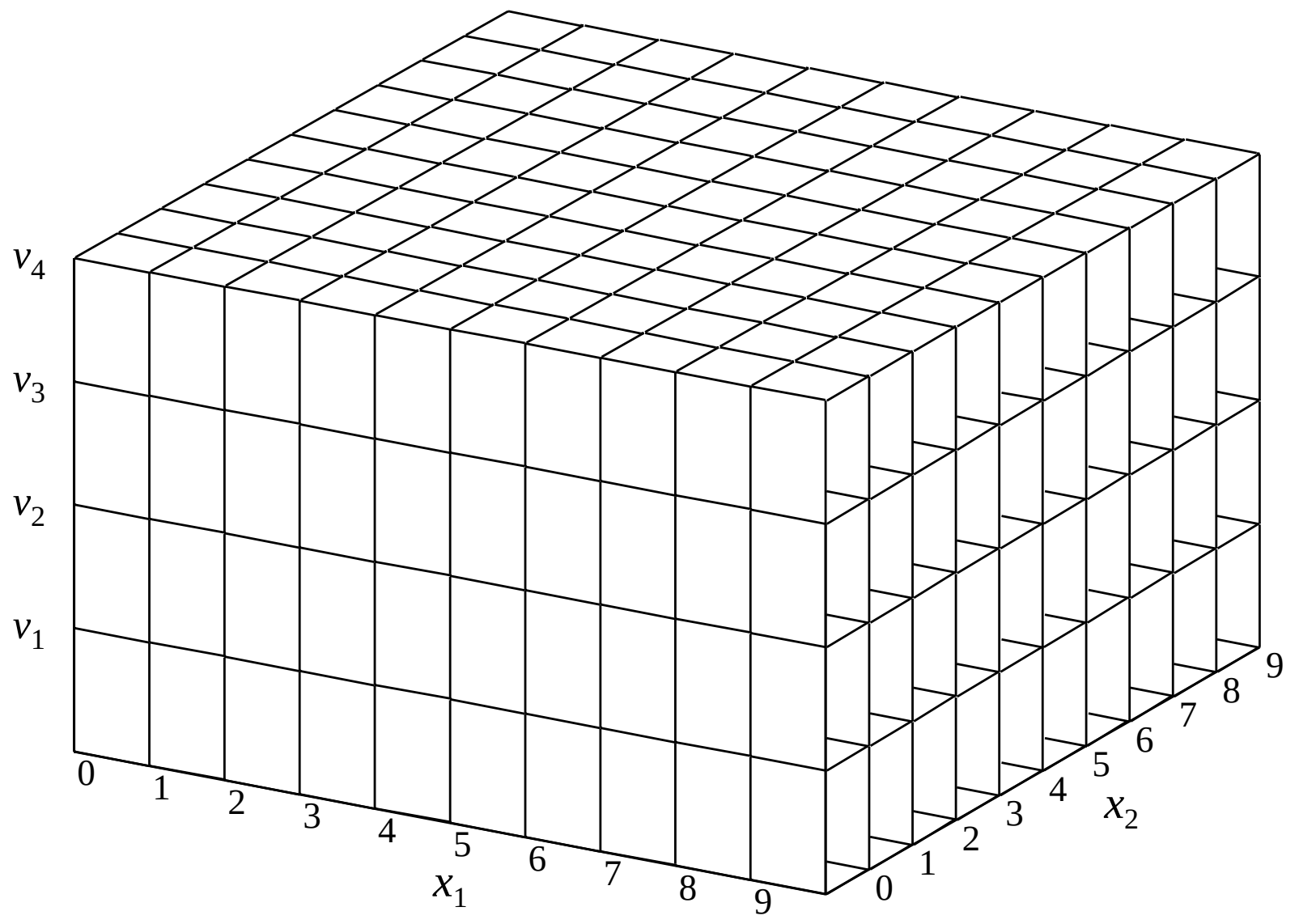

**Figure 3.6:** Structure of the phase space of single particle moving with any one of the velocities $\boldsymbol{v}_1$, $\boldsymbol{v}_2$, $\boldsymbol{v}_3$, and $\boldsymbol{v}_4$ in physical systems existing over square spatial lattice in terms of the coordinate systems describing its location in the physical position space and the momentum space.

A simpler and clearer visualization of the phase space of such systems is possible by ordering the velocity vectors in an arbitrary sequence and assigning unique number to them, *e.g.*, as $\boldsymbol{v}_i$, $1 \leq i \leq \mathcal{N}_{\mathrm{v}}$. If these numbers are interpreted as "height" or "levels" above surface of the spatial lattice, the phase space can be visualized as a three-dimensional bar chart. In this bar chart, a bar of height $i$ drawn at a lattice site on the spatial lattice means that a particle located at this lattice site can have the velocity $\boldsymbol{v}_i$. If more than one bar is drawn at some lattice site, it means that particle located at this lattice site can have any one of the velocity vectors corresponding to the heights of the bars.

For discrete systems in which only one particle can occupy a lattice site at a time step and the particles are allowed to move with four velocity vectors $\boldsymbol{v}_1$, $\boldsymbol{v}_2$, $\boldsymbol{v}_3$, and $\boldsymbol{v}_4$ over square spatial lattice, the phase space visualized using the procedure outlined above should appear as shown in Fig. 3.6 if the system has been described correctly. In this figure there are four bars at every lattice site on the spatial lattice. This shows that particles moving with any one of the four velocities can occupy any lattice site and that the location of particles in the position space and velocity (or, momentum) space are not correlated.

Structure of the phase space of both the HPP and TM gases in PSLR, in view of the description of these lattice gases given in Secs. 2.2.6.2 and 2.2.6.4, is as shown in Fig. 3.7.

8) Here, the momentum space and velocity space are equivalent.

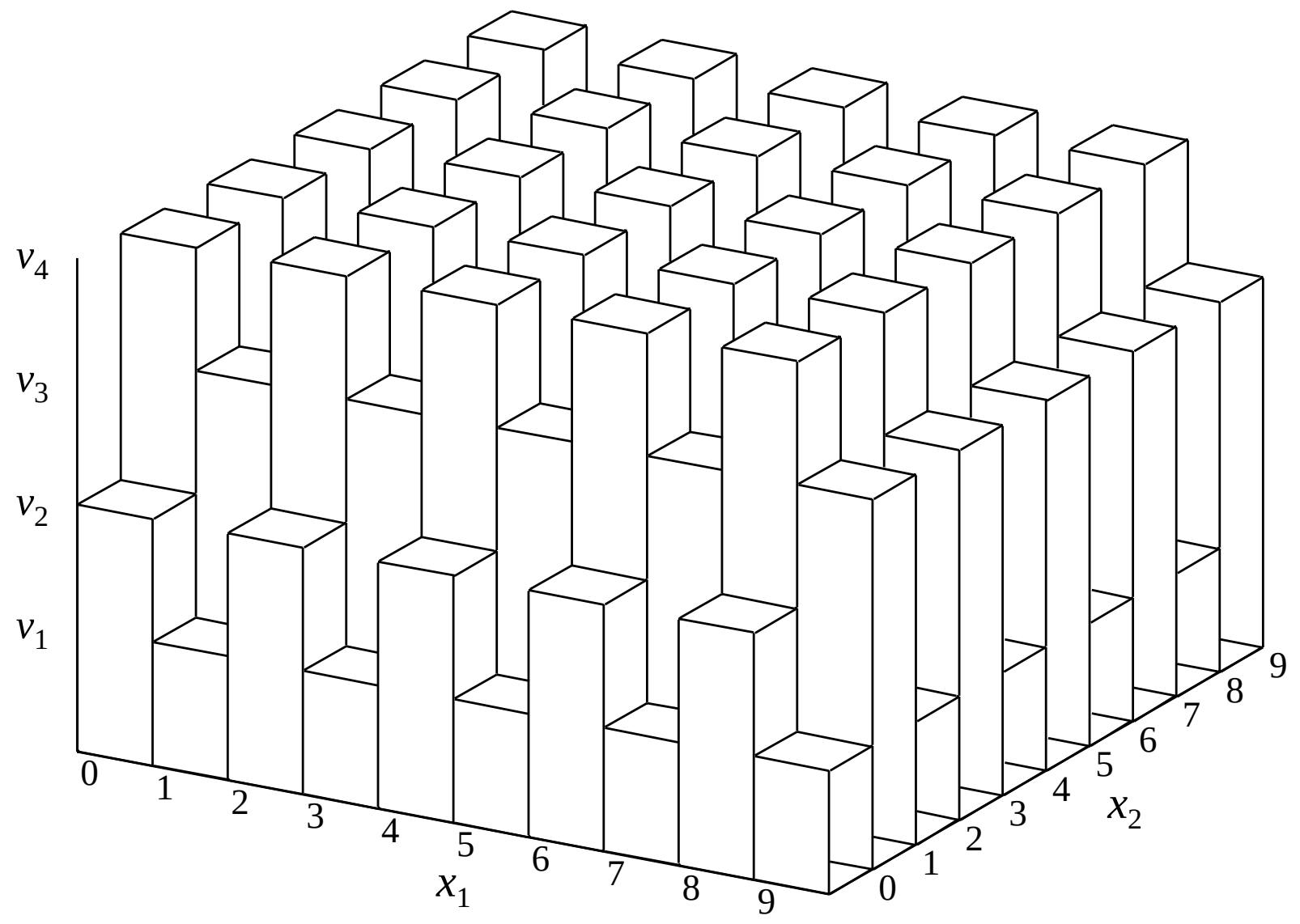

**Figure 3.7:** Structure of the phase space of single particle in HPP and TM gases in PSLR in terms of coordinate systems describing its location in the physical position space and momentum space. The ordering of velocity vectors used here is shown in table 3.1.

| $i$ | $\boldsymbol{v}_i$ | HPP gas both partitions | TM gas solid partition | TM gas dashed partition |
|---|---|---|---|---|
| 1 | $\boldsymbol{v}_1$ | $(-1,1)$ | $(-1,0)$ | $(0,1)$ |
| 2 | $\boldsymbol{v}_2$ | $(1,1)$ | $(0,1)$ | $(1,0)$ |
| 3 | $\boldsymbol{v}_3$ | $(-1,-1)$ | $(0,-1)$ | $(-1,0)$ |
| 4 | $\boldsymbol{v}_4$ | $(1,-1)$ | $(1,0)$ | $(0,-1)$ |

**Table 3.1:** Ordering of velocity vectors for the HPP and TM gases used in Fig. 3.7 corresponding to their evolution rules in PSLR shown in Figs. 2.10 and 2.13, respectively.

The ordering of velocity vectors for the HPP and TM gases used in this figure is given in table 3.1. An important observation from Fig. 3.7 is that in both the HPP and TM gases in PSLR particles having specific velocities can occupy only specific cells. Alternatively, in these lattice gases in PSLR the location of particles in position and velocity spaces are correlated in that the particle occupying the cell located at $(i, j)$ on the spatial lattice can only (or, must necessarily) have the velocity $\boldsymbol{v}_k$, where $k = 2 - (i \bmod 2) + 2(j \bmod 2)$. This correlation makes the structure of phase space in these lattice gases different from that of a correctly described system as is seen clearly from Figs. 3.6 and 3.7.

The above implies that systems in which particles move with four velocities over square spatial lattice and only one particle occupies a cell (or, lattice site) at a time step are not described completely by both the HPP and TM gases in PSLR. An obvious consequence of this incompleteness is that the dynamics produced by the HPP and TM gases will necessarily be incomplete and different from the correct dynamics of such systems (with particle velocities corresponding to those in the HPP and TM gases).

The above analysis and its conclusion holds good for all multiparticle lattice gases described using PSLR. It shows that description of lattice systems in multiparticle lattice gases using PSLR, though computationally efficient in terms of memory requirements, is physically inconsistent if the cells are distinguished and assigned separate identities on the basis of their coordinates. This inconsistency can be overcome only if all the cells comprising each block in each partition are considered together as a single unit whose location in the position space (or, on the spatial lattice) is identified by a single coordinate point.

#### 3.1.2.2 Value of Impact Parameter

It was mentioned in Sec. 3.1.1 and also shown in Sec. 3.1.2.1 that in PSLR each block must be considered as a single unit whose location must be identified by a single coordinate point, *e.g.*, the coordinate of the centroid of the block. As a result, in any analysis all the cells comprising each block must be considered together and the coordinates of cells comprising a block become meaningless. In the coordinate system for the spatial lattice used in PSLR, however, each cell (or, lattice site) comprising the spatial lattice is assigned unique coordinate. As a result, the irrelevance of the coordinates of cells in PSLR is usually ignored and each cell is considered as a single entity independent of other cells. A peculiarity which arises because of this incorrect treatment of cells and blocks in multiparticle lattice gases in PSLR is that it becomes possible to devise evolution rules in PSLR which cannot be translated to MPR keeping all the system parameters unchanged. One particular system parameter which cannot be preserved during transformation such evolution rules from PSLR to MPR is the impact parameter of particles during collisions. This occurs as follows:

In MPR interactions occur among particles occupying the same lattice site. Since all the particles occupying the same lattice site have the same coordinates, in MPR the impact parameter is always zero during collisions among particles. In PSLR also, if all the cells comprising a block are considered together as a single unit, the impact parameter is always zero. If, however, the cells in PSLR are considered as separate units it is possible to devise evolution rules in which impact parameter is non-zero. A peculiarity of such evolution rules is that if one treats the cells as separate units during simulations, the results obtained from simulations also reproduce the behavior of a system with non-zero impact parameter. An example of such evolution rules in PSLR is the evolution rules of the TM gas. These rules, naturally, cannot be transformed into MPR because in MPR the impact parameter will always be zero because of the very nature of the description.

It is noteworthy that a description of the TM gas in MPR is not available in the literature. In the following paragraphs an attempt has been made to transform TM gas from PSLR to MPR using the procedure outlined in Sec. 3.1.1. During this transformation the impact parameter of the TM gas changes to zero and in MPR it becomes identical with the HPP gas. This illustrates that (i) if the cells comprising a block in PSLR are treated as separate units based on their coordinates the TM gas cannot be transformed from PSLR to MPR, and (ii) if the cells comprising a block in PSLR are treated as a single unit then the dynamics of the TM gas in PSLR, which should have been equivalent to that of the HPP gas in PSLR (*i.e.*, with zero impact parameter), is physically inconsistent.

The above brings out an important point that extreme care should be exercised while using PSLR for implementing multiparticle lattice gases on computer for simulation studied. Despite the above mentioned peculiarities in PSLR, its use for implementing multiparticle lattice gases on computer might be desirable because it is computationally efficient compared to MPR in terms of memory requirements.

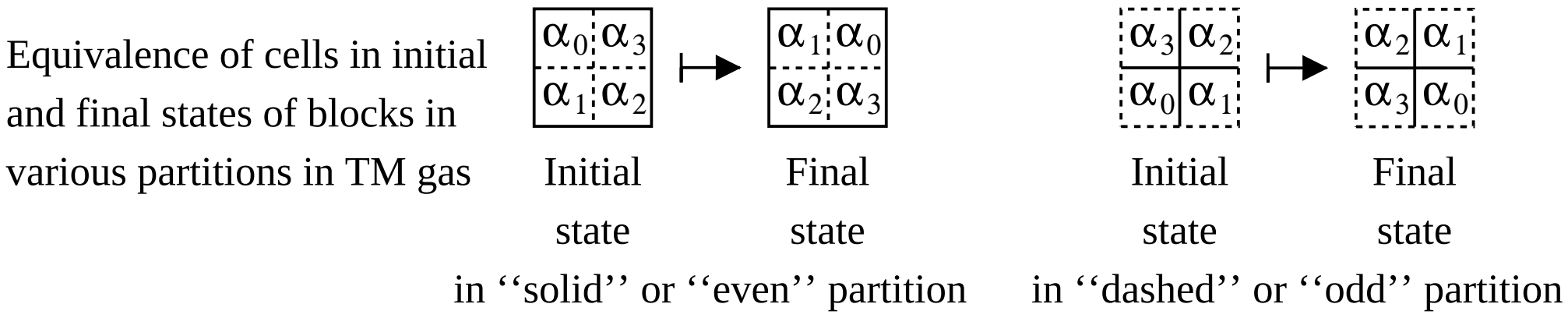

**Figure 3.8:** Equivalence of cells in the initial and final states of blocks in partitioned spatial lattice representation of the TM gas on the basis of equivalence of velocity of particles occupying them. Cells marked with the same symbol in the initial and final states of the block are equivalent in that the state (here, mass and velocity) of the particle occupying them is the same.

**Transformation of the TM gas from PSLR to MPR:** The equivalence of cells in the initial and final states of blocks in the evolution rules for both the partition for TM gas is shown in Fig. 3.8. This figure also shows cross equivalence of cells in the evolution rules for both the partitions. The equivalence of cells is evaluated in terms of the states (mass and velocity) of particles occupying them. Using this equivalence and the transformation procedure outlined in Sec. 3.1.1, the evolution rules for both the partitions in PSLR have been transformed to collision rules in MPR. The transformation is shown in Fig. 3.9. In this figure, the evolution rules in PSLR are taken from Fig. 2.13 and the collision rules in MPR have been written using the notation described in Sec. 2.2.6.2. Comparison of these collision rules with the collision rules for the HPP gas outlined in table 2.7 shows that both are identical. This has occurred, as expected and outlined in the previous paragraph, because the impact parameter of particles becomes zero during the transformation. It is worthwhile to note that although the evolution rules for both the partitions are different in PSLR, they result in identical collision rules in MPR. The numbers written by the side of the collision rules in MPR correspond to those in table 2.7 for the HPP gas in MPR.

## 3.2 Role of Collision Rules in the Dynamics of Multiparticle Lattice Gases

In the following sections doubts arise on the role of collision rules in the micro- and macrodynamics of multiparticle lattice gases. This happens because of ambiguity in the meaning of the phrase *"different collision rules"* which causes misinterpretation of certain findings from literature. The doubts are resolved if differences between usage of this phrase in the literature and in the present investigation are understood. These are as follows:

In multiparticle lattice gases, many a times, multiple final states are possible for the same initial state during collisions. Different final states for the same initial state give rise to different deterministic cellular automata rules. As a result, collision rules of multiparticle lattice gases are constructed by mixing all the cellular automata rules together. The mixing is usually done by assigning probabilities to various cellular automata rules with which they operate on lattice sites at different time steps, *e.g.*, as in the FHP gas. By controlling these probabilities, the cellular automata rules can be mixed in different desired proportions. In the lattice gas literature [1,2] collision rules obtained by mixing the cellular automata rules in various proportions are called as *"different collision rules"*. These *"different collision rules"* have different microscopic behavior, and thus, lead to different coefficients of viscosity. As far as macroscopic behavior of such collision rules is concerned, investigations carried out by Wolfram [1] and Frisch *et. al.* [2] on single speed multiparticle lattice gases show that if the rules satisfy the condition of semi-detailed balance they have identical macrodynamics in that their coarse graining leads to identical equations in the hydrodynamic limit.

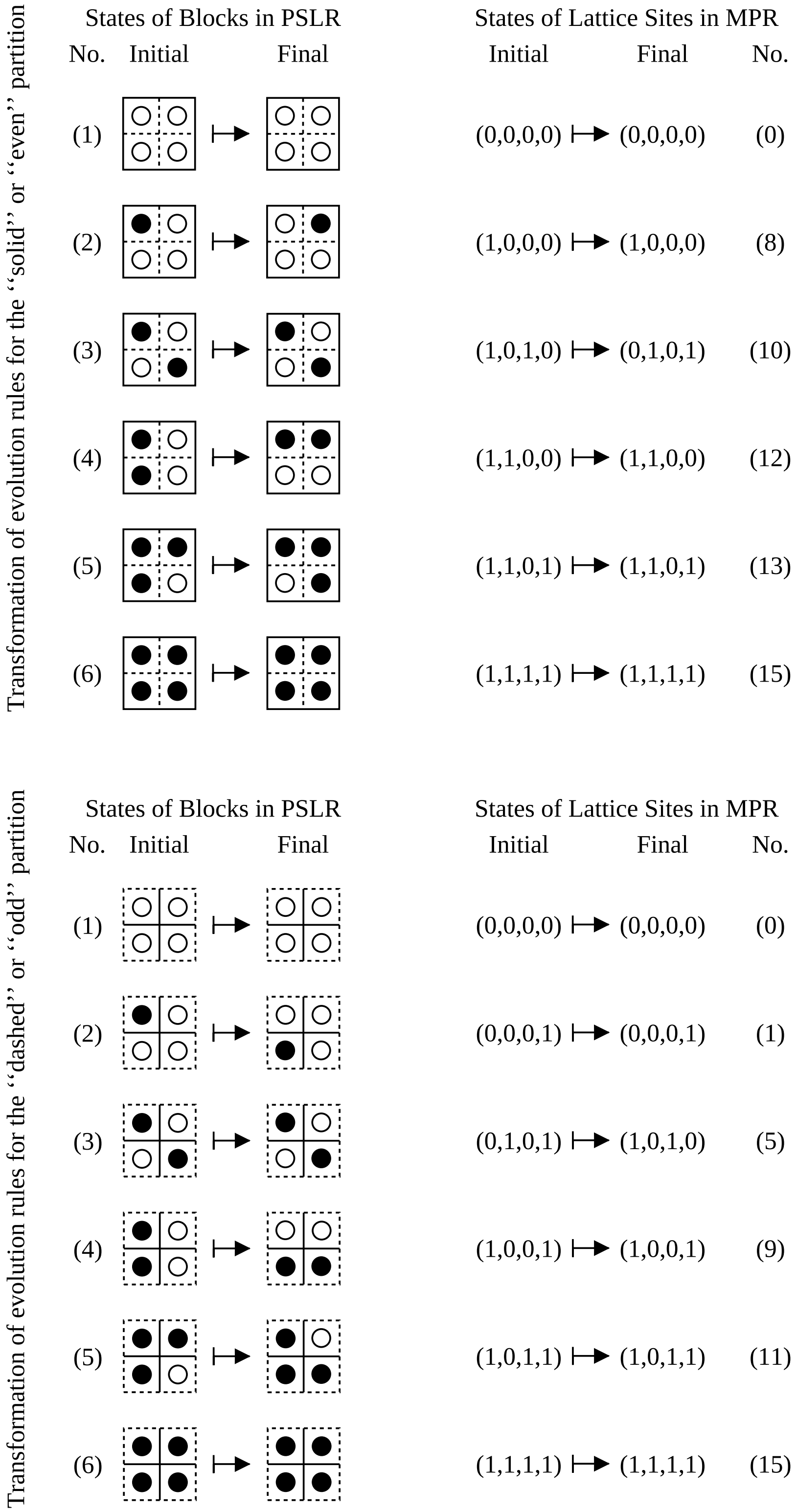

**Figure 3.9:** Transformation of block rules in partitioned spatial lattice representation to collision rules in multiple particle representation for the TM gas. The evolution rules in partitioned spatial lattice representation are reproduced from Fig. 2.13. The collision rules in multiple particle representation correspond with those for the HPP gas and are written as in table 2.7.

The above illustrates the sense in which the phrase *"different collision rules"* is used in the lattice gas literature. In the present investigation the fraction with which various cellular automata rules are mixed for constructing collision rules is also considered as a parametric element of the collision rules themselves. As a result, changes which occur in collision rules because of change in the mixing fraction are not viewed as giving rise to *"different collision rules"*. They are considered only as giving rise to different coefficients of viscosity. Because of this, in the present investigation the phrase *"different collision rules"* is used in exactly the same sense in which the phrase *"different lattice gases"* has been used in the lattice gas literature. Various aspects of this usage are described in detail below.

Two important elements in the definition of lattice gases and their collision rules are the interaction neighborhood and velocity set. Lattice gases whose interaction neighborhoods and/or velocity sets differ have *"different collision rules"* and are called as *"different lattice gases"* in the literature. Thus, the phrases *"different collision rules"* and *"different lattice gases"* are essentially equivalent and have been treated as such in the present investigation.

Different multiparticle lattice gases cannot be distinguished on the basis of their interaction neighborhood because by definition it is restricted to one lattice site in all multiparticle lattice gases (*c.f.*, Sec. 2.3). As a result, in the literature multiparticle lattice gases are distinguished solely on the basis of differences in their collision rules. In multiparticle lattice gases one velocity set results in only one set of collision rules. Thus, each velocity set is specific to exactly one multiparticle lattice gas. Moreover, all the lattice gases available in the literature are multiparticle lattice gases. As a result, current understanding is that *"different lattice gases"* must necessarily have *"different velocity sets"*.

In view of the above and in view of the fact that all the lattice gases available in the literature are multiparticle lattice gases, presently it appears meaningless to talk about *"different lattice gases"* having the same velocity set or equivalently about *"different collision rules"* developed over the same velocity set. In principle, however, it is possible to vary the size of interaction neighborhood and develop *"different lattice gases"* having *"different collision rules"* but the same velocity set. With this view, it is quite possible that collision rules affect the micro- and macrodynamics of lattice gases even though they are developed over the same velocity set.

## 3.3 Investigation on the Cause of non-Galilean Invariance and Incompressibility of Multiparticle Lattice Gases

The cause of non-Galilean invariance and incompressibility of multiparticle lattice gases has been investigated in this section. The literature (*c.f.*, chapter 2) does not suggest any appropriate starting point for these investigations. As a result, an *ab initio* analysis of lattice gases has been carried out. In this analysis, the basic elements and process of construction of lattice gases (not just multiparticle lattice gases) have been scrutinized to identify their relationship with the dynamics of lattice gases. Following this, the role played by these elements in the dynamics of multiparticle lattice gases has been analyzed to determine the exact cause of non-Galilean invariance and incompressibility of multiparticle lattice gases.

### 3.3.1 Basic Elements and Process of Construction of Lattice Gases

Lattice gases are constructed by associating specific forms and structures with a number of basic elements or components. These elements are: (i) spatial lattice, (ii) symbol set for particles of each species, (iii) velocity set for particles of each species, (iv) interaction neighborhood, and (v) evolution rule defined over (i)–(iv). If evolution during one time step is decomposed into collision and translation steps, as in multiparticle lattice gases, construction of the evolution rules involves construction of the collision and translation rules. Such a decomposition, however, is not necessary and many times may not be possible. In terms of the basic elements of cellular automata, the symbol set and velocity set for particles of each species allow construction of symbols representing the states of lattice sites.

Construction of lattice gases in terms of the five basic elements mentioned above can be viewed as exact and detailed specification of the elements. This process can be described as an algorithmic process consisting of five major steps. A broad outline of these steps showing their interrelationship (and relationship with some basic elements of physical systems) is as follows:

**(i)** Structure of the spatial lattice is selected according to the requirements of the system to be simulated. The specification of the spatial lattice structure involves specification of the basic geometrical arrangement of lattice sites, *e.g.*, in two-dimensions it could be square, triangular, hexagonal, or like the one shown in Fig. 3.2(C), *etc.* The connectivity of lattice sites is not specified at this stage. The specification of connectivity involves specification of the number and location of lattice sites to which each lattice site is connected. As a result, it gets specified when velocity sets for particles of each species are specified (which is done in the step (iii) below).

**(ii)** Necessary symbols for representing particles of each species are selected as required by the system to be simulated. For this one species is associated with all the particles having the same properties, *i.e.*, the same mass, charge, spin, *etc.*

**(iii)** Velocity sets are constructed for each particle species. This is done by specifying the velocity vectors with which particles of each species can move over the spatial lattice. The construction of velocity sets is necessarily carried out within the constraints imposed by the structure of the spatial lattice. For example, for square spatial lattice the permissible elements of the velocity set must necessarily be members of the set $\{(\pm i, \pm j)\}$, where $i$ and $j$ are integers. Specification of the velocity set automatically leads to specification of the connectivity of lattice sites. For example, if $(i, j)$ is an element in the velocity set then a lattice site $(x_0, y_0)$ is necessarily connected to the lattice sites $(x_0 + i, y_0 + j)$ and $(x_0 - i, y_0 - j)$. This is because of Newton's first law which requires that a particle in free motion with velocity $(i, j)$ occupying the lattice site $(\alpha, \beta)$ must be able to move to the lattice site $(\alpha + i, \beta + j)$ in the next evolution.

**(iv)** The interaction neighborhood of particles, *i.e.*, the spatial zone around a lattice site within which the particle positioned at that lattice site interacts with other particles, is specified. In general this is coupled with velocity sets of particles. This coupling, however, can be removed by imposing external constraints. For example, for any velocity set the interaction neighborhood can be reduced to one lattice site by allowing multiple particles to occupy the same lattice site simultaneously or by partitioning the spatial lattice appropriately as in multiparticle lattice gases. As a result, the specification of the interaction neighborhood is largely a matter of choice.

**(v)** The evolution rule through which the lattice gas comprising of the above components evolves in time is constructed.[9] The construction of evolution rule is carried out within the constraints imposed by all the laws governing the dynamics of the system to be simulated. At this point it is noteworthy that after specifying the elements (i)–(iv) many times it may not be possible to construct an evolution rule that leads to a dynamics which is consistent with physical observations and at times it may not be possible to construct an evolution rule at all, especially when the constraints imposed by the conservation laws—specifically the laws of conservation of mass, momentum, and energy—must necessarily be satisfied. If such a thing happens, entire exercise has to be repeated starting from the step (iii) onwards (this is because the output of the steps (i) and (ii) is dictated solely by requirements of the system to be simulated).

### 3.3.2 Elements to be Investigated and Method of Investigation

From Sec. 3.3.1 it is clear that among the five basic elements of lattice gases the externally controllable elements are the velocity set, interaction neighborhood, and evolution rule. The nature of the other elements, *viz.*, (structure of) the spatial lattice and symbol set for particles of each species, is dictated solely by the simulation requirements and the nature of system to be modeled. As a result, in order to develop lattice gases for a given system, one has freedom to play around only with the velocity set, interaction neighborhood, and evolution rule for making the dynamics of the lattice gases conform with that of experimental observations. This implies that problems with multiparticle lattice gases must also be related in some or the other way to these three elements only.[10] Thus, to find the cause of the problems with multiparticle lattice gases, the role of these three elements in the dynamics of multiparticle lattice gases needs to be investigated rigorously.

The parameter space provided by the velocity set, interaction neighborhood, and evolution rule is too large to be amenable to systematic analysis. More so because no rigorous method for addressing this entire parameter space is available at present. Thus, the problem needs to be simplified further. For this note that if one desires to develop lattice gases for specific purposes, like simulation of athermal systems, the selection of the velocity set

---

[9] If evolution through one time step is decomposed in many sub-steps, then construction of evolution rule implies that the evolution rules for each sub-step are constructed. Mathematically, if the operator leading to evolution through one time step $\mathcal{E}$ is decomposed in $N$ suboperators $\mathcal{E}_i$, $i = 1, \dots, N$, as $\mathcal{E} = \mathcal{E}_1\mathcal{E}_2 \cdots \mathcal{E}_N$, then construction of evolution rules for $\mathcal{E}$ implies that rules for all $\mathcal{E}_i$, $i = 1, \dots, N$, are constructed.

Here it is assumed that the method of construction of evolution rules for $\mathcal{E}$ or $\mathcal{E}_i$, as the case may be, is known; which, except for multiparticle lattice gases, is not the case at present. It has been seen in chapter 2 that multiparticle lattice gases are problem laden and as yet have not proved to be good enough models of physical systems. Thus, rigorous method of construction of evolution rules (or, lattice gases) for a given dynamics (or, physical system) is not known at present. In fact, to devise a rigorous method of construction of evolution rules (or, lattice gases) for given dynamics is the objective of this investigation (*c.f.*, Sec. 1.4).

[10] This is under the assumption that the structure of spatial lattice and symbol set for particles of each species have been selected as dictated by the system.

can also be taken to be dictated by the simulation requirements alone. This leaves the interaction neighborhood and the evolution rule as the only elements with which one can play around to develop lattice gases with desired dynamical behavior.

In multiparticle lattice gases the evolution rule is decomposed into two subrules known as the *"collision rule"* and *"translation rule"* (*c.f.*, Sec. 2.3). The translation rule, by definition, leads to pure streaming of particles and can either precede or succeed the collision rule without any effect on the resulting macrodynamics although the microdynamics is different in both the cases. Furthermore, it is well known that pure streaming of particles leaves their distribution function unchanged [3]. This implies that for studying the effect of evolution rule on the macrodynamics of lattice gases it is not necessary to consider the translation rule part of the evolution rule at all.

The above implies that the cause of non-Galilean invariance and incompressibility of multiparticle lattice gases must be related to the structure of the interaction neighborhood and collision rules. Since, in general, collision rules have to be defined over some preselected interaction neighborhood only, these two cannot be said to be independent of each other. As a result, in any investigation the effect of both of them must be analyzed simultaneously. This task is undoubtedly very complex if attempted in the obvious manner wherein all the possible collision rules defined over all the possible interaction neighborhoods are analyzed. In multiparticle lattice gases, however, this task becomes simplified because the interaction neighborhood is always restricted to one lattice site.

Thus, I propose the hypothesis that the structure and construction of collision rules and interaction neighborhood has some role to play in the incompressibility and non-Galilean invariance of multiparticle lattice gases. In view of this hypothesis, the role of interaction neighborhood and collision rules in the dynamics of multiparticle lattice gases has been analyzed from the next section onwards. The results of this analysis will, naturally, also serve as a check of validity of this hypothesis.

### 3.3.3 Collision Dynamics in Physical Systems and Multiparticle Lattice Gases

Problems with multiparticle lattice gases reflect in the form of departure in their macrodynamics from that observed in physical systems. Furthermore, it has been hypothesized in Sec. 3.3.2 that this departure is probably because of the structure of collision rules and interaction neighborhood in multiparticle lattice gases. Collision rules encode the dynamics of collisions. This indicates that collision dynamics in multiparticle lattice gases is likely to be different from that in physical systems. In view of this, the dynamics of collisions in physical systems and in multiparticle lattice gases has been analyzed and compared below (in Secs. 3.3.3.1 and 3.3.3.2). This analysis reveals that collision dynamics in multiparticle lattice gases is drastically different from that in physical systems. The relationship of these differences with non-Galilean invariance and incompressibility of multiparticle lattice gases has been analyzed in Secs. 3.3.4 and 3.3.5.

#### 3.3.3.1 Collision Dynamics in Physical Systems

In physical systems collisions occur among particles occupying different locations (points) in physical space. Here, for simplicity, the location of a particle is being identified with the coordinate of its center of mass. Consider a system of identical spherical particles with binary interaction potential $\phi(r; a_0, a_1, \cdots, a_k)$, where $r$ is the distance between two particles and $a_i$, $i = 0, \ldots, k$, are parameters dependent on nature of particles. For known binary interactions potentials $\phi \to \infty$ as $r \to 0$ and $\phi \to 0$ as $r \to \infty$ [4]. Since the range of interaction potential is infinite, the trajectory of particles separated by any finite distance will be deflected due to mutual interaction. As a result, the exact dynamics (or, trajectory) of a particle in this system can be computed only by considering the influence of all the other particles on it. This, however, is not feasible because collision cross sections diverge for interaction potentials which decay to zero only in the limit $r \to \infty$ [3]. Because of this it becomes necessary to limit the range of interactions within finite bounds. For this, note that the deflection in the trajectory of particles will be infinitesimally small if their distance is more than some critical distance, say, $r_c$. As a result, the finite range of interactions is usually achieved by imposing a reasonable cutoff on the deflection angle. Typically $r_c$ is of the order of few particle diameters. Under this localization, particles interact with each other only if their distance becomes less than or equal to $r_c$ and not otherwise.

Now, for simplicity, consider a binary collision between two particles, say, $\alpha$ and $\beta$. The distance between the center of mass of these particles at any instant of time is $r_{\alpha\beta} = |\boldsymbol{r}_\alpha - \boldsymbol{r}_\beta|$. When the particles are not interacting $r_{\alpha\beta} > r_c$ and during collisions $r_{\alpha\beta} \leq r_c$. In fact, closer look at the collision process shows that the trajectories of the particles start deflecting from their original path as soon as the particles approach each other within $r_c$ and the trajectories continue to deflect till the particle rebound from each other and move away at a distance greater than $r_c$. During this entire process the particles move with finite velocities. As a result, in physical systems collisions between particles take finite time $t_c$, $t_c > 0$, for completion. Also, at the beginning of collision the distance between particles $r_{\alpha\beta}$ decreases from $r_c$ to a distance of closest approach $r_{\min}$ and then increases to $r_c$ at the end of the collision. It is important to note that $r_{\min}$ is a function of relative velocity and impact parameter of particles and *never* goes to zero. $r_{\min}$ will go to zero *iff* the relative velocity of particles is infinite (because $\phi = \infty$ at $r = 0$) and their trajectories lead to a head-on collision, which is physically impossible since the relative velocity can never be infinite. Hence, in physical systems $r_{\min} = 0$, or equivalently $r_{\alpha\beta} = 0$, *is a physical impossibility*, and during collisions one always has $0 < r_{\alpha\beta} \leq r_c$. This entire dynamics along with its approximation using the hard sphere model of particles is shown in Fig. 3.10. The distances $r_{\alpha\beta}$ and $r_{\alpha'\beta'}$ shown in this figure are the minimum distances between particles (or, distance of closest approach of particles) during the collision process.

#### 3.3.3.2 Collision Dynamics in Multiparticle Lattice Gases

In multiparticle lattice gases, by definition, multiple particles are allowed to occupy the same lattice site simultaneously, interaction neighborhood is restricted to one lattice site, and collisions occur among particles occupying the same lattice site (*c.f.*, Sec. 2.3). Since particles occupying the same lattice site have the same coordinates, the distance between the center of mass of particles during collisions is *always* zero in multiparticle lattice gases.[11)]

---

[11)] This observation makes the basis of all the conclusions that follow. As a result, attempts have been made to present arguments to contradict this statement and its conclusion. These arguments are based on partitioned spatial lattice representation of multiparticle lattice gases and are incorrect. The arguments and the reasons of their incorrectness are as follows:

**Argument 1:** To show that the footmarked statement is incorrect, one asserts that particles occupying the same lattice site do not occupy the same point in space. Instead, they are located in boxes (seen on the on the dual of the lattice) around the lattice sites and occupy different locations inside these boxes.

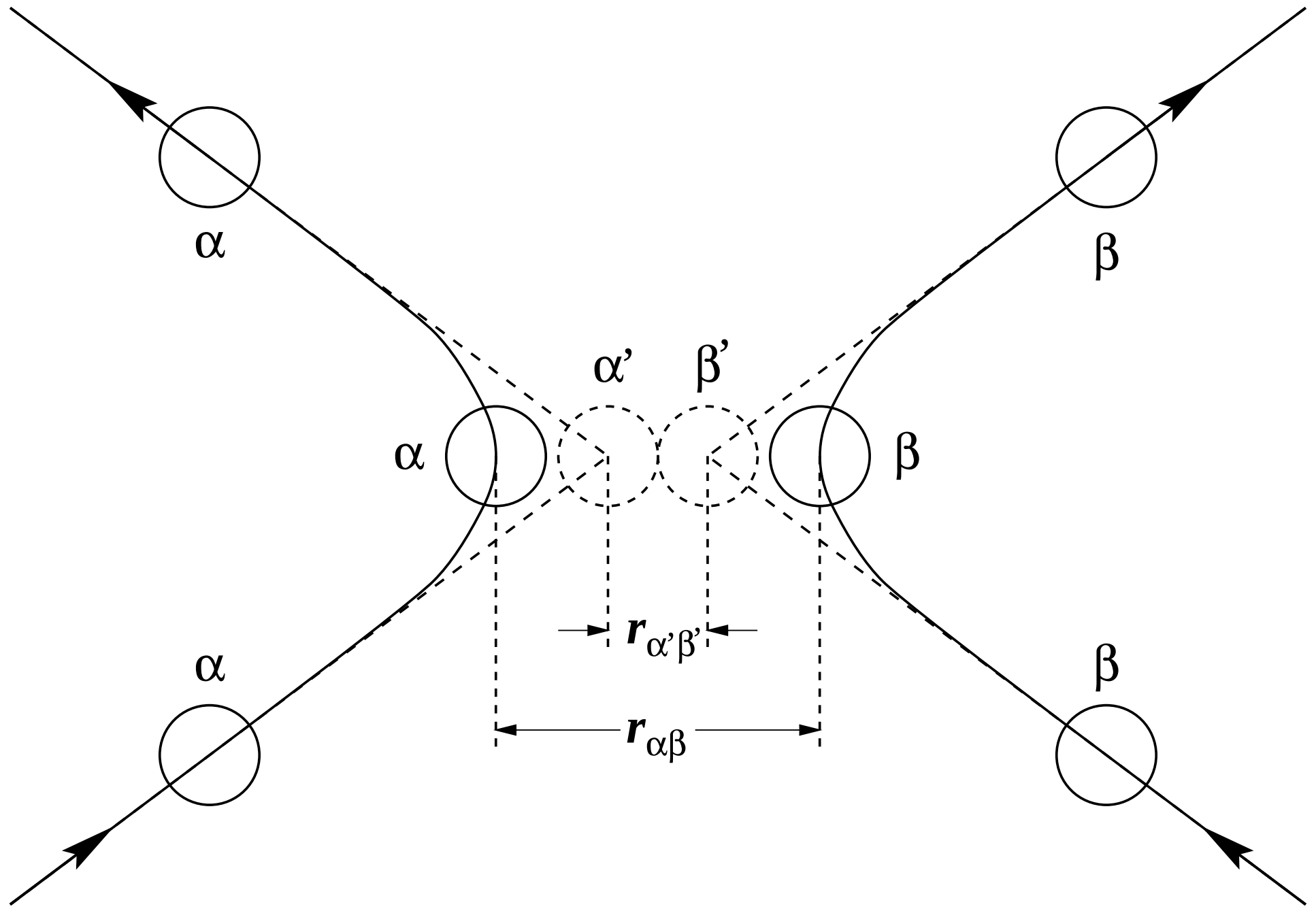

**Figure 3.10:** Dynamics of binary collisions in physical systems (curved trajectories; solid lines) and its approximation using hard sphere model of particles (straight trajectories; dotted lines).

Since the distance between particles is always zero during collisions, associating interaction potentials similar to those in physical systems described in Sec. 3.3.3.1 with particles in multiparticle lattice gases leads to a collision dynamics which cannot occur in physical systems and their models. As a result, no meaningful interaction potential can be associated with particles in multiparticle lattice gases. In fact, in order to obtain a consistent collision dynamics in multiparticle lattice gases, the only interaction potential that can be attributed to particles is a delta function. This implies that the model of particles encoded into multiparticle lattice gases (or, the model of particles on which multiparticle lattice gases are based) is that of *rigid point particles*. This model, though mathematically consistent with the dynamics of multiparticle lattice gases, is known to be physically incorrect and problematic because of occurrence of infinite potentials and forces during collisions.

Comparison of the collision dynamics in multiparticle lattice gases described above with that in physical systems shows that in multiparticle lattice gases particle collisions lead to a physically inconsistent collision dynamics and that the model of particles encoded

---

The above argument is incorrect because in simulations and also during analysis of multiparticle lattice gases, *e.g.*, for arriving at the coarse grained equations using Chapman-Enskog procedure (*c.f.*, Sec. 2.4), the coordinates of all the particles inside a box are taken to be the same as the coordinate of the box itself. Thus, for all practical purposes and physical analysis, the distance between particles occupying the same lattice site is always zero in multiparticle lattice gases implying that they occupy the same point in space irrespective of whether the lattice sites are treated as points or as boxes around the points.

**Argument 2:** In the second argument one asserts that the correct representation of the system is given by partitioned spatial lattice representation and in it no two particles occupy the same point in space.

The apparent correctness of the above argument is based on the assumption that cells inside boxes in PSLR can be distinguished and assigned separate identities on the basis of their coordinates. This assumption is incorrect as has been rigorously argued in Sec. 3.1.2 (more specifically, in Sec. 3.1.2.1).

in multiparticle lattice gases is no where near to that in physical systems or to that used in classical theories of physical systems. As a result, it is to be expected that multiparticle lattice gases will not be able to reproduce the dynamical behavior of physical systems correctly. That it is so is clearly seen from Sec. 2.5. Analysis of the exact role played by these differences in the macrodynamics of multiparticle lattice gases, specifically in their non-Galilean invariance and incompressibility, is carried out in the following sections.

### 3.3.4 Non-Galilean Invariance of Multiparticle Lattice Gases

Mathematically, non-Galilean invariance of multiparticle lattice gases shows up in the form of a multiplicative density dependent factor $g(\rho)$, the Galilean invariance breaking parameter, in the nonlinear terms of coarse grained hydrodynamic equations (*c.f.*, Sec. 2.5.1). Thus, the coarse grained equations of multiparticle lattice gases differ from the true Navier-Stokes equations primarily in the form of the coefficient of the nonlinear terms. On substituting $g(\rho) = 1$ in the coarse grained equations of multiparticle lattice gases, they unconditionally reduce to true Navier-Stokes equations. In view of this and also in view of of Secs. 3.3.2 and 3.3.3, the cause of non-Galilean invariance of multiparticle lattice gases can be determined by analyzing and understanding the dynamical process and phenomena in the microscopic world which give rise to nonlinear terms in coarse grained equations of physical systems and of multiparticle lattice gases.

#### 3.3.4.1 Process Represented by Nonlinear Terms of Navier-Stokes Equation

Nonlinear terms of Navier-Stokes equation in nonconservative from [5–7] can be written as

$$\rho v_\beta \frac{\partial v_\alpha}{\partial x_\beta}$$

for $\alpha$-component of momentum.

The differentials in these terms give the spatial gradient of $\alpha$-component of velocity in $\beta$-direction. Each complete term gives the net rate of change of $\alpha$-momentum due to influx/efflux of $\beta$-momentum which occurs because of spatial gradient of $\alpha$-velocity in $\beta$-direction. This is because $\rho v_\beta$ is component of the local momentum in $\beta$-direction and hence the net rate of change of $\beta$-momentum to $\alpha$-momentum is the magnitude of $\beta$-momentum multiplied by the gradient of $\alpha$-velocity in $\beta$-direction. Thus, the nonlinear terms of the Navier-Stokes equation are, in fact, mathematical statements representing the mechanism of conversion of various components of momentum into each other because of spatial gradients of various components of local fluid velocity in various directions. Alternatively, it can also be said that the nonlinear terms of the Navier-Stokes equation give the mechanism of convection of various components of momentum into various directions (or, equivalently, the mechanism of spatial momentum redistribution).

When fluids are considered as continuum, the appearance of the nonlinear terms can be explained on the basis of existence of spatial gradients of each component of velocity and consequent change of various components of momentum into each other as done above. The macroscopic observations made above, however, cannot be used directly for reconstructing the processes and phenomena occurring in the underlying microscopic world. This is because in fluids, continuum arises as a limiting case from the coarse graining of the discrete microscopic world and in this process entire information about the dynamics of the microscopic world is lost (in completely unrecoverable manner). Nevertheless, since the nonlinear terms of Navier-Stokes equation relate to spatial momentum redistribution, their origin from the microscopic world must necessarily be related to the microscopic process and phenomena which cause spatial momentum redistribution. The processes and phenomena which cause spatial momentum redistribution in physical systems and multiparticle lattice gases and give rise to nonlinear terms in their coarse grained equations have been determined in the following sections (Secs. 3.3.4.2 and 3.3.4.3).

#### 3.3.4.2 Microscopic Processes Causing Spatial Momentum Redistribution in Physical Systems

The dynamics of classical microscopic world is fully determined by two basic processes, namely, translation of particles and collisions among them. Thus, in the microscopic world momentum transfer/redistribution can occur only through these two processes. Translation of particles causes transfer of momentum associated with the particles from one location to another. This process, however, leads to redistribution of momentum only in space and there is no redistribution of momentum among the particles, *i.e.*, the net momentum of each particle remains unchanged.[12)] Redistribution of momentum among the particles occurs when particles interact (or, collide) and momentum is transferred from one particle to another. Interactions among particles in physical systems, however, do not merely lead to redistribution of momentum among them. They also cause redistribution of momentum in space over distances spanning the interaction zone of particles.[13)] This spatial momentum redistribution during interparticle interactions occurs because particles interact at a distance (*c.f.*, Sec. 3.3.3.1). The time duration over which this redistribution occurs is the duration for which the collisions last. To summarize, in physical systems there are two different microscopic processes which cause redistribution of momentum in space and give rise to the nonlinear terms of the Navier-Stokes equation, *viz.*, (i) translation (or, free motion) of particles from one location to another, and (ii) interparticle interactions.

#### 3.3.4.3 Microscopic Processes Causing Spatial Momentum Redistribution in Multiparticle Lattice Gases

Multiparticle lattice gases evolve in discrete time steps. Their dynamics during one time step is decomposed into two sub-steps, *viz.*, interparticle interactions and particle translation (*c.f.*, Sec. 2.3). Translation of particles causes transfer of momentum associated with the particles from one lattice site to another. This process, however, leads to redistribution of momentum only in space and there is no redistribution of momentum among the particles, *i.e.*, the net momentum of each particle remains unchanged. Redistribution of momentum among the particles occurs in the collision step (or, interaction step) when particles interact and momentum is transferred from one particle to another. Unlike in physical systems, interactions among particles in multiparticle lattice gases do not cause spatial momentum redistribution. This is because in multiparticle lattice gases the model of particles is that of *hard point particles* (*c.f.*, Sec. 3.3.3.2) and interactions among them

---

[12)] Because of this, one-particle distribution function remains unchanged in systems in which particles move without any type of mutual interaction. This type of particle motion is called streaming. It can occur only if all particles have the same velocity or if all particles are moving in the same direction and their velocity increases in a non-decreasing manner along that direction. Systems in which there is streaming of particles do not show interesting dynamics.

[13)] Usually, these distances are of the order of few molecular diameters.

are *contact interactions* which occur only when the distance between particles becomes zero. As a result, in multiparticle lattice gases redistribution of momentum among particles during interparticle interactions is not accompanied with redistribution of momentum in space, *i.e.*, momentum is not transferred from one lattice site to another during interparticle interactions (irrespective of whether the processes occur in zero time or non-zero time). To summarize, in multiparticle lattice gases there is one microscopic processes which causes redistribution of momentum in space and gives rise to the nonlinear terms in coarse grained hydrodynamic equations, *viz.*, translation of particles from one lattice site to another.

#### 3.3.4.4 Cause of non-Galilean Invariance of Multiparticle Lattice Gases

Secs. 3.3.4.2 and 3.3.4.3 show that the processes causing spatial momentum redistribution in multiparticle lattice gases differ from those in physical systems. Specifically, in multiparticle lattice gases no spatial momentum redistribution occurs during interparticle interactions. Because of this lack of one process, insufficient spatial momentum redistribution occurs in multiparticle lattice gases. As a result, a multiplicative factor, signifying this insufficiency, appears in the nonlinear terms (and in some other terms also) of the coarse grained equations of multiparticle lattice gases and leads to violation of Galilean invariance.

From and Sec. 2.4.5.2 and Appendix A it is seen that the multiplicative factor appears in the form of the $\nabla \cdot (g\rho\boldsymbol{v}\boldsymbol{v})$ and not in the term of the form $g\rho(\boldsymbol{v} \cdot \nabla)\boldsymbol{v}$. This is because the term signifying spatial momentum redistribution which arises during coarse graining in multiparticle lattice gases is $\nabla \cdot (g\rho\boldsymbol{v}\boldsymbol{v})$ and not $g\rho(\boldsymbol{v} \cdot \nabla)\boldsymbol{v}$. The corresponding terms in physical systems are $\nabla \cdot (\rho\boldsymbol{v}\boldsymbol{v})$ and $\rho(\boldsymbol{v} \cdot \nabla)\boldsymbol{v}$. In physical system the term

$$\frac{\partial(\rho\boldsymbol{v})}{\partial t} + \nabla \cdot (\rho\boldsymbol{v}\boldsymbol{v})$$

can be rewritten in the form

$$\rho\frac{\partial\boldsymbol{v}}{\partial t} + \rho(\boldsymbol{v} \cdot \nabla)\boldsymbol{v}$$

by using the continuity equation and the expression $\nabla \cdot (\rho\boldsymbol{v}\boldsymbol{v}) = \rho(\boldsymbol{v} \cdot \nabla)\boldsymbol{v} + \boldsymbol{v}\nabla \cdot (\rho\boldsymbol{v})$. Such rearrangement, however, is not possible in multiparticle lattice gases because the continuity equation of multiparticle lattice gases is the same as that of physical systems and does not involve the factor $g$ in its convective term. In multiparticle lattice gases, the term $\nabla \cdot (g\rho\boldsymbol{v}\boldsymbol{v})$ can be rewritten as $g\rho(\boldsymbol{v} \cdot \nabla)\boldsymbol{v}$ only in the incompressible limit by using Eq. (A.12) and invoking appropriate approximations for neglecting irrelevant terms as done by Wolfram [1] and Frisch *et. al.* [2].

The analysis presented above gives the exact cause of non-Galilean invariance of multiparticle lattice gases and shows that this violation occurs because the microdynamics of multiparticle lattice gases differs from that in physical systems. Specifically, it shows that in multiparticle lattice gases violation Galilean invariance occurs because the dynamics of interparticle interactions encoded in them lacks some elements, *viz.*, the elements causing spatial momentum redistribution, which are actually present in all physical systems.

### 3.3.5 Incompressibility of Multiparticle Lattice Gases

The second problem of interest in the present investigation that continues to persist with multiparticle lattice gases despite all the development they have seen (*c.f.*, Sec. 2.2.3) is their inability in modeling and simulation of physical systems and phenomena involving compressibility. It is hard to argue that positive progress has been made at resolving this problem because indications to this effect are not available in the literature. It is worthwhile to note that for long the incompressibility of multiparticle lattice gases has been assumed to be related to their non-Galilean invariance. The findings of Sec. 2.5.2, however, show that it is not so and that the recovery of Galilean invariance is only a necessary (and not sufficient) condition for correct simulation of compressible phenomena. Thus, it seems that the exact reason of incompressibility of multiparticle lattice gases is not known yet.

Secs. 3.3.2 and 3.3.3 show that the primary difference, besides discreteness, in the microdynamics of physical systems and multiparticle lattice gases lies in the collision dynamics of the two. In view of this, in the following sections an attempt has been made to identify the exact cause of incompressibility of multiparticle lattice gases and relate it to their collision dynamics. This relationship, however, cannot be established in as simple a manner as was sufficient for relating the non-Galilean invariance of multiparticle lattice gases with their collision dynamics (*c.f.*, Sec. 3.3.4). For associating the incompressibility of multiparticle lattice gases with their collision dynamics it is, first of all, necessary to understand how the compressible phenomena originate in physical systems and exactly what happens at the microscopic level at the onset of these phenomena.

#### 3.3.5.1 Origin of Compressible Phenomena in Physical Systems

The usual mechanical definition of compressibility associates a numerical measure with this quantity. It does not give any explicit indication about the (dynamical) mechanism of origin of compressible phenomena. Understanding of this mechanism, however, is necessary for understanding the cause of incompressibility of multiparticle lattice gases. Consequently, in the following paragraphs the dynamical mechanism of origin of compressible phenomena in physical systems has been established at the microscopic level of their description.

Little observation is needed to note that compressible phenomena in fluids arise from (nearly) instantaneous response of groups of particles to motion of obstructions relative to the fluid.[14] The obstructions, usually, are solid objects present into the fluid and their motion relative to the fluid may or may not be impulsive. In the microscopic world, the response of particles to the motion of obstructions relative to them reflects in the form of change in the velocity of particles present in the neighborhood of the obstructions. This change, if caused by the motion of obstructions into the fluid, is such that it makes the affected particles move away from the obstructions and, if caused by the motion of obstructions away from the fluid, is such that it makes the affected particles move towards

[14] Also from a similar response of particles to changes in the externally applied fields. Study of the response of particles to fields, *e.g.*, magnetic field, however, is beyond the scope of the present investigation because presently it is not clear how to incorporate fields into lattice gases. Hence, the discussion in the present investigation will be restricted only to the study of appearance of compressible phenomena in fluids because of motion of obstructions relative to the fluid.

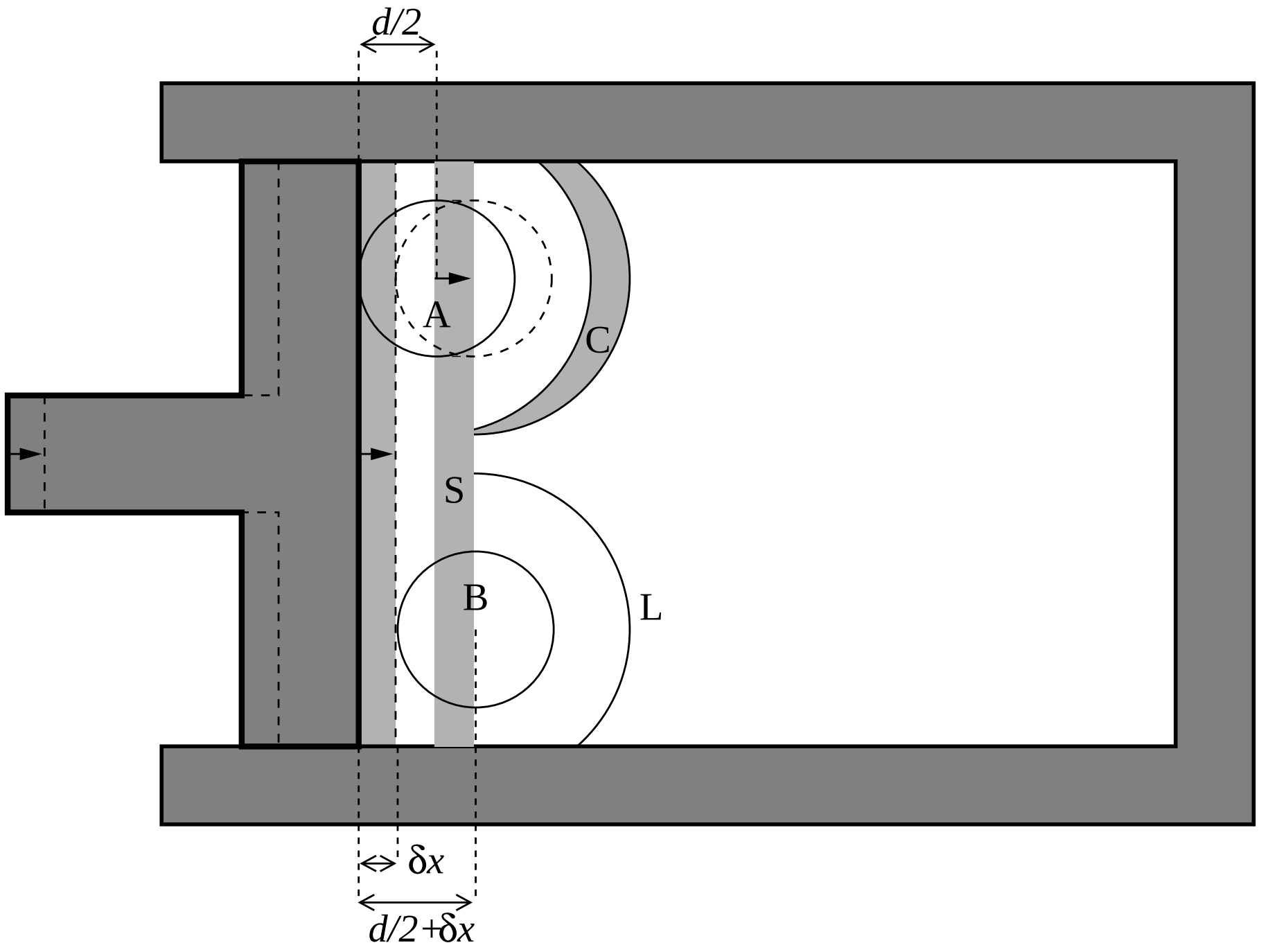

**Figure 3.11:** Schematic representation of the piston-cylinder setup and the dynamics which occurs in it at the moment of instantaneously pushing the piston. All the length scales have been disproportionately enlarged for clarity of representation and only few particles are shown.

the space that was previously occupied by the obstructions. The processes which lead to this response of particles can be illustrated through an idealized experiment as follows:

Consider an ideal friction-less piston-cylinder assembly in which a gas of hard sphere particles of diameter $d$ is enclosed between the piston and walls of the cylinder as shown in Fig. 3.11. Let, collisions of gas particles with the piston, walls of the cylinder, and other gas particles be perfectly elastic and instantaneous. Let, the spatial location of the particles at any time be specified by the location of their center of mass. Let, initially the system be at equilibrium. Now consider that the piston is pushed into the cylinder instantaneously, *i.e.*, in the time interval $\delta t$ in the limit $\delta t \to 0$, by an infinitesimally small distance $\delta x$ in the limit $\delta x \to 0$. Observations on this system reveal that as soon as the piston is pushed there is an instantaneous information propagation into the gas and a wave is formed which originates from the piston and propagates away from it into the gas. The processes involved in this complete dynamics are best studied by dividing them into two parts as: (I) the origin or formation of the wave, and (II) propagation of the wave into the gas.

**(I) Formation of Wave:** Since the gas consists of hard sphere particles, formation of wave is an instantaneous event which occurs as soon as the piston is pushed. Furthermore, it is well known that waves can be formed either because of change in the local density of gas or because of change in the local momentum of gas. In the present example, the piston is pushed by infinitesimally small distance $\delta x$ which does not cause changes in the local density of gas and thus formation of the wave is not because of density change. Instead, the wave is formed because of change in the momentum of particles in the neighborhood of the piston due to transfer of momentum from the piston to particles. This transfer process occurs in two basic steps leading to formation of the wave. These steps, in the order of their occurrence, are: (A) transfer of momentum from piston to particles in its neighborhood, and (B) transfer of momentum from these particles to other particles in their neighborhood.

**(A) First Step (Transfer of Momentum from the Piston to Particles):** As the piston is pushed, it collides and transfers momentum to all the particles located in the strip of width $\delta x$ which starts at a distance of $d/2$ from the initial location of the piston. In Fig. 3.11 this strip is marked with the letter S. During the collision, the piston instantaneously pushes all these particles by infinitesimally small distance $\delta x$ to the far edge of the strip, *i.e.*, at the location $d/2 + \delta x$. For example, in Fig. 3.11 the particle A, which is initially located at the inner edge of the strip S, is instantaneously pushed to the far edge of the strip by the piston. In the figure, the initial and final locations of this particle are shown by continuous and dashed circles, respectively. Similarly, momentum is also transferred to all the particles located just inside the far edge of the strip, *e.g.*, to particle B in Fig. 3.11, with the difference that the displacement of these particles is zero. This is because, although the piston collides with these particles, it does not push (displace) these particles. Instead, it just comes in their contact and transfers momentum to them.

**(B) Second Step (Transfer of Momentum from Particles to Other Particles):** The particles that have been pushed by the piston to the edge of the strip S in turn collide with their neighbors and transfer momentum to them. For example, in Fig. 3.11, the particle A collides and transfers momentum to all the particles located in the curved strip marked with the letter C and pushes them to the far (outer) edge of this strip. The particle B, however, collides and transfers momentum only to particles located on the curved line marked with the letter L. Unlike the particles located in strip C, the particles located on the line L are not displaced because the displacement of particle B itself is zero.[15)] These particles (the particles located in the strip C and on the line L) in turn transfer momentum to their neighbors through a similar process. This process continues till a stage is reached wherein all the strips (similar to strip C and line L) are empty so that no more collisions occur.

The processes elaborated in the step (A) above lead to instantaneous readjustment of momentum vectors of all the particles located within the strip S. The processes outlined in the step (B) above lead to instantaneous readjustment of momentum vectors of particles which, being located beyond the strip S, do not come in direct contact with the piston. These readjustments are such that the velocity vectors of the affected particles point in the half-plane away from the piston. The consequence of completion of both these steps, *viz.*, the steps (A) and (B), is readjustment of momentum vectors of particles comprising the gas and formation of wave; which then propagates into the cylinder.

It is worthwhile to note that in the steps (A) and (B) outlined above all the processes are instantaneous and synchronous (because collisions occur instantaneously and the piston is also pushed instantaneously). That is, the processes occur and are completed exactly at the instant at which the piston is pushed. Furthermore, at all non-zero and finite gas densities there is non-zero and finite probability that a particle will be found *"within a small zone around a point"* in space. As a result, there is non-zero probability

[15)] This also implies that the particle B must already be in contact with the particles located on the line L even before the piston is pushed. This is because all the events are occur in infinitesimally small time (*i.e.*, are instantaneous) and the displacement of particles during this time interval is necessarily zero. Thus, it is not possible that while the piston is being pushed some particles move over to the line L.

that particles will be present in the strip C. The probability that a particle will be found *"at a point"* in space, however, is always zero. As a result, the probability that particles will be present on the line L is also always zero.

**(II) Propagation of Wave:** Once the wave is formed as described above, it propagates into the cylinder. The propagation of the wave into the cylinder exactly follows the step (B) outlined above with the difference that once the wave is formed, the particles whose momentum vectors have changed move to new locations before colliding with other particles.[16)] During these collisions the particles do not get displaced like particle A in Fig. 3.11. Instead, the momentum transfer occurs only through contact as with particle B in Fig. 3.11. Through this process the momentum gets transferred from one location in space to another and also from one particle to another leading to propagation of the wave into the cylinder.

In the idealized system considered above, the process of momentum transfer via collisions has a finite probability of never terminating, especially if the density of particles is large enough. Experience, however, shows that in all real physical systems this process terminates after finite number of collisions. In physical systems the maximum distance to which momentum gets transferred through this process depends upon the physical properties of the particles comprising the medium. Furthermore, it is not possible to instantaneously move an obstruction through non-zero distances in a physical medium of massive (as opposed to mass-less) particles. Thus, in physical systems a delay is introduced between all consecutive interactions of a particle. As a result, one has to look at the events which occur in an infinitesimally small time interval $\delta t$ ($\delta t \to 0^+$ and not $\delta t \to 0$), rather than at the events occurring at an instant. Analysis of such realistic situations does not alter the arguments given above in any significant way, other than introducing an additional step of motion of particles coupled with collisions through infinitesimally small distance between collisions. In addition to this one also has to allow for a very small time for completion of collisions and an associated displacement of particles during this interval. In the present investigation, however, the time required for completion of a collision will always be taken to be zero as is usually done in the literature, *e.g.*, in the derivation of the Boltzmann equation, because it is negligible compared to the *mean free time*.[17)] Note that neglecting *collision time*[18)] in comparison with the mean free time is an approximation. More details on this approximation and its validity in the present context follow in chapter 5.

#### 3.3.5.2 Microscopic Processes Causing Compressible Phenomena in Physical Systems

The discussion outlined in Sec. 3.3.5.1 brings out the microscopic processes responsible for the origin of compressible phenomena as well as for propagation of compressibility

[16)] The time for which each particle remains in free motion and the distance which it covers in free motion will, in general, be different for different particles.

[17)] Mean free time of a particle is the mean time for which it is in free motion, alternatively, it also the mean time between two consecutive collisions of a particle. It is also known as relaxation time.

[18)] In some investigations, in the kinetic theory literature, the term *"collision time"* is taken to be synonymous with the term *"mean free time"* since the later is the mean time between two consecutive collision of a particle. In some other investigations these two terms are strongly differentiated and the term *"collision time"* is used to refer to the time which is required for completion of a collision. This differentiation is maintained in the present investigation also and the term *"collision time"* is used with the later meaning.

effects in system of particles. It shows that the origin of compressible phenomena at microscopic level is due, primarily, to redistribution of momentum in space which occurs through interparticle interactions.[19)] It also shows that role of spatial momentum redistribution due to translation of particles is negligible (or, irrelevant) compared to that due to interparticle interactions in *giving rise to* compressible phenomena. This is because the instantaneous displacement of particles is negligible compared to the distances over which spatial momentum redistribution occurs during the onset of compressible phenomena. A straight forward way of visualizing the irrelevance of particle translation in *giving rise to* compressible phenomena is to consider a system of particles in which there are no interparticle interactions and thus momentum redistribution through interparticle interactions does not occur (either in space or among the particles). In such a system, naturally, all particles move with the same velocity without accelerating and the distance between any two particles remains same at all times. As a result, compressible phenomena do not (and, cannot) originate in this system even though there is spatial momentum redistribution due to particle translation.

Regarding propagation of compressibility effects, the discussion outlined in Sec. 3.3.5.1 shows that at the microscopic level the propagation of compressibility effects involves spatial redistribution of momentum due to particle translation as well as interparticle interactions. It also shows that both these processes work in unison leading to propagation of compressibility effects in the medium. As a result, neither of these processes can be assigned greater importance (and neglected) in comparison with the other. A (more or less) qualitative comparison of the role played by these processes in different systems, however, is possible in terms of the distance over which momentum is redistributed by these process in unit time. In this comparison, in solids, spatial momentum redistribution by interparticle interactions gains predominance over that by particle translation. On the other hand, in gases the inverse is true. Although the contribution of spatial momentum redistribution due to interparticle interactions, in gases, is very small in comparison to that due to particle translations, the former cannot be neglected. This is because propagation of waves from one location to the other involves formation of waves at the new location for which spatial momentum redistribution (as brought out above) plays the central role.

The analysis of the mechanism of formation and propagation of waves outlined above is fully in terms of redistribution of momentum. It is, however, customary and (possibly) more intuitively appealing to look at compressibility effects in terms of density variations in the system. For correlating density variations with the dynamics of spatial momentum redistribution outlined above, note that the spatial momentum redistribution processes invoke collective response from groups of particles[20)] and cause them to move towards (or,

---

[19)] As explained in Sec. 3.3.4.2, spatial momentum redistribution due to interparticle interactions necessarily involves (or, is accompanied by) redistribution of momentum among particles. Because of this one might be tempted to rephrase the footmarked assertion as

> *The origin of compressible phenomena at microscopic level is due, primarily, to redistribution of momentum among particles which occurs through interparticle interactions.*

This rephrasal, although correct in the context of physical systems, is incorrect in the wider context of models of physical systems. This is because it is not necessary that in models of physical systems momentum redistribution among particles will always be accompanied by spatial momentum redistribution as, for example, is the case with multiparticle lattice gases.

[20)] These are the particles whose momentum has changed in the spatial momentum redistribution processes due to interaction with the piston or with other similar particles.

away from) some zone. This causes crowding of particles in the zone towards which they are moving and gives rise to density variations in the system.

In view of the above, it can also be said that compressible phenomena in fluids originate from collective response of particles which leads to mutual adjustment in their momentum vectors through collisions and makes them move either towards or away from some zone and causes transfer of momentum from one location in space to another. Here, it is important to note that energy is also transferred along with the momentum. This, in most of the systems (model systems are included), is true. At this point, however, it is necessary to note that in certain model systems known as *athermal systems*[21)] the kinetic energy of all particles remains same at all times. As a result, to take care of the most general situations that are possible, it is necessary to restrict the analysis to be only in terms of *momentum* transfer. In all classical systems, this restriction is acceptable because the mass $m$, momentum $\boldsymbol{p}$, and kinetic energy $E$ of particles are interrelated via $E = \boldsymbol{p}^2/(2m)$.

#### 3.3.5.3 Cause of Incompressibility of Multiparticle Lattice Gases

The findings of Sec. 3.3.5.2 show that spatial momentum redistribution caused by particle translations as well as interparticle interactions is the primary factor responsible for the origin and propagation of compressible phenomena in physical systems. In multiparticle lattice gases, however, there is no spatial momentum redistribution due to interparticle interactions (*c.f.*, Sec. 3.3.4.3). As a result, multiparticle lattice gases are not able to simulate compressible systems correctly. The formation and propagation of waves which is observed in multiparticle lattice gases, *e.g.*, as in [8], is because of initial density perturbations or because of density perturbations caused by obstructions placed in the path of particles. Although, these waves also propagate into the medium, they do not represent phenomena occurring in physical systems correctly because spatial momentum redistribution during interparticle interactions is absent in multiparticle lattice gases.

Another argument that also leads to the above conclusion is as follows: In the multiparticle lattice gases collisions are completely local to the lattice sites and by definition the net momentum is conserved during collisions. As a result, while resolving collisions on a lattice site there is no interaction (momentum transfer) among particles occupying different lattice sites, and each lattice site acts as an isolated system whose net momentum flux vector remains unchanged before and after collision resolution. As a result, in multiparticle lattice gases it is impossible to obtain collective response from particles occupying different lattice sites. Since, such a response is essential for correct simulation of compressible phenomena, it can be concluded that it is impossible to correctly simulate compressible systems using the multiparticle lattice gases.

## 3.4 A Note on Interrelationship of non-Galilean Invariance and Incompressibility of Multiparticle Lattice Gases

It was argued in Sec. 2.5.2 that non-Galilean invariance and incompressibility of multiparticle lattice gases are not interrelated in that it is not necessary that overcoming

[21)] *Athermal systems* are those systems in which all particles move with the same *speed* at all times, *e.g.*, the HPP gas (*c.f.*, Sec. 2.2.6.2) and the FHP gas (*c.f.*, Sec. 2.2.6.5).

the problem of non-Galilean invariance is sufficient to overcome the problem of incompressibility as well. The arguments were in contradistinction to the literature wherein the incompressibility of multiparticle lattice gases has been attributed to appearance of Galilean invariance breaking parameter in the nonlinear terms of coarse grained momentum equation. The findings of Secs. 3.3.4 and 3.3.5, however, show that in multiparticle lattice gases, both, non-Galilean invariance and incompressibility are because of absence of spatial momentum redistribution during interparticle interactions. This seems to contradict the argument presented in Sec. 2.5.2. Note, however, that there is no contradiction here. This is because of the following:

The fact that both these problems arise because of the same deficiency in the microdynamics multiparticle lattice gases does not imply that both these problems are interrelated in that overcoming the problem of non-Galilean invariance is sufficient to overcome the problem of incompressibility as well. This is because it might be possible to overcome the problem of non-Galilean invariance in ways[22)] through which the problem of incompressibility still remains. One such method is to artificially alter spatial momentum redistribution, *e.g.*, by addition of rest particles,[23)] such that the Galilean invariance breaking parameter either becomes unity or a constant.

## 3.5 Conclusions

The analysis presented in this chapter brings out the following conclusions:

**1)** All multiparticle lattice gases can be represented using either multiple particle representation or partitioned spatial lattice representation. Both these representations are equivalent for multiparticle lattice gases and can be transformed in each other.

**2)** In multiple particle representation the impact parameter of particles is always zero because collisions occur among particles occupying the same lattice site.

**3)** Partitioned spatial lattice representation shows number of peculiarities when the cells comprising the spatial lattice are treated as independent units. These peculiarities are: (i) the spatial location and velocity of particles become correlated, (ii) the structure of the phase space (of each particle; thus also of the entire system) becomes different from that in physical systems, and (iii) it becomes possible to define lattice gases whose evolution rules have non-zero impact parameter (these lattice gases cannot be transformed correctly to multiple particle representation because in multiple particle representation the impact parameter is necessarily zero). The peculiarities (i) and (ii) make this representation physically inconsistent.

**4)** The peculiarities of partitioned spatial lattice representation can be bypassed by treating the blocks, rather than the cells (or, lattice sites), comprising the spatial lattice as independent units whose location is identified with one coordinate point each. This treatment, in essence, transforms the partitioned spatial lattice representation into multiple particle representation which is a physically consistent representation.

[22)] Of course, without removing the deficiency (*i.e.*, of lack of spatial momentum redistribution during interparticle interactions) that has been found in the microdynamics of multiparticle lattice gases. If this deficiency is removed then, naturally, both these problems will be overcome simultaneously.

[23)] This has been attempted frequently in the literature [2,9] with partial success.

**5)** Lattice gases whose collision rules have been developed using either different velocity sets or different interaction neighborhoods must be considered as different lattice gases. Different lattice gases having the same velocity set but different interaction neighborhoods are likely to show different micro- and macrodynamical behavior.

**6)** Non-Galilean invariance and incompressibility of multiparticle lattice gases is because of the structure and construction of their collision rules (interaction neighborhood, being always restricted to one lattice site, need not be treated separately).

**7)** The dynamics of interparticle interactions in physical systems differs from that in multiparticle lattice gases in that in physical systems interactions occur among particles occupying different points in space and the distance among particles never goes to zero during interactions, whereas in multiparticle lattice gases interaction occur among particles occupying the same lattice site (*i.e.*, the same point in space) and the distance among particles is always zero during interactions. As a result, the nature of the model of particles encoded in multiparticle lattice gases also differs from that in physical systems (and thus, from that used in classical models of physical systems). Specifically, the model of particles encoded in multiparticle lattice gases is that of *hard point particles* (*i.e.*, zero range of interaction), whereas in physical systems the model of particles (in most general terms) is that of *point centers of forces with a hard core* (*i.e.*, non-zero range of interaction).

**8)** Spatial momentum redistribution occurs during interparticle interactions in physical systems but not in multiparticle lattice gases. Spatial momentum redistribution during particle translation and redistribution of momentum among particles during interparticle interactions occur in both physical systems and multiparticle lattice gases.

**9)** The nonlinear terms of Navier-Stokes equation encode the overall dynamics of spatial momentum redistribution. The coefficient of nonlinear terms of Navier-Stokes equation arises from combined effect of spatial momentum redistribution during both particle translation as well as interparticle interactions.

**10)** Spatial momentum redistribution due to both particle translation as well as interparticle interactions is necessary for origin and propagation of compressible phenomena and compressibility effects in physical systems. Thus, the presence of both these mechanism is necessary in models of physical systems for correct simulation of compressible systems and phenomena.

**11)** Absence of spatial momentum redistribution during interparticle interactions leads to insufficient redistribution of momentum in space in multiparticle lattice gases which reflects in the form of appearance of the multiplicative Galilean invariance breaking parameter in the nonlinear terms of coarse grained momentum equation. In essence, violation of Galilean invariance in multiparticle lattice gases occurs because there is no spatial momentum redistribution during interparticle interactions.

**12)** Multiparticle lattice gases are not able to correctly simulate compressible systems and phenomena because in them spatial momentum redistribution does not occur during interparticle interactions. Alternatively, the problem of incompressibility in multiparticle lattice gases is because it is not possible to invoke collective response from particles occupying different lattice sites in them.

**13)** Although, both non-Galilean invariance and incompressibility of multiparticle lattice gases originate from the same deficiency in the microdynamics of multiparticle lattice gases, they are neither interrelated nor consequences of each other in that it is not necessary that overcoming the problem of non-Galilean invariance is sufficient for overcoming the problem of incompressibility as well. This is because it is possible to (partially) overcome the problem of non-Galilean invariance in ways, *e.g.*, by addition of rest particles, which leave the problem of incompressibility unresolved.

The above conclusions give not only the cause of the problems of non-Galilean invariance and incompressibility of multiparticle lattice gases but also corroborate the findings and conclusions of chapter 2. Furthermore, these conclusions also suggest that the problems can be overcome by incorporating a mechanism which leads to spatial momentum redistribution during collisions in multiparticle lattice gases. If such a mechanism can be devised the objective of this investigation (*c.f.*, Sec. 1.4) would be fulfilled easily. Thus, further questions related to possibility, impossibility, and the method of overcoming these problems in lattice gases (not just multiparticle lattice gases) have been addressed in chapter 4.

# Chapter 4

# Overcoming non-Galilean Invariance and Incompressibility of Lattice Gases

. . . and those who insist on talking about problems must explain their cause.
*Yes, sire.*
Those who show the cause of problems must tell how to eliminate them.
*Yes, sire.*
Even if they find it obvious.
. . .

Following the conclusions of chapter 3, the investigations presented in this chapter are directed towards finding out the necessary condition(s) and method of developing lattice gases which are not constrained by the problems of non-Galilean invariance and incompressibility observed in multiparticle lattice gases. In addition to this, whether or not multiparticle lattice gases can be made Galilean invariant and capable of simulating compressible systems and phenomena has also been discussed.

## 4.1 Overcoming non-Galilean Invariance and Incompressibility in Lattice Gases

The analysis presented in chapter 3 shows that the problems of non-Galilean invariance and incompressibility of multiparticle lattice gases arise because the dynamics of collisions in them differs from that in physical systems. Specifically, Secs. 3.3.4 and 3.3.5 bring out that the cause of both these problems in multiparticle lattice gases is the lack of spatial momentum redistribution during interparticle interactions in them. This suggests that for lattice gases (not just multiparticle lattice gases) to be free from these problems, their collision rules (or, more generally their evolution rules) should be such that spatial momentum redistribution occurs during interparticle interactions in them.

The method of restoring spatial momentum redistribution during interparticle interactions in lattice gases, whether or not spatial momentum redistribution during interparticle interactions can be restored in multiparticle lattice gases, and the necessary conditions for the same and their consequences, have been discussed in the following sections.

### 4.1.1 Restoring Spatial Momentum Redistribution During Interparticle Interactions in Lattice Gases

Secs. 3.3.3 and 3.3.4.3 show that the absence of spatial momentum redistribution during interparticle interactions in multiparticle lattice gases is because in them interactions occur only among particles occupying the same lattice site and distance between particles is zero during interactions. This implies that spatial momentum redistribution during interparticle interactions can be restored by defining the evolution rules in such a way that interactions occur among particles occupying different lattice sites. This is because if interactions occur among particles occupying different lattice sites, the distance between them will necessarily be non-zero during interactions; which will cause transfer/redistribution of momentum over non-zero distance in space during interparticle interactions.

### 4.1.2 On Restoring Spatial Momentum Redistribution During Interparticle Interactions in Multiparticle Lattice Gases

From the previous section (Sec. 4.1.1) it is clear that in multiparticle lattice gases spatial momentum redistribution during interparticle interactions can be restored only by incorporating interactions among particles occupying different lattice sites in their collision rules. From the definition of multiparticle lattice gases (*c.f.*, Sec. 2.3) it appears that interactions among particles occupying different lattice sites can be easily incorporated in them by expanding their interaction neighborhood from one lattice site to more than one lattice sites and appropriately redefining the collision rules over the expanded interaction neighborhood. Considerations involved in the expansion of interaction neighborhood are not relevant here. They can either be *ad hoc* or based on rigorous physical reasoning. In either case, however, the interaction neighborhood must be expanded symmetrically in all the directions because all the directions are equivalent. Whether or not collision rules can be defined in a physically consistent manner over the expanded interaction neighborhood is, however, not clear. As a result, the considerations involved in defining collision rules over the expanded interaction neighborhood need to be addressed and analyzed explicitly. This is done below for single species multiparticle lattice gases[1)] under separate cases. The cases are as follows:

**1)** Multiparticle lattice gases in which no two particles occupying the same lattice site simultaneously can have the same velocity.

**2)** Multiparticle lattice gases in which two or more particles occupying the same lattice site simultaneously are allowed to have the same velocity. The maximum number of particles with the same velocity that can occupy a lattice site simultaneously is constrained to be finite. The maximum number of particles for each velocity vector can be different. The maximum number of particles must be greater than or equal to 2 for at least one velocity vector.

**3)** Multiparticle lattice gases in which two or more particles occupying the same lattice site simultaneously are allowed to have the same velocity. The maximum number of particles with the same velocity that can occupy a lattice site simultaneously is not constrained to be finite for at least one velocity vector.

[1)] The restriction to single species multiparticle lattice gases is only for the sake of simplifying the overall analysis and discussion. Generalizations to other cases proceed along similar lines and are straight forward.

**Case 1:** Consider multiparticle lattice gases in which interaction neighborhood of particles extends to more than one lattice site and no two particles occupying the same lattice site simultaneously are permitted to have the same velocity. For these multiparticle lattice gases collision rules cannot be defined over the expanded interaction neighborhood in physically consistent manner. As a result, spatial momentum redistribution during interparticle interactions cannot be restored in them. This is because of the following:

At any non-zero density the probability that there are some fully occupied lattice sites is non-zero in lattice gases. As a result, in multiparticle lattice gases with expanded interaction neighborhood the probability that during collision among particles occupying two (or, more) lattice sites at least one lattice site will be fully occupied is non-zero. In such a collision, the post collision velocity vector of the particle occupying the fully occupied lattice site will become same as that of another particle occupying the same lattice site. This leads to an inconsistent situation since by definition no two particles having the same velocity are permitted to occupy the same lattice site simultaneously. This inconsistency can be bypassed only by enforcing, in definition, the condition that collisions do not occur in such situations. This condition is the same as that in the HPP and FHP gases and other multiparticle lattice gases (*c.f.*, Sec. 2.3). Enforcement of this condition, however, leads to a problem that as the density of particles increases above some critical value the probability of collisions among particles starts decreasing monotonically and becomes zero at the maximum density.[2)] This behavior is observed in other multiparticle lattice gases also and has been said to occur because of a peculiar phenomena called *particle-hole duality*[3)] which does not have a counterpart in the physical reality [1]. In physical reality, the probability of collisions (or, the collision frequency $\nu$) varies linearly with density ($n$), *i.e.*, $\nu \propto n$.

**Case 2:** In this case also the collision rules cannot be defined consistently over expanded interaction neighborhood for restoring spatial momentum redistribution during interparticle interactions. The arguments proceed along the same lines as in the case 1 above.

**Case 3:** In this case it is possible to define collisions rules over the expanded interaction neighborhood so that spatial momentum redistribution during interparticle interactions is restored. This is because at least one channel can have infinitely many particles. This case, however, is not of interest because theoretically it does not lead to true cellular automata models (because the total number of possible states for the lattice sites is no longer finite) and practically the resulting lattice gases cannot be implemented on digital computers (because digital computers have finite memory; as a result neither can the symbols representing all possible states be created nor can the rule table be stored).

**Conclusion:** The analysis outlined above shows that interactions among particles occupying different lattice sites cannot be incorporated in a physically consistent manner in

[2)] This problem, like the problems on non-Galilean invariance and incompressibility, is also well known in the literature. It, however, has received considerably less attention because its presence was noted in the early developments and it was found that this problem surfaces above certain critical density only (the critical density, in all multiparticle lattice gases, is typically 50% of the maximum possible density). As a result, from the very beginning the usage of lattice gases (multiparticle lattice gases) has been constrained to be below the critical density [1] to that the problem is not encountered.

[3)] Here, space or unoccupied positions are termed as holes. This concept has been adopted from condensed matter physics wherein it is used extensively, *e.g.*, while studying flow of charge in semiconductors. Taking the example of single species athermal multiparticle lattice gases, *particle-hole duality* refers to and is a consequence of the fact that collision rules remain invariant if particles and holes are interchanged.

multiparticle lattice gases. As a result, it is not possible to restore spatial momentum redistribution during interparticle interactions in multiparticle lattice gases.

### 4.1.3 Necessary Condition for Restoring Spatial Momentum Redistribution During Interparticle Interactions in Lattice Gases: The Single Particle Exclusion Principle

An important result from Sec. 4.1.2 is that incorporating collisions among particles occupying different lattice sites (in multiparticle lattice gases; in fact, in lattice gases, in general) is not *sufficient* for restoring spatial momentum redistribution during interparticle interactions in a physically consistent manner. It seems that additional conditions are needed for obtaining physical consistency.[4)] For finding out these conditions, note that in multiparticle lattice gases physical inconsistency arises because the situations in which interactions can occur among particles need to be restricted through explicitly imposed external constraints. These restrictions are needed to ensure that the basic defining constraints of multiparticle lattice gases on the maximum number of particles having the same state that can occupy a lattice site simultaneously are not violated during interparticle interactions.

Restricting the situations in which particles can interact, however, is not the only method for ensuring that the basic defining constraints of lattice gases[5)] are not violated during interparticle interactions. A few other new methods can also be devised. Two such methods have been outlined and discussed in the following. The additional advantage common to both these methods is that the resulting lattice gases can be believed to be physically consistent in that the collision frequency is expected to increase monotonically with density.[6)] The disadvantage of both these methods is that the modeling philosophy contained in them departs drastically from that contained in multiparticle lattice gases.

The first method, of relevance in the context on multiparticle lattice gases, is completely removing the defining constraints on the maximum number of particles having the same state that can occupy the same lattice site simultaneously. This method, as pointed out in Sec. 4.1.2, leads to impractical multiparticle lattice gases[7)] violating the basic definition of cellular automata, and thus, is not practically useful.

The second method focuses on modification of basic defining constraints of lattice gases[5)] to bring them closer to observations on physical systems. The new constraints are determined as follows: Note that in multiparticle lattice gases[7)] constraints on the maximum number of particles that can occupy a lattice site simultaneously are imposed by fixing the maximum number of particles with the same state that can occupy a lattice site simultaneously. Because of this, in these lattice gases[7)] more than one particles can occupy the same lattice site simultaneously irrespective of whether or not interactions

[4)] Here, the lack of physical consistency relates to peculiar behavior of collision frequency with density; and, thus, also to the behavior of all the quantities which depend upon collision frequency.

[5)] Here *defining constraints* refers, primarily, to constraints on the maximum number of particles having the same state that can (or, are allowed to) occupy the same lattice site simultaneously. With sufficient caution, however, one may include other constraints, if any, also.

[6)] This statement, at this stage, appears to be intuitive. It, in fact, is based on inspection of collision rules of certain new types of lattice gases that have been introduced in the following chapter. That the collision frequency will increase monotonically with density in these lattice gases can be seen with little observation and will also become clear in the following chapters. The exact variation in the form of a rigorous functional mathematical relationship between the two, however, cannot be ascertained easily.

[7)] This includes multiparticle lattice gases with expanded interaction neighborhood (*c.f.*, Sec. 4.1.2) also.

occur among them. Furthermore, maximum number of particles with the same state that can occupy the same lattice site simultaneously is invariably less than the maximum number of particles possible on it. Observations on *classical physical systems*, however, affirm that no two particles can ever occupy the same location in physical position space simultaneously irrespective of their states. In addition to it, in physical systems a particle located at a point in the physical position space can be in any state (that is possible for it). These observation, in the context of defining constraints of lattice gases, bring out that constraints on the maximum number of particles that can occupy the same lattice site simultaneously should not be related to the states of particles. Instead, they should be imposed without any reference to the states of particles as in physical systems. In addition, only one particle should be allowed to occupy a lattice site at any time step as in classical physical systems.

The last statement above arises directly from observations on physical systems. Thus, it has a valid physical basis. In addition to it there are some finer points also. These get clarified by the following elaboration: If only one particle is allowed to occupy a lattice site at any time step then interactions can be defined only among particles occupying different lattice sites. In this case the resulting model will be equivalent to molecular dynamics in all respects other than discreteness of space. On the other hand, if multiple particles are allowed to occupy the same lattice site simultaneously then evolution rules must necessarily be defined to take care of their interactions with particles occupying neighboring lattice sites (*c.f.*, Sec. 4.1.1) and optionally to take care of interactions among them. If interactions are defined to occur among particle occupying the same lattice site then the collision dynamics of the resulting model suffers from all the problems discussed in Sec. 3.3.3.[8] To eliminate these problems it is necessary that particles occupying the same lattice site should not interact. In this case, the resulting model merely simulates superimposition of many replicas of the system which is simulated by *an equivalent model*[9] in which only one particle is allowed to occupy a lattice site at any time step. Such models, however, appear to have an advantage in that simulations carried out using them will be less noisy compared to those carried out using the models in which at most one particle is allowed to occupy a lattice site at any time step. The overall cost incurred, however, belittles this advantage because the simulation time and storage memory requirements grow (at least) exponentially with the maximum number of particles that can occupy a lattice site simultaneously. Thus, instead of using such models it will always be advantageous to use a model in which only one particle is allowed to occupy a lattice site at any time step and to study ensemble averaged quantities by simulating many (statistically equivalent) realizations of the system.

The discussion outlined in previous paragraphs of this section brings out one condition that must necessarily be incorporated into the definition of lattice gases to ensure that interactions occur only among particles occupying different lattice sites, spatial momentum redistribution occurs during interparticle interactions, and resulting lattice gases are free from physical inconsistencies. The condition is that *only one particle should be allowed to*

[8] Note that the problem of absence of spatial momentum redistribution during interparticle interactions does not come up because interactions among particles occupying different lattice sites are also incorporated. Thus, interaction among particles occupying the same lattice site work simply as randomization enhancing process. The cost of incorporating this process, however, is very high because the number of states per lattice site increases exponentially with the maximum number of particles that can occupy the same lattice site simultaneously. The size of the evolution rule table increases still more rapidly.

[9] Here, equivalence is in terms of equivalence between the number of particle species, possible states for particles of each species (including velocity vectors, *etc.*), and the topology of the interaction neighborhood.

*occupy a lattice site at any time step irrespective of its state*. This condition, in addition to the above, also ensures that collision frequency varies monotonically with density and physical inconsistencies related to it do not arise in the resulting lattice gases. It is noteworthy that this condition is an *exclusion principle* based on the total number of particles (restricted to one here) without reference to their states, rather than on total number of particles for each state (as in multiparticle lattice gases), that can occupy a lattice site simultaneously. In general, this exclusion principle can be relaxed and more than one particles can be allowed to occupy the same lattice site simultaneously. This, however, is disadvantageous in terms of simulation time and memory requirements as discussed in the previous paragraph. As a result, in the statement of the exclusion principle given above the maximum number of particles has been deliberately constrained to be one.[10)] Henceforth, this exclusion principle will be referred to in a more descriptive way as the *single particle exclusion principle* and lattice gases developed using it will be collectively referred to as *single particle lattice gases*.

Some consequence of this exclusion principle have been pointed out and discussed in the following section. The method of developing lattice gases by incorporating this exclusion principle in their definition and the analysis of the resulting lattice gases is a subject of the remaining chapters of the present investigation.

## 4.2 Consequences of the Single Particle Exclusion Principle

Incorporating the single particle exclusion principle in the definition of lattice gases makes the resulting lattice gases fully discrete counterparts of molecular dynamics. This is because these lattice gases and molecular dynamics become identical in all respects except that space and time are discrete in the former.[11)] The exact consequences of incorporating the single particle exclusion principle in the definition of lattice gases are as follows:

---

[10)] Inquisitive readers might like to relax the maximum number of particles in the exclusion principle and investigate the behavior of the resulting lattice gases. Data obtained from comparative studies allowing different number of particles will definitely substantiate the discussion outlined in this section and is welcome. The inquisitive, however, is being cautioned that overwhelming difficulties will arise in trying to develop evolution rules when the exclusion principle is relaxed to permit more than one particles to occupy the same lattice site simultaneously. It should be noted, that for such data to be useful no constraints should be imposed on the mutual states of particles occupying the same lattice site simultaneously. The intricacies of the procedure involved in developing such lattice gases are much the same as those detailed in the following chapters about developing lattice gases by allowing at the most one particle.

[11)] In fact, to be exact, in molecular dynamics *simulations* also space and time are discrete. This happens because of the inherent nature of computations on digital computers. The details are as follows: Theoretically space and time are continuous in molecular dynamics. Actual computer simulations, however, are carried out in small steps of non-zero time interval and algorithms are developed for the same. While developing the algorithms space is still treated as continuous and implemented as such using real (instead of integer) numbers. The digital computers, however, suffer from finiteness of precision because of which the spatial domain of computation automatically gets discretized in large number of small finite intervals. For example, consider computer implementation of molecular dynamics simulation in one-dimensional space in which coordinates of particles can vary continuously in the closed interval $[x_1, x_2]$. Let, on the computer, the coordinates be stored in a data type having $N$ bits. Then, the continuous domain $[x_1, x_2]$ gets discretized in $2^N$ discrete intervals each of size $(x_2 - x_1)/2^N$. As a result, the simulation, instead of occurring in the continuous domain $[x_1, x_2]$ having infinitely many points, actually occurs in a discrete domain consisting of the set of finitely many points $\{y_i : y_i = x_1 + i(x_2 - x_1)/2^N, i = 0, \ldots, 2^N - 1\}$.

The discreteness of space and time in lattice gases, however, differs in a very fundamental way from the above in that in lattice gases space and time are discrete *by definition* and are treated as such even in theoretical analysis of the models and their dynamics. In theoretical analysis of these models continuum arises

**Interaction Potential:** One of the most important consequence of incorporating the single particle exclusion principle into the definition of lattice gases is that desired realistic interaction potentials (albeit discretized ones) can be easily associated with the particles. This association is possible only because in single particle lattice gases the distance between particles is always non-zero during interactions. This, in turn, is possible only because in these lattice gases at most one particle is allowed to occupy a lattice site at a time step and interactions necessarily occur among particles occupying different lattice sites.

If the single particle exclusion principle is relaxed and more than one particles are allowed to occupy the same lattice site simultaneously, then realistic interactions potentials (other than the repulsive delta function form) cannot be associated with particles in physically consistent and unambiguous manner. This is because the distance between particles occupying the same lattice site, which is the case in these lattice gases, is necessarily zero for all practical purposes and thus repulsion between particles occupying the same lattice site becomes infinite. Similarly, in multiparticle lattice gases also interactions occur among particles occupying the same lattice site and the distance between them is always zero. As a result, in multiparticle lattice gases also it is not possible to associate realistic interaction potentials (other than the repulsive delta function form) with particles without introducing physical inconsistencies and violation of conservation laws in the resulting models.[12)]

**Interactions, Interaction Neighborhood, and Evolution Rules:** Some consequences of incorporating the single particle exclusion principle in the definition of lattice gases surface in the form of changes in the method of identification and definition of interparticle interactions, interaction neighborhood, and evolution rules. Some important differences are as follows:

In multiparticle lattice gases precise identification of "interactions" among particles and construction of "interaction rules" (with wider generalization "evolution rules") is straight forward and goes as follows. At any time step, interactions are said to occur among particles on each lattice site which is occupied by at least two particles. Such a clear identification of interactions is possible only because the interaction neighborhood of particles is restricted to one lattice site and interacting particles are the particles occupying the same lattice site. The outcome of interactions, usually, is change in the state of interacting particles (or, equivalently, lattice sites; because the state of each lattice site is obtained by superimposing the states of all the particles occupying it), subject to the constraints imposed by conservation laws. Since the constraints imposed by the conservation laws must necessarily be satisfied, interaction need not necessarily change the states of interacting particles.

In single particle lattice gases identification of "interactions" among particles and construction of "interaction rules" is not as straight forward as in multiparticle lattice gases. This is because in single particle lattice gases the interaction neighborhood of particles necessarily extends beyond the lattice sites occupied by them.[13)] This necessitates identification of the interaction neighborhood of particles prior to development of interaction

---

as a limiting case wherein the length and time intervals used in simulation become negligible compared to length and time scales of interest and over which variation in properties of interest is small.

[12)] Here, it is assumed that the laws of conservation of mass, momentum, and energy must necessarily be satisfied by all the physical systems taken as a whole and also by their individual parts.

[13)] This happens because the single particle exclusion principle embedded in the definition of single particle lattice gases constraints at most one particle to occupy a lattice site at any time step and forces interactions to occur only among particles occupying different lattice sites.

rules, thus adding an additional step which is absent in multiparticle lattice gases. Identification of the interaction neighborhood of particles poses many complications in the development of single particle lattice gases because the geometry and size of the interaction neighborhood depend not only on the velocity sets of particles but also on their interaction potentials. The procedure of identification of the interaction neighborhood and construction of interaction/evolution rules for desired velocity sets and interaction potentials and various considerations involved therein will be described in details in the following chapters.

## 4.3 Another Interpretation of Conclusions of Secs. 3.3.4 and 3.3.5 and its Consequences

The arguments given in Secs. 3.3.4 and 3.3.5 can also be looked at as implying that for lattice gases to be Galilean invariant and capable of simulating compressible phenomena correctly the net momentum of *lattice sites* on which interactions occur should, in general, change following the interactions. Note that this statement does not mean that the initial and final momentum of lattice sites on which interactions occur should necessarily be different after the interactions have been computed. Instead, it means that the probability that initial and final momentum of the lattice sites will be different is not identically zero.[14)]

The above point of view suggests that to overcome the problems of non-Galilean invariance and incompressibility of lattice gases one only needs to find and incorporate into the definition of lattice gases a mechanism through which change in momentum, in the way mentioned above, can be achieved at lattice sites at which interaction occur among particles. One such mechanism, *viz.*, the single particle exclusion principle, has already been outlined in Sec. 4.1 (more specifically in Sec. 4.1.3). It seems that other alternative mechanisms, not rooted in some way at the single particle exclusion principle, which can lead to lattice gases that are as simple as the multiparticle lattice gases and at the same time have physically consistent dynamics, do not exist. This is in view of the argument given below.

Let us assume that an alternative mechanism, say $\mathcal{M}$, of the type mentioned above exists. Let the lattice gases having the mechanism $\mathcal{M}$ built into them be called as $\mathcal{M}$-lattice gases. Now, note that the simplicity of multiparticle lattice gases arises from the simplicity of their evolution rules which, in turn, are simple only because the interaction neighborhood is restricted to one lattice site, *i.e.*, because collisions occur among particles occupying the same lattice site. Thus, if $\mathcal{M}$-lattice gases are as simple as multiparticle lattice gases, the interaction neighborhood employed in them should also be restricted to one lattice site.[15)] As a result, multiple particles must be allowed to occupy the same lattice site in these lattice gases (otherwise interactions will not occur among particles). This implies that the primary difference and the cause of all other differences between $\mathcal{M}$-lattice gases and multiparticle lattice gases is the presence or absence of $\mathcal{M}$. Thus, $\mathcal{M}$-lattice gases can be developed by incorporating the mechanism $\mathcal{M}$ into multiparticle lattice gases. The remaining part of the argument showing nonexistence of the mechanism $\mathcal{M}$ is as follows:

---

[14)] Note that this probability is identically zero in multiparticle lattice gases.

[15)] That simplicity of the type seen in multiparticle lattice gases is lost when interaction neighborhood is extended to more than one lattice site has already be seen in Secs. 4.1.2 and 4.1.3.

If one tries to incorporate the mechanism $\mathcal{M}$ into multiparticle lattice gases the law of conservation of momentum will necessarily be violated in the resulting lattice gases (the $\mathcal{M}$-lattice gases) at the lattice sites at which interactions occur among particles. This is because in these lattice gases the interaction neighborhood of particles is restricted to one lattice site and consequently each lattice site behaves as an isolated system during interparticle interactions. As a result, in order that the law of conservation of momentum be satisfied at each lattice site the net momentum of each lattice site should remain unchanged before and after interactions have been computed. In $\mathcal{M}$-lattice gases, however, the net momentum of lattice sites at which interactions occur among particles will necessarily change with non-vanishing probability after interactions have been computed. As a result, lattice gases obtained by incorporating the mechanism $\mathcal{M}$ into multiparticle lattice gases will not be physically consistent. This contradicts the starting assumption that the mechanism $\mathcal{M}$ will lead to physically consistent lattice gases. Therefore, $\mathcal{M}$ does not exist.

## 4.4 Lattice Gases Violating Conservation Laws

Arguments presented in Sec. 4.3 show that it is not possible to develop physically consistent lattice gases that are as simple as multiparticle lattice gases by incorporating, in multiparticle lattice gases, a mechanism which, as a consequence of interactions among particles, leads to change in the net momentum of lattice sites at which the interactions occur. Trying to extend multiparticle lattice gases in this way leads to violation of conservation laws in the resulting lattice gases at the lattice sites at which interactions occur among particles. Thus, such an extension is undesirable because the conservation laws must be satisfied at each lattice site at each stage of evolution for ensuring physical consistency.

Despite the above observation, in the literature certain lattice gases have been developed and studied wherein the law of conservation of momentum is violated at the lattice sites during interactions (see [2] for examples). Violation of conservation laws in a model makes the model unacceptable for all practical purposes. As a result, in these lattice gases alternative views regarding conservation of required quantities are employed. Specifically, in these lattice gases the conservation laws are said to be satisfied in one of the two ways, *viz.*, (i) statistically over the entire spatial lattice, *e.g.*, the net momentum over the entire spatial lattice remains unchanged at each stage of evolution, and (ii) in the ensemble at each lattice site (or, for the entire system), *e.g.*, the mean momentum at each lattice site (or, the entire system) in an ensemble remains unchanged at each stage of evolution. The first one of these views can be accepted as correct (though with some reservations; see below). The second view, however, is incorrect. This is because if conservation laws are violated at the lattice sites (or, by the entire system) in a simulation then the dynamical behavior reproduced in that simulation will be unphysical. As a result, the overall dynamical behavior observed after ensemble averaging will also become unphysical.

The first point of view, being statistical in nature, renders the lattice gases based on it as purely statistical (rather than dynamical) tools. Close comparison of ideas behind these lattice gases with those employed in the *direct simulation Monte Carlo* method [3] reveals that these lattice gases are essentially completely discrete analogs of the direct simulation Monte Carlo method; apparently, without additional advantages. The major different between the two being that in the former the results of all interactions are precomputed whereas in the later they are computed as and when required.

The above approach for developing simple lattice gases, however, is unacceptable in view of the following observations: (i) This approach permits development of unphysical lattice gases violating Newton's third law of motion and the laws of thermodynamics during interparticle interactions. This casts an undesirable shadow of doubt over physical validity of simulation results because the consequences of these violations are unpredictable. One might attempt to avoid these violations by explicitly incorporating some (deterministic) mechanism which, for example, ensures that the net momentum over the entire spatial lattice before and after interactions have been computed remains unaltered. This cannot be allowed because it is equivalent to incorporating a global mechanism for verification of conservation laws and leads to loss of the desired simplicity. (ii) These lattice gases are *inherently* physically inconsistent because the dynamics of interactions among particles necessarily remains physically inconsistent because of the violation of conservation laws.

The above observations show that if it is desired to develop physically consistent lattice gases without violating the conservation laws then the only consistent method is to incorporate the single particle exclusion principle in the definition of lattice gases. This rewinds the discussion to the beginning of Sec. 4.2 and then into complexities of developing collision rules for single particle lattice gases. These have been discussed in the following chapters.

## 4.5 Conclusions

The primary conclusions of the analysis presented in this chapter are the following:

1) It is not possible to restore spatial momentum redistribution during interparticle interactions in multiparticle lattice gases in a physically consistent manner. Thus, it is not possible to overcome the problems of non-Galilean invariance and incompressibility of multiparticle lattice gases.

2) Spatial momentum redistribution during interparticle interactions can be restored by incorporating *single particle exclusion principle* in the definition of lattice gases. The resulting lattice gases, termed *single particle lattice gases*, will be physically consistent, Galilean invariant, and capable of correctly simulating compressible systems.

The single particle lattice gases can be viewed as discrete analogs of molecular dynamics based on the formalism of cellular automata. These lattice gases, being radically different in their underlying philosophy from multiparticle lattice gases, cannot be developed like multiparticle lattice gases. Thus, to fulfill the objectives this investigation (*c.f.*, Sec. 1.4), further investigations on the method of development of these lattice gases are required.

## Chapter 5

# Construction of Single Particle Lattice Gases

. . . and remember, as humans we have many peculiarities.
*Yes, sire.*
Having faith is good and usually necessary in life; but it cannot be forced.
*Yes, sire.*
It is unfortunate that as scientists we are faithless.
*A doubt, sire! After all this analysis!*
Pointers are just indicative.
. . .

In view of the conclusions of chapter 4, the investigations presented in this chapter are directed towards formalizing a systematic method for construction of single particle lattice gases. Specifically, in this chapter various considerations involved in construction of single particle lattice gases and a systematic procedure for their construction have been outlined.

Since single particle lattice gases are fully discrete analogs of molecular dynamics, many of the basic considerations involved in their construction correspond closely to those in molecular dynamics and are equally indispensable. The inherent discreteness of these lattice gases, however, makes them fundamentally different from molecular dynamics in many ways. This necessitates certain additional considerations, hitherto unknown and unnecessary, in these lattice gases. These considerations are addressed in the following sections. Thereafter, the systematic procedure of construction of single particle lattice gases is outlined.

## 5.1 Necessary Considerations

Mathematical models of physical systems are developed under a number of assumptions for simplifying the task of modeling. Simplifications are needed also because the efficiency of available methods for dealing with the complexity of resulting models and the time required for producing intelligible results from the models are usually constrained in investigations. Although on one hand simplifying assumptions reduce the complexity of models and speed up the task of modeling, on the other hand they impose limits (usually not sharply defined) on the range of variation of input parameters of the models. As a result, the models stand valid and produce intelligible results as long as they are used within the constraints imposed by the simplifying assumptions that were invoked during modeling. The results produced by the models start departing from experimental observations as soon as the constraints imposed by the simplifying assumptions are compromised and finally after a limit the models break down, *i.e.*, become invalid or no longer describe the system correctly.

The simplifying assumptions are made such that spurious and uninteresting information about the physical system is eliminated from the model while the information of interest remains unaffected and is reproduced correctly by the model. As far as models of particle dynamical systems are concerned, the simplifying assumptions are usually based on the length and time scales of interest. Depending on the problem's requirements and the modeling methodology that has been adopted, assumptions involving other parameters might also be made as and when required. One such assumption, particularly in microscopic models like molecular dynamics, Monte-Carlo methods, and the single particle lattice gases being investigated herein, relates to the nature of interactions among particles which is specified through various interaction potentials.[1)] It should, however, be noted and I wish to emphasize this fact, that assumptions involving length scales and time scales of interest lie at the core of all the classical and semiclassical descriptions of physical systems. In fact, it does not seem possible to develop an intelligible microscopic model of particle dynamical systems within the classical and semiclassical frameworks without making assumptions, in some form or the other, about the length and time scales of interest and also about the nature of interaction potential. It is these assumptions and the way and form in which they enter into single particle lattice gases that is of primary concern in this section.

### 5.1.1 Length Scales, Time Scales, and related Assumptions

While modeling physical systems, especially fluid dynamical systems, one basic observation that one makes is that there exist many different length scales and time scales on which the systems can be described. The choice of the length and time scales on which a system should be described in a modeling exercise depends largely on the problem's requirements. In the present investigation the scope of inquiry will be restricted to phenomena occurring at length scales of the order of particle diameter or larger. Addressing smaller length scales goes beyond the scope of this investigation because of predominance of quantum effects.

#### 5.1.1.1 Conventional Description of Physical Systems

Above and at length scales larger than particle diameter there exist three different characteristic lengths at which different phenomena occur in physical systems. These characteristic lengths are, (i) the range of interactions $R_{\mathrm{I}}$, (ii) the mean free path $\lambda$, and (iii) the length over which a macroscopic property changes by a finite amount $L_{\mathrm{h}}$. The characteristic times associated with these characteristic lengths are, (i) the time duration of a collision $t_{\mathrm{c}}$, (ii) the time between two successive collisions of a particle $t_{\mathrm{k}}$, and (iii) the macroscopic time $t_{\mathrm{h}}$, respectively. The correspondence between these characteristic lengths and times is

$$\begin{aligned} t_{\mathrm{c}} &\simeq R_{\mathrm{I}}/c_{\mathrm{mts}} \\ t_{\mathrm{k}} &\simeq \lambda/c_{\mathrm{mts}} \\ t_{\mathrm{h}} &\simeq L_{\mathrm{h}}/c_{\mathrm{s}} \end{aligned}$$

[1)]Usually binary interaction potentials are employed [1] which suffice for studying weakly correlated systems, *e.g.*, gases. For studying properties of strongly correlated systems, *e.g.*, bulk and surface properties of liquids and condensed matter, many-body interaction potentials are needed and should be used [2].

where $c_{\text{mts}}$ is the mean thermal speed of particles and $c_{\text{s}}$ is the speed of sound.

The time scales corresponding to the characteristic times mentioned above are known as, (i) the collision time scale, (ii) the kinetic time scale, and (iii) the hydrodynamic time scale, respectively. These time scales were first formally introduced and used for studying irreversible processes occurring in gases in the theory of Bogoliubov and later in the theory of Prigogine and Balescu and also in the theory of Frieman and Sandri [3]. In gases at ordinary densities these time scales are widely different from each other. This difference permits accurate description of gases at many different levels (of approximation), *e.g.*, the Navier-Stokes equation exists at and above the hydrodynamic time scale, the Boltzmann equation exists at and above the kinetic time scale, and the kinetic equations obtained from the theories of Bogoliubov, and Prigogine and Balescu, and Frieman and Sandri exist at and above the collision time scale.

In calculus based descriptions of particle dynamical systems, assumptions involving length and time scales of interest have to be incorporated explicitly into the models as is done in the theories mentioned above. This, rather than being a shortcoming, is a welcome flexibility of such descriptions because it facilitates development of models at the desired level of approximation and complexity as elaborated earlier in Sec. 5.1 on page 119.

#### 5.1.1.2 Single Particle Lattice Gases

As in the usual calculus based descriptions of physical systems, the flexibility of choosing length and time scales of interest is available in single particle lattice gases also. It, however, is somewhat constrained because of discreteness of space-time in which these lattice gases exist. In addition to this, in these lattice gases the single particle exclusion principle also imposes some constraints on the length and time scales of interest. Because of these constraints, the way and form in which the assumptions regarding length and time scales of interest enter into single particle lattice gases becomes very different from that in which they enter into the usual calculus based descriptions. In single particle lattice gases, the assumptions regarding the length and time scales of interest enter as elaborated below.

The discreteness of space-time in single particle lattice gases necessitates that the length of links connecting two neighboring lattice sites $\Delta x$ and the duration of each evolution (or, the time step) $\Delta t$ be known so that particle velocities can be quantized appropriately. Furthermore, if one wants to associate finite diameter $d$ with the particles, the single particle exclusion principle requires that the condition $\Delta x > d$ should necessarily be satisfied, *i.e.*, the dimensions of each cell should be large enough to contain a particle in its entirety. This is under the assumption that the particles have a rigid core of diameter $d$. For more generalized particle models, *e.g.*, point center of force particles, the lower bound of $d$, naturally, is the smallest distance of closest approach $r_{c_{\min}}$ of particles in the system; which, in a simple gas system, is achieved in the most energetic head-on collision. The upper bound on $d$ and the interrelationship between $\Delta x$ and $r_{c_{\min}}$ is determined by certain additional considerations on interaction potentials and discrete velocity sets outlined in Secs. 5.1.3.8 and 5.1.4.

In order to determine the relationship of the evolution time $\Delta t$ with other characteristic times of the system, further considerations on the assumptions and strategy that will be used for evolving the lattice gas system by one time step are required. Many different strategies can be devised. Description of all the possible strategies, however, is not within the scope of this investigation. Brief description of the evolution strategy that will be used in single particle lattice gases being proposed and addressed herein is as follows:

> **Evolution Strategy in Single Particle Lattice Gases:** *The evolution during one time step is decomposed into two sub-steps, namely, interaction step and particle translation step. Separate rules are developed for each of these sub-steps. In the interaction step particles only interact without moving from their locations and in the translation step particles move to new locations without interaction with each other. For evolving the system by one time step the rules for the interaction step and the particle translation step are applied consecutively and in the same order, i.e., first the interaction rules are applied on (the given state of) the system and a new intermediate state of the system is obtained and then the translation rules are applied on this intermediate state to obtain the final state of the system; which completes one evolution.*

This strategy of decomposition of evolution into two consecutive sub-steps has been adopted directly from multiparticle lattice gases (*c.f.*, Sec. 2.3). It must be noted that adoption of this strategy is *not* a necessity for developing evolution rules in either single particle or multiparticle lattice gases, *i.e.*, in both the types of lattice gases evolution rules can be developed without decomposing the evolution during one time step into interaction and particle translation steps, also. This a decomposition, however, simplifies the task of development of evolution rules considerably, and thus, has been adopted.

The decomposition of evolution during one time step into two consecutive sub-steps as detailed above requires that the duration of each sub-step be known before proceeding with other considerations involved in the actual construction of the lattice gas. For the evolution strategy described above, the duration of each sub-step is determined as follows:

Let the duration of the collision resolution step[2)] be $\Delta t_{\mathrm{C}}$ and that of the particle translation step be $\Delta t_{\mathrm{T}}$. With this the interrelationship between $\Delta t$, $\Delta t_{\mathrm{C}}$, and $\Delta t_{\mathrm{T}}$ is

$$\Delta t = \Delta t_{\mathrm{C}} + \Delta t_{\mathrm{T}} \tag{5.1}$$

Let us change the perspective slightly and view $\Delta t_{\mathrm{C}}$ and $\Delta t_{\mathrm{T}} = \Delta T - \Delta t_{\mathrm{C}}$ to be times elapsed during the collision and translation processes for individual particles. The total time available to each particle for undergoing through these two processes during one evolution (or, time step) being $\Delta T$. Now note that in single particle lattice gases being addressed herein, at any time step a particle either does not collide or it undergoes exactly one collision with one or many of its neighbors. One consequence of this constraint is that in these lattice gases $\Delta t_{\mathrm{C}}$ is not identical for all the particles even if the particles are identical. For particles which do not undergo collision $\Delta t_{\mathrm{C}} = 0$ and for particles which undergo a collision $\Delta t_{\mathrm{C}} = t_{\mathrm{c}}$. Because of this, during each evolution, the time available to each particle during which it remains in free motion is different. As a result, displacement of each particle during one time step is different even if they are moving with the same speed during free motion. In single particle lattice gases the validity of this statement rests on the fact that collisions change only the velocity of particles. This is true in non-reacting systems. In reacting systems, however, one or many of the reacting particles might get

[2)] In general, the interaction step. All interactions, however, can be viewed as collisions and their consequences; which is customary in the kinetic theory literature. Hence, the terms *collisions* and *interactions* have been used interchangeably throughout in the present investigation.

displaced to new locations during reactions. Thus, for achieving exactness, in general, it is necessary to consider displacement of particles during interactions also.

As far as the objective of this investigation (*c.f.*, Sec. 1.4) is concerned, consideration of the displacement of particles during collisions for developing lattice gases becomes too elaborate and will turn out to be inconsequential as has been shown by developments in classical kinetic theory. Thus, the assumption that displacement of particles during collisions is negligible compared to their displacement during free motion is invoked. This assumption, since the displacement of particles during the time interval $\Delta t_{\mathrm{C}}$, in general, is $\simeq \Delta t_{\mathrm{C}} c_{\mathrm{mts}}$ and that during the time interval $\Delta t_{\mathrm{T}}$ is $= \Delta t_{\mathrm{T}} c_{\mathrm{mts}}$, implies that collisions have been assumed to be instantaneous, *i.e.*, it has been assumed that the condition

$$\Delta t_{\mathrm{C}} \ll \Delta t_{\mathrm{T}} \tag{5.2}$$

or, equivalently the condition

$$\Delta t_{\mathrm{C}} \ll \Delta t \tag{5.3}$$

holds for the system.

Another consequence of the constraint that particles undergo either zero or exactly one collision at any time step is that in single particle lattice gases the condition

$$\Delta t \leq t_{\mathrm{mft}} \tag{5.4}$$

always holds; where $t_{\mathrm{mft}}$ is the mean free time. This is because, statistically a particle undergoes one collision in the time interval $t_{\mathrm{mft}}$.

The meaning and implications of Eqs. (5.3) and (5.4) in terms of classical kinetic theory are revealed by combining them and substituting $\Delta t_{\mathrm{C}} \leq t_{\mathrm{c}}$. This gives that in single particle lattice gases the condition

$$t_{\mathrm{c}} \ll t_{\mathrm{mft}} \tag{5.5}$$

is always satisfied. This condition is the first assumption in Boltzmann's analysis of gases and also one of the basic ideas in Bogoliubov's theory [3]. This implies that in the single particle lattice gases the assumption given by Eq. (5.3) along with the constraint that a particle undergoes either zero or exactly one collision at any time step is equivalent to the Boltzmann's basic assumption given by Eq. (5.5).

The recovery of Boltzmann's basic assumption in single particle lattice gases implies that results obtained from simulations carried out using these lattice gases should be at least as accurate as those obtained by solving the Boltzmann equation. Thus, it can be expected that single particle lattice gas simulations will be able to capture and show all those details of a system which can be seen by solving the Boltzmann equation for it; with, perhaps, some additional features. In fact, it will be seen from simulation results presented in chapters 7 and 8 that this expectation is correctly fulfilled (within the limits in which behavior and capabilities of a simple single particle lattice gas be generalized). The exact relationship of results obtained from single particle lattice gas simulations with the results obtained from Bogoliubov's theory varies with the details of the lattice gas. Some elaborations of this aspect of single particle lattice gases will be furnished later as and when needed.

In the *natural units* of lattice gas systems the length of links, duration of each evolution (or, time step), and duration of the collision resolution step are denoted by $\delta x$, $\tau$, and $\delta\tau$, respectively. The transition from $\Delta x$, $\Delta t$, and $\Delta t_{\mathrm{C}}$ to $\delta x$, $\tau$, and $\delta\tau$ can be viewed to be due to invocation of a non-dimensionalization procedure wherein the lengths are represented in the units of $\Delta x$ and the time is represented in the units of $\Delta t$. Thus, in the natural units of lattice gas system

$$\delta x = \Delta x / \Delta x = 1$$

$$\tau = \Delta t / \Delta t = 1$$

$$\delta \tau = \Delta t_{\mathrm{C}} / \Delta t \simeq 0$$

for the single particle lattice gases being considered in the present investigation.

### 5.1.2 Interaction Potentials and related Assumptions

A variety of interaction potentials are available for use in continuum and partially discrete (*i.e.*, when at least one out of the space, time, or dependent variables is continuous and at least one of them is discrete) simulation methods [1,2]. These interaction potentials, however, cannot be employed directly in fully discrete simulation models, *e.g.*, single particle lattice gases. This is because in these models particles occupy only specific locations marked as lattice sites on the spatial lattice and thus interactions among particles occur at well defined discrete distances. As a result, discretized versions of interaction potentials are needed for computing and analyzing interparticle interactions in such methods.

The method of arriving at discrete forms of interaction potentials which are equivalent to given forms in continuum will be described in Sec. 5.1.2.4. Before proceeding with the description, however, it is necessary to look into certain important considerations which surface during development of discretized versions of interaction potentials. More so because these considerations have important consequences in the construction of lattice gases also.

#### 5.1.2.1 Range of Interactions

It is well known in kinetic theory that collision cross sections diverge for interaction potentials which decay to zero only in the limit $r \to \infty$ [4]. To overcome this problem it becomes necessary to limit the range of interactions $R_{\mathrm{I}}$ to be within finite bounds. In the methods of classical kinetic theory, this finiteness is usually achieved in an indirect manner by imposing a reasonable cutoff on the scattering (or, deflection) angle [4].

Problems arise in single particle lattice gases also, though in a slightly different form, if the range of interaction potential is infinite. This is because if $R_{\mathrm{I}} = \infty$, every particle in the system interacts with all the other particles in the system. This gives rise to a kind of global coupling among all the particles in the system which cannot be overcome or eliminated in any (known) way without constraining $R_{\mathrm{I}}$ to be finite. Henceforth, this kind of global coupling will be referred to as *"potential global coupling"* because it arises from the nature of the interaction potential. This term is required, and thus has been coined, for distinction because later on it will be seen that one more kind of global coupling, which does not arise from the nature of the interaction potential, is also present in single particle lattice gases.

The consequences of potential global coupling are extremely severe in that it prevents even the development of single particle lattice gases. This is because presently no method is known by which evolution rules can be written for single particle lattice gases with infinite range interaction potentials. In fact, it appears that no such method exists at all. This is because in general the computational cost of writing the evolution rules for such a lattice gas will be infinite both in terms of time and space (storage memory) requirements.

In order to bypass the occurrence of potential global coupling in single particle lattice gases, it is essential to limit the range of interactions within finite bounds. This is done before discretizing the continuum interaction potentials. Thus, the *"locality hypothesis"*[3] is implicitly invoked for developing discretized versions of continuum interaction potentials. This implies that ones discretized interaction potentials are available, locality hypothesis need never be reinvoked while constructing lattice gas analog of the physical system.

In view of the above, the first step for developing discretized versions of continuum interaction potentials is to fix the range of interactions $R_{\mathrm{I}}$. The criteria for fixing $R_{\mathrm{I}}$ is to be chosen by the investigator and it could be *any* appropriate criteria. Its exact nature or description is of no consequence here except for the constraint that it should be such that it results in a finite and non-zero value of $R_{\mathrm{I}}$. Let $\tilde{R}_{\mathrm{I}}$ be the range of interactions in natural units of the lattice system. The transformation of $R_{\mathrm{I}}$ into $\tilde{R}_{\mathrm{I}}$ is achieved by invoking a non-dimensionalization procedure wherein all the lengths are represented in the units of $\Delta x$ as has been done earlier in Sec. 5.1.1.2. Thus,

$$\tilde{R}_{\mathrm{I}} = R_{\mathrm{I}}/\Delta x \tag{5.6}$$

#### 5.1.2.2 Finite Range Interactions in Continuum Space

To understand the interaction among particles with finite $R_{\mathrm{I}}$ in continuum space consider, for the sake of simplicity, a system of identical particles so that $R_{\mathrm{I}}$ is the same for all the particles. Let us select an arbitrary particle A in this system and look at its interaction with other particles located in its neighborhood in a reference frame fixed at A. Thus, A can be treated as the source of potential. The force experienced by other particles due to their presence in the neighborhood of A at a distance of $r$ from A will be zero if $r > R_{\mathrm{I}}$. Thus, such particles do not interact with A. However, if the other particles are located at a distance of $r \leq R_{\mathrm{I}}$ from A, they interact with A and the force felt by them could either be attractive, repulsive, or zero depending on $r$ and the exact form of the interaction potential. Note that the probability that two particles separated by a distance $r \leq R_{\mathrm{I}}$ will experience zero force *approaches* zero because usually zero force occurs on a surface of thickness zero at a distance of $r < R_{\mathrm{I}}$ around the particles; most of attractive-repulsive interaction potentials are of this type. This straight forward scenario of interparticle interactions in continuum space is implied by the finiteness of $R_{\mathrm{I}}$. The equivalent scenario in discrete space, however, becomes quite different because of the discreteness of the spatial lattice.

#### 5.1.2.3 Finite Range Interaction in Discrete Space

To understand the interaction among particles with finite $R_{\mathrm{I}}$ in discrete space consider a system of identical particles existing over a discrete spatial lattice subject to the single particle exclusion principle. In this system, particles occupy only those locations in space which coincide with the lattice sites. As a result, interparticle interactions occur only at well defined discrete distances. However, $R_{\mathrm{I}}$ is obtained from considerations in continuum.

[3] **Locality Hypothesis:** All interactions are local in space. Alternatively, all the events occurring at a point in space depend only on the states of a (or, events occurring in a) finite zone around that point.

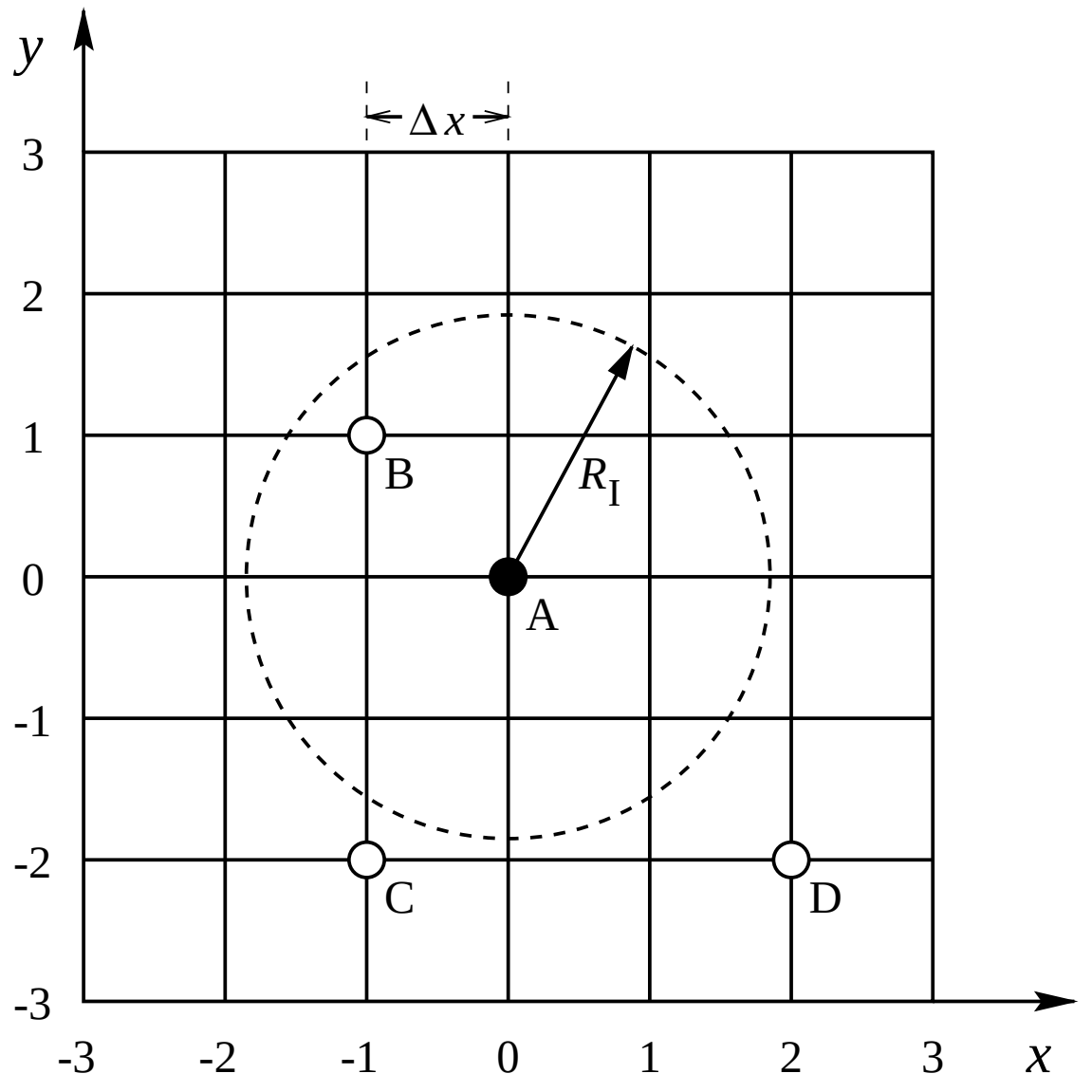

**Figure 5.1:** Interaction among particles on square spatial lattice (solid lines). Small black circle represents the particle at which potential is centered. Big dashed circle represents the interaction zone of this particle in continuum space. Small hollow circles represent other particles in the system.

This leads to complications in deciding whether or not two particles interact with each other. The way and form in which these complications arise and the method of their resolution is illustrated clearly through the following example:

Let us assume, for the sake of simplicity, that the system exists over square spatial lattice and look at the interaction of an arbitrary particle (say, A) with the other particles (say, B, C, ...) in this system. Consider an instance of this system with the reference frame fixed to the particle A as shown in Fig. 5.1 on page 126. In this figure only four particles A, B, C, and D have been shown. Particle A, being the reference particle, has been shown with black circle and the particles B, C, and D have been shown with hollow circles. The dashed circle of radius $R_{\mathrm{I}}$ drawn around the particle A shows the zone of interaction of this particle with other particles in its neighborhood in *continuum* space. Now, for this system the question *"Which of the particles B, C, and D interact with A, and why?"* is addressed below.

If the analysis that was used in continuum space (*c.f.*, Sec. 5.1.2.2) is employed without alterations for answering the above question, one arrives at the conclusion that particle B interacts with particle A and that particles C and D do not interact with particle A. This conclusion, however, is incorrect. This is because the spatial lattice of systems which exist in discrete position space, supposedly, is an equivalent discrete representation of the continuum position space.[4] This equivalence is correctly established when the discrete space is viewed as having a densely packed cellular structure in which the lattice sites coincide with the centroid of each cell; and not when it is viewed as an array of discrete points connected to each other with infinitely thin links. Note that mathematically both these concepts are equivalent in that a collection of points is known as a *"lattice"* or *"lattice of sites"* with the points making the *"lattice sites"* and the collection of closely packed boxes whose centroids coincide with the lattice sites is known as the *"dual of the lattice"*.

[4]The present knowledge of physics does not tell with certainty whether the *position space* is discrete or continuous in physical reality. In calculus based models of physical systems, however, the position space has been treated as a continuum since long. Whether this treatment is an outcome of the invention of real numbers or whether the position space is actually a continuum in physical reality is subject to debate, which I bypass. For the existence of calculus based models it is, however, necessary that position space (and other relevant variables) be continuous. This is because of the very definition of derivative as a limit wherein the denominator approach zero [5] (as a consequence of this, discreteness vanishes).

The above implies that for obtaining discrete equivalents of continuum properties (which includes interparticle interactions and the like) in particle dynamical systems existing in discrete position space one should work on the dual of the spatial lattice and not directly on the spatial lattice itself. This requires further elaboration because as far as mathematical analysis is concerned, spatial lattice and its dual are equivalent concepts. As a result, the natural expectation is that any mathematical analysis of a particle dynamical system existing in discrete space should yield identical results irrespective of whether the analysis has been carried out on spatial lattice or its dual. This expectation, however, contradicts the earlier conclusion; thus raising the doubt *"why should the results obtained from an analysis carried out over (a given) spatial lattice be different from those obtained from a similar analysis carried out over the dual of the spatial lattice?"*

When some aspect of particle dynamical systems which exist in discrete space is being analyzed on the dual of the spatial lattice, the particles need to be viewed as residing inside the cells rather than as located on the lattice sites. Since the dimensions of each cell are finite and non-zero (as opposed to the dimensions of the lattice sites which are universally zero) and since the smallest distinguishable length scale is restricted to the dimensions of each cell, the particles are free to occupy any location inside a cell with equal probability. Thus, on the dual of a spatial lattice the particles are (necessarily) viewed as delocalized inside the cells, *i.e.*, having finite uncertainty in position. In contradistinction to this, note that on a spatial lattice the particles are usually viewed as fixed on the lattice sites, *i.e.*, having zero uncertainty in position. Thus, the uncertainty associated with the position of particles on a spatial lattice differs from that on its dual. This difference is the cause of difference in results when an exercise seeking the discrete equivalent of a continuum property is carried out on a spatial lattice and on its dual.

The above shows that analysis for ascertaining which of the particles B, C, and D interact with particle A in the example system must be carried out after transforming the spatial lattice of the system to its dual. As shown in Fig. 5.1, the example system exists on square spatial lattice. The dual of square spatial lattice is shown in Fig. 5.2 and new configuration of the example system on dual of square spatial lattice is shown in Fig. 5.3. For analyzing the system in its new configuration it shall be assumed, for the sake of simplicity, that interaction potentials of particles are centered at the centroids of cells (*i.e.*, the lattice sites) occupied by them rather than at the particles themselves.

---

Since the currently existent view point about the position space being a continuum has *worked* for time immemorial and since there have not been any acceptable indications which hint at the possibility of position space being discrete (see Bacry [6] and Penrose [7] for some arguments, counter arguments, and hypotheses), I prefer to look at the spatial lattice of lattice gases as an equivalent discrete representation of the continuum space rather than as the correct representation of a physical reality in which the position space is actually discrete. If it so turns out some time in future that the position space is indeed discrete in physical reality, the present investigation will be affected only in that the statements wherein the spatial lattice has been taken to be an equivalent discrete representation of continuum space will have to be discarded.

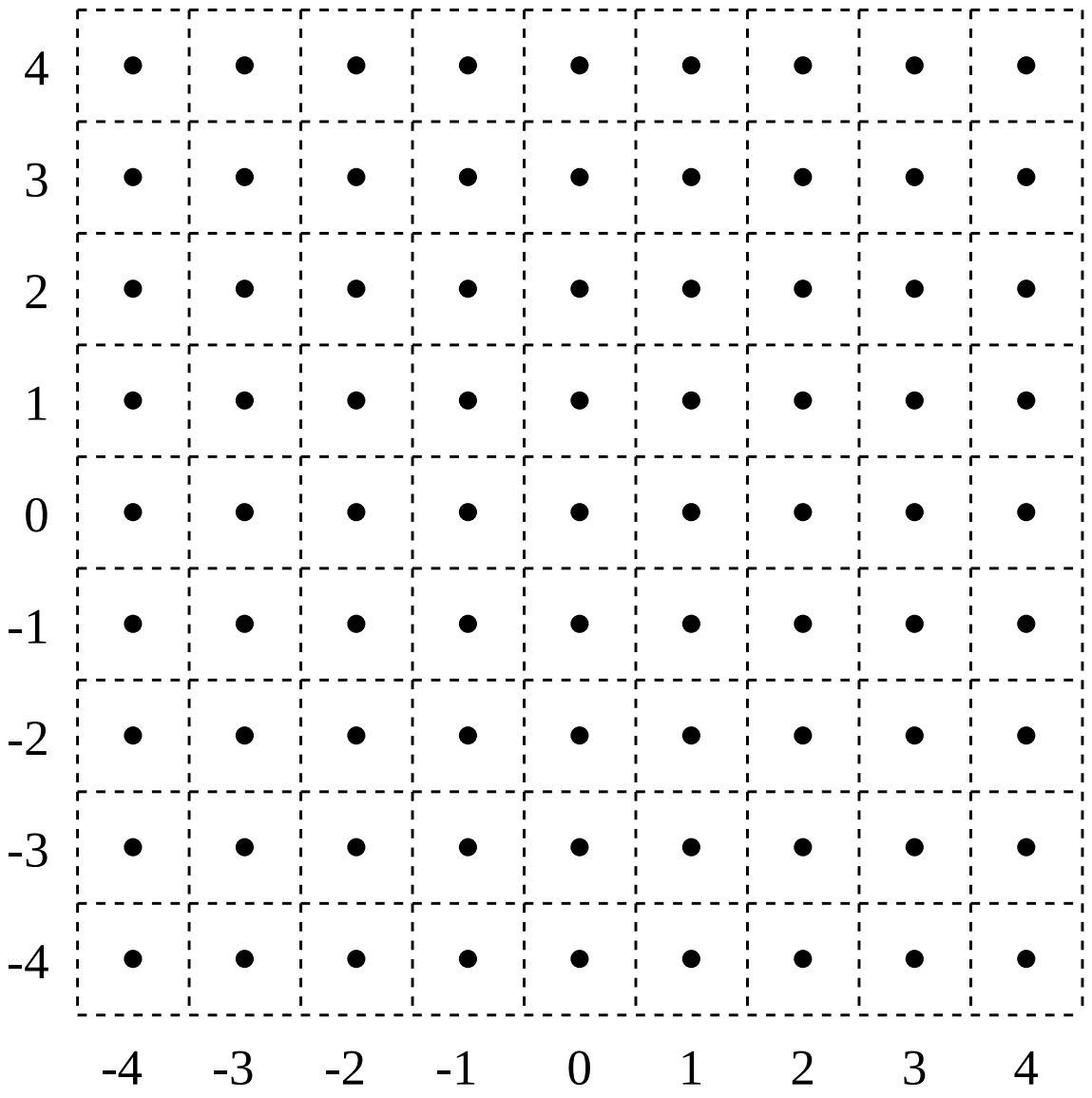

**Figure 5.2:** Schematic of square spatial lattice and its dual. Black circles represent the lattice sites. The boxes made by dotted lines around each lattice site form the dual of the lattice. The distance between two consecutive nearest neighbor lattice sites is $\Delta x$. The numbers at the bottom and left are coordinates of the lattice sites (in units of $\Delta x$) in an arbitrary reference frame fixed at the lattice site located at $(0, 0)$.

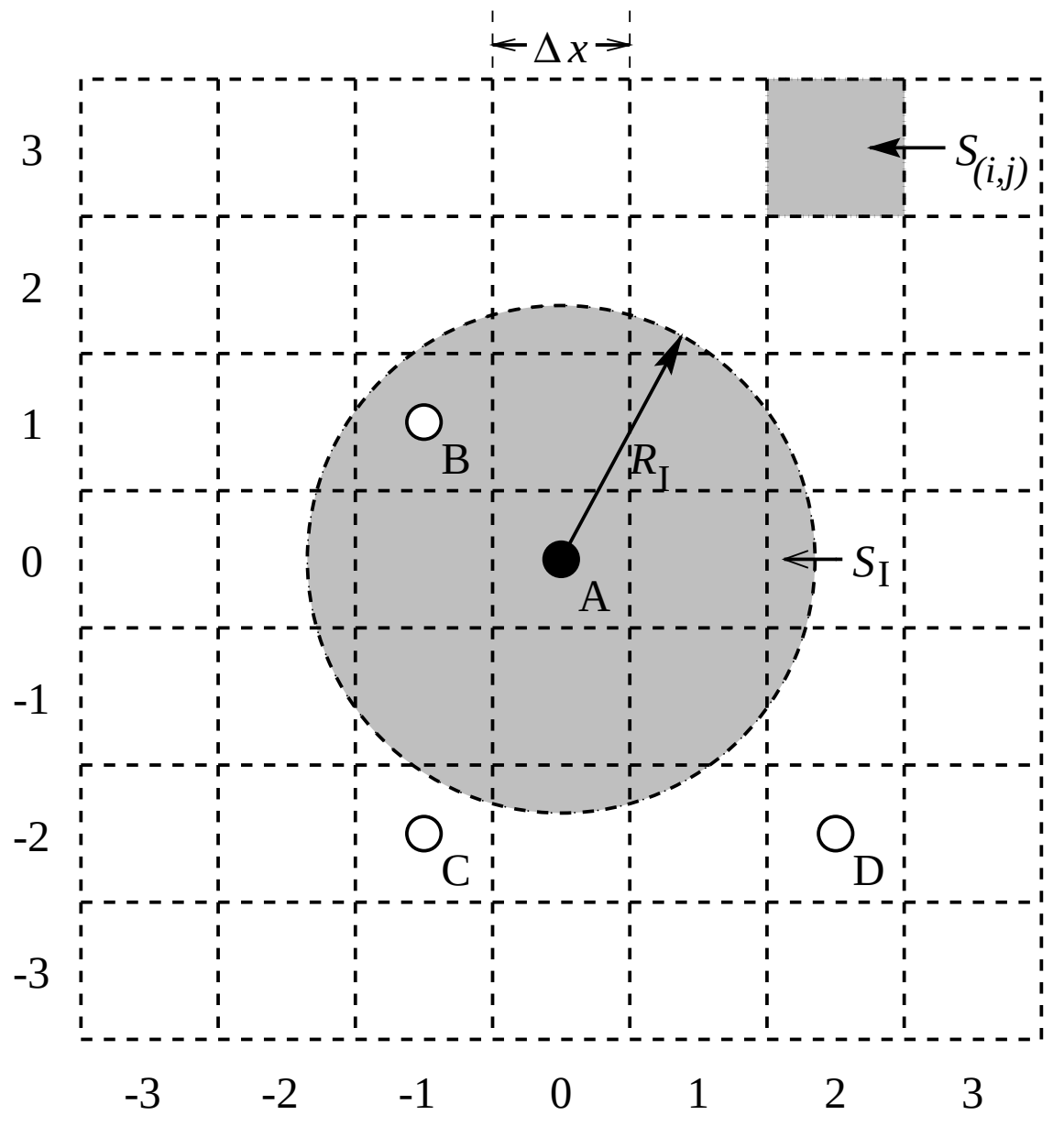

**Figure 5.3:** Interaction among particles on the dual of square spatial lattice.

The above assumption might appear to be somewhat counterintuitive because in usual analysis procedures (*e.g.*, as in molecular dynamics, kinetic theory, and Monte-Carlo methods) the interaction potentials of particles are always taken to be centered at the particles themselves. This doubt is amenable to easy clarification because the situation in which the above assumption has been invoked happens to be totally different from the one that exists in the usual analysis procedures. In the usual analysis procedures, locations of particles are taken to be known with zero uncertainty. As a result, it becomes necessary that interaction potentials of particles be viewed as centered at the particles themselves. Whereas, in particle dynamical systems existing in discrete position space, the particles are delocalized inside the cells and it is in this state that they interact with other particles. In delocalized state, the particles can be viewed to be simultaneously present at all the points inside their respective cells because the probability of finding the particles at any point inside their cells is the same. Alternatively, the delocalization of particles can be viewed to be giving rise to continuum of mass within the cells occupied by them (this situation can also be viewed as if the particles were "filling" their respective cells). Thus, the mean location of particles, in the center of force model, lies at the centroid of the cells (*i.e.*, lattice sites) occupied by them. As a result, the interaction potentials of particles can be safely taken to be centered at the centroids of cells occupied by them whereas the particles themselves are delocalized inside their respective cells; as has been assumed above.

With the above insights, I revert back to the analysis of the example system. Since the particles are delocalized inside their cells, the particle A will interact with all the particles whose *cells* lie, either fully or partially, within the range of interaction potential of the particle A. For mathematical formulation of this statement, let the zone occupied by the cell around lattice site $(i, j)$ be $S_{(i,j)}$ and the zone affected by the interaction potential of particle A be $S_{\mathrm{I}}$ (see the shaded regions in Fig. 5.3). Then, mathematically the above statement says that particle A will interact with a particle located at the site $(i, j)$ *iff* the condition

$$\left(S_{(i,j)} \cap S_{\mathrm{I}}\right) \supset \emptyset \tag{5.7}$$

is satisfied, and not otherwise; where, $\emptyset$ represents the empty set or null set. This condition reveals that in the example system particle A will interact with particles B and C and not with particle D. In general, the particle A will interact with particles occupying any one of the lattice sites $(\pm 1, 0)$, $(0, \pm 1)$, $(\pm 1, \pm 1)$, $(\pm 2, 0)$, $(0, \pm 2)$, $(\pm 2, \pm 1)$, or $(\pm 1, \pm 2)$ and not with particles occupying other lattice sites.

#### 5.1.2.4 Discretization of Continuum Interaction Potentials

Information furnished in previous sections makes it possible to proceed with the description of the procedure of discretization of continuum interaction potentials. The discretized forms of interaction potentials, thus developed, can be employed for constructions of single particle lattice gases. The details of the discretization procedure are as follows:

Let, $\Phi(\boldsymbol{r})$ be the continuum interaction potential with range of interactions $R_{\mathrm{I}}$ that is to be discretized. Let, the potential be centered at the origin of coordinate system being used. Then the mean potential $\bar{\Phi}(S)$ experienced by a particle delocalized in some region $S$ is

$$\bar{\Phi}(S) = \frac{1}{A} \int_S \Phi(\boldsymbol{r}) ds \tag{5.8}$$

where $A$ is the area of $S$ in two-dimensional space, volume of $S$ in three-dimensional space, *etc.* Generalized schematic representation depicting physical meanings of various variables appearing in the above equation is shown in Fig. 5.4.

On square spatial lattice having links of size $\Delta x$, the mean potential $\bar{\Phi}(x_i, y_j)$ experienced by a particle delocalized in the cell around the lattice site $(i, j)$ located at the point $(x_i, y_j)$, $x_i = i\Delta x$, $y_j = j\Delta x$, in a coordinate system centered at the source of potential, is

$$\begin{aligned} \bar{\Phi}(x_i, y_j) &= \frac{1}{A(x_i, y_j)} \int_{S(x_i, y_j)} \Phi(x, y)\, ds \\ &= \frac{1}{(\Delta x)^2} \int_{y_j - \Delta x/2}^{y_j + \Delta x/2} \int_{x_i - \Delta x/2}^{x_i + \Delta x/2} \Phi(x, y)\, dx\, dy \end{aligned} \tag{5.9}$$

where $S(x_i, y_j)$ represents the region occupied by the cell around the lattice site $(i, j)$ and $A(x_i, y_j) = (\Delta x)^2$ is area of the cell.

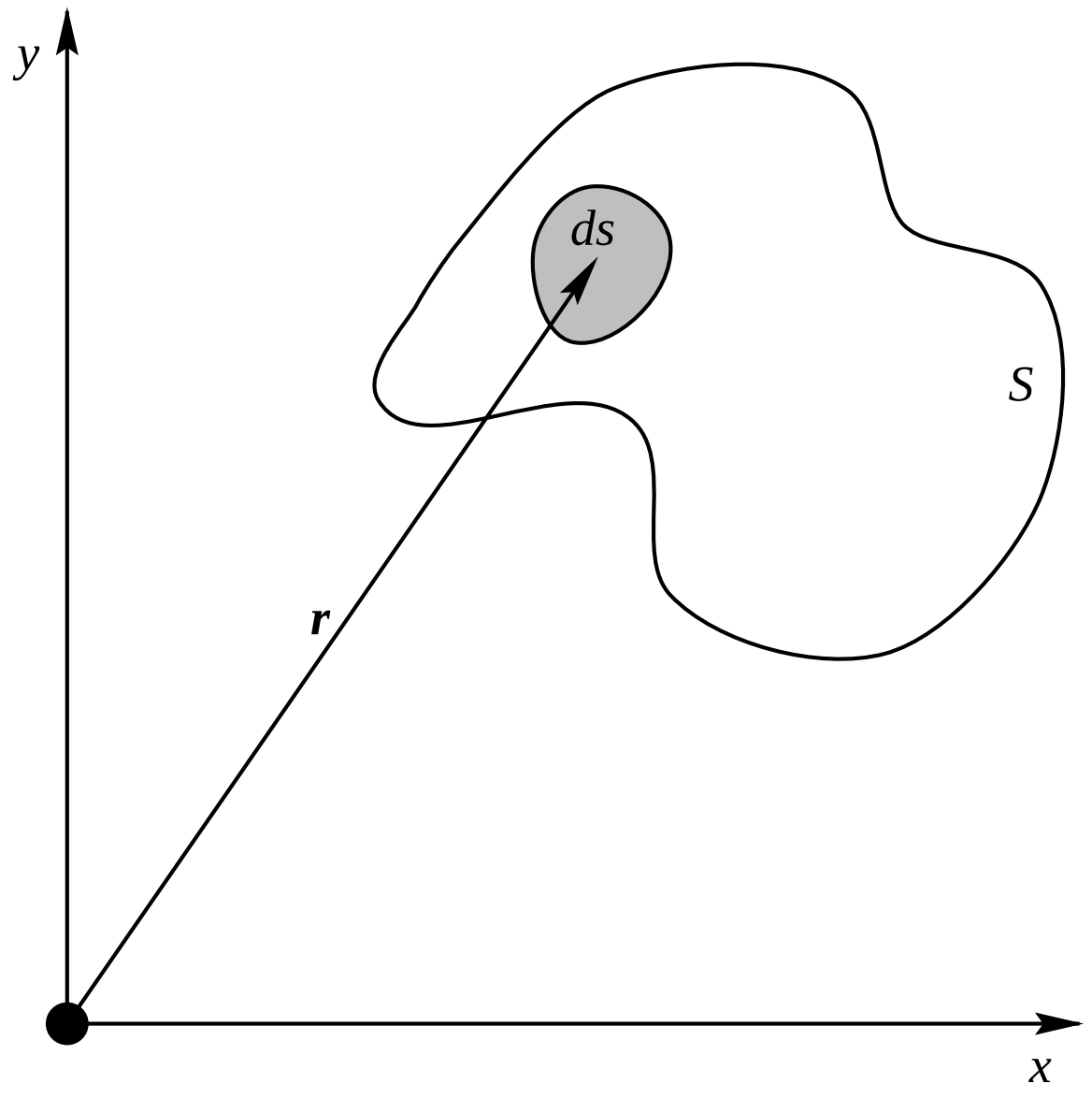

**Figure 5.4:** Generalized schematic representation in two-dimensional Cartesian position space depicting physical meanings of various variables appearing in Eq. (5.8) for development of discretized forms of interaction potentials which are equivalent to given forms in continuum.

In the natural units of lattice system, *i.e.*, in the system of units wherein all the distances are expressed in units of lattice links $\Delta x$, Eq. (5.9) gives

$$\bar{\Phi}(i, j) = \int_{j-1/2}^{j+1/2} \int_{i-1/2}^{i+1/2} \Phi(\tilde{x}, \tilde{y})\, d\tilde{x}\, d\tilde{y} \tag{5.10}$$

where the dummy variables $\tilde{x}$, $\tilde{y}$, $d\tilde{x}$ and $d\tilde{y}$ have also been non-dimensionalized by $\Delta x$. Note that $\bar{\Phi}(x_i, y_j) \equiv \bar{\Phi}(i, j)$ because only the lengths (and not the height of the potential function) have been non-dimensionalized.

In two-dimensional space, the case when $\Phi(x, y)$ is circularly symmetric and $\Phi(0, 0) = \infty$, deserves special attention because of its simplified nature. In this case, Eq. (5.10) implies that on any two-dimensional spatial lattice the mean potential at the lattice site $(0, 0)$ is

$$\bar{\Phi}(0, 0) = \infty \tag{5.11}$$

This equation asserts that no two particles can occupy the same lattice site, which is same as the assertion of the single particle exclusion principle. This, once again, shows that single particle exclusion principle, rather than being an arbitrary constraint, is demanded by the dynamics of interparticle interactions (so that it can be physically consistent).

In addition to Eq. (5.11), circular symmetry of $\Phi(x, y)$ implies that on the square spatial lattice, the relation

$$\bar{\Phi}(\pm i, \pm j) = \bar{\Phi}(\pm j, \pm i)$$

always holds good. Furthermore, for a lattice site $(i, j)$, $\bar{\Phi}(i, j)$ will be zero if Eq. (5.7) is not satisfied. If, however, Eq. (5.7) is satisfied, $\bar{\Phi}(i, j)$ could be zero, positive, or negative depending on the exact form of $\Phi(x, y)$ and the length of links $\Delta x$.

The discretized interaction potentials obtained as above can be used for computing the mean force experienced by particles and thus employed in construction of interaction rules. Since the force is

$$\boldsymbol{F}(\boldsymbol{r}) = -\boldsymbol{\nabla}\Phi(\boldsymbol{r}) \tag{5.12}$$

and the mean force is

$$\bar{\boldsymbol{F}}(\boldsymbol{r}) = -\boldsymbol{\nabla}\bar{\Phi}(\boldsymbol{r}) \tag{5.13}$$

it can be computed from the discretized interaction potential using finite difference formulae. Some examples being

$$\bar{\boldsymbol{F}}(\boldsymbol{r}) = \frac{\bar{\Phi}(\boldsymbol{r}+\Delta\boldsymbol{r}) - \bar{\Phi}(\boldsymbol{r})}{\Delta\boldsymbol{r}} + \mathcal{O}(\Delta\boldsymbol{r}) \qquad \text{Forward difference} \tag{5.14}$$

$$= \frac{\bar{\Phi}(\boldsymbol{r}) - \bar{\Phi}(\boldsymbol{r}-\Delta\boldsymbol{r})}{\Delta\boldsymbol{r}} + \mathcal{O}(\Delta\boldsymbol{r}) \qquad \text{Backward difference} \tag{5.15}$$

$$= \frac{\bar{\Phi}(\boldsymbol{r}+\Delta\boldsymbol{r}) - \bar{\Phi}(\boldsymbol{r}-\Delta\boldsymbol{r})}{2\Delta\boldsymbol{r}} + \mathcal{O}(\Delta\boldsymbol{r})^2 \qquad \text{Central difference} \tag{5.16}$$

Computation of force from discretized interaction potential using finite difference formulae as outlines above, however, is subject to inaccuracies arising from truncation of the Taylor series used for arriving at the finite difference formulae. As a result, it is desirable to compute the discretized values of mean force directly from Eq. (5.12). The procedure is similar to that employed for interaction potentials. Thus,

$$\bar{\boldsymbol{F}}(S) = -\frac{1}{A}\int_S \boldsymbol{\nabla}\Phi(\boldsymbol{r})ds \tag{5.17}$$

On square spatial lattice the above equation gives,

$$\begin{aligned}\bar{\boldsymbol{F}}(x_i, y_j) &= -\frac{1}{A(x_i, y_j)}\int_{S(x_i,y_j)} \boldsymbol{\nabla}\Phi(x, y)\, ds \\ &= -\frac{1}{(\Delta x)^2}\int_{y_j-\Delta x/2}^{y_j+\Delta x/2}\int_{x_i-\Delta x/2}^{x_i+\Delta x/2} \boldsymbol{\nabla}\Phi(x, y)\, dx\, dy \end{aligned} \tag{5.18}$$

where $S(x_i, y_j)$ represents the region occupied by the cell around the lattice site $(i, j)$ and $A(x_i, y_j) = (\Delta x)^2$ is area of the cell.

In the natural units of lattice system Eq. (5.18) gives

$$\bar{\boldsymbol{F}}(i, j) = -\int_{j-1/2}^{j+1/2}\int_{i-1/2}^{i+1/2} \boldsymbol{\nabla}\Phi(\tilde{x}, \tilde{y})\, d\tilde{x}\, d\tilde{y} \tag{5.19}$$

where the dummy variables $\tilde{x}$, $\tilde{y}$, $d\tilde{x}$ and $d\tilde{y}$ have also been non-dimensionalized by $\Delta x$.

In two-dimensional space, for circularly symmetric $\Phi(x, y)$ with $\Phi(0, 0) = \infty$, Eq. (5.19) implies that the mean force at the lattice site $(0, 0)$ on any spatial lattice is

$$\bar{\boldsymbol{F}}(0, 0) = \infty \tag{5.20}$$

This equation is merely reformulation of Eq. (5.11) and has identical implications.

In addition to Eq. (5.20), circular symmetry of $\Phi(x, y)$ implies that on the square spatial lattice, the relation

$$\bar{\boldsymbol{F}}(\pm i, \pm j) = \bar{\boldsymbol{F}}(\pm j, \pm i)$$

always holds good. Furthermore, for a lattice site $(i, j)$, $\bar{\boldsymbol{F}}(i, j)$ will be zero if Eq. (5.7) is not satisfied. If, however, Eq. (5.7) is satisfied, $\bar{\boldsymbol{F}}(i, j)$ could be zero, positive (repulsive), or negative (attractive) depending on the exact form of $\Phi(x, y)$ and the length of links $\Delta x$.

### 5.1.3 Interaction Mechanisms and Interaction Neighborhoods

The elaborations on length and time scales in single particle lattice gases and their relationship with those in physical systems furnished in Sec. 5.1.1.2 are sufficient for determining the time step to be used for evolving single particle lattice gases. The length scales and in particular the length of links $\Delta x$, however, cannot be determined at present. Knowledge of $\Delta x$, however, being necessary for discretization of continuum interaction potentials (*c.f.*, Sec. 5.1.2.4), is essential for construction of single particle lattice gases. Determination of $\Delta x$ as well as construction of interaction rules of single particle lattice gases requires certain considerations on interaction among particles. Thus, rigorous analysis of various mechanisms of interparticle interactions and associated interaction neighborhoods in single particle lattice gases is necessary. These details are furnished in the following sections.

In single particle lattice gases, particles interact with each other through two radically different mechanisms. These mechanisms are: (i) interaction due to the mutual interaction potential of particles, and (ii) interaction due to physical contact between particles. Henceforth, the interactions which occur due to the interaction potential of particles will be called as *field interactions* and the interactions which occur due to physical contact between particle will be called as *contact interactions* or *impact interactions*. Field interactions are well known in classical theories of classical physics. In classical physics, the contact interactions are known to appear in collision between hard spheres.

In single particle lattice gases both the field and contact interactions of particles occur under specific conditions when other particles are located in a finite neighborhood around them. Henceforth, the neighborhoods associated with these interactions will be referred to as *field interaction neighborhood* and *contact interaction neighborhood*, respectively. Both these interaction neighborhoods play important role in the development of single particle lattice gases. They, however, are not used directly. Instead, an *overall interaction neighborhood*, which subsumes both of them and is obtained by merging them, is employed. These interactions, their mechanisms, interactions neighborhoods, and various interrelationships among them are described in the following sections (Secs. 5.1.3.1–5.1.3.8).

#### 5.1.3.1 Field Interactions in Continuum and Discrete Space

*Field interactions* are the interactions which occur among particles at a distance through their mutual interaction potential. In these interactions, the force exerted by the particles on each other depends only on their mutual orientation and distance of separation and has no dependence on the velocities of the particles, *e.g.*, in a binary interaction between identical spherical particles in three-dimensions, any two particles separated by the same distance exert the same force on each other irrespective of their velocity vectors. These interactions take finite time for completion which is equal to the time interval between the instant at which the particles enter into each other's range of influence and the instant at which they go out of each other's range of influence. In these interactions the distance between particles changes continuously throughout the duration of interaction. Furthermore, the distance of closest approach of particles, depending on the energy of particles and other parameters (*e.g.*, the impact parameter, *etc.*) at the instant when the particles start influencing each other, can become arbitrarily small during interactions but can never become zero because the classical interaction potentials $\Phi(\boldsymbol{r})$ have a pole at $\boldsymbol{r} = 0$. Field interactions occur only in those models of physical systems in which particles are treated as centers of forces,[5)] irrespective of whether the models exist in continuum space or discrete space.

If the range of interaction potential is finite, which is necessarily true in single particle lattice gases, the field interactions of particles occur locally with other particles located in a small neighborhood around them. This neighborhood is referred to as the *field interaction neighborhood* or the *neighborhood of field interactions*. The method of determining topology of the field interaction neighborhood in single particle lattice gases is detailed in Sec. 5.1.3.2.

#### 5.1.3.2 Field Interaction Neighborhood in Discrete Space

For developing evolution rules of single particle lattice gases it is necessary to know the topology of field interaction neighborhood of particles. The topology is known if the exact form of discretized interaction potential is known. To know just the topology, however, it is not necessary to compute the exact form of the discretized interaction potential. It can be determined using Eq. (5.7) if the range of interactions, structure of spatial lattice, and lattice parameters are known. This is because topology of any domain in discrete position space is known if coordinates of all the lattice sites comprising that domain are known.

It has been pointed out in Sec. 5.1.3 that information furnished till now does not permit determination of lattice parameters (*i.e.*, length of links $\Delta x$, *etc.*).[6)] As a result, topology of the field interaction neighborhood cannot be determined at this stage. In order to bypass the need of knowing the lattice parameters, in the following, the topology of the field interaction neighborhood will be discussed in a generalized fashion based only on the range of interactions in the natural units of the lattice system. In this discussion, for the sake of simplicity, it will be assumed that the interaction potential is circularly symmetric and the system exists over square spatial lattice. Generalizations to other spatial lattices and other forms of interaction potentials are straight forward.

The range of interactions is a parameter derived from considerations in continuum. Whereas, in single particle lattice gases interparticle interactions occur only at well defined discrete distances because of discreteness of position space. As a result, topology of the field interaction neighborhood of particles remains identical for a range of values of $\tilde{R}_{\mathrm{I}}$. Let

[5)]For field interactions to occur, the particles must necessarily have a non-rigid potential around them; they may or may not have a rigid core. In general, their interaction potentials should be of the form

$$\Phi(\boldsymbol{r}) = \begin{cases} \infty & \text{when } f(\boldsymbol{r}) \leq 0 \\ \phi(\boldsymbol{r}) & \text{when } f(\boldsymbol{r}) > 0 \end{cases}$$

where $f(\boldsymbol{r}) = 0$ is the equation of the rigid core, $f(\boldsymbol{r}) \leq 0$ specifies the interior of the rigid core having infinite potential, and $f(\boldsymbol{r}) > 0$ specifies the exterior of the rigid core having soft (*i.e.*, finite) potential.

[6)]Additional constraints which are necessary for determination of lattice parameters come from considerations on the discrete velocity set and the topology of the *contact interaction neighborhood* of particles. These are given in Secs. 5.1.3.8 and 5.1.4.

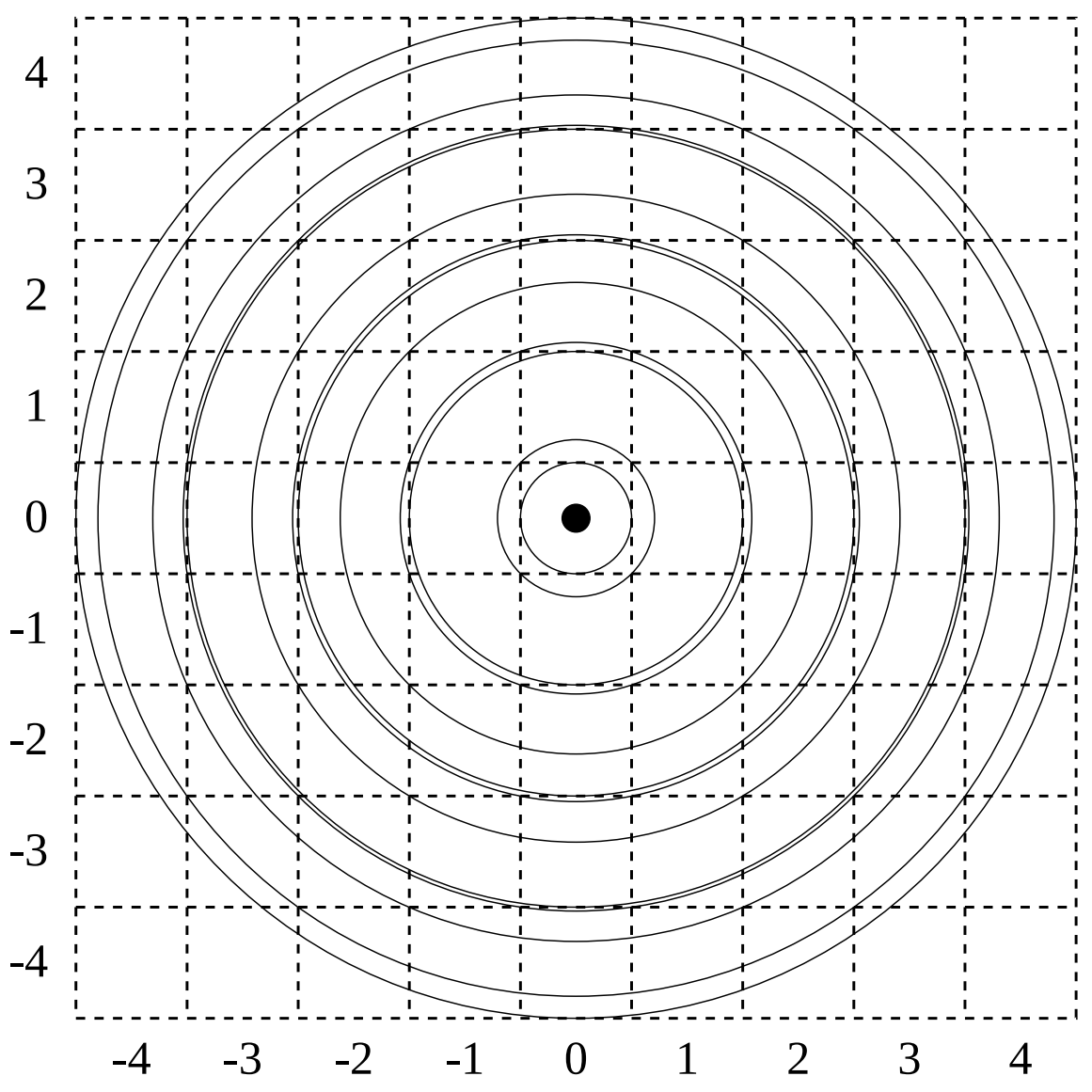

**Figure 5.5:** The bands of $\tilde{R}_{\mathrm{I}}$ for symmetric interaction potentials in which topology of the field interaction neighborhood remains unchanged over square spatial lattice. Each band is made by two consecutive circles. The inner circle for the inner most band has zero radius. All the lattice sites which are covered, either fully or partially, by the outer most circle for a band make the field interaction neighborhood for values of $\tilde{R}_{\mathrm{I}}$ lying within that band. The origin of the coordinate system is fixed to lattice site at which interaction potential is centered (marked with solid circle).

| Range of $4\tilde{R}^2_{\mathrm{I}}$ | | Coordinates of lattice sites lying in the field interaction neighborhood | Number of sites |
|---|---|---|---|
| $4\tilde{R}^2_{\mathrm{I_{min}}}$ | $4\tilde{R}^2_{\mathrm{I_{max}}}$ | | |
| 0 | 1 | $(0,0)$ | 1 |
| 1 | 2 | $\oplus$ $(\pm1,0)$, $(0,\pm1)$ | 5 |
| 2 | 9 | $\oplus$ $(\pm1,\pm1)$ | 9 |
| 9 | 10 | $\oplus$ $(\pm2,0)$, $(0,\pm2)$ | 13 |
| 10 | 18 | $\oplus$ $(\pm2,\pm1)$, $(\pm1,\pm2)$ | 21 |
| 18 | 25 | $\oplus$ $(\pm2,\pm2)$ | 25 |
| 25 | 26 | $\oplus$ $(\pm3,0)$, $(0,\pm3)$ | 29 |
| 26 | 34 | $\oplus$ $(\pm3,\pm1)$, $(\pm1,\pm3)$ | 37 |
| 34 | 49 | $\oplus$ $(\pm3,\pm2)$, $(\pm2,\pm3)$ | 45 |
| 49 | 50 | $\oplus$ $(\pm4,0)$, $(0,\pm4)$ | 49 |
| 50 | 58 | $\oplus$ $(\pm4,\pm1)$, $(\pm1,\pm4)$, $(\pm3,\pm3)$ | 61 |
| 58 | 74 | $\oplus$ $(\pm4,\pm2)$, $(\pm2,\pm4)$ | 69 |
| 74 | 81 | $\oplus$ $(\pm4,\pm3)$, $(\pm3,\pm4)$ | 77 |

**Table 5.1:** Coordinates of lattice sites lying within the field interaction neighborhood of particles on square spatial lattice for various ranges of variation of $\tilde{R}_{\mathrm{I}}$. The sign "$\oplus$" should be read as "all the above lattice sites and".

$\tilde{R}_{\mathrm{I_{min}}}$ be the minimum and $\tilde{R}_{\mathrm{I_{max}}}$ be the maximum value of $\tilde{R}_{\mathrm{I}}$ for which the topology of the field interaction neighborhood remain unaltered. The values of $\tilde{R}_{\mathrm{I_{min}}}$ and $\tilde{R}_{\mathrm{I_{max}}}$ depend on the structure of spatial lattice. Various ranges of variation of $\tilde{R}_{\mathrm{I}}$ (up to $\tilde{R}_{\mathrm{I}} \leq 4.5$) over the square spatial lattice within which the topology of field interaction neighborhood remains unaltered are shown in Fig. 5.5. The coordinates of lattice sites comprising the field interaction neighborhood for some of the ranges of variation of $\tilde{R}_{\mathrm{I}}$, $\tilde{R}_{\mathrm{I_{min}}} < \tilde{R}_{\mathrm{I}} \leq \tilde{R}_{\mathrm{I_{max}}}$, are given in table 5.1. Schematic visualization of field interaction neighborhood for some of the ranges of variation of $\tilde{R}_{\mathrm{I}}$, $0 \leq \tilde{R}_{\mathrm{I}} \leq 3.5$, over square spatial lattice is shown in Fig. 5.6.

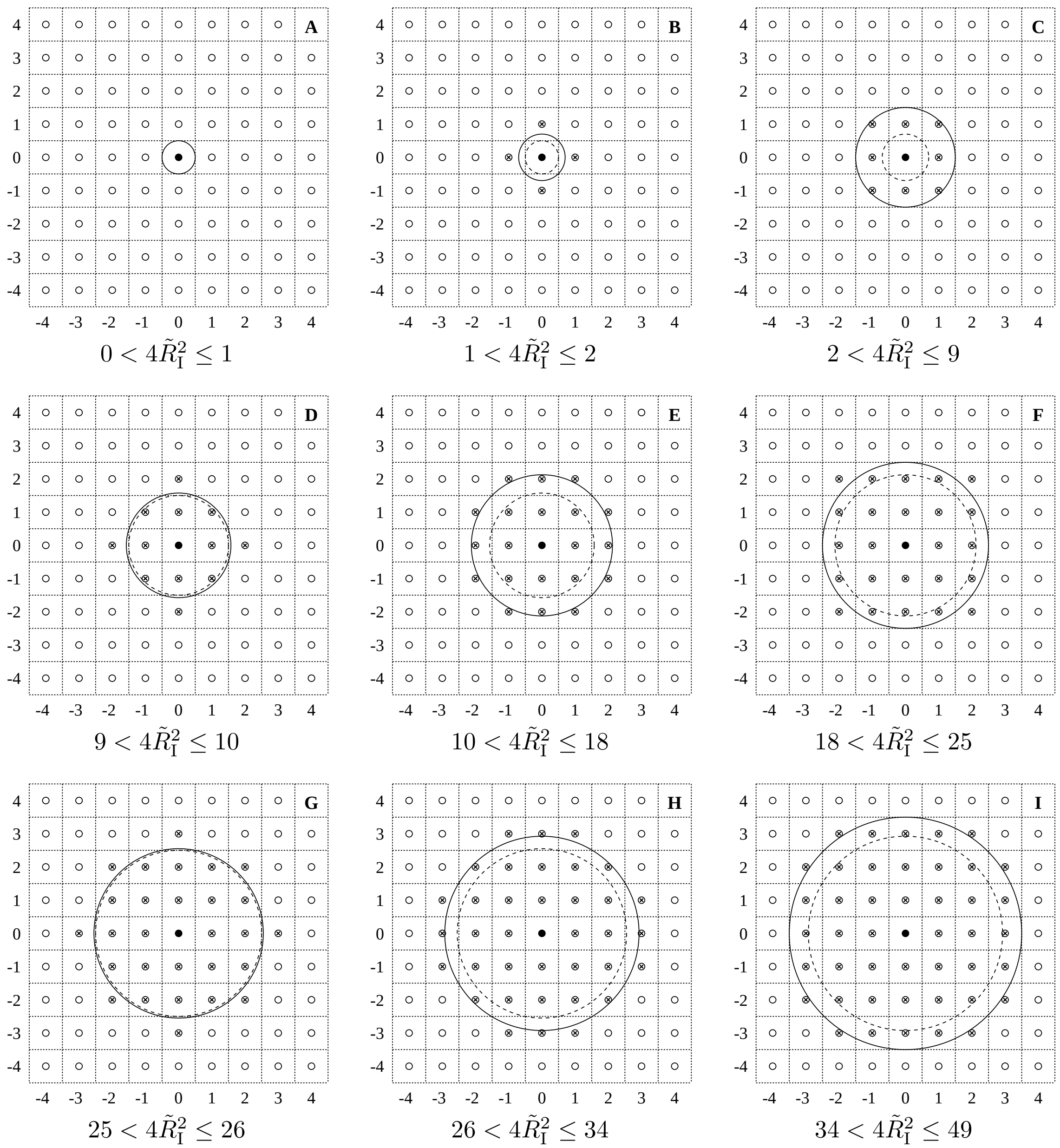

**Figure 5.6:** Topology of the field interaction neighborhood over square spatial lattice for circularly symmetric interaction potentials in some ranges of variation of $\tilde{R}_{\mathrm{I}}$ (shown with big solid and dashed circles). Solid circles ($\bullet$) represent the lattice site at which the potential is centered. Crossed circles ($\otimes$) and solid circles represent the lattice sites comprising the field interaction neighborhood.

The format of table 5.1 is somewhat unusual and should be understood as follows: In the first and second columns of this table the range of variation of $4\tilde{R}_{\mathrm{I}}^2$, instead of $\tilde{R}_{\mathrm{I}}$, has been given. This is because, $4\tilde{R}_{\mathrm{I_{min}}}^2$ and $4\tilde{R}_{\mathrm{I_{max}}}^2$ are always non-negative integers. The sign "$\oplus$", which appears in the third column of the table, indicates that all the lattice sites listed in the previous rows are included in this field interaction neighborhood also.

#### 5.1.3.3 Contact Interactions in Continuum and Discrete Space: Vertex and Edge Interactions

*Contact interactions* are the interactions which occur among particles that influence each other through physical contact. In these interactions, the force exerted by the particles on each other depends on their momentum at the time of contact and not on the distance of separation between them. Occurrence of these interactions depends strongly on velocity vectors of particles because for these interactions to occur the particles should be moving with velocities such that they reach at the same location in space at same time and thus come into contact with each other. These interactions are truly instantaneous interactions because the information transfer between rigid bodies occurs instantaneously. In continuum space, these interactions are said to occur only when the distance between the centroids of particles equals a *fixed* value which depends on the geometry and orientation of the particles at the time of contact. The distance at the time of contact is *fixed* also in the sense that it does not change with time and that the interaction occurs *only when* the particles moving relative to each other are separated exactly by this distance. Contact interactions occur only in those models of particle dynamical systems in which particles are treated as having a rigid core of finite and non-zero dimensions, *e.g.*, hard spheres, irrespective of whether the models exist in continuum space or discrete space.

In single particle lattice gases, particles necessarily have a rigid core of finite and non-zero dimensions. The rigid core cannot be eliminated, *i.e.*, its dimensions cannot be made zero, because its origin is a natural consequence of discreteness of position space and single particle exclusion principle. As a result, contact interactions appear naturally in all single particle lattice gases and no special effort is needed to incorporate them into the models. Field interactions, on the other hand, have to be incorporated explicitly in single particle lattice gases. This is done by superimposing an interaction potential of the desired form over the rigid core of particles. Thus, in single particle lattice gases particles can be treated either as *rigid bodies* undergoing only contact interactions or as *centers of forces with rigid core of finite and non-zero dimensions* undergoing both contact and field interactions but not as *point centers of forces* which undergo only field interactions. The impossibility of modeling particles as *point centers of forces* is a boon for single particle lattice gases because it completely eliminates the troublesome singularity which occurs when the distance between particles which are point centers of forces becomes zero.

In single particle lattice gases contact interactions among particles, like field interactions, also occur locally because particles move with discrete velocities of finite magnitude. The neighborhood around particles in which they can undergo contact interactions with other particles is referred to as the *contact interaction neighborhood*. For constructing single particle lattice gases it is necessary to determine the topology of contact interaction neighborhood because these interactions necessarily occur in all single particle lattice gases.

The mechanism by which contact interactions occur in models existing in continuum position space is as explained above. In continuum position space these interaction are subsumed within field interactions and treated in an identical way as field interaction by assigning infinite potential to the rigid core (if any) of particles. In models existing in discrete position space, however, major differences arise in the mechanism of occurrence and treatment of these interactions because of discreteness of position space (and other constraints, if any, *e.g.*, the single particle exclusion principle in the case of single particle lattice gases). As a result, before attempting to understand the method of determining the topology of the contact interaction neighborhood, it is necessary to understand the exact nature of interactions which are being termed as *"contact interactions"* and the mechanism by which they occur in single particle lattice gases being addressed herein.

To understand various types of contact interactions and their mechanism of occurrence in single particle lattice gases consider a system of *rigid particles* existing in discrete position space subject to the single particle exclusion principle. The system evolves in discrete time steps. During the evolution, particles move over the spatial lattice with discrete velocities and interact with each other. These interactions occur through two radically different mechanisms, *viz.*, (i) when two or more particles are headed for the same lattice site (or, vertex) simultaneously, and (ii) when two or more particles moving parallel to each other and headed for different lattice sites cross each other while passing through a common link (or, edge). Interactions occurring through mechanism (i) are termed as *"vertex interactions"* and those occurring through mechanism (ii) are termed as *"edge interactions"*. Collectively, these interactions are termed as *"contact interactions"*. The exact nature of these interactions, the detailed mechanism of their occurrence, and the method of determining topology of contact interaction neighborhood are elaborated in Secs. 5.1.3.4–5.1.3.7 below.

#### 5.1.3.4 Nature and Mechanism of Vertex Interactions

In a given configuration of particles over the spatial lattice, *vertex interactions* occur when there exists at least one group of two or more particles in which the particles are competing for the same lattice site. Thus, for these interactions to occur the particles, depending on their velocities, must be located in very specific geometric arrangements relative to each other. In general, if vertex interaction occurs among $N$ particles located at the lattice sites $\boldsymbol{x}_i$, $i = 1, \ldots, N$, and moving with the velocities $\boldsymbol{v}_i$, $i = 1, \ldots, N$, then the condition

$$\boldsymbol{x}_i + \boldsymbol{v}_i \Delta t = \boldsymbol{x}_j + \boldsymbol{v}_j \Delta t = \boldsymbol{x}_k \tag{5.21}$$

is necessarily satisfied; where, $i, j = 1, \ldots, N$, $i \neq j$, and $\boldsymbol{x}_k$ is the lattice site which is the common target of all the $N$ particles. In addition to this, exactly one of the following three conditions regarding the state of the lattice site $\boldsymbol{x}_k$ also holds: (i) The lattice site $\boldsymbol{x}_k$ is empty, *i.e.*, $\boldsymbol{x}_k \notin \{\boldsymbol{x}_1, \ldots, \boldsymbol{x}_N\}$. (ii) Velocity of the particle occupying the lattice site $\boldsymbol{x}_k$ is zero, *i.e.*, $\boldsymbol{x}_k \in \{\boldsymbol{x}_1, \ldots, \boldsymbol{x}_N\}$ and $\boldsymbol{v}_k = 0$. This also implies that the particle occupying the lattice site $\boldsymbol{x}_k$ is one of the $N$ particles among which vertex interaction occurs. (iii) Velocity of the particle occupying the lattice site $\boldsymbol{x}_k$ is non-zero and it does not undergo an *edge interaction*[7)] (to be defined below) with any of the $N$ particles under question, *i.e.*, $\boldsymbol{x}_k \notin \{\boldsymbol{x}_1, \ldots, \boldsymbol{x}_N\}$ and $\boldsymbol{v}_k \neq 0$. Some examples of configurations over square spatial lattice in which vertex interactions occur among particles are shown in Fig. 5.7.

#### 5.1.3.5 Nature and Mechanism of Edge Interactions

In a given configuration of particles over the spatial lattice, *edge interactions* occur when there exists at least one group of two or more particles in which the particles, in order to reach at their targeted lattice sites, have to pass through the same link simultaneously. These interactions occur because it is *possible* that while passing through a common link

[7)] If the velocity of the particle occupying the lattice site $\boldsymbol{x}_k$ is non-zero, it will never undergo vertex interaction with any of the $N$ particles under question because of Eq. (5.21).

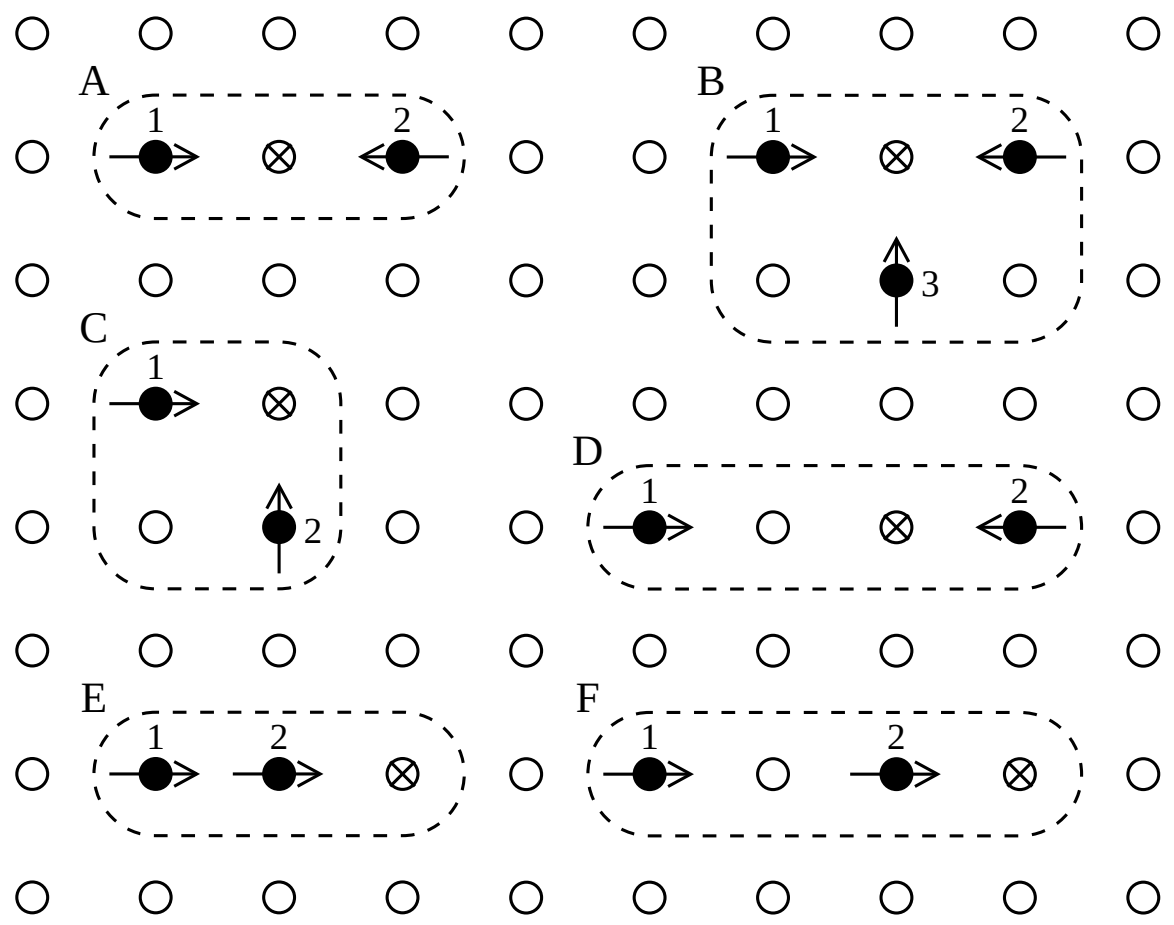

**Figure 5.7:** Examples of vertex interactions over square spatial lattice in single particle lattice gases. Each configuration is enclosed in a dashed oval and marked with a capital letter. Various symbols represent: • particles, ∘ unoccupied lattice sites, →↑← direction of motion of particles, and ⊗ the lattice site which is common target of all the particles in each configuration. Particles in each configuration are labeled 1, 2, 3. Velocities of various particles are: $\boldsymbol{v}_{\mathrm{A1}} = (1,0)$, $\boldsymbol{v}_{\mathrm{A2}} = (-1,0)$, $\boldsymbol{v}_{\mathrm{B1}} = (1,0)$, $\boldsymbol{v}_{\mathrm{B2}} = (-1,0)$, $\boldsymbol{v}_{\mathrm{B3}} = (0,1)$, $\boldsymbol{v}_{\mathrm{C1}} = (1,0)$, $\boldsymbol{v}_{\mathrm{C2}} = (0,1)$, $\boldsymbol{v}_{\mathrm{D1}} = (2,0)$, $\boldsymbol{v}_{\mathrm{D2}} = (-1,0)$, $\boldsymbol{v}_{\mathrm{E1}} = (2,0)$, $\boldsymbol{v}_{\mathrm{E2}} = (1,0)$, $\boldsymbol{v}_{\mathrm{F1}} = (3,0)$, and $\boldsymbol{v}_{\mathrm{F2}} = (1,0)$, where $\boldsymbol{v}_{\alpha\beta}$ represents velocity of particle $\beta$ in configuration $\alpha$ and the pairs $(v_x, v_y)$ represent the x- and y- components of the velocity vector.

the particles might collide (or, interact) with each other. The probability of occurrence of such a collision depends on the size of the particles relative to each other and relative to the size of the link (or, equivalently, the size of the cells occupied by them). For these interactions to occur among a group of particles, the particles must be located on a straight line and the velocity vectors of the particles must be parallel to this line. In addition to this, the particles must be located within specific distance from each other depending on their velocities (see Eqs. (5.22)–(5.24) below). Edge interactions among particles can be of varying complexity (defined later) and can occur in different ways in different types of geometric configurations. These are elaborated below.

**Binary Edge Interactions:** Since the lattice links are one-dimensional, edge interactions are primarily binary in nature (multiparticle edge interactions can also occur and will be discussed later). The binary edge interactions can occur only in one of the following three ways: (i) When two particles are moving in opposite directions and cross each other while passing through the same link. (ii) When two particles are moving in the same direction and one of them overtakes the other while passing through the same link. (iii) When one moving particle passes over a stationary particle. For a binary edge interaction to occur in any one of these three ways, the conditions

$$\boldsymbol{x}_1 - \boldsymbol{x}_2 \quad \| \quad \boldsymbol{v}_1 \qquad \text{if} \quad \boldsymbol{v}_1 \neq 0 \tag{5.22}$$

$$\boldsymbol{x}_1 - \boldsymbol{x}_2 \quad \| \quad \boldsymbol{v}_2 \qquad \text{if} \quad \boldsymbol{v}_2 \neq 0 \tag{5.23}$$

$$|\boldsymbol{x}_2 - \boldsymbol{x}_1| \quad < \quad |\boldsymbol{v}_1 - \boldsymbol{v}_2|\Delta t \tag{5.24}$$

must necessarily be satisfied; where $\boldsymbol{x}_1$ and $\boldsymbol{x}_2$ are coordinates of the particles, $\boldsymbol{v}_1$ and $\boldsymbol{v}_2$ are velocities of the particles both of which cannot be zero simultaneously, and $\boldsymbol{a}\|\boldsymbol{b}$ denotes that the vectors $\boldsymbol{a}$ and $\boldsymbol{b}$ are parallel. That both $\boldsymbol{v}_1$ and $\boldsymbol{v}_2$ are not zero simultaneously is rigorously ensured by Eq. (5.24) combined with the single particle exclusion principle.

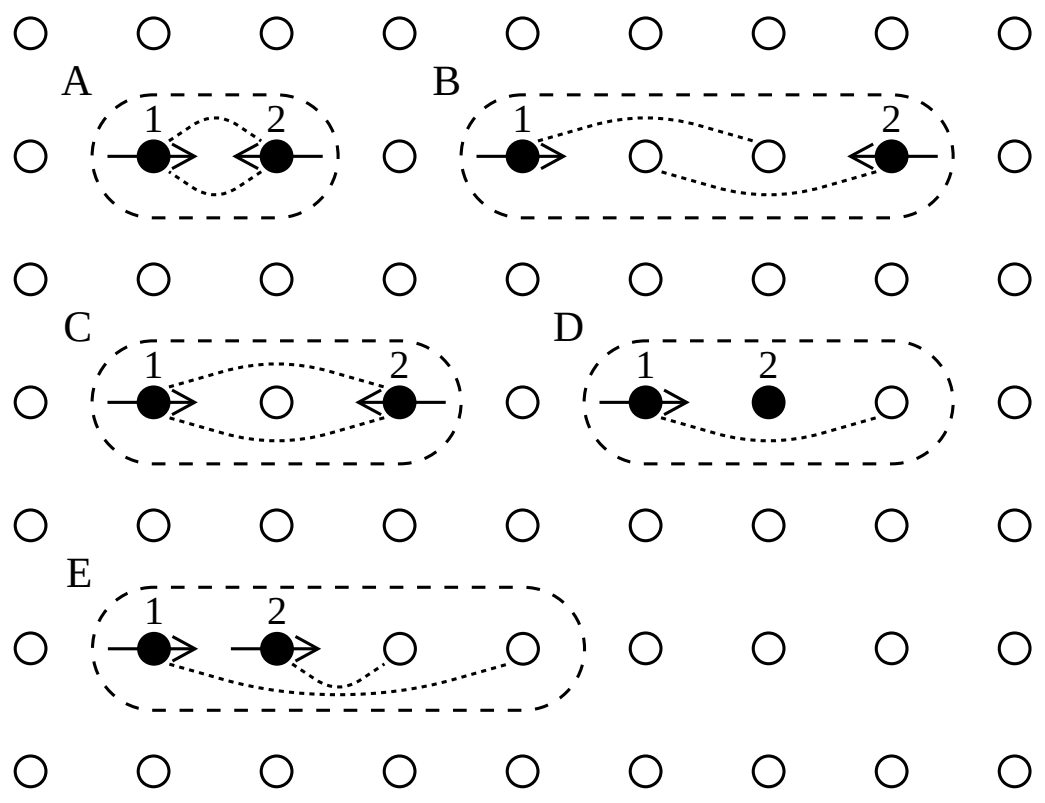

**Figure 5.8:** Examples of configurations in which binary edge interactions occur among particles over square spatial lattice in single particle lattice gases. The dotted lines are drawn to pictorially bring out the cross-over which causes interaction; they originate from the lattice sites occupied by the particles and go up to the lattice sites targeted by the particles. Velocities of various particle are: $\boldsymbol{v}_{\mathrm{A}1} = (1,0)$, $\boldsymbol{v}_{\mathrm{A}2} = (0,-1)$, $\boldsymbol{v}_{\mathrm{B}1} = (2,0)$, $\boldsymbol{v}_{\mathrm{B}2} = (-2,0)$, $\boldsymbol{v}_{\mathrm{C}1} = (2,0)$, $\boldsymbol{v}_{\mathrm{C}2} = (-2,0)$, $\boldsymbol{v}_{\mathrm{D}1} = (2,0)$, $\boldsymbol{v}_{\mathrm{D}2} = (0,0)$, $\boldsymbol{v}_{\mathrm{E}1} = (3,0)$, and $\boldsymbol{v}_{\mathrm{E}2} = (1,0)$. Symbols and notations are as in Fig. 5.7.

Some example configurations over square spatial lattice in which binary edge interactions occur among particles are shown in Fig. 5.8. Although these configurations are shown over square spatial lattice, they are inherently one-dimensional because of Eqs. (5.22)–(5.24); in fact, all configurations involving only edge interactions are inherently one-dimensional. As a result, the configurations involving only edge interactions shown in this figure (and other such figures which follow) can, in general, exist over any $\mathcal{D}$-dimensional spatial lattice in appropriate $\mathcal{D}$-dimensional single particle lattice gases.

The probability $p$ of occurrence of a binary edge interaction (through any of the three mechanisms mentioned above) is, in general, given by

$$p = \begin{cases} \kappa \left( \dfrac{d_1 + d_2}{\Delta x} \right)^2 & \text{if} \quad d_1 + d_2 \leq \Delta x \\ 1 & \text{if} \quad d_1 + d_2 > \Delta x \end{cases} \tag{5.25}$$

where $\kappa \in (0, 1]$ is a fine tuning parameter whose value depends on the geometry of particles and lattice links (usually $\kappa = 1$), $d_1$ and $d_2$ are characteristic sizes of the particles, and $\Delta x$ is the size of the lattice link on which the collision is likely to occur (usually same as the size of the boxes occupied by the particles).

**Complexity of Edge Interactions:** In the elaborations which follow, a term *complexity of edge interactions* will be required and used. It is defined as follows:

> **Complexity of Edge Interactions:** In single particle lattice gases, the *complexity of edge interactions* in a configuration of particles with interlinked edge interactions is the total number of two-particle configurations with binary edge interactions into which the original configuration can be decomposed.

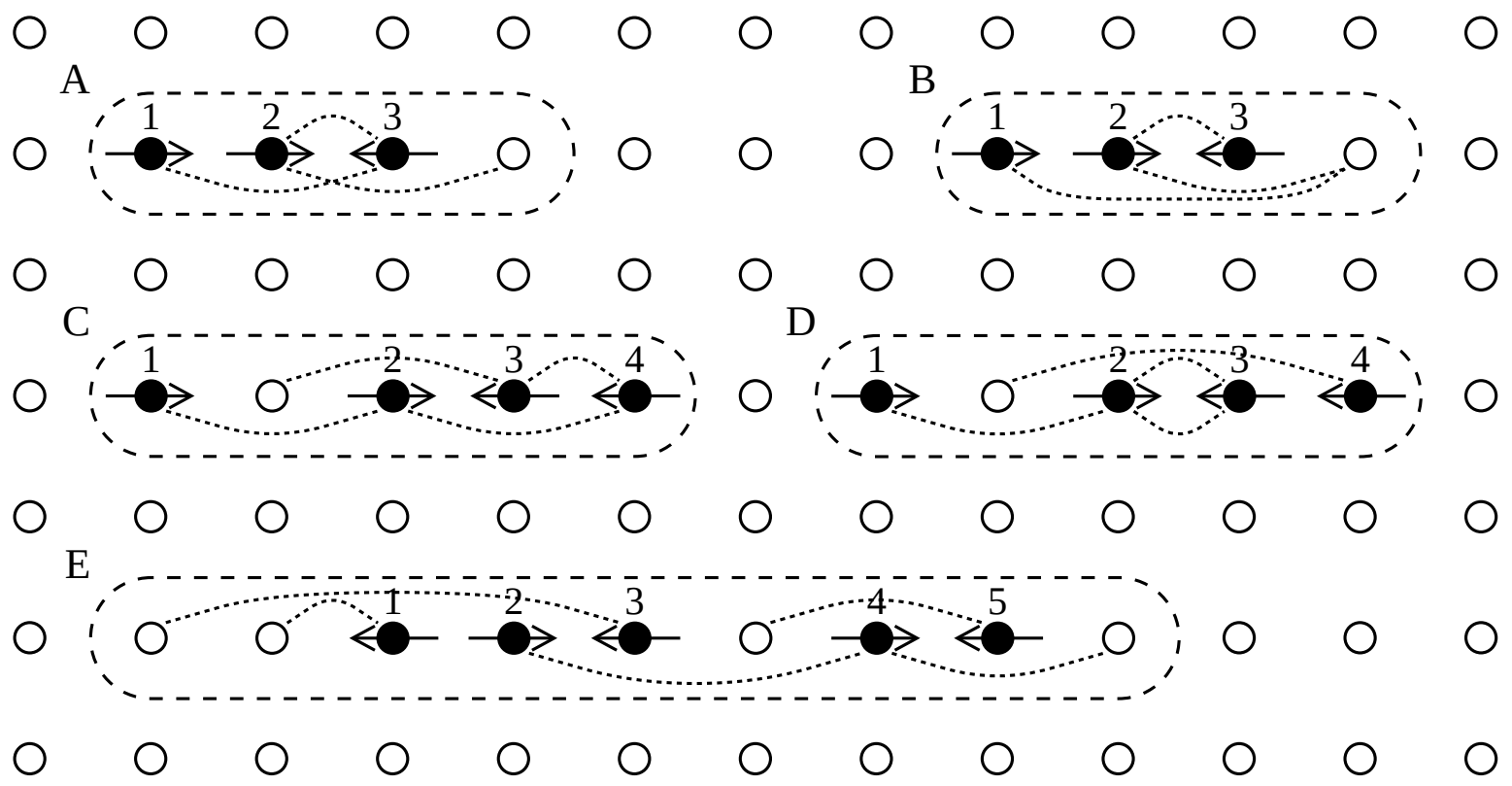

**Figure 5.9:** Examples of configurations in which multiparticle edge interactions occur among particles over square spatial lattice in single particle lattice gases. Velocities of various particle are: $\boldsymbol{v}_{\rm A1} = (2,0)$, $\boldsymbol{v}_{\rm A2} = (2,0)$, $\boldsymbol{v}_{\rm A3} = (-1,0)$, $\boldsymbol{v}_{\rm B1} = (3,0)$, $\boldsymbol{v}_{\rm B2} = (2,0)$, $\boldsymbol{v}_{\rm B3} = (-1,0)$, $\boldsymbol{v}_{\rm C1} = (2,0)$, $\boldsymbol{v}_{\rm C2} = (2,0)$, $\boldsymbol{v}_{\rm C3} = (-2,0)$, $\boldsymbol{v}_{\rm C4} = (-1,0)$, $\boldsymbol{v}_{\rm D1} = (2,0)$, $\boldsymbol{v}_{\rm D2} = (1,0)$, $\boldsymbol{v}_{\rm D3} = (-1,0)$, $\boldsymbol{v}_{\rm D4} = (-3,0)$, $\boldsymbol{v}_{\rm E1} = (-1,0)$, $\boldsymbol{v}_{\rm E2} = (3,0)$, $\boldsymbol{v}_{\rm E3} = (-4,0)$, $\boldsymbol{v}_{\rm E4} = (2,0)$, and $\boldsymbol{v}_{\rm E5} = (-2,0)$. Symbols and notations are as in Fig. 5.8.

According to this definition the complexity of all binary edge interactions is unity and the complexity of multiparticle edge interactions (defined and described below) is necessarily greater than unity. Furthermore, the complexity is always a natural number.

**Multiparticle Edge Interactions:** In addition to the binary edge interactions outlined in previous paragraphs, multiparticle edge interactions, *i.e.*, edge interactions involving three or more particles, also occur in single particle lattice gases. Some such configurations involving three, four, and five particles are shown in Fig. 5.9.

The complexity of multiparticle edge interactions and whether or not they will actually occur in a single particle lattice gas depends primarily on the velocity vectors with which particle are permitted to move over the spatial lattice in the lattice gas. For example, multiparticle edge interactions, *i.e.*, edge interactions whose complexity is more than one, cannot occur in single particle lattice gases in which particles are constrained to move with *unit speed* over the spatial lattice, *i.e.*, if $|\boldsymbol{v}_i| = 1$, $i = 0, \cdots, N-1$, where $N$ is the number of elements in the discrete velocity set (or, the connectivity of each lattice site) and $\boldsymbol{v}_i$'s are the elements of the discrete velocity set $\mathcal{V} \equiv \{\boldsymbol{v}_0, \cdots, \boldsymbol{v}_{N-1}\}$.

The complexity of multiparticle edge interactions increases rapidly as the number of elements in the discrete velocity set of particles increases. In fact, it should be clear from configurations B, D, and E shown in Fig. 5.9 that for (almost) all the discrete velocity sets which contain two velocity vectors $\boldsymbol{v}_1$ and $\boldsymbol{v}_2$ such that $|\boldsymbol{v}_1| \geq 3$, $|\boldsymbol{v}_2| \geq 3$, $\boldsymbol{v}_1 \cdot \boldsymbol{v}_2 \leq -9$, and $\boldsymbol{v}_1 \times \boldsymbol{v}_2 = 0$ (*i.e.*, $\boldsymbol{v}_1 \| \boldsymbol{v}_2$), it is possible to have (arbitrarily) long chain like geometrical configurations of particles in which multiparticle edge interactions occur. Because of this, analysis of multiparticle edge interactions for most of discrete velocity sets turns out to be considerably more complicated than that of binary edge interactions. In fact, if the discrete velocity set contains two velocity vectors $\boldsymbol{v}_1$ and $\boldsymbol{v}_2$ of the type described above, multiparticle edge interactions, in general, cannot be analyzed at all.

Complications that appear in the analysis of multiparticle edge interactions due to formation of (arbitrarily) long chains of interacting particles pose severe difficulties in the development of single particle lattice gases. These difficulties surface in the form of increase in the computational complexity of algorithms that have to be used for identification of these configurations, and then in their treatment, during simulations.[8)] Thus, to keep the task of development of single particle lattice gases within the bounds of feasibility the analysis of multiparticle edge interactions must be simplified.[9)] This simplification can be achieved by decomposing the multiparticle configurations having edge interactions into two-particle configurations with binary edge interactions[10)] and then resolving these binary edge interactions in an appropriate manner. That such a decomposition is feasible and that in single particle lattice gases all the multiparticle edge interactions can be decomposed in terms of binary edge interactions will become clear soon. It will also be seen that decomposition of most of multiparticle edge interactions, in any single particle lattice gas, can be carried out without the aid of additional simplifying assumptions and thus such a decomposition can be considered to be exact. For decomposing some special types of multiparticle edge interactions, however, some additional simplifying assumptions need to be invoked; in these cases, naturally, the exactness is lost. It is important to remember at this point that this decomposition is essential and needed during simulations only if it is desired to incorporate multiparticle edge interactions into single particle lattice gases.

**Decomposition of Multiparticle Edge Interactions into Binary Edge Interactions:** The decomposition of multiparticle edge interactions into binary edge interactions is carried out by decomposing the multiparticle configurations into all possible two particle configurations and then selecting only those two particle configurations in which binary edge interactions occur. This decomposition, if the multiparticle configuration consists of $N$ particles, gives a total of $\binom{N}{2}$ different two particle configurations. From these two particle configurations, the configurations in which binary edge interactions occur are identified by checking, for every two particle configuration, whether the coordinates and velocities of particles in the configuration satisfy Eqs. (5.22)–(5.24). These equations, as brought out earlier, are the necessary and sufficient conditions for occurrence of binary edge interactions.

**Resolving Multiparticle Edge Interactions:** During simulations multiparticle edge interactions are resolved by resolving binary edge interactions in various two particle configurations obtained by decomposing the multiparticle configurations. Decomposition of a

---

[8)] For an overview of the extent of difficulties that appear in treatment of multiparticle edge interactions during simulations, assume that the collision rules are to be used in the form of a rule table. The fact that for most of discrete velocity sets multiparticle edge interactions can occur among arbitrarily large numbers of particles (the numbers of particle can go up to infinity if the size of system is infinite) implies that in general the collision rule table will have infinite elements. As a result, it is impossible to construct the collision rule table and store it in the memory. Such attempts, thus, should not be made. Instead, algorithms which encode the entire rule table should be developed and used for resolving interparticle interactions online. Since these algorithms must necessarily be able to reproduce all the elements of the (hypothetical) rule table, they, in general, will have high computational complexity and thus might be impractical.

[9)] One can carry out exact analysis of multiparticle edge interactions provided that the discrete velocity set is such that arbitrarily long chains of particles with multiparticle edge interactions are not formed. Such an analysis is expected to yield more accurate model. But, whether the gain in accuracy will be significant and useful and whether such an attempt at gaining accuracy is meaningful in view of Eqs. (5.3) and (5.4), remains questionable. This question, if one wants to be exact, can be resolved only by making at least one such attempt and analyzing the results. No attempt of this kind has been made in the present investigation. This is because I think that in view of Eqs. (5.3) and (5.4), the accuracy that might be gained by exact analysis of multiparticle edge interactions will be of no practical significance.

[10)] This is equivalent to viewing the multiparticle configurations with edge interactions as consisting of (or, as superimposition of) many two-particle configurations with only binary edge interactions.

multiparticle configuration with only edge interactions into two particle configurations will always yield at least 2 and at most $\binom{N}{2}$ different two particle configurations with binary edge interactions, where $N$ is number of particles in the multiparticle configuration. This raises the following doubt: *"In which order should binary edge interactions in various two particle configurations be resolved during simulations?"*

The above doubt is resolved as follows: Note that multiparticle edge interactions occur because of formation of configurations in which at least one particle, instead of interacting with exactly one other particle, interacts with two or more particles. Also note that edge interactions occur when the particles collide with each other while passing through a common link. Since the links are one-dimensional, a particle can collide with at the most two more particles *simultaneously*—one approaching it from the front and the other approaching it from the rear. In case two or more particles approach a particle from the same side (front or rear), the particle first forms a configuration having binary edge interaction with the particle which is nearest to it and then, if no edge interaction occurs between the two, it forms another configuration having binary edge interaction with the particle that is farther away, and so on. This is illustrated in the multiparticle configuration A shown in Fig. 5.9. In this configuration, particle 3 forms a configurations with binary edge interaction first with the particle 2 and then with the particle 1. This implies that there is a natural sequence (on the event horizon) in which edge interactions will occur in multiparticle configurations.

The above implies that binary edge interactions in two particle configurations obtained by decomposing multiparticle configurations should be addressed in the same sequence in which they (are likely to) occur in time. The sequence in which the binary edge interactions are likely to occur is determined by computing the time at which the particles in various two particle configurations cross each other while passing through the common links. This is easily accomplished by solving the equation

$$t^i - t_0 = \frac{\boldsymbol{x}_1^i - \boldsymbol{x}_2^i}{\boldsymbol{v}_2^i - \boldsymbol{v}_1^i} \tag{5.26}$$

for every two particle configuration; where $t_0$ is current time, $t^i$ is time at which particles collide in the $i^{\text{th}}$ two particle configuration (assuming that they are point hard spheres[11)]), $\boldsymbol{x}_1^i$ and $\boldsymbol{x}_2^i$ are coordinates of particles at time $t_0$ in the $i^{\text{th}}$ two particle configuration, $\boldsymbol{v}_1^i$ and $\boldsymbol{v}_2^i$ are velocities of particle at time $t_0$ in the $i^{\text{th}}$ two particle configuration (the velocities remain unchanged till collision occurs). Particles in each two particle configuration are numbered 1 and 2. The two particle configurations are numbered $1, \ldots, N_{\text{tpc}}$, where $N_{\text{tpc}} \in \left[2, \binom{N}{2}\right]$ is the total number of two particle configurations into which a multiparticle configuration of $N$ particles has been decomposed.

Eq. (5.26) can be used for ascertaining the occurrence of vertex interactions also. The occurrence or otherwise of various interactions and the types of the interactions are indicated by the values of $t^i - t_0$ as follows: (i) $t^i - t_0 = 0$ is impossible because of single particle exclusion principle and finiteness of velocities. (ii) $0 < t^i - t_0 < 1$ means that an edge interaction occurs between particles in the current evolution. (iii) $t^i - t_0 = 1$ means that vertex interaction occurs between particles in the current evolution. (iv) $t^i - t_0 < 0$ indicates *possible* situation in the past. (v) $t^i - t_0 > 1$ indicates *possible* situation in the future following the current evolution. The indications about the past and future in the

[11)]This approximation is reasonable as long as it is used only for computing the time of collision between particles for deciding the sequence in which the binary edge interactions should be treated.

points (iv) and (v) above should be judged along the lines indicated in the points (ii) and (iii) above. In the above and in Eq. (5.26) $t^i - t_0$ is the time interval relative to the current time step. In the following it will be expressed only through $t^i$ since all the discussion pertains to the next evolution and thus, is relative to the current time step.

Use of Eq. (5.26) for computing the time of occurrence[12)] of various binary edge interactions in multiparticle configurations reveals that they are of two basic types or combinations of these types. The basic types are: (I) multiparticle configurations in which all binary edge interactions occur at different time instants, and (II) multiparticle configurations in which $N$, $N \geq 2$, binary edge interactions occur at the same time instant and total number of particles involved in these interactions is less than $2N$. Permutations and combinations of these two basic types of multiparticle configurations give rise to many complex situations. Some examples of such situations are: (III)[13)] multiparticle configurations in which $N$, $N \geq 2$, binary edge interactions occur at the same time instant and total number of particles involved in these interactions is $2N$, and (IV) multiparticle configurations which are mixture of multiparticle configurations of the types (I)–(III) defined above. Note that multiparticle configurations of type (III) are mixtures of many multiparticle configurations of the basic type (I). Examples of each of these types of multiparticle configurations along with their decomposition two particle configurations with binary edge interactions, method of ascertaining the sequence in which various binary edge interactions occur in the multiparticle configurations, and finally the method of resolving interactions in the multiparticle configurations during simulations are described in the following.

The symbols used in the following descriptions are: Multiparticle configurations with edge interactions are labeled as A. Two-particle configurations with binary edge interactions obtained by decomposing the multiparticle configurations are labeled as B, C, D, E, F, G, H, *etc.* Time instants given by Eq. (5.26) at which binary edge interactions occur in these two-particle configurations are denoted by $t^{\mathrm{B}}$, $t^{\mathrm{C}}$, $t^{\mathrm{D}}$, $t^{\mathrm{E}}$, $t^{\mathrm{F}}$, $t^{\mathrm{G}}$, $t^{\mathrm{H}}$, *etc.*

**(I) Multiparticle Configurations in Which All Binary Edge Interactions Occur at Different Time Instants:** Some examples of this type of multiparticle configurations with edge interactions are shown in Figs. 5.10 and 5.11. The situations depicted in these figures, and those conceptually similar to these, are analyzed as described below.

Configuration A in Fig. 5.10 shows multiparticle edge interaction among three particles numbered 1, 2, and 3 over square spatial lattice. Velocities of the particles are $\boldsymbol{v}_1 = (2, 0)$, $\boldsymbol{v}_2 = (2, 0)$, and $\boldsymbol{v}_3 = (-1, 0)$. This configuration can be decomposed into three two-particle configurations formed by the pairs 1–2, 2–3, and 1–3 of particles. Out of these three pairs, the pairs 1–3 and 2–3 satisfy Eqs. (5.22)–(5.24) but the pair 1–2 does not. This implies that binary edge interactions occur in the pairs 2–3 and 1–3 and not in the pair 1–2. As a result, the configuration A can be decomposed into two two-particle configurations, formed by the pairs 2–3 and 1–3, with binary edge interaction. These pairs are shown in the figure and labeled as configurations B and C, respectively. The time instants given by Eq. (5.26), in natural units of the lattice system, at which binary edge interactions occur in these two-particle configurations are: $t^{\mathrm{B}} = 1/3$ and $t^{\mathrm{C}} = 2/3$. These time instants are shown above the respective two-particle configurations within the figure. These time instants, being all different, give an unambiguous sequences in which

12) Recall, from the elaborations give earlier (on page 139), that binary edge interactions are probabilistic events and the phrase *"occurrence of binary edge interactions"* in the footmarked sentence and elsewhere in the current section has been used in this sense only. In a given instance, unless the probability is unity, binary edge interactions may or may not actually occur.

13) The numbering used in the previous lines has been continued.

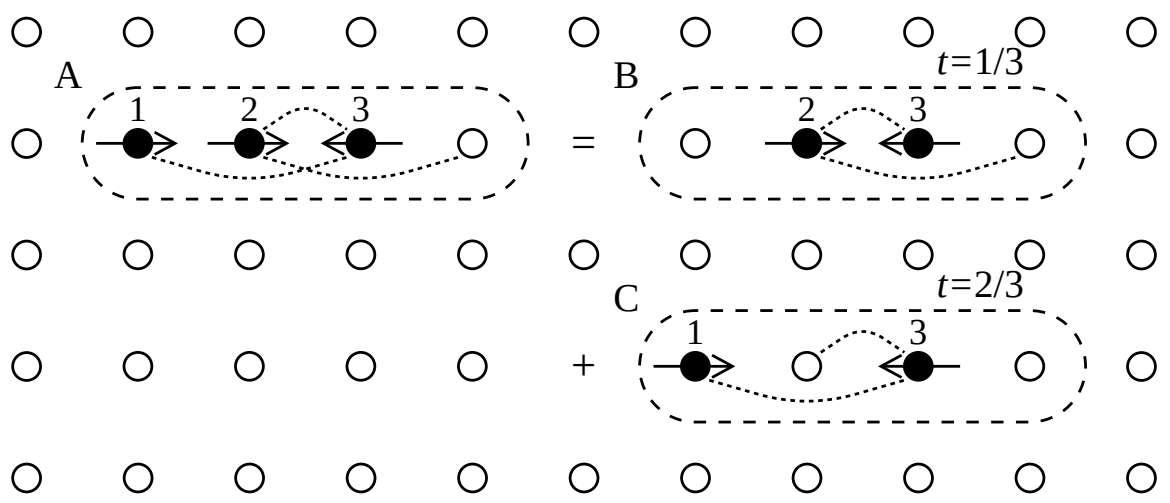

**Figure 5.10:** A three particle configuration with multiparticle edge interactions and its decomposition into two-particle configurations with binary edge interactions. Note that all binary edge interactions occur at different time instants. Velocities of various particles are: $\boldsymbol{v}_1 = (2,0)$, $\boldsymbol{v}_2 = (2,0)$, and $\boldsymbol{v}_3 = (-1,0)$. Symbols and notations are as in Fig. 5.8.

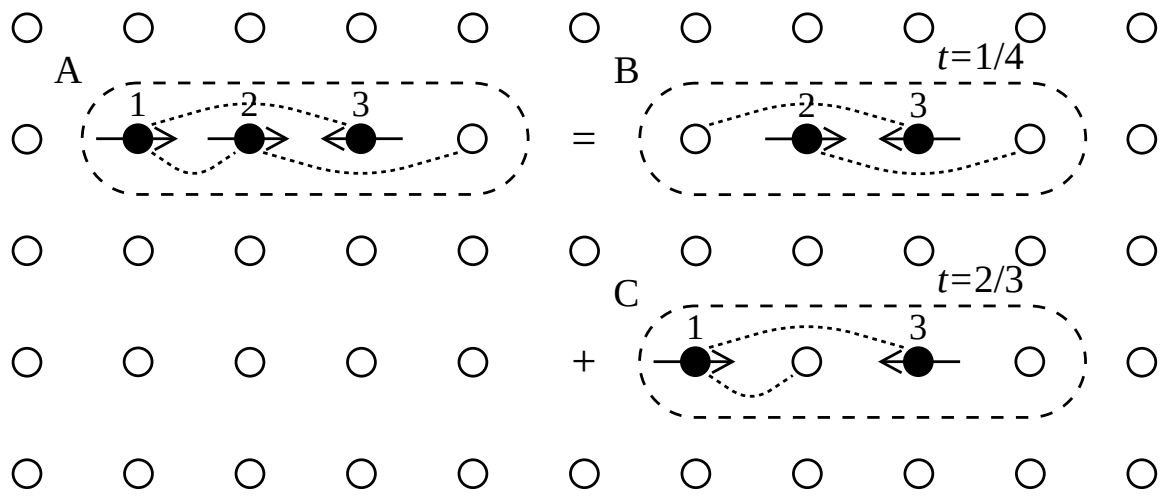

**Figure 5.11:** A three particle configuration with multiparticle edge interactions and its decomposition into two-particle configurations with binary edge interactions. Note that all binary edge interactions occur at different time instants. Velocities of various particles are: $\boldsymbol{v}_1 = (1,0)$, $\boldsymbol{v}_2 = (2,0)$, and $\boldsymbol{v}_3 = (-2,0)$. Symbols and notations are as in Fig. 5.8.

the binary edge interactions occur, *viz.*, (B, C), *i.e.*, binary edge interaction occurs first in configuration B and then in configuration C.[14] During simulations this sequence is used for resolving edge interactions in the multiparticle configuration A. The procedure of resolving edge interactions is as follows: The occurrence of binary edge interaction in configuration B is checked first.[15] If the interaction occurs, it is resolved and then the process stops. If the interaction does not occur, the occurrence of binary edge interaction in configuration C is checked. If this interaction occurs, it is resolved and then the process stops. If this interaction does not occur then also the process stops because all two-particle configurations have been checked.

The sequence in which various binary edge interactions occur in the multiparticle configuration shown in Fig. 5.11 can be obtained in the same way as above. Decomposition of this multiparticle configuration into two-particle configurations with binary edge interactions and the time intervals after which binary edge interactions occur in these two-particle configurations are shown within the figure. The edge interactions in the multiparticle configurations shown in this figure can be resolved using the same procedure that has been outlined for the multiparticle configuration shown in Fig. 5.10. This is because the number of binary edge interactions and the sequence in which they occur in each of this multiparticle configuration are identical with those for the configuration shown in Fig. 5.10.

[14]Note the notation used here. The binary edge interaction which occur in a sequence are written in the descending order separated by commas. and all binary edge interactions in a multiparticle configuration are enclosed within parenthesis. This notation will be used in the following also.

[15]The occurrence of a binary edge interaction, it being a probabilistic event occurring with *a-priori* known probability $p$, is checked by generating a random number $r$ between 0 and 1 (pseudo-random number in the computer) and comparing it with $p$. The interaction occurs if $r < p$, and not otherwise.

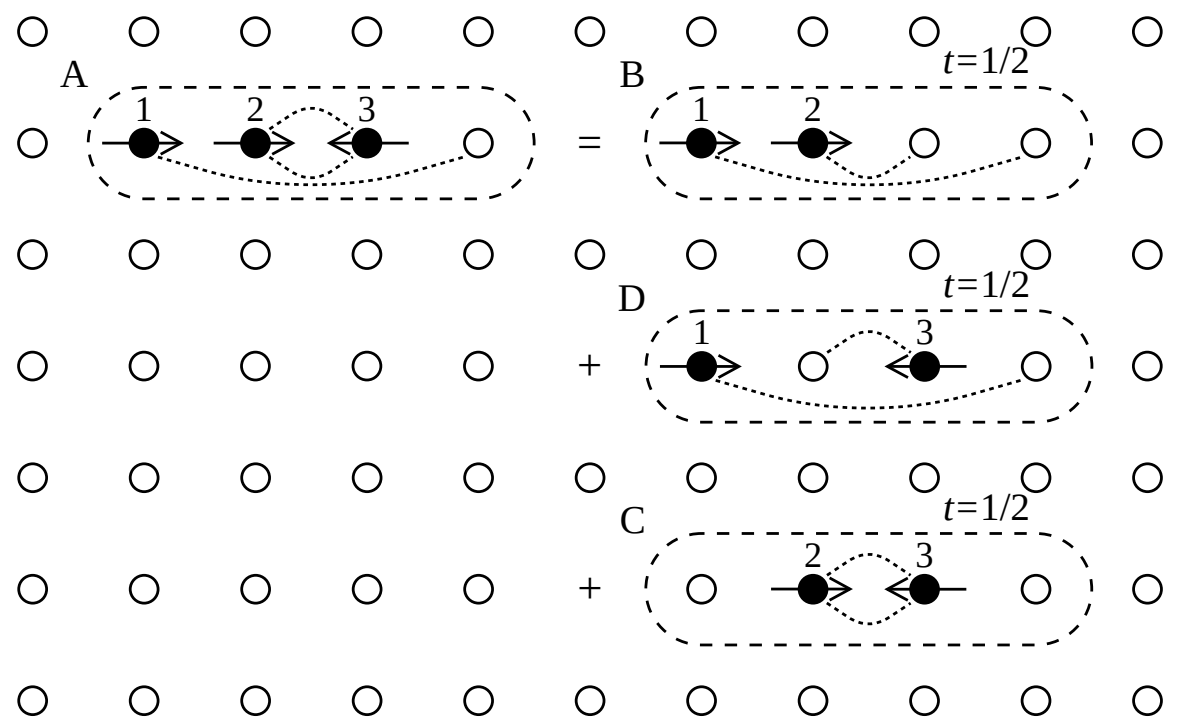

**Figure 5.12:** A three particle configuration with multiparticle edge interactions and its decomposition into two-particle configurations with binary edge interactions. Note that 3 binary edge interactions occur at the same time instant and total number of particles involved in these interactions, being 3, is less than twice the number of interactions. Velocities of various particles are: $\boldsymbol{v}_1 = (3, 0)$, $\boldsymbol{v}_2 = (1, 0)$, and $\boldsymbol{v}_3 = (-1, 0)$. Symbols and notations are as in Fig. 5.8.

The sequence of occurrence of binary edge interactions in multiparticle configurations of type (I) can be ascertained as done above. Once the sequence has been ascertained, it can be used for resolving the edge interactions during simulations. The general procedure of resolving edge interactions in multiparticle configurations of type (I) is as follows: To begin with, the two-particle configuration that comes first in the sequence is selected and whether or not binary edge interaction occurs in it is checked.[15)] If the interaction occurs, it is *resolved* and then the process stops, *i.e.*, no more checking is done.[16)] If the interaction does not occur, the two-particle configuration that comes next in the sequence is selected and the same procedure that was used for the first two-particle configuration is repeated. This process continues till either an edge interaction occurs in some two-particle configuration (following which the process stops) or all two-particle configurations have been checked.

**(II) Multiparticle Configurations in Which $N$, $N \geq 2$, Binary Edge Interactions Occur at the Same Time Instant and Total Number of Particles Involved in These Interactions is Less Than $2N$:** Some examples of this type of multiparticle configurations with edge interactions are shown in Figs. 5.12 and 5.13. Situations depicted in these figures, and those conceptually similar to these, are analyzed as described below.

In Fig. 5.12, a three particle configuration, A, with multiparticle edge interactions and its decomposition (carried out as explained on page 141) into two-particle configurations, B, C, and D, with binary edge interactions is shown. The time instant relative to the current time step at which binary edge interactions occur in these two-particle configurations are: $t^{\mathrm{B}} = 1/2$, $t^{\mathrm{C}} = 1/2$, and $t^{\mathrm{D}} = 1/2$. In the figure, these time instants are shown above

---

[16)] When binary edge interaction in a two-particle configuration is *resolved*, the states of particles comprising this configuration change. Because of this, the status of edge interactions in all the two-particle configurations formed by either of these two particles is affected. In general, the new states of particles might eliminate many of the edge interactions that were present earlier and might give to many new edge interactions. Unless, these situations are reviewed the process cannot be continued and thus is stopped.

the respective configurations. These time instants are computed using Eq. (5.26) under the assumption that the particles are rigid point particles. Since $t^{\mathrm{B}} = t^{\mathrm{C}} = t^{\mathrm{D}}$, binary edge interactions in all these two-particle configurations (appear to) occur simultaneously and it does not appear possible to decide the sequence in which they should be resolved. The sequence, however, can be decided by computing the time interval after which binary edge interactions occur in these two-particle configurations using Eq. (5.26) under the assumption that the particles have a finite size, say $2\delta$, $0 < 2\delta < 1$. In this computation, the coordinates employed in Eq. (5.26) are the coordinates of the edges of particles facing each other. For the two-particle configuration B shown in Fig. 5.12 this gives

$$t^{\mathrm{B}} = \frac{\boldsymbol{x}_1^{\mathrm{B}} - \boldsymbol{x}_2^{\mathrm{B}}}{\boldsymbol{v}_2^{\mathrm{B}} - \boldsymbol{v}_1^{\mathrm{B}}} = \frac{\delta - (1-\delta)}{1-3} = 1/2 - \delta$$

Similarly, for the configurations C and D one gets: $t^{\mathrm{C}} = 1/2 - \delta/2$ and $t^{\mathrm{D}} = 1/2 - \delta$. These values, for the entire range of variation of $\delta$, indicate that binary edge interactions in the configurations B and D occur before that in the configuration C. Furthermore, in the configurations B and D the interactions occur simultaneously. From this information it is clear that binary edge interaction in the configuration C is to be resolved after that in configurations B and D. As far as the configurations B and D are concerned, two different sequences, *viz.*, B then D or D then B, are possible. Which of these two sequences is correct cannot be decided.[17] It, however, is necessary to assign a sequence to the configurations B and D since one particle is common among them. This is done by selecting, either randomly or on the basis of some arbitrary criteria, the configuration which should be first and the one which should be second. I recommend that in such cases the choice should be random with equal probability for each configuration, rather than being based on arbitrary criteria. This is because random choice with equal probabilities would, on an average, favor all the configurations equally and ensure maximum randomization of particle trajectories. Thus, there are two possible sequences in which binary edge interactions can be resolved in the multiparticle configuration shown in Fig. 5.12. These are: (B, D, C) and (D, B, C). Both these sequences are equally probable and, during simulations, exactly one of them is selected and used for resolving edge interactions in the multiparticle configuration. Introducing a more compact and convenient notation (explained in the footnote accompanying this sentence), the sequence in which binary edge interactions occur in the multiparticle configuration shown in Fig. 5.12 is written as: (B $\oplus$ D, C).[18]

---

[17] This is under the assumption that the particles comprising the configurations B and D are identical. If the particles were not identical, as would be the case in multispecies mixtures, they would have different sizes and the results of the analysis carried out above could have been different.

[18] Here the notation used for writing the sequence of binary edge interactions multiparticle configurations of type (I) (described earlier on page 144 in footnote 15) has been augmented and used. The augmentation is introduction of a new symbol $\oplus$. The properties of this symbol are as follows: It is used in place of commas between two interlinked simultaneous binary edge interactions. In general, if there are $N$ interlinked simultaneous binary edge interactions, $\mathrm{B}_i$, $i = 1, \ldots, N$, in a multiparticle configuration then, using this symbol, they are written as $\mathrm{B}_1 \oplus \mathrm{B}_2 \oplus \ldots \oplus \mathrm{B}_{N-1} \oplus \mathrm{B}_N$. The order of writing the binary edge interactions is not important, *i.e.*, if $N = 2$ then $\mathrm{B}_1 \oplus \mathrm{B}_2$ and $\mathrm{B}_2 \oplus \mathrm{B}_1$ are equivalent. Since the binary edge interactions are interlinked, a sequence must necessarily be assigned to them during simulations. In general, for $N$ interlinked simultaneous binary edge interactions, there can be total of $N!$ different sequences because each binary edge interaction can take any one of the $N$ places in the sequence with equal probability, *i.e.*, there are $N$ choices for the first place, $N-1$ choices for the second, and so on. The process of assigning a unique sequence is as explained in the paragraph containing the footmarked statement. Once as sequence is assigned, the symbol $\oplus$ is replaced by comma and the order in which binary edge interaction are written becomes important. Thus, the sequence (B $\oplus$ D, C) in the footmarked

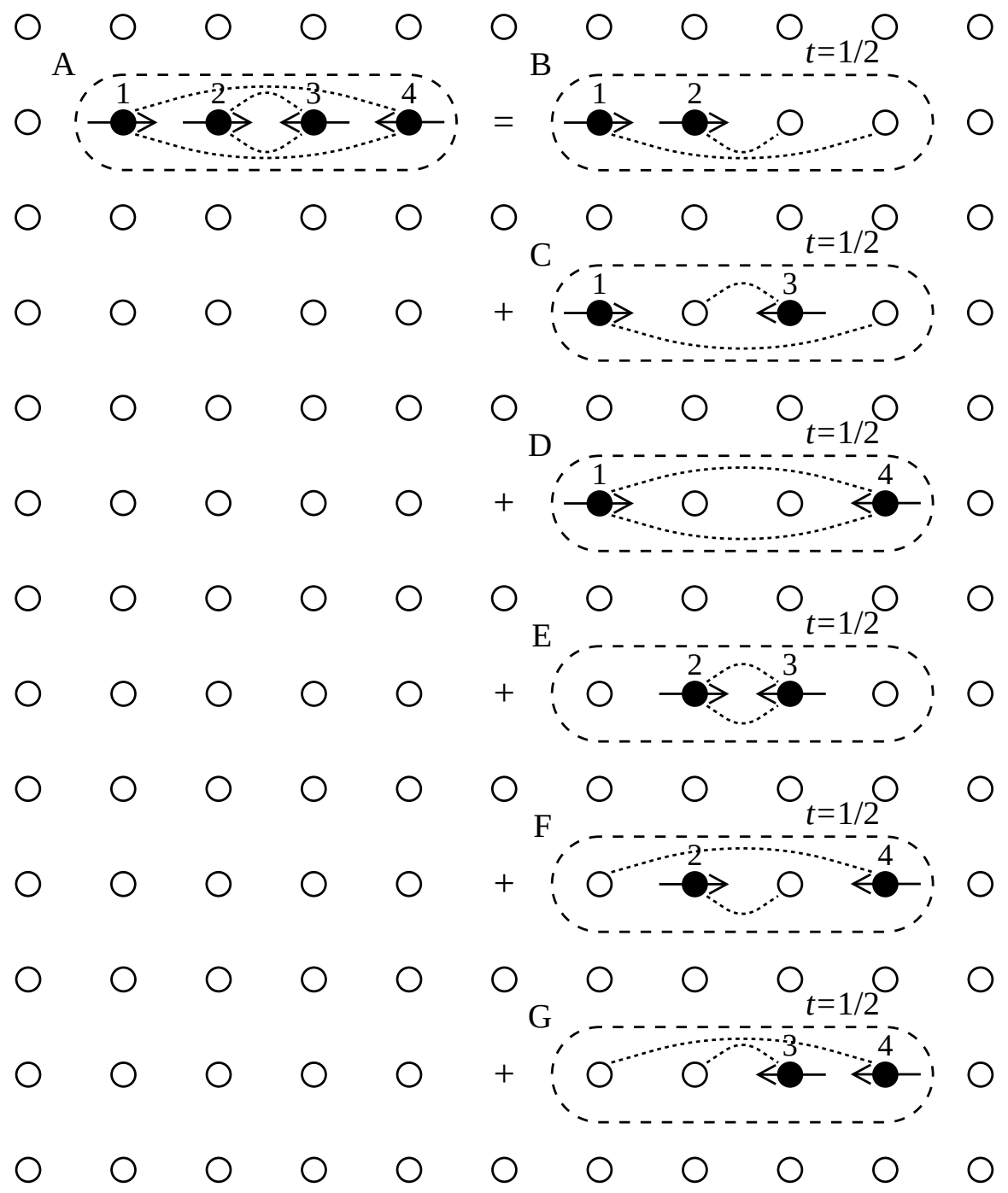

**Figure 5.13:** A four particle configuration with multiparticle edge interactions and its decomposition into two-particle configurations with binary edge interactions. Note that 6 binary edge interactions occur at the same time instant and total number of particles involved in these interactions, being 4, is less than twice the number of interactions. Velocities of various particles are: $\boldsymbol{v}_1 = (3, 0)$, $\boldsymbol{v}_2 = (1, 0)$, $\boldsymbol{v}_3 = (-1, 0)$, and $\boldsymbol{v}_4 = (-3, 0)$. Symbols and notations are as in Fig. 5.8.

The mulitparticle configuration shown in Fig. 5.13 can also be analyzed in the same way as above. The time instants relative to the current time step at which binary edge interactions occur in this multiparticle configuration, calculated using Eq. (5.26) under rigid point particle approximations, are: $t^{\mathrm{B}} = 1/2$, $t^{\mathrm{C}} = 1/2$, $t^{\mathrm{D}} = 1/2$, $t^{\mathrm{E}} = 1/2$, $t^{\mathrm{F}} = 1/2$, and $t^{\mathrm{G}} = 1/2$. Thus, under rigid point particle approximation all the binary edge interactions occur simultaneously. Recomputation of the time instants assuming the particles to be of finite size gives: $t^{\mathrm{B}} = 1/2-\delta$, $t^{\mathrm{C}} = 1/2-\delta/2$, $t^{\mathrm{D}} = 1/2-\delta/3$, $t^{\mathrm{E}} = 1/2-\delta$, $t^{\mathrm{F}} = 1/2-\delta/2$, and $t^{\mathrm{G}} = 1/2-\delta$. Thus, the sequence in which binary edge interactions should be resolved is: (B $\oplus$ E $\oplus$ G, C $\oplus$ F, D). This means that the first, second, and third choices are from B, E, G, fourth and fifth choices are from C and F and the sixth choice is D. The choices, *e.g.*, of which is the first, second, and third among the configurations B, E, and G, can be random or based on some arbitrary criteria. If the selection is random with equal probability, there are total of 12 possible sequences in which binary edge interactions can be resolved. These are: (i) (B, E, G, C, F, D), (ii) (B, G, E, C, F, D), (iii) (E, B, G, C, F, D), (iv) (E, G, B, C, F, D), (v) (G, B, E, C, F, D), (vi) (G, E, B, C, F, D), (vii) (B, E, G, F, C, D), (viii) (B, G, E, F, C, D), (ix) (E, B, G, F, C, D), (x) (E, G, B, F, C, D), (xi) (G, B, E, F, C, D), and (xii) (G, E, B, F, C, D). All these sequences are equally probable and only one of them is selected and used for resolving edge interactions during simulations.

---

statement will necessarily get translated with equal probability either into the sequence (B, D, C) or (D, B, C) during simulations.

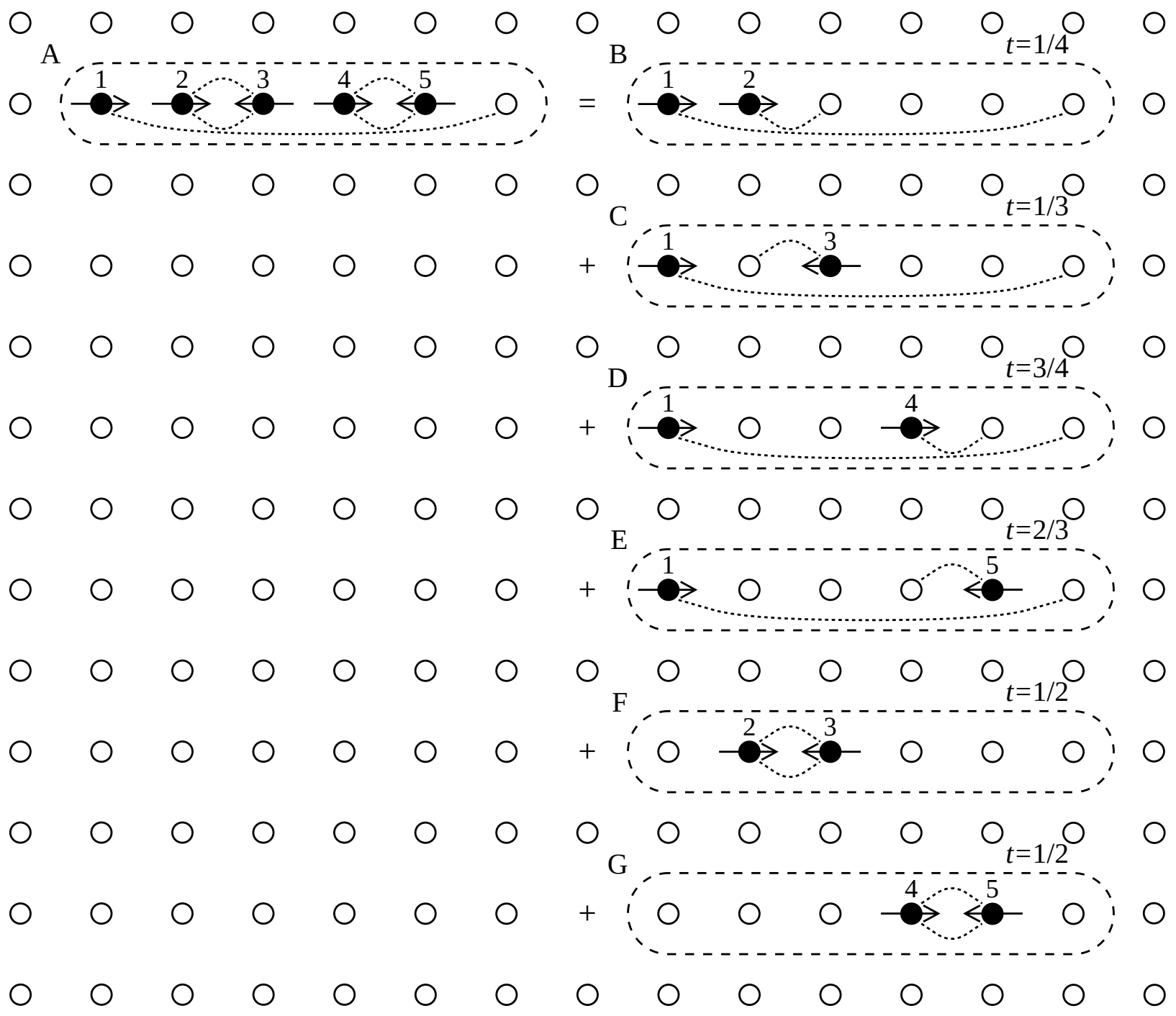

**Figure 5.14:** A five particle configuration with multiparticle edge interactions and its decomposition into two-particle configurations with binary edge interactions. Note that 2 binary edge interactions occur at the same time instant and total number of particles involved in these interactions, being 4, is equal to twice the number of interactions. Velocities of various particles are: $\boldsymbol{v}_1 = (5,0)$, $\boldsymbol{v}_2 = (1,0)$, $\boldsymbol{v}_3 = (-1,0)$, $\boldsymbol{v}_4 = (1,0)$, and $\boldsymbol{v}_5 = (-1,0)$. Symbols and notations are as in Fig. 5.8.

The analysis presented above brings out that in multiparticle configurations of type (II) there can, in general, be many sequences in which binary edge inetarctions occur. Once the sequences have been ascertained as described above, one of them must be selected for resolving edge interactions in the mulitparticle configuration. Selection is done during simulations and edge interactions in the mulitparticle configuration are resolved in the selected sequence using the procedure outlined earlier for multiparticle configurations of type (I).

**(III) Multiparticle Configurations in Which $N$, $N \geq 2$, Binary Edge Interactions Occur at the Same Time Instant and Total Number of Particles Involved in These Interactions is $2N$:** Some examples of this type of mulitparticle configurations with edge interactions are shown in Figs. 5.14 and 5.15. The situations depicted in these figures, and those conceptually similar to these, are analyzed as described below.

In Fig. 5.14, a five particle configuration with multiparticle edge interactions and its decomposition (carried out as explained on page 141) into two-particle configurations with binary edge interactions is shown. The five particle configuration is labeled A and the two-particle configurations are labeled B, C, D, E, F, and G. The time instant relative to the current time step at which binary edge interactions occur in these two-particle configurations are: $t^{\mathrm{B}} = 1/4$, $t^{\mathrm{C}} = 1/3$, $t^{\mathrm{D}} = 3/4$, $t^{\mathrm{E}} = 2/3$, $t^{\mathrm{F}} = 1/2$, and $t^{\mathrm{G}} = 1/2$. In the figure, these time instants are shown above the respective configurations. These time instants are computed using Eq. (5.26) under the assumption that the particles are rigid point

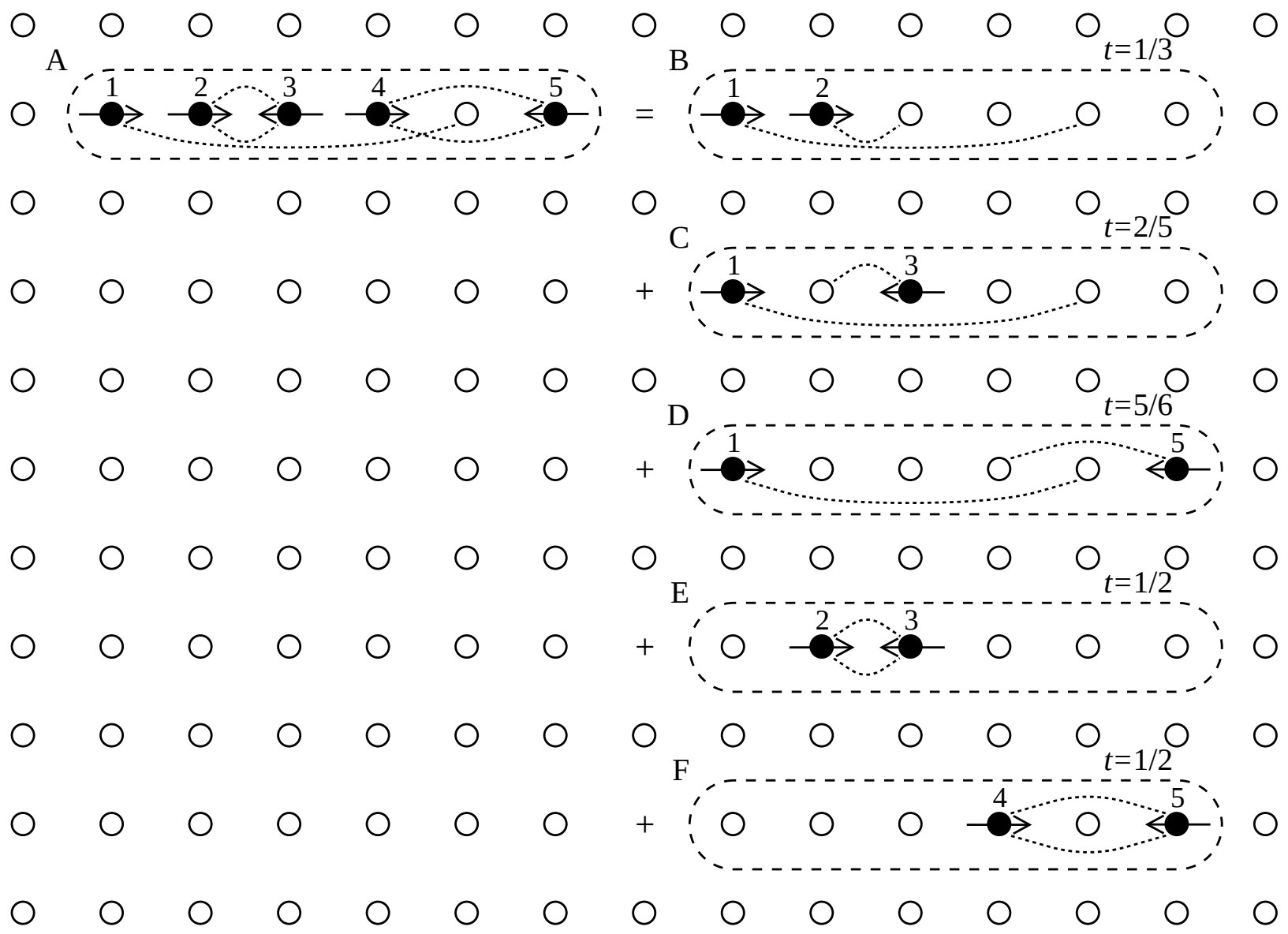

**Figure 5.15:** A five particle configuration with multiparticle edge interactions and its decomposition into two-particle configurations with binary edge interactions. Note that 2 binary edge interactions occur at the same time instant and total number of particles involved in these interactions, being 4, is equal to twice the number of interactions. Velocities of various particles are: $\boldsymbol{v}_1 = (4, 0)$, $\boldsymbol{v}_2 = (1, 0)$, $\boldsymbol{v}_3 = (-1, 0)$, $\boldsymbol{v}_4 = (2, 0)$, and $\boldsymbol{v}_5 = (-2, 0)$. Symbols and notations are as in Fig. 5.8.

particles. From these time instants the sequence in which binary edge interactions occur in most of the two-particle configurations is evident except for the configurations F and G. In both these configurations binary edge interactions (appear to) occur simultaneously. Recomputation of the time instant, assuming particles to be of finite size, at which binary edge interactions occur in these two-particle configurations gives: $t^{\mathrm{F}} = 1/2 - \delta$ and $t^{\mathrm{G}} = 1/2 - \delta$. This implies that in these configurations binary edge interactions occur simultaneously. Since no particle is common to both these configurations, it is not necessary to assign a specific sequence to them. Binary edge interactions occurring in these configurations can be treated simultaneously or on after the other in any arbitrary sequence with identical results. These binary binary edge interactions must be treated as if they were occurring in two different two-particle configurations. Thus, introducing a new compact notation (explained in the footnote accompanying this sentence), the sequence in which binary edge interactions occur in the mulitparticle configuration shown in Fig. 5.14 is: (B, C, F ∥ G, E, D).[19)]

The five particle configuration with multiparticle edge interactions shown in Fig. 5.15 can also be analyzed along the same lines as above. The time instants at which various

[19)] Note that the notation used earlier has been augmented further by introducing a new symbol ∥. The properties and usage of this symbol are as follows: The symbol ∥ is used in place of commas between two simultaneous binary edge interactions which are not interlinked. In general, if there are $N$ noninterlinked simultaneous binary edge interactions, $\mathrm{B}_i$, $i = 1, \ldots, N$, in a multiparticle configuration then, using this symbol, they are written as $\mathrm{B}_1 \parallel \mathrm{B}_2 \parallel \ldots \parallel \mathrm{B}_{N-1} \parallel \mathrm{B}_N$. The order of writing the binary edge interactions is not important, *i.e.*, if $N = 2$ then $\mathrm{B}_1 \parallel \mathrm{B}_2$ and $\mathrm{B}_2 \parallel \mathrm{B}_1$ are equivalent. Since the binary edge interactions are noninterlinked, no specific sequence need to be assigned to them during simulations. During simulations all of them all must be checked and resolved. Since they are noninterlinked, they can be processed in parallel.

binary edge interactions occur in this multiparticle configuration, computed using Eq. (5.26) under rigid point particle assumption, are: $t^{\mathrm{B}} = 1/3$, $t^{\mathrm{C}} = 2/5$, $t^{\mathrm{D}} = 5/6$, $t^{\mathrm{E}} = 1/2$, and $t^{\mathrm{F}} = 1/2$. From these time instants, the sequence of occurrence of almost all binary edge interactions is evident except for the binary edge interactions E and F which (appear to) occur simultaneously. Recomputation of the time instants, assuming particles to be of finite size, gives: $t^{\mathrm{E}} = 1/2 - \delta$ and $t^{\mathrm{F}} = 1/2 - \delta/2$. This gives an unambiguous sequence, the sequence being (B, C, E, F, D), of occurrence to all the binary edge interactions. This sequence, however, should not be used for resolving edge interactions during simulations. This is because binary edge interactions in two-particle configurations E and F are non-interlinked and the occurrence of any one of them does not affect the occurrence of the other. Hence, both the binary edge interactions E and F must be resolved in parallel. Thus, the sequence in which edge interactions in the multiparticle configuration shown in Fig. 5.15 should be resolved during simulations is: (B, C, E $\|$ F, D).

Note that the situations described above are different from those arising in multiparticle configurations of type (II) described earlier. In multiparticle configurations of type (II), a sequence is necessarily needed because the simultaneous binary edge interactions are interlinked and the occurrence of one affects the occurrence of the other(s). Whereas in the multiparticle configurations of type (III) being addressed currently, *e.g.*, like the ones shown in Figs. 5.14 and 5.15, the occurrence of any one of simultaneous binary edge interactions does not affect the occurrence of the other(s) because the two-particle configurations are not interlinked. As a result, such simultaneous binary edge interactions must be treated as if they were occurring in different two-particle configurations and during simulations all of them must be checked and resolved, *i.e.*, the process should not be stopped unless all simultaneous binary edge interactions of this type have been checked. Rest of the process remains same as explained earlier for multiparticle configurations of type (I).

**(IV) Multiparticle Configurations Which are Mixture of Multiparticle Configurations of Types (I)–(III):** An example of this type of multiparticle configurations with edge interactions is shown in Fig. 5.16. Since these multiparticle configurations are mixture of all the three types of multiparticle configurations (types (I)–(III)) discussed and described earlier, their analysis requires elaborate considerations wherein all the three methods of analysis outlined earlier for multiparticle configurations of types (I)–(III) have to be used. This multiparticle configuration, being a mixture of all the three types, serves as a simple example of the most general type of multiparticle configurations. The situation depicted in this figure, and those conceptually similar to it, are analyzed as described below.

The time instants relative to the current time step at which various binary edge interactions occur in the five particle configuration with multiparticle edge interactions shown in Fig. 5.16 are: $t^{\mathrm{B}} = 1/3$, $t^{\mathrm{C}} = 1/3$, $t^{\mathrm{D}} = 3/4$, $t^{\mathrm{E}} = 4/7$, $t^{\mathrm{F}} = 1/3$, $t^{\mathrm{G}} = 3/4$, and $t^{\mathrm{H}} = 1/3$. These time instants are computed using Eq. (5.26) under rigid point particle assumption. In the figure, they are shown above the respective configurations. These time instants point out the existence two different groups of simultaneous binary edge interactions in the multiparticle configuration. The first group being formed by the configurations B, C, F, and H and the second group being formed by the configurations D and G. Recomputation of time instants for these groups, assuming particles to be of finite size, gives: $t^{\mathrm{B}} = 1/3 - 2\delta/3$, $t^{\mathrm{C}} = 1/3 - \delta/3$, $t^{\mathrm{F}} = 1/3 - 2\delta/3$, and $t^{\mathrm{H}} = 1/3 - 2\delta/3$ for the first group and $t^{\mathrm{D}} = 3/4 - \delta/2$ and $t^{\mathrm{G}} = 3/4 - \delta/2$ for the second group. This implies that in the first group binary edge interactions in the configurations B, F, and H occur simultaneously followed by binary edge interaction in the configuration C. Among the configurations B, F, and H, the configurations B and F are interlinked but the configuration H is not interlinked with either of them. As a result, the sequence of binary edge interactions in the first group is: (B $\oplus$ F) $\|$ H, C.[20] In the second group, binary edge interactions in both the configurations D and G occur simultaneously and are not interlinked. As a result, the sequence of binary edge interactions in the second groups is: D $\|$ G. Thus, the overall sequence in which edge interaction should be resolved during simulations is: ((B $\oplus$ F) $\|$ H, C, E, D $\|$ G).[20]

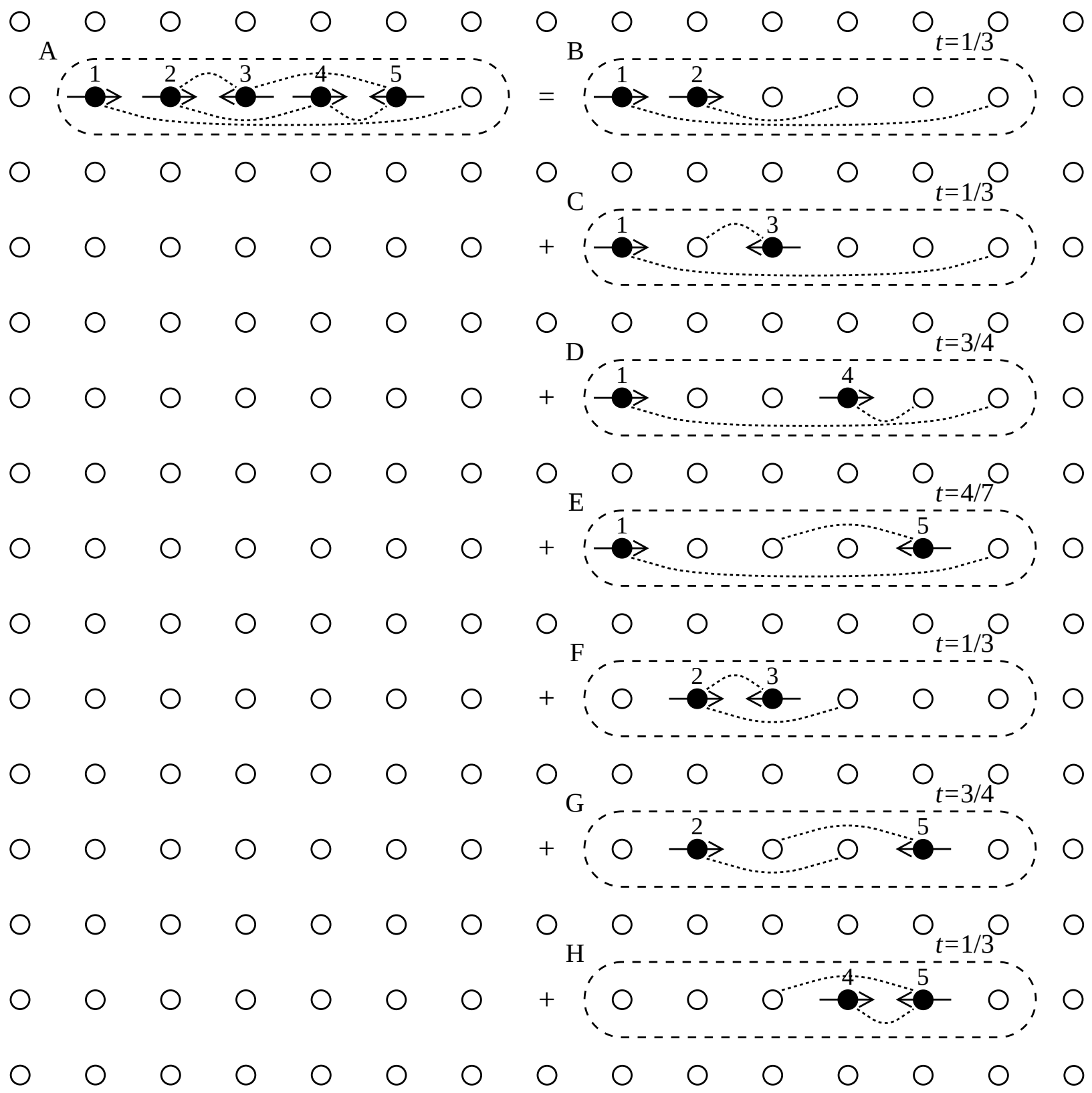

**Figure 5.16:** A five particle configuration with multiparticle edge interactions and its decomposition into two-particle configurations with binary edge interactions. Note the mixture of situations which occur in multiparticle configurations of type (I)–(III). Velocities of various particles are: $\boldsymbol{v}_1 = (5, 0)$, $\boldsymbol{v}_2 = (2, 0)$, $\boldsymbol{v}_3 = (-1, 0)$, $\boldsymbol{v}_4 = (1, 0)$, and $\boldsymbol{v}_5 = (-2, 0)$. Symbols and notations are as in Fig. 5.8.

#### 5.1.3.6 Mixed Vertex and Edge Interactions

At times, vertex interactions will also be present along with edge interactions in multiparticle configurations. An example of such multiparticle configurations is shown in Fig. 5.17 (this configuration is also shown in Fig. 5.9). The situation depicted in this figure, and those conceptually similar to it, are analyzed as described below.

In Fig. 5.17, a three particle configuration, A, with mixed vertex and edge interactions and its decomposition into two-particle configurations, B, C, and D, is shown. Time instants relative to the current time step at which various (vertex and binary edge) interactions occur in these two-particle configurations, computed using Eq. (5.26) under rigid point particle assumption, are: $t^{\mathrm{B}} = 1$, $t^{\mathrm{C}} = 1/2$, and $t^{\mathrm{D}} = 1/3$. In the figure, these time instants are shown above the respective configurations. From these time instants it is clear that vertex interaction occurs in configuration B and binary edge interactions occur in configurations C and D.[21)] Furthermore, these time instants give an unambiguous sequence of occurrence of the interactions, the sequence being: (D, C, B). From this sequence it is clear that in the multiparticle configuration vertex interaction occurs only if neither of the edge interactions occur. As a result, for resolving various interactions in the multiparticle configuration the same procedure that was outlined for resolving edge interactions in multiparticle configurations of type (I) can be employed. The difference is that the last interaction, being a vertex interaction, necessarily occurs if none of the edge interactions occur.

[20] Note that here parenthesis are being used for grouping also.

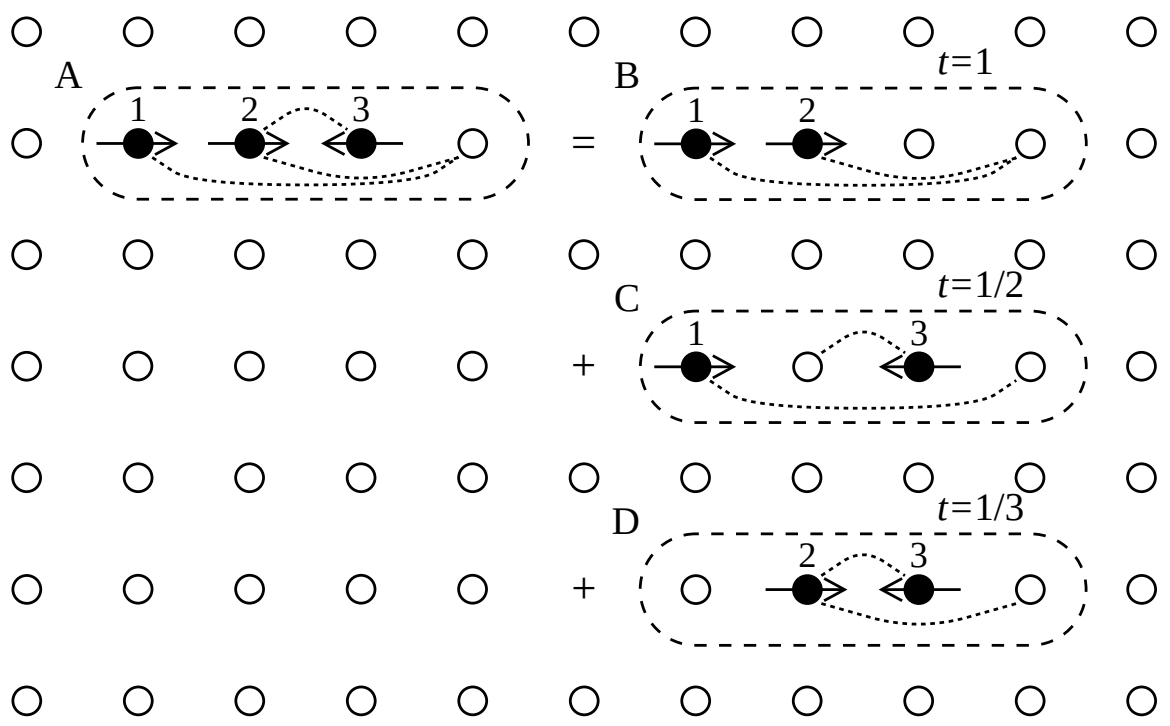

**Figure 5.17:** A three particle configuration with mixed vertex and edge interactions and its decomposition into two-particle configurations. Velocities of various particles are: $\boldsymbol{v}_1 = (3, 0)$, $\boldsymbol{v}_2 = (2, 0)$, and $\boldsymbol{v}_3 = (-1, 0)$. Symbols and notations are as in Fig. 5.8.

Since the time instant relative to the current time step at which vertex interactions occur is necessarily unity (*c.f.*, paragraph following Eq. (5.26)), they will always appear last in the sequence in which various interactions in multiparticle configurations with mixed vertex and edge interactions occur. As a result, all multiparticle configurations with mixed vertex and edge interactions can be processed by first processing the edge interactions and then, if none of the edge interactions occur, processing the vertex interaction(s). The edge interactions can be processed as outlined in Sec. 5.1.3.5.

---

[21)] That vertex interaction occurs in configuration B and binary edge interactions occur in configurations C and D is *visually* seen in Fig. 5.17. This same fact is evident from the computed values of the time instants at which the interactions occur. As elaborated earlier in the paragraph following Eq. (5.26), if the computed time instant equals unity vertex interaction occurs between particles in the current evolution and if it lies in $(0, 1)$ edge interaction occurs. Hence, $t^{\mathrm{B}} = 1$ implies that vertex interaction occurs in configuration B, and $t^{\mathrm{C}} = 1/2$ and $t^{\mathrm{D}} = 1/3$ imply that binary edge interactions occur in configurations C and D. The advantage of arriving at these inferences from computed values of time instants, instead of through visual perception, is that the deduction procedure can be formulated as robust and error free mechanical/logical algorithm suitable for implementation on computers; whereas the procedure of deduction through visual perceptions is difficult, if not impossible, to formulate as robust and error free mechanical/logical algorithm.

#### 5.1.3.7 Topology of Contact Interaction Neighborhood

The topology of the contact interaction neighborhood depends on the discrete velocity set and structure of the spatial lattice (connectivity, lattice parameters, *etc.*). Determination of the discrete velocity set requires the knowledge of lattice parameters (particularly, length of links $\Delta x$) and time step. The information furnished till now, however, does not permit determination of the lattice parameters (see also Secs. 5.1.3 and 5.1.3.2). As a result, velocity vectors which comprise the elements of the discrete velocity set can also not be determined. Thus, in the following, the method of determining the topology of the contact interaction neighborhood will be outlined in a generalized fashion and exemplified through some specific discrete velocity sets over square spatial lattice.

The contact interaction neighborhood of particles is the zone around the particles within which they can interact with other the particles through physical contact of their rigid core. In single particle lattice gases, the contact interaction neighborhood of particles can be determined as follows: Let the set $\mathcal{V} \equiv \{\boldsymbol{v}_i : i = 1, \cdots, N_{\rm v}\}$, where $N_{\rm v}$ is the number of elements (discrete velocity vectors) in the set, be the overall discrete velocity set of particles in a single particle lattice gas.[22] In this single particle lattice gas, consider a particle A moving with velocity $\boldsymbol{v}_{\rm A}$, $\boldsymbol{v}_{\rm A} \in \mathcal{V}$. This particle can undergo contact interactions with particles occupying some lattice sites around it provided that they are moving with appropriate velocities. These lattice sites comprise the contact interaction neighborhood of any particle moving with the velocity $\boldsymbol{v}_{\rm A}$. Thus, in a reference frame fixed at particles, the coordinates, $\boldsymbol{x}_k$, of the lattice sites which comprise the contact interaction neighborhood of particles[23] in a single particle lattice gas can be obtained by solving the equations

$$\boldsymbol{x}_k = \boldsymbol{v}_i - \boldsymbol{v}_j \qquad \forall\, i, j \in [1, N_{\rm v}] \quad : \quad i \neq j \tag{5.27}$$

$$|\boldsymbol{x}_k| < |\boldsymbol{v}_i - \boldsymbol{v}_j| \qquad \forall\, i, j \in [1, N_{\rm v}] \quad : \quad i \neq j \quad \text{and} \quad \boldsymbol{v}_i \| \boldsymbol{v}_j \tag{5.28}$$

and eliminating all but one of the repeated coordinates. The lattice site $(0, 0)$ is, by definition, contained in the contact interaction neighborhood. In the above equations, Eq. (5.28) comes from the conditions of occurrence of vertex interactions and Eq. (5.28) comes from the conditions of occurrence of binary edge interactions. Note that in the above multiparticle edge interactions have not been considered separately for computing the coordinates of lattice sites comprising the contact interaction neighborhood. This is because multiparticle edge interactions are resolved by decomposing them into binary edge interactions (*c.f.*, Sec. 5.1.3.5) and thus do not need to be considered separately. In the following, the contact interaction neighborhood of particles will be treated as a set, $\mathcal{L}$, whose elements are lattice site coordinates in a reference frame fixed at the targeted particle.

Consider a system of particles existing over square spatial lattice. For this system, the topology of the contact interaction neighborhood depends only upon the discrete velocity set of particles. For the sake of illustrating the method outlined above, consider five different discrete velocity sets $\mathcal{V}_1$, $\mathcal{V}_2$, $\mathcal{V}_3$, $\mathcal{V}_4$, and $\mathcal{V}_5$ enlisted in table 5.2. The coordinates of lattice sites comprising the contact interaction neighborhood of particle for each of these discrete velocity sets, computed using Eqs. (5.27) and (5.28), are enlisted in table 5.3. The topology of the contact interaction neighborhood for these discrete velocity sets is shown in Fig. 5.18.

[22] In general, there can be more than one species of particles in single particle lattice gases and discrete velocity sets for particles of each species can be different. The overall discrete velocity set of particles being used in the footmarked statement is union of the discrete velocity sets of particles of all the species comprising a single particle lattice gas. For example, if a single particle lattice gas consists of particles of $N_{\rm p}$ different species and $\mathcal{V}_i$, $i = 1, \cdots, N_{\rm p}$, is the discrete velocity set for particles of species $i$, then the overall discrete velocity set $\mathcal{V}$ for the single particle lattice gas is $\mathcal{V} = \mathcal{V}_1 \cup \mathcal{V}_2 \cup \cdots \cup \mathcal{V}_{N_{\rm p}}$.

[23] Note that the particles can be moving with any velocity vector belonging to $\mathcal{V}$.

| Some discrete velocity sets over square spatial lattice | | | |
|---|---|---|---|
| # | Name | Elements | Number of elements |
| 1 | $\mathcal{V}_1$ | $\{(\pm1,0),(0,\pm1)\}$ | 4 |
| 2 | $\mathcal{V}_2$ | $\mathcal{V}_1 \cup \{(\pm1,\pm1)\}$ | 8 |
| 3 | $\mathcal{V}_3$ | $\mathcal{V}_1 \cup \{(\pm2,0),(0,\pm2)\}$ | 8 |
| 4 | $\mathcal{V}_4$ | $\mathcal{V}_2 \cup \{(\pm2,0),(0,\pm2)\}$ | 12 |
| 5 | $\mathcal{V}_5$ | $\mathcal{V}_4 \cup \{(\pm2,\pm1),(\pm1,\pm2)\}$ | 20 |

**Table 5.2:** Some discrete velocity sets over square spatial lattice.

| # | Discrete Velocity Set | Contact interaction neighborhood of particles over square spatial lattice | | |
|---|---|---|---|---|
| | | Name | Elements | Number of sites |
| 1 | $\mathcal{V}_1$ | $\mathcal{L}_1$ | $\{(0,0),(\pm1,0),(0,\pm1),(\pm1,\pm1),(\pm2,0),(0,\pm2)\}$ | 13 |
| 2 | $\mathcal{V}_2$ | $\mathcal{L}_2$ | $\mathcal{L}_1 \cup \{(\pm2,\pm1),(\pm1,\pm2),(\pm2,\pm2)\}$ | 25 |
| 3 | $\mathcal{V}_3$ | $\mathcal{L}_3$ | $\mathcal{L}_2 \cup \{(\pm3,0),(0,\pm3),(\pm4,0),(0,\pm4)\}$ | 33 |
| 4 | $\mathcal{V}_4$ | $\mathcal{L}_4$ | $\mathcal{L}_3 \cup \{(\pm3,\pm1),(\pm1,\pm3)\}$ | 41 |
| 5 | $\mathcal{V}_5$ | $\mathcal{L}_5$ | $\mathcal{L}_4 \cup \{(\pm3,\pm2),(\pm2,\pm3),(\pm3,\pm3),$ $(\pm4,\pm1),(\pm1,\pm4),(\pm4,\pm2),(\pm2,\pm4)\}$ | 69 |

**Table 5.3:** Coordinates of lattice sites comprising the contact interaction neighborhood of particles over square spatial lattice for the discrete velocity sets enlisted in table 5.2.

From Eqs. (5.27) and (5.28) and the elaboration furnished above, it is evident that if a discrete velocity set is subset of the other then the same relationship must hold between the contact interaction neighborhoods corresponding to them also, *i.e.*, if $\mathcal{V}_\alpha \subseteq \mathcal{V}_\beta$ then $\mathcal{L}_\alpha \subseteq \mathcal{L}_\beta$. The converse, however, need not be true, *i.e.*, although $\mathcal{V}_\alpha \subseteq \mathcal{V}_\beta$ implies that $\mathcal{L}_\alpha \subseteq \mathcal{L}_\beta$, $\mathcal{L}_\alpha \subseteq \mathcal{L}_\beta$ does not, in general, imply that $\mathcal{V}_\alpha \subseteq \mathcal{V}_\beta$. These relationships are readily illustrated through the information contained in tables 5.2 and 5.3.

#### 5.1.3.8 Overall Interaction Neighborhood

Secs. 5.1.3.3–5.1.3.6 show that the mechanism of occurrence of contact interactions in single particle lattice gases is drastically different from that in the physical systems. The difference is that in physical systems the rigid cores of particles come in actual physical contact with each other during contact interactions and the velocities of particles change immediately after the physical contact (which is an instantaneous event), whereas in single particle lattice gases the rigid cores of particles never come in actual physical contact with each other and the velocities of particles are changed in response to an anticipated contact interaction.[24)] On the other hand, the mechanism of occurrence of field interactions in

[24)]Note from Secs. 5.1.3.3–5.1.3.6 that in single particle lattice gases whether or not contact interactions will occur in the next evolution is anticipated in the current evolution and all the contact interactions which will occur are resolved, *i.e.*, the velocities (in general, states) of the particles involved in these interactions are changed in accordance with the conservation laws such that they do not occur. As a result, contact

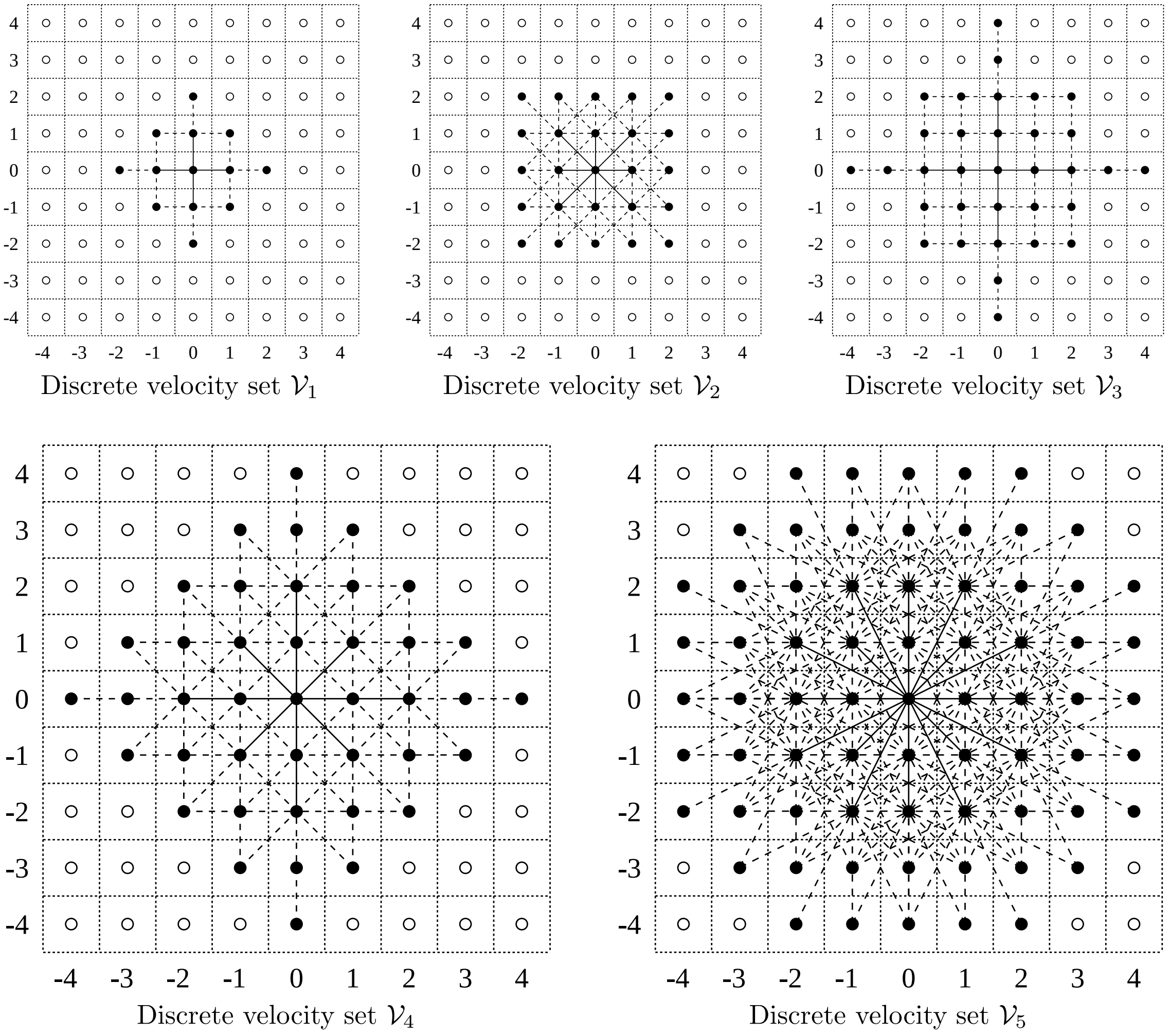

**Figure 5.18:** Topology of contact interaction neighborhood over square spatial lattice (dotted lines) for discrete velocity sets enlisted in table 5.2. Solid circles (•) represent the lattice sites which comprise the contact interaction neighborhood. Hollow circles (◦) represent other lattice sites. Reference point is lattice site $(0, 0)$. Solid lines connecting lattice site $(0, 0)$ to other lattice sites represent the possible velocities which a particle occupying this lattice site can have. Dashed lines indicate the possible paths along which particles occupying other lattice sites can move and interact with the particle occupying the lattice site $(0, 0)$.

---

interactions never actually occur (rather, they are not allowed to occur) in single particle lattice gases. This evolution procedure is necessitated by the single particle exclusion principle so that conservation laws do not get violated during simulations. If the contact interactions (particularly vertex interactions) were left unresolved during simulations, more than one particles would reach the same lattice site and get *annihilated* (explained below) causing violation of conservation laws. Such violation would not occur if edge interactions were left unresolved. This is because edge interactions occur during cross over of particles which are headed for different lattice sites. Since edge interactions are probabilistic in nature, leaving them unresolved is equivalent to choosing the probability of their occurrence to be zero.

To illustrate the above, consider a situation in space in which there are only two particles headed for the same lattice site so that a vertex interaction occurs between them in the next evolution. If this vertex interaction is left unresolved, then both the particles reach at the same lattice site in the next evolution and only one of them remains and the other gets *annihilated* (explained below). This occurs because enforcement of the single particle exclusion principle permits only one particle to occupy a lattice site at single particle lattice gases is similar to that in physical systems in that field interactions, in both, occur among particles located at a distance from each other. The only difference in the nature of field interactions in single particle lattice gases and physical systems is that in the former, because of discreteness of space, field interactions occur among particles separated by very specific distances whereas in the later arbitrary distances are possible (*c.f.*, Sec. 5.1.3.1).

The above mentioned characteristic of contact interactions in single particle lattice gases does not have any (known) counterpart in physical reality and is a consequence of discreteness of the spatial lattice combined with the single particle exclusion principle. As a result, in simulations of systems comprising of rigid particles the dynamics of particles will become approximate (and, perhaps, somewhat unphysical). This problem will surface even if the particles have a non-rigid potential because in all single particle lattice gases the particles necessarily have a rigid core (*c.f.*, Sec. 5.1.3). In this case, however, the problem can be eliminated by choosing a field interaction neighborhood that is large enough so that it subsumes the contact interaction neighborhood within it. This, in essence, is merging of the field interaction neighborhood and contact interaction neighborhood to arrive at one interaction neighborhood which, henceforth, will be referred to as the *overall interaction neighborhood*. It is evident that the overall interaction neighborhood is identical to the field interaction neighborhood. This merging is also necessitated by the fact that development of evolution rules of single particle lattice gases (in fact, cellular automata, in general) involves only one interaction/evolution neighborhood (*c.f.*, Sec. 2.1.2.4).

The problems arising out of unphysical characteristics of contact interactions get eliminated by merging the field interaction neighborhood and the contact interaction neighborhood in the way mentioned above because field interactions guide the result of all possible contact interactions and thus eliminate their unphysical effect.[25)] This makes the dynamics observed in single particle lattice gas simulations similar to that of physical systems because the mechanism of occurrence of field interactions in single particle lattice gases is similar to that in physical systems (*c.f.*, Sec. 5.1.3.1).

The requirement that the contact interaction neighborhood be subsumed inside the field interaction neighborhood for the dynamics observed in simulations to be similar to that in physical systems imposes some constraints on $\tilde{R}_{\mathrm{I}}$ (the range of interactions in

---

any time step. On the other hand, if the vertex interaction is resolved then conservation laws will not be violated.

In the above paragraph, *annihilation* is a property of the evolution algorithm and means irrecoverable loss of information. In a simulation on a sequential computer, if multiple particles try to reach at the same lattice site then the one which reaches last remains on the lattice site and the rest get annihilated. This is because, out of two particles which reach the same lattice site consecutively, the one which reaches last overwrites entire information about the one which had reached first. On sequential computers, the sequence in which particles reach a lattice site, *e.g.*, in an unresolved vertex interaction, is dictated by the evolution algorithm (*i.e.*, by the way in which the evolution algorithm has been implemented on the computer; which is specific to the computer programmer since algorithms can be implemented in may ways). This sequence is of no relevance at any time in the simulation or its results except in the event of errors (*i.e.*, when violation of conservation laws is observed). In simulations on parallel or massively parallel computers, two or more particles can reach the same lattice site simultaneously, *i.e.*, information about two or more particles can reach the processor simultaneously, which of these will be accepted and which ones will be dropped depends upon the hardware and software implementation of information transfer mechanisms and the simulation algorithm (in computer science terms this is called as *collision*). Although this process is, in general, more involved than that on sequential computers, its end result is that only one (or, possibly even zero) particle remains on the lattice site and the rest get annihilated.

[25)] The exact process through which it happens will be illustrated and elaborated upon soon.

| # | Discrete Velocity Set | $4\|\boldsymbol{r}'\|^2$ | Number of lattice sites in contact interaction neighborhood | Number of lattice sites in smallest field interaction neighborhood | Maximum value of $4\tilde{R}_{\mathrm{I}}^2$ for smallest field interaction neighborhood |
|---|---|---|---|---|---|
| 1 | $\mathcal{V}_1$ | 9 | 13 | 13 | 10 |
| 2 | $\mathcal{V}_2$ | 18 | 25 | 25 | 25 |
| 3 | $\mathcal{V}_3$ | 49 | 33 | 49 | 50 |
| 4 | $\mathcal{V}_4$ | 49 | 41 | 49 | 50 |
| 5 | $\mathcal{V}_5$ | 58 | 69 | 69 | 74 |

**Table 5.4:** The minimum permissible range of (field) interactions $|\boldsymbol{r}'|$ over square spatial lattice for the discrete velocity sets shown in table 5.2 so that the field interaction neighborhood subsumes the contact interaction neighborhood. The values of $|\boldsymbol{r}'|$ are in the natural units of the lattice system.

natural units of the lattice system). These constraints depend solely on the discrete velocity set because the topology of the contact interaction neighborhood depends solely on the discrete velocity set (for given spatial lattice). These constraints can be determined algorithmically as demonstrated below for systems existing over $\mathcal{D}$-dimensional *hypercubic* spatial lattices. Generalization of the procedure to other spatial lattices are straight forward.

Let the set $\mathcal{V} \equiv \{\boldsymbol{v}_i : i = 1, \cdots, N_{\mathrm{v}}\}$ containing $N_{\mathrm{v}}$ discrete velocity vectors be the discrete velocity set of particles in a single particle lattice gas. Let $\tilde{R}_{\mathrm{CI}}$ be *range of contact interactions* in natural units of the lattice system defined as

$$\tilde{R}_{\mathrm{CI}} = \max[|\boldsymbol{v}_i - \boldsymbol{v}_j|] \qquad \forall\, i, j \in [1, N_{\mathrm{v}}] \tag{5.29}$$

By this definition $\tilde{R}_{\mathrm{CI}}$ is the maximum distance between particles at which contact interactions can occur among them.[26)] Let $\boldsymbol{r} \equiv (r_1, \ldots, r_{\mathcal{D}})$ be a vector in $\mathcal{D}$-dimensional space such that all of its components are positive, *i.e.*, $r_i > 0$, $i = 1, \ldots, \mathcal{D}$, and $|\boldsymbol{r}| = \tilde{R}_{\mathrm{CI}}$. Let $\boldsymbol{r}' \equiv (r'_1, \ldots, r'_{\mathcal{D}})$ be another vector in $\mathcal{D}$-dimensional space with its components defined in terms of components of $\boldsymbol{r}$ as

$$r'_i = \begin{cases} r_i - \frac{1}{2} & \text{if} \quad r_i \neq 0 \\ r_i & \text{if} \quad r_i = 0 \end{cases}$$

With the above, for systems existing over $\mathcal{D}$-dimensional hypercubic spatial lattices the range of (field) interactions $\tilde{R}_{\mathrm{I}}$, in natural units of lattice system, should be

$$\tilde{R}_{\mathrm{I}} > |\boldsymbol{r}'| \tag{5.30}$$

This means that $|\boldsymbol{r}'|$ is the minimum permissible range of (field) interactions in natural units of the lattice system for the discrete velocity set $\mathcal{V}$. It is fully determined by the discrete velocity set and structure of the spatial lattice.

In Sec. 5.1.3.7, the topology of contact interaction neighborhood of particles was exemplified for systems existing over square spatial lattice using discrete velocity sets enlisted

[26)] It is not necessary that contact interactions will occur between particles if their distance is less than or equal to $\tilde{R}_{\mathrm{CI}}$ because the occurrence of contact interactions depends upon the velocity vectors of particles as well as their positions relative to each other in addition to their separation.

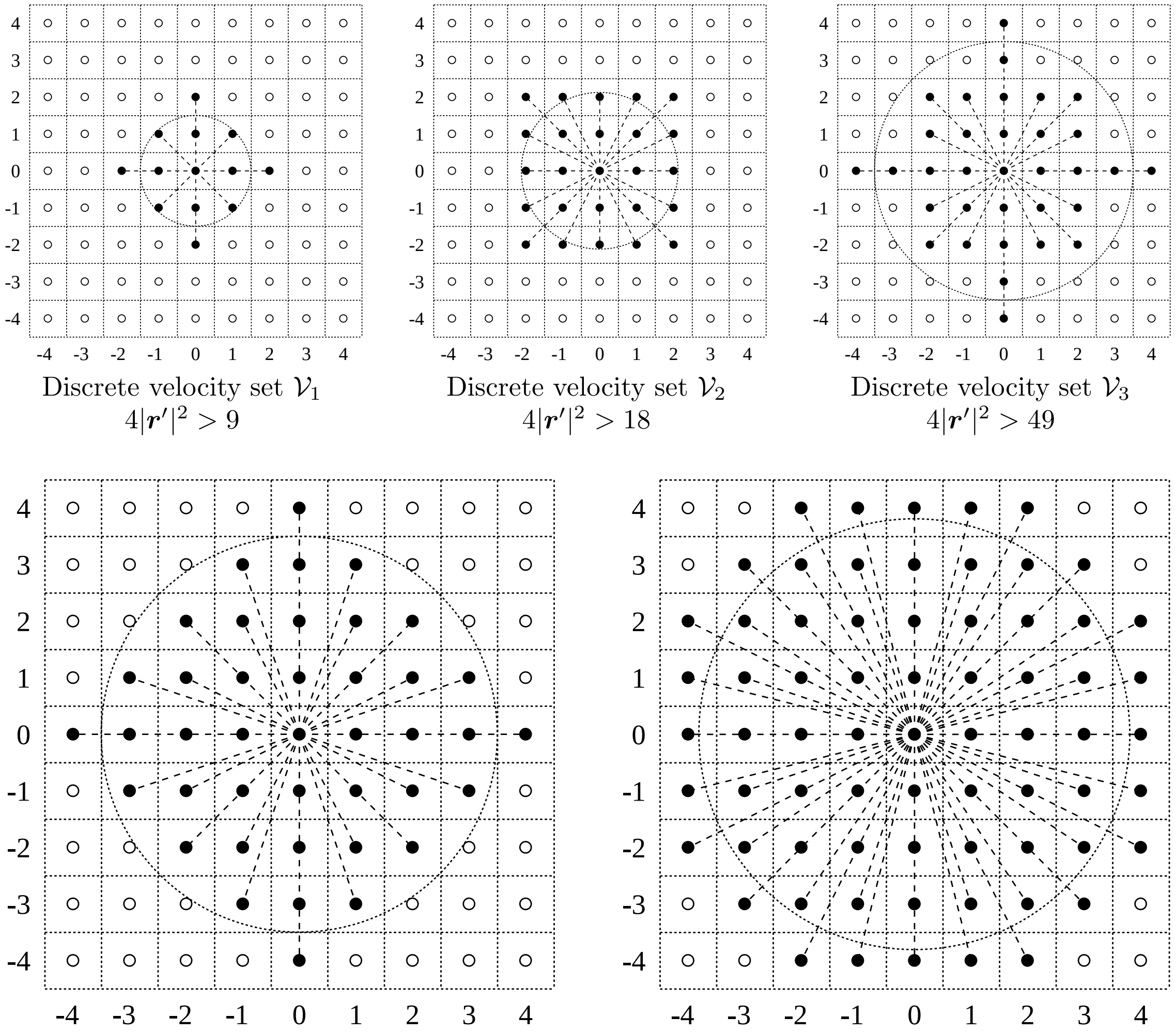

Discrete velocity set $\mathcal{V}_1$
$4|\boldsymbol{r}'|^2 > 9$

Discrete velocity set $\mathcal{V}_2$
$4|\boldsymbol{r}'|^2 > 18$

Discrete velocity set $\mathcal{V}_3$
$4|\boldsymbol{r}'|^2 > 49$

Discrete velocity set $\mathcal{V}_4$
$4|\boldsymbol{r}'|^2 > 49$

Discrete velocity set $\mathcal{V}_5$
$4|\boldsymbol{r}'|^2 > 58$

**Figure 5.19:** Subsumption of the contact interaction neighborhood by the field interaction neighborhood over square spatial lattice (dotted lines) for discrete velocity sets enlisted in table 5.2. Reference point is lattice site $(0, 0)$. Solid circles ($\bullet$) represent the lattice sites which comprise the contact interaction neighborhood. Hollow circles ($\circ$) circles represent other lattice sites. Radius of the dotted circle centered at the reference lattice site is $|\boldsymbol{r}'|$. For $\tilde{R}_{\mathrm{I}} > |\boldsymbol{r}'|$ the field interaction neighborhood subsumes the contact interaction neighborhood. Lattice sites comprising the contact interaction neighborhood have been connected to the reference lattice site with straight dashed lines showing the possible paths of particles, relative to the path of the particle occupying the reference lattice site, along which contact interactions can occur.

in table 5.2. The values of $|\boldsymbol{r}'|$ for these example systems, computed using the procedure outlined above, are enlisted in table 5.4. In this table the maximum value of $\tilde{R}_{\mathrm{I}}$ for which the field interaction neighborhood subsuming the contact interaction neighborhood is the smallest possible, and the number of lattice sites contained in each one of these interaction neighborhoods is also given. The subsumption is illustrated in Fig. 5.19.

#### 5.1.3.9 Selection of Discrete Velocity Set

For single particle lattice gases being discussed in the present investigation selection of the discrete velocity is, in essence, selection of the elements comprising it in the natural units of the lattice system. This selection is essentially a matter of free choice. This is because (i) single particle lattice gases can be developed for *almost* any discrete velocity set that one chooses, and (ii) in single particle lattice gases the system dependent parameters, *e.g.*, the interaction potential, manifest only during field interactions and the velocity set dependent parameters manifest only during contact interactions and, the system dependent parameters guide and determine the overall effect of the velocity set dependent parameters in the dynamics of the model (this will get clarified soon). Since interaction rules must satisfy the conservations laws, some constraints appear on the discrete velocity set. These constraints must be accounted for while selecting the discrete velocity set otherwise construction of physically consistent interaction rules might not be possible. These constraints are brought out and elaborated upon in Sec. 5.2.3.2 (specifically on page 177).

Nevertheless, the choice of the discrete velocity set, in essence, establishes a compromise between the effort required for developing evolution rules and the level of accuracy desired in approximating the velocity and speed distribution functions. This compromise comes up because in terms of statistical mechanics, the elements of the discrete velocity set provide an approximation to the velocity and speed distribution functions. Thus, more the number of velocity vectors and speeds in the discrete velocity set better is the accuracy of approximation of the velocity and speed distribution functions in the single particle lattice gas. Increasing the number of elements comprising the discrete velocity set, however, increases the size of the overall interaction neighborhood as well as the number of possible states of lattice sites. Because of this the size of the evolution rule table and the effort required for developing the evolution rules increases rapidly with increase in the size of the discrete velocity set. As a result, the compromise involved between the effort required for developing evolution rules and the level of accuracy desired in approximating the velocity and speed distribution functions must be considered while selecting the discrete velocity set.

### 5.1.4 Determination of Lattice Parameters and Time Step

It is clear from Secs. 5.1.3.7 and 5.1.3.2 that topology of the contact interaction neighborhood depends on the velocity vectors comprising the discrete velocity set and that of the field interaction neighborhood depends on the lattice parameters $\Delta x$ (for a given range of interactions $R_{\mathrm{I}}$ or, equivalently, interaction potential). Furthermore, the discrete velocity set imposes constraints on $\tilde{R}_{\mathrm{I}}$, given by Eq. (5.30), in order that the contact interaction neighborhood be subsumed inside the field interaction neighborhood. As a result, Eq. (5.30) combined with Eq. (5.6) gives constraints on the lattice parameters $\Delta x$ as

$$\Delta x < \frac{R_{\mathrm{I}}}{|\boldsymbol{r}'|} \tag{5.31}$$

for specified discrete velocity sets. This constraint permits unambiguous selection of the lattice parameters. In fact, within the scope of this constraint, selection of lattice parameters is essentially free except for some system specific constraints. These additional constraints, for specific types of physical systems, are as follows:

If the particles comprising a system are point centers of forces, then an additional constraint on $\Delta x$ is (*c.f.*, Sec. 3.3.3.1)

$$\Delta x > 0 \tag{5.32}$$

This constraint is also required by the fact that single particle lattice gases exist in discrete space and the extent of the spatial lattice (in terms of lattice sites) becomes unbounded for $\Delta x = 0$ making it impossible to map it back to any finite spatial domain in continuum.

If the particles comprising a physical system whose single particle lattice gas analog is to be (or, being) developed have a rigid core of characteristic dimension $d$ (they may or may not have a non-rigid potential around them), then an additional constraint

$$\Delta x > d \tag{5.33}$$

also appears on $\Delta x$ (*c.f.*, Sec. 5.1.1.2). This constraint is necessitated by the requirement that the rigid cores of particles should be contained in their entirety within the cells occupied by the particles. This constraint subsumes the one given by Eq. (5.32) for point center of force particles (for which $d = 0$).

Although the above constraints permit determination of lattice parameters $\Delta x$, they do not facilitate determination of the time step $\Delta t$. Since, $\Delta x$ and $\Delta t$ are related through elements of the discrete velocity set, it is imperative that the constraints on them cannot be independent of each other. This, in fact, is true since a more rigorous constraint on $\Delta x$ which also permits determination of $\Delta t$ exists. It is as follows:

Consider a single particle lattice gas in which particles move with velocities belonging to the discrete velocity set $\mathcal{V} \equiv \{\boldsymbol{v}_i : i = 1, \cdots, N_\text{v}\}$ containing $N_\text{v}$ discrete velocity vectors. Let $r^{ij}_{\text{c}_\text{min}}$ be the distance of closest approach of particles in continuum space corresponding to the velocity vectors $\boldsymbol{v}_i \in \mathcal{V}$ and $\boldsymbol{v}_j \in \mathcal{V}$ of the single particle lattice gas. Let $r^{\text{min}}_{\text{c}_\text{min}}$ and $r^{\text{max}}_{\text{c}_\text{min}}$ be the minimum and maximum values of $r^{ij}_{\text{c}_\text{min}}$. This gives the constraint on $\Delta x$ as

$$\Delta x > r^{\text{max}}_{\text{c}_\text{min}} \tag{5.34}$$

This constraint means that the cells should be large enough to contain interacting particles with largest separation in their entirety. This constraint appears because in single particle lattice gases, by definition, states of the particles and the lattice sites which they can occupy are not correlated, this in turn means that interacting particles occupying neighboring lattice sites can have any velocity vector. This constraint is also necessitated by the facts that in single particle lattice gases being addressed presently (i) particles move over to new lattice sites pointed out by their velocity vectors in one hop, and (ii) interparticle interactions are resolved by resolving the contact interactions and the field interactions only guide the outcome of the contact interactions (the method will be elaborated soon).

Combining Eqs. (5.31) and (5.34) gives overall constraints on lattice parameters as

$$r^{\text{max}}_{\text{c}_\text{min}} < \Delta x < \frac{R_\text{I}}{|\boldsymbol{r}'|} \tag{5.35}$$

This inequality, in order to give non-empty range for selection of $\Delta x$, requires that $r^{\text{max}}_{\text{c}_\text{min}} < R_\text{I}/|\boldsymbol{r}'|$ must necessarily hold. The value of $r^{\text{max}}_{\text{c}_\text{min}}$, however, depends upon the velocity vectors comprising the discrete velocity set, besides the mutual interaction potential of particles. In natural units of the lattice system, the velocity vectors $\boldsymbol{v}_i$ comprising the discrete velocity set are selected freely (*c.f.*, Sec. 5.1.3.9). Each one of these vectors $\boldsymbol{v}_i$, in physical units, corresponds to the vector $\boldsymbol{v}_i \Delta x/\Delta t$ which are used for computing the distance of closest approach of particles. This constrains the minimum value of $\Delta x/\Delta t$ because the distance of closest approach of particles decreases with increase in their velocity, and thus also with increase in $\Delta x/\Delta t$. This constraint is determined as follows:

Let $r^{\max}_{c_{\min}}$ correspond to the velocity vectors $\boldsymbol{v}_\alpha \in \mathcal{V}$ and $\boldsymbol{v}_\beta \in \mathcal{V}$, $\boldsymbol{v}_\alpha \neq \boldsymbol{v}_\beta$, in natural units of the lattice system. In physical units the velocity vectors corresponding to these velocity vectors are $\boldsymbol{v}_\alpha \Delta x/\Delta t$ and $\boldsymbol{v}_\beta \Delta x/\Delta t$. With this, to ensure that $r^{\max}_{c_{\min}} < R_{\mathrm{I}}/|\boldsymbol{r}'|$, the constraint on the value of $\Delta x/\Delta t$ is

$$\frac{\Delta x}{\Delta t} > \sqrt{\frac{2\Phi(R_{\mathrm{I}}/|\boldsymbol{r}'|)}{m\left[\boldsymbol{v}_\alpha^2 - (\boldsymbol{v}_\alpha \cdot \hat{\boldsymbol{v}}_{\mathrm{cm}})^2 + \boldsymbol{v}_\beta^2 - (\boldsymbol{v}_\beta \cdot \hat{\boldsymbol{v}}_{\mathrm{cm}})^2\right]}} \tag{5.36}$$

for systems comprising of identical particles of mass $m$ interacting via interaction potential $\Phi(r)$ with particles moving in a plane on intersecting undisturbed trajectories, where $\hat{\boldsymbol{v}}_{\mathrm{cm}} = \boldsymbol{v}_{\mathrm{cm}}/|\boldsymbol{v}_{\mathrm{cm}}|$ and $\boldsymbol{v}_{\mathrm{cm}} = (\boldsymbol{v}_\alpha + \boldsymbol{v}_\beta)/2$ (see also Appendix B). To determine the values of $\Delta x$ and $\Delta t$, a value of $\Delta x/\Delta t$ satisfying Eq. (5.36) is selected and the value of $r^{\max}_{c_{\min}}$ is computed. The value of $r^{\max}_{c_{\min}}$ is the solution of the equation

$$\Phi(r) = \frac{1}{2} m \left[\boldsymbol{v}_\alpha^2 - (\boldsymbol{v}_\alpha \cdot \hat{\boldsymbol{v}}_{\mathrm{cm}})^2 + \boldsymbol{v}_\beta^2 - (\boldsymbol{v}_\beta \cdot \hat{\boldsymbol{v}}_{\mathrm{cm}})^2\right] \left[\frac{\Delta x}{\Delta t}\right]^2 \tag{5.37}$$

because when $r = r^{\max}_{c_{\min}}$, the potential energy of the system of particles is equal to the sum of initial kinetic energy of the particles minus sum of the kinetic energy of the particles at the point of closest approach (see Appendix B for details).

Once the value of $r^{\max}_{c_{\min}}$ has been obtained, a value of $\Delta x$ satisfying Eq. (5.35) is selected. With this, the value of $\Delta t$ also becomes known since the value of $\Delta x/\Delta t$, selected for computing the value of $r^{\max}_{c_{\min}}$ through Eq. (5.37), is already known.

## 5.2 Definition, Evolution Strategy, and Evolution Rules of Single Particle Lattice Gases

Various important aspects of particle dynamics which need to be considered for developing single particle lattice gas analogs of physical systems have been brought out in Sec. 5.1 and subsections therein. The elaborations furnished in these sections give a general overview of the essential nature of single particle lattice gases and permit detailed description of the structure and construction of various elements comprising them. These details, including the formal definition of single particle lattice gases, are furnished below.

### 5.2.1 Definition

Single particle lattice gases exist over an underlying regular $\mathcal{D}$-dimensional spatial lattice and operate in discrete time steps $\tau$. In general, particles of $\mathcal{N}_p$ different species exist over the spatial lattice. Each lattice site can be occupied by either zero or at the most one particle of any species at any time step. Particles of each species move over the spatial lattice with finitely many discrete velocity vectors. The velocity vectors for particles of each species form a discrete velocity set $\mathcal{V}^{(i)} \equiv \{\boldsymbol{v}_j^{(i)} : j = 1, \ldots, \mathcal{N}_{\mathrm{v}}^{(i)}\}$ containing $\mathcal{N}_{\mathrm{v}}^{(i)}$ discrete elements (velocity vectors), where the superscript $(i)$ refers to particles of species $i$ and $i = 1, \ldots, \mathcal{N}_{\mathrm{p}}$. The overall discrete velocity set $\mathcal{V}$ for the lattice gas is defined as $\mathcal{V} \equiv \mathcal{V}^{(1)} \cup \mathcal{V}^{(2)} \cup \cdots \cup \mathcal{V}^{(\mathcal{N}_{\mathrm{p}})}$ and contains a total of $\mathcal{N}_{\mathrm{v}}$ discrete elements. The velocities of particles at any time step are neither correlated to nor restricted by the coordinates of lattice sites occupied by them. The particles, as a result of their motion, spontaneously interact with each other or with solid surfaces and their velocities (or, more generally, states) change after each interaction. The system evolves in discrete time steps as described in Sec. 5.2.2 below.

### 5.2.2 Evolution Strategy

In single particle lattice gases being proposed herein, the dynamics of particles over spatial lattice during one time step is decomposed into two steps, namely (i) *particle translation*, and (ii) *interparticle interactions*. In the particle translation step the particles move to the lattice sites pointed out by their respective velocity vectors. The motion is said to be completed in time $(\tau - \delta\tau)$ in the limit $\delta\tau \to \epsilon$, where $0 \leq \epsilon \ll 1$. In the interparticle interaction step various interactions of the particles with their neighbors and solid surfaces are processed in accordance with predefined *interaction rules*. The interactions are said to occur in time $\delta\tau$ in the limit $\delta\tau \to \epsilon$. The interaction rules are defined to conserve mass, momentum, and energy (and other relevant conservation laws, if any) for each group of interacting particles at each time step. In the single particle lattice gases being proposed and studied in the present investigation interactions are assumed to be instantaneous in compliance with Eq. (5.3), *i.e.*, $\delta\tau = \epsilon = 0$. It is noteworthy that in the above decomposition the particle translation step necessarily follows the interparticle interaction step (the cause of appearance of this constraint will be explained in Sec. 5.2.2.2).

The generalized mathematical formulation of the evolution strategy outlined above is as follows: Let $\mathcal{E}$ be the operator for overall evolution during one time step, $\mathcal{T}$ be the operator for evolution during translation step, $\mathcal{C}$ be the operator for evolution during interparticle interaction step, and $\Gamma^{(\tau)}$ be the state of the system at the end of time step $\tau$ (or, equivalently, at the beginning of time step $\tau + 1$), then the overall evolution of the system during one time step (from the end of time step $\tau$ to the end of time step $\tau + 1$) is

$$\Gamma^{(\tau+1)} = \mathcal{E}\left(\Gamma^{(\tau)}\right) \tag{5.38}$$

Introducing an intermediate step at time $\tau + \delta$, step wise evolution during one time step is

$$\Gamma^{(\tau+\delta)} = \mathcal{C}\left(\Gamma^{(\tau)}\right) \tag{5.39}$$

$$\Gamma^{(\tau+1)} = \mathcal{T}\left(\Gamma^{(\tau+\delta)}\right) \tag{5.40}$$

where $\Gamma^{(\tau+\delta)}$ is the intermediate state of the system. Combining the evolution in these two sub-steps gives an alternate form of overall evolution during one time step as

$$\Gamma^{(\tau+1)} = \mathcal{T}\left(\mathcal{C}\left(\Gamma^{(\tau)}\right)\right) \tag{5.41}$$

This equation, along with Eq. (5.38), gives decomposition of the operator for overall evolution during one time step as

$$\mathcal{E} = \mathcal{T}\mathcal{C} \tag{5.42}$$

It is important to note that the above decomposition of the overall evolution operator is carried out in multiparticle lattice gases also. The difference, however, is that in multiparticle lattice gases the translation step need not necessarily follow the interaction step, *i.e.*, $\mathcal{E}$ can be decomposed as given by Eq. (5.42) and also as $\mathcal{CT}$. Whereas in single particle lattice gases $\mathcal{E}$ can be decomposed only as given by Eq. (5.42) and not as $\mathcal{CT}$. In fact, in single particle lattice gases it is impossible to decompose $\mathcal{E}$ in the form $\mathcal{CT}$ (the cause of this constraint will become evident in Sec. 5.2.2.2).

In single particle lattice gases the mechanism of particle translation is straight forward and as described in the first paragraph of this section and the mechanism of interparticle interactions is as outlined and discussed in Sec. 5.1 and various subsections therein. During computer simulations, particle translation and interparticle interactions have to be processed algorithmically. The structure and method of development of these algorithms are as described in Secs. 5.2.2.1 and 5.2.2.2 which follow.

#### 5.2.2.1 Processing of Particle Translation

During simulations translation of particles is accomplished by repositioning the particles from the lattice sites occupied by them to the lattice sites pointed out by their velocity vectors. On sequential machines, this can be accomplished in two different ways, *viz.*, (I) non-table lookup lattice site scanning algorithm, and (II) non-table lookup particle scanning algorithm.[27] Although both these algorithm accomplish the same task, their programming complexities and execution efficiencies differ. As a result, their usefulness for different types of systems (different in terms of particle density) also differs. Development of parallel and massively parallel counterparts of these algorithms is straight forward because translation of each particle from the occupied to the targeted lattice site is independent of translation of other particles. The details of these algorithms for sequential machines are given below. In the following descriptions, the symbols defined in Sec. 5.2.1 are used.

**(I) Non-Table Lookup Lattice Site Scanning Algorithm:** In this algorithm each lattice site is selected sequentially and processed for its new state. The lattice sites can be selected by sequentially scanning the spatial lattice in any desired manner. Let $\boldsymbol{x}_{\mathrm{c}}$ be the selected lattice site whose new state is to be determined. The coordinates of lattice sites from where particles can move over to the lattice site $\boldsymbol{x}_{\mathrm{c}}$ are $\boldsymbol{y}_i = \boldsymbol{x}_{\mathrm{c}} - \boldsymbol{v}^{(i)}$, $\boldsymbol{v}^{(i)} \in \mathcal{V}$, $i = 1, \ldots, \mathcal{N}_{\mathrm{v}}$. From the lattice sites $\boldsymbol{y}_i$ a particle will move over to the lattice site $\boldsymbol{x}_{\mathrm{c}}$ provided that a lattice site, say $\boldsymbol{y}_j$, $\boldsymbol{y}_j \in \{\boldsymbol{y}_i : i = 1, \ldots, \mathcal{N}_{\mathrm{v}}\}$, $j \in [1, \mathcal{N}_{\mathrm{v}}]$, is occupied by a particle moving with velocity $\boldsymbol{v}^{(j)} \in \mathcal{V}$. That at most one or zero particle, out of all the particles occupying the lattice sites $\boldsymbol{y}_i$, will be headed for and move over to the lattice site $\boldsymbol{x}_{\mathrm{c}}$ is guaranteed by the interparticle interaction step which precedes the particle translation step. This is because in the interparticle interaction step interactions among particles are processed such that at the end of processing vertex interactions do not occur among particles, *i.e.*, no two particles are headed for the same lattice site. With this, the new state of the lattice site $\boldsymbol{x}_{\mathrm{c}}$ is determined as follows: If out of the lattice sites $\boldsymbol{y}_i$ a lattice site $\boldsymbol{y}_j$ is occupied by particle moving with velocity $\boldsymbol{v}^{(j)}$ then the new state of the lattice site $\boldsymbol{x}_{\mathrm{c}}$ is same as the state of the lattice site $\boldsymbol{y}_j$, else the new state of the lattice site $\boldsymbol{x}_\mathrm{c}$ is empty (irrespective of its old state).

[27] Note that table lookup algorithms are not being considered here. This is because these algorithms not only require extra memory for storage of lookup tables but also require many operations for construction of symbols to be looked up into the table before the new states of lattice sites can be determined.

**(II) Non-Table Lookup Particle Scanning Algorithm:** In this algorithm all particle occupied lattice sites are identified and a list of particles is made by scanning the lattice sites sequentially in any desired order. Following this, the new state of all the lattice sites is initialized to empty. Finally, the particles are selected sequentially, one at a time, and repositioned on the lattice site pointed out by their velocity vector. Repositioning of the $i^{\mathrm{th}}$ particles is carried out as follows: Let the lattice site occupied by a particle be $\boldsymbol{x}_i$ and its velocity be $\boldsymbol{v}^{(i)}$. Then, coordinate of the new lattice site at which this particles moves over is $\boldsymbol{x}_i + \boldsymbol{v}^{(i)}$. That each particle will be headed for a different lattice site is guaranteed as explained in the previous paragraph.

On sequential machines both the algorithms outlined above require two different buffers for storing the states of the system before and after particle translation (or, the current and the new states, respectively). Let $\Gamma^{(1)}$ and $\Gamma^{(2)}$ be the buffers storing the current and the new states of the system. Then, in the algorithms described above, input is taken from $\Gamma^{(1)}$ and alterations are done on $\Gamma^{(2)}$. Once the processing is over, $\Gamma^{(1)}$ is replaced by $\Gamma^{(2)}$. Thus, both these algorithms require a temporary buffer, namely $\Gamma^{(2)}$, for successful execution.

Both the algorithms outlined above require only additions, comparisons, and assignments. Total number of additions carried out in the first algorithm is $2\mathcal{N}_{\mathrm{sites}}\mathcal{N}_{\mathrm{v}}$ and that in the second algorithm is $\mathcal{N}_{\mathrm{sites}} + 3\mathcal{N}_{\mathrm{tp}}$, where $\mathcal{N}_{\mathrm{sites}}$ is the total number of lattice sites comprising the spatial lattice, $\mathcal{N}_{\mathrm{v}}$ is the total number of velocity vectors comprising the overall discrete velocity set $\mathcal{V}$ of the system, and $\mathcal{N}_{\mathrm{tp}}$ is the total number of particles posited over the spatial lattice. Note that the value of $\mathcal{N}_{\mathrm{tp}}$ can vary from 0 to $\mathcal{N}_{\mathrm{sites}}$. Furthermore, in physically consistent single particle lattice gases which can be employed for meaningful simulations the discrete velocity set must contain at least two elements, *i.e.*, $\mathcal{N}_{\mathrm{v}} \geq 2$.[28) This, assuming additions to be substantially more expensive compared to comparisons and assignments, implies that the second algorithm is more efficient compared to the first algorithm. The second algorithm, however, requires additional memory for storing the list of particles. As a result, for large and dense systems it might be desirable to use the first algorithm instead of the second one in order to meet constraints on the available memory.

#### 5.2.2.2 Processing of Interparticle Interactions

The basic mechanisms of interparticle interaction in single particle lattice gases have been described in detail in Sec. 5.1.3 and subsections therein. These descriptions show that the mechanism of field interactions in single particle lattice gases is similar to that in physical systems. As a result, field interactions in single particle lattice gases can be treated as in physical systems. The mechanism of contact interactions (which are divided into vertex interactions and edge interactions) in single particle lattice gases, however, differs from that in physical systems. It is also evident that whether or not edge interactions should be incorporated in single particle lattice gases is a matter of free choice.[29) The

[28) If the discrete velocity set contains only one element, no meaningful simulations can be carried out. This is because such a velocity set can only reproduce pure streaming of particles in one direction.

[29) This is because edge interactions, whether resolved or not, do not cause violation of conservation laws in single particle lattice gases. Not incorporating edge interaction in single particle lattice gases means that they are not processed during simulations or, equivalently, that their processing is not defined/incorporated in the interaction rules of the single particle lattice gas. Incorporating the elements for processing of edge vertex interactions, however, are necessarily present in all single particle lattice gases. These interactions, if left unresolved, cause annihilation of particles and lead to violation of conservation laws as explained in footnote 24 on page 156. This necessitates special considerations for processing vertex interactions to ensure that conservation laws remain intact during simulations. These considerations and the way in which they are arrived at is described below.

**Vertex Interactions and Exclusion Global Coupling:** Consider a system of identical rigid particles existing over square spatial lattice within the framework of single particle exclusion principle. For the sake of simplicity consider that the particles move over the spatial lattice with velocity vectors belonging to the discrete velocity set $\mathcal{V} \equiv \{(\pm 1, 0), (0, \pm 1)\}$, *i.e.*, the speed of all the particles is unity. Since the particles comprising this system are rigid bodies, only contact interactions occur among them. Let the dynamics of particles in this system obey (or, be constrained by) the laws of conservation of mass, momentum, and energy.

Now consider one realization of possible states of this system shown as the initial state or the state (1) in Fig. 5.20. In this initial state a vertex interaction occurs between particles B and C moving with velocities $(1, 0)$ and $(-1, 0)$, respectively. This vertex interaction, within the constraints imposed by the conservation laws, can be resolved in three different ways by changing the velocities of particles B and C to $(-1, 0)$ and $(1, 0)$, or to $(0, -1)$ and $(0, 1)$, or to $(0, 1)$ and $(0, -1)$, respectively. During simulations only one of these three ways is selected in an appropriate manner. The new states of the system obtained by resolving the vertex interaction in each one of these three possible ways are shown as the states (2a), (2b), and (2c) in the figure. Among these new states, no contact interaction occurs in the state (2c) whereas fresh vertex interactions arise in the states (2a) and (2b). As a result, if one of these two states, *viz.*, state (2a) or (2b), is obtained during simulations, interparticle interactions have to be processed again. The results of further processing of states (2a) and (2b), till states in which no contact interactions occur have been obtained, are shown as states (3a), (3b), and (4a) within the figure.

The important point to be noted in the above is that in the states (4a) and (3b) velocity of particle D has changed despite the fact that in the initial state it does not undergo contact interaction with any particle (lying within its contact interaction neighborhood). This change of state has occurred because particle C undergoes a vertex interaction with particle B which, in turn, lies within the contact interaction neighborhood of particle D. One more example illustrating this phenomena is shown in Fig. 5.21.

The above discussion illustrates that in single particle lattice gases resolution of vertex interactions (in general, any type of interactions) among particles can give rise to fresh vertex interactions of these particles with other particles lying within their contact interaction neighborhood. This gives rise to a kind of *global coupling* among particles which cannot be bypassed like the potential global coupling (*c.f.*, Sec. 5.1.2.1). This global coupling, just like the vertex interactions themselves, is a consequence of discreteness of spatial lattice combined with the single particle exclusion principle. It, henceforth, will be referred to as *"exclusion global coupling"*. The effect of this global coupling is that the new state of a particle depends not only on the state of particles lying within its contact interaction neighborhood but also on the state of many other particles lying outside it. In fact, if

---

interaction in the interaction rules of single particle lattice gases only changes some transport properties, *e.g.*, viscosity, of the system. It does not affect the dynamical behavior of the system in any other way. This is because the edge interactions alter only the trajectory of particles.

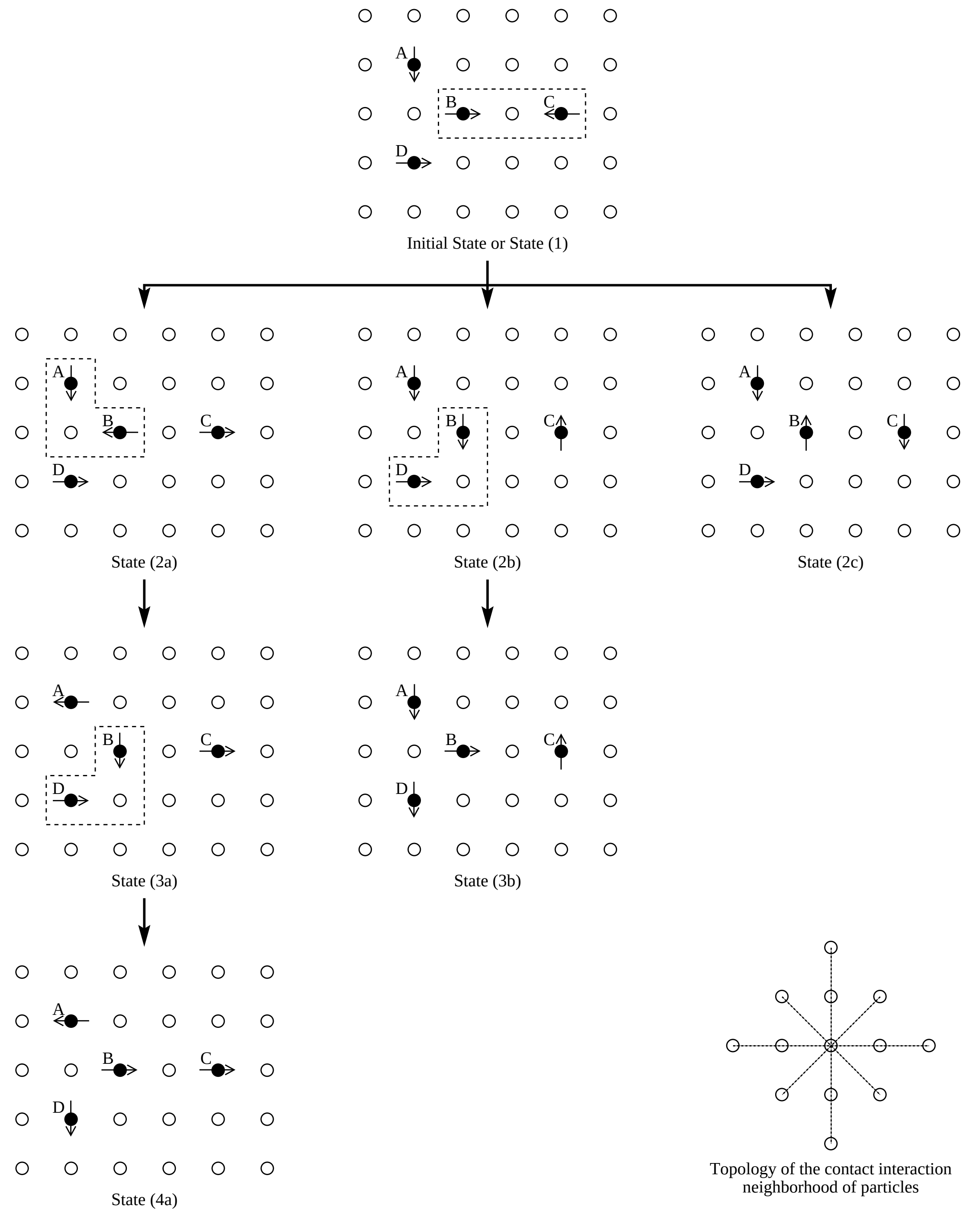

**Figure 5.20:** Illustration of exclusion global coupling in single particle lattice gases in a four particle system. Topology of the contact interaction neighborhood of particles in the system is shown in the lower right corner. Configurations in which particles undergo vertex interactions are enclosed in dashed boxes. All possible processing paths are shown. Various symbols represent: ○ unoccupied lattice sites, • particles, →↑↓← direction of motion of particles. Speed of all the particles is unity.

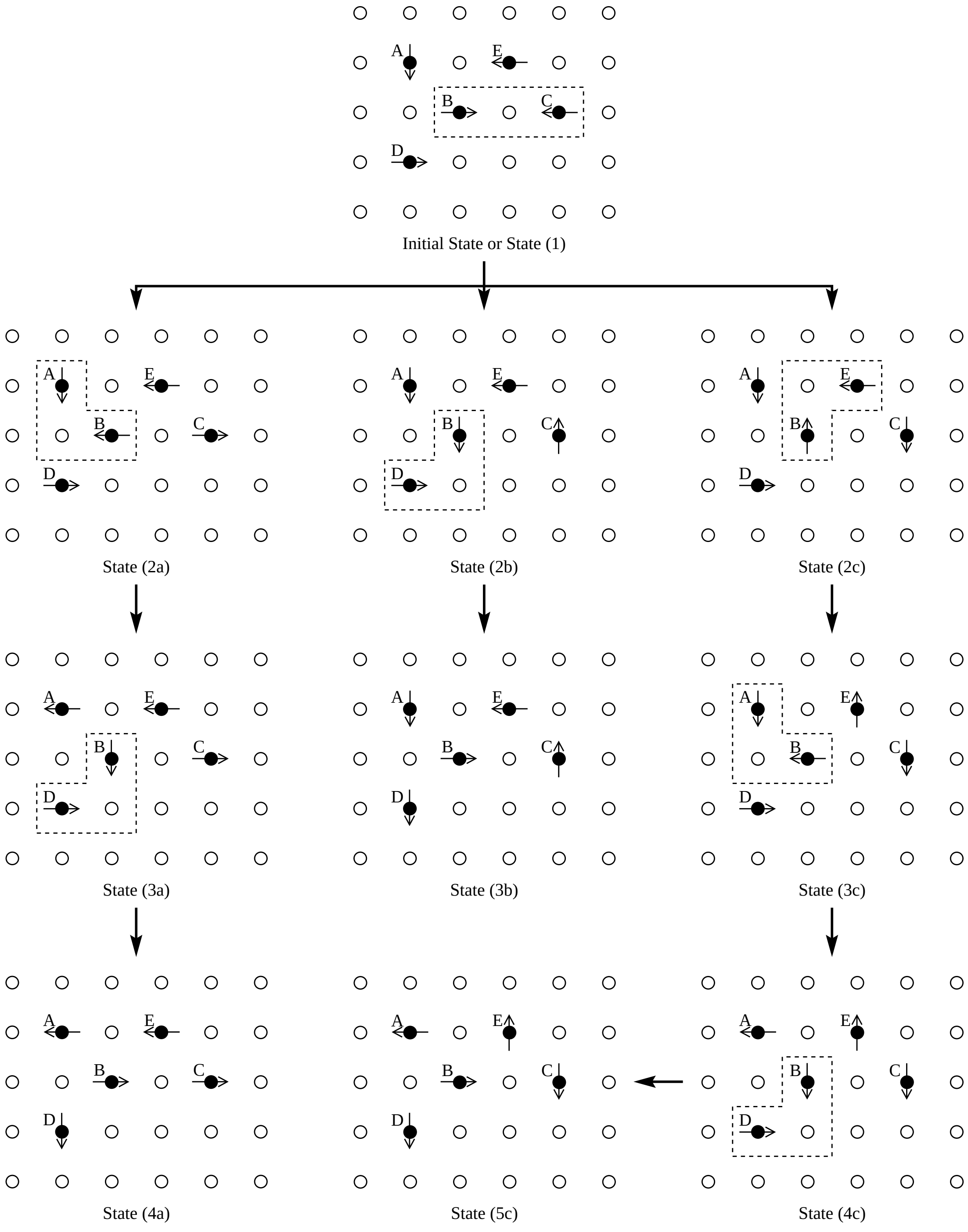

**Figure 5.21:** Illustration of exclusion global coupling in single particle lattice gases in a five particle system. Topology of the contact interaction neighborhood of particles in the system is as shown in Fig. 5.20. Configurations in which particles undergo vertex interactions are enclosed in dashed boxes. All possible processing paths are shown. Various symbols represent: ∘ unoccupied lattice sites, • particles, →↑↓← direction of motion of particles. Speed of all the particles is unity.

particle density is sufficiently high new state of each particle may depend on the current state of all the particles positioned over the spatial lattice. Thus, in general, the new state of a particle is governed by the current states of all the particles positioned over the spatial lattice.

**Consequence of Exclusion Global Coupling—Iterative Processing of Interparticle Interactions:** The occurrence of exclusion global coupling in single particle lattice gases necessitates that interaction rules (or, evolution rules for processing interparticle interactions) be developed by considering the current state of all the lattice sites comprising the spatial lattice. This, however, is not possible because (i) separate interaction rules will be required for spatial lattices of different dimensions, and (ii) the complexity of the interaction rules (or the rule table), in terms of total number of elements comprising them, becomes overwhelmingly large. This dependence of interaction rules on dimensions of the spatial lattice can be bypassed by developing interaction rules over interaction neighborhoods of fixed dimensions and processing the interparticle interactions iteratively as done in the examples shown in Figs. 5.20 and 5.21. The fixed size interaction neighborhood to be employed for developing the interaction rules is the overall interaction neighborhood (*c.f.*, Sec. 5.1.3.8).

The iterative processing of interparticle interactions mentioned above, in order to be terminated, requires that the possibility of occurrence of vertex interactions among particles be checked at the end of each iteration. If vertex interactions are found, another iteration must be carried out and all the interparticle interactions be processed once again. If no vertex interactions are found, processing of interparticle interactions is said to be over and no more iterations need to be carried out.[30] The exact number of iterations that will be required for processing all the interparticle interactions in a given state of system depends on the size of the system, states of particles, and the density and distribution of the particles. This number, however, cannot be calculated in advance. This is because the interaction rules, in general, contain probabilistic elements (this will be seen in Sec. 5.2.3).

**Mathematical Formulation of the Method of Processing Interparticle Interactions:** The generalized mathematical formulation of the method of processing interparticle interaction in single particle lattice gases outlined above is as follows: Let $\Gamma^{(\tau)}$ be the state of the system at the end of time step $\tau$ or at the beginning of processing of interparticle interactions for time step $\tau + 1$. Since processing of interparticle interactions precedes that of particle translation for evolution in each time step, $\Gamma^{(\tau)}$ is same as the state of the system at the beginning of time step $\tau + 1$. Let $\Gamma^{(\tau+\delta)}$ be the state of the system at the end of processing of interparticle interactions, where $\tau + \delta$ is an intermediate step between the time steps $\tau$ and $\tau + 1$. Since processing of particle translation follows processing of interparticle interactions, $\Gamma^{(\tau+\delta)}$ is the state of the system at the beginning of particle translation step. Let $\mathcal{C}$ be the operator for overall evolution of the system during interparticle interaction step resulting from iterative processing of interparticle interactions. In terms of these definitions, overall evolution of the system resulting from processing of

[30] If one wants one can carry out further iterations which might alter the state of the system and give rise to fresh vertex interactions. Appearance of fresh vertex interactions, in order that they may be resolved, will necessitate further iterations which also will have to be carried out till a state in which no vertex interactions exist is obtained. This process can be continued as long as desired. The iterations, however, can be stopped only when a state has been obtained in which no vertex interactions exist. This is because the iterative processing is required only to ensure that no vertex interaction exists over the spatial lattice so that particles do not get annihilated during translation step and the conservations laws remain intact.

interparticle interactions is

$$\Gamma^{(\tau+\delta)} = \mathcal{C}\left(\Gamma^{(\tau)}\right) \tag{5.43}$$

In single particle lattice gases interparticle interactions have to be processed iteratively as brought out in previous paragraphs. This implies that the operator for overall evolution in interaction step is decomposed in terms of fractional operators (or, operators for fractional evolution in each iteration). Let $\mathcal{C}'$ be the operator for evolution of the system during one iteration carried out for processing interparticle interactions. In terms of $\mathcal{C}'$, evolution of the system in each iteration carried out for processing interparticle interactions is

$$\Gamma^{(\tau+\delta_1)} = \mathcal{C}'\left(\Gamma^{(\tau)}\right) \tag{5.44}$$

$$\Gamma^{(\tau+\delta_2)} = \mathcal{C}'\left(\Gamma^{(\tau+\delta_1)}\right) \tag{5.45}$$

$$\vdots$$

$$\Gamma^{\left(\tau+\delta_{\mathrm{N_{iter}}-1}\right)} = \mathcal{C}'\left(\Gamma^{\left(\tau+\delta_{\mathrm{N_{iter}}-2}\right)}\right) \tag{5.46}$$

$$\Gamma^{(\tau+\delta)} = \mathcal{C}'\left(\Gamma^{\left(\tau+\delta_{\mathrm{N_{iter}}-1}\right)}\right) \tag{5.47}$$

where $\tau+\delta_i$, $i = 1, \ldots, \mathrm{N_{iter}}$, are intermediate steps lying sequentially between the steps $\tau$ and $\tau+\delta$ ($\equiv \tau+\delta_{\mathrm{N_{iter}}}$), $\Gamma^{(\tau+\delta_i)}$ is the intermediate state of the system at the end of the intermediate step $\tau+\delta_i$, $\Gamma^{(\tau+\delta)} \equiv \Gamma^{\left(\tau+\delta_{\mathrm{N_{iter}}}\right)}$, and $\mathrm{N_{iter}}$ is the total number of iterations required for processing interparticle interactions in the state $\Gamma^{(\tau)}$ till a state in which vertex interactions do not occur, *viz.*, the state $\Gamma^{(\tau+\delta)}$, has been obtained. Combining Eqs. (5.44)–(5.47)—which represent fractional evolution of the system in each iteration carried out for processing interparticle interactions—gives an alternate expression for overall evolution of the system in the interparticle interaction step as

$$\Gamma^{(\tau+\delta)} = \underbrace{\mathcal{C}'\left(\mathcal{C}'\left(\cdots\left(\mathcal{C}'\left(\Gamma^{(\tau)}\right)\right)\right)\right)}_{\mathrm{N_{iter}}\ \text{times}} \tag{5.48}$$

In the above, the exact value of $\mathrm{N_{iter}}$ depends on the form of $\Gamma^{(\tau)}$. This value cannot be computed in advance, *i.e.*, without carrying out the iterations, even if $\Gamma^{(\tau)}$ is known. This is because the fractional evolution operator $\mathcal{C}'$ (and thus, the overall evolution operator $\mathcal{C}$ also) contains probabilistic elements whose values vary from simulation to simulation.

Comparison of Eqs. (5.43) and (5.48) gives decomposition of the overall evolution operator $\mathcal{C}$ in terms of fractional evolution operator $\mathcal{C}'$, in single particle lattice gases, as

$$\mathcal{C} = \underbrace{\mathcal{C}'\mathcal{C}'\cdots\mathcal{C}'}_{\mathrm{N_{iter}}\ \text{times}} \tag{5.49}$$

**Relationship Between Iterative and Non-Iterative Procedures of Processing Interparticle Interactions:** The iterative procedure of processing interparticle interactions in single particle lattice gases outlined above has been devised as a substitute for the single step procedure necessitated by exclusion global coupling.[31] The size of the interaction neighborhood employed in the iterative procedure is, in general, very small compared to the dimensions of the spatial lattice which makes the interaction neighborhood for the single step procedure. From this it is evident that the iterative procedure provides only an approximate substitute of the single step procedure. The approximation involved here is in terms of the number of final states that can be obtained by the two procedure for the same initial state. Specifically, for a given initial state of the system the total number of final states that can be obtained using the iterative procedure is, in general, only a small subset of the total number of final states that can be obtained by using the single step procedure. That this is so can be readily seen from the example initial state studied earlier and shown in Fig. 5.20. As shown in this figure the iterative procedure gives three possible final states for the example initial state. Whereas, a total of 14 different final states, shown in Fig. 5.22, are possible for this initial state and can be obtained using single the step procedure.

[31] In the single step procedure the interaction rules have to be defined by considering the current states of all the lattice sites comprising the spatial lattice.

Mathematical formulation of the above can be carried out as follows: Let $\widetilde{\mathcal{C}}$ be the operator for evolution of the system during interparticle interaction step when the interactions have been processed using the single step procedure. Let $\left\{\mathcal{C}\left(\Gamma^{(\tau)}\right)\right\}$ be the set of all possible final states for the initial state $\Gamma^{(\tau)}$ that can be obtained by application of $\mathcal{C}$ over $\Gamma^{(\tau)}$ or by iterative application of $\mathcal{C}'$ over $\Gamma^{(\tau)}$. Let $\left\{\widetilde{\mathcal{C}}\left(\Gamma^{(\tau)}\right)\right\}$ be the set of all possible final states for the initial state $\Gamma^{(\tau)}$ that can be obtained by application of $\widetilde{\mathcal{C}}$ over $\Gamma^{(\tau)}$. In terms of these definitions, various relationships pointed out above can be written as

$$\mathcal{C} \neq \widetilde{\mathcal{C}} \tag{5.50}$$

$$\left\{\mathcal{C}\left(\Gamma^{(\tau)}\right)\right\} \subseteq \left\{\widetilde{\mathcal{C}}\left(\Gamma^{(\tau)}\right)\right\} \tag{5.51}$$

If the operators $\mathcal{C}$ and $\widetilde{\mathcal{C}}$ are written in the form of interaction rule tables, then more precise form of Eq. (5.50) is

$$\mathcal{C} \subseteq \widetilde{\mathcal{C}} \tag{5.52}$$

### 5.2.3 Method of Construction of Interaction Rules

From Sec. 5.2.2.2 it is clear that in single particle lattice gases interparticle interactions need to be processed iteratively. The interaction rules, thus, refer to and are developed for processing to be carried out during each iteration.

Since single particle lattice gases are fully discrete counterparts of molecular dynamics, it appears that their interactions rules can be constructed easily by adopting procedure used in molecular dynamics for processing interparticle interactions. The need of iterations for resolving interparticle interactions in single particle lattice gases, however, arises because of occurrence of vertex interactions (*c.f.*, Sec. 5.2.2.2). This makes adoption of the molecular dynamics procedure a failure. The nature of this failure is outlined in Sec. 5.2.3.1. This failure necessitates use of an alternative method for construction of interaction rules of single particle lattice gases. This alternative method is outlined in Sec. 5.2.3.2.

#### 5.2.3.1 Failure of Adoption of Molecular Dynamics Procedure

Consider a system of particles existing in discrete space subject to the single particle exclusion principle. The system evolves in discrete time steps $\tau$ of duration $\Delta t$. Let $\Delta x$ be the length of links connecting consecutive lattice sites. Let $\mathcal{L}_\mathrm{O}$ be the overall interaction neighborhood of particles in the system. $\mathcal{L}_\mathrm{O}$ is a set containing coordinates of all the lattice sites comprising the overall interaction neighborhood of particles in system. The coordinates are relative to the coordinate of the reference lattice site or the lattice site occupied by the particle whose overall interaction neighborhood is desired. Let $\Gamma^{(\tau)} \equiv \Gamma^{(\tau+\delta_0)}$ be the state of the system at the end of time step $\tau$ or at the beginning of time step $\tau+1$. Let $\Gamma^{(\tau+\delta)}$ be the state of the system obtained after processing interparticle interactions in the state $\Gamma^{(\tau)}$. Let $\Gamma^{(\tau+1)}$ be the state of the system after processing particle translation in the state $\Gamma^{(\tau+\delta)}$ or at the end of time step $\tau+1$. Let a particle occupying the lattice site $\boldsymbol{x}_i$ in the system be referred to as the particle $i$. Let the mass of the particle $i$ be $m_i$. Let $\mathcal{V}_i$ be the discrete velocity of particle $i$ and all other particles whose species is identical to that of particle $i$. Let $\boldsymbol{v}_i^{(\alpha)} \in \mathcal{V}_i$ be the velocity of particle $i$ in the state $\Gamma^{(\alpha)}$. Then, $\boldsymbol{v}_i^{(\tau+\delta)} = \boldsymbol{v}_{i'}^{(\tau+1)}$, where $\boldsymbol{x}_{i'} = \boldsymbol{x}_i + \boldsymbol{v}_i^{(\tau+\delta)}$ is the coordinate of lattice site occupied by particle $i$ after translation; this is because translation does not alter the velocity of particles.

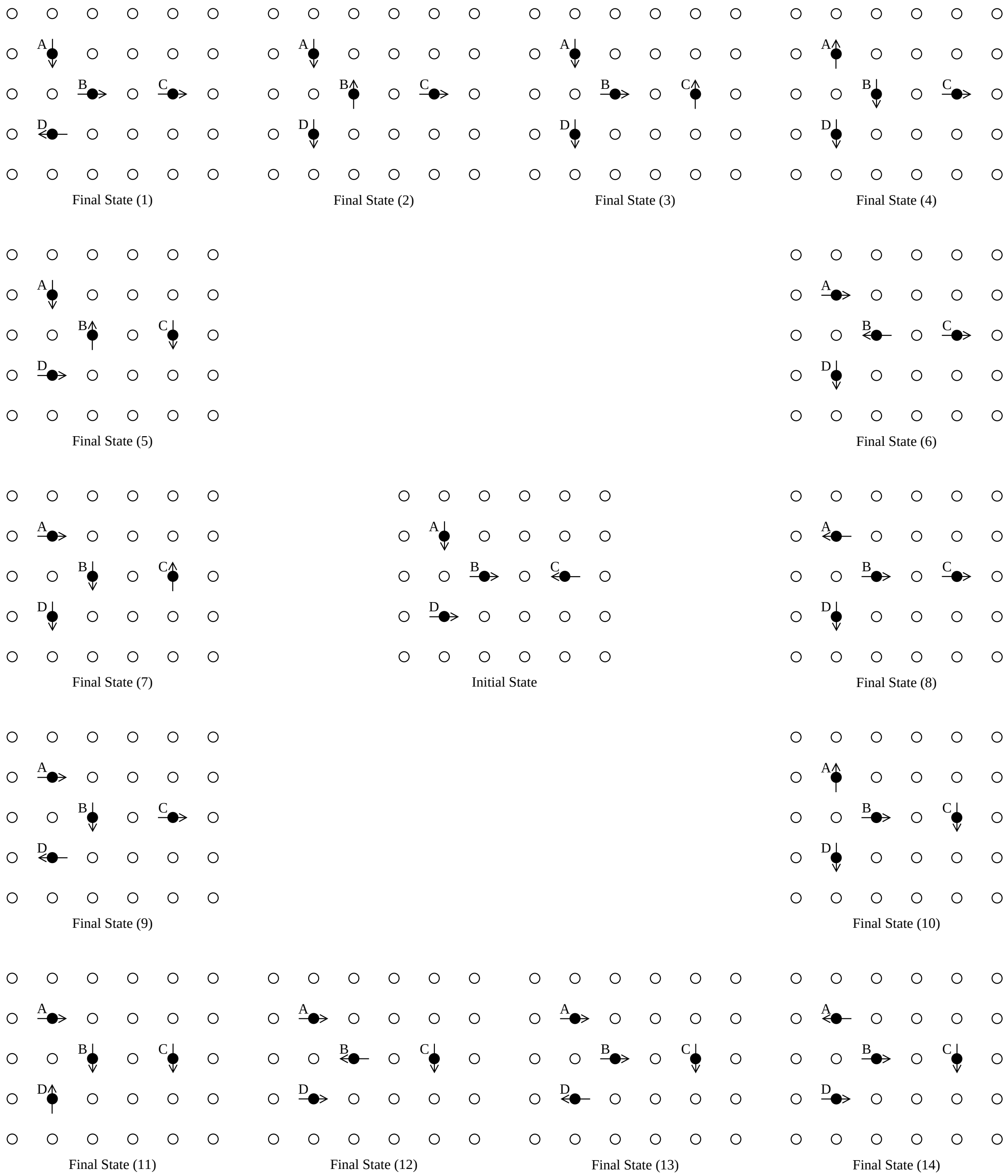

**Figure 5.22:** Possible final states for the initial state shown in Fig. 5.20 which can be obtained using interaction rules developed by considering the state of all the lattice sites comprising the spatial lattice. Note that the mass, momentum, and energy of the system is same in the initial and all final states. Various symbols represent: • particles, ∘ unoccupied lattice sites, →↑↓← direction of motion of particles. Speed of all the particles is unity.

Let the force exerted on particle $i$ by another particle $j$ be $\bar{\boldsymbol{F}}_{ij}$. Then, the net force $\bar{\boldsymbol{F}}_i$ experienced by the particle $i$ due to other particles lying in its interaction neighborhood is

$$\bar{\boldsymbol{F}}_i = \sum_{\substack{\forall\,(j-i)\in\mathcal{L}_\mathrm{O}\\ j\neq i}} \bar{\boldsymbol{F}}_{ij} \tag{5.53}$$

This force, if the system of particles described above were existing in continuum position space, would change the velocity of particle $i$ by an amount $\Delta\boldsymbol{v}_i$

$$\Delta\boldsymbol{v}_i = \frac{\bar{\boldsymbol{F}}_i}{m_i}\Delta t_i \tag{5.54}$$

in accordance with Newton's second law, where $\Delta t_i$ is the time for which particle $i$ experiences force $\bar{\boldsymbol{F}}_i$.

The net mean force $\bar{\boldsymbol{F}}_i$ used above is the mean force experienced by the particle $i$ (due to presence of other particles in its overall interaction neighborhood) at the beginning of time step $\tau+1$. It is, thus, an instantaneous approximation of the mean force that would actually be experienced by the particle in continuum space during the time interval $\Delta t$. The mean force $\bar{\boldsymbol{F}}_i$ would be experienced by the particle for time duration for which it remains within the box formed on the dual of the spatial lattice around the lattice site $i$. This time interval $\Delta t_i$ can vary from 0 to $\Delta t/\boldsymbol{v}_i^{(\tau)}$, *i.e.*, $0 < \Delta t_i \le \Delta t/\boldsymbol{v}_i^{(\tau)}$, depending upon the position of particle within the box at the beginning of the time step.[32] Thus, on the average, the particle would experience the net mean force $\bar{\boldsymbol{F}}_i$ for time duration of $\Delta t_i = \Delta t/(2\boldsymbol{v}_i^{(\tau)})$. This approximation of $\Delta t_i$ becomes exact in the limit $\Delta x \to 0$ and $\Delta t \to 0$.

If exact molecular dynamics procedure were being used, $\Delta t_i$ would be equal to $\Delta t$ and $\bar{\boldsymbol{F}}_i$ would be average of the force experienced by the particle during the time interval $\Delta t$ from the beginning of the time step $\tau+1$ till its end. In continuum space, this force would have to be computed in the limit $\Delta t \to 0$ because particles are constantly under motion. This computation, however, is not feasible. As a result, the approximation outlined in the previous paragraph has to be employed. Note, however, that this approximation does not conform to (in fact, it subsumes or is the superset of) the basic assumption regarding time scales, *viz.*, $\Delta t_{\mathrm{C}} \ll \Delta t$, employed in single particle lattice gases (*c.f.*, Sec. 5.1.1.2). This is because $\Delta t_i$ in Eq. (5.54) corresponds to $\Delta t_{\mathrm{C}}$ in the basic assumption.

[32] Note that $\boldsymbol{v}_i^{(\tau)}$ is the velocity of particle in natural units of the lattice system. In physical units, the velocity of the particle is $\boldsymbol{v}_i^{(\tau)}\Delta x/\Delta t$. Thus, the maximum time for which the particle remains inside a box of dimensions $\Delta x$ is $\Delta x/(\boldsymbol{v}_i^{(\tau)}\Delta x/\Delta t) = \Delta t/\boldsymbol{v}_i^{(\tau)}$.

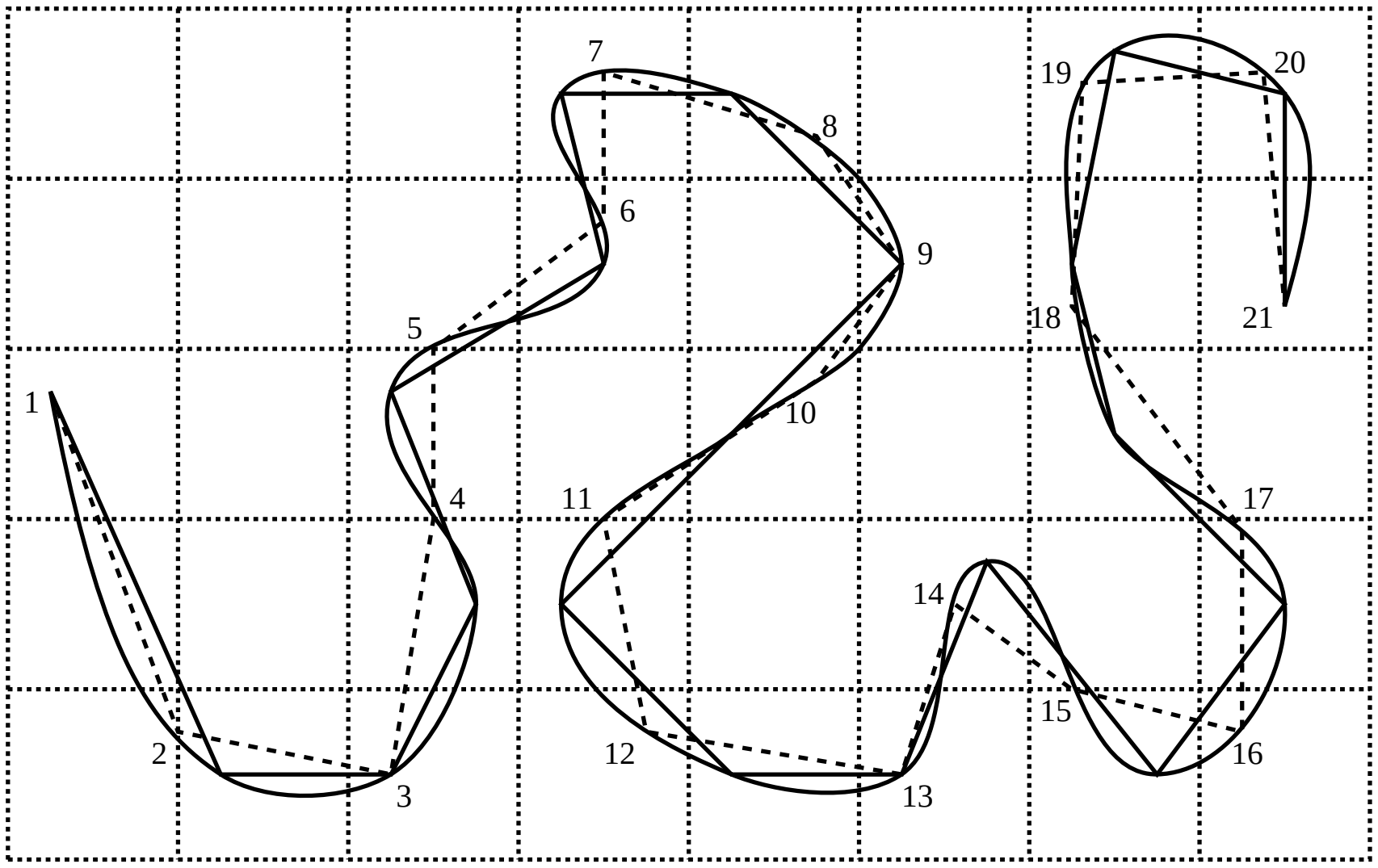

**Figure 5.23:** The exact trajectory of a particle (curved solid line) in two-dimensional position space and its possible approximations (straight solid and dashed lines) using the molecular dynamics procedure outlined in Sec. 5.2.3.1. A sample grid (dotted lines) is superimposed to indicate nature of approximation that would be involved in single particle lattice gases.

The above procedure would approximate the actual trajectory of particle as shown in Fig. 5.23. Depending upon the time step, many different approximations are possible as shown in Fig. 5.23 by two different line styles. Note that in these approximations the velocity of particle as it arrives at a station (marked by non-differentiable points on the approximate trajectories) is different from the one with which it leaves that station. The incoming and outgoing velocities of particles at different stations on the approximate trajectory will, in general, be different. The duration of each time step may be kept fixed or made variable as desired. Usually it is kept fixed. Another important aspect of the molecular dynamics procedure is that various stations on approximate trajectories would, in general, not coincide with the exact trajectory. This is because of the approximation involved in estimating the mean force experienced by the particle in each time step as well as because of numerical inaccuracies which creep in due to round-off errors on finite precision machines.

Irrespective of the method employed for estimating $\Delta t_i$ and $\bar{\boldsymbol{F}}_i$, $\Delta \boldsymbol{v}_i^{(\tau)}$ is finite and if the system were evolving from time step $\tau$ to $\tau + 1$ in continuum position space, would change the velocity of particle $i$ from $\boldsymbol{v}_i^{(\tau)}$ to $\boldsymbol{v}_{i,\mathrm{cont}}^{(\tau+\delta)}$ as

$$\boldsymbol{v}_{i,\mathrm{cont}}^{(\tau+\delta)} = \boldsymbol{v}_i^{(\tau)} + \Delta \boldsymbol{v}_i^{(\tau)} \tag{5.55}$$

$$= \boldsymbol{v}_i^{(\tau)} + \frac{\bar{\boldsymbol{F}}_i}{m_i} \Delta t_i \tag{5.56}$$

where the additional subscript in $\boldsymbol{v}_{i,\text{cont}}^{(\tau+\delta)}$ is to emphasize the fact that this velocity would be the outgoing velocity[33] of the particle (it will also be the velocity of the particle at the end of time step $\tau+1$) if the system were existing in continuum position space.[34]

The system, however, exists in discrete position space and not in continuum position space. As a result, it will, in general, not be possible to have $\boldsymbol{v}_{i,\text{cont}}^{(\tau+\delta)}$ as the outgoing velocity $\boldsymbol{v}_i^{(\tau+\delta)}$ of the particle $i$ in the system. The impossibility of having $\boldsymbol{v}_i^{(\tau+\delta)} = \boldsymbol{v}_{i,\text{cont}}^{(\tau+\delta)}$ in single particle lattice gases surfaces because (i) the constraints imposed on the velocity of particles by the discrete velocity set $\mathcal{V}_i$ necessitate that $\boldsymbol{v}_i^{(\tau+\delta)} \in \mathcal{V}_i$ and $\boldsymbol{v}_{i,\text{cont}}^{(\tau+\delta)}$ might not be (in fact, in general, will not be) contained in $\mathcal{V}_i$ since $\mathcal{V}_i$ contains only finitely many discrete velocity vectors pointing from one lattice site to another, and (ii) it is necessary that $\boldsymbol{v}_i^{(\tau+\delta)}$ be such that vertex interactions do not occur[35] among particles in the final state of the system.

A method of determining the the outgoing velocity $\boldsymbol{v}_i^{(\tau+\delta)}$ that appears to be feasible is to approximate $\boldsymbol{v}_{i,\text{cont}}^{(\tau+\delta)}$ by a velocity vector contained in $\mathcal{V}_i$ and use it as the value of $\boldsymbol{v}_i^{(\tau+\delta)}$. For this approximation to be as good as possible, the vector to be selected from $\mathcal{V}_i$ should be as close to $\boldsymbol{v}_{i,\text{cont}}^{(\tau+\delta)}$ as possible. Although this method appears to be appealing, it cannot be used in actual simulations. This is because of two problems. Firstly, the approximation involved in this method leads to violation of conservation laws, particularly, of the laws of conservation of momentum and energy. This violation occurs because the outgoing velocity of each particle is computed by considering its field interactions with other particles and hence the conservation laws are pivoted to field interactions of particles. Thus, the actual value of the outgoing velocity dictated by the conservation laws is $\boldsymbol{v}_{i,\text{cont}}^{(\tau+\delta)}$. As a result, approximating it by any other value causes violation of conservation laws. This violation cannot be overcome because it is a consequence of the approximation process. Secondly, this method does not guarantee that vertex interactions will not occur among particles in the final state of the system; which, as pointed out earlier, is a necessary constraint on the velocities of particles in the final state of the system. The second problem can be overcome easily by employing iterations if the second best, third best, *etc.*, approximations to $\boldsymbol{v}_{i,\text{cont}}^{(\tau+\delta)}$ are also permitted (possibly, in an appropriate probabilistic manner) instead of only the best approximation. Although the second problem can be overcome easily, the first problem persists and renders the method unusable.

From the discussion outlined above it is clear that molecular dynamics procedure cannot be adopted as it is for developing interaction rules for processing interparticle interactions in single particle lattice gases. An important consequence of this impossibility is that the microscopic dynamics reproduced in single particle lattice gas simulations will differ from that of an identical system existing in continuum space and evolving in discrete time step. This is because the velocity, and thus the trajectory, of particles differ in both these systems. In order that single particle lattice gas simulations be of use for studying the dynamical behavior of physical systems it, however, is necessary that the velocity and trajectory of particles in single particle lattice gas simulations be as close as possible, if not identical, to the actual once. In addition, the final velocity of particles has to be determined in such a way that all the conservation laws of the system remain intact during simulations and vertex interactions do not occur in the final state of the system. An alternate method of constructing interaction rules of single particle lattice gases which takes care of all these aspects is outlined in the following sections (*i.e.*, Sec. 5.2.3.2–5.2.3.6).

---

[33] In continuum position space, $\boldsymbol{v}_i^{(\tau)}$ is the incoming velocity of the particle at station $i$ and $\boldsymbol{v}_{i,\text{cont}}^{(\tau+\delta)}$ is the outgoing velocity of the particle.

[34] Note that if the system were existing in continuum position space particles, in general, can move with any velocity without restrictions like in single particle lattice gases. Alternatively, if the system were existing in continuum position space the discrete velocity set of particles will, in general, have infinitely many elements and not finitely many discrete elements as in single particle lattice gases.

[35] Whether or not having $\boldsymbol{v}_i^{(\tau+\delta)} = \boldsymbol{v}_{i,\text{cont}}^{(\tau+\delta)}$ will lead to vertex interactions in the final state of the system is debatable. It, in fact, might not give rise to vertex interactions. But then, the constraints imposed by the discrete velocity set, described as the first cause in point (i), also have to be taken into account.

#### 5.2.3.2 The Alternate Method of Constructing Interaction Rules

The failure of molecular dynamics procedure for computing the outgoing velocity of particles occurs because $\boldsymbol{v}_{i,\mathrm{cont}}^{(\tau+\delta)}$ may, in general, not belong to $\mathcal{V}_i$ and also because approximating it by a vector contained in $\mathcal{V}_i$ causes violation of conservation laws. This in turn happens because in this procedure, while computing the outgoing velocity of particles, the primary stress lies on field interactions among particles. In single particle lattice gases, however, two different types of interparticle interactions occur, *viz.*, field interactions and contact interactions (*c.f.*, Sec. 5.1.3 and subsection therein). As a result, another procedure for computing the outgoing velocity can be devised wherein the primary stress lies on contact interactions, instead of field interactions, among particles. The advantages of this procedure are that in it the computed value of the outgoing velocity $\boldsymbol{v}_i^{(\tau+\delta)}$ necessarily belongs to the discrete velocity set $\mathcal{V}_i$ and the problem of violation of conservation laws does not surface. This procedure is an iterative procedure. It, along with some definitions in addition to those outlined in the first paragraph of Sec. 5.2.3.1, is described below.

Let $\mathcal{V}$ be the overall discrete velocity set or the discrete velocity set obtained by combining the discrete velocity sets for particles of all the species comprising the system. Let $\mathcal{L}_\mathrm{C}$ be the contact interaction neighborhood corresponding to $\mathcal{V}$. Let $\Gamma^{(\tau)} = \Gamma^{(\tau+\delta_0)}$ and $\Gamma^{(\tau+\delta)} = \Gamma^{\left(\tau+\delta_{\mathrm{N_{iter}}}\right)}$, where $\mathrm{N_{iter}}$ is the total number of iterations required for processing interparticle interactions in the state $\Gamma^{(\tau)}$. Let interparticle interactions be processed in the state $\Gamma^{(\tau)}$ resulting in the new state $\Gamma^{(\tau+\delta_k)}$ after $k$ iterations are over. In this new state, contact interactions (more specifically, vertex interactions) may or may not occur among particles. If vertex interactions do not occur, then processing of interparticle interactions is over, *i.e.*, $\mathrm{N_{iter}} = k$ and $\Gamma^{(\tau+\delta)} = \Gamma^{\left(\tau+\delta_{\mathrm{N_{iter}}}\right)} = \Gamma^{(\tau+\delta_k)}$, and particle translation can be carried out to complete the evolution for time step $\tau+1$. If, on the other hand, vertex interactions occur, then they will necessarily involve at least two particles and further processing of interparticle interactions would be required. Let there be vertex interactions in the state $\Gamma^{(\tau+\delta_k)}$. Let $\Gamma^{(\tau+\delta_{k+1})}$ be the new state of the system after processing interactions among particles in the state $\Gamma^{(\tau+\delta_k)}$. This processing involves computing the velocity of particles[36)] in the state $\Gamma^{(\tau+\delta_{k+1})}$ of the system and is carried out as follows.

In the state $\Gamma^{(\tau+\delta_k)}$ particles can be found in one of the two possible situations, *viz.*, (i) particle does not undergo contact interactions with any particle lying within its contact interaction neighborhood, and (ii) particle undergoes contact interactions (*i.e.*, vertex or edge interactions) with one or many particles lying within its contact interaction neighborhood. If a particle does not undergo contact interactions with any of its neighbors,

[36)] Only velocity needs to be computed because the system is considered to be a classical system consisting of neutral particles and there are no chemical reactions. In more generalized case, charge and other state parameters may also be associated with the particles. In such a case, one computes the *state* of particles instead of just the velocity. The state subsumes the velocity of particles also.

then its velocity remains unchanged in the state $\Gamma^{(\tau+\delta_{k+1})}$, *i.e.*, $\boldsymbol{v}_i^{(\tau+\delta_{k+1})} = \boldsymbol{v}_i^{(\tau+\delta_k)}$, where the subscript $i$ denotes that the velocities refer to the particle $i$ (*c.f.*, Sec. 5.2.3.1). On the other hand, if a particle undergoes contact interactions with one or many of its neighbors, then its velocity in the state $\Gamma^{(\tau+\delta_{k+1})}$ will, in general, not be same as that in the state $\Gamma^{(\tau+\delta_k)}$. In such a situation, the velocity in the state $\Gamma^{(\tau+\delta_{k+1})}$ is determined as described below.

Let there be contact interaction among $\mathcal{N}_{\text{PC}}$ particles located at lattice sites $\boldsymbol{x}_{i_1}, \ldots, \boldsymbol{x}_{i_{\mathcal{N}_{\text{PC}}}}$. Each one of these particles, following the convention outlined in Sec. 5.2.3.1, will be referred to as particle $i_j$, $j = 1, \ldots, \mathcal{N}_{\text{PC}}$. In this contact interaction, in the absence of chemical reactions, the species of particles does not change, *i.e.*,

$$m_{i_j}^{(\tau+\delta_{k+1})} = m_{i_j}^{(\tau+\delta_k)} \qquad \forall\, j \in [1, \mathcal{N}_{\text{PC}}] \tag{5.57}$$

where $m_{i_j}^{(\alpha)}$ is the mass of particle $i_j$ in the state $\Gamma^{(\alpha)}$ of the system. Note that the notation used in Sec. 5.2.3.1 for representing the mass of particles has been augmented to incorporate possible change in mass (*i.e.*, species of particles) between consecutive time step and consecutive iterations in the same time step. In the following, for the sake of simplicity, only non-reacting systems will be considered.

In view of Eq. (5.57), the law of conservation of mass

$$\sum_{j=1}^{\mathcal{N}_{\text{PC}}} m_{i_j}^{(\tau+\delta_{k+1})} = \sum_{j=1}^{\mathcal{N}_{\text{PC}}} m_{i_j}^{(\tau+\delta_k)} \tag{5.58}$$

remains satisfied during the interaction. The laws of conservation of momentum and energy require that the equations

$$\sum_{j=1}^{\mathcal{N}_{\text{PC}}} m_{i_j}^{(\tau+\delta_{k+1})} \boldsymbol{v}_{i_j}^{(\tau+\delta_{k+1})} = \sum_{j=1}^{\mathcal{N}_{\text{PC}}} m_{i_j}^{(\tau+\delta_k)} \boldsymbol{v}_{i_j}^{(\tau+\delta_k)} \tag{5.59}$$

$$\sum_{j=1}^{\mathcal{N}_{\text{PC}}} m_{i_j}^{(\tau+\delta_{k+1})} \left[\boldsymbol{v}_{i_j}^{(\tau+\delta_{k+1})}\right]^2 = \sum_{j=1}^{\mathcal{N}_{\text{PC}}} m_{i_j}^{(\tau+\delta_k)} \left[\boldsymbol{v}_{i_j}^{(\tau+\delta_k)}\right]^2 \tag{5.60}$$

hold during the interaction. In both these equations, all parameters in the right hand side are known and through Eq. (5.57), in the absence of chemical reactions, $m_{i_j}^{(\tau+\delta_{k+1})}$ in the left hand side are also known. In $\mathcal{D}$-dimensional position space, these two equations are equivalent to $\mathcal{D}+1$ scaler algebraic equations in $\mathcal{D}\mathcal{N}_{\text{PC}}$ scalar unknowns. The unknowns are components of $\mathcal{D}$-dimensional vectors $\boldsymbol{v}_{i_j}^{(\tau+\delta_{k+1})}$, $j = 1, \ldots, \mathcal{N}_{\text{PC}}$, which appear in the left hand side of these equations. Note that $\mathcal{D} \geq 1$ and, since contact interactions necessarily involve at least two particles, $\mathcal{N}_{\text{PC}} \geq 2$. Details on existence and uniqueness of solutions for this system of equations for various values of $\mathcal{N}_{\text{PC}}$ and $\mathcal{D}$ for various types of contact interactions (*i.e.*, vertex and edge interactions) are described below.

**Edge Interactions:** For all edge interactions, $\mathcal{N}_{\text{PC}} = 2$. This is because all configurations with edge interactions are processed by decomposing them into two particle configurations with binary edge interactions (*c.f.*, Sec. 5.1.3.5). For binary edge interactions, irrespective of the value of $\mathcal{D}$, $\boldsymbol{v}_{i_j}^{(\tau+\delta_{k+1})} = \boldsymbol{v}_{i_j}^{(\tau+\delta_k)}$ is a solution of the system of equations because the occurrence of these interactions is probabilistic (*c.f.*, Sec. 5.1.3.5). In addition to this solution, the system of equations will always have at least one more solution. As a result, the system of equations always has at least two solutions. Let, for a given binary edge interaction, $\mathcal{N}_{\text{sol}}$, $\mathcal{N}_{\text{sol}} \geq 2$, be the total number of solutions for the system of equations. The exact value of $\mathcal{N}_{\text{sol}}$ depends upon $\mathcal{D}$. All these solutions, however, may not be *acceptable* because of the constraints imposed on $\boldsymbol{v}_{i_j}^{(\tau+\delta_{k+1})}$ by the discrete velocity set $\mathcal{V}_{i_j}$, $j = 1, \ldots, \mathcal{N}_{\text{PC}}$, of particles participating in the edge interaction.[37] Let the total number of acceptable solutions be $\mathcal{N}_{\text{acc}}$. Then, $\mathcal{N}_{\text{acc}} \geq 1$ because $\boldsymbol{v}_{i_j}^{(\tau+\delta_{k+1})} = \boldsymbol{v}_{i_j}^{(\tau+\delta_k)}$ is also a solution for the system of equations. If $\mathcal{N}_{\text{acc}} = 1$, then this solution unambiguously gives the velocities of particles in the state $\Gamma^{(\tau+\delta_{k+1})}$. On the other hand, if $\mathcal{N}_{\text{acc}} > 1$, then only one of these solutions has to be selected for the velocities of particles in the state $\Gamma^{(\tau+\delta_{k+1})}$. The selection should be carried out probabilistically because all the solutions are valid solutions as far as conservation laws of the system are concerned. The probabilities can be selected either in some *ad hoc* manner or on the basis of some desired criteria.

**Vertex Interactions:** In vertex interactions, unlike edge interactions, $\boldsymbol{v}_{i_j}^{(\tau+\delta_{k+1})}$ and $\boldsymbol{v}_{i_j}^{(\tau+\delta_k)}$ must necessarily be different otherwise the interaction remains unprocessed. As a result, the number of acceptable solutions $\mathcal{N}_{\text{acc}}$ can vary from 0 to the maximum number of possible solutions $\mathcal{N}_{\text{sol}}$. If $\mathcal{N}_{\text{acc}} > 0$, one solution is selected as described in the previous paragraph for edge interactions. On the other hand, if $\mathcal{N}_{\text{acc}} = 0$, physically consistent interaction rules cannot be constructed for the single particle lattice gas.[38] This is because vertex interactions which do not have a solution cannot be processed and cause violation of conservation laws due to annihilation of particles during simulations. As a result, selection of particle species and discrete velocity set for particles of each species must be done carefully so that such a situation does not arise for any possible vertex interaction in the system. For single species systems, the occurrence of such situations can be eliminated by ensuring that the discrete velocity set contains elements in pairs of the form $\pm\boldsymbol{v}$, $\forall\ |\boldsymbol{v}| \neq 0$.

The method outlined above can be used for construction of interaction rules of single particle lattice gases with desired particle species and discrete velocity sets (selected within the constraints outlined in Sec. 5.1.3.9 and in the previous paragraph). Note that in this method the interaction neighborhood of particles is restricted to their contact interaction neighborhood because only contact interactions of particles are considered. As a result, construction of interaction rules in the usual form of lookup tables, essentially, involves tabulation of all possible initial states of the contact interaction neighborhood and the possible final states for each one of these initial states. The construction of lookup tables, however, is unwieldy because the site of the contact interaction neighborhood (*i.e.*, the number of lattice sites $\mathcal{N}_{\mathcal{L}_{\text{C}}}$ comprising the contact interaction neighborhood) increases rapidly with $\mathcal{D}$, $\mathcal{N}_{\text{v}}$, and the magnitude of largest velocity vector $\max[|\boldsymbol{v}_1|, \ldots, |\boldsymbol{v}_{\mathcal{N}_{\text{v}}}|]$ contained in $\mathcal{V}$. Because of this the total number of elements in the lookup table (*i.e.*, the total number of states $\mathcal{N}_{\text{s,cin}}$ of the contact interaction neighborhood and all possible final states corresponding to them) becomes very large even for simplest single species two-dimensional systems with smallest discrete velocity sets. The value of $\mathcal{N}_{\text{s,cin}}$ for the discrete velocity sets shown in table 5.2 and corresponding contact interaction neighborhoods shown in table 5.3 for single species systems existing over two-dimensional square spatial lattice is given in table 5.5.

Since contact interactions are directional, states of most of the lattice sites comprising the contact interaction neighborhood become irrelevant for the outcome of contact interactions among particles. The lattice sites whose states are important in contact interactions depends upon the velocity of particles involved in the interaction. Furthermore, since the occurrence of contact interactions depends upon the velocity of particles, they can occur only in very specific configurations among particles. These configurations will henceforth be referred to as *"basic collision configurations"* (BCCs). The structure and number of BCCs depends upon the overall discrete velocity set of particles and the structure of spatial lattice. For single species single particle lattice gases with discrete velocity set containing $\mathcal{N}_{\text{v}}$ discrete velocity vectors, total number of BCCs arising out of vertex interactions $\mathcal{N}_{\text{bcc,vi}}$, after ignoring the state of the commonly targeted lattice site, is

$$\mathcal{N}_{\text{bcc,vi}} = \sum_{i=2}^{\mathcal{N}_{\text{v}}} \binom{\mathcal{N}_{\text{v}}}{i} = 2^{\mathcal{N}_{\text{v}}} - \mathcal{N}_{\text{v}} - 1 \tag{5.61}$$

[37] A solution is acceptable only if $\boldsymbol{v}_{i_j}^{(\tau+\delta_{k+1})} \in \mathcal{V}_{i_j}$, $j = 1, \ldots, \mathcal{N}_{\text{PC}}$, and not otherwise.

[38] This means that single particle lattice gases corresponding to particle species and discrete velocity sets for which vertex interactions without any acceptable solution of Eqs. (5.59) and (5.60) occur, do not exist.

| $i$ | $\mathcal{V}$ | $\mathcal{N}_{\text{v}}$ | $\mathcal{N}_{\text{s}}$ | $\mathcal{N}_{\mathcal{L}_\text{C}}$ | $(\mathcal{N}_{\text{s}})^{\mathcal{N}_{\mathcal{L}_\text{C}}} = \mathcal{N}_{\text{s,cin}}$ |
|---|---|---|---|---|---|
| 1 | $\mathcal{V}_1$ | 4 | 5 | 13 | $5^{13} \approx 1.22 \times 10^9$ |
| 2 | $\mathcal{V}_2$ | 8 | 9 | 25 | $9^{25} \approx 7.18 \times 10^{23}$ |
| 3 | $\mathcal{V}_3$ | 8 | 9 | 33 | $9^{33} \approx 3.09 \times 10^{31}$ |
| 4 | $\mathcal{V}_4$ | 12 | 13 | 41 | $13^{41} \approx 4.70 \times 10^{45}$ |
| 5 | $\mathcal{V}_5$ | 20 | 21 | 69 | $21^{69} \approx 1.71 \times 10^{91}$ |

**Table 5.5:** Number of states of the contact interaction neighborhood for some single species single particle lattice gases existing over square spatial lattice with discrete velocity sets shown in table 5.2 and corresponding contact interaction neighborhoods shown in table 5.3. $\mathcal{N}_{\text{s}}$ is the possible number of states for each lattice site, $\mathcal{N}_{\mathcal{L}_\text{C}}$ is the number of lattice sites comprising the contact interaction neighborhood, and $\mathcal{N}_{\text{s,cin}}$ is the number of states of the contact interaction neighborhood.

This number will always be smaller (in fact, orders of magnitude smaller) compared to the number of states $\mathcal{N}_{\text{s,cin}}$ of the contact interaction neighborhood. This is because the number of states of contact interaction neighborhood is given by $\mathcal{N}_{\text{s,cin}} = (\mathcal{N}_{\text{s}})^{\mathcal{N}_{\mathcal{L}_\text{C}}}$, where the conditions $\mathcal{N}_{\text{s}} > 2$ and $\mathcal{N}_{\mathcal{L}_\text{C}} > \mathcal{N}_{\text{v}}$ always hold for all single particle lattice gases. For computing the number of BCCs arising out of edge interactions $\mathcal{N}_{\text{bcc,ei}}$ it is necessary to know the elements comprising the discrete velocity set (for particles of each species). As a result, general expressions for $\mathcal{N}_{\text{bcc,ei}}$, like Eq. (5.61), cannot be given. This number, however, will also be orders of magnitude smaller compared to $\mathcal{N}_{\text{s,cin}}$. The BCCs for a one-dimensional single species single particle lattice gas with discrete velocity set $\{(\pm 1)\}$ and a two-dimensional single species single particle lattice gas existing over square spatial lattice with discrete velocity set $\{(\pm 1, 0), (0, \pm 1)\}$ are shown in Fig. 5.24.[39)] It is worth comparing the values of $\mathcal{N}_{\text{s,cin}}$ and $\mathcal{N}_{\text{bcc}} = \mathcal{N}_{\text{bcc,vi}} + \mathcal{N}_{\text{bcc,ei}}$ in these single particle lattice gases. In the one-dimensional single particle lattice gas $\mathcal{N}_{\text{s,cin}} = 3^5 = 243$ and $\mathcal{N}_{\text{bcc}} = 2$, and in the two-dimensional single particle lattice gas $\mathcal{N}_{\text{s,cin}} = 5^{13} = 1220703125$ and $\mathcal{N}_{\text{bcc}} = 13$. The orders of magnitude difference in the values of $\mathcal{N}_{\text{s,cin}}$ and $\mathcal{N}_{\text{bcc}}$ is noteworthy.

[39)] Note that in this figure the BCCs are shown using a compressed notation wherein all the BCCs having the same state of interacting particles but many different states of the targeted lattice site are represented as one BCC using the symbol $\otimes$ which means as explained within the figure. Eq. (5.61) counts only the number of compressed BCCs arising out of vertex interactions. Although the total number of compressed BCCs $\mathcal{N}_{\text{bcc}}$ shown in Fig. 5.24 for the two-dimensional case is 13, the total number of (uncompressed) BCCs obtained by expanding the symbol $\otimes$ becomes $2 + 3 \times 6 + 2 \times 4 + 1 = 29$. The advantage of using this compressed notation is that the size of the BCC lookup table reduces considerably.

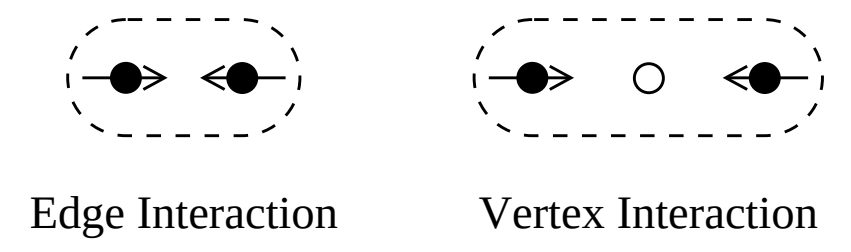

**(A)** Contact interactions in one-dimensional systems with $\mathcal{V} = \{(\pm 1)\}$.

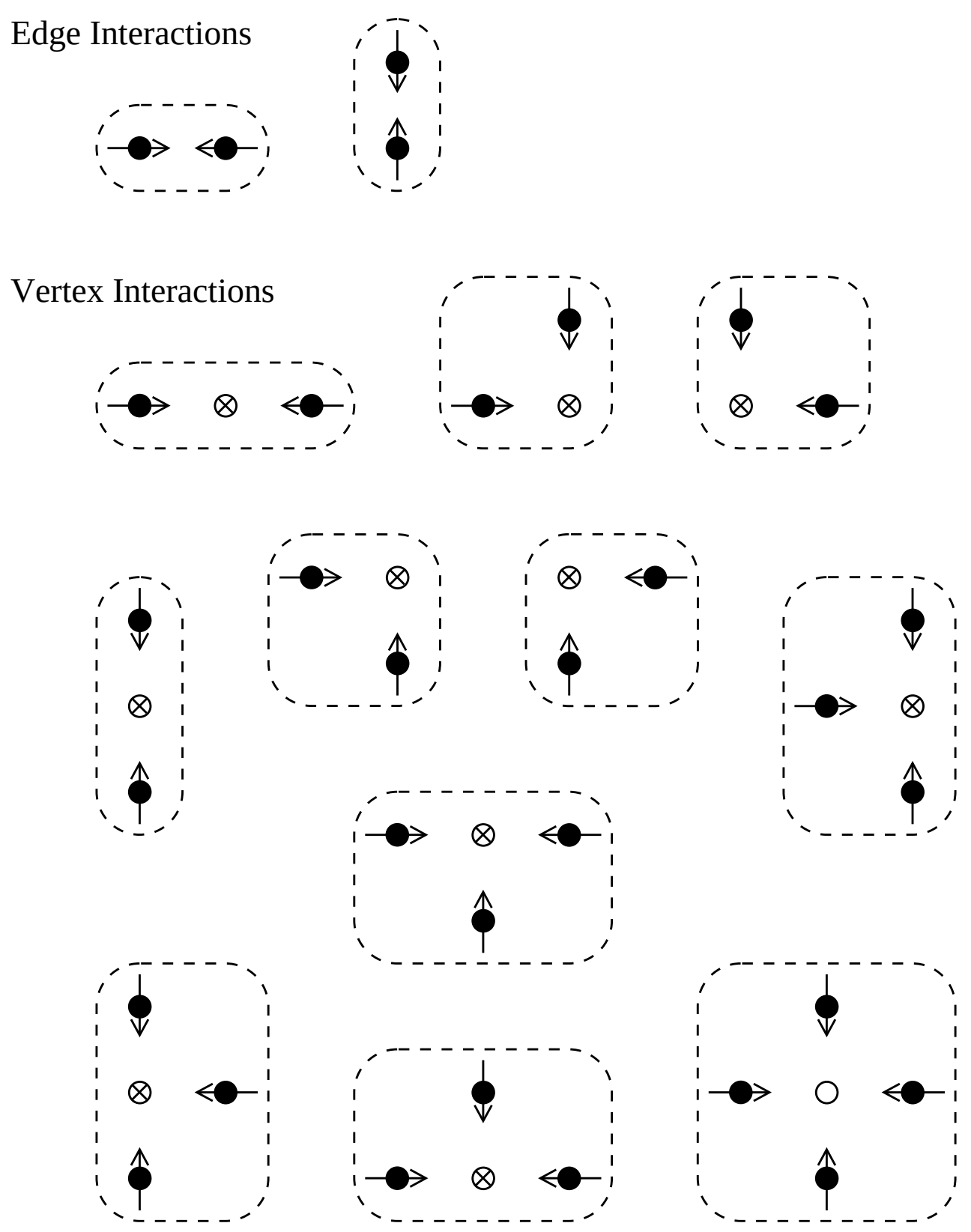

**(B)** Contact interactions in two-dimensional systems with $\mathcal{V} = \{(\pm 1, 0), (0, \pm 1)\}$ existing over square spatial lattice.

**Figure 5.24:** The basic collision configurations for contact interactions among particles in (A) one-dimensional systems with overall discrete velocity set $\mathcal{V} = \{(\pm 1)\}$, and (B) two-dimensional systems existing over square spatial lattice with the overall discrete velocity set $\mathcal{V} = \{(\pm 1, 0), (0, \pm 1)\}$. Each configuration is enclosed in dashed ovals. Various symbols mean: $\bullet$ particle, $\rightarrow\uparrow\downarrow\leftarrow$ direction of motion of particles, $\circ$ empty lattice site, and $\otimes$ lattice sites which are empty or occupied by a particle which does not undergo contact interaction with any particle in the configuration. In each configuration, lattice sites whose state is irrelevant for the interaction are not shown.

From the above it is evident that the total number of BCCs arising out of contact interactions will, in general, be orders of magnitude smaller compared to the total number of states of the contact interaction neighborhood. Because of this it is advantageous to tabulate the BCCs and their possible final states instead of constructing the usual lookup tables based on the states of the entire contact interaction neighborhood. The tables of BCCs and their possible acceptable solutions will henceforth be referred to as *"BCC lookup tables"*. The BCC lookup tables will, in general, contain many isometrically equivalent

BCCs as can be seen from Fig. 5.24. As a result, the size of the BCCs lookup tables can be reduced by keeping only one of such BCCs and its acceptable solutions. The other BCCs and their solutions can be constructed using simple isometric transformations of the one in the table. Thus, for the two-dimensional case shown in Fig. 5.24, the BCC lookup tables need to contain only 5 BCCs and their acceptable solutions. The BCC lookup tables can be used for processing contact interactions among particles during simulations in much the same manner as the usual lookup tables (further details are given in Sec. 5.2.3.3).

The method of construction of interaction rules outlined above is fully logical/mechanical method and does not contain *intuitive* elements requiring abilities beyond those of presently available computing machines. As a result, it can be implemented on digital computers for constructing the interaction rules in the desired form, *e.g.*, as lookup tables based on the state of the entire contact interaction neighborhood, or as BCCs and their solutions, or as algorithms based on any of these. Construction of lookup tables based on the state of the entire contact interaction neighborhood and algorithms based on them is not recommended because of excessive memory requirements, instead, algorithms based on BCCs and their solutions should be constructed and used. In fact, even from the point of view of complexity of algorithms to be employed for using these two types of lookup tables, use of the BCC lookup tables is advantageous, simpler, and more straight forward compared to using the lookup tables based on the state of lattice sites comprising the entire contact interaction neighborhood.

#### 5.2.3.3 Processing Interparticle Interactions

The method of using the BCC lookup tables for processing interparticle interactions (specifically, contact interactions) during simulations is as follows: For each iteration at any time step the spatial lattice is scanned for the presence of BCCs arising from both edge and vertex interactions. In the first iteration for any time step, further processing is carried out if at least one BCC of any of these types is found otherwise the iterations are terminated. In the subsequent iterations, further processing is carried out if at least one BCC arising out of vertex interactions is found otherwise the iterations are terminated.[40) Each one of the BCCs, thus found over the spatial lattice, is processed as follows: The BCC is looked up into the BCC lookup table and its solution is identified. If the BCC has only one solution, then that solution is selected. If, however, the BCC has more than one solution then one of the solutions is chosen in accordance with preselected *"state transition probabilities"* (STPs) (further details on STPs are given in Sec. 5.2.3.4). Following this, states of all the interacting particles comprising the BCC are substituted by the states of corresponding particles in the selected solution. This completes the processing of the BCC.

[40)] Note that it is not necessary that further processing be carried out only if at least one BCC arising out of vertex interactions is found. If desired, processing can be carried out even if BCCs arising out of vertex are not present and at least one BCC arising out of edge interactions is found. The iterations, however, can be terminated only if no BCCs arising out of vertex interactions are found over the spatial lattice. A more strict condition for terminating iterations is to ensure that no BCCs of any type are found over the spatial lattice. This condition is computationally expensive and not required for simulating fluid dynamic systems. It, however, can be used for studying domain formation and the like in dense systems.

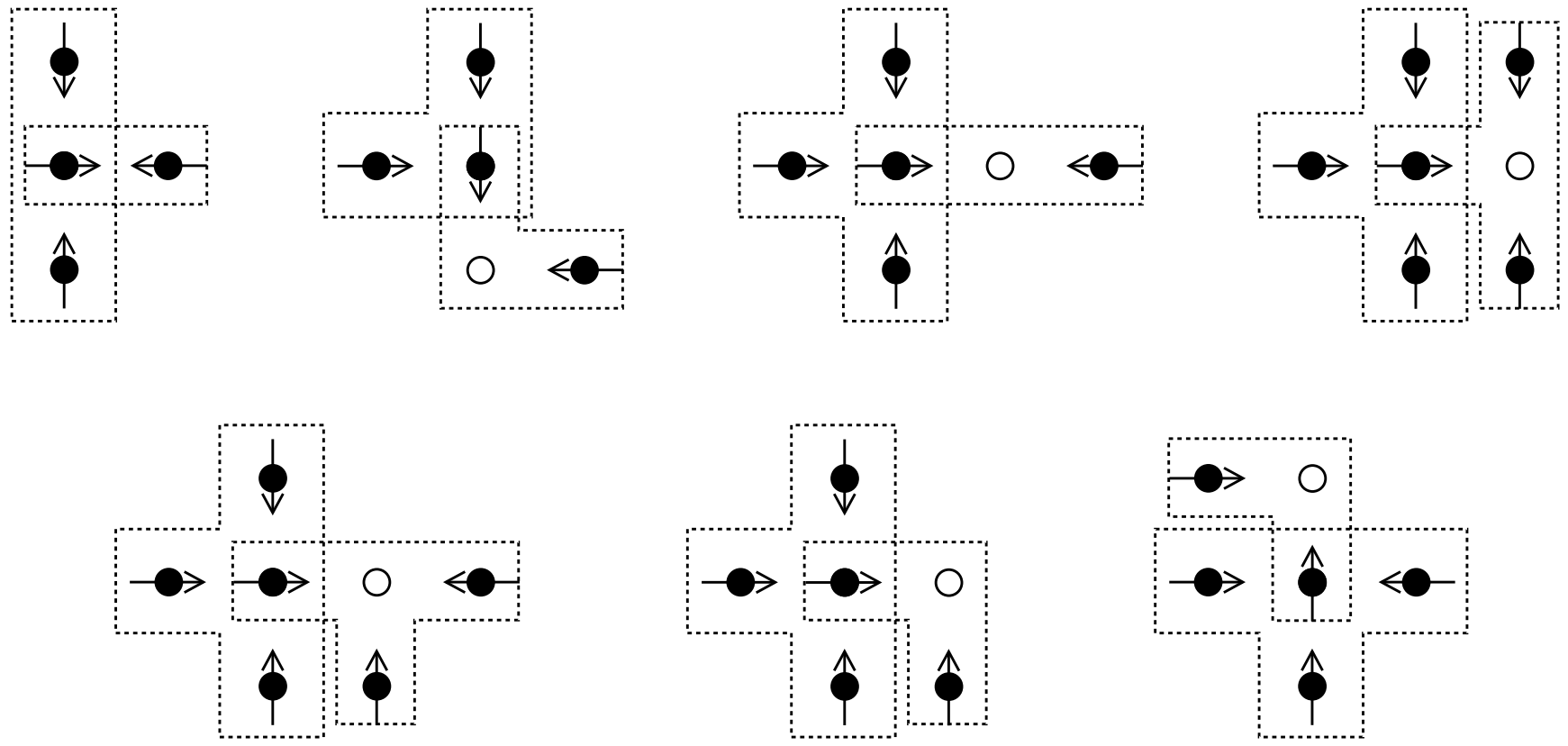

**Figure 5.25:** Examples of overlapping basic collision configurations in two-dimensional systems existing over square spatial lattice with the overall discrete velocity set $\mathcal{V} = \{(\pm 1, 0), (0, \pm 1)\}$. Each configuration is enclosed in dotted boxes. The symbols mean as in Fig. 5.24.

The BCCs and BCC lookup tables have two key features, *viz.*, (i) the BCC occurring over the spatial lattice can overlap[41)] (see Fig. 5.25 for some examples), and (ii) some of the BCCs contained in the BCC lookup tables can be supersets of other BCCs, *i.e.*, some BCCs in the BCC lookup tables can be decomposed in terms of other BCCs enlisted therein (see Fig. 5.26 for an example). These features, unless appropriate precautions are taken, can lead to incorrect identification of BCCs over the spatial lattice and give rise to undesirable problems during simulations. As a result, while searching for the BCC over the spatial lattice care must be taken to ensure that the occurrence of every BCC is identified correctly. The precautions that should be taken are described below.

The possibility of incorrect identification of BCCs due to their decomposability (*viz.*, the point (ii) above) can be eliminated by searching for them in the order of decreasing number of particles comprising them during simulations. This can be done by arranging the BCCs in the order of decreasing number of particles in the lookup tables and checking them in the order of arrangement while searching during simulations. The BCCs with the same number of particles can be arranged consecutively in the tables. The same effect can be achieved by assigning search priorities to the BCCs in the order of decreasing number of particles comprising them and then searching for them in the order of assigned priorities. The BCCs with the same number of particles can be assigned consecutive search priorities.

The possibility of incorrect identification of BCCs due to their overlap over spatial lattice (*viz.*, the point (i) above) can be eliminated checking for possible formation of BCCs for all the particles comprising the system and excluding (from further checking) only those lattice sites which are occupied by interacting particles comprising the BCCs. The lattice sites which are part of the BCCs and occupied by particles which do not take part in the interaction represented by the BCCs (*e.g.*, the lattice sites marked $\otimes$ in Fig. 5.24) should be checked further for formation of BCCs containing them.

---

[41)]This happens because most of the BCCs are not densely packed structures in that they contain lattice sites for which more than one state is permissible for the same state of interacting particles comprising the BCC, *e.g.*, see the BCCs marked containing lattice sites marked with the symbol $\otimes$ in Fig. 5.24.

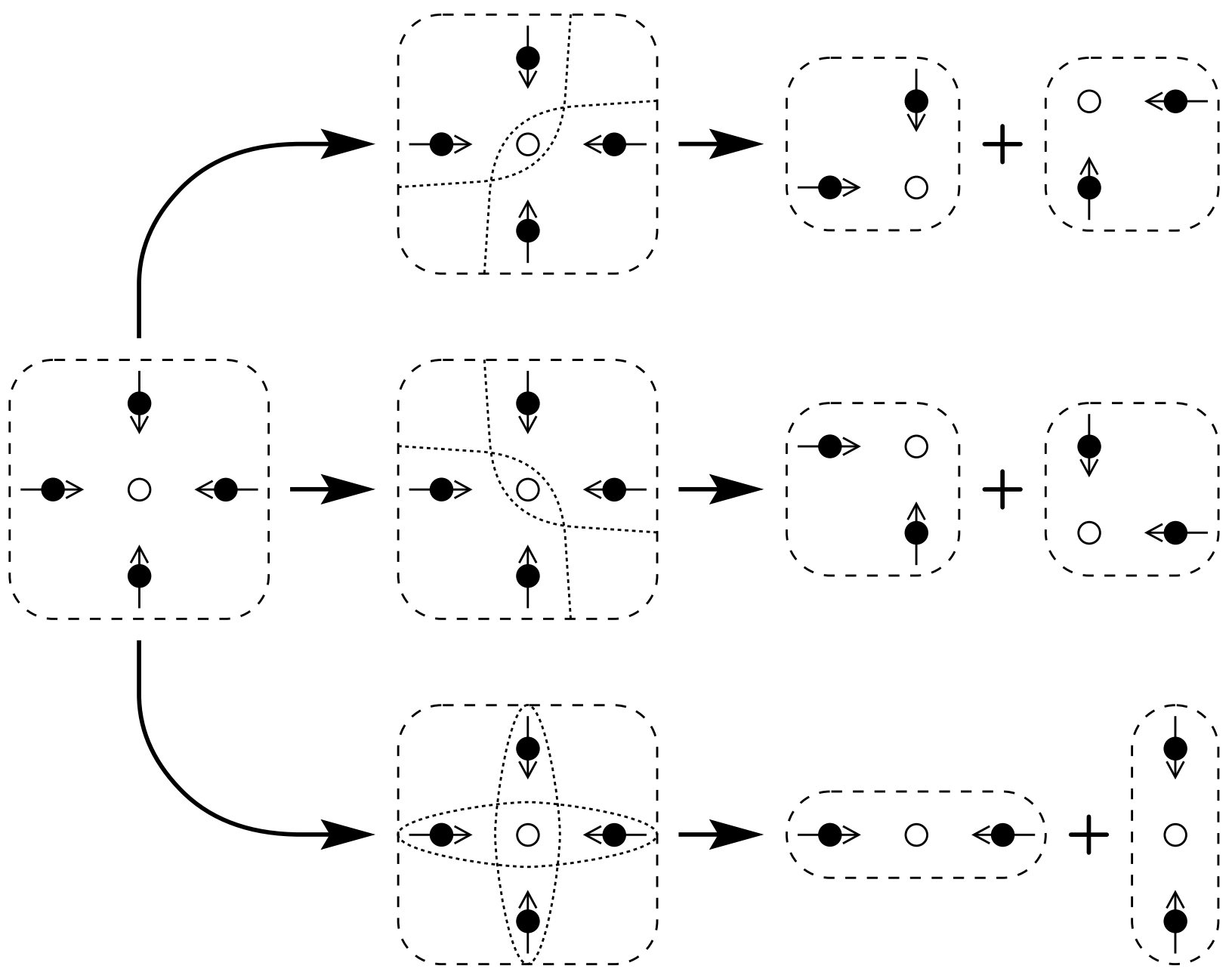

**Figure 5.26:** Various possible ways of decomposing a four-particle BCC into two-particle BCCs in two-dimensional systems existing over square spatial lattice with the overall discrete velocity set $\mathcal{V} = \{(\pm 1, 0), (0, \pm 1)\}$. The BCCs are enclosed in dashed ovals. The decomposition is shown by dotted lines. The symbols mean as in Fig. 5.24.

#### 5.2.3.4 State Transition Probability Table

It is clear from Sec. 5.2.3.2 that BCCs, in general, can have multiple solutions. During simulations, however, only one solution is used as the final state of particles comprising such a BCC. As a result, when a BCC having multiple solutions is identified during search, decision problem regarding solution to be selected arises. This problem is resolved (and can only be resolved) by assigning *probabilities* to all the BCC $\mapsto$ SOLUTION state transition pairs. These probabilities are termed as *"state transition probabilities"* (STPs). The STPs form a table termed as the *"state transition probability table"* (STP table) corresponding to the BCC lookup table. This table is needed during simulations and thus, must be constructed before the simulation starts. The entries in the table can either remain invariant or vary with time (and/or, possibly, lattice site coordinates). In case the table varies, it must be constructed afresh before processing for each evolution (and, possibly, lattice site) starts. In the following, for the sake of simplicity, the STP tables have been taken to be invariant through out the simulations.[42] The values of STPs must satisfy some constraints which arise from symmetry requirements and conservation laws of the system. Within these constraints the STPs can be chosen freely. These constrains are described below.

Let $\mathcal{S}_i$ represent the $i^{\text{th}}$ BCC in a single particle lattice gas, $i = 1, \ldots, \mathcal{N}_{\text{BCC}}$, where $\mathcal{N}_{\text{BCC}}$ is the total number of BCCs. Let the total number of solutions for $\mathcal{S}_i$ be $\mathcal{N}_{\text{sol},i}$. From Sec. 5.2.3.2 it is clear that $\mathcal{N}_{\text{sol},i} \geq 1, \forall\, i \in [1, \mathcal{N}_{\text{BCC}}]$. Let $\mathcal{S}_{i,j}$ be the $j^{\text{th}}$ solution of

[42] In fact, properties of the STP tables which vary with time (and/or, possibly, lattice site coordinates) are extremely difficult to analyze either theoretically or through simulations. As a result, such tables have not been addressed in the present investigation.

$\mathcal{S}_i$ in the BCC lookup table. Let $p_{i,j}$ be the STP for $\mathcal{S}_i \mapsto \mathcal{S}_{i,j}$ state transition. With this notation, the constraint on $p_{i,j}$ for BCCs which cannot be decomposed into other BCCs is

$$\sum_{j=1}^{\mathcal{N}_{\text{sol},i}} p_{i,j} = 1 \tag{5.62}$$

This condition is necessitated by the fact that every BCC which can possibly be found over the spatial lattice must have at least one solution otherwise annihilation of particles will occur causing violation of conservation laws (*c.f.*, Sec. 5.2.3.2). The BCCs arising from edge interactions are their own solutions, *i.e.*, $\mathcal{S}_i \in \{\mathcal{S}_{i,j} : j = 1, \ldots, \mathcal{N}_{\text{sol},i}\}$. As a result, Eq. (5.62) is unconditionally satisfied for all BCC arising from edge interactions.

The constraint on $p_{i,j}$ for BCCs which can be decomposed into other BCCs is

$$\sum_{j=1}^{\mathcal{N}_{\text{sol},i}} p_{i,j} = \mathcal{P}_i \qquad 0 \leq \mathcal{P}_i \leq 1 \tag{5.63}$$

This is because these BCCs can be resolved by decomposing them into other BCCs which, if decomposable and need to be decomposed, finally leads to non-decomposable BCCs for which Eq. (5.62) holds. Thus, the decomposable BCCs, even if $\mathcal{P}_i < 1$, will always have a solution. The number of state transitions for such BCCs in the case $\mathcal{P}_i < 1$ may or may not be different from that in the case $\mathcal{P}_i = 1$. Whether or not the number of state transitions for such BCCs will be different under these two conditions is dictated by the exact details of the BCC lookup table and the corresponding STP table, *i.e.*, the values of $p_{i,j}$.

In addition to the constraints outlined above, one more constraint arises from consideration of the law of conservation of angular momentum which requires that the angular momentum of particles undergoing contact interactions must remain conserved. This is because during contact interactions the particles behave as rigid bodies and the system constituted by them, in effect, remains isolated from other particles. Thus, the angular momentum of a BCC and its solutions should be same. The acceptable solutions of BCCs contained in the BCC lookup tables, however, are obtained by solving the Eqs. (5.58)–(5.60). Many of these solutions may not have the same angular momentum as the corresponding BCCs. As a result, the law of conservation of angular momentum will be violated during simulations. This violation can be overcome in two ways, *viz.*, (i) by adding an equation for conservation of angular momentum, which, using the notation of Sec. 5.2.3.2, is

$$\sum_{j=1}^{\mathcal{N}_{\text{PC}}} m_{i_j}^{(\tau+\delta_{k+1})} \left[\boldsymbol{x}_{i_j}^{(\tau+\delta_{k+1})} \times \boldsymbol{v}_{i_j}^{(\tau+\delta_{k+1})}\right] = \sum_{j=1}^{\mathcal{N}_{\text{PC}}} m_{i_j}^{(\tau+\delta_k)} \left[\boldsymbol{x}_{i_j}^{(\tau+\delta_k)} \times \boldsymbol{v}_{i_j}^{(\tau+\delta_k)}\right] \tag{5.64}$$

to the equation set given by Eqs. (5.58)–(5.60) and then solving this new system of equations for obtaining acceptable solutions of the BCCs, and (ii) by assigning STPs in such a way that the angular momentum remains conserved statistically (over the spatial lattice during each simulation as well as in an ensemble).

For one-dimensional systems, Eq. (5.64) becomes irrelevant because $\boldsymbol{x}_{i_j} \times \boldsymbol{v}_{i_j}$ is always zero. For $\mathcal{D}$-dimensional systems, $\mathcal{D} > 1$, use of Eq. (5.64) along with the Eqs. (5.58)–(5.60) for obtaining acceptable solutions of BCCs, in essence, reduces the number of acceptable solutions (compared to those obtained by solving only the Eqs. (5.58)–(5.60)). In fact, the new system formed by the Eqs. (5.58)–(5.60) and (5.64) might not have any acceptable solution for many BCCs. In the cases in which this happens (which will be many, especially for multispecies systems), physically consistent interaction rules cannot be constructed. As a result, it is not advisable to add the equation for conservation of angular momentum to the system of equations to be used for obtaining acceptable solutions for BCCs. Instead, the second method pointed out above, *viz.*, ensuring statistical conservation of angular momentum by appropriate selection of the STPs, should be used. The justification of this method is based on the fact that in single particle lattice gas simulations the properties of interest have to be sampled either as ensemble averages (or, as time averages). As a result, statistical conservation of angular momentum in the ensemble (or, in the time domain used for averaging) is sufficient to ensure physical consistency. The method of selecting the STPs to ensure statistical conservation of angular momentum is as described below.

Consider, for the sake of simplicity, a non-reacting system. Let the angular momentum of a BCC $\mathcal{S}_i$ be $L_i$ and that of its solutions $\mathcal{S}_{i,j}$ be $L_{i,j}$, $j = 1, \ldots, \mathcal{N}_{\mathrm{sol},i}$. Let $\mathcal{S}_i$ consist of $\mathcal{N}_{\mathrm{PC},i}$ particles of mass $m_{i,k}$, moving with velocities $\boldsymbol{v}_{i,k}$, and located at lattice sites $\boldsymbol{x}_{i,k}$, $k = 1, \ldots, \mathcal{N}_{\mathrm{PC},i}$. Let the velocities of the particles in the solution $\mathcal{S}_{i,j}$ of the BCC be $\boldsymbol{v}_{i,j,k}$, $k = 1, \ldots, \mathcal{N}_{\mathrm{PC},i}$. Then $L_i$ and $L_{i,j}$ are given by

$$L_i = \sum_{k=1}^{\mathcal{N}_{\mathrm{PC},i}} m_{i,k} \boldsymbol{x}_{i,k} \times \boldsymbol{v}_{i,k} \tag{5.65}$$

$$L_{i,j} = \sum_{k=1}^{\mathcal{N}_{\mathrm{PC},i}} m_{i,k} \boldsymbol{x}_{i,k} \times \boldsymbol{v}_{i,j,k} \tag{5.66}$$

In general, $L_i$ and $L_{i,j}$ may be different for the same $i$. The conservation of angular momentum, however, may be ensured by choosing $p_{i,j}$'s (the STPS) in a way such that

$$L_i = \sum_{j=1}^{\mathcal{N}_{\mathrm{PC},i}} p_{i,j} L_{i,j} \tag{5.67}$$

To ensure that STPs satisfying this condition exist, it is sufficient to ensure (in single species single particle lattice gases) that the discrete velocity set contains velocity vectors in pairs of the form $\pm\boldsymbol{v}$. General conditions for multispecies multiparticle lattice gases can also be derived; this, being irrelevant to the concepts presented herein, will not be attempted here.

On symmetric spatial lattices and for symmetric discrete velocity sets many isometrically equivalent BCCs occur over the spatial lattice (*e.g.*, see two-dimensional case in Fig. 5.24). In such cases, for single species systems, STPs can always be chosen such that Eq. (5.67) is satisfied. For multispecies systems, however, problems might arise in choosing STPs satisfying Eq. (5.67). In these cases, STPs can be chosen such that the condition

$$\sum_{\mathcal{I}[\mathcal{S}_i]} L_i = \sum_{\mathcal{I}[\mathcal{S}_i]} \sum_{j=1}^{\mathcal{N}_{\mathrm{PC},i}} p_{i,j} L_{i,j} \tag{5.68}$$

is satisfied, where the subscript $\mathcal{I}[\mathcal{S}_i]$ in summations means that summations must be carried out over all BCCs which are isometrically equivalent to $\mathcal{S}_i$. This condition arises from the fact that all the isometrically equivalent BCCs have same probability of occurring any where over the spatial lattice. This condition, unlike the condition given by Eq. (5.67), can be satisfied for any and all systems existing over symmetric spatial lattices with symmetric discrete velocity sets (for particles of each species). It is evident that the condition given by Eq. (5.67) is stronger compared to that given by Eq. (5.68) and that STPs satisfying the former condition will necessarily satisfy the later condition also.

**Example:** Within the constraints outlined above, the STPs can be selected in a very simple and straight forward manner if all $\mathcal{S}_i \mapsto \mathcal{S}_{i,j}$ state transitions are permitted and all $\mathcal{P}_i$ are taken to be unity and the discrete velocity set (for particles of each species) has appropriate symmetries. For this, note that all the solutions of various BCCs (having unique or multiple solutions) contained in the BCC lookup tables are acceptable and valid within the constraints imposed by the conservation laws. This, in the absence of any other constraint, implies that each solution of a BCC should be taken to be equally likely, *i.e.*,

$$p_{i,j} = p_{i,k} \qquad \forall\, j,k \in [1, \mathcal{N}_{\mathrm{sol},i}] \tag{5.69}$$

This gives the STPs as

$$p_{i,j} = \frac{1}{\mathcal{N}_{\mathrm{sol},i}} \qquad \forall\, j \in [1, \mathcal{N}_{\mathrm{sol},i}] \tag{5.70}$$

for all the BCCs, *i.e.*, for all $i \in [1, \mathcal{N}_{\mathrm{BCC}}]$. These values of $p_{i,j}$ satisfy the condition given by Eqs. (5.62) and (5.67) (and thus, also the condition given by Eq. (5.68)).

### 5.2.3.5 Remarks on Usage of Interaction Rules

The method of construction of interaction rules for single particle lattice gases described in Sec. 5.2.3.2 (with additional details furnished in Secs. 5.2.3.3 and 5.2.3.4) is based only on processing of contact interactions among particles; field interactions among particles are ignored in this processing. Since contact interactions occur only among particles having rigid core (of finite and non-zero dimensions), the interaction rules constructed using this method are applicable only for simulation of dynamics of rigid particle systems. Since the discrete velocity set contains finitely many elements[43)] and the interaction rules contain probabilistic elements, the microdynamics observed in these simulations will not be a good replica of the microdynamics of an identical system of particles existing in continuum space.[44)] As a result, in single particle lattice gas simulations, properties of interest must be sampled as ensemble averages or as time averages. For sampling of macroscopic properties (*e.g.*, pressure, temperature, density, *etc.*), this averaging is appropriate and desired because macroscopic properties are defined as ensemble/time averages of microscopic properties. Thus, in single particle lattice gas simulations, macroscopic property $\xi_i^{(\tau)}$ should be sampled at the location $\boldsymbol{x}_i$ at time step $t = \tau\Delta t$ using ensemble averaging as $\left\langle \xi_i^{(\tau)} \right\rangle$ in large enough ensemble or using time averaging as $\bar{\xi}_{\boldsymbol{i}}^{(\tau)}$ in large enough time domain.

For ideal ensemble averaging sampling should be carried out as

$$\left\langle \xi_i^{(\tau)} \right\rangle = \lim_{\mathcal{N}_{\mathrm{ens}}\to\infty} \frac{1}{\mathcal{N}_{\mathrm{ens}}} \sum_{k=1}^{\mathcal{N}_{\mathrm{ens}}} \left[\xi_i^{(\tau)}\right]^k \tag{5.71}$$

43) This is in addition to the discreteness of space and time.

44) Irrespective of whether it is evolving in discrete time or continuous time.

where $\mathcal{N}_{\text{ens}}$ is number of realizations of the system contained in the ensemble, and $\left[\xi_i^{(\tau)}\right]^k$ is the value sampled at the time step $\tau$ in the $k^{\text{th}}$ realization of the system, $k = 1, \ldots, \mathcal{N}_{\text{ens}}$. In actual simulations, the limit is approximated by taking a finite but large value for $\mathcal{N}_{\text{ens}}$. In the above equation, the method of sampling of $\left[\xi_i^{(\tau)}\right]^k$ is irrelevant; it may be sampled at $\boldsymbol{x}_i$ or in some appropriate domain around $\boldsymbol{x}_i$ as required.

For ideal time averaging sampling should be carried out as

$$\bar{\xi}_i^{(\tau)} = \lim_{\mathcal{N}_\tau \to \infty} \frac{1}{\mathcal{N}_\tau} \sum_{k=1}^{\mathcal{N}_\tau} \xi_i^{(\tau-k)} \tag{5.72}$$

where $\mathcal{N}_\tau$ is number of time steps over which averaging is to be carried out, and $\xi_i^{(\tau-k)}$ is the value sampled at the time step $\tau - k$, $k = 1, \ldots, \mathcal{N}_\tau$. In actual simulations, the limit is approximated by taking a finite but large value for $\mathcal{N}_\tau$. In the above equation, the method of sampling of $\xi_i^{(\tau-k)}$ is irrelevant; it may be sampled at $\boldsymbol{x}_i$ or in some appropriate domain around $\boldsymbol{x}_i$ as required.

Since the effect of field interactions is not incorporated in the interaction rules constructed using the method described in Sec. 5.2.3.2, doubts might arise regarding the choice of transport properties. This is because they depend upon interaction potentials [8]. In this regard, it should be noted that dependence of transport properties on interaction potentials arises from their dependence on scattering of particles during interparticle interactions which occurs differently for different interaction potentials. In the interaction rules constructed as described in Sec. 5.2.3.2, the scattering of particles (and thus, their nature and the nature of interaction among them) can be controlled as desired by appropriately choosing the STPs for various edge and vertex interactions. Since the STPs vary continuously in $[0, 1]$, a very precise fine tuning of transport properties is possible by fine tuning the STPs.

The probabilistic nature of interaction rules gives rise to another doubt which relates to sampling of particle trajectories. This is because a single simulation of a given initial state of a system, being probabilistic, cannot (and will not) reproduce the trajectories of particles which conform to those of an equivalent system existing in continuum space.[44)] This, however, does not imply that trajectories of particles cannot sampled from single particle lattice gas simulations. The trajectories of particles can be sampled as ensemble average of large number of simulations carried out using the same initial state of the system with the same set of STPs. Let $\boldsymbol{y}_i^{(\tau)}$ be the location of particle $i$ at time step $\tau$. Then, the ideal trajectory of particle $i$ can be sampled as $\left\langle \boldsymbol{y}_i^{(\tau)} \right\rangle$ using

$$\left\langle \boldsymbol{y}_i^{(\tau)} \right\rangle = \lim_{\mathcal{N}_{\text{sim}} \to \infty} \frac{1}{\mathcal{N}_{\text{sim}}} \sum_{k=1}^{\mathcal{N}_{\text{sim}}} \left[\boldsymbol{y}_i^{(\tau)}\right]^k \tag{5.73}$$

where $\mathcal{N}_{\text{sim}}$ is number of simulations carried out using the same initial state of the system, and $\left[\boldsymbol{y}_i^{(\tau)}\right]^k$ is the location of particle $i$ at the time step $\tau$ in the $k^{\text{th}}$ simulation of the system, $k = 1, \ldots, \mathcal{N}_{\text{rep}}$. In actual simulations, the limit is approximated by taking a finite but large value for $\mathcal{N}_{\text{sim}}$. For sampling the trajectory of particles it is important that the initial state of the system should not be changed. The STPs, as they govern the nature of particles and the nature of interparticle interactions, should also not be changed. The assumption underlying this procedure is that $\left\langle \boldsymbol{y}_i^{(\tau)} \right\rangle \to \left[\boldsymbol{y}_i^{(\tau)}\right]_{\text{C}}$ in the limit $\mathcal{N}_{\text{sim}} \to \infty$, where $\left[\boldsymbol{y}_i^{(\tau)}\right]_{\text{C}}$ is the trajectory of particle $i$ when the system with identical initial state exists in continuum space and evolves in discrete time steps of duration $\Delta t$.

A key point in the elaborations furnished above, that makes single particle lattice gases capable of simulating systems with desired transport coefficients or desired interaction potentials, is that in single particle lattice gases the nature of particles (and thus, the nature of interactions among them) can be controlled by appropriately choosing the STPs. This implies that STPs are linked to interaction potentials. The method choosing STPs for desired interaction potentials or equivalently, the method of incorporating the effect of field interactions into interaction rules constructed using the procedure outlined in Sec. 5.2.3.2 is described in the following section (Sec. 5.2.3.6).

#### 5.2.3.6 Incorporating Field Interactions

It is evident from Sec. 5.2.3.1 that in single particle lattice gases $\boldsymbol{v}_i^{(\tau+\delta)} = \boldsymbol{v}_{i,\text{cont}}^{(\tau+\delta)}$ is, in general, not possible. In fact, in general, $\boldsymbol{e}_i^{(\tau+\delta)} = \boldsymbol{e}_{i,\text{cont}}^{(\tau+\delta)}$ is also not possible, where $\boldsymbol{e}_\alpha^{(\beta)}$

$$\boldsymbol{e}_\alpha^{(\beta)} = \frac{\boldsymbol{v}_\alpha^{(\beta)}}{\left|\boldsymbol{v}_\alpha^{(\beta)}\right|} \tag{5.74}$$

is *unit vector* in the direction of $\boldsymbol{v}_\alpha^{(\beta)}$. As a result, the influence of field interactions cannot be incorporated exactly into interaction rules of single particle lattice gases. Alternatively, the trajectory of particles obtained in single particle lattice gases simulations cannot be made to correspond exactly to the *exact trajectory* (*i.e.*, trajectory observed in an identical system existing in continuum space and evolving in discrete time steps of duration $\Delta t$). A method, however, can be devised wherein the trajectory obtain in single particle lattice gas simulations follows the exact trajectory closely. The basic insight underlying this method is that in single particle lattice gas simulations the scattering of particles during interparticle interactions depends upon the STPs. Thus, by appropriately selecting the STPs, the scattering of particles can be controlled in such a way that it becomes equivalent to that of the desired interaction potential. The details of the method are as described below.

For incorporating the effect of field interactions into interaction rules constructed using the method described in Sec. 5.2.3.2, it is necessary to compute $\boldsymbol{v}_{i,\text{cont}}^{(\tau+\delta)}$ for all the particles comprising the system. This computation is carried out as described in Sec. 5.2.3.1, *i.e.*, by considering the field interactions of each particle with all the particles lying in its overall interaction neighborhood, with the difference that $\Delta t_i$ is taken to be very small compared to $\Delta t$ (say, at least two or three orders of magnitude smaller) in order to stay within the assumptions outlined in Sec. 5.1.1.2.[45] With this, the *unit vector* $\boldsymbol{e}_{i,\text{cont}}^{(\tau+\delta)}$ in the direction of $\boldsymbol{v}_{i,\text{cont}}^{(\tau+\delta)}$ can be computed. This unit vector gives the direction in which particle will move following interactions if the system were existing in continuum position space. In single particle lattice gases, in general, particles cannot be made to move exactly along this direction. Instead, they can be made to move close to this direction. In fact, their

[45] It, however, is not necessary that these constraints should be satisfied. If desired, one can take $\Delta t_i$ to be of the same order of magnitude as $\Delta t$ or even equal to $\Delta t$. If such a choice is made, considerations and assumptions outlined in Sec. 5.1.1.2 will no longer be applicable; in fact, they become subset of the new considerations and assumptions that will come up.

motion can be controlled in such a way that the ensemble averaged value of their direction of motion is identical to that of $\boldsymbol{e}_{i,\text{cont}}^{(\tau+\delta)}$. This is done as follows:

For processing a BCC $\mathcal{S}_n$ comprising of $\mathcal{N}_{\text{PC}}$ particles, the units vectors $\boldsymbol{e}_{n,i_j,\text{cont}}^{(\tau+\delta)}$, $i_j = 1, \ldots, \mathcal{N}_{\text{PC}}$, are computed. Let there be $\mathcal{N}_{\text{sol},n}$ solutions $\mathcal{S}_{n,k}$, $k = 1, \ldots, \mathcal{N}_{\text{sol},n}$ for this BCC. Let the velocities of particles in these solutions be $\boldsymbol{v}_{n,k,i_j}$. Each of these velocity vectors depart by an angle $\theta_{n,k,i_j}$ from $\boldsymbol{e}_{n,i_j,\text{cont}}^{(\tau+\delta)}$, where

$$\cos\theta_{n,k,i_j} = \frac{\sum_{l=1}^{\mathcal{D}} \left(\boldsymbol{e}_{n,i_j,\text{cont}}^{(\tau+\delta)}\right)_l \left(\boldsymbol{v}_{n,k,i_j}\right)_l}{\left|\boldsymbol{v}_{n,k,i_j}\right|} \tag{5.75}$$

From this equation, the angular departures are computed in the closed interval $[-\pi, \pi]$. With this, the sum $\Theta_{n,k}$ of the absolute values of these angular departures for all the particles comprising a BCC are computed as

$$\Theta_{n,k} = \sum_{j=1}^{\mathcal{N}_{\text{PC}}} \left|\theta_{n,k,i_j}\right| \tag{5.76}$$

The STP $p_{n,k}$ for $\mathcal{S}_n \mapsto \mathcal{S}_{n,k}$ state transitions can, now, be decided as

$$p_{n,k} = f_{n,k} \tag{5.77}$$

where $f_{n,k}$, $f_{n,k} \geq 0$, is monotonic function of $\Theta_{n,k}$ which decays with increase in $\Theta_{n,k}$ and it is such that $p_{n,k}$ satisfies the condition given by Eq. (5.62) or by Eq. (5.63), which ever may be appropriate. The $f_{n,k}$ can be chosen freely. Other constraints on $f_{n,k}$ are identical to and rise from those on $p_{n,k}$. It should also be noted that the STPs obtained using the above procedure will, in general, not satisfy the condition given by Eq. (5.67). They, however, will satisfy the condition given by Eq. (5.68) if the discrete velocity set has appropriate symmetries pointed out in Sec. 5.2.3.4.

In the method outlined above, the unit vector $\boldsymbol{e}_{i,\text{cont}}^{(\tau+\delta)}$ depends upon the state of all the lattice sites comprising the overall interaction neighborhood of particle located at the lattice site $\boldsymbol{x}_i$ at the time step $\tau$. As a result, a lookup table which gives $\boldsymbol{e}_{i,\text{cont}}^{(\tau+\delta)}$ for all the possible states of the overall interaction neighborhood can be constructed. This lookup table, for an overall interaction neighborhood comprising of $\mathcal{N}_{\mathcal{L}_{\text{O}}}$ lattice sites with $\mathcal{N}_{\text{s}}$ states for each lattice site, will contain $\mathcal{N}_{\text{etab}} = \mathcal{N}_{\text{s}}^{\mathcal{N}_{\mathcal{L}_{\text{O}}}}$ elements. The value of $\mathcal{N}_{\mathcal{L}_{\text{O}}}$ is, in general, large and $\mathcal{N}_{\text{etab}}$ increases exponentially with the size of the overall interaction neighborhood. As a result, the memory required for storing the lookup table is, usually, overwhelmingly large. In view of this, constructing and storing the lookup table at the start of simulations is not recommended. Instead, it is more appropriate and advantageous to compute the unit vector $\boldsymbol{e}_{i,\text{cont}}^{(\tau+\delta)}$ online as and when required as done in the procedure outlined above. When this online procedure is adopted, it appears that STP tables appear to be dependent on the lattice site coordinates. This apparent dependence is only a false dependence which arises because of contraction of the overall interaction neighborhood of the particles undergoing contact interactions to the BCC itself. If the BCCs and their solutions are enlisted by expanding the entire combined overall interaction neighborhood of particles undergoing contact interactions, the STPs will not show dependence on lattice site coordinates.

Use of the procedure outlined above for incorporating desired interaction potentials in single particle lattice gases makes the simulations computationally very expensive both in terms of storage memory and computation time requirements. Still, this is the only feasible procedure by which desired interaction potentials can be incorporated into single particle lattice gases. This procedure, being computationally expensive, should preferably be used only when microscopic properties, *e.g.*, trajectory of particles, need to be sampled. For sampling macroscopic properties, especially in the hydrodynamic limit and for system under equilibrium or departing only slightly from equilibrium, this procedure is not needed. Under such conditions, the procedure outlined in Secs. 5.2.3.2–5.2.3.4 is sufficient.

## 5.3 Conclusions

This chapter relates to various aspects of construction of lattice gases within the constraints of the single particle exclusion principle. These lattice gases are termed as *"single particle lattice gases"*. Some important conclusions of this chapter are the following:

**1)** Single particle lattice gases are similar to multiparticle lattice gases only in the sense that the evolution of the system during one time step, in both, is decomposed into two sub-steps, *viz.*, (i) particle translation, and (ii) interparticle interactions. The important difference between the two is that the method of processing interparticle interactions in multiparticle lattice gases is a single step procedure whereas in single particle lattice gases it is an iterative (or, multistep) procedure.

**2)** The method of construction of single particle lattice gases existing in the limit of negligible collision time (negligible compared to time step) has been discovered and outlined. The method of construction of single particle lattice gases differs drastically from that of multiparticle lattice gases because in this method several considerations, in addition to those in multiparticle lattice gases, are required. These additional considerations surface because of enforcement of the single particle exclusion principle.

**3)** The method of construction of single particle lattice gases, though quite involved, is fully algorithmic and can be programmed on conventional digital computers for constructing desired type of single particle lattice gases.

**4)** Because of enforcement of single particle exclusion principle, in these lattice gases desired interaction potentials can be associated with particles. As a result, single particle lattice gases are fully discrete analogs of molecular dynamics based on the formalism of cellular automata. These lattice gases exist in the limit of collision time being negligible compared to the time step.

The overall conclusion is that a systematic and algorithmic method for construction of single particle lattice gases has been conceived and formalized. All the relevant details are outlined and discussed in this chapter. The details furnished in this chapter indicate that the space of single particle lattice gases existing in the limit of collision time being negligible compared to the time step cannot be empty. This can be proved only through construction of a single particle lattice gas which will be done in chapter 6.

# Chapter 6

# Construction of an Example Single Particle Lattice Gas over Square Spatial Lattice

. . . and you must observe those inside blind alleys.
*Yes, sire.*
They refuse to believe the truth asking for its cause. Then they ask for the way out and many doubts follow.
*Yes, sire.*
To tell them how to make way is not enough.
*Yes, sire. Why so, sire?*
If they could follow procedures, would they be in?
. . .

In this chapter the method of construction of single particle lattice gases formalized in chapter 5 will be demonstrated by constructing the evolution rules for a single species single velocity single particle lattice gas over square spatial lattice. This construction, in addition to giving a concrete and usable single particle lattice gas, will also serve to prove that the space of single particle lattice gases described in chapter 5 is not empty.

## 6.1 Definition of the Example Lattice Gas: The SPLG-1 Model

The example lattice gas to be constructed in the following sections is a single species single speed single particle lattice gas which exists over an underlying two-dimensional square spatial lattice. In this lattice gas all the lattice sites are identical and can either be empty or occupied by exactly one particle, irrespective of the state of the particle,[1)] at any time step (the *"single particle exclusion principle"*). All particles are identical and move over the spatial lattice with the same speed. The velocities of the particles belong to the discrete velocity set $\mathcal{V} \equiv \{(\pm 1, 0), (0, \pm 1)\}$ containing four discrete velocity vectors (*i.e.*, $\mathcal{N}_v = 4$). During their motion over the spatial lattice, the particles interact with each other (and with solid boundaries, if any) and are scattered according to predefined interaction rules. The interactions among particles are instantaneous. The dynamics of particles during each time step is decomposed into two sequential sub-steps, *viz.*, (i) interparticle interaction step, and (ii) particle translation step, in the same sequence. Thus, during simulations, processing for evolving the system of particles during each time

[1)] This means that velocities (in general, states) of particles are not correlated with lattice sites coordinates, time step, or states of neighboring particles. This is a necessary necessary defining constraint in single particle lattice gases and a part of the *"single particle exclusion principle"* (*c.f.*, Sec. 4.1.3).

step is carried out in two sub-steps, *viz.*, (i) processing of interparticle interactions, which is followed by (ii) processing of particle translation. The rules for processing to be carried out in these two sub-steps are described in Secs. 6.2 and 6.3. This lattice gas, being the first single particle lattice gas, will henceforth be referred to as the *"SPLG-1 Model"* or the *"SPLGA-1 Model"*.

## 6.2 Translation Rules for the SPLG-1 Model

Since all particles move with unit speed in the SPLG-1 model, the translation rules simply reposition the particles on the nearest neighbor lattice sites pointed out by the velocity vectors of the particles. During translation of particles, the boundary conditions have to be taken care of. Three different types of boundary conditions are possible, *viz.*, (i) free or open boundaries, (ii) toroidal or periodic boundaries, and (iii) solid boundaries.

On free or open boundaries the density and velocity distributions of particles must be specified. Particles positioned on the lattice site adjacent to these boundaries, if moving out of the boundary, vanish from the system. Depending upon the boundary conditions specified, new particles are generated on lattice sites adjacent to these boundaries.

On toroidal or periodic boundaries, particles are neither generated nor vanish from the system. Instead, if they be moving out of the boundary, they are repositioned on the lattice site next to the corresponding opposite boundary. Thus, these boundary conditions always occur in pairs. Consider a rectangular system comprising of $(N_x, N_y)$ lattice sites along the $x$- and $y$-axes with toroidal boundary conditions along all the four sides. Let the origin $(0, 0)$ of the system be located at the lower-left corner as shown in Fig. 6.1. Then, particles positioned at boundary lattice sites, if moving out of the boundaries, are repositioned as

$$\begin{aligned}
(0, y) &\mapsto (N_x - 1, y) && \text{if velocity is } (-1, 0) \\
(N_x - 1, y) &\mapsto (0, y) && \text{if velocity is } (1, 0) \\
(x, 0) &\mapsto (x, N_y - 1) && \text{if velocity is } (0, -1) \\
(x, N_y - 1) &\mapsto (x, 0) && \text{if velocity is } (0, 1)
\end{aligned}$$

On lattice sites near solid boundaries no special processing needs to be carried out because particles positioned on these lattice sites will always be moving into the system; either away from the boundary or along the boundary depending upon fluid-particle–boundary-particle interaction rules. Fluid particles moving into solid boundaries, or interacting with solid boundaries, will not be found in the system during particle translation step because all the interactions are processed during interparticle interaction step.

## 6.3 Interaction Rules for the SPLG-1 Model

The interaction rules of SPLG-1 model can be constructed either by considering the effect of only contact interactions or by considering the effect of both contact and field interactions. If only contact interactions are considered, the construction of interaction rules involves tabulation of all possible BCCs and their solutions. The STPs are selected at the time of simulations depending upon the simulation requirements. If the effect of both contact and field interactions is considered, then also the interaction rules need to be developed by tabulating the BCCs and their solutions; following which the effect of field interactions for desired interaction potentials can be incorporated via STPs. In this case, if the interaction rules are to be used in the form of compressed BCC lookup tables, the STPs have to determined at the time of simulations for the desired interaction potentials with appropriate range of interactions (so that contact interaction neighborhood gets subsumed with in field interaction neighborhood). Instead, if expanded BCC lookup tables are used the STPs can be determined before starting the simulations.

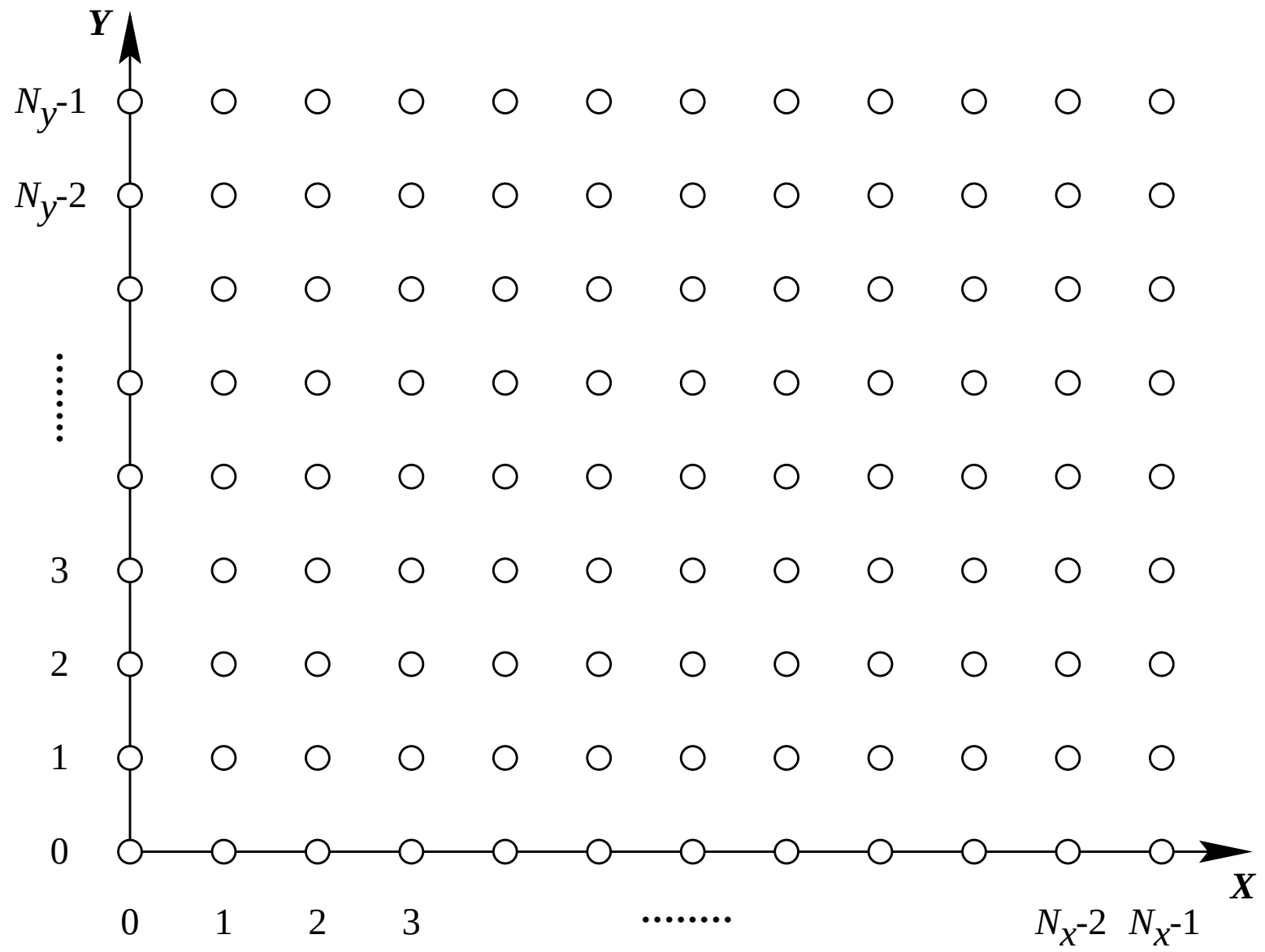

**Figure 6.1:** The coordinate system and convention of numbering lattice sites in two-dimensional rectangular systems.

The construction of interaction rules by considering only the contact interactions, *i.e.*, construction of compressed BCC lookup tables (or simply, the BCC lookup tables), is described in Sec. 6.3.1. The interaction rules based only on contact interactions are sufficient for simulation of systems in the hydrodynamic limit since the exact particles trajectories are not important in this limit. As a result, selection of STPs by considering the interaction potentials will not demonstrated; some comments, however, shall be made on construction of complete BCC lookup tables incorporating the effect of field interactions in Sec. 6.3.2.

### 6.3.1 Construction of BCC Lookup Tables for the SPLG-1 Model

The contact interaction neighborhood for the SPLG-1 model is shown in Fig. 6.2. This contact interaction neighborhood is for particles irrespective of their velocity. It is obtained by superimposing the contact interaction neighborhoods for particles of different velocities. The contact interaction neighborhood for particles of a given velocity is smaller compared to this. The contact interaction neighborhoods for particles of different velocities in the SPLG-1 model are shown in Fig. 6.3. One aspect that is evident from this figure is that the topology of contact interaction neighborhoods for particles of different velocities is identical for all velocity vectors belonging to the discrete velocity. That this must be so is evident from the fact that the largest and most complex contact interaction (which is a vertex interaction) involves $\mathcal{N}_{\mathrm{v}}$ particles posited on appropriate lattice sites relative to each other; each particle in this contact interaction moves with a unique velocity vector belonging to the discrete velocity set. Since this contact interaction involves particles of all velocities, its topology is necessarily contained in the contact interaction neighborhood for particles of any velocity belonging to the discrete velocity set. Furthermore, since this contact interaction is the largest possible contact interaction, its topology dictates and normalizes the topology of the contact interaction neighborhood for particles of any velocity belonging to the discrete velocity set. This contact interaction neighborhood will, henceforth, be referred to as the *"reduced contact interaction neighborhood"* and denoted by $\mathcal{L}_{\mathrm{R}}$.

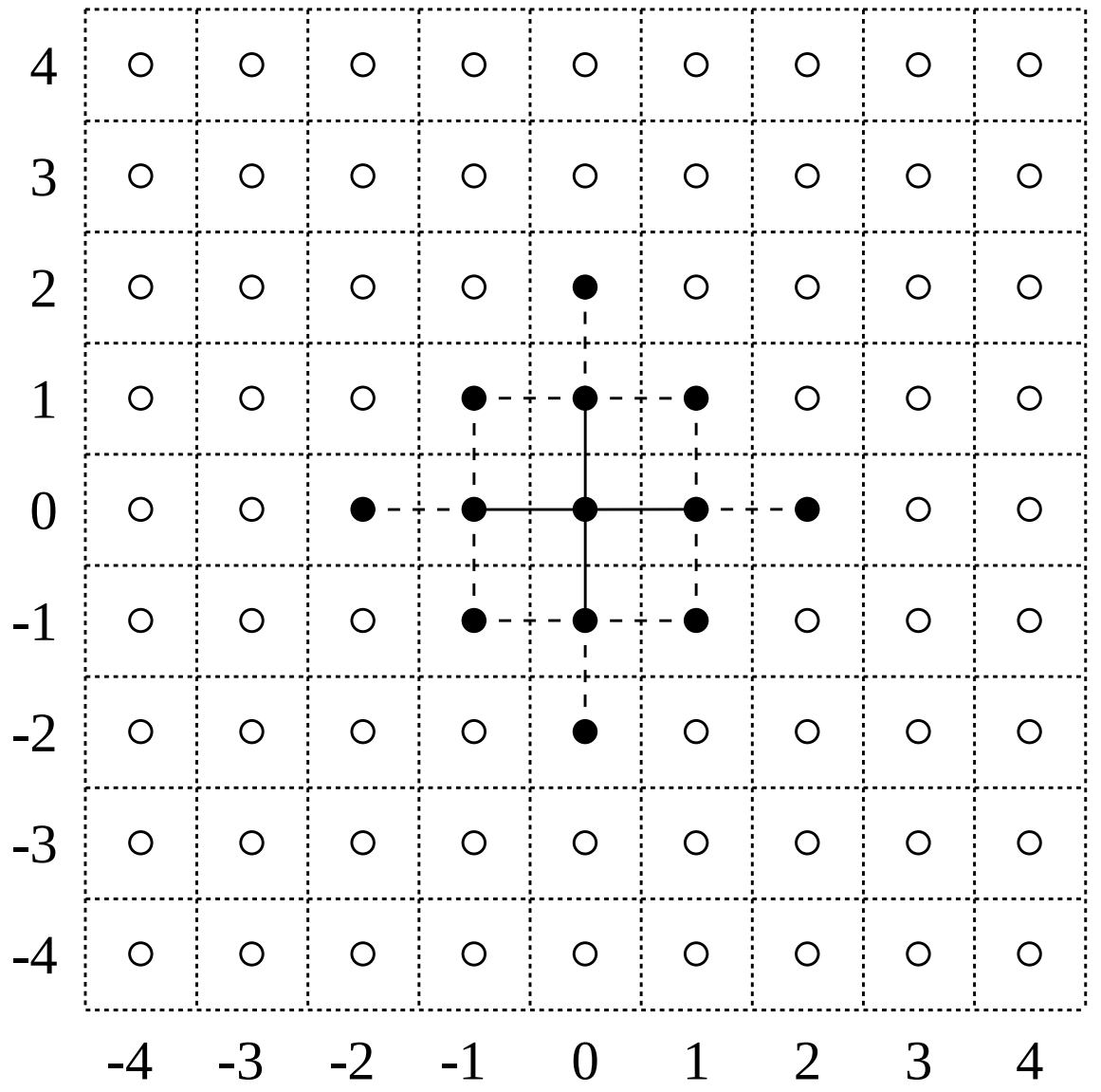

**Figure 6.2:** Topology of contact interaction neighborhood of particles in the SPLG-1 model. Reference frame is fixed at the particle. Solid circles (•) represent lattice sites comprising the contact interaction neighborhood. Hollow circles (○) represent other lattice sites. Solid lines connecting lattice site $(0, 0)$ to other lattice sites represent possible paths along which particle positioned at this lattice site can move. Dashed lines (and solid lines) represent possible paths along which particles occupying other lattice sites can move and interact with the particle occupying the lattice site $(0, 0)$.

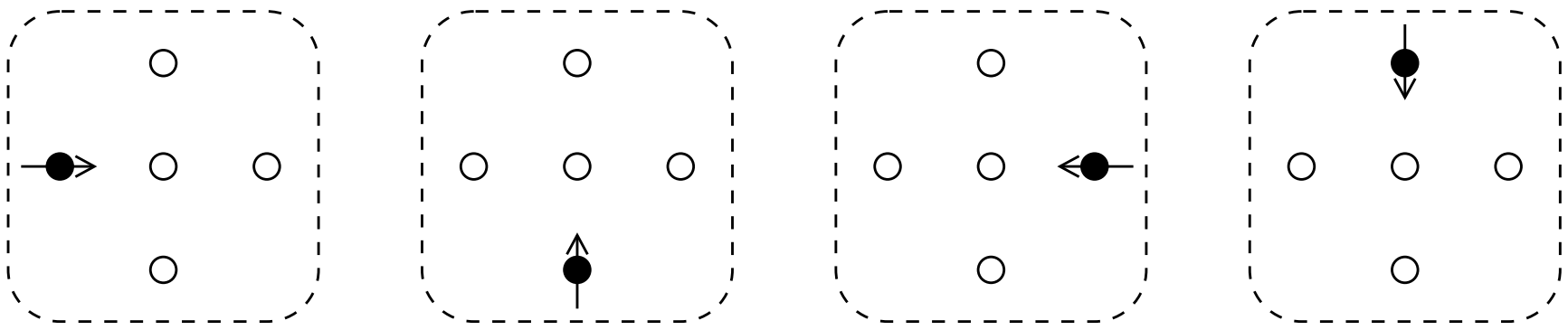

**Figure 6.3:** Contact interaction neighborhoods for particles with different velocities in the SPLG-1 model. Symbols represent: ○ and • lattice sites comprising the contact interaction neighborhood of particle positioned at lattice site shown with • and moving along the arrows →↑↓←.

The BCCs for the SPLG-1 model can be tabulated by tabulating all possible states of the reduced contact interaction neighborhood in which contact interactions occur and removing those lattice sites whose states are irrelevant for the contact interaction. The number of these BCCs can be reduced further by using a new symbol, *e.g.*, $\otimes$, for all those lattice sites which can have only some of the possible states but otherwise are not part of the BCC. The BCCs in the SPLG-1 model, tabulated in this manner, are shown in Fig. 6.4. Many of the BCCs shown in this figure are isometrically equivalent, *i.e.*, they can be obtained from each other through isometric transformations. For constructing the BCC lookup table only one out of each group of isometrically equivalent BCCs needs to be kept. This reduces the size of the BCC lookup table considerably. The unique BCCs that will be kept in the BCC lookup table to be constructed in the following paragraph are shown in Fig. 6.5.

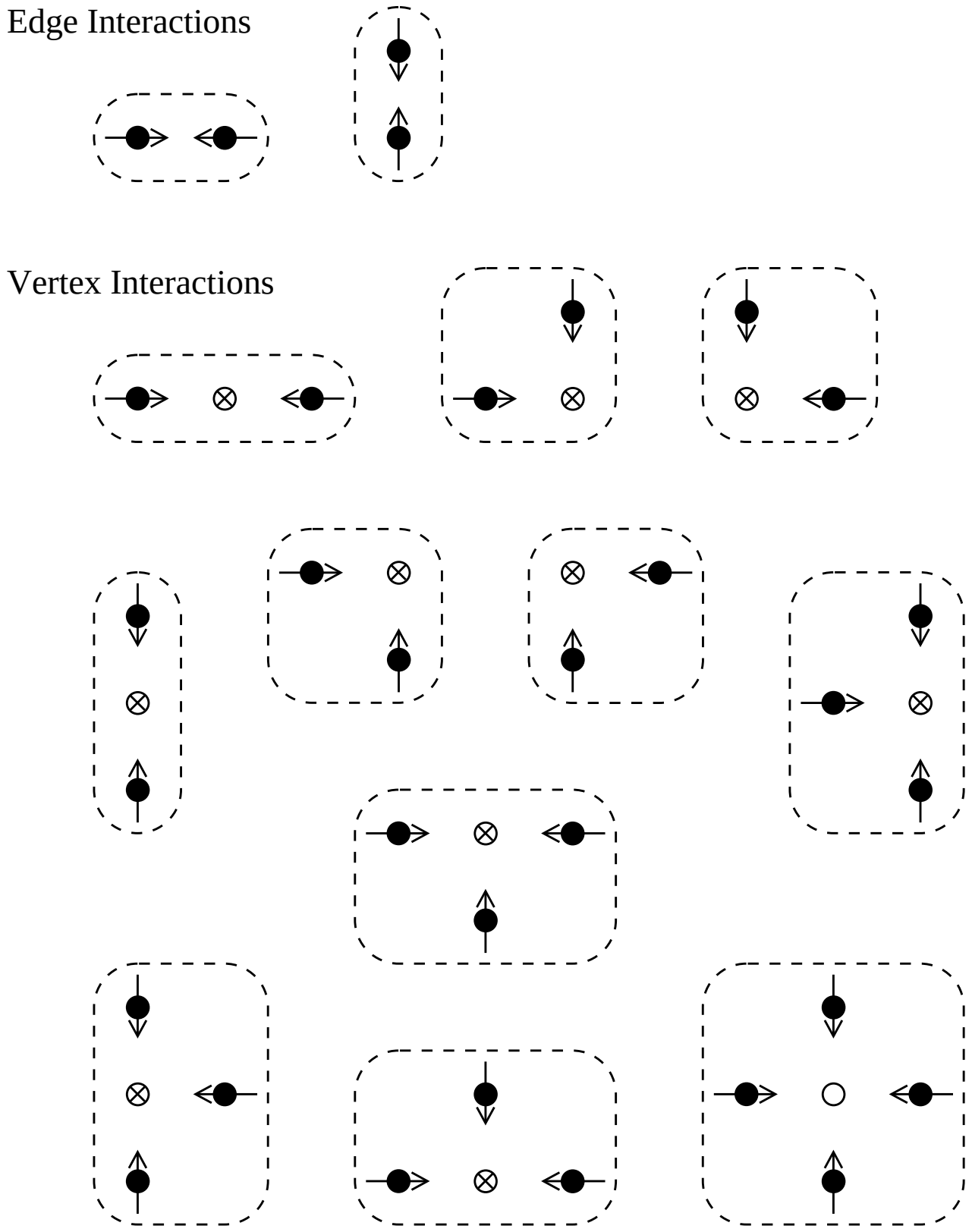

**Figure 6.4:** Basic collision configurations in the SPLG-1 model. Symbols mean: ○ empty lattice site, • particle, →↑↓← direction of motion of particles, and ⊗ lattice sites which are empty or occupied by a particle which does not undergo contact interaction with any particle in the configuration.

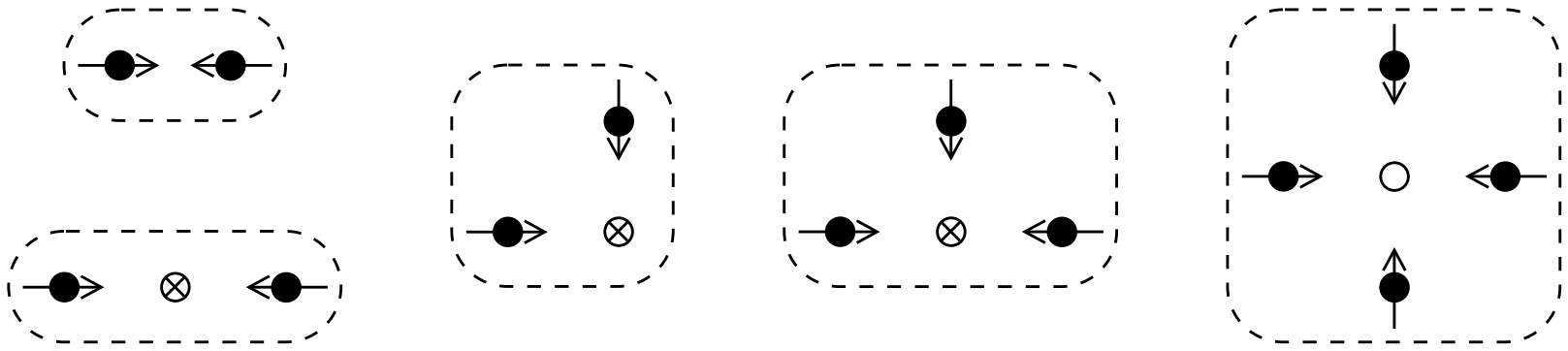

**Figure 6.5:** Unique basic collision configurations in the SPLG-1 model. Symbols are as in Fig. 6.4.

For constructing the BCC lookup table, all possible solutions need to be obtained for the BCCs shown in Fig. 6.5. The solutions for each of these BCCs can be obtained using the procedure described in Sec. 5.2.3.2. These solutions are shown in Fig. 6.6 which is the required BCC lookup table of the SPLG-1 model for processing interactions among fluid particles during simulations. In this table, BCC (1) arises from edge interactions and BCCs (2)–(5) arise from vertex interactions. This table is in reduced from and contains only 5 BCCs. Whereas, if the interaction rules were constructed by considering the states of all lattice sites in the entire reduced contact interaction neighborhood, then they will contain $\mathcal{N}_{\rm s}^{\mathcal{N}_{\mathcal{L}_{\rm R}}} = 5^5 = 3125$ different configurations and all possible solutions corresponding to each one of them, where $\mathcal{N}_{\rm s} = \mathcal{N}_{\rm v} + 1 = 5$ is the number of possible states for each lattice site and $\mathcal{N}_{\mathcal{L}_{\rm R}} = 5$ is the number of lattice sites comprising the reduced contact interaction neighborhood. The contact interaction neighborhood (shown in Fig. 6.2) for the SPLG-1 model comprises of 13 lattice sites. As a result, if the interaction rules were constructed by considering the state of the entire contact interaction neighborhood (which, in fact, is not necessary), then they will contain $5^{13} = 1220703125 \approx 1.22 \times 10^9$ different configurations and all possible solutions corresponding to each one of them. The reduction in the memory required for storing the interaction rules through the use of BCCs is noteworthy.

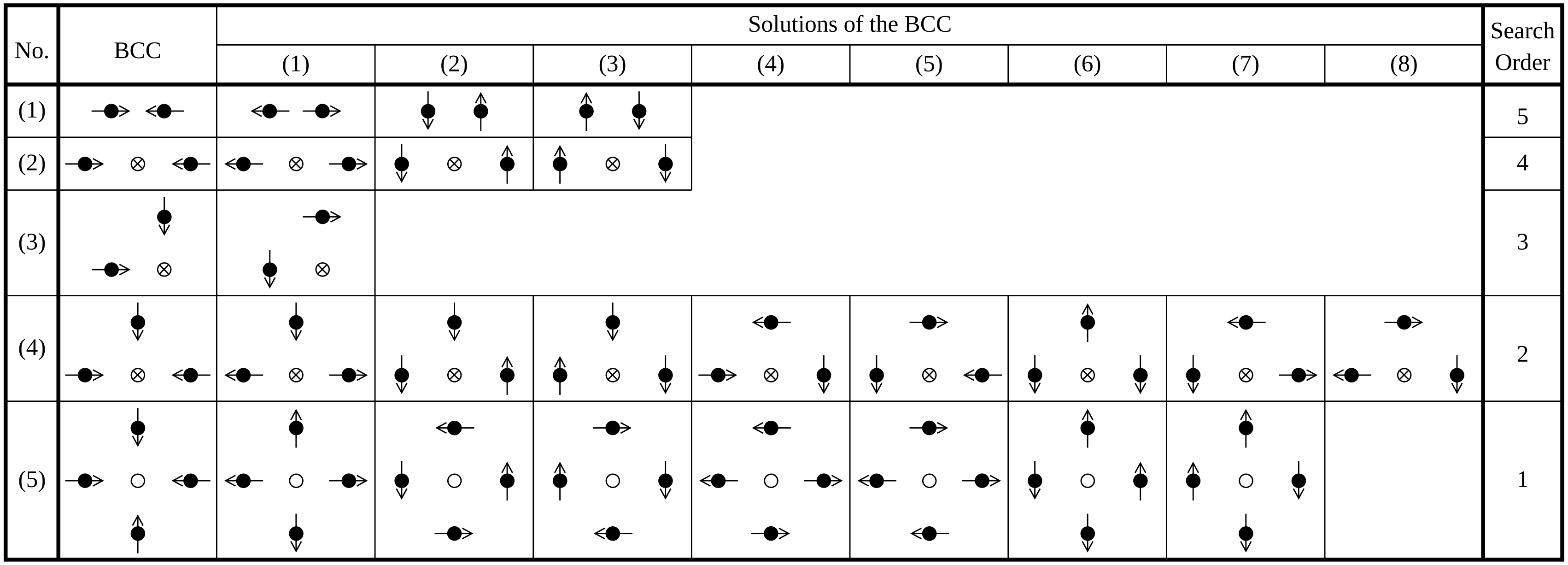

**Figure 6.6:** BCC lookup table of the SPLG-1 model for processing interaction among fluid particles during simulations. Symbols are as in Fig. 6.4.

In systems containing solid boundaries, fluid particles interact with the boundary particles also. As a result, for simulation of such systems, interaction rules (or, BCC lookup table) for processing interactions of fluid particles with boundary particles, in addition to the those for processing interactions among fluid particles, are also required. These rules are constructed by tabulating all possible configurations in which fluid particles can interact with boundary particles. In these rules, it is possible to take care of the thermal velocity of boundary particles along with their mean mass motion velocity. Incorporating the thermal velocity of boundary particles, however, is not only difficult but also unnecessary for simulation of macroscopic systems. As a result, the interaction rules should be constructed by considering only the mean mass motion velocity of boundary particles. The rules for processing interaction of fluid particles with stationary boundary particles in SPLG-1 model are tabulated in Fig. 6.7. The rules for moving boundaries will also be similar to these and can be constructed along similar lines.

Desired interaction potentials can be incorporated in the rules for interaction of fluid particles with stationary boundary particles, also. This, however, is not required for simulation of fluid dynamic systems or for sampling properties in the hydrodynamic limit, *i.e.*, macroscopic properties. The interaction of fluid particles with boundary particles can be specified without using the interaction potentials by selecting STPs corresponding to the desired scattering, energy and normal and tangential momentum accommodation coefficients.

Let $\alpha_{\rm en}$, $\alpha_{\rm nm}$, and $\alpha_{\rm tm}$ be the energy, normal momentum, and tangential momentum accommodation coefficients, respectively. These are defined as

$$\alpha_{\rm en} = \frac{\boldsymbol{v}_{\rm i}^2 - \boldsymbol{v}_{\rm r}^2}{\boldsymbol{v}_{\rm i}^2 - \boldsymbol{v}_{\rm w}^2} \tag{6.1}$$

$$\alpha_{\rm nm} = \frac{\boldsymbol{v}_{\rm n,i} - \boldsymbol{v}_{\rm n,r}}{\boldsymbol{v}_{\rm n,i} - \boldsymbol{v}_{\rm n,w}} \tag{6.2}$$

$$\alpha_{\rm tm} = \frac{\boldsymbol{v}_{\rm t,i} - \boldsymbol{v}_{\rm t,r}}{\boldsymbol{v}_{\rm t,i} - \boldsymbol{v}_{\rm t,w}} \tag{6.3}$$

where $\boldsymbol{v}_{\rm i}$ is velocity of particles incident on the surface, $\boldsymbol{v}_{\rm r}$ is velocity of particles reflected from the surface, $\boldsymbol{v}_{\rm w}$ is velocity of particles in complete equilibrium with the surface, and $\boldsymbol{v}_{\rm n,a}$ and $\boldsymbol{v}_{\rm t,a}$ are normal and tangential components of $\boldsymbol{v}_{\rm a}$ relative to the surface. Usually, the tangential velocity of particles in equilibrium with the surface $\boldsymbol{v}_{\rm t,w}$ is zero [1].

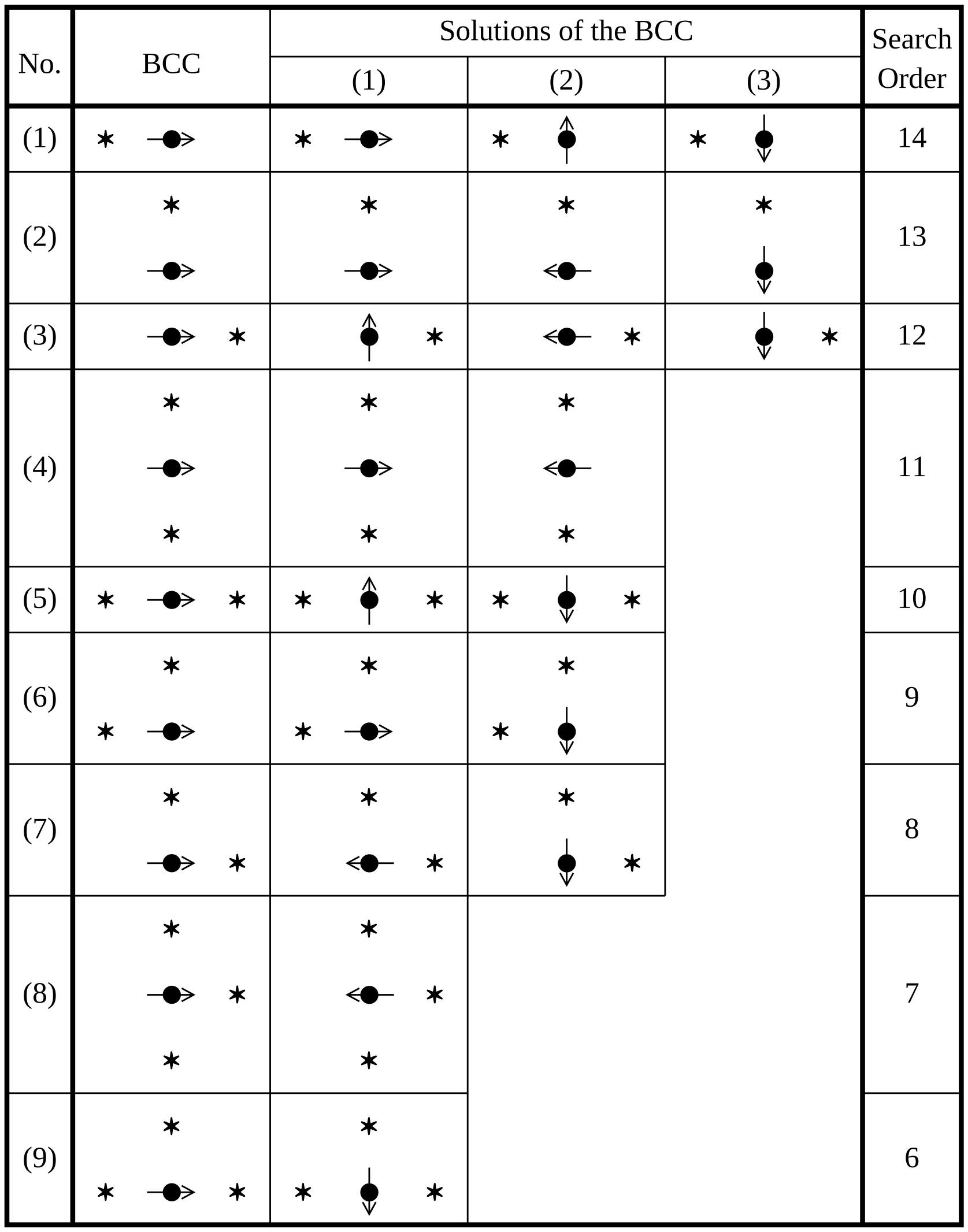

**Figure 6.7:** BCC lookup table of the SPLG-1 model for processing interaction of fluid particles with stationary boundary particles ($*$) during simulations. Symbols are as in Fig. 6.4.

In addition to the above, for simulation of fluid dynamic systems, STPs should be selected in such a manner that no slip boundary condition is satisfied. For this, the STPs should satisfy the condition

$$\sum_{r=1}^{\mathcal{N}_\mathrm{v}} p_{i,r} \boldsymbol{v}_{\mathrm{t},r} = \boldsymbol{v}_{\mathrm{t,w}} \tag{6.4}$$

where $p_{i,r}$ is the probability that a particle incident with velocity $\boldsymbol{v}_i$ will be reflected with velocity $\boldsymbol{v}_r$, $\boldsymbol{v}_{\mathrm{t},r}$ is the tangential component of the reflected velocity (relative to the surface), and $\boldsymbol{v}_{\mathrm{t,w}}$ is the tangential component of the mean mass motion velocity of the surface. For stationary surface, $|\boldsymbol{v}_\mathrm{w}| = 0$ and $|\boldsymbol{v}_{\mathrm{t,w}}| = 0$.

The condition given by Eq. 6.4 demands that statistically the tangential component of reflected velocity for each incident velocity vector should be equal to the tangential component of the mean mass motion velocity of the surface. It is, thus, a very strict condition. It, however, can always be satisfied if the discrete velocity set contains velocity vectors in pairs of the form $\pm\boldsymbol{v}$. A more relaxed condition, which also ensures no-slip boundary conditions, is

$$\sum_{i=1}^{\mathcal{N}_{\rm v}} \sum_{r=1}^{\mathcal{N}_{\rm v}} p_{i,r}\boldsymbol{v}_{{\rm t},r} = \boldsymbol{v}_{\rm t,w} \tag{6.5}$$

This condition subsumes that given by Eq. (6.4). Thus, it will necessarily be satisfied if the condition given by Eq. (6.4) is satisfied.

### 6.3.2 Incorporating Interaction Potentials in the SPLG-1 Model

For incorporating interaction potentials in SPLG-1 model, the desires interaction potential must be selected and its range must be determined. Following this, lattice parameters must be determined as explained in Sec. 5.1.4 in such a way that the resulting field interaction neighborhood subsumes the contact interaction neighborhood. This can be done as explained in Sec. 5.1.3.8. The steps for the SPLG-1 models are described below.

The range of contact interactions in natural units of the lattice system $\tilde{R}_{\rm CI}$ for the SPLG-1 model is 2. A vector $\boldsymbol{r}$ in the first quadrant of the two-dimensional position space with the same magnitude as $\tilde{R}_{\rm CI}$ is $(2,0)$ (*c.f.*, Sec. 5.1.3.8). The vector $\boldsymbol{r}'$ corresponding to this vector is $(\frac{3}{2},0)$. Thus, in order that contact interaction neighborhood be subsumed within the field interaction neighborhood, the range of field interactions in natural unit of the lattice system $\tilde{R}_{\rm I}$ should be

$$\tilde{R}_{\rm I} > \frac{3}{2}$$

With this, the smallest field interaction neighborhood that subsumes the contact interaction neighborhood is obtained for

$$\frac{3}{2} < \tilde{R}_{\rm I} \leq \frac{\sqrt{10}}{2}$$

The field interaction neighborhood for this range of $\tilde{R}_{\rm I}$ is shown in Fig. 5.6. It is seen from this figure that for SPLG-1 model the smallest field interaction neighborhood is identical to the contact interaction neighborhood which is shown in Fig. 6.2.

Using the details given above, an appropriate field (or, overall) interaction neighborhood can be selected. Following this, field interactions can be incorporated into the SPLG-1 model by appropriately defining the STPs as explained in Sec. 5.2.3.6. It was pointed out in Sec. 5.2.3.6, on the grounds of overwhelmingly large storage memory requirements, that construction of the complete BCC lookup tables (and corresponding STP tables) incorporating the effect of field interactions should not be attempted. A concrete example, illustrating this, is given below. In this example the smallest overall interaction neighborhood is used.

It is explained in Sec. 5.2.3.6, after incorporating the desired interaction potential in SPLG-1 model, the full BCC lookup table and the corresponding STP table will contain solutions of all the BCCs and their STPs depending on the state of overall interaction neighborhood for all the particles undergoing contact interaction in each BCC. Since the topology of the largest BCC subsumes that of all other BCCs, the reduced contact interaction neighborhood can be used for constructing the complete BCC lookup table. The topology of the domain whose state becomes important and must be enlisted in this BCC lookup table can be obtained by superimposing the reference lattice site (or, vertex) of the overall interaction neighborhood over all the vertices in the reduced contact interaction neighborhood. This domain will, henceforth, be referred to as the *"joint interaction neighborhood"* and denoted by $\mathcal{L}_\mathrm{J}$. The joint interaction neighborhood in the SPLG-1 model for the smallest overall interaction neighborhood is shown in Fig. 6.8.

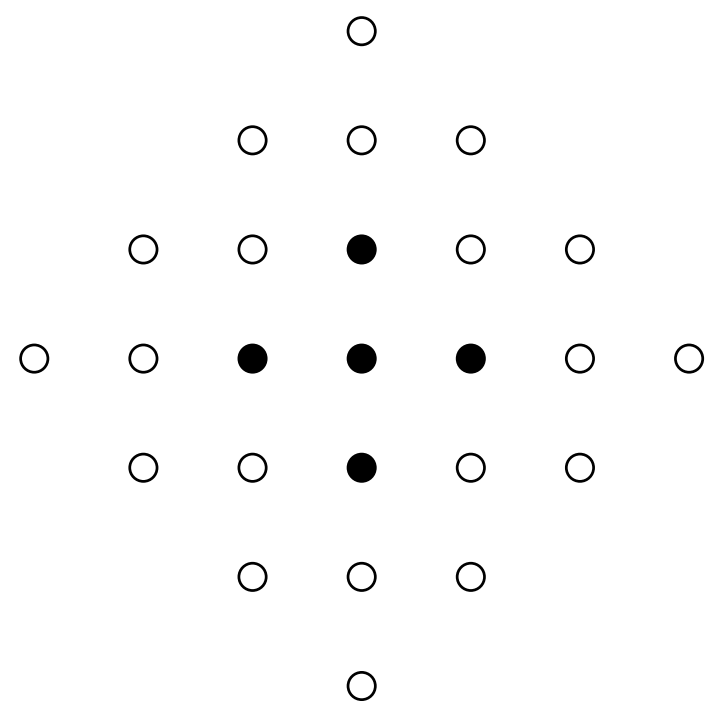

**Figure 6.8:** The joint interaction neighborhood in the SPLG-1 model for the smallest overall interaction neighborhood which subsumes the contact interaction neighborhood. Symbols represent: • lattice sites comprising the reduced contact interaction neighborhood, ∘ and • lattice sites comprising the joint interaction neighborhood.

It is seen from Fig. 6.8 that the joint interaction neighborhood comprises of 25 lattice sites, *i.e.*, $\mathcal{N}_{\mathcal{L}_\mathrm{J}} = 25$. As a result, the full BCC lookup table after incorporating field interactions using the smallest overall interaction neighborhood will contain

$$\mathcal{N}_\mathrm{s}^{\mathcal{N}_{\mathcal{L}_\mathrm{J}}} = 5^{25} = 298023223876953125 \approx 2.98 \times 10^{17}$$

configurations and all possible solutions corresponding to each one of them. This lookup tables requires immense amount of memory for storage and thus is not worth constructing.

## 6.4 Key Aspects of Interaction Rules for the SPLG-1 Model

In the SPLG-1 model, all particles move with unit speed. As a result, they are always reflected from surfaces with the same speed. Because of this, the thermal energy of solid boundaries does not convect into the fluid. Alternatively, it means that in this model it is meaningless to make the thermal energy of solid boundaries different from that of the fluid. Thus, the energy accommodation coefficient $\alpha_\mathrm{en}$ does not play any role in this model.

In the BCC lookup table for processing interaction of fluid particles with boundary particles in the SPLG-1 model (*c.f.*, Fig. 6.7), configuration in which a fluid particle is surrounded by boundary particles on all the four nearest neighbor lattice sites is not shown. This is because, the occurrence of such a configuration in the system during simulations will cause processing of interparticle interaction to end up into infinite loop (*i.e.*, to never terminate). This happens because the fluid particle, for all possible velocities, will undergo vertex interaction with one of the four boundary particles. This, if absent in the initial state of the system, cannot be produced during simulations in systems containing only stationary solid boundaries. Thus, care should be taken to ensure that the initial condition (or, the initial state) of the system does not contain such configurations. In order that this problem does not surface, it is necessary that stationary velocity vectors be contained in the discrete velocity sets for particles of each species in single particle lattice gases.

Inspection of the BCC lookup table for processing interactions among fluid particles in the SPLG-1 model (*c.f.*, Fig. 6.6) reveals that several *fixed states* are possible in this model. These states, during processing of interparticle interactions, form a closed repeating sequence of states with vertex interactions. As a result, if any of these states occurs in the system, the iterations carried out for processing interparticle interactions will never terminate. When the boundaries are toroidal or periodic, these states occur in pairs which lead to each other. These states can occur only when the spatial lattice comprises of even number of lattice sites along both $x$- and $y$-axes and either 50% or 100% of the lattice sites are occupied. When the spatial lattice if 50% full, these states are formed when particles with velocity $\boldsymbol{v}_1$ are positioned at odd lattice sites in odd rows (or, odd lattice sites in even rows) and particles with velocity $\boldsymbol{v}_2$ are positioned at even lattice sites in even rows (or, even lattice sites in odd rows) and $\boldsymbol{v}_1 \cdot \boldsymbol{v}_2 = 0$. Four such state pairs are shown in Fig. 6.9. When the spatial lattice if 100% full, these states are formed when particles with velocity $\boldsymbol{v}_1$ are positioned at all lattice sites in odd rows and particles with velocity $\boldsymbol{v}_2$ are positioned at all lattice sites in even rows and $\boldsymbol{v}_1 \cdot \boldsymbol{v}_2 = 0$. Examples of these states can be constructed in a straight forward manner as shown for the 50% full case in Fig. 6.9. Isometric transforms of the fixed states are also fixed states.

An important property of these fixed states, when the boundary conditions are toroidal or periodic, is that they can occur during simulations only if the initial state itself is fixed state. This is because if evolution of the system in these states is retraced by one time step, two particles appear to be coming from the same lattice site. Thus, the state preceding the fixed states is inconsistent in view of the single particle exclusion principle and cannot occur in single particle lattice gases (during simulations). As a result, if simulations do not start with a fixed state, the system, by itself, cannot enter into a fixed state during simulations. Thus, these fixed states are not strange attractors of the system.

Fixed states of the type mentioned above can occur in open infinite systems, also. These situations, however, are of not practical interest because simulation of infinite systems is not possible. Fixed states can occur under solid boundary conditions also. In this condition, however, their occurrence depends upon the STP table corresponding to rules for processing interaction of fluid particles with boundary particles. The occurrence of these states, when the system has solid boundaries, can be avoided by carefully selecting the STPs, ensuring all closed spaces are of dimensions $(2, 2)$ or larger, and all closed spaces of dimensions $(1, N)$ or $(N, 1)$ are neither fully filled nor contain particles on all alternate lattice sites.

## 6.5 Analysis of the SPLG-1 Model

Detailed mathematical analysis of the SPLG-1 model[2)] is not feasible because interparticle interactions are processed iteratively using fractional interaction rules. In the following sections, however, some important details about SPLG-1 model—which include the probability of occurrence of BCCs, mean free time (*i.e.*, mean time between two consecutive collisions of a particle), mean free path, and speed of sound—are provided for spatially uniform systems in equilibrium at mean particle density of $n$ particles per lattice site.

[2)] In fact, of any single particle lattice gas.

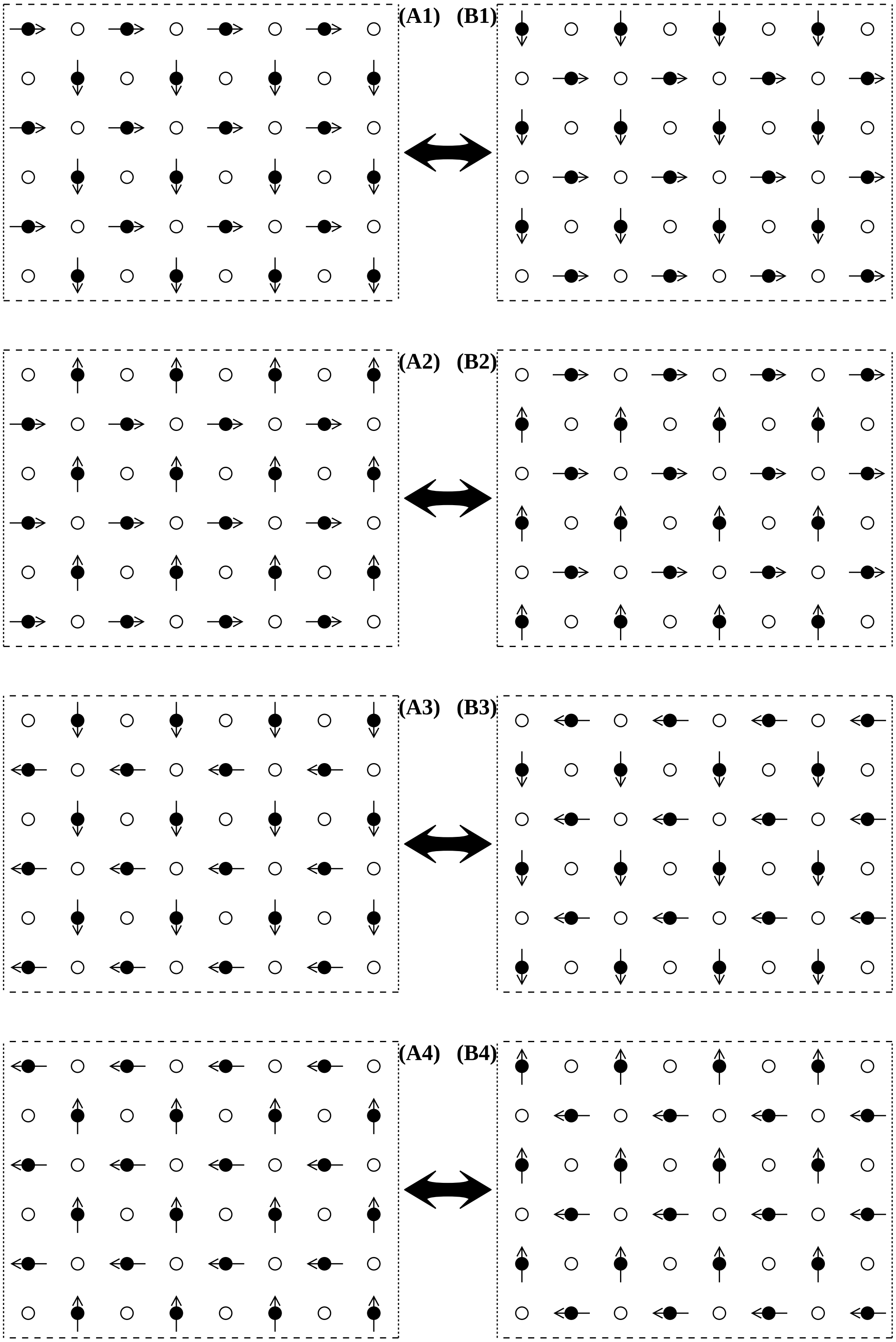

**Figure 6.9:** Four possible fixed states with 50% particle density in the SPLG-1 model. Symbols are as in Fig. 6.4.

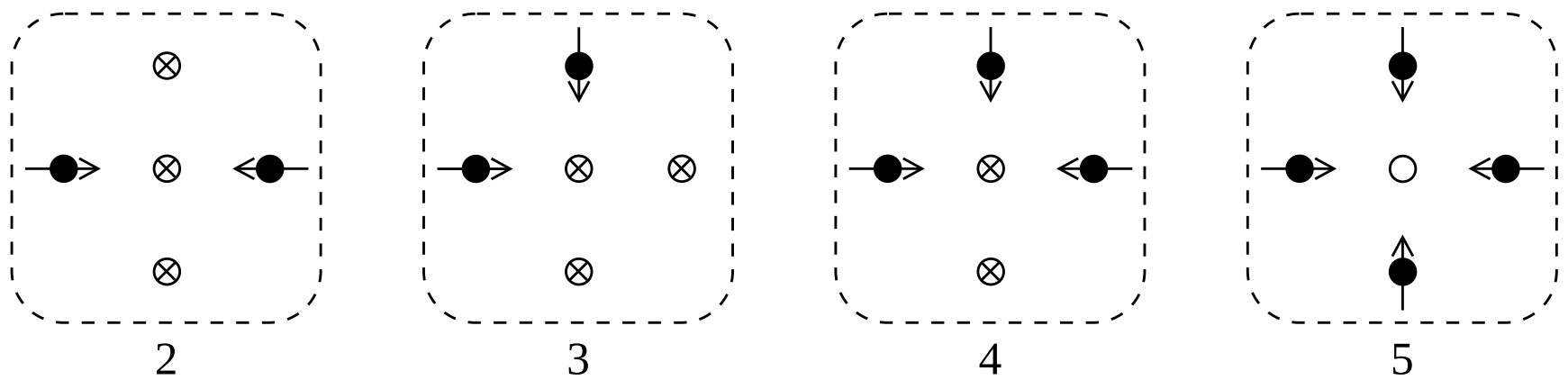

**Figure 6.10:** Basic collision configurations in the SPLG-1 model redrawn using the entire reduced contact interaction neighborhood. Symbols are as in Fig. 6.4.

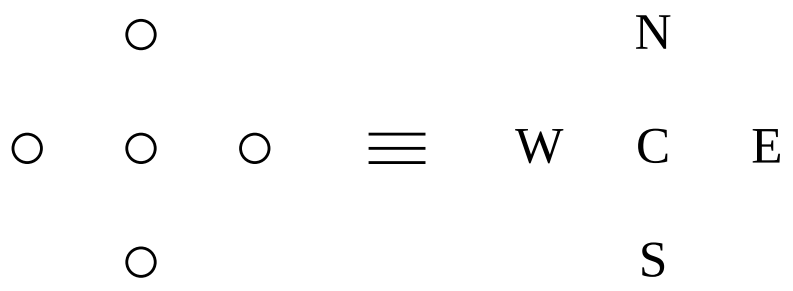

**Figure 6.11:** Lattice site labels in the reduced contact interaction neighborhood of SPLG-1 model.

### 6.5.1 Probability of Occurrence of BCCs

In Figs. 6.4, 6.5, and 6.6 the BCCs are tabulated in a vary compact fashion. This form, although permits unambiguous detection of the BCCs in as ordered search (the order is specified in Fig. 6.6) during simulations, does not permit straight forward calculation of their probability of occurrence over the spatial lattice (except for the BCC arising out of edge interactions; shown as BCC 1 in Fig. 6.6). As a result, the BCC 2–5 have been redrawn in Fig. 6.10 using the entire reduced contact interaction neighborhood. The probability of occurrence of these BCCs over spatial lattice in spatially uniform systems at equilibrium with mean particle density of $n$ particles per lattice site is calculated as described below.

Let the lattice sites in the reduced contact interaction neighborhood be labeled E, S, C, W, and N as shown in Fig. 6.11. Let $\rightarrow\uparrow\downarrow\leftarrow$ denote that a lattice site is occupied by a particle moving with velocity $(1,0)$, $(0,1)$, $(0,-1)$, and $(-1,0)$, respectively. Let $\circ$ denote that a lattice site is empty. Let probability that a lattice site S is in state R be denoted by $P_{\mathrm{X}}(\mathrm{Y})$, where $\mathrm{X} \in \{\mathrm{E},\mathrm{W},\mathrm{N},\mathrm{S},\mathrm{C}\}$ and $\mathrm{Y} \in \{\rightarrow,\uparrow,\downarrow,\leftarrow,\circ\}$. Then, the probabilities that a lattice site X is in various possible states are

$$\left.\begin{array}{lcl} P_{\mathrm{X}}(\rightarrow) & = & n/4 \\ P_{\mathrm{X}}(\uparrow) & = & n/4 \\ P_{\mathrm{X}}(\downarrow) & = & n/4 \\ P_{\mathrm{X}}(\leftarrow) & = & n/4 \\ P_{\mathrm{X}}(\circ) & = & 1-n \end{array}\right\} \tag{6.6}$$

and the probabilities that a lattice site X is *not* in various possible states are

$$\left.\begin{array}{lcl} Q_{\mathrm{X}}(\rightarrow) & = & 1-n/4 \\ Q_{\mathrm{X}}(\uparrow) & = & 1-n/4 \\ Q_{\mathrm{X}}(\downarrow) & = & 1-n/4 \\ Q_{\mathrm{X}}(\leftarrow) & = & 1-n/4 \\ Q_{\mathrm{X}}(\circ) & = & n \end{array}\right\} \tag{6.7}$$

where $Q_{\mathrm{X}}(\mathrm{Y}) = 1 - P_{\mathrm{X}}(\mathrm{Y})$.

| BCC # $i$ | $\mathcal{N}_{\text{iso}}^{(i)}$ | $\hat{P}^{(i)}$ | $P^{(i)} = \mathcal{N}_{\text{iso}}^{(i)} \hat{P}^{(i)}$ |
|---|---|---|---|
| 1 | 2 | $(n/4)^2$ | $2(n/4)^2$ |
| 2 | 2 | $(n/4)^2(1-n/4)(1-3n/4+n^2/4)$ | $2(n/4)^2(1-n/4)(1-3n/4+n^2/4)$ |
| 3 | 4 | $(n/4)^2(1-n/4)(1-3n/4+n^2/4)$ | $4(n/4)^2(1-n/4)(1-3n/4+n^2/4)$ |
| 4 | 4 | $(n/4)^3(1-n/2)^2$ | $4(n/4)^3(1-n/2)^2$ |
| 5 | 1 | $(n/4)^4(1-n)$ | $(n/4)^4(1-n)$ |

**Table 6.1:** Number of isometries $\mathcal{N}_{\text{iso}}^{(i)}$ and the probability of occurrence of various BCCs, $\hat{P}^{(i)}$ and $P^{(i)}$, over spatial lattice in the SPLG-1 model in spatially uniform system at equilibrium with mean particle density of $n$ particles per lattice site.

Let the probability of occurrence of a BCC $i$ over spatial lattice including all its isometries be $P^{(i)}$ and the that of just the BCC itself (in the form shown in Fig. 6.6 and 6.10 and excluding its isometries) be $\hat{P}^{(i)}$. Let the number of isometries for a BCC $i$ be $\mathcal{N}_{\text{iso}}^{(i)}$. Then

$$P^{(i)} = \mathcal{N}_{\text{iso}}^{(i)} \hat{P}^{(i)} \tag{6.8}$$

With this $\hat{P}^{(i)}$ for various BCCs are calculated as follows:

$$\begin{aligned}
\hat{P}^{(1)} &= P_{\text{W}}(\rightarrow)P_{\text{C}}(\leftarrow) \\
&= (n/4)^2 && (6.9) \\
\hat{P}^{(2)} &= P_{\text{W}}(\rightarrow)P_{\text{E}}(\leftarrow)\left[P_{\text{C}}(\uparrow)Q_{\text{S}}(\uparrow) + P_{\text{C}}(\downarrow)Q_{\text{N}}(\downarrow) + P_{\text{C}}(\circ)Q_{\text{S}}(\uparrow)Q_{\text{N}}(\downarrow)\right] \\
&= (n/4)^2\left[(n/4)(1-n/4) + (n/4)(1-n/4) + (1-n)(1-n/4)^2\right] \\
&= (n/4)^2(1-n/4)(1-3n/4+n^2/4) && (6.10) \\
\hat{P}^{(3)} &= P_{\text{W}}(\rightarrow)P_{\text{N}}(\downarrow)\left[P_{\text{C}}(\rightarrow)Q_{\text{S}}(\uparrow) + P_{\text{C}}(\downarrow)Q_{\text{E}}(\leftarrow) + P_{\text{C}}(\circ)Q_{\text{S}}(\uparrow)Q_{\text{E}}(\leftarrow)\right] \\
&= (n/4)^2\left[(n/4)(1-n/4) + (n/4)(1-n/4) + (1-n)(1-n/4)^2\right] \\
&= (n/4)^2(1-n/4)(1-3n/4+n^2/4) && (6.11) \\
\hat{P}^{(4)} &= P_{\text{W}}(\rightarrow)P_{\text{N}}(\downarrow)P_{\text{E}}(\leftarrow)\left[P_{\text{C}}(\downarrow) + P_{\text{C}}(\circ)Q_{\text{S}}(\uparrow)\right] \\
&= (n/4)^3\left[(n/4) + (1-n)(1-n/4)\right] \\
&= (n/4)^3(1-n/2)^2 && (6.12) \\
\hat{P}^{(5)} &= P_{\text{W}}(\rightarrow)P_{\text{N}}(\downarrow)P_{\text{E}}(\leftarrow)P_{\text{E}}(\uparrow)P_{\text{C}}(\circ) \\
&= (n/4)^4(1-n) && (6.13)
\end{aligned}$$

The above results along with number of isometries for each BCC and their net probability of occurrence over spatial lattice in the SPLG-1 model are summarized in table 6.1.

### 6.5.2 Collision Probability, Mean Free Time, and Mean Free Path

Let $\mathcal{P}$ be the probability that a particle undergoes a collision at any time step (in SPLG-1 model) in spatially uniform systems at equilibrium with mean density of $n$ particles per lattice site. Then, the mean time that a particle spends per collision (or, the mean free time $\tau_{\text{mf}}$ of particles) in the SPLG-1 model is

$$\tau_{\text{mf}} = 1/\mathcal{P} \tag{6.14}$$

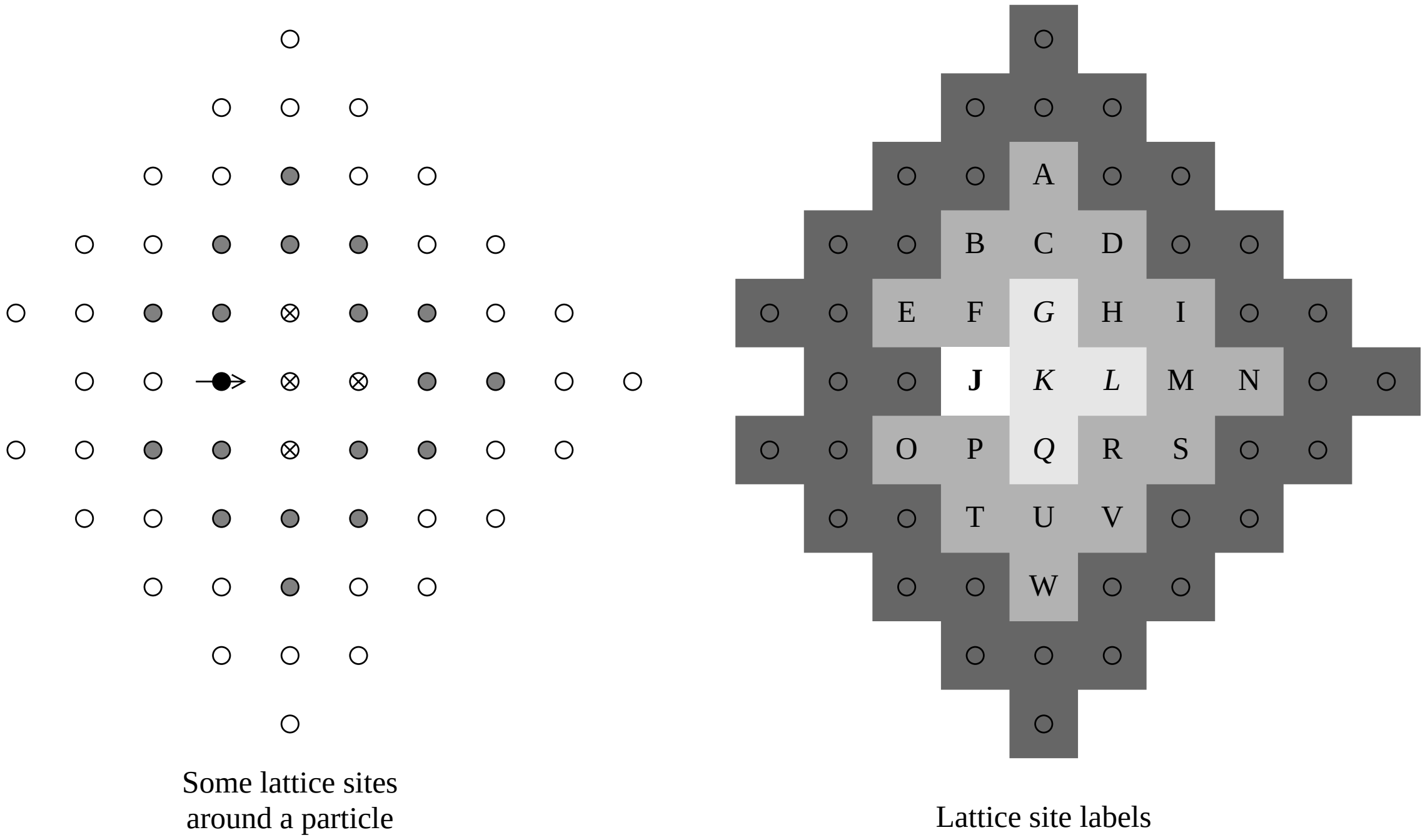

**Figure 6.12:** Some lattice sites and their labels in the neighborhood of a particle, the particle J, moving with velocity $(1, 0)$ in the SPLG-1 model. Different groups of neighbors of particle J are shown differently. The outer most neighbors are not labeled.

Let the mean speed of particles between consecutive collisions be $v_{\text{mf}}$. With this, the mean distance traveled by a particle between two consecutive collisions (or, the mean free path $\lambda_{\text{mf}}$ of particles) is

$$\lambda_{\text{mf}} = v_{\text{mf}} \tau_{\text{mf}} \tag{6.15}$$

In the SPLG-1 model particles always move with the speed of one lattice site per time step, *i.e.*, $v_{\text{mf}} = 1$. This, combined with Eqs. (6.14) and (6.15), gives

$$\lambda_{\text{mf}} = \tau_{\text{mf}} = 1/\mathcal{P} \tag{6.16}$$

for the SPLG-1 model.

In single particle lattice gases, interactions among particles are processed iteratively using fractional interaction rules and the number of iterations required for processing depends upon density and velocity distribution functions of particles and the STP tables. As a result, in single particle lattice gases $\mathcal{P}$ cannot be estimated easily as it requires incorporating the effect of all the iterations. In the following, however, an approximate attempt is made at estimating $\mathcal{P}$ for the SPLG-1 model.

#### 6.5.2.1 Collision Events in Various Iteration: Collision Tree

Consider a small domain of a system as shown in Fig. 6.12 with lattice sites as labeled therein. The topology of the domain shown in this figure is irrelevant and its peculiarity should not distract. It is shown because only these lattice sites are relevant in the following derivations. Consider that this state is at the beginning of a time step and processing of interparticle interactions is to be carried out (which is an iterative process). In this domain consider a particle located at lattice site marked J (henceforth, particle J, and so on) and moving with velocity $(1, 0)$ as shown by the arrow in the figure. During processing of interparticle interactions, particle J can either collide or not collide in various iterations. This gives rise to a collision tree for each particle as shown in Fig. 6.13. The events that can happens with particle J in the first and second iteration are described below.

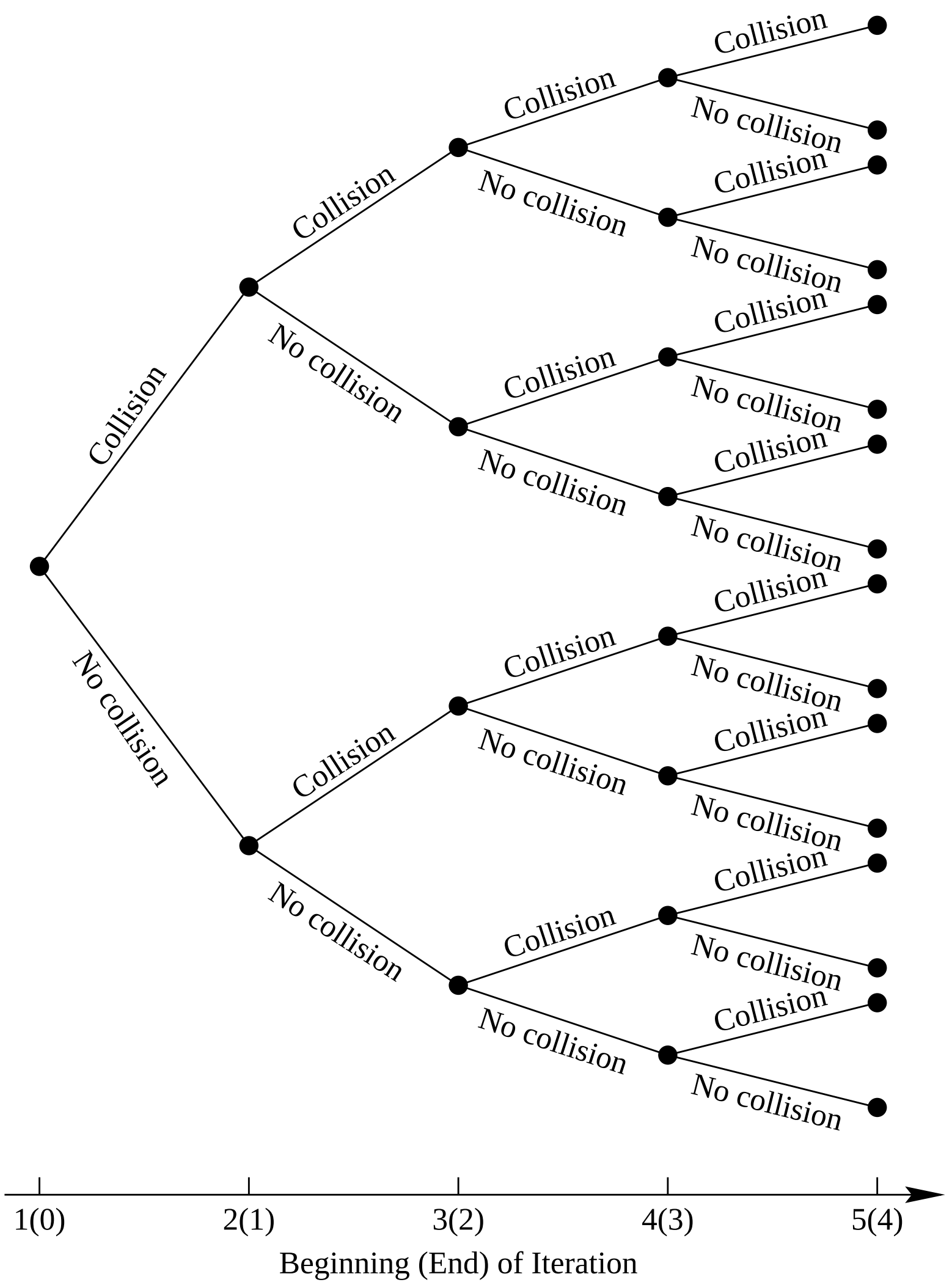

**Figure 6.13:** Collision tree that is formed for each particle while processing interparticle interactions at each time step in single particle lattice gases. First iteration begins with the time step.

**Events with Particle J in First Iteration:** In the first iteration the particle J can collide only with particles G, K, Q, and L. These will be called as *"nearest interaction neighbors"* (NINs) of particle J. If a collision occurs, the state of particle J can change to any of the four possible states $\rightarrow$, $\uparrow$, $\downarrow$, and $\leftarrow$. If collision does not occur, the state of the particle J remains unchanged at the end of first iteration.

**Events with Particle J in Second Iteration:** If particle J does not collide in the first iteration, it can collide in the second iteration. This collision can occur only if at least one of the particles G, K, Q, and L (*i.e.*, NINs of particle J) collides with its neighbors (other then particle J) in the first iteration and then goes to an appropriate state facilitating collision with particle J in the second iteration. The particles G, K, Q, and L can collide within themselves or any of the particle A-F, H, I, M-P, R-W. These will be called as *"next nearest interaction neighbors"* (NNINs) of particle J. If particle J has collided in the first iteration, then also it can collide in the second iteration. If the state of particle J, following the collision in the first iteration, is ↓, it can collide with particle O, P, Q, and T, if it is ↑, it can collide with B, E, F, and G, similarly for ← and →. Occurrence of a collision in the second iterations when the state is one of ↓, ↑, and ← does not have constraints. If, however, the state is →, particle J can collide with its NINs only if they undergo a collision with one of their neighbors, *i.e.*, among themselves or with NNINs of particle J. This is because of the nature of the interaction rules, in which, as can be seen from Fig. 6.6, the state of at most one particle remains unchanged following interactions.

**Events with Particle J in Third and Higher Iterations:** The tree of collision events widens in third and higher iterations and rapidly becomes extremely complex. Except for the complexity, the events that can happen with particle J in the third and higher iterations can also be analyzed along lines similar to those outlined above. At present, however, the analysis of third and higher iteration is being skipped because of the complexity involved.

#### 6.5.2.2 Estimating Collision Probability

**Mathematical Notations:** The mathematical notation employed in the analysis outlined below is as follows: $p_{\mathrm{X}}^{(i)}(\mathrm{Y})$ represents the probability that particle X in state Y at the beginning of $i^{\mathrm{th}}$ iteration collides in the $i^{\mathrm{th}}$ iteration. $p_{\mathrm{X}}^{(i)}$ represents the probability that particle X (irrespective of its state) collides in the $i^{\mathrm{th}}$ iteration. $p_{\mathrm{X}}^{(i)-(j)}(\mathrm{Y})$ represents the probability that particle X in state Y at the beginning of $i^{\mathrm{th}}$ iteration collides at least once between the $i^{\mathrm{th}}$ and $j^{\mathrm{th}}$ iterations, both inclusive. $p_{\mathrm{N^kIN,X}}^{(i)}$ represents the probability that $\mathrm{N^kIN}$ of particle X collide with their neighbors (which includes $\mathrm{N^{k+1}IN}$ of X and $\mathrm{N^kIN}$ themselves) in the $i^{\mathrm{th}}$ iteration, where $\mathrm{N^kIN}$ means $k-1$ times removed nearest interaction neighbors, *e.g.*, $\mathrm{N^1IN}$ = NIN, $\mathrm{N^2IN}$ = NNIN, and $\mathrm{N^3IN}$ = NNNIN. Finally, $\mathcal{N}_{\mathrm{N^kIN,X}}$ represents the number of particles comprising the $\mathrm{N^kIN}$ of particle X.

In the above, $p_{\mathrm{X}}^{(i)}(\mathrm{Y})$ and $p_{\mathrm{X}}^{(i)}$ are related through

$$p_{\mathrm{X}}^{(i)} = \frac{1}{\mathcal{N}_{\mathrm{v}}} \sum_{j=1}^{\mathcal{N}_{\mathrm{v}}} p_{\mathrm{X}}^{(i)}(\boldsymbol{v}_j) \tag{6.17}$$

where $\boldsymbol{v}_j \in \mathcal{V}$, $j = 1, \ldots, \mathcal{N}_{\mathrm{v}}$, represents possible state(s) of particle X, $\mathcal{V}$ is the discrete velocity set (for particle X) and $\mathcal{N}_{\mathrm{v}}$ is the number of elements in $\mathcal{V}$. This equation is an approximation because in it $p_{\mathrm{X}}^{(i)}(\boldsymbol{v}_j)$ has been assigned equal weight for all $j$. This weighting scheme, in general, will not be applicable. It, however, must suffice as a first approximation. For the SPLG-1 model $\mathcal{V} \equiv (\pm 1, 0), (0, \pm 1) \equiv \{\rightarrow, \uparrow, \downarrow, \leftarrow\}$ and $\mathcal{N}_{\mathrm{v}} = 4$.

In the first iteration, $p_{\mathrm{X}}^{(1)}(\boldsymbol{v}_i) = p_{\mathrm{X}}^{(1)}(\boldsymbol{v}_j)$, for all $\boldsymbol{v}_i, \boldsymbol{v}_j \in \mathcal{V}$. This is because the system is spatially uniform and at equilibrium and in single particle lattice gases only one particle can occupy a lattice site at any time step. With Eq. (6.17), this gives

$$p_{\mathrm{X}}^{(1)}(\boldsymbol{v}_i) = p_{\mathrm{X}}^{(1)} = p_{\mathrm{X}}^{(1)}(\bullet) = p^{(1)}(\bullet) = p^{(1)} \qquad \forall\, i \in \mathcal{V} \tag{6.18}$$

where new notation is introduced in the last three equalities and • represents a particle irrespective of its state. In the last two equalities the subscript representing lattice site has been dropped because the system is spatially uniform and at equilibrium and all lattice sites are equivalent to each other. In the last equality the state dependence is also dropped because in single particle lattice gases only one particle can occupy a lattice site at any time step and it can have only one state. For the SPLG-1 model, Eq. (6.18) gives

$$p_{\mathrm{X}}^{(1)}(\rightarrow) = p_{\mathrm{X}}^{(1)}(\uparrow) = p_{\mathrm{X}}^{(1)}(\downarrow) = p_{\mathrm{X}}^{(1)}(\leftarrow) = p_{\mathrm{X}}^{(1)} = p_{\mathrm{X}}^{(1)}(\bullet) = p^{(1)}(\bullet) = p^{(1)} \tag{6.19}$$

for spatially uniform systems at equilibrium. These notations will be useful below.

In single particle lattice gases particles do not change lattice sites during processing of interparticle interactions, *i.e.*, a particle occupying lattice site X in the first iteration continues to occupy the same lattice site in the subsequence iterations also. This means that subscript X can be dropped and $p_{\mathrm{X}}^{(1)-(i)}$ can be written as $p^{(1)-(i)}$.

**Estimating Collision Probability in SPLG-1 Model:** Sec. 6.5.2.1 suggests that $\mathcal{P}$ can be evaluated as sum of infinite series written using the above notation as

$$\mathcal{P} = p^{(1)-(\infty)}(\mathrm{Y}) = \sum_{i=1}^{\infty} p_{\mathrm{X}}^{(i)} \tag{6.20}$$

The exact form of all the terms of this infinite series, and thus the exact sum of this series, cannot be determined easily. As a result, in the following this series will be truncated and $\mathcal{P}$ will be estimated in various approximations as

$$\mathcal{P}^{(i)} = p^{(1)-(i)} \tag{6.21}$$

where $p^{(1)-(i)} = p_{\mathrm{X}}^{(1)-(i)} = p_{\mathrm{X}}^{(1)-(i)}(\mathrm{Y})$ because of Eq. (6.18) and $p^{(1)-(1)} = p^{(1)}$.

For evaluating $p^{(1)-(i)}$ all the branches of the collision tree shown in Fig. 6.13, except for the lower most branch, must be summed up till $i^{\text{th}}$ iteration. This summation can be simplified by evaluating the probability that a particle does not collide in any iteration from the first till the $i^{\text{th}}$, which involves only the lower most branch of the collision tree. This probability will be denoted as $q^{(1)-(i)}$ and is related to $p^{(1)-(i)}$ through

$$p^{(1)-(i)} = 1 - q^{(1)-(i)} \tag{6.22}$$

where the subscript denoting lattice site has been dropped from $q^{(1)-(i)}$ following the same notation as that of $p^{(1)-(i)}$, *i.e.*, $q^{(1)-(i)} = q_{\mathrm{X}}^{(1)-(i)}$, and $q^{(1)-(1)} = q^{(1)}$.

In the following, various probabilities defined above will be evaluated exactly as well as approximately for the SPLG-1 mode. Specifically, the exact form of $q^{(1)}$, $p^{(1)}$, $\mathcal{P}^{(1)}$, $q^{(1)-(2)}$, $p^{(1)-(2)}$, and $\mathcal{P}^{(2)}$ will be evaluated and used for arriving at the approximate forms of some other terms. Approximate forms of $q^{(1)-(2)}$, $p^{(1)-(2)}$, and $\mathcal{P}^{(2)}$ will also be evaluated and compared with the corresponding exact forms for estimating the accuracy of the approximation. The approximate forms of $q^{(1)-(i)}$, $p^{(1)-(i)}$, and $\mathcal{P}^{(i)}$ will be denoted by adding the subscript "app" as $q_{\text{app}}^{(1)-(i)}$, $p_{\text{app}}^{(1)-(i)}$, and $\mathcal{P}_{\text{app}}^{(i)}$, respectively.

**Exact Evaluation of $q^{(1)}$, $p^{(1)}$, and $\mathcal{P}^{(1)}$:** In the SPLG-1 model Eqs. (6.19) and (6.22) gives $p^{(1)} = p_{\mathrm{J}}^{(1)}(\rightarrow)$ and thus $q_{\mathrm{X}}^{(1)} = q_{\mathrm{J}}^{(1)}(\rightarrow)$ for X = J. The corresponding configuration is shown in Fig. 6.12. While processing interparticle interactions, in the first iteration, the particle J shown in this figure can collide only with particles positioned at lattice site G,

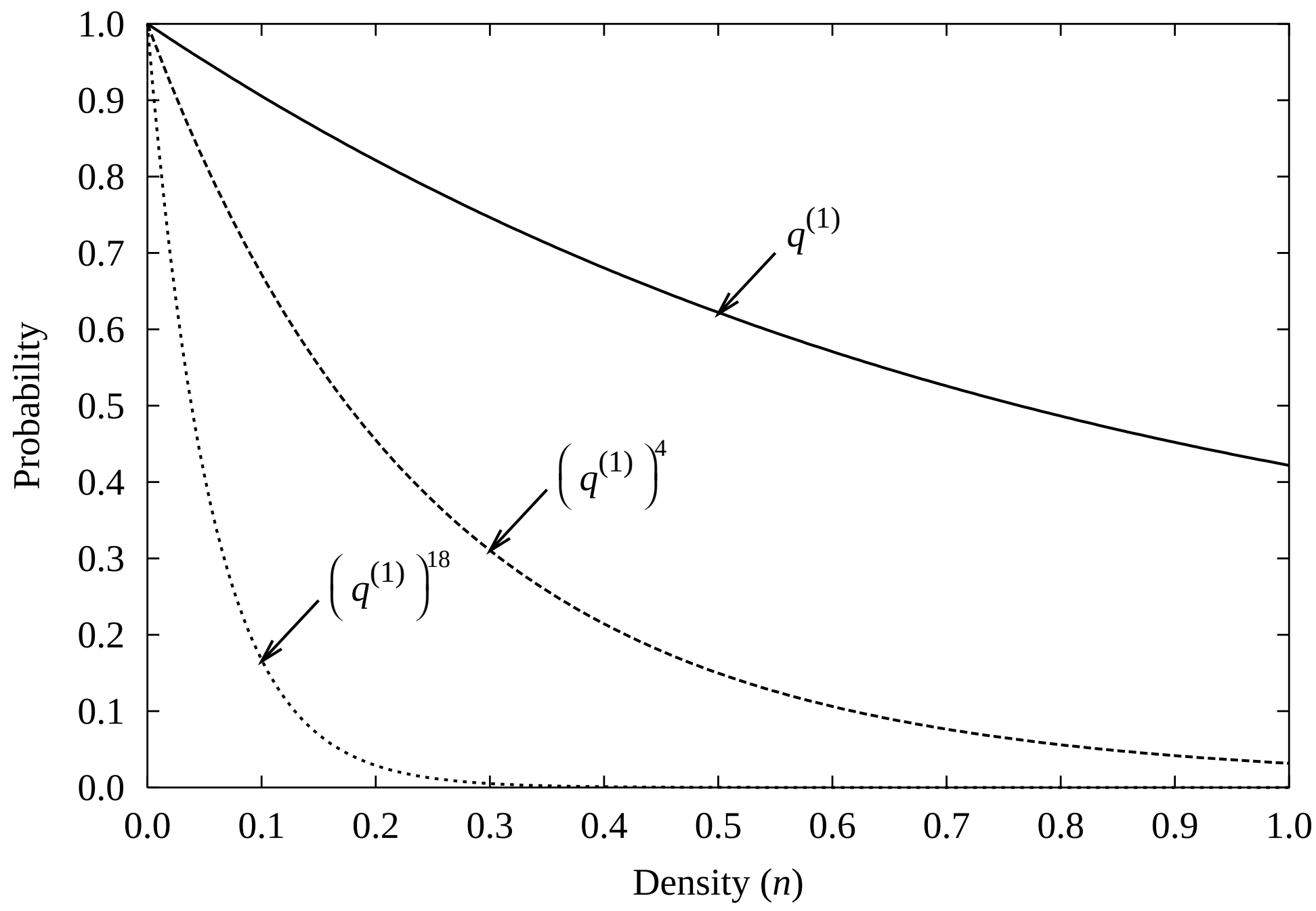

**Figure 6.14:** Variation of $q^{(1)}$ given by Eq. (6.23) and its 4$^{\text{th}}$ and 18$^{\text{th}}$ powers.

K, L, and Q. With this, the probability that a particle will not collide in the first iteration $q^{(1)}$ is

$$\begin{aligned} q^{(1)} &= q_{\mathrm{J}}^{(1)}(\rightarrow) \\ &= P_{\mathrm{K}}(\rightarrow)Q_{\mathrm{G}}(\downarrow)Q_{\mathrm{Q}}(\uparrow) + \\ &\quad P_{\mathrm{K}}(\uparrow)Q_{\mathrm{L}}(\leftarrow)Q_{\mathrm{Q}}(\uparrow) + \\ &\quad P_{\mathrm{K}}(\downarrow)Q_{\mathrm{L}}(\leftarrow)Q_{\mathrm{G}}(\downarrow) + \\ &\quad P_{\mathrm{K}}(\circ)Q_{\mathrm{G}}(\downarrow)Q_{\mathrm{L}}(\leftarrow)Q_{\mathrm{Q}}(\uparrow) \\ &= 3(n/4)(1-n/4)^2 + (1-n)(1-n/4)^3 \\ &= 1 - 4(n/4) + 9(n/4)^2 - 10(n/4)^3 + 4(n/4)^4 \end{aligned} \tag{6.23}$$

This function and its 4$^{\text{th}}$ and 18$^{\text{th}}$ powers (needed later) are plotted in Fig. 6.14.

Eqs. (6.21)–(6.23), when combined, give collision probability in the first iteration as

$$\mathcal{P}^{(1)} = p^{(1)} = 1 - q^{(1)} = 4(n/4) - 9(n/4)^2 + 10(n/4)^3 - 4(n/4)^4 \tag{6.24}$$

This expression for the probability of collision in the first iteration can be verified only by recomputing $p^{(1)}$ directly using the first principles. For this, again, consider particle J with its state and the domain shown in Fig. 6.12. Various steps and the result of this recomputation are as follows:

$$\begin{aligned} p^{(1)} &= p_{\mathrm{J}}^{(1)}(\rightarrow) \\ &= P_{\mathrm{K}}(\leftarrow) + \\ &\quad P_{\mathrm{K}}(\rightarrow)\,[\,P_{\mathrm{G}}(\downarrow)Q_{\mathrm{Q}}(\uparrow) + Q_{\mathrm{G}}(\downarrow)P_{\mathrm{Q}}(\uparrow) + P_{\mathrm{G}}(\downarrow)P_{\mathrm{Q}}(\uparrow)\,] + \\ &\quad P_{\mathrm{K}}(\uparrow)\,[\,P_{\mathrm{L}}(\leftarrow)Q_{\mathrm{Q}}(\uparrow) + Q_{\mathrm{L}}(\leftarrow)P_{\mathrm{Q}}(\uparrow) + P_{\mathrm{L}}(\leftarrow)P_{\mathrm{Q}}(\uparrow)\,] + \end{aligned}$$

$$
\begin{aligned}
& P_{\mathrm{K}}(\downarrow)\,[\,P_{\mathrm{L}}(\leftarrow)Q_{\mathrm{G}}(\downarrow) + Q_{\mathrm{L}}(\leftarrow)P_{\mathrm{G}}(\downarrow) + P_{\mathrm{L}}(\leftarrow)P_{\mathrm{G}}(\downarrow)\,] + \\
& P_{\mathrm{K}}(\circ)\,[\,P_{\mathrm{L}}(\leftarrow)Q_{\mathrm{G}}(\downarrow)Q_{\mathrm{Q}}(\uparrow) + Q_{\mathrm{L}}(\leftarrow)P_{\mathrm{G}}(\downarrow)Q_{\mathrm{Q}}(\uparrow) + \\
& \qquad Q_{\mathrm{L}}(\leftarrow)Q_{\mathrm{G}}(\downarrow)P_{\mathrm{Q}}(\uparrow) + P_{\mathrm{L}}(\leftarrow)P_{\mathrm{G}}(\downarrow)Q_{\mathrm{Q}}(\uparrow) + \\
& \qquad P_{\mathrm{L}}(\leftarrow)Q_{\mathrm{G}}(\downarrow)P_{\mathrm{Q}}(\uparrow) + Q_{\mathrm{L}}(\leftarrow)P_{\mathrm{G}}(\downarrow)P_{\mathrm{Q}}(\uparrow) + \\
& \qquad P_{\mathrm{L}}(\leftarrow)P_{\mathrm{G}}(\downarrow)P_{\mathrm{Q}}(\uparrow)\,] \\
= \; & (n/4) + 3(n/4)\left[2(n/4)(1-n/4) + (n/4)^2\right] + \\
& (1-n)\left[3(n/4)(1-n/4)^2 + 3(n/4)^2(1-n/4) + (n/4)^3\right] \\
= \; & 4(n/4) - 9(n/4)^2 + 10(n/4)^3 - 4(n/4)^4 \qquad (6.25)
\end{aligned}
$$

which is identical to that given by Eq. (6.24). This recomputation of $p^{(1)}$ not only verifies the expression for $p^{(1)}$ but also illustrates that direct computation of $p^{(1)}$ is more involved and lengthy process compared to computation of $q^{(1)}$ and then computation of $p^{(1)}$ as $1 - q^{(1)}$. In fact, in general, the complexity of computation of $p^{(i)}$ increases more rapidly with $i$ compared to the complexity of computation of $q^{(i)}$. As a result, in the following, $p^{(i)}$ will be computed only as $1 - q^{(i)}$ and not directly.

**Exact Evaluation of $q^{(1)-(2)}$, $p^{(1)-(2)}$, and $\mathcal{P}^{(2)}$:** For estimating the probability of collision of a particle in the second iteration, the effect of collision of NINs of the particle (among themselves or with NNINs of the particle) also have to be considered. These considerations for particle J shown in Fig. 6.12 give the expression for $q^{(1)-(2)}$ as

$$
q^{(1)-(2)} = q_{\mathrm{J}}^{(1)-(2)} = q_{\mathrm{J}}^{(1)}\left[q_{\mathrm{NIN,J}}^{(1)} + p_{\mathrm{NIN,J}}^{(1)} q_{\mathrm{J}}^{(2)}\right] \qquad (6.26)
$$

The term $q_{\mathrm{J}}^{(1)}$ in this equation appears because J does not collide in the first iteration. In the second iteration J will not collide if either the NINs of J do not collide in the first iteration or, if they do, then—subject to this condition—J does not collide in the second iteration. If the NINs of J do not collide in the first iteration (represented by the term $q_{\mathrm{NIN,J}}^{(1)}$), then their states and the state of J remain unchanged in the second iteration and J automatically does not collide in the second iteration. If the NINs of J collide in the first iteration (represented by the term $p_{\mathrm{NIN,J}}^{(1)}$), then J may or may not collide with its NINs in the second iteration and hence the probability that J does not collide with its NINs in the second iteration must be incorporated explicitly (represented by the term $q_{\mathrm{J}}^{(2)}$).

In Eq. (6.26), the probability that J does collide in the second iteration is given by the expression in brackets "$[\cdots]$" and not by the term $q_{\mathrm{J}}^{(2)}$ alone. The term $q_{\mathrm{J}}^{(2)}$ only accounts for the probability of J not colliding in the second iteration provided that its NINs have collided in the first iteration. Furthermore, in this equation the terms $q_{\mathrm{NIN,J}}^{(1)}$, $p_{\mathrm{NIN,J}}^{(1)}$, and $q_{\mathrm{J}}^{(2)}$ must be evaluated subject to the condition that particle J has not collided in the first iteration. This is because this condition or the event of J not colliding in the first iteration, gives information about the possible states of NINs and thus alters their collision probabilities. Furthermore, $q_{\mathrm{J}}^{(2)}$ must be evaluated under an additional condition that the NINs of J have collided (with particle other than J) in the first iteration.

In Eq. (6.26) $q_{\mathrm{J}}^{(1)}$ is as given by Eq. (6.23) and

$$
\begin{aligned}
q_{\mathrm{NIN,J}}^{(1)} &= q_{\mathrm{G}}^{(1)} q_{\mathrm{K}}^{(1)} q_{\mathrm{Q}}^{(1)} q_{\mathrm{L}}^{(1)} \qquad (6.27) \\
p_{\mathrm{NIN,J}}^{(1)} &= 1 - q_{\mathrm{NIN,J}}^{(1)} \qquad (6.28)
\end{aligned}
$$

where $q_{\mathrm{X}}^{(1)}$, $\mathrm{X} \in \{\mathrm{G}, \mathrm{K}, \mathrm{Q}, \mathrm{L}\}$, is given by

$$q_{\mathrm{X}}^{(1)} = q_{\mathrm{X}}^{(1)}(\circ) + \frac{1}{4}\left[q_{\mathrm{X}}^{(1)}(\rightarrow) + q_{\mathrm{X}}^{(1)}(\uparrow) + q_{\mathrm{X}}^{(1)}(\downarrow) + q_{\mathrm{X}}^{(1)}(\leftarrow)\right] \tag{6.29}$$

because the states of lattice sites other than J are not known. The factor of 1/4 appears because a particle positioned on a lattice site can only be in one state at any iteration in a time step. As a result, the probability that a particle occupying a lattice site will not collide at any iteration in a time step is the weighted mean of the particle being in various possible states and not colliding. In the first iteration, all weights are equal because the system is spatially uniform and at equilibrium.

For evaluating the above probabilities note that

$$q_{\mathrm{X}}^{(i)}(\circ) = 1 - n \qquad \forall\, \mathrm{X} \in \{\mathrm{G}, \mathrm{K}, \mathrm{Q}, \mathrm{L}\} \tag{6.30}$$

for all $i$. This is because the occurrence or non-occurrence of interactions among particles does not change the probability that a lattice site is empty since the particle no not move during processing of interparticle interactions.

Also note that

$$\begin{aligned} q_{\mathrm{K}}^{(1)}(\rightarrow) = q_{\mathrm{K}}^{(1)}(\uparrow) = q_{\mathrm{K}}^{(1)}(\downarrow) = q_{\mathrm{L}}^{(1)}(\rightarrow) = q_{\mathrm{L}}^{(1)}(\uparrow) = q_{\mathrm{L}}^{(1)}(\downarrow) = \\ q_{\mathrm{G}}^{(1)}(\rightarrow) = q_{\mathrm{G}}^{(1)}(\uparrow) = q_{\mathrm{Q}}^{(1)}(\rightarrow) = q_{\mathrm{Q}}^{(1)}(\downarrow) \end{aligned} \tag{6.31}$$

This is because the non-occurrence of collision of particle J with its neighbors G, K, Q, and L does not give any information about the states of particle in the reduced contact interaction neighborhood of these particles if their states are the once out in Eq. (6.31).

Symmetry of the system shown in Fig. 6.12 gives

$$q_{G}^{(1)}(\leftarrow) \;=\; q_{Q}^{(1)}(\leftarrow) \tag{6.32}$$

$$q_{G}^{(1)}(\downarrow) \;=\; q_{Q}^{(1)}(\uparrow) \tag{6.33}$$

Eqs. (6.31)–(6.33), in view of the geometry shown in Fig. 6.12, imply that $q_{\mathrm{NIN,J}}^{(1)}$ can be determined by evaluating $q_{\mathrm{G}}^{(1)}(\circ)$, $q_{\mathrm{K}}^{(1)}(\uparrow)$, $q_{G}^{(1)}(\leftarrow)$, $q_{K}^{(1)}(\leftarrow)$, $q_{G}^{(1)}(\downarrow)$, and $q_{L}^{(1)}(\leftarrow)$. $q_{\mathrm{G}}^{(1)}(\circ)$ has already been evaluated above and is as given by Eq. (6.30). The others are as follows:

$$\begin{aligned} q_{\mathrm{K}}^{(1)}(\uparrow) \;&=\; P_{\mathrm{K}}(\uparrow)\,[\,P_{\mathrm{G}}(\circ)Q_{\mathrm{C}}(\downarrow)Q_{\mathrm{F}}(\rightarrow)Q_{\mathrm{H}}(\leftarrow) + \\ &\qquad P_{\mathrm{G}}(\uparrow)Q_{\mathrm{F}}(\rightarrow)Q_{\mathrm{H}}(\leftarrow) + \\ &\qquad P_{\mathrm{G}}(\rightarrow)Q_{\mathrm{C}}(\downarrow)Q_{\mathrm{F}}(\rightarrow) + \\ &\qquad P_{\mathrm{G}}(\leftarrow)Q_{\mathrm{C}}(\downarrow)Q_{\mathrm{H}}(\leftarrow)\,] \\ &=\; (n/4)\left[(1-n)(1-n/4)^3 + 3(n/4)(1-n/4)^2\right] \end{aligned} \tag{6.34}$$

$$\begin{aligned} q_{\mathrm{G}}^{(1)}(\leftarrow) \;&=\; P_{\mathrm{G}}(\leftarrow)\,[\,P_{\mathrm{F}}(\circ)Q_{\mathrm{E}}(\rightarrow)Q_{\mathrm{B}}(\downarrow) + \\ &\qquad P_{\mathrm{F}}(\downarrow)Q_{\mathrm{E}}(\rightarrow)Q_{\mathrm{B}}(\downarrow) + \\ &\qquad P_{\mathrm{F}}(\uparrow)Q_{\mathrm{E}}(\rightarrow) + \\ &\qquad P_{\mathrm{F}}(\leftarrow)Q_{\mathrm{B}}(\downarrow)\,] \\ &=\; (n/4)\left[(1-n)(1-n/4)^2 + (n/4)(1-n/4)^2 + 2(n/4)(1-n/4)\right] \end{aligned} \tag{6.35}$$

$$q_{\mathrm{K}}^{(1)}(\leftarrow) \;=\; 0 \tag{6.36}$$

$$q_{\mathrm{G}}^{(1)}(\downarrow) \;=\; 0 \tag{6.37}$$

$$q_{\mathrm{L}}^{(1)}(\leftarrow) \;=\; 0 \tag{6.38}$$

The probabilities given by Eqs. (6.36)–(6.38) are noteworthy since they are all zero and require further explanation. For $q_{\rm G}^{(1)}(\downarrow) = 0$, the explanation is that if the particle G is in the state ↓ then, in order that it does not collide with J (since it is known that J does not collide with any of G, K, Q, and L), it must necessarily collide with K which should necessarily be in the state ↑. Identical explanations hold for $q_{\rm L}^{(1)}(\leftarrow) = 0$. For $q_{\rm K}^{(1)}(\leftarrow) = 0$, the explanation is that particle K cannot be in the state ← because in this state it will necessarily collide with J, which should not happen since J does not collide in first iteration.

Eqs. (6.34)–(6.38), when combined with Eqs. (6.29)–(6.33), give the expression for $q_{\rm X}^{(1)}$, $\rm X \in \{G, K, Q, L\}$, as

$$q_{\rm G}^{(1)} = \frac{1}{4}\left[4 - 13(n/4) - 11(n/4)^2 + 23(n/4)^3 - 23(n/4)^4 + 8(n/4)^5\right] \quad (6.39)$$

$$q_{\rm K}^{(1)} = \frac{1}{4}\left[4 - 13(n/4) - 12(n/4)^2 + 36(n/4)^3 - 30(n/4)^4 + 12(n/4)^5\right] \quad (6.40)$$

$$q_{\rm Q}^{(1)} = \frac{1}{4}\left[4 - 13(n/4) - 11(n/4)^2 + 23(n/4)^3 - 23(n/4)^4 + 8(n/4)^5\right] \quad (6.41)$$

$$q_{\rm L}^{(1)} = \frac{1}{4}\left[4 - 13(n/4) - 12(n/4)^2 + 36(n/4)^3 - 30(n/4)^4 + 12(n/4)^5\right] \quad (6.42)$$

The expressions for $q_{\rm NIN,J}^{(1)}$ and $q_{\rm NIN,J}^{(1)}$ can be easily obtained by substituting the expressions for $q_{\rm X}^{(1)}$, $\rm X \in \{G, K, Q, L\}$, given above in Eqs. (6.27) and (6.28). The results are

$$q_{\rm NIN,J}^{(1)} = \frac{1}{4^4}\left[4 - 13(n/4) - 12(n/4)^2 + 36(n/4)^3 - 30(n/4)^4 + 12(n/4)^5\right]^2 \left[4 - 13(n/4) - 11(n/4)^2 + 23(n/4)^3 - 23(n/4)^4 + 8(n/4)^5\right]^2 \quad (6.43)$$

$$p_{\rm NIN,J}^{(1)} = 1 - \frac{1}{4^4}\left[4 - 13(n/4) - 12(n/4)^2 + 36(n/4)^3 - 30(n/4)^4 + 12(n/4)^5\right]^2 \left[4 - 13(n/4) - 11(n/4)^2 + 23(n/4)^3 - 23(n/4)^4 + 8(n/4)^5\right]^2 \quad (6.44)$$

The exact evaluation of $q_{\rm J}^{(2)}$ requires estimation of probabilities for various possible states of NINs of J. This estimation is required because the probabilities change since particle J does not collide in the first iteration and NINs of J collide among themselves and with NNINs of J in the first iteration. This estimation requires detailed considerations wherein the STPs need to be accounted for. It is, thus, an extremely involved task and will not be undertaken. Instead, it will be assumed that the probabilities for NINs of J being in various possible states in the second iteration do not differ significantly from those in the first iteration.[3)] With this assumption, $q_{\rm J}^{(2)}$ can be approximated as

$$q_{\rm J}^{(2)} \approx q_{\rm J}^{(1)} \quad (6.45)$$

which is as given by Eq. (6.23).

Substituting from Eqs. (6.23), (6.43), (6.44), and (6.45) into Eq. (6.26) gives the expression for $q^{(1)-(2)}$ as

$$q^{(1)-(2)} = \left[1 - 4(n/4) + 9(n/4)^2 - 10(n/4)^3 + 4(n/4)^4\right]$$

3) In fact, from this point onwards, the assumption is that the probability of a lattice site being in various possible states in any iteration is not vary different from that in the first iteration.

$$\left\{ \frac{1}{4^4} \left(4 - 13(n/4) - 12(n/4)^2 + 36(n/4)^3 - 30(n/4)^4 + 12(n/4)^5\right)^2 \right.$$
$$\left(4 - 13(n/4) - 11(n/4)^2 + 23(n/4)^3 - 23(n/4)^4 + 8(n/4)^5\right)^2 +$$
$$\left[ 1 - \frac{1}{4^4} \left(4 - 13(n/4) - 12(n/4)^2 + 36(n/4)^3 - 30(n/4)^4 + 12(n/4)^5\right)^2 \right.$$
$$\left. \left(4 - 13(n/4) - 11(n/4)^2 + 23(n/4)^3 - 23(n/4)^4 + 8(n/4)^5\right)^2 \right]$$
$$\left. \left[1 - 4(n/4) + 9(n/4)^2 - 10(n/4)^3 + 4(n/4)^4\right] \right\} \tag{6.46}$$

It must be noted that this evaluation of $q^{(1)-(2)}$ is exact except for the approximation expressed through Eq. (6.45). This expression can be substituted into Eqs. (6.21) and (6.22) to obtain the expressions for $p^{(1)-(2)}$ and $\mathcal{P}^{(2)}$.

**Approximate Evaluation of $q^{(1)-(2)}$, $p^{(1)-(2)}$, and $\mathcal{P}^{(2)}$:** As mentioned earlier, the approximate forms of various probabilities are denoted by adding the subscript "app" to the notations for the exact forms. Thus, in the following, the approximate forms of $q^{(1)-(2)}$, $p^{(1)-(2)}$, and $\mathcal{P}^{(2)}$ are denoted by $q_{\text{app}}^{(1)-(2)}$, $p_{\text{app}}^{(1)-(2)}$, and $\mathcal{P}_{\text{app}}^{(2)}$, respectively.

In the exact evaluation of $q^{(1)-(2)}$ carried out earlier, all terms of Eq. (6.26) were evaluated exactly except for the term $q_{\text{J}}^{(2)}$ which was approximated by $q_{\text{J}}^{(1)}$ for simplifying the analysis. Despite this approximation the overall analysis for obtaining $q^{(1)-(2)}$ turned out to be extremely involved and lengthy. This analysis, if extended for evaluating $q^{(1)-(i)}$, rapidly assumes overwhelming proportions for all $i \geq 3$. As a result, it is desirable to find out simple and good approximations which can be used while evaluating $q^{(1)-(i)}$ for $i \geq 3$. These approximations relate to evaluation of $q_{\text{N}^k\text{IN,J}}^{(i)}$. The assumption involved in these approximations is that the occurrence or non-occurrence of collisions among particles in an iteration does not alter the probability of occurrence or non-occurrence of collision of other particles in the next iteration significantly. Thus, in these approximations the information about states of particles provided by previous iterations is ignored (or not used) for computing the probabilities in the next iteration. In the context of evaluation of $q_{\text{NIN,J}}^{(1)}$, these approximations imply replacement of $q_{\text{X}}^{(1)}$, for all $\text{X} \in \{\text{G}, \text{K}, \text{Q}, \text{L}\}$, by $q_{\text{J}}^{(1)}$ or, equivalently, by $q^{(1)}$. This gives $q_{\text{NIN,J,app}}^{(1)}$ and $p_{\text{NIN,J,app}}^{(1)}$—the approximate forms of $q_{\text{NIN,J}}^{(1)}$ and $p_{\text{NIN,J}}^{(1)}$—as

$$q_{\text{NIN,J,app}}^{(1)} = \left(q^{(1)}\right)^4 \tag{6.47}$$
$$p_{\text{NIN,J,app}}^{(1)} = 1 - \left(q^{(1)}\right)^4 \tag{6.48}$$

which correspond to Eqs. (6.27) and (6.28), respectively. Similarly, $q^{(1)-(2)}$ in Eq. (6.26) becomes $q_{\text{app}}^{(1)-(2)}$ and the equation takes the approximate form

$$q_{\text{app}}^{(1)-(2)} = q^{(1)} \left\{ \left(q^{(1)}\right)^4 + \left[1 - \left(q^{(1)}\right)^4\right] q^{(1)} \right\} \tag{6.49}$$

which on further reduction becomes

$$q_{\text{app}}^{(1)-(2)} = q^{(1)} \left[q^{(1)} + p^{(1)} \left(q^{(1)}\right)^4\right] \tag{6.50}$$

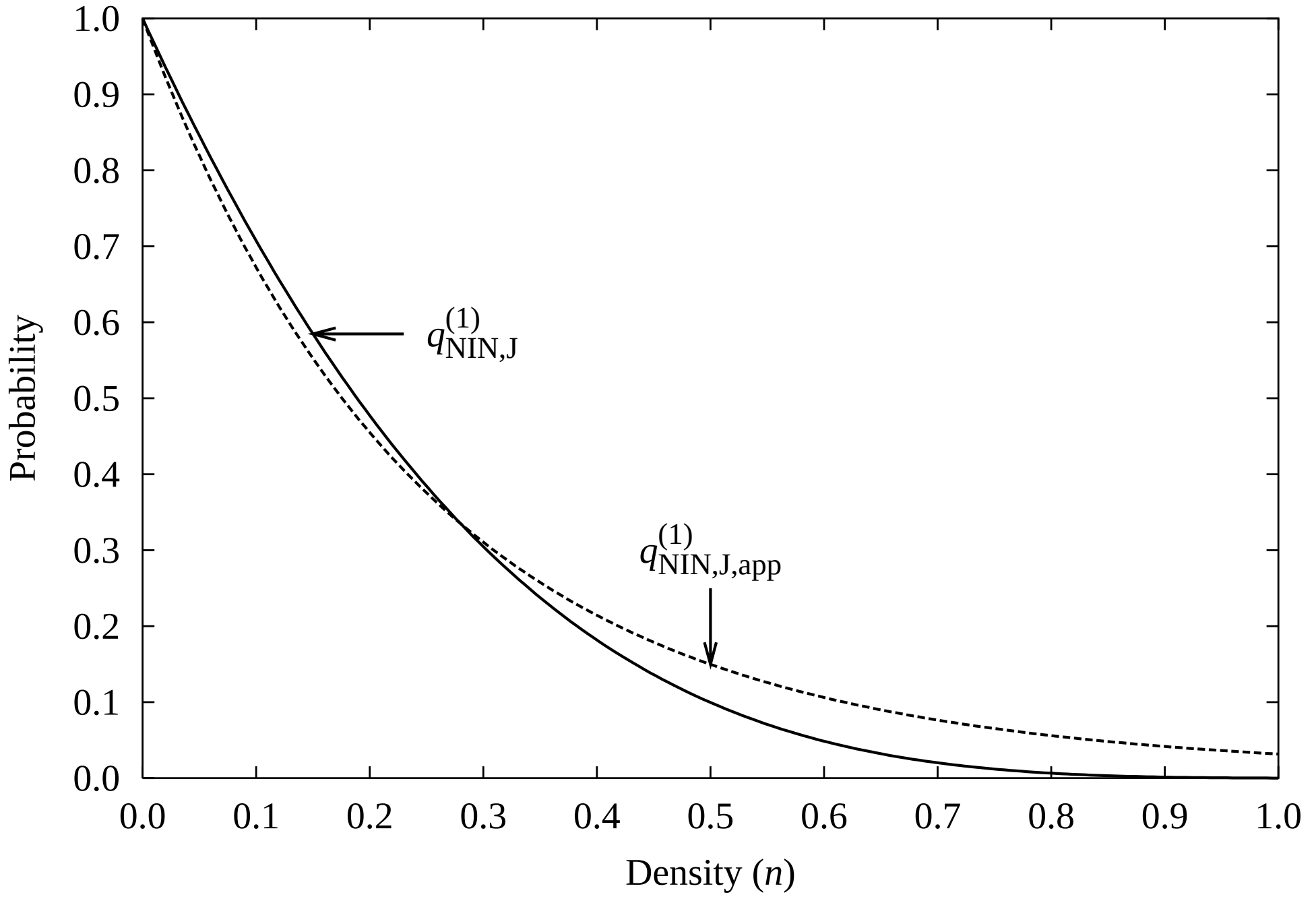

**Figure 6.15:** Comparison of the exact value of $q^{(1)}_{\mathrm{NIN,J}}$ given by Eq. (6.43) and the corresponding approximate value $q^{(1)}_{\mathrm{NIN,J,app}}$ given by Eq. (6.51).

Substituting for $q^{(1)}$ and $p^{(1)}$ from Eqs. (6.23) and (6.24) in Eqs. (6.47) and (6.50) gives

$$q^{(1)}_{\mathrm{NIN,J,app}} = \Big[1 - 4(n/4) + 9(n/4)^2 - 10(n/4)^3 + 4(n/4)^4\Big]^4 \tag{6.51}$$

$$\begin{aligned} q^{(1)-(2)}_{\mathrm{app}} = {} & \Big[1 - 4(n/4) + 9(n/4)^2 - 10(n/4)^3 + 4(n/4)^4\Big] \\ & \Big\{ \Big[1 - 4(n/4) + 9(n/4)^2 - 10(n/4)^3 + 4(n/4)^4\Big] + \\ & \Big[4(n/4) - 9(n/4)^2 + 10(n/4)^3 - 4(n/4)^4\Big] \\ & \Big[1 - 4(n/4) + 9(n/4)^2 - 10(n/4)^3 + 4(n/4)^4\Big]^4 \Big\} \end{aligned} \tag{6.52}$$

Graphical comparison of the exact and approximate forms of $q^{(1)}_{\mathrm{NIN,J}}$ given by Eqs. (6.43) and (6.51), respectively, is shown in Fig. 6.15. The departure of $q^{(1)}_{\mathrm{NIN,J,app}}$ from $q^{(1)}_{\mathrm{NIN,J}}$ is shown in Fig. 6.16. The figures show that $q^{(1)}_{\mathrm{NIN,J,app}}$ departs considerably from $q^{(1)}_{\mathrm{NIN,J}}$, especially at high densities, with the maximum departure of $\approx 0.0579211$ at $n \approx 0.631$.

Graphical comparison of the exact and approximate forms of $q^{(1)-(2)}$ given by Eqs. (6.46) and Eq. (6.52), respectively, is shown in Fig. 6.17. The departure of $q^{(1)-(2)}_{\mathrm{app}}$ from $q^{(1)-(2)}$ is shown in Fig. 6.18. The functional forms for $q^{(1)-(2)}_{\mathrm{app}}$ and $q^{(1)-(2)}$ shown in Fig. 6.17 are almost indistinguishable from each other till $n \approx 0.4$, following which they start departing slowly. The departure, however, is almost insignificant and has a maximum of $\approx 0.0143181$ at $n \approx 0.645$ as seen in Fig. 6.18. It is noteworthy that although $q^{(1)}_{\mathrm{NIN,J,app}}$ departs considerably from $q^{(1)}_{\mathrm{NIN,J}}$, especially at high densities, the effect of this departure is not much evident in the corresponding probabilities $q^{(1)-(2)}_{\mathrm{app}}$ and $q^{(1)-(2)}$. It should also be noted that the approximate form gives lower value of collision probability (*i.e.*, higher value of probability of non-occurrence of interactions as seen in the figure) especially at high densities. This is consistent with the nature of approximation wherein some spurious interactions—the interactions whose probability of non-occurrence is zero or otherwise smaller compared to $q^{(1)}$—also get accounted for, especially at higher densities.

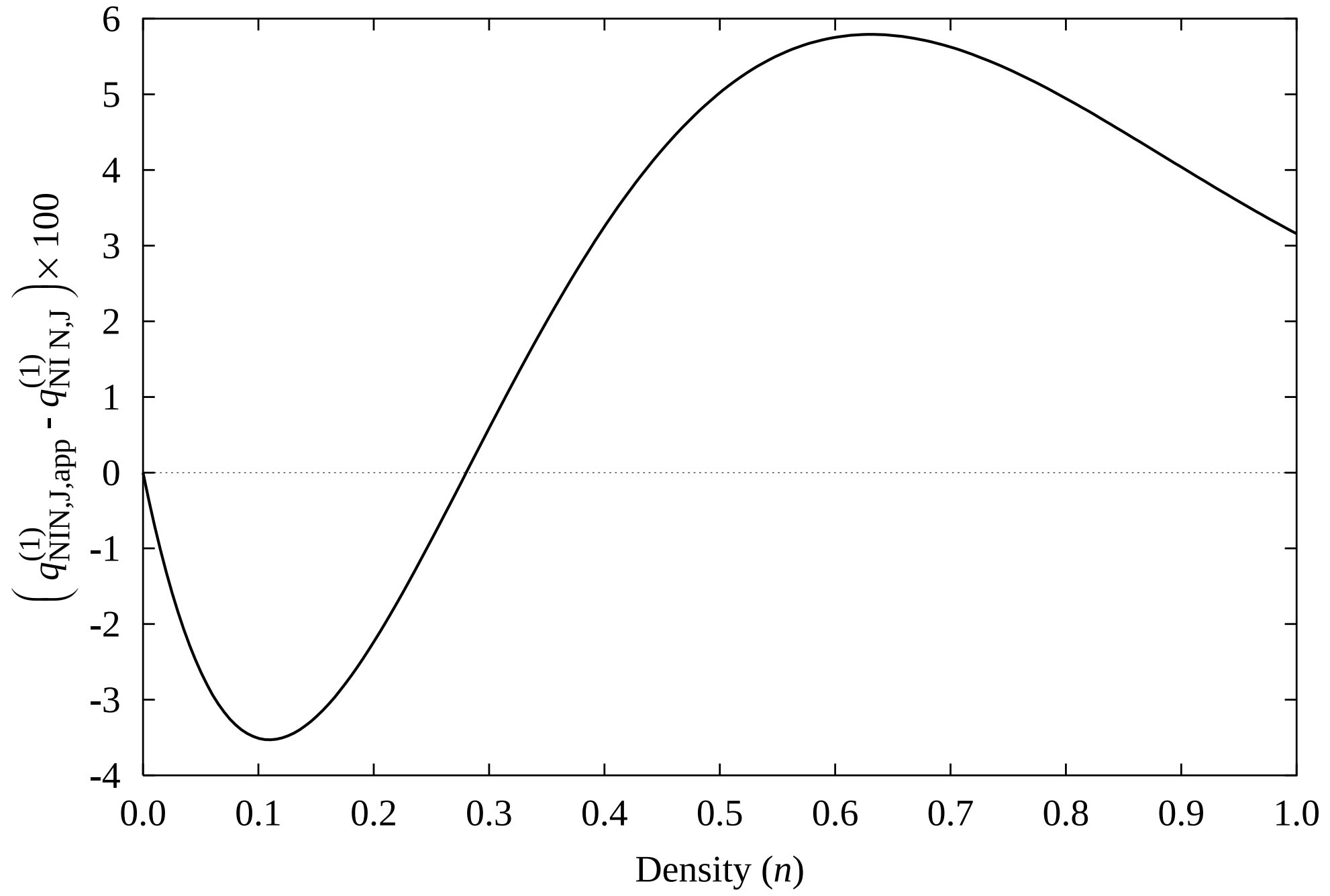

**Figure 6.16:** Departure of the approximate value $q^{(1)}_{\mathrm{NIN,J,app}}$ given by Eq. (6.51) from the exact value $q^{(1)}_{\mathrm{NIN,J}}$ given by Eq. (6.43).

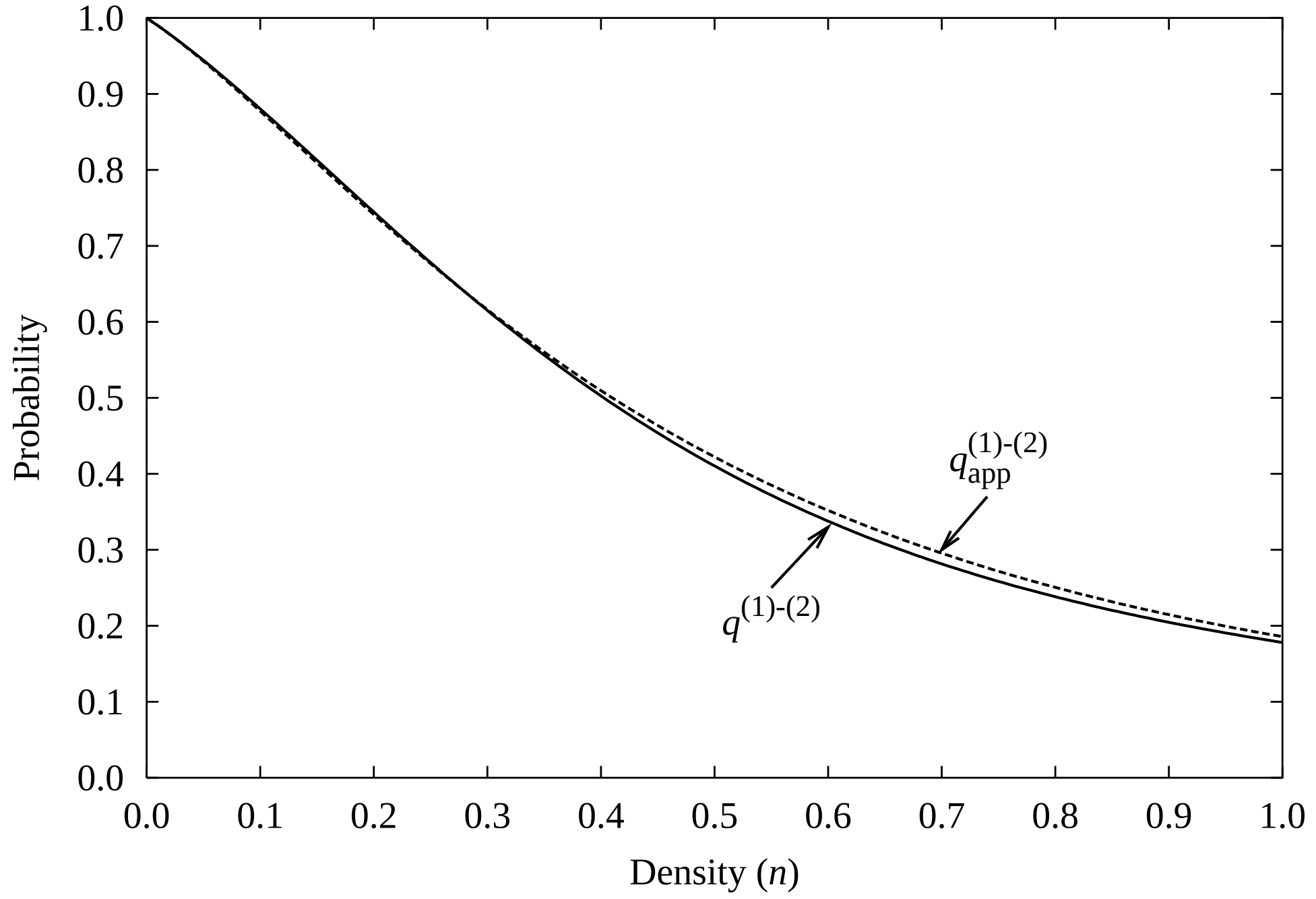

**Figure 6.17:** Comparison of the exact value of $q^{(1)-(2)}$ given by Eq. (6.46) and the corresponding approximate value $q^{(1)-(2)}_{\mathrm{app}}$ given by Eq. (6.52).

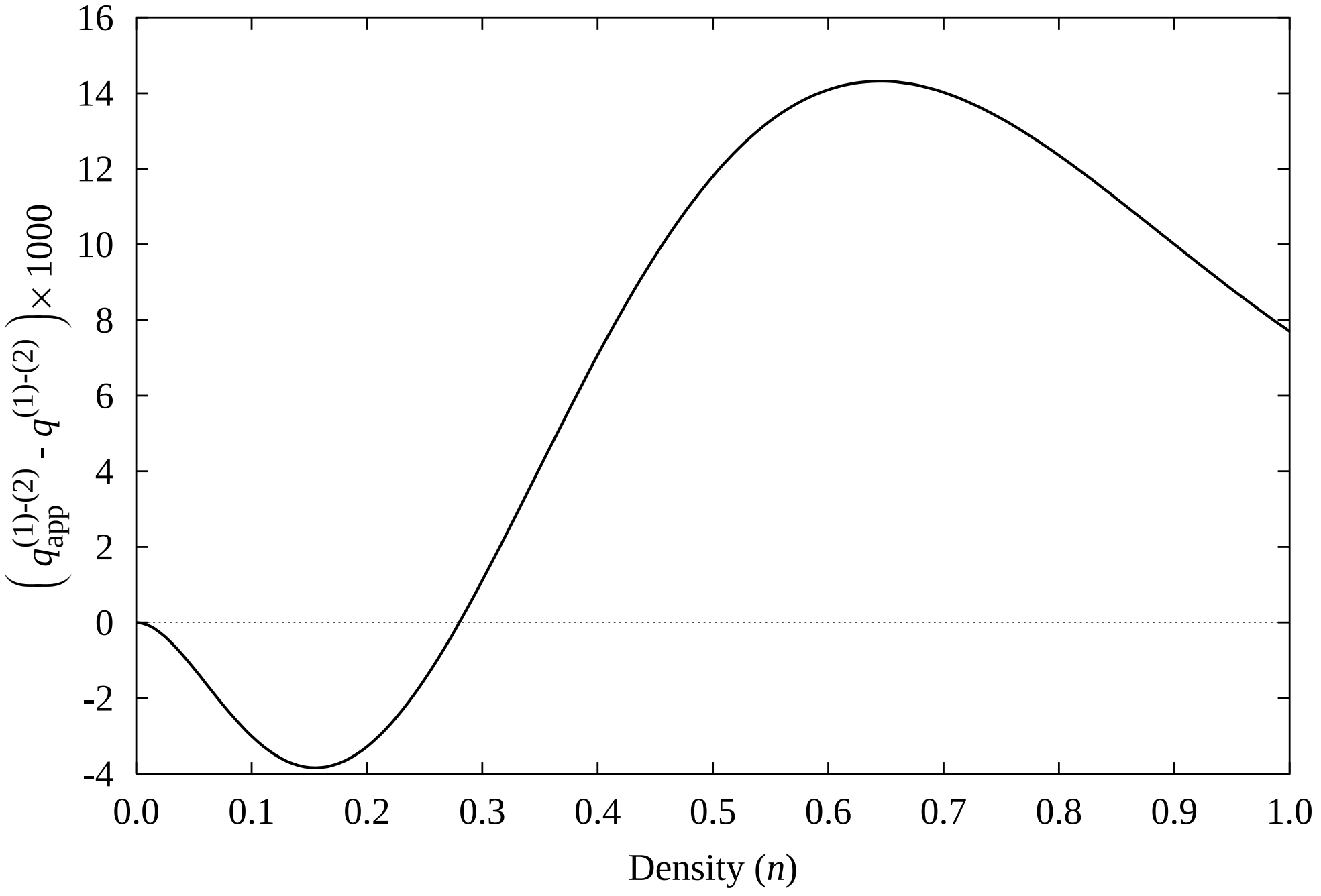

**Figure 6.18:** Departure of the approximate value $q_{\text{app}}^{(1)-(2)}$ given by Eq. (6.52) from the exact value $q^{(1)-(2)}$ given by Eq. (6.46).

The results of the comparison carried out above indicate that the simplifying assumptions made for obtaining the approximate form $q^{(1)}_{\text{NIN,J,app}}$ of $q^{(1)}_{\text{NIN,J}}$ give good approximation of $q^{(1)-(2)}$. These simplifying assumptions can, in general, also be employed for obtaining approximate forms for $q^{(i)}_{\text{N}^k\text{IN,J}}$ and $q^{(1)-(i+1)}$. It must, however, be noted that as the approximation becomes crude as $k$ and $i$ increase. Thus, as $i$ increases, the approximate form of $q^{(1)-(i)}$ is not expected to be good representative of the sum of the infinite series given by Eq. (6.20). In fact, it is expected that as $i$ increases, $q^{(1)-(i)}$ will initially approach towards the exact sum of the series and then start diverging rapidly.

**Approximate Evaluation of $q^{(1)-(3)}$, $p^{(1)-(3)}$, and $\mathcal{P}^{(3)}$:** The exact expression for $q^{(1)-(3)}$, similar to that for $q^{(1)-(2)}$ given by Eq. (6.26), can be obtained along the same lines as done for $q^{(1)-(2)}$. This expression involved consideration of interactions of NNINs of J and is

$$\begin{aligned} q^{(1)-(3)} \;=\; & q^{(1)}_{\text{J}} \Big\{ q^{(1)}_{\text{NIN,J}} \Big[ q^{(1)}_{\text{NNIN,J}} + p^{(1)}_{\text{NNIN,J}} \Big( q^{(2)}_{\text{NIN,J}} + p^{(2)}_{\text{NIN,J}} q^{(3)}_{\text{J}} \Big) \Big] + \\ & p^{(1)}_{\text{NIN,J}} q^{(2)}_{\text{J}} \Big[ q^{(2)}_{\text{NNIN,J}} + p^{(2)}_{\text{NNIN,J}} \Big( q^{(2)}_{\text{NIN,J}} + p^{(2)}_{\text{NIN,J}} q^{(3)}_{\text{J}} \Big) \Big] \Big\} \end{aligned} \qquad (6.53)$$

The exercise for exact evaluation of this expression is very difficult and involved because interaction probabilities have to be considered for NNINs and NINs of J at both first and second iterations and the number of NINs and NNINs is $\mathcal{N}_{\text{NIN,J}} = 4$ and $\mathcal{N}_{\text{NNIN,J}} = 18$, respectively. The evaluation of this expression can be simplified using the approximations

$$\left.\begin{aligned} q^{(2)}_{\text{NNIN,J}} &\approx q^{(1)}_{\text{NNIN,J}} \\ p^{(2)}_{\text{NNIN,J}} &\approx p^{(1)}_{\text{NNIN,J}} \\ q^{(2)}_{\text{NIN,J}} &\approx q^{(1)}_{\text{NIN,J}} \\ p^{(2)}_{\text{NIN,J}} &\approx p^{(1)}_{\text{NIN,J}} \\ q^{(3)}_{\text{J}} &\approx q^{(2)}_{\text{J}} \approx q^{(1)}_{\text{J}} \end{aligned}\right\} \quad (6.54)$$

which are similar to those employed earlier for approximate evaluation of $q^{(1)-(2)}$. With these approximations, Eq. (6.53) becomes

$$\begin{aligned} q^{(1)-(3)}_{\text{app}} &= q^{(1)}_{\text{J}} \Big\{ q^{(1)}_{\text{NIN,J}} \left[q^{(1)}_{\text{NNIN,J}} + p^{(1)}_{\text{NNIN,J}} \left(q^{(1)}_{\text{NIN,J}} + p^{(1)}_{\text{NIN,J}} q^{(1)}_{\text{J}}\right)\right] + \\ &\qquad p^{(1)}_{\text{NIN,J}} q^{(1)}_{\text{J}} \left[q^{(1)}_{\text{NNIN,J}} + p^{(1)}_{\text{NNIN,J}} \left(q^{(1)}_{\text{NIN,J}} + p^{(1)}_{\text{NIN,J}} q^{(1)}_{\text{J}}\right)\right] \Big\} \\ &= q^{(1)}_{\text{J}} \left[q^{(1)}_{\text{NIN,J}} + p^{(1)}_{\text{NIN,J}} q^{(1)}_{\text{J}}\right] \left\{q^{(1)}_{\text{NNIN,J}} + p^{(1)}_{\text{NNIN,J}} \left[q^{(1)}_{\text{NIN,J}} + p^{(1)}_{\text{NIN,J}} q^{(1)}_{\text{J}}\right]\right\} \end{aligned} \quad (6.55)$$

For evaluating this expression $q^{(1)}_{\text{NNIN,J}}$, and $p^{(1)}_{\text{NNIN,J}}$ can be approximated in terms of $q^{(1)}$ as

$$q^{(1)}_{\text{NNIN,J}} \approx q^{(1)}_{\text{NNIN,J,app}} = \left(q^{(1)}\right)^{\mathcal{N}_{\text{NNIN,J}}} = \left(q^{(1)}\right)^{18} \quad (6.56)$$

$$p^{(1)}_{\text{NNIN,J}} \approx p^{(1)}_{\text{NNIN,J,app}} = 1 - q^{(1)}_{\text{NNIN,J,app}} = 1 - \left(q^{(1)}\right)^{18} \quad (6.57)$$

These approximations are similar to those employed for approximating $q^{(1)}_{\text{NIN,J}}$ and $p^{(1)}_{\text{NIN,J}}$ as given in Eqs. (6.47) and (6.48), respectively.

For $q^{(1)}_{\text{NIN,J}}$ and $p^{(1)}_{\text{NIN,J}}$ appearing in Eq. (6.55), either exact expressions given by Eqs. (6.43) and (6.43) can employed or the approximate expressions given by Eqs. (6.47) and (6.48) can be employed. Also note that the first two terms in the product on the left hand side of Eq. (6.55) are identical to $q^{(1)-(2)}$ if $q^{(1)}_{\text{NIN,J}}$ and $p^{(1)}_{\text{NIN,J}}$ are not approximated and are identical to $q^{(1)-(2)}_{\text{app}}$ if they are approximated. Under these two conditions, two different approximate expressions result for Eq. (6.55). These will be denoted by $q^{(1)-(3)}_{\text{app1}}$ and $q^{(1)-(3)}_{\text{app1}}$, respectively, and are

$$q^{(1)-(3)}_{\text{app1}} = q^{(1)-(2)} \left\{q^{(1)}_{\text{NNIN,J,app}} + p^{(1)}_{\text{NNIN,J,app}} \left[q^{(1)}_{\text{NIN,J}} + p^{(1)}_{\text{NIN,J}} q^{(1)}_{\text{J}}\right]\right\} \quad (6.58)$$

$$q^{(1)-(3)}_{\text{app2}} = q^{(1)-(2)}_{\text{app}} \left\{q^{(1)}_{\text{NNIN,J,app}} + p^{(1)}_{\text{NNIN,J,app}} \left[q^{(1)}_{\text{NIN,J,app}} + p^{(1)}_{\text{NIN,J,app}} q^{(1)}_{\text{J}}\right]\right\} \quad (6.59)$$

Also note that the correction arising out of NNINs becomes significant at high densities and at these densities the probability that none of the NNINs of J will collide is expected to be negligible. This is evident from graph of $\left(q^{(1)}\right)^{18}$, which approximates $q^{(1)}_{\text{NNIN,J}}$ and $q^{(2)}_{\text{NNIN,J}}$, shown in Fig. 6.14. As a result, two more approximations of $q^{(1)-(3)}_{\text{app}}$, along the lines of $q^{(1)-(3)}_{\text{app1}}$ and $q^{(1)-(3)}_{\text{app2}}$, can also be obtained by neglecting $q^{(1)}_{\text{NNIN,J}}$ in Eq. (6.55). These approximations will be denoted by $q^{(1)-(3)}_{\text{app1b}}$ and $q^{(1)-(3)}_{\text{app2b}}$, respectively, and are

$$q^{(1)-(3)}_{\text{app1b}} = q^{(1)-(2)} \left[q^{(1)}_{\text{NIN,J}} + p^{(1)}_{\text{NIN,J}} q^{(1)}_{\text{J}}\right] \quad (6.60)$$

$$q^{(1)-(3)}_{\text{app2b}} = q^{(1)-(2)}_{\text{app}} \left[q^{(1)}_{\text{NIN,J,app}} + p^{(1)}_{\text{NIN,J,app}} q^{(1)}_{\text{J}}\right] \quad (6.61)$$

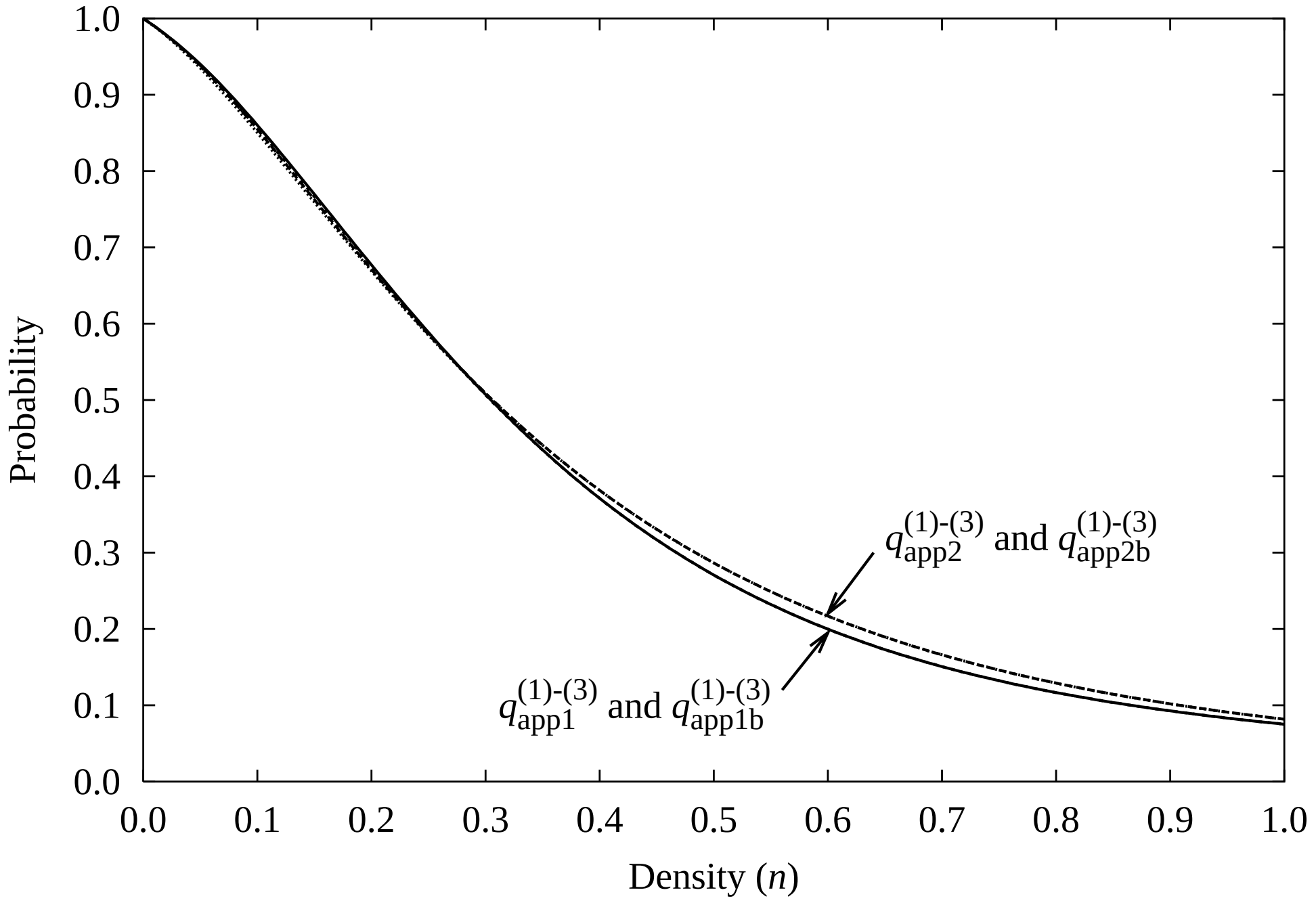

**Figure 6.19:** Comparison of various approximations of $q^{(1)-(3)}$ given by Eqs. (6.58)–(6.61).

Comparison of various approximations of $q^{(1)-(3)}$ given by Eqs. (6.58)–(6.61) is shown in Fig. 6.19. This figure shows that curves for $q^{(1)-(3)}_{\text{app1}}$ and $q^{(1)-(3)}_{\text{app1b}}$ are indistinguishable and those for $q^{(1)-(3)}_{\text{app2}}$ and $q^{(1)-(3)}_{\text{app2b}}$ are indistinguishable, especially at high densities. From this it is clear that the correction arising out of NNINs is, indeed, negligible. Thus, the approximation involved in neglecting the effect of NNINs (in general, N$^{\text{k}}$INs, $k \geq 2$) is quite accurate and can be freely employed for estimating $q^{(1)-(i)}$, $i \geq 3$, without incurring significant errors. In the figure it is also seen that the $q^{(1)-(3)}_{\text{app2}}$ and $q^{(1)-(3)}_{\text{app2b}}$ are higher compared to $q^{(1)-(3)}_{\text{app1}}$ $q^{(1)-(3)}_{\text{app1b}}$. This trend and its reason is identical to that seen in Fig. 6.17 and is because $q^{(1)}_{\text{NIN,J}}$ has been approximated in $q^{(1)-(3)}_{\text{app2}}$ and $q^{(1)-(3)}_{\text{app2b}}$ but not in $q^{(1)-(3)}_{\text{app1}}$ and $q^{(1)-(3)}_{\text{app1b}}$. Both the later probabilities, thus, are more accurate compared to the former ones. In general, among various possible approximations for $q^{(1)-(i)}$ (approximations carried out along the lines described above) the one which gives lower values at high densities will be more accurate compared to the others. This is because of the nature of approximations as explained earlier and also because $q^{(1)-(i)}$ must go to zero at $n = 1$ in the limit $i \to \infty$.

**Comparison of Various Approximations of $q^{(1)-(i)}$ and $p^{(1)-(i)}$ ($= \mathcal{P}^{(i)}$) and Corresponding Mean Free Paths and Times:** Various exact and approximate expressions for $q^{(1)-(i)}$, $i \in [1, 3]$, obtained in previous paragraphs are plotted together in Fig. 6.20 for comparison. From this figure it is seen that these expressions fall in three distinct groups which can be identified as having the same $i$. The variation of probabilities among different groups differ significantly from each other. This is as expected because different number of iterations are considered in different groups. Various approximations of $q^{(1)-(i)}$ within the same group, *i.e.*, for the same $i$, differ only slightly from each other. Among these expressions, the most accurate one is that for $q^{(1)-(3)}_{\text{app1}}$ given by Eq. 6.58. This is because in

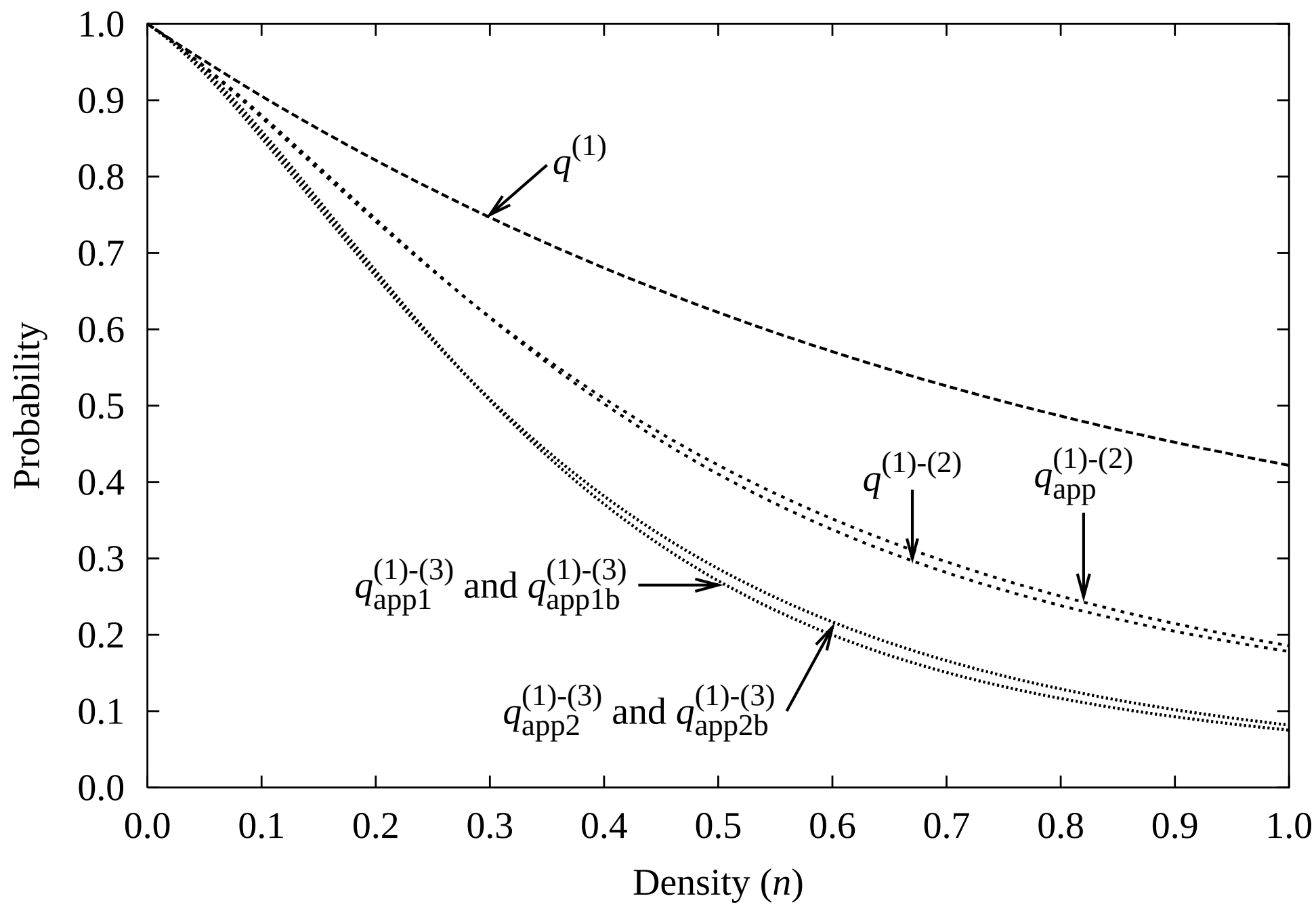

**Figure 6.20:** Comparison of various approximations of $q^{(1)-(i)}$, $i \in [1, 3]$, given by Eqs. (6.23), (6.46), (6.52), (6.58)–(6.61).

it three iterations have been considered and only moderate approximations are involved. The interaction probabilities corresponding to various approximations of $q^{(1)-(i)}$, $i \in [1, 3]$, are plotted in Fig. 6.21. The $p_\alpha^\beta$ shown in this figure correspond to $q_\alpha^\beta$ and are given by

$$p_\alpha^\beta = 1 - q_\alpha^\beta \tag{6.62}$$

where $\alpha$ and $\beta$ are appropriate subscripts and superscripts, *e.g.*, $\alpha \equiv (1) - (3)$ and $\beta \equiv$ app1.

The mean free paths and mean free times corresponding to the various approximations of $q(1) - (i)$, $i \in [1, 3]$, given by Eqs. (6.23), (6.46), (6.52), and (6.58)–(6.61), can computed using Eqs. (6.16), (6.21), and (6.22). The mean free path and mean free time corresponding to $q_\alpha^\beta$ are given by

$$\lambda_{\text{mf},\alpha}^\beta = \tau_{\text{mf},\alpha}^\beta = \frac{1}{p_\alpha^\beta} = \frac{1}{1 - q_\alpha^\beta} \tag{6.63}$$

where $\alpha$ and $\beta$ are appropriate subscripts and superscripts, *e.g.*, $\alpha \equiv (1) - (3)$ and $\beta \equiv$ app1. The mean free paths and mean free times calculated as above, corresponding to various approximations of $q(1) - (i)$, $i \in [1, 3]$, are plotted in Fig. 6.22. In this figure a closeup view in the density range $0.04 \leq n \leq 0.12$ is also shown. As seen in this figure, the mean free paths and mean free times for three different groups depending upon the number of iterations considered. Various approximations within each group do not differ significantly, especially at high densities. At low densities, various approximations within each group differ, though not much, as is also seen in the closeup. It is seen that at low densities the mean free path and time corresponding to $q_{\text{app1}}^{(1)-(3)}$—the most accurate approximate estimation of $q^{(1)-(3)}$—is higher compared to other approximations of $q^{(1)-(3)}$. At high densities, all the approximations of $q^{(1)-(3)}$ give mean free path and time close to unity. This is expected because in the SPLG-1 model all particle move with unit speed, and thus, at any time step they will always move by one lattice site.

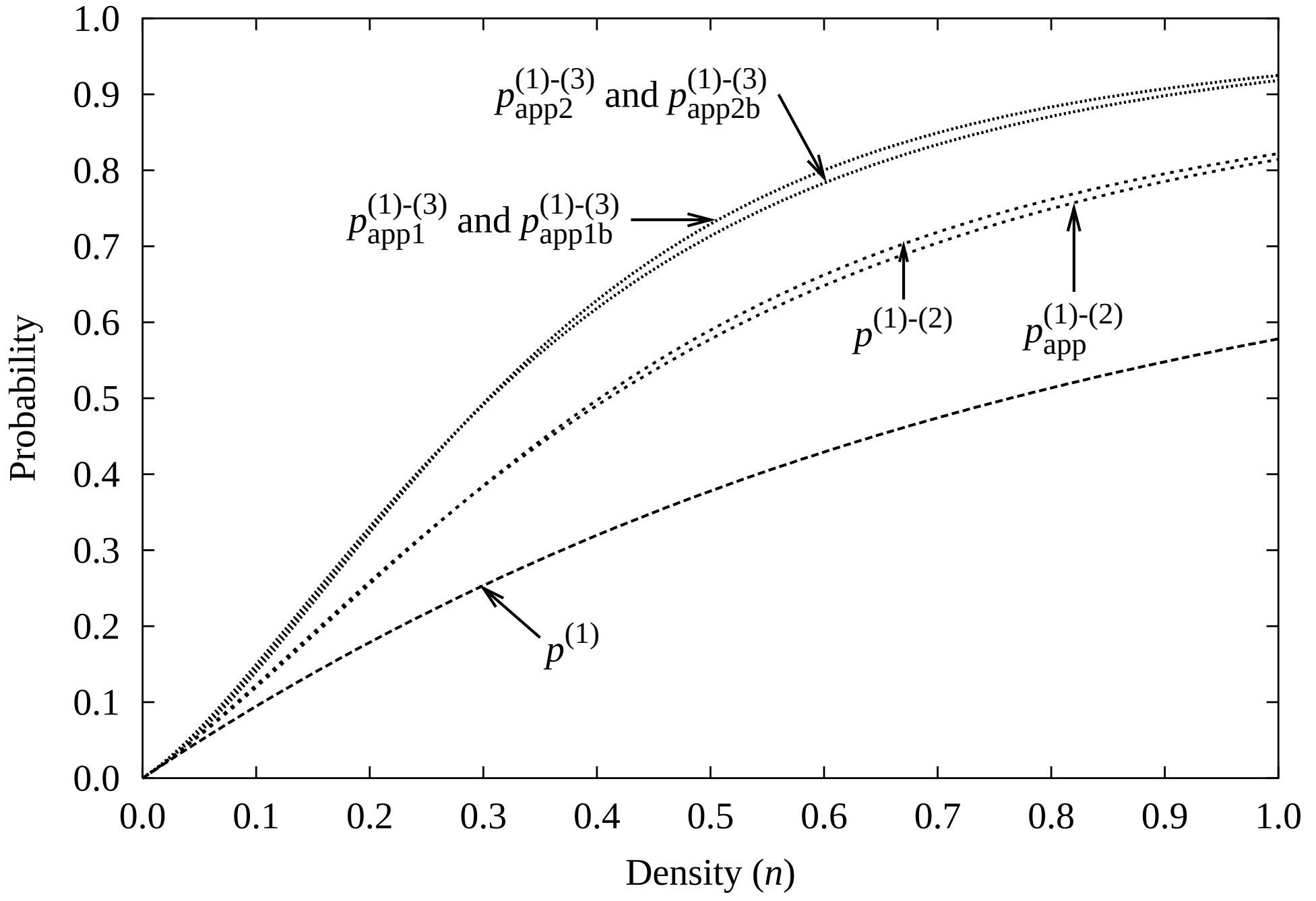

**Figure 6.21:** Comparison of interaction probabilities corresponding to various approximations of $q^{(1)-(i)}$, $i \in [1, 3]$, given by Eqs. (6.23), (6.46), (6.52), (6.58)–(6.61).

### 6.5.3 Speed of Sound

The speed of sound is defined as [2,3]

$$c_{\text{s}}^2 = \frac{1}{\mathcal{D}}\langle \boldsymbol{v}(0) \cdot \boldsymbol{v}(0)\rangle = \frac{1}{\mathcal{D}}\sum_{i=1}^{\mathcal{N}_{\text{v}}} f_i(\boldsymbol{v}_i \cdot \boldsymbol{v}_i) \tag{6.64}$$

where $\langle \boldsymbol{v}(0) \cdot \boldsymbol{v}(0)\rangle$ is the velocity autocorrelation function at $t = 0$, $\boldsymbol{v}(0)$ is the velocity of particle at $t = 0$, and $\mathcal{D}$ is number of dimensions. If $f_i$ is the velocity distribution function for particles of velocity $\boldsymbol{v}_i$ such that $\sum_{i=1}^{\mathcal{N}_{\text{v}}} f_i = 1$, then $\langle \boldsymbol{v}(0) \cdot \boldsymbol{v}(0)\rangle$ can be calculated as

$$\langle \boldsymbol{v}(0) \cdot \boldsymbol{v}(0)\rangle = \sum_{i=1}^{\mathcal{N}_{\text{v}}} f_i(\boldsymbol{v}_i \cdot \boldsymbol{v}_i) \tag{6.65}$$

where $\boldsymbol{v}_i \in \mathcal{V}$, $i = 1, \ldots, \mathcal{N}_{\text{v}}$, are velocity vectors comprising the discrete velocity set $\mathcal{V}$. Combining Eqs. (6.64) and (6.65) gives the speed of sound as

$$c_{\text{s}}^2 = \frac{1}{\mathcal{D}}\sum_{i=1}^{\mathcal{N}_{\text{v}}} f_i(\boldsymbol{v}_i \cdot \boldsymbol{v}_i) \tag{6.66}$$

In the SPLG-1 model all particle move with unit speed. Thus, for the SPLG-1 model, for all possible velocity distribution functions, Eq. (6.66) gives

$$c_{\text{s}} = \frac{1}{\sqrt{2}} \tag{6.67}$$

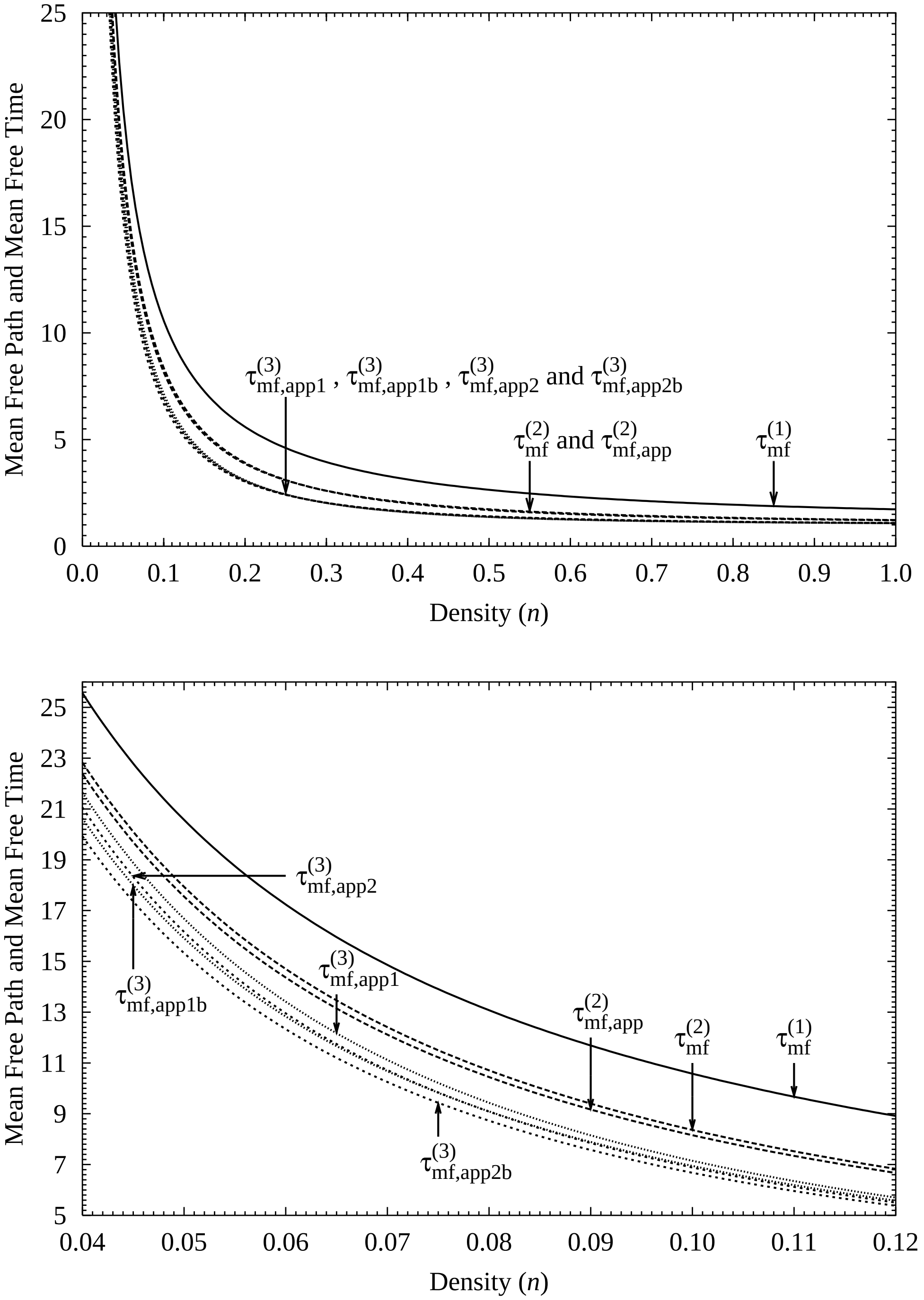

**Figure 6.22:** Comparison of the mean free paths and mean free times corresponding to various approximations of collision probability shown in Fig. 6.21. Labels for mean free time are shown in the figure. The labels for mean free path are same as those for mean free time *via* Eq. (6.16). The bottom view is closeup of the top in the density range $0.04 \leq n \leq 0.12$.

## 6.6 Conclusions

This chapter relates to construction of a single species single speed single particle lattice gas over square spatial lattice. This chapter, in addition to giving a concrete single particle lattice gas, elucidates the generalized method of construction of single particle lattice gases outlined in chapter 5. The important conclusions of this chapter are as follows:

**1)** A single species single speed single particle lattice gas existing over square spatial lattice has been constructed. This model is called as the *"SPLG-1"* model. In this model, interactions occur in negligible time compared to the time step; this is equivalent to collision time being negligible compared to the mean free time. In this model evolution of the system during one time step is decomposed into two sub-steps, *viz.*, (i) particle translation, and (ii) interparticle interactions. Interparticle interactions are processed iteratively using fractional interaction rules.

**2)** Fractional interaction rules (the BCC lookup tables) have been constructed both for processing interactions among fluid particles as well for processing interactions of fluid particles with stationary solid boundary particles in the SPLG-1 model. The rules are constructed in compressed form. In this form, the rules for processing interactions among fluid particles contain only 5 BCCs. On the other hand, if the rules were constructed in uncompressed form using the reduced contact interaction neighborhood they would contain 3125 configurations and if the entire contact interaction neighborhood was used they would contain $1220703125 \approx 1.22 \times 10^9$ configurations.

**3)** In the SPLG-1 model desired interaction potentials can be incorporated by appropriately selecting the STPs. Using this model, systems with various types of boundary conditions, *viz.*, (i) free or open boundaries, (ii) toroidal or periodic boundaries, and (iii) solid boundaries, can be simulated. Method of application of various types on boundary conditions is described.

**4)** The SPLG-1 model has several fixed states under specific boundary conditions which form a closed repeating sequence of states during processing of interparticle interactions. These states cannot occur during simulations unless the starting state itself is a fixed states with appropriate boundary conditions. These states are not strange attractors for the system.

**5)** Exact mathematical analysis of the SPLG-1 model leading to various coarse grained equations is not possible because of iterative (or, recursive) decomposition of interaction operator in terms of fractional interaction operators.

**6)** Analytical expressions for the probability of occurrence of BCCs, probability of collision of particles, mean free path, mean free time, and speed of sound in the SPLG-1 model have been derived. It is found that both the mean free part and mean free time become unity as particle density becomes unity (*i.e.*, when the lattice is fully occupied). In this model, the speed of sound is a constant $(1/\sqrt{2})$ independent of density and other model parameters.

Finally, the construction of SPLG-1 model conclusively shows that the space of single particle lattice gases existing in the limit of collision time being negligible compared to the time step (or, alternatively, in the limit of collision time being negligible to the mean free time) is not empty. Some simulations using this model are described in chapters 7 and 8.

# Chapter 7

# Particles Enclosed in a Box at Equilibrium

. . . plans, tools, and materials are necessary for mechanical constructions.
*Yes, sire.*
They are only suggestive of an end.
*Yes, sire.*
Where is the end?
*Here it is . . . , sire!*
Ah! What about a demonstration?
. . .

In this chapter a system of particles enclosed in a box at equilibrium has been simulated and studied using the SPLG-1 model constructed in chapter 6. Various aspects that have been studied, simulation setup, and simulation results are described in the following sections.

## 7.1 Aspects to be Studied

The literature related to multiparticle lattice gases [1–5] suggests that usability of a lattice gas for accurate simulation of physical systems can be determined by studying its basic properties of isotropy, Galilean invariance, and geometrical and physical dimensionality, *etc.* It is well known that isotropy is essentially a geometrical property that depends upon the structure of the underlying spatial lattice and discrete velocity set [3,6]. Consequently, its presence in a lattice gas is determined strictly by simulation requirements. Similarly geometrical dimensionality is also determined by simulation requirements, alone. It is worth noting that the SPLG-1 model is geometrically two-dimensional and, because it exists over square spatial lattice with single speed discrete velocity set, lacks isotropy. Furthermore, this model is expected to be Galilean invariant as argued for single particle lattice gases in chapter 4. The presence or absence of Galilean invariance, however, is not important for simulations of equilibrium systems which have zero mean mass motion (or, flow) velocity.

### 7.1.1 Geometrical and Physical Dimensionality and Dynamical Exponents

Various studies on multiparticle lattice gases [1–3,7] show that the geometrical dimensionality of lattice gases does not have one-to-one correspondence with the dimensionality of their (physical) dynamical behavior as indicated by the dynamical exponent. For example, it is well established that the geometrically two-dimensional HPP, TM, and FHP gases have dynamical behaviors of different dimensionality [4]. This difference arises from the fact that the physical dimensionality of a lattice gas depends upon its collision rules and the impact parameter embedded therein. Since the equivalence of physical and geometrical dimensionality of lattice gases dictates whether simulations carried out using them are physically consistent and meaningful, it is necessary to determine their physical dimensionality.

The physical dimensionality of a lattice gas is given by its dynamical exponent ($\alpha$) that shows the behavior of the long time tail of *Time Correlation Function for Velocity* (TCFV) at a lattice site in macroscopically stationary dynamical systems at equilibrium. Since the dynamical exponent of a model depend upon the structure of its collision rules, it is not purely geometrical property and cannot be determined easily. It can only be determined either through rigorous mathematical analysis of the model or through extensive computer simulations. The SPLG-1 model has been found to be too complex to be amenable to rigorous mathematical analysis, hence computer simulations have been carried out for determining its dynamical exponent ($\alpha$).

### 7.1.2 Randomness and Diffusive Behavior

The second important exercise for understanding the dynamical behavior of lattice gases relates to study of randomness in the modeled dynamics. This study is also necessary for establishing whether the rules of the model introduce any bias in the motion of particles and for determining the nature of bias, if any. For lattice gases that are not easily amenable to rigorous mathematical treatment, *e.g.*, the single particle lattice gases, randomness properties in modeled dynamics can be studied by sampling the mean displacement and the mean square displacement of tagged particles in a macroscopically stationary dynamical system at equilibrium. The sampled results can be compared with the available theoretical results, *e.g.*, that from the Langevin equation, for verification.

### 7.1.3 Effect of Model Parameters

While trying to analyze a simulation model, it is necessary to spawn the complete range of parameters on which its dynamics may depend. In the case of SPLG-1 model, the simulation parameters are: (i) spatial lattice dimensions, (ii) boundary conditions, (iii) initial particle density and velocity distributions, and (iv) STP sets corresponding to interaction rules given in Figs. 6.6 and 6.7. Out of these parameters, (i)–(iii) are problem dependent parameters and are dictated by the problem to be simulated, and the STP set (iv) is the model parameter which dictates the dynamical behavior of the model. Consequently, for analysis of the model, study of the effect of STP set on dynamics becomes very important. Spawning the complete range of STP sets, however, is not feasible because selection of a STP sets presents infinitely many possibilities due to continuous variation of probabilities in $[0, 1]$. Consequently, for demonstrating and getting a feel of the STP set dependent nonlinearities of the SPLG-1 model some sample STP sets have been chosen arbitrarily. These STP sets are enlisted in tables 7.1 and 7.2 which correspond to Figs. 6.6 and 6.7, respectively.

| | FCC # | Solutions (1) | (2) | (3) | (4) | (5) | (6) | (7) | (8) |
|---|---|---|---|---|---|---|---|---|---|
| STP Set-1 | (1) | 1 | 1 | 1 | - | - | - | - | - |
| | (2) | 1 | 1 | 1 | - | - | - | - | - |
| | (3) | 1 | - | - | - | - | - | - | - |
| | (4) | 1 | 1 | 1 | 1 | 1 | 1 | 1 | 1 |
| | (5) | 1 | 1 | 1 | 1 | 1 | 1 | 1 | - |
| STP Set-2 | (1) | 1 | 1 | 1 | - | - | - | - | - |
| | (2) | 1 | 1 | 1 | - | - | - | - | - |
| | (3) | 1 | - | - | - | - | - | - | - |
| | (4) | 1 | 1 | 1 | 1 | 1 | 1 | 1 | 1 |
| | (5) | 0 | 0 | 0 | 0 | 0 | 0 | 1 | - |
| STP Set-3 | (1) | 1 | 1 | 1 | - | - | - | - | - |
| | (2) | 1 | 1 | 1 | - | - | - | - | - |
| | (3) | 1 | - | - | - | - | - | - | - |
| | (4) | 0 | 0 | 0 | 0 | 0 | 0 | 0 | 1 |
| | (5) | 1 | 1 | 1 | 1 | 1 | 1 | 1 | - |
| STP Set-4 | (1) | 1 | 1 | 1 | - | - | - | - | - |
| | (2) | 1 | 1 | 1 | - | - | - | - | - |
| | (3) | 1 | - | - | - | - | - | - | - |
| | (4) | 0 | 0 | 0 | 0 | 0 | 0 | 0 | 1 |
| | (5) | 0 | 0 | 0 | 0 | 0 | 0 | 1 | - |
| STP Set-5 | (1) | 1 | 1 | 1 | - | - | - | - | - |
| | (2) | 1 | 1 | 1 | - | - | - | - | - |
| | (3) | 1 | - | - | - | - | - | - | - |
| | (4) | 1 | 1 | 1 | 1 | 1 | 0 | 0 | 0 |
| | (5) | 1 | 1 | 1 | 1 | 1 | 1 | 1 | - |

**Table 7.1:** Unnormalized state transition probability sets corresponding to the rule table shown in Fig. 6.6. Normalization can be carried out by dividing each entry by the sum of all the entries in its row.

| FCC # | Solutions (1) | (2) | (3) |
|---|---|---|---|
| (1) | 1 | 1 | 1 |
| (2) | 1 | 1 | 1 |
| (3) | 1 | 1 | 1 |
| (4) | 1 | 1 | - |
| (5) | 1 | 1 | - |
| (6) | 1 | 1 | - |
| (7) | 1 | 1 | - |
| (8) | 1 | - | - |
| (9) | 1 | - | - |

**Table 7.2:** Unnormalized state transition probability set corresponding to the rule table shown in Fig. 6.7. Normalization can be carried out by dividing each entry by the sum of all the entries in its row.

In the STP sets 2, 3, 4, and 5 shown in table 7.1 certain BCC $\mapsto$ SOLUTION state transition pairs have been assigned zero transition probability, *i.e.*, these state transitions are excluded from the dynamics or forbidden during processing of interparticle interactions during simulations. This allows very precise control over the dynamics because all undesired state transitions can be eliminated by lowering their transition probabilities to zero and the importance of other transitions can be fixed relative to each other by appropriately adjusting their transition probabilities. It, however, should also be noted that all the state transition pairs of a BCC cannot be assigned zero probability simultaneously. This is because all the BCC can occur over the spatial lattice at any time step in an unrestricted manner and their occurrence cannot be controlled externally. Hence, for resolving all possible collisions that can occur over the spatial lattice, each BCC should necessarily be allowed to undergo at least one state transition, *i.e.*, each BCC should necessarily have at least one BCC $\mapsto$ SOLUTION state transition pair with non-zero transition probability.

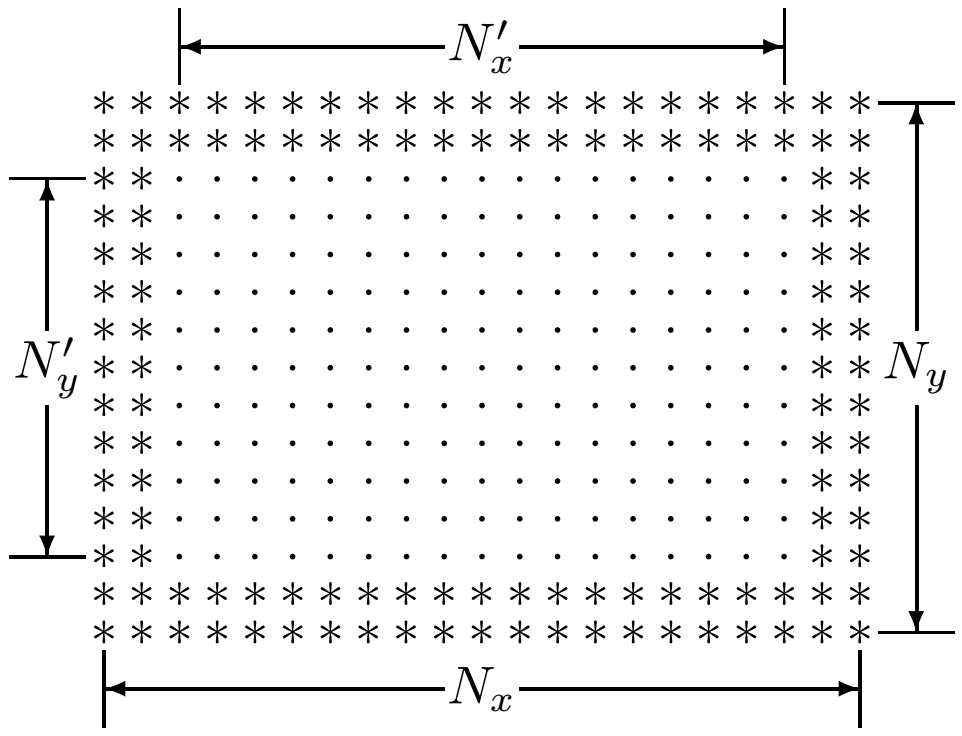

**Figure 7.1:** Schematic representation of the simulation setup used for studying the dynamical behavior of the SPLG-1 model for various state transition probability sets. $*$ represents stationary boundary walls and dots $(\cdot)$ show the lattice sites that may be empty or occupied by fluid particles.

## 7.2 Simulation Setup

For studying the dynamical behavior of SPLG-1 model, simulations have been carried out on a macroscopically stationary system of particles at equilibrium at density of $n$ particles per lattice site enclosed in a container of dimensions $(N_x, N_y) = (516, 516)$ with 2 lattice site thick stationary boundary walls (the thickness of walls, though specified for the sake of completeness, is irrelevant for results). A schematic representation of the simulation setup is shown in Fig. 7.1. In these simulations various STP sets shown in table 7.1 have been used in combination with the STP set shown in table 7.2. In the following analysis, results obtained for different STP set combinations have been labeled with the number of the STP set given in table 7.1 for distinction.

Since the SPLG-1 model is primarily intended for simulation of fluid dynamical systems, all the simulations presented in the following sections have been carried out for moderate and low particle densities only. Specifically, the values of $n = 0.05$ and $0.10$ have been used. Simulations of systems with high values of $n$ $(n > 0.15)$ were also attempted, but had to be abandoned because in all the trial simulations available computer time turned out to be a sever limiting factor. This happened because too many iterations were required for processing interparticle interactions at high densities.

## 7.3 Sampling Procedure

During simulations the mean displacement and mean square displacement of tagged particles and time correlation function for velocity at lattice sites were sampled. The tagged particle data were sampled by tagging $N_{\text{tag}} = 253$ particles lying nearest to the center of the box $(258, 258)$ and tracing them for $t = 1000$ time steps. The time correlation was sampled at all the lattice sites excluding those occupied by the stationary boundary particles. The sampled data were ensemble averaged over $N = 25$ realizations of the system, rather than in the limit $N \to \infty$ as required ideally.

The mean displacement $\langle \boldsymbol{r}(t) \rangle$ and the mean square displacement $\langle r^2(t) \rangle$ were sampled using

$$\langle \boldsymbol{r}(t) \rangle = \lim_{N\to\infty} \frac{1}{N} \sum_{k=1}^{N} \left( \frac{1}{N_{\text{tag}}} \sum_{i=1}^{N_{\text{tag}}} \left[ \boldsymbol{r}_i^k(t) - \boldsymbol{r}_i^k(0) \right] \right) \tag{7.1}$$

$$\begin{aligned} \left\langle r^2(t) \right\rangle &\equiv \left\langle [\boldsymbol{r}(t) - \boldsymbol{r}(0)]^2 \right\rangle \\ &= \lim_{N\to\infty} \frac{1}{N} \sum_{k=1}^{N} \left( \frac{1}{N_{\text{tag}}} \sum_{i=1}^{N_{\text{tag}}} \left[ \boldsymbol{r}_i^k(t) - \boldsymbol{r}_i^k(0) \right]^2 \right) \end{aligned} \tag{7.2}$$

where $\boldsymbol{r}_i^k(t)$ is the location of $i^{\text{th}}$ tagged particle at time step $t$ for $k^{\text{th}}$ realization of the system.

The time correlation function for velocity at lattice sites $\phi_s(t)$ was sampled using

$$\begin{aligned} \phi_s(t) &\equiv \langle \boldsymbol{v}(t) \cdot \boldsymbol{v}(0) \rangle_s \\ &= \lim_{N\to\infty} \frac{1}{N} \sum_{k=1}^{N} \left( \frac{1}{N'_x N'_y} \sum_{j=1}^{N'_y} \sum_{i=1}^{N'_x} \left[ \boldsymbol{v}_{ij}^k(t) \cdot \boldsymbol{v}_{ij}^k(0) \right] \right) \end{aligned} \tag{7.3}$$

where $N'_x$ and $N'_y$ refer to the region available for motion of fluid particles (see Fig. 7.1), $\boldsymbol{v}_{ij}^k(t)$ is the velocity of particles occupying the lattice site $(i, j)$ at time step $t$ for the $k^{\text{th}}$ realization of the system or zero if the lattice site is unoccupied, and the coordinates $(i, j)$ refer to lattice sites inside the region demarcated by $N'_x$ and $N'_y$.

## 7.4 Theoretical Results for Comparison with Simulations

Rigorous theoretical results are not available for the SPLG-1 model that has been used for simulations presented herein. The diffusive behavior and randomness properties observed in the SPLG-1 model, however, can be compared with the solution of Langevin equation [8,9]. In what follows, results from the Langevin analysis and the mean free time results obtained in chapter 6 for the SPLG-1 model have been combined to arrive at the quantities that are required and used for comparison with the simulation results.

In the absence of external forces the dynamical behavior of a neutral particle in random (Brownian) motion in a neutral fluid at equilibrium is Gaussian Markov process. It is described by the Langevin equation [8–10]

$$m\ddot{\boldsymbol{r}}(t) = -\frac{m}{\tau}\dot{\boldsymbol{r}}(t) + \boldsymbol{F}(t) \tag{7.4}$$

where $m$ is the mass of the particle, $\boldsymbol{r} \equiv (x, y)$ is the coordinate of the particle, $\tau$ is the mean collision time or the mean free time, $\boldsymbol{F}(t)$ is the random force due to collisions with the surrounding fluid medium, and the dots over quantities represent their time derivatives.

Solution of Eq. (7.4) gives time variation of velocity autocorrelation function for particles in random motion, which may be tagged particles, as

$$\langle \boldsymbol{v}(t) \cdot \boldsymbol{v}(0) \rangle = \left\langle v^2(0) \right\rangle e^{-t/\tau} \tag{7.5}$$

where $\langle v^2(0) \rangle = kT/m$ by the equipartition theorem, $k$ is the Boltzmann constant, and $T$ is temperature; and $\langle v^2(0) \rangle = 1$ because the system is athermal (see also Eq. (6.65)), *i.e.*, all the fluid particles move with the same speed at all time steps ($|\boldsymbol{v}(t)| = 1$). Note that Eqs. (7.3) and (7.5) are different from each other in that Eq. (7.3) is used for sampling the time correlation function for velocity at lattice sites whereas Eq. (7.5) gives the velocity autocorrelation function for tagged particles.

Since the modeled system is Newtonian, VACF and mean square displacement of the tagged particles satisfy the identity [8,11]

$$\frac{\partial^2 \left\langle [\boldsymbol{r}(t) - \boldsymbol{r}(0)]^2 \right\rangle}{\partial t^2} \equiv 2\langle \boldsymbol{v}(t) \cdot \boldsymbol{v}(0) \rangle \tag{7.6}$$

which can be easily verified. Eq. (7.6) on integration with Eq. (7.5) and after substituting for $\langle v^2(0) \rangle$ gives the variation of mean square displacement of particles with time as

$$\left\langle r^2(t) \right\rangle = 2\tau \left[ t - \tau \left( 1 - e^{-t/\tau} \right) \right] \tag{7.7}$$

In the above equation, the mean free time $\tau$ of particles is computed as explained in Sec. 6.5.2 and 6.5.2.2 using Eqs. (6.16), (6.21), (6.22), and an appropriate approximation of $q^{(1)-(i)}$. The most accurate approximate estimate of $q^{(1)-(i)}$ obtained in Sec. 6.5.2.2 is given by Eq. (6.58). For this approximate estimation, the expression for mean free time is

$$\tau = \frac{1}{1 - q_{\text{app1}}^{(1)-(3)}} \tag{7.8}$$

which has been employed in the following comparisons.

## 7.5 Analysis of Simulation Results

### 7.5.1 Mean Displacement

The sampled variation of mean displacement of tagged particles is shown in Fig. 7.2. For analysis of this data note that in an equilibrium system in the absence of external force fields tagged particles should perform random walk with zero mean displacement. In the presence of externally imposed fields and also in the event of violation conservation laws (including the law of conservation of angular momentum), the motion of tagged particles becomes biased resulting in non-zero mean displacement. The figure shows that the STP sets 1, 2, and 5 introduces less bias in the random motion of particles compared to the STP sets 3 and 4. Among all these STP sets, the STP set 1 introduces the least amount of bias in the motion of particles.

The shape of curves seen in Fig. 7.2 indicates the nature of bias introduced by STP sets and can be related to collision rules. The counter-clockwise (CCW) motion of tagged particles reflected in the mean displacement curve for $n = 0.10$ and STP set 3 is a direct consequence of the CCW motion introduced in the center of mass of colliding particles by the solution of BCC 4 in STP set 3. The behavior of tagged particle for $n = 0.10$ and STP set 4 also has the same explanation with an addition that the combined effect of CCW motion introduced by the solutions of both the BCC 4 and 5 leads to more circular path of tagged particle compared to that obtained in STP set 3. No CCW motion of tagged particle is seen for $n = 0.05$ because at this density the probability of occurrence of BCC 4 and 5 around any lattice site is negligible as can be seen from table 7.3. Similarly, since the probability of occurrence of BCC 5 for both $n = 0.05$ and 0.10 is negligible, the bias introduced by the STP set 2 in the motion of tagged particle is not visible for either $n = 0.05$ or 0.10. The bias is introduced because of violation of the law of conservation of angular momentum during interparticle interactions.

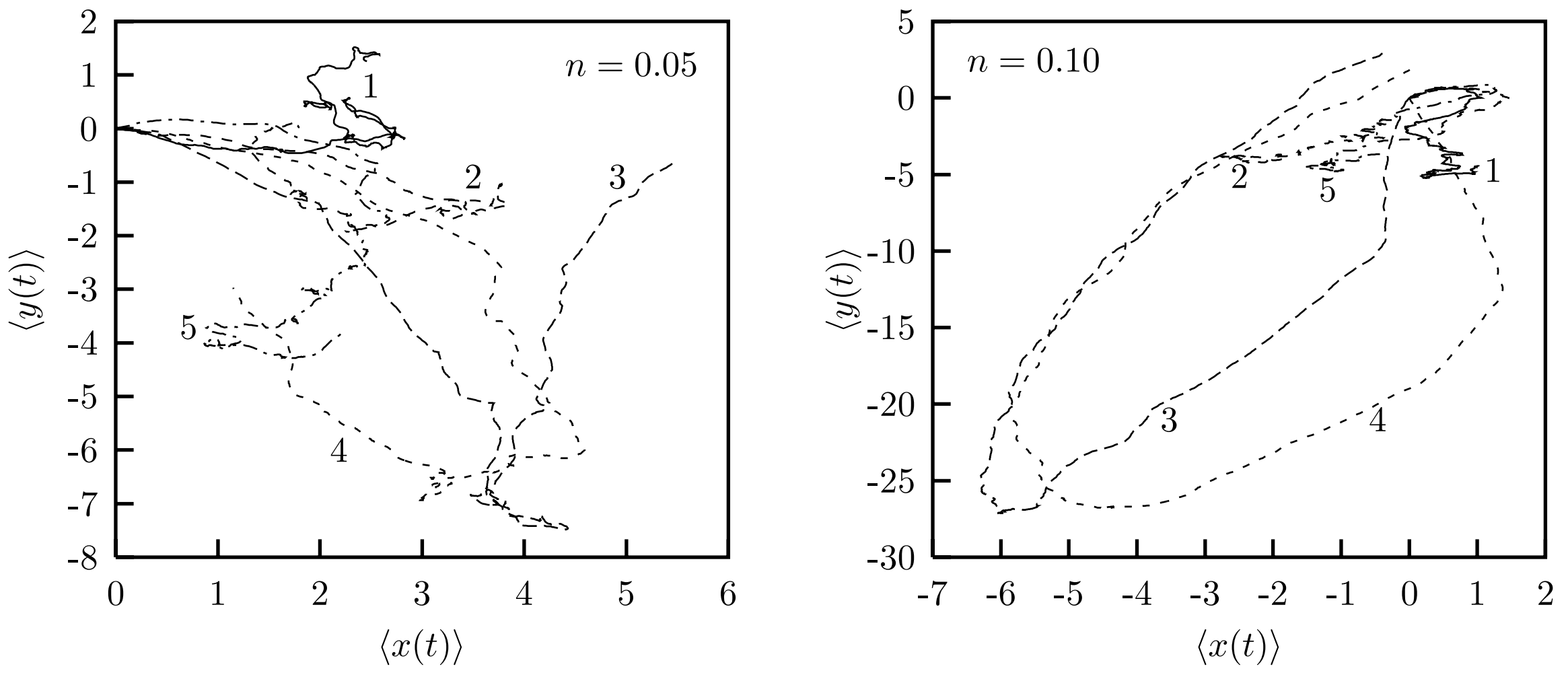

**Figure 7.2:** Time variation of the mean displacement $\langle \boldsymbol{r}(t)\rangle \equiv (\langle x(t)\rangle, \langle y(t)\rangle)$. Numerals near the curves correspond to number of the STP set (from table 7.1) that has been used during these simulation. All the curves are for $t \in [0, 1000]$.

| BCC | Probability of occurrence of BCCs including all isometries for mean particle density of | | |
|---|---|---|---|
| # | $n$ | $n = 0.05$ | $n = 0.10$ |
| 1 | $2(n/4)^2$ | $3.125 \times 10^{-4}$ | $1.25 \times 10^{-3}$ |
| 2 | $2(n/4)^2(1-n/4)(1-3n/4+n^2/4)$ | $2.9721435546875 \times 10^{-4}$ | $1.130390625 \times 10^{-3}$ |
| 3 | $4(n/4)^2(1-n/4)(1-3n/4+n^2/4)$ | $5.944287109375 \times 10^{-4}$ | $2.26078125 \times 10^{-3}$ |
| 4 | $4(n/4)^3(1-n/2)^2$ | $7.4267578125 \times 10^{-6}$ | $5.640625 \times 10^{-5}$ |
| 5 | $(n/4)^4(1-n)$ | $2.3193359375 \times 10^{-8}$ | $3.515625 \times 10^{-7}$ |

**Table 7.3:** Probability of occurrence of various BCC shown in Fig. 6.6 (including their isometries) around a lattice site in an equilibrium system at various particle densities.

For STP sets 1 and 2 no bias should be visible in an ideal ensemble because in these STP sets the net bias introduced by solutions of each BCC is zero, *i.e.*, the law of conservation angular momentum is satisfied during interparticle interactions. Furthermore, due to presence of additional solutions for BCC 4 in STP set 1, the STP set 1 should cause more randomization in the system compared to the STP set 5. This implies that the mean displacement of tagged particle for STP set 1 should be less compared to that for STP set 5. This is readily verified from the trajectories shown in Fig. 7.2.

## 7.5.2 Mean Square Displacement

The sampled mean square displacement of tagged particle is shown in Fig. 7.3 for $n = 0.05$ and 0.10 along with the variation predicted by Langevin solution, Eq. (7.7). The mean free times at these densities are 16.6801947269498 and 7.1497088570489, respectively.

Fig. 7.3 shows that for $n = 0.05$ the variation of mean square displacement for all the STP sets is in excellent agreement with that predicted by the Langevin solution till $t \approx 350$. The departure seen after $t \approx 350$ is because of finite size effects and can be explained as follows: "Since the size of the simulated system is finite, the tagged particle diffuses and collides with the system boundaries at $t \approx 350$. After the collision the normal path of tagged particle is altered and it is reflected back into the system, *i.e.*, it cannot move any farther than the limits imposed by the system boundaries. As a result, the rate of increase of its mean square displacement goes down. Consequently, the sampled mean square displacement of tagged particle starts departing from the predicted one from the time of collision of the tagged particle with system boundaries, as seen in Fig. 7.3."

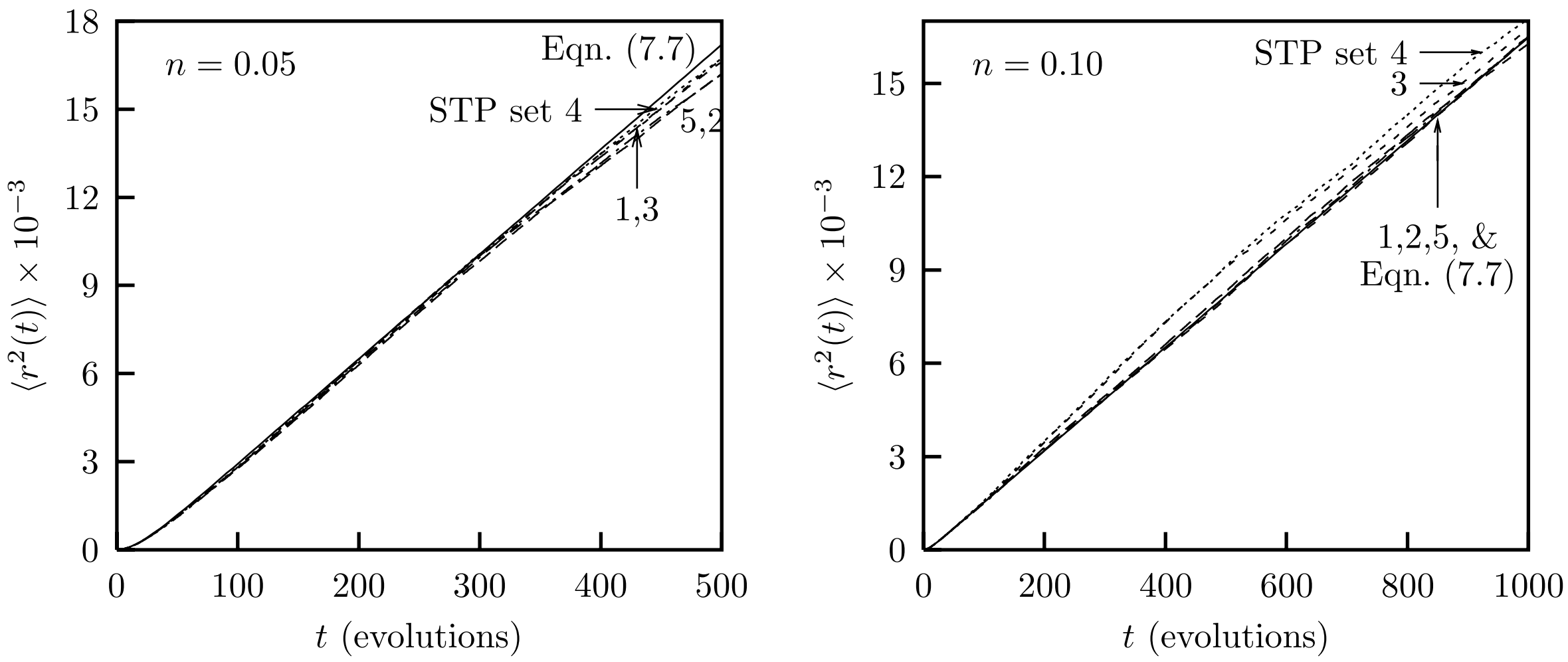

**Figure 7.3:** Time variation of the mean square displacement $\langle r^2(t)\rangle$. Eq. (7.7) (solid) and sampled data (various line styles). Numerals near the curves correspond to number of the STP set (from table 7.1) that has been used during these simulation.

For $n = 0.05$ no significant effect of differences in the STP sets is visible on the variation of mean square displacement either before or after collision of the tagged particle with system boundaries. This is because for $n = 0.05$ the bias introduced by different STP sets in the motion of particle is negligible as discussed in Sec. 7.5.1.

Fig. 7.3 shows that for $n = 0.10$ and STP sets 1, 2, and 5 the variation of mean square displacement is in excellent agreement with the Langevin prediction through out the time domain of simulation. This is because the STP sets 1, 2, and 5 do not introduce any significant bias in the motion of particles (as discussed in Sec. 7.5.1 and also seen from table 7.3). Furthermore, for $n = 0.10$ the tagged particle does not experience the confinement due to system boundaries, either. This is because the collision frequency is higher for $n = 0.10$ compared to that for $n = 0.05$. As a result, for $n = 0.10$ the tagged particle takes more time to reach the system boundaries compared to that for $n = 0.05$. In fact, the simulations show that for $n = 0.10$ the time taken by the tagged particle to reach the system boundaries is more than the total simulation time. In Fig. 7.3, the variation of mean square displacement for $n = 0.10$ and STP sets 3 and 4 show significant departure from the Langevin prediction. This departure is not because of collision of tagged particle with system boundaries because for $n = 0.10$ the collision occurs only after $t = 1000$. The departure is because of the bias introduced by the STP sets 3 and 4 in the motion of tagged particle. The nature of this bias has already been discussed in Sec. 7.5.1 and is clearly seen in Fig. 7.2. Furthermore, the departure shown by the data/curve for STP set 4 is more compared to that for STP set 3. This is in conformity with the fact that the bias introduced by the STP set 4 is more compared to the bias introduced by the STP set 3.

| $n$ | STP set | $t_1$ | $t_2$ | $b$ | $\alpha$ | Sum of square residuals |
|---|---|---|---|---|---|---|
| 0.05 | 1 | 20 | 450 | 0.103017(±0.00834741) | 0.977219(±0.0206897) | $6.26323 \times 10^{-5}$ |
| | 2 | 20 | 450 | 0.123940(±0.00986215) | 1.024910(±0.0206068) | $5.80601 \times 10^{-5}$ |
| | 3 | 20 | 450 | 0.109639(±0.00891527) | 0.990357(±0.0208452) | $6.38264 \times 10^{-5}$ |
| | 4 | 20 | 450 | 0.112269(±0.00897831) | 0.992581(±0.0205145) | $6.35078 \times 10^{-5}$ |
| | 5 | 20 | 450 | 0.114008(±0.00996523) | 1.005730(±0.0225099) | $6.98894 \times 10^{-5}$ |
| 0.10 | 1 | 15 | 450 | 0.1653850(±0.00857383) | 1.064610(±0.0145795) | $5.63381 \times 10^{-5}$ |
| | 2 | 15 | 450 | 0.1460890(±0.00695378) | 1.031970(±0.0132400) | $4.83276 \times 10^{-5}$ |
| | 3 | 15 | 450 | 0.0986275(±0.00391530) | 0.877855(±0.0104187) | $5.36523 \times 10^{-5}$ |
| | 4 | 15 | 450 | 0.1020970(±0.00408149) | 0.887570(±0.0105332) | $5.38821 \times 10^{-5}$ |
| | 5 | 15 | 450 | 0.1619370(±0.00829087) | 1.054900(±0.0143522) | $5.70097 \times 10^{-5}$ |

**Table 7.4:** Parameters obtained by least-square curve fitting Eq. (7.9) through sampled velocity time correlation function data at various particle densities and STP set combinations. Variation of the parameters in 68.3% confidence interval is given in parentheses.

### 7.5.3 Time Correlation Function for Velocity at Lattice Sites

The sampled variation of time correlation function for velocity set lattice sites for $n = 0.05$ and 0.10 and for all the STP sets is shown in Figs. 7.4 and 7.5. In these figures each data set has been visualized on both linear and log-log scales superimposed on the same graph with the lower curve corresponding to the linear axes at bottom and left, and the upper curve corresponding to the log-log axes at top and right. The raw data curves on log-log axes are *"approximate"* recast of the data plotted on linear axes because while making the log-log curves all non-positive values in the raw data were been neglected.

The solid lines in Figs. 7.4 and 7.5 correspond to the least-square curve

$$\frac{\phi_s(t)}{\phi_s(0)} = \begin{cases} 1 & \text{for } t = 0 \\ bt^{-\alpha} & \text{for } t > 0 \end{cases} \tag{7.9}$$

fitted through the sampled data in the time domain $[t_1, t_2]$, $0 < t_1 < t_2 < 1000$. The selection of time domain $[t_1, t_2]$ for curve fitting is a compromise between very long times and very short times because at very long times the data becomes too noisy (see the log-log curves in Figs. 7.4 and 7.5) and extracting meaningful information about long time tail from it becomes very difficult, and at very short times the extremely slow decay of correlation function considerably alters the computed long time behavior. The particular form of $\phi_s(t)/\phi_s(0)$ given by Eq. 7.9 has been chosen for curve fitting because earlier studies suggest that in the long time limit $\phi_s(t)$ varies as $t^{-\alpha}$ [4], where $\alpha$ is the dynamical exponent and $\alpha = D/2$ for a $D$ dimensional system ($D \geq 2$). The values of $t_1$ and $t_2$ and the parameters $b$ and $\alpha$ for various STP sets and particle densities are given in table 7.4.

Fig. 7.4 shows that for $n = 0.05$ curves for all the STP sets follow nearly identical trend. This is expected because at low density the STP sets do not introduce significant bias in the motion of particles and hence in an ideal ensemble velocity time correlation functions for them should not differ. The curves show that in the long time limit the dynamical exponent ($\alpha$) for all the STP sets is nearly 1. The exact value of $\alpha$ obtained by least square curve fitting Eq. (7.9) through the raw data is give in table 7.4 along with the uncertainty in 68.3% confidence interval. These $\alpha$ values shows that the dynamical behavior of SPLG-1 model is truly two-dimensional for $n = 0.05$ for all the STP sets given in table 7.1. In fact, this conclusion is also applicable to dynamical behavior of the SPLG-1 model for all $n \leq 0.05$ because the bias introduced by the STP sets becomes more and more insignificant as the mean particle density decreases.

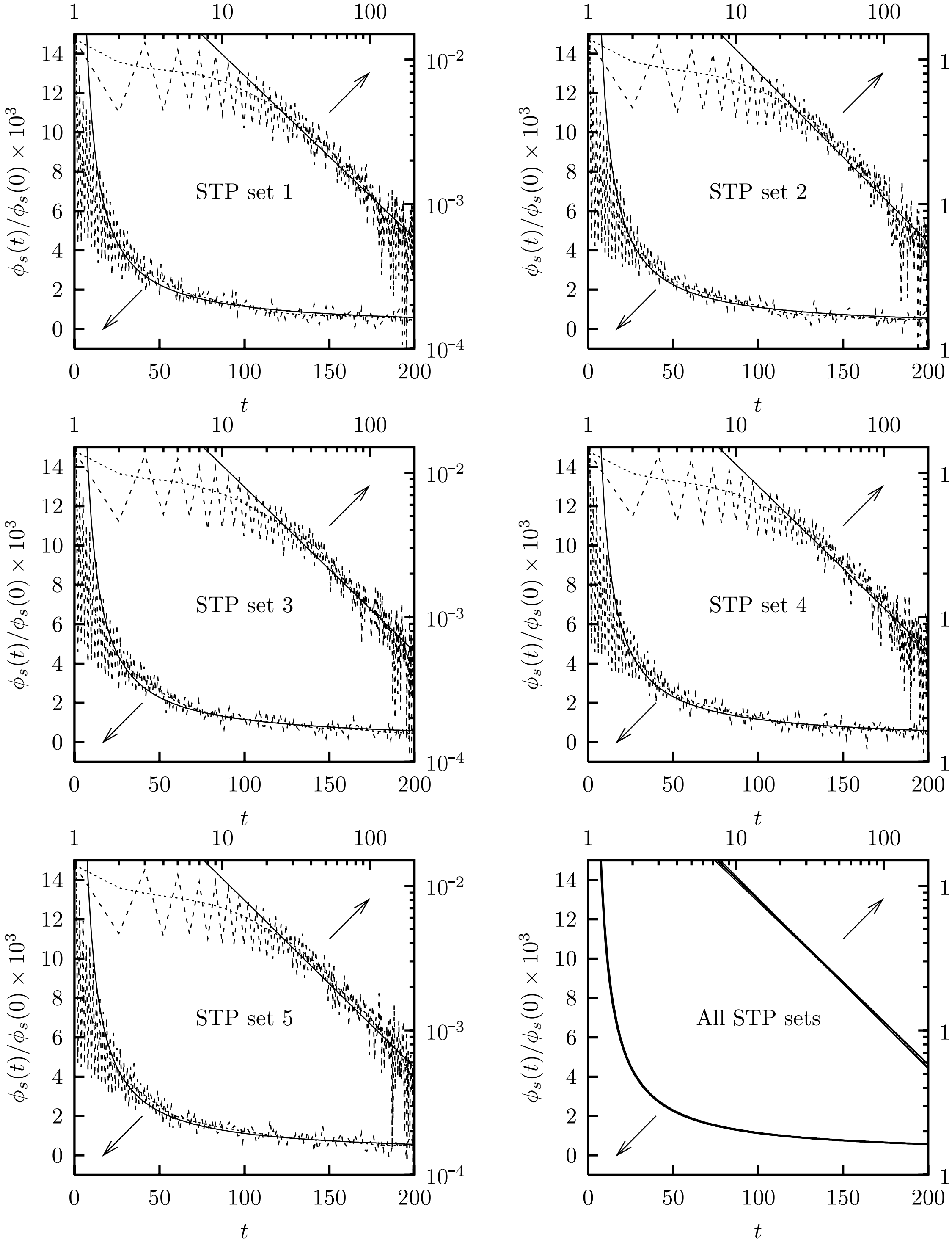

**Figure 7.4:** Time variation of time correlation function for velocity at lattice sites for $n = 0.05$ for various STP sets. Sampled data (big dash), after Bezier smoothing (small dash), and curve fit (solid).

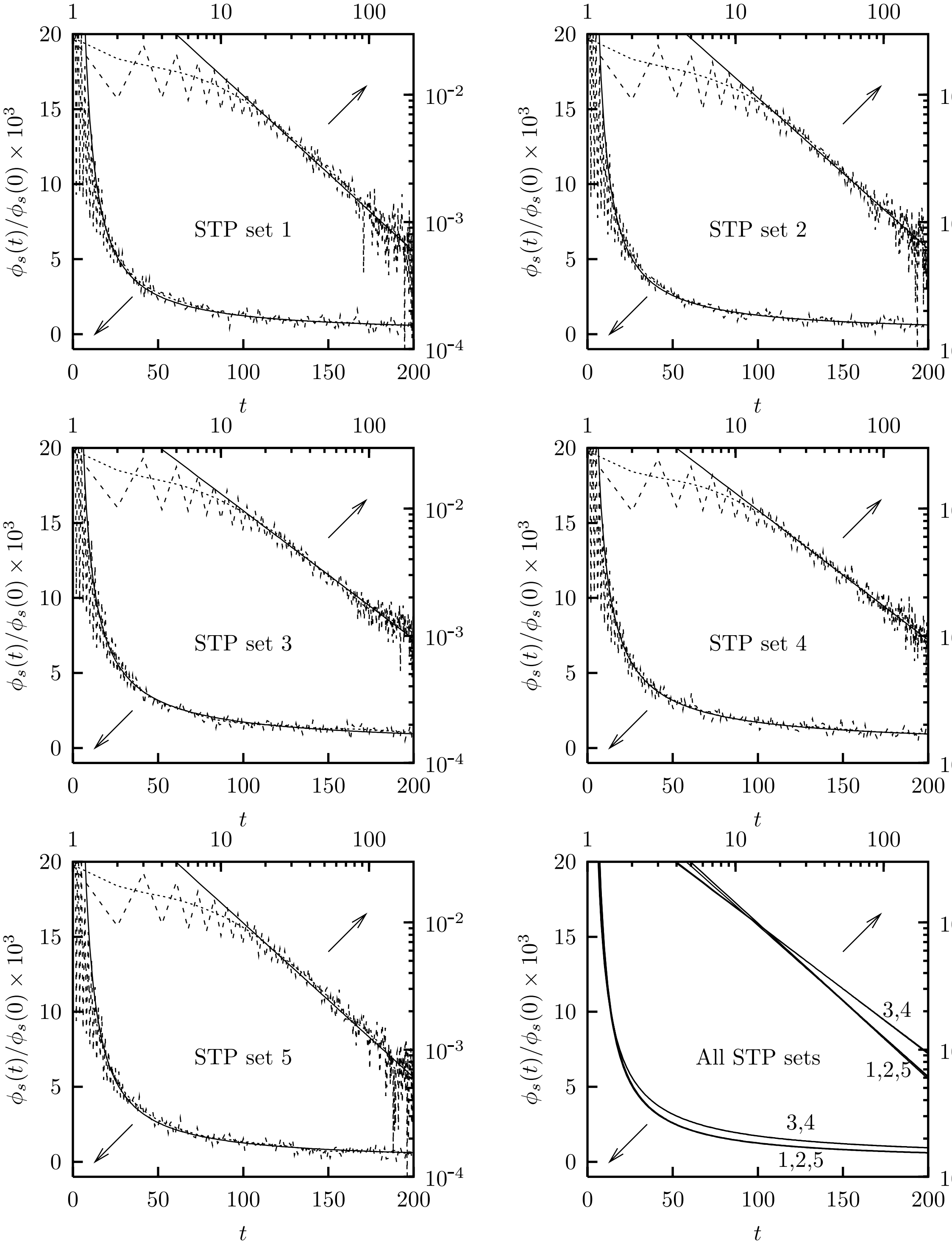

**Figure 7.5:** Time variation of time correlation function for velocity at lattice sites for $n = 0.10$ for various STP sets. Sampled data (big dash), after Bezier smoothing (small dash), and curve fit (solid).

In Fig. 7.5, the velocity time correlation curves for $n = 0.10$ show a behavior dependent on STP sets. The dependence stands out in the last graph of this figure wherein the curves for STP sets 3 and 4 group together and separate out from those for STP sets 1, 2, and 5. The $\alpha$ values given in table 7.4 for the curves in these two groups also differ considerably indicating that for $n = 0.10$ the dynamical behavior of the SPLG-1 model is considerably different for the STP sets in these groups. The $\alpha$ values show that the dynamical behavior of the SPLG-1 model for STP sets 1, 2, and 5 is truly two-dimensional for $n = 0.10$ (and less), whereas for STP sets 3 and 4 the dimensionality lies somewhere between one and two. The grouping of curves and corresponding difference in $\alpha$ values is because of the bias introduced by the STP sets 3 and 4 in the motion of particles. It is interesting that for the biased STP sets the dynamical exponent is less than 1 implying that biases introduced by violation of conservation laws (specifically, the law of conservation of angular momentum) generate dynamical behavior of lesser dimensionality compared to that in the absence of biases. In these simulations, this happens because the biased STP sets enforce order in the motion of particles—making them move in CCW direction as discussed in Sec. 7.5.1—and thus lead to dynamical behavior of lesser dimensionality.

## 7.6 Conclusions

This chapter relates to simulation of a system of particles enclosed in a box at equilibrium using the SPLG-1 model. The results indicate that the SPLG-1 model is physically consistent and suitable for simulation of diffusive systems. The major conclusions that can be obtained from the simulation results presented in this chapter are as follows:

**1)** The motion of particles in the SPLG-1 model forms a Gaussian Markov process for STP sets which satisfy all the conservation laws including the law of conservation of angular momentum. The time variation of mean and mean square displacements of particle is as expected physically. Since the diffusive behavior of this model is identical to that reproduced by the Langevin equation, its the diffusive properties can be obtained by solving the Langevin equation.

**2)** The STP sets in which the law of conservation angular momentum is violated, introduce bias in the motion of particles and force them to move in preferred directions. This results in dynamical behavior of lesser dimensionality compared to two as indicated by the dynamical exponent which is less than unity. The exact nature of bias can be determined by analyzing the STP sets.

**3)** At long times the time correlation function for velocity decays as expected for two-dimensional systems. The dynamical exponents being unity indicate that the dynamical behavior of the model is two-dimensional. This is expected for a physically consistent two-dimensional model. Compared to this, the dynamical exponents of HPP and FHP gases indicate dynamical behavior of dimensionality other than two (between one and two for the HPP gas, and between two and three for the FHP gas [4]).

**4)** The theoretical estimation of mean free path in the SPLG-1 model carried out in chapter 6 (Secs. 6.5.2 and 6.5.2.2), and as given by Eq. (7.8), is strikingly accurate despite many simplifications and approximations involved.

# Chapter 8

# Relaxation of Strong Density Perturbation in Finite Length Tube

. . . that was neat and simple. Isn't it petty, too?
*It is fundamental and necessary, sire.*
True. Still, petty.
*It shows feasibility and applicability, sire.*
Of? To?
*More involved constructions, sire.*
Like?
. . .

In this chapter relaxation of strong density perturbation in a system of particles enclosed in a finite length tube has been simulated and studied using the SPLG-1 model constructed in chapter 6. Various aspects that have been studied, simulation setup, and simulation results are described in the following sections.

## 8.1 Description of the Problem and Simulation Setup

For studying the relaxation of strong density perturbations in a fluid enclosed in a finite length tube, simulations were carried out using a closed tube of dimensions $(N_x, N_y) = (516, 255)$ with 2 lattice site thick boundary walls (the thickness of walls, though specified for the sake of completeness, is irrelevant for results). Initially the tube was partitioned into left ($x \leq 52$) and right ($x > 52$) chambers and which were filled with particles at equilibrium at density of $n_1$ and $n_2$ particles per site, respectively. In actual simulations these initial densities were taken to be $n_1 = 0.11$ and $n_2 = 0.04$ particles per lattice site. The mean free time (and also the mean free path) of particles at these densities is 6.35035616632152 and 21.6488386218452, respectively. During simulations the chambers were merged by instantaneously removing the partition at $t = 0$ and the behavior of the system was observed for 1000 time steps. At each time step the mean particle density ($n$), the mean $x$-momentum ($M_x$), and the mean $y$-momentum ($M_y$) were sampled at all $x$ stations in the tube using a $(1, 251)$ window centered at $y = 127$. The sampled data were ensemble averaged over 50 realizations of the system for reducing noise. The noise remaining after ensemble averaging was removed by Bezier filtering/smoothing.

## 8.2 Selection of the STP Set

In the simulations presented in this chapter the STP set 1 shown in table 7.1 was used in combination with the STP set shown in table 7.2. The selection of STP set (from table 7.1) was based on the results of studies conducted in Sec. 7. These studies suggest that the STP sets 1, 2, or 5 are suitable for simulation of neutral system of particles because they do not introduce bias in the motion of particles till $n = 0.10$. Theoretically, however, the STP set 2 is known to have bias, though negligible and not effecting simulations till $n = 0.10$. As a result, it cannot be used for simulation of systems having compressible phenomena because its behavior at high densities ($n > 0.10$) is not known. This leaves STP sets 1 and 5 as suitable for simulation of neutral system of particles involving compressible phenomena. From these, the STP set 1 has been selected. This selection, though essentially arbitrarily, is influenced by the following differences observed between the STP sets 1 and 5—(i) The number of solutions of the BCC 4 is more in the STP set 1 compared to that in the STP set 5 and hence the STP set 1 causes more randomization in the motion of particles than the STP set 5. (ii) As a consequence of (i) the mean displacement of the tagged particle at any time is smaller in the simulations carried out using STP set 1 compared to that in the simulations carried out using STP set 5.

## 8.3 Analysis of Simulation Results

### 8.3.1 Transformation of the Sampled Data

In the following analysis non-dimensionalized particle density has been used for studying the spatio-temporal variation of density in the tube. For extracting various relevant parameters in these studies, the sampled particle density $n(x,t)$ has been non-dimensionalized using the following two transformations

$$\rho(x,t) = \frac{n(x,t) - n_2}{n_1 - n_2} \tag{8.1}$$

$$\rho'(x,t) = \frac{n(x,t) - \min[n(x)](t)}{\max[n(x)](t) - \min[n(x)](t)} \tag{8.2}$$

Among these two transformations, the transformation given by Eq. (8.1) is a global transformation because it non-dimensionalizes the sampled particle density at all time steps using global parameters, *i.e.*, parameters which are invariant in the complete space-time domain of simulation. Whereas, the transformation given by Eq. (8.2) is a (temporally) local transformation because it non-dimensionalizes the sampled particle density at a particular time step using parameters which are local to that time step, *i.e.*, the non-dimensionalization parameters are different for different time steps. Specifically, Eq. (8.1) employs the global maxima $n_1$ and global minima $n_2$ of density in the complete space-time domain of simulation for non-dimensionalization, whereas Eq. (8.1) employs the global maxima $\max[n(x)](t)$ and global minima $\min[n(x)](t)$ of density in the tube at each time step. Consequently, $\rho(x,t)$ will henceforth be referred to as *globally non-dimensionalized particle density* and $\rho'(x,t)$ will be referred to as *locally non-dimensionalized particle density*. In view of Eqs. (8.1) and (8.2) the mutual relationship between $\rho(x,t)$ and $\rho'(x,t)$ is

$$\rho'(x,t) = \frac{\rho(x,t) - \min[\rho(x)](t)}{\max[\rho(x)](t) - \min[\rho(x)](t)}$$

In studies involving spatio-temporal variation of momentum and velocity, a functional called *extrema* has been used frequently. Usually, the term “extrema” collectively refers to the “minima” and “maxima” of a function, however, in this study *extrema* has been used as functional defined as follows: The extrema of a quantity $A(x)$, $\text{ext}[A(x)]$, is defined as the value of $A(x)$ having the largest absolute value. Alternatively, $\text{ext}[A(x)]$ is the value of $A(x)$ at the maxima of $|A(x)|$, *i.e.*,

$$\text{ext}[A(x)] = \begin{cases} \max[A(x)], & \text{if} \quad \max[A(x)] = \quad \max[|A(x)|] \\ \min[A(x)], & \text{if} \quad \min[A(x)] = -\max[|A(x)|] \end{cases} \tag{8.3}$$

In view of this equation, the functional *extrema* is a hybrid of the functionals *minima* and *maxima*. Henceforth, all the references to “extrema” in this study will only be as the functional defined by the equation given above.

For studying the spatio-temporal variation of velocity ($\boldsymbol{V}$) of particles in the tube, the mean velocity of particles at a given location in the tube at a given time is computed from the sampled density and momentum data using the equation

$$\boldsymbol{V}(x,t) = \frac{\boldsymbol{M}(x,t)}{\rho(x,t)} \tag{8.4}$$

where $\boldsymbol{V} \equiv (V_x, V_y)$ and $\boldsymbol{M} \equiv (M_x, M_y)$.

### 8.3.2 Some New Terminology

The basis of the following analysis of the spatio-temporal variation of density, momentum, and velocity lies in the $\rho$-$x$-$t$, $M_x$-$x$-$t$, and $V_x$-$x$-$t$ surfaces generated from the sampled data. These surfaces show the propagation of various quantities in the tube in time after removing the diaphragm, *e.g.*, the $M_x$-$x$-$t$ surface shows propagation of $x$-momentum in the tube in time. These quantities propagate in the form of waves showing complex dynamics. Consequently, in the following analysis of the sampled data, the waves seen on these three surfaces have been referred to by the following names—(i) *density wave:* wave seen on the $\rho$-$x$-$t$ surface, (ii) *$x$-momentum wave:* wave seen on the $M_x$-$x$-$t$ surface, and (iii) *$x$-velocity wave:* wave seen on the $V_x$-$x$-$t$ surface (here $V_x$ is $x$-velocity of particles). Whether these waves are identical or not, various similarities and differences between these waves, and mutual relationship among these waves and their properties are revealed in the course of the analysis presented in the following section. The only motivation of introducing these three intuitively appealing terms is to simplify the analysis and bring out some important aspects of dynamics in closed systems, and nothing else.

### 8.3.3 Variation of Density

The sampled spatio-temporal variation of density in the tube is shown in Fig. 8.1. A planar projection of this figure on the $x$-$\rho$–plane for time steps $t = 0(20)1000$ is given in Fig. 8.2. These figures show a density wave (shock wave) propagating in the tube. The wave moves from left to right, collides with the right wall, reflects, and then starts moving from right to left. During propagation from left to right in the tube, the wave gets dissipated (looses strength) monotonically till it reaches and collides with the right wall at $t \approx 530$. On collision with the right wall, strength of the wave increases and its velocity starts reversing. Full reversal of velocity occurs at $t \approx 600$. At this time the density (or, strength of the wave) attains a new maxima with the non-dimensionalized density at the peak of the wave rising to $\approx 2$ times its value at the beginning of collision. Full reversal of velocity marks the end of collision of the density wave with the wall. Following this, the wave starts moving from right to left showing almost similar dissipation trend as in its motion from left to right at long times. The rate of dissipation at the beginning of right to left motion of the density wave is slower compared to that observed during its left to right motion. The complete dynamics of dissipation of the density wave in the time duration $[0, 1000]$ is summarized in Fig. 8.3 wherein the variation of maximum density in the wave with time has been shown. In this figure the slope (decay rate) in the region of almost linear decay in density, *i.e.*, in $90 \leq t \leq 520$, is $\partial n/\partial t \approx -2.130 \times 10^{-5}$ particles per time-step or $\partial \rho/\partial t = 3.042 \times 10^{-4}$ per time-step.

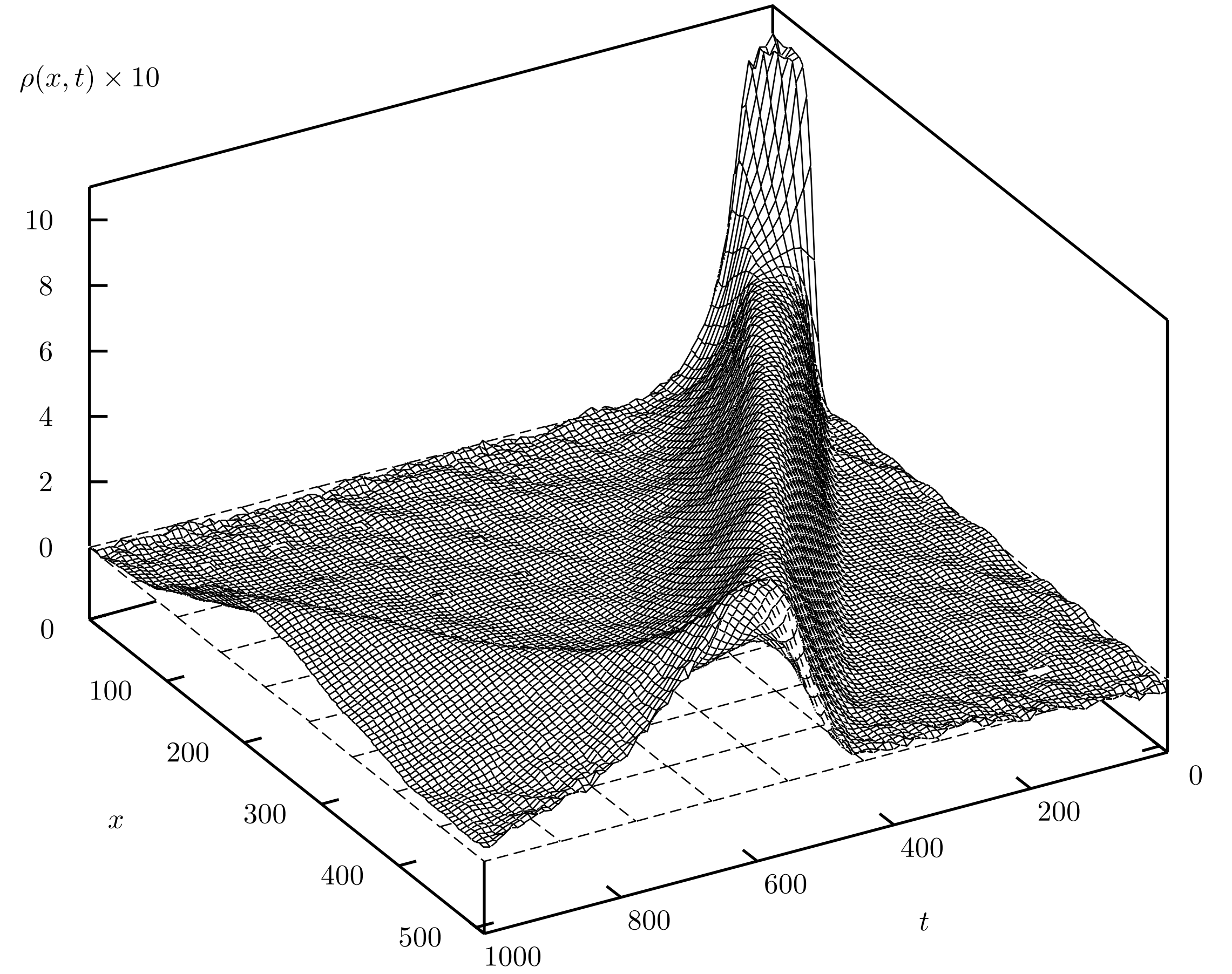

**Figure 8.1:** The spatio-temporal variation of density in the tube for $n_1 = 0.11$, $n_2 = 0.04$, and STP set 1. Data has been Bezier smoothed.

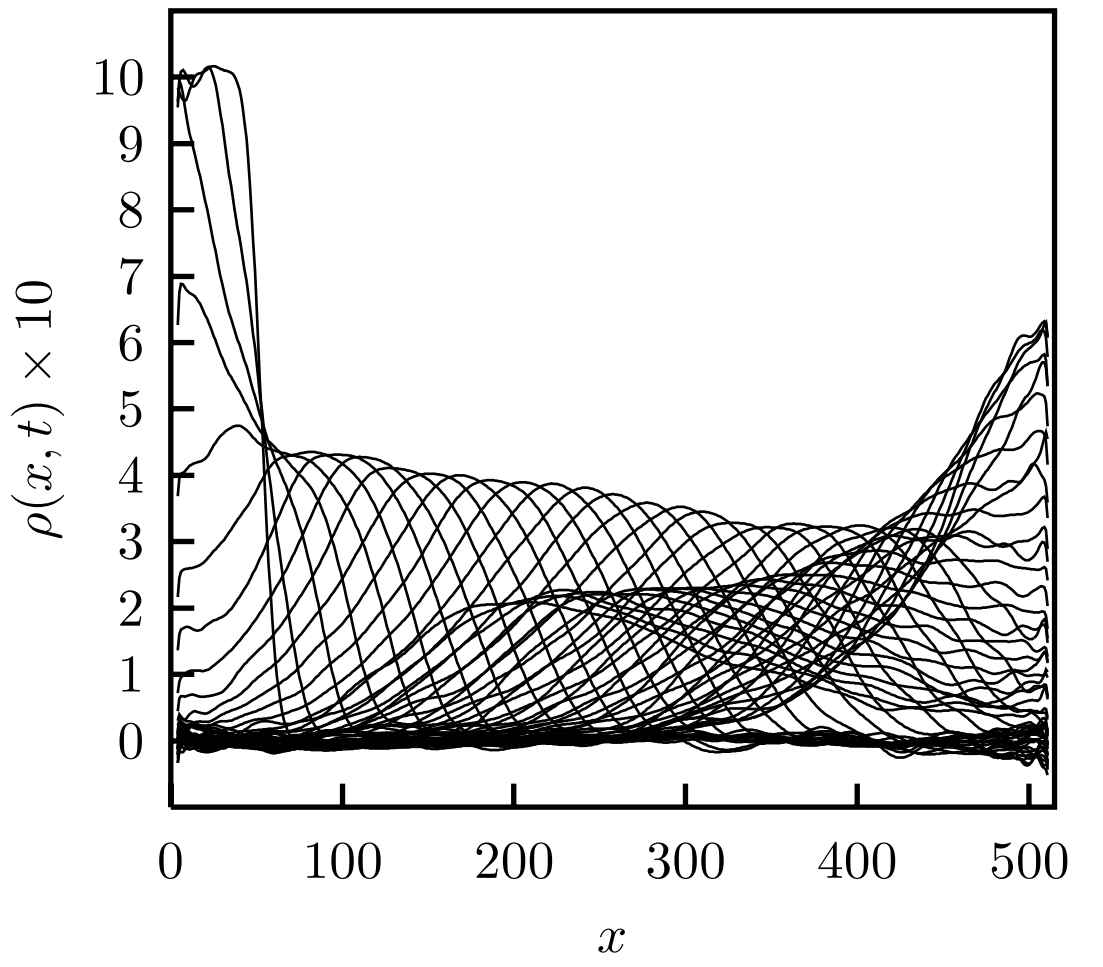

**Figure 8.2:** Spatial variation of density in the tube at time steps $t = 0(20)1000$ for $n_1 = 0.11$, $n_2 = 0.04$, and STP set 1. This figure is projection of Fig. 8.1 on the $x$-$\rho$–plane. Data has been Bezier smoothed.

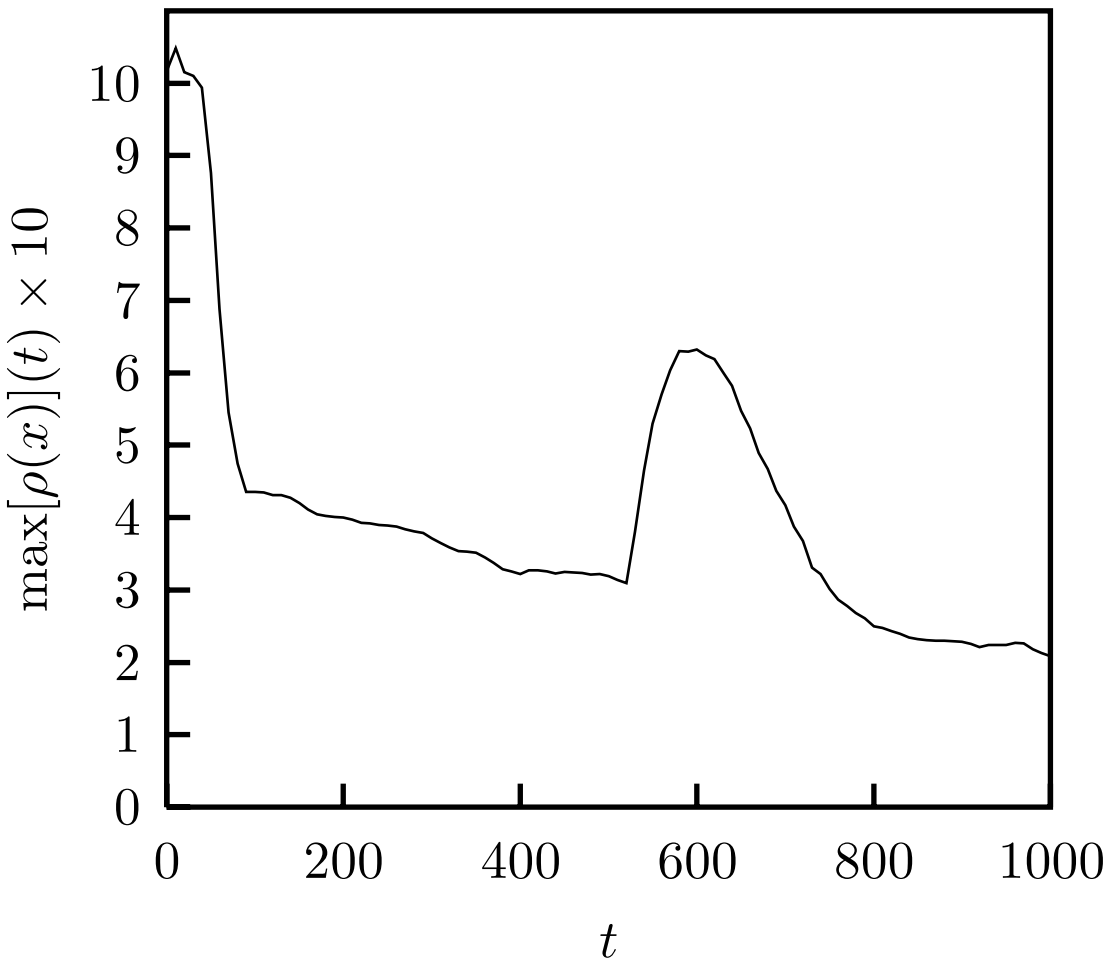

**Figure 8.3:** Variation of maximum density in the tube with time for $n_1 = 0.11$, $n_2 = 0.04$, and STP set 1. Maxima have been extracted from the Bezier smoothed data.

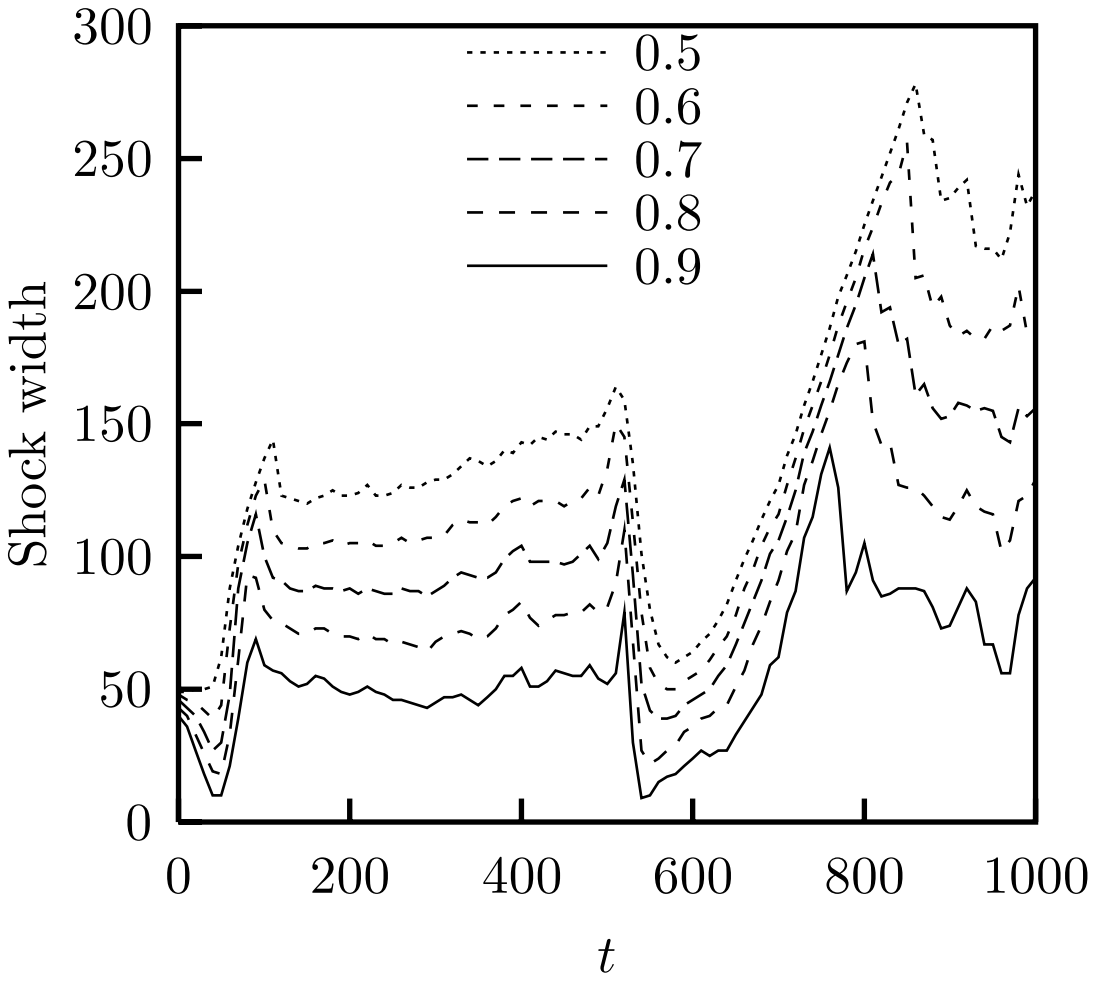

**Figure 8.4:** Variation of the width of the density wave with time for $f = 0.5$, 0.6, 0.7, 0.8, and 0.9. The legends give $f$ value associated with each line style.

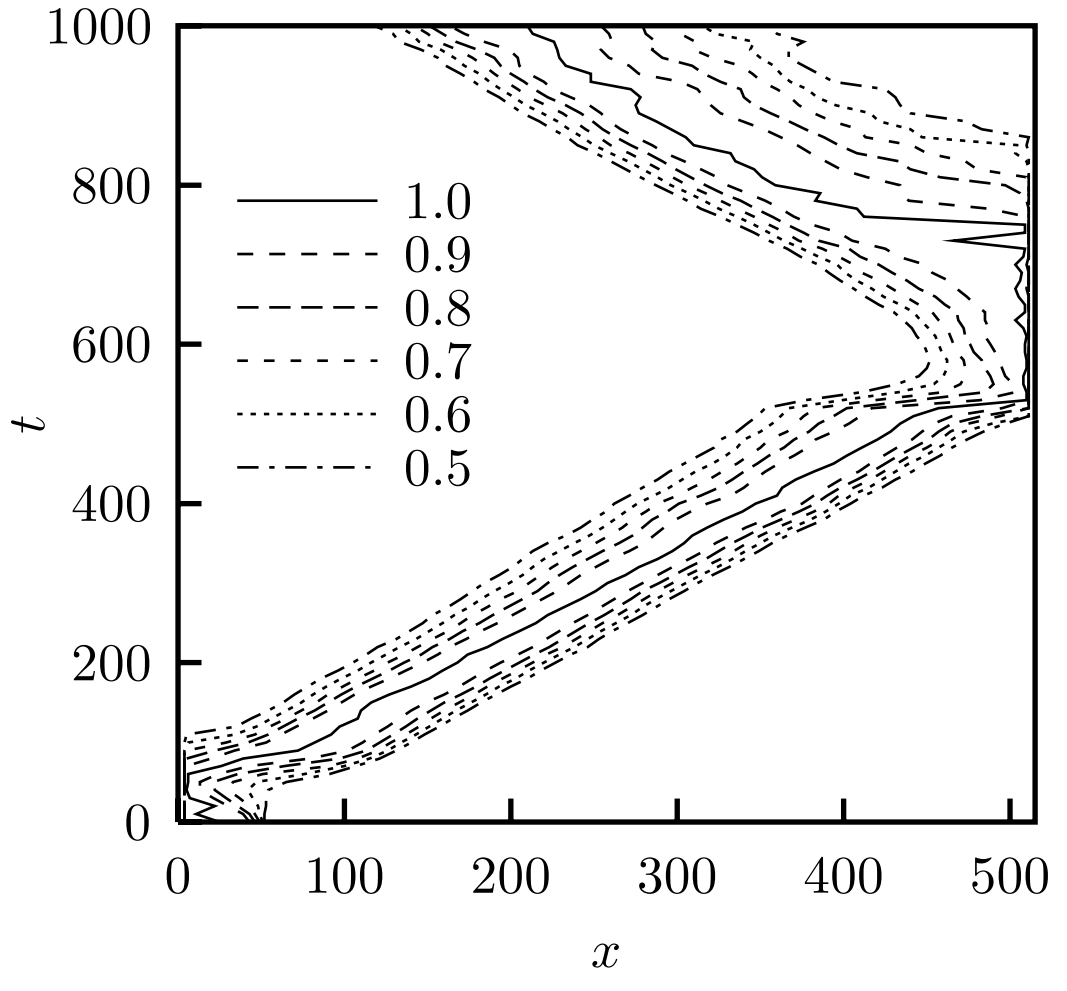

**Figure 8.5:** Path take by the density wave and the edges of its width defining sections located at $f = 0.5$, 0.6, 0.7, 0.8, and 0.9. The legends give $f$ value associated with various line styles.

As the density wave dissipates, energy and momentum are transferred to neighboring particles and its width increases. Here the width of the density wave at a given time is defined as the width of the section taken at $\rho' = f$ around the peak of the density wave in the locally non-dimensionalized density $\rho'(x,t)$ curve, $1 \leq \rho'(x,t) \leq 0$ at a given time step $t$. The section location parameter $f$ varies linearly in $[0, 1]$, with $f = 0$ at the base of the wave and $f = 1$ at the peak of the wave. The variation of width of the density wave (shock width) with time is shown in Fig. 8.4 for $f = 0.5$, 0.6, 0.7, 0.8, and 0.9. The figure shows that as the wave moves away from the wall after collision, its width increases very steeply for some time and then the rate of increase of width shows a sudden drastic drop. As the wave approaches wall, its width decreases very rapidly and goes to a minimum. At this time the wave can be said to be in full contact with the wall or wetting the wall. This minima is maintained for some time steps till the wave recoils from the wall and starts moving away from it. The dynamics described above is very clearly visible in Fig. 8.5. This figure shows the locus of the density wave in the $x$-$t$–plane along with the locus of the left and right edges of its width defining sections located at $f = 0.5$, 0.6, 0.7, 0.8, and 0.9.

| Position relative to peak of shock | $f$ | Velocity in the time interval [120, 490] |
|---|---|---|
| Left | 0.5 | $0.804051(\pm 2.73304 \times 10^{-3})$ |
| | 0.6 | $0.815798(\pm 2.55853 \times 10^{-3})$ |
| | 0.7 | $0.833323(\pm 2.95727 \times 10^{-3})$ |
| | 0.8 | $0.851996(\pm 3.57283 \times 10^{-3})$ |
| | 0.9 | $0.871771(\pm 4.34943 \times 10^{-3})$ |
| Peak | 1.0 | $0.901751(\pm 3.46226 \times 10^{-3})$ |
| Right | 0.9 | $0.882482(\pm 2.98834 \times 10^{-3})$ |
| | 0.8 | $0.877436(\pm 2.70390 \times 10^{-3})$ |
| | 0.7 | $0.876289(\pm 2.41452 \times 10^{-3})$ |
| | 0.6 | $0.878140(\pm 2.08874 \times 10^{-3})$ |
| | 0.5 | $0.885965(\pm 2.04651 \times 10^{-3})$ |

**Table 8.1:** Velocity of the peak and width defining edges of the density wave in the time interval $[120, 490]$. Numbers in parentheses give the variation of velocity in 68.3% confidence interval.

Fig. 8.5 shows that the density wave and its edges take a straight line path in the time interval $[120, 490]$. Figs. 8.1 and 8.2 show that during this time interval the wave is detached from both the left and the right walls. The velocity of the wave and its edges, computed from their path in the $x$-$t$–plane in this time interval, is given in table 8.1 along with the variation in the computed values in 68.3% confidence interval. This table shows that the speed of the density wave, when it is away from the walls, is less than the maximum speed with which a fluid particle can move at any time step. Furthermore, as one moves away from the peak towards the trailing edge (left) of the density wave the velocity decreases monotonically. On the other hand, when one moves from the peak towards the leading edge (right) of the wave the decay of velocity shows a different trend. The velocity first decreases as expected till $f = 0.7$ and then starts increasing again. This trend is not explicable.

Fig. 8.5 also shows that as the density wave approaches the right wall its velocity increases very sharply till it collides with the wall and becomes (almost) stationary. The velocity of the density wave and its edges computed from their path in $x$-$t$–plane in the time interval of their approach towards the right wall is given in table 8.2. One important aspect visible from this table is that while approaching the wall the density wave and its edges move with a velocity that is much greater than the maximum speed with which the fluid particles can move at any time step. This is because in the SPLG-1 model (in fact, in all single particle lattice gases, in general) momentum transfer can occur independent of mass transfer. This a consequence of the single particle exclusion principle which causes an additional mode of momentum transfer, *viz.*, momentum transfer during interparticle interactions, to appear in single particle lattice gases. Since interparticle interactions are processed iteratively and, in general, multiple iterations might be required at each time step momentum can be transferred over arbitrarily long distances in one time step through interparticle interactions. In fact, the speed of momentum transfer during interparticle interactions is always more than the maximum speed of mass transfer. This is because, even if one iteration is required, momentum transfer can occur across distances of 1, $\sqrt{2}$, or 2 lattice sites depending upon the configuration of interacting particles whereas in a time step mass transfer can occur only across the distance of 1 lattice site. Whether or not the net speed of momentum transfer will be more than the speed of mass transfer, depends upon the density of particles and their distribution functions. Since particle density increases rapidly as the shock approaches the wall and interparticle interactions also become more frequent because of reflection of particles from the wall, the rapid momentum transfer during interparticle interactions becomes the major mode of information transfer and leads to the observed behavior of shock near the wall.

| Position relative to peak of shock | $f$ | $t_1$ | $t_2$ | Velocity in the time interval $[t_1, t_2]$ |
|---|---|---|---|---|
| Left | 0.5 | 520 | 560 | 2.44485(±0.1658080) |
| | 0.6 | 520 | 550 | 2.97814(±0.1455680) |
| | 0.7 | 520 | 540 | 3.85541(±0.0835339) |
| | 0.8 | 520 | 540 | 4.15020(±0.0166841) |
| | 0.9 | 520 | 540 | 4.67519(±0.7981170) |
| Peak | 1.0 | 520 | 530 | 5.30000(±0.7502830) |
| Right | 0.9 | 510 | 530 | 2.37101(±0.2407780) |
| | 0.8 | 500 | 520 | 2.33623(±0.1694610) |
| | 0.7 | 490 | 520 | 1.61750(±0.0846657) |
| | 0.6 | 490 | 510 | 1.96068(±0.0838737) |
| | 0.5 | 490 | 510 | 1.35062(±0.0166910) |

**Table 8.2:** Velocity of the peak and width defining edges of the density wave during its approach towards right wall in the time interval $[t_1, t_2]$. Numbers in parentheses give the variation of velocity in 68.3% confidence interval.

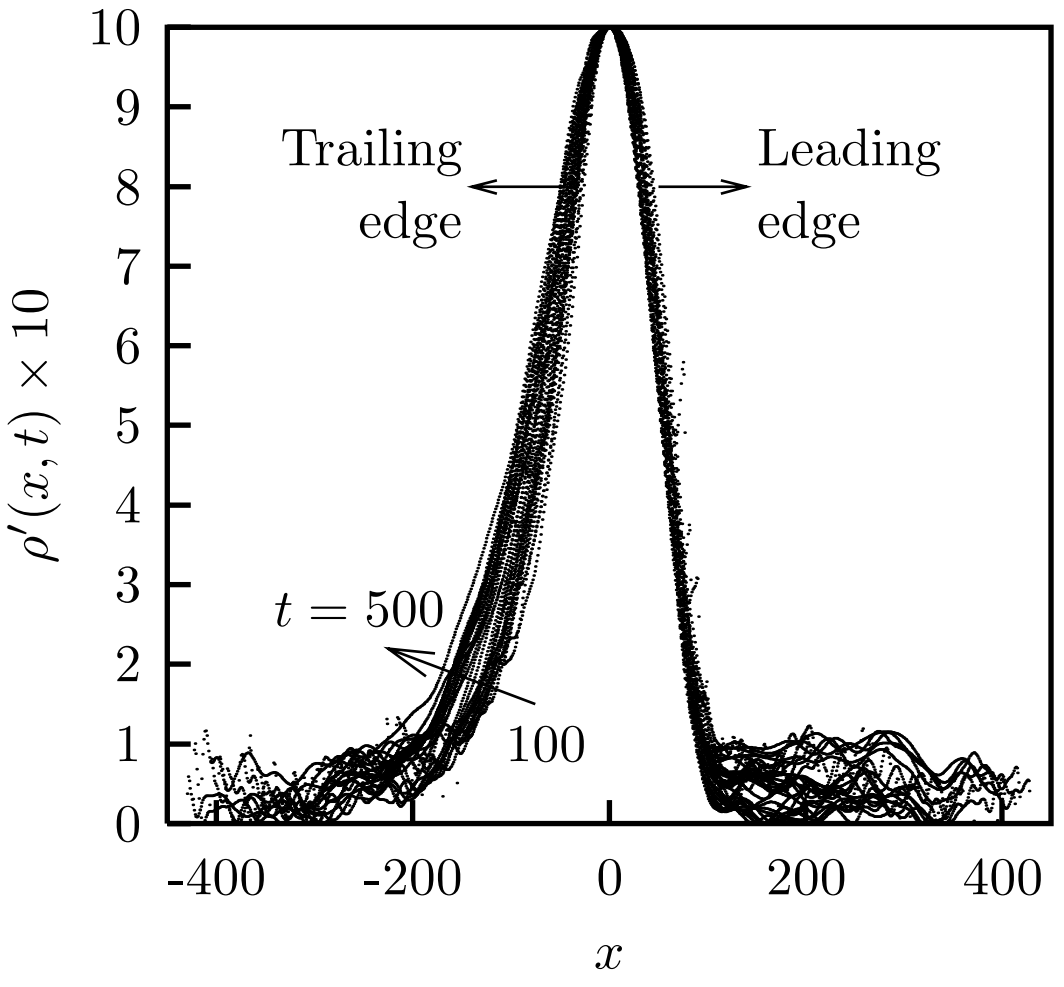

**Figure 8.6:** Collapse of the locally non-dimensionalized density $\rho'(x,t)$ curves for various time steps in the time interval $[100, 500]$ for $n_1 = 0.11$, $n_2 = 0.04$, and STP set 1.

In the time interval of almost linear decay in density at the peak of the density wave seen in Fig. 8.2, *i.e.*, in the time interval $90 \leq t \leq 520$, the locally non-dimensionalized density $\rho'(x,t)$ curves for each time step collapse together as shown in Fig. 8.6. In this figure, the $x$-origin of all the curves has been shifted at the point of maximum density in them, *i.e.*, for the curve at a given time $t$, the origin is shifted to a location $x$ such that $\rho'(x,t) = 1$. As noted previously, this figure also shows that the spread at the leading edge of the density wave is much small compared to that at its trailing edge. This behavior, including the collapse, has also been observed elsewhere in the context of free expansion of gas clouds and other related problems solved analytically under various constraints and/or computed using the direct simulation Monte-Carlo method [1]. At present, however, it is not possible to furnish quantitative comparative details because as yet results using other methods are not available for the problem simulated herein or even for a "similar" problem.

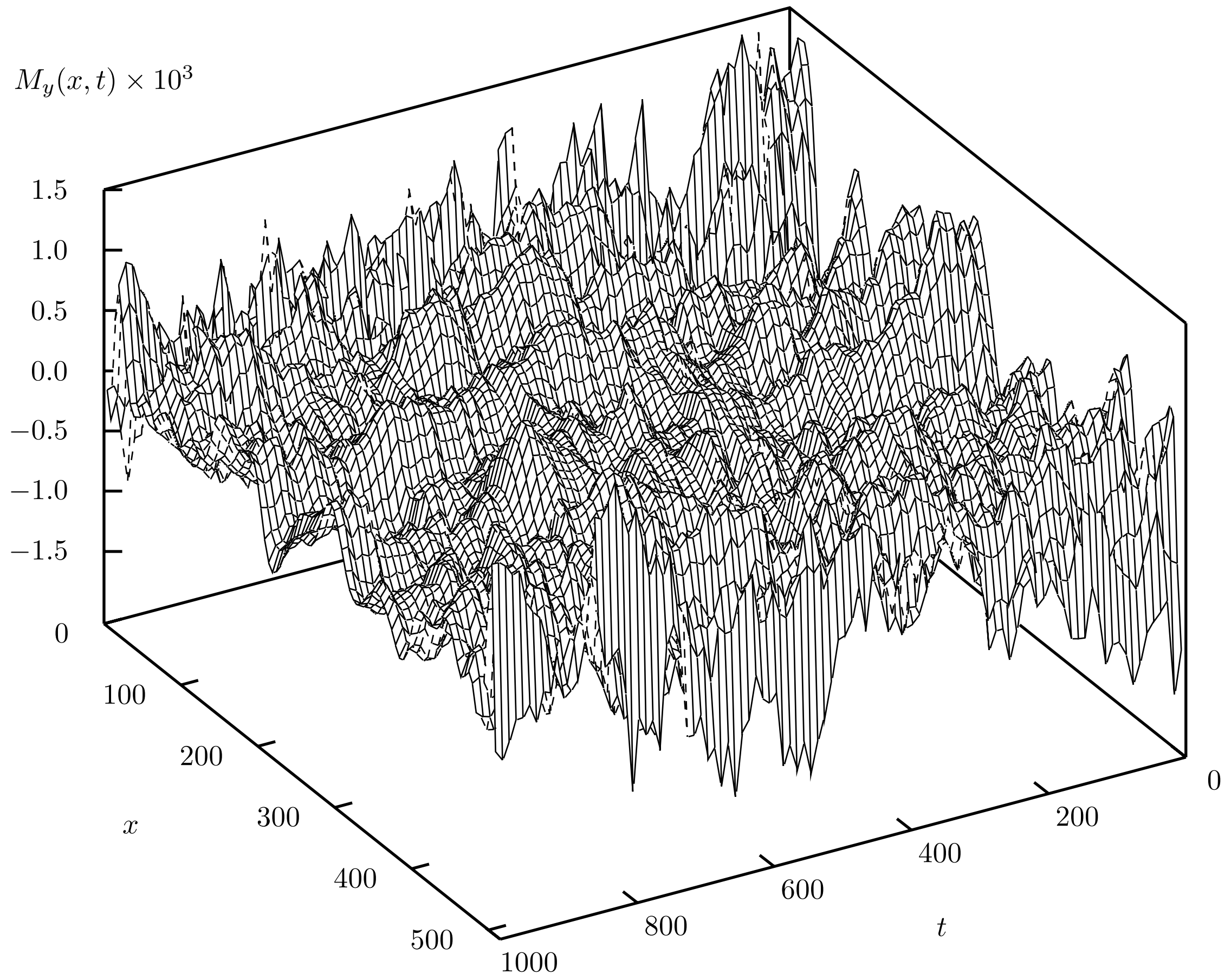

**Figure 8.7:** Spatio-temporal variation of $y$-momentum in the tube for $n_1 = 0.11$, $n_2 = 0.04$, and STP set 1. Data has been Bezier smoothed.

### 8.3.4 Variation of Momentum

In the simulations presented herein, the density front or wave propagates only along the $x$-axis. As a result, the cross flow of momentum must either be uniformly zero or limited to *thermal fluctuations*[1)] only. This is evident from the sampled $y$-momentum data, $M_y(x,t)$,

[1)]Here the term *thermal fluctuations* has been used with many reservations because the SPLG-1 model is a single speed or an athermal model, *i.e.*, in this model the energy equation is not/will not be independent of the mass and momentum equations because all particles move with the same speed. As a result, there is no well defined temperature in this model. The only thing that can be said about temperature is that

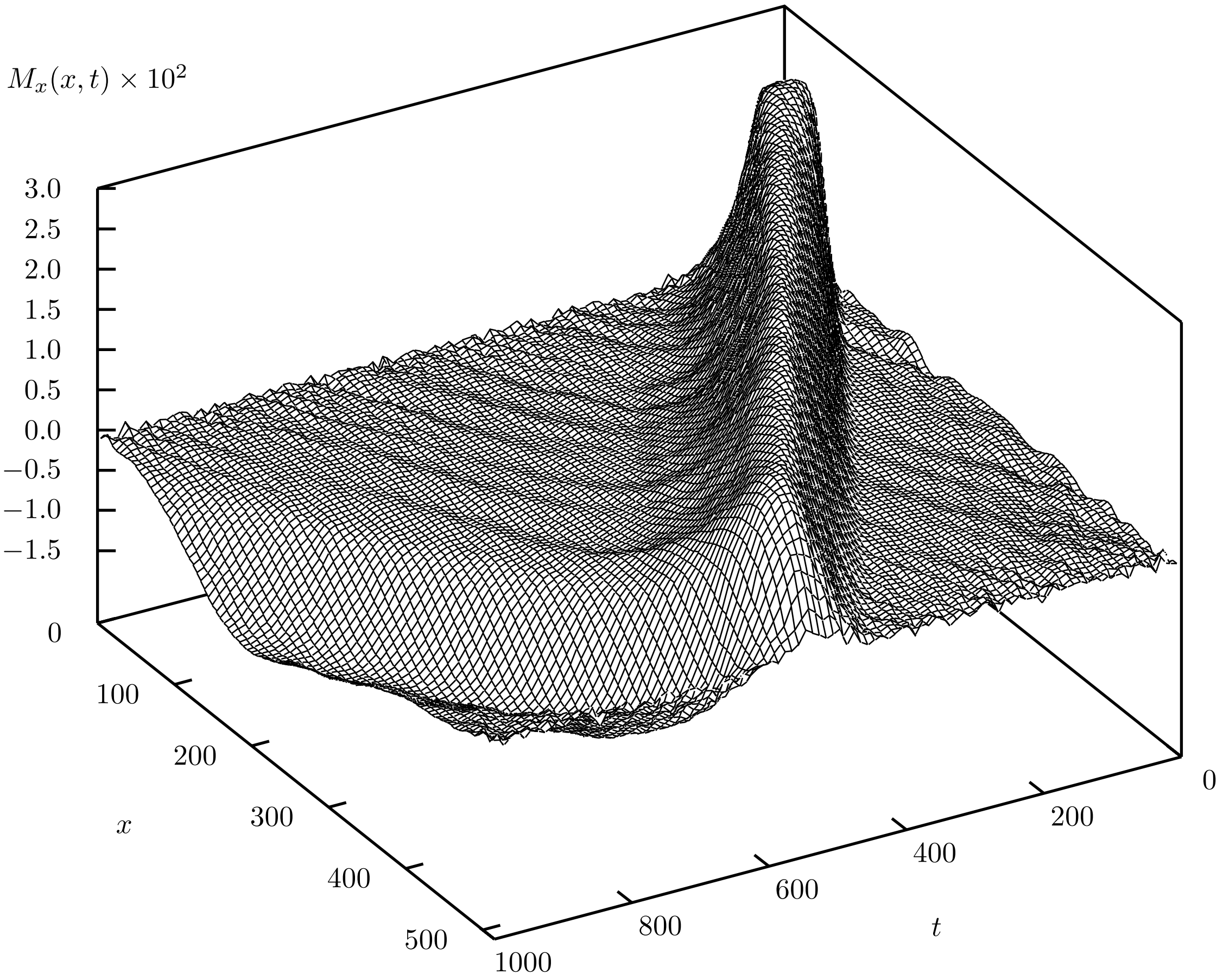

**Figure 8.8:** Spatio-temporal variation of $x$-momentum in the tube for $n_1 = 0.11$, $n_2 = 0.04$, and STP set 1. First view. Data has been Bezier smoothed.

whose spatio-temporal variation has been shown in Fig. 8.7. This figure shows that the $y$-momentum is not uniformly zero but fluctuates (almost randomly) in the bounded domain $[-1.5 \times 10^{-3}, 1.5 \times 10^{3}]$. The issue, whether these fluctuations are *thermal fluctuations* or *statistical fluctuations* continues to remain unresolved in the literature including the present investigation. The exploration and explanation of this issue goes beyond the scope this study. The resolution of this issue in the present study is complicated because of the fact that if the fluctuations are statistical they would vanish in an ideal ensemble consisting of vary large number of simulations (limit $N \to \infty$, $N$ is the number of simulations). Carrying out so many simulations, however, is not feasible because statistically the fluctuations will decay at the best as $N^{-1/2}$. As a result, in the present study usage of the term *thermal fluctuations* will be continued throughout. In case it is found that these fluctuations are actually *statistical fluctuations*, the new term can be substituted throughout without substantial impact on the findings which are not related to these fluctuations.

Spatio-temporal variation of the sampled $x$-momentum is shown in Figs. 8.8 and 8.9. These figures represent the same data set from two different perspectives. Both these it is always above the critical temperature so that solid-liquid, solid-gas, and liquid-gas phase transitions cannot occur in this model. This statement holds good for all other single speed lattice gases, including the single speed multiparticle lattice gases, *e.g.*, the HPP, FHP, and TM gases, also.

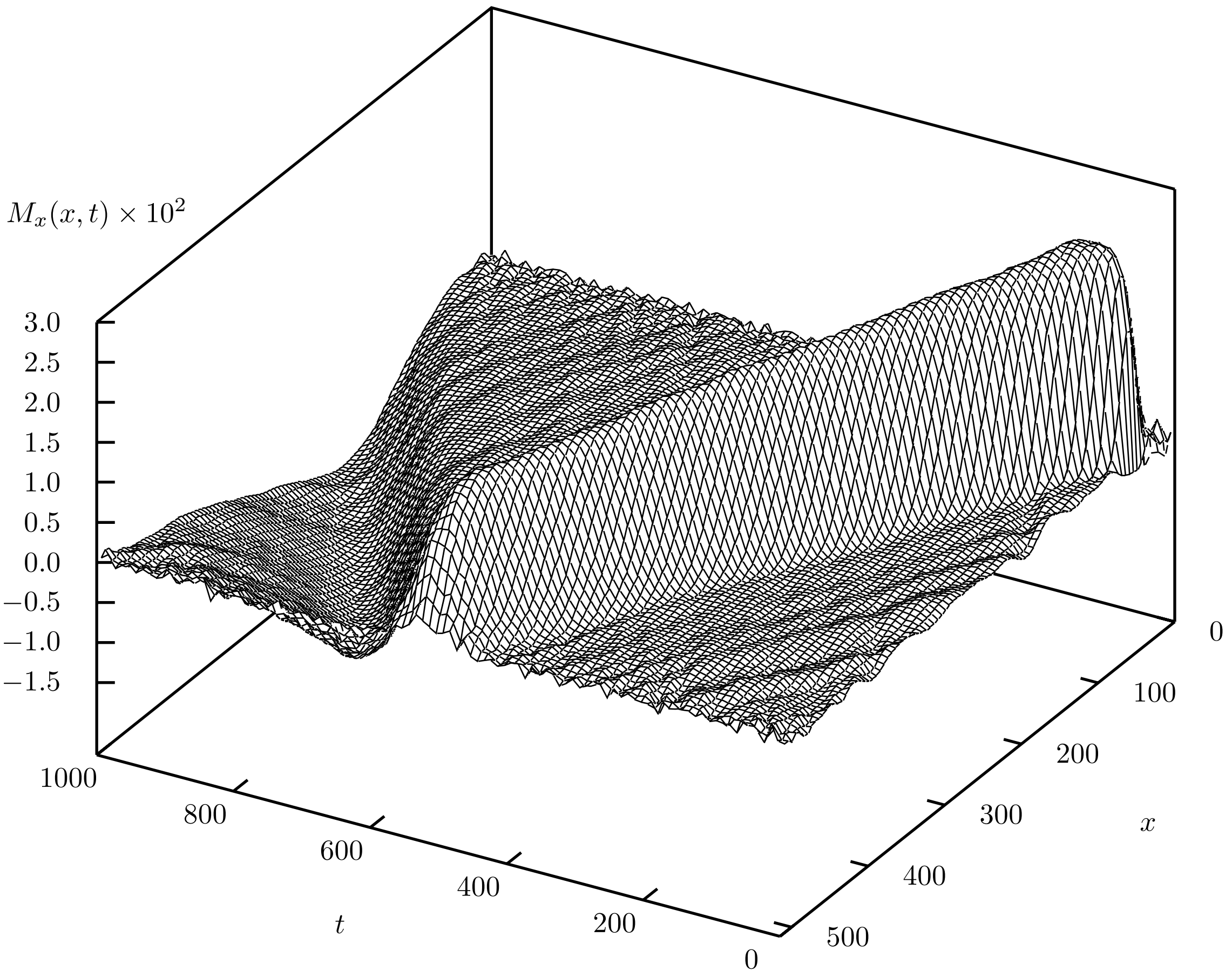

**Figure 8.9:** Spatio-temporal variation of $x$-momentum in the tube for $n_1 = 0.11$, $n_2 = 0.04$, and STP set 1. Second view. Data has been Bezier smoothed.

perspectives are shown because all the information contained in the $M_x(x,t)$ surface cannot be represented fully by a single perspective. A planar projection of these figures on the $M_x$-$t$–plane that shows the spatial variation of $x$-momentum at time steps $t = 0(30)1000$ is given in Fig. 8.10.

Figs. 8.8, 8.9, and 8.10 show that after the left and right chambers are merged at $t = 0$ the $x$-momentum of particles at the interface of the chambers starts increasing and a *momentum wave* propagates into the tube. This wave must be distinguished from the wave seen in Fig. 8.1 which is referred to as the *density wave*. For some time steps after $t = 0$ the $x$-momentum of the momentum wave increases very rapidly, attains a maxima, and then starts decaying. The decay is slow and almost linear till the momentum wave comes in contact with the right wall. On contact with the wall the $x$-momentum starts decaying very rapidly. The decay continues till $t \approx 620$. At this time the $x$-momentum of particles becomes indistinguishable from *thermal fluctuations* but still continues to reverse. The reversal is completed at $t \approx 650$.

As noted in Sec. 8.3.3, the density wave starts moving away from the wall after $t \approx 600$ with its trailing edge still in contact with the wall. The trailing edge remains in contact with the wall till $t \approx 650$. During this time interval, *i.e.*, in $[600, 650]$, two events occur simultaneously in the momentum wave, *viz.*, (i) in the zone away from the wall $x$-momentum continues to decay rapidly and goes to zero, and (ii) in the zone near the wall $x$-momentum increases rapidly along the $-ve$ $x$-direction progressing towards a fresh maxima. These two events have been summarized in Fig. 8.11. The fresh maxima in the $x$-momentum of momentum wave occurs at $t \approx 820$. Following this, the momentum wave starts detaching from the wall and moves away showing almost linear decay in $x$-momentum. The complete dynamics of change in $x$-momentum of the momentum wave that occurs in the time interval $[0, 1000]$ has been summarized in Fig. 8.12 through curves showing the variation of minima, maxima, and extrema (as defined in Sec. 8.3.1) of $x$-momentum in the tube with time.

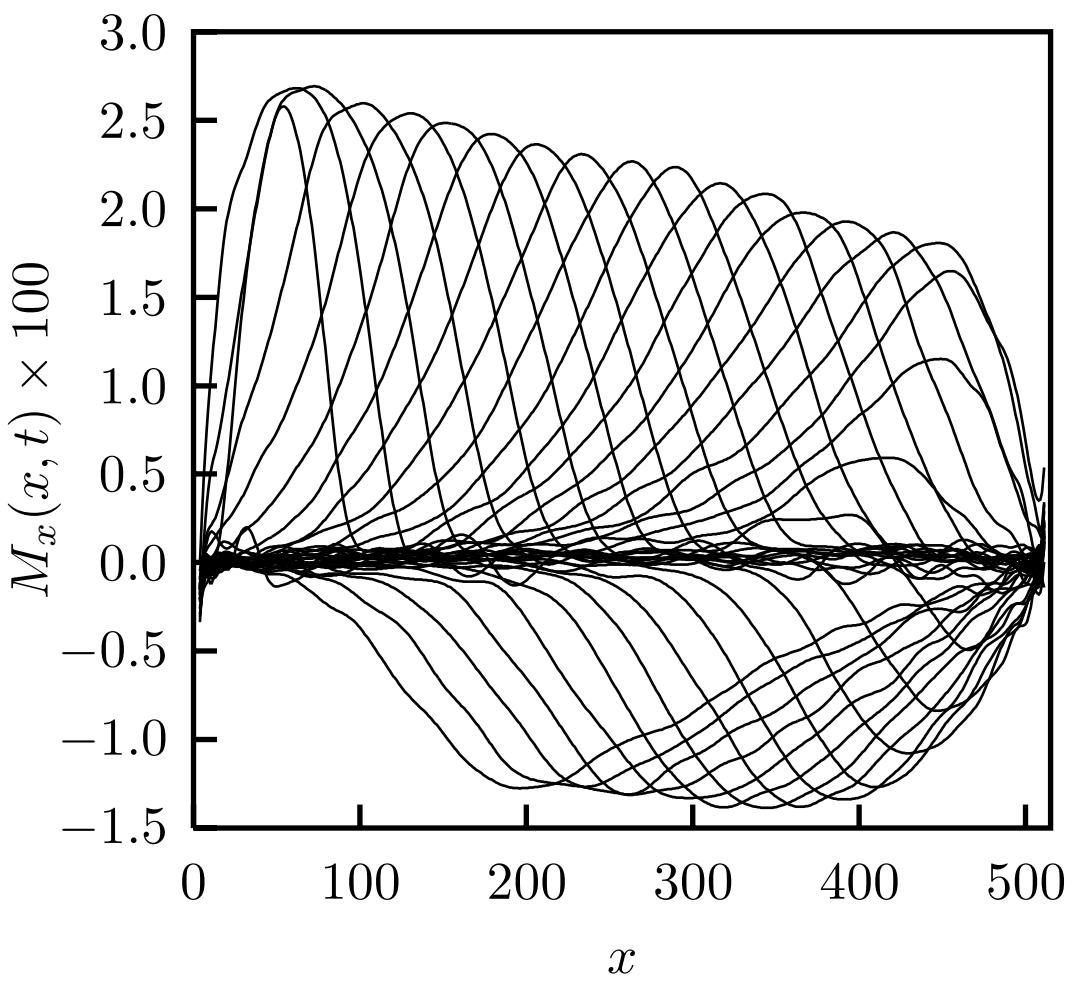

**Figure 8.10:** Spatial variation of $x$-momentum in the tube at time steps $t = 0(30)1000$ for $n_1 = 0.11$, $n_2 = 0.04$, and STP set 1. This figure is planar projection of Figs. 8.8 and 8.9 at the $x$-$M_x$–plane. Data has been Bezier smoothed.

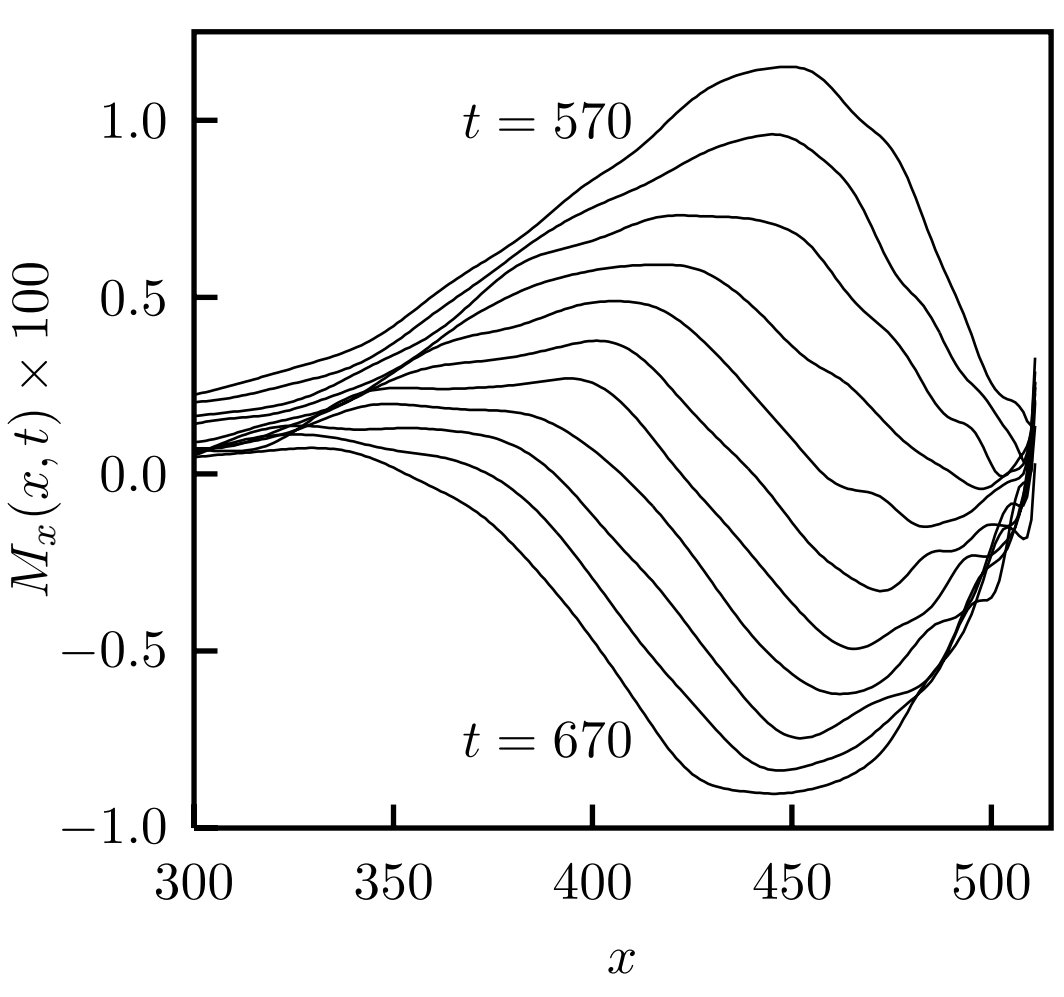

**Figure 8.11:** Spatial variation of $x$-momentum in the tube at time steps $t = 570(10)670$ for $n_1 = 0.11$, $n_2 = 0.04$, and STP set 1. This figure is expansion of Fig. 8.10 for enhancing details.

In the following analysis it is assumed that the momentum wave and its path can be accurately represented by its $x$-component alone and that the extrema of momentum wave coincides with the extrema of $x$-momentum wave. This assumption seems to be quite accurate and well justified because the global extrema of $y$-momentum varies in less than 5% of the global extrema of $x$-momentum, *i.e.*,

$$|\,\text{ext}[M_y(x,t)]| \le 0.05|\,\text{ext}[M_x(x,t)]|$$

or,

$$\max[|M_y(x,t)|] \le 0.05 \max[|M_x(x,t)|]$$

The temporal variation of $x$-momentum at the peak of density wave, *i.e.*, at locations where density is maximum at a given time, is shown in Fig. 8.13 along with the variation of extrema of $x$-momentum (or the peak of momentum wave). Fig. 8.13 shows that near the wall the behavior of $x$-momentum at the peak of density wave is drastically different from the behavior of $x$-momentum at the peak of momentum wave. For some initial time steps after $t = 0$, the $x$-momentum at the peak of momentum wave increases very rapidly attaining a maxima and then (almost) stabilizes at this value for some time. Whereas, during the same time interval the $x$-momentum at the peak of density wave shows an almost non-increasing random behavior and then increases very rapidly equalizing itself with the $x$-momentum at the peak of momentum wave at $t \approx 90$.

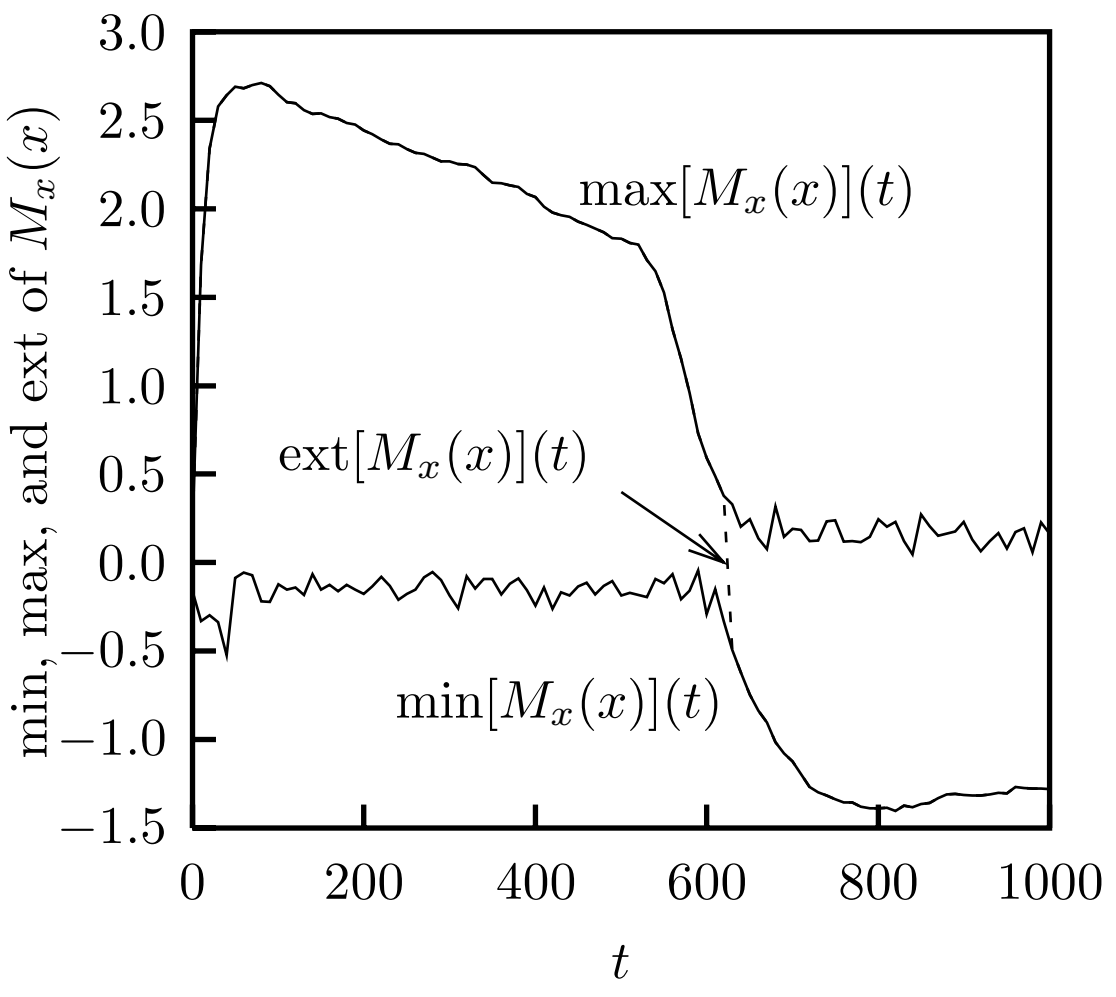

**Figure 8.12:** Variation of minima, maxima, and extrema of $M_x(x)$ in the tube with time for $n_1 = 0.11$, $n_2 = 0.04$, and STP set 1. The extrema have been extracted from the Bezier smoothed data.

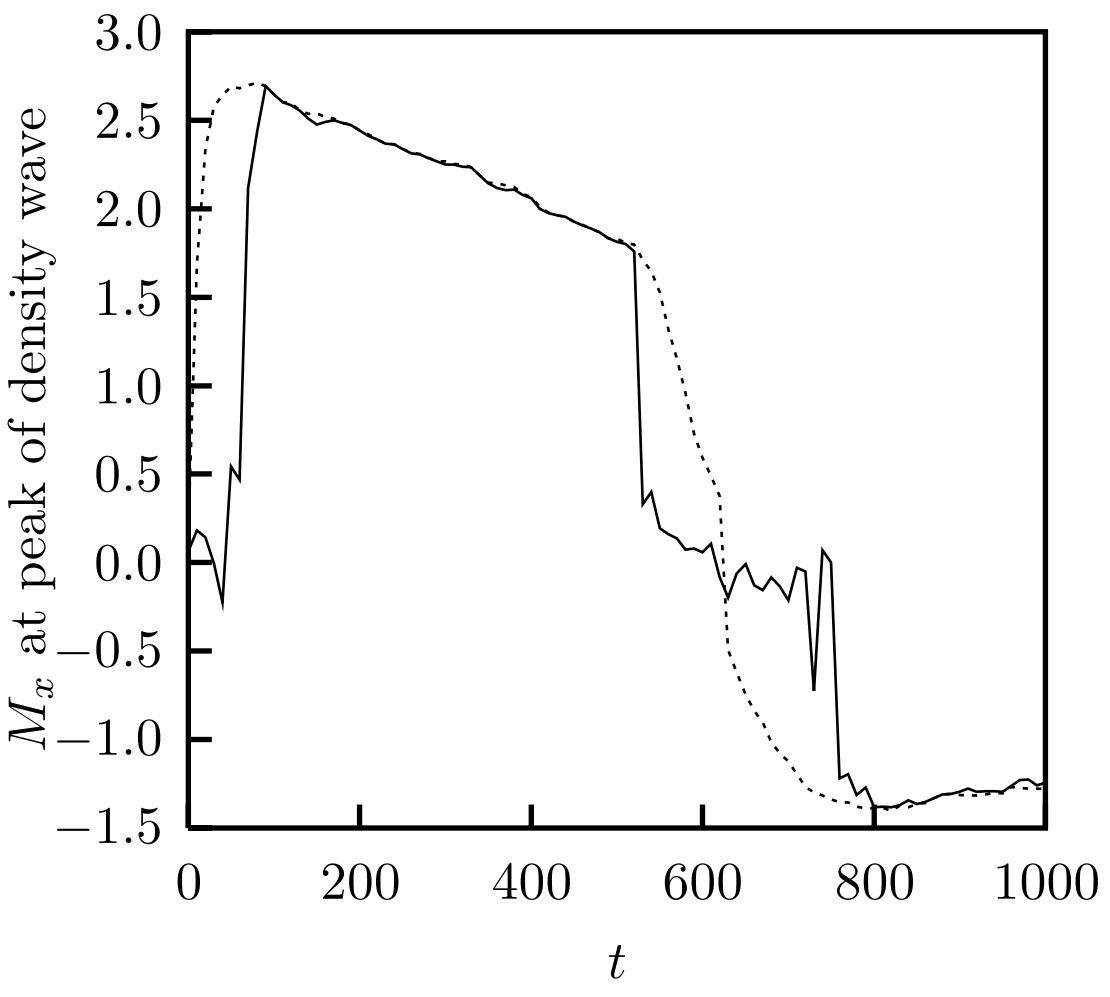

**Figure 8.13:** Variation of $M_x$ at the peak of density wave in the tube with time for $n_1 = 0.11$, $n_2 = 0.04$, and STP set 1. Broken lines show the time variation of extrema of $x$-momentum, ext[$M_x(x)$].

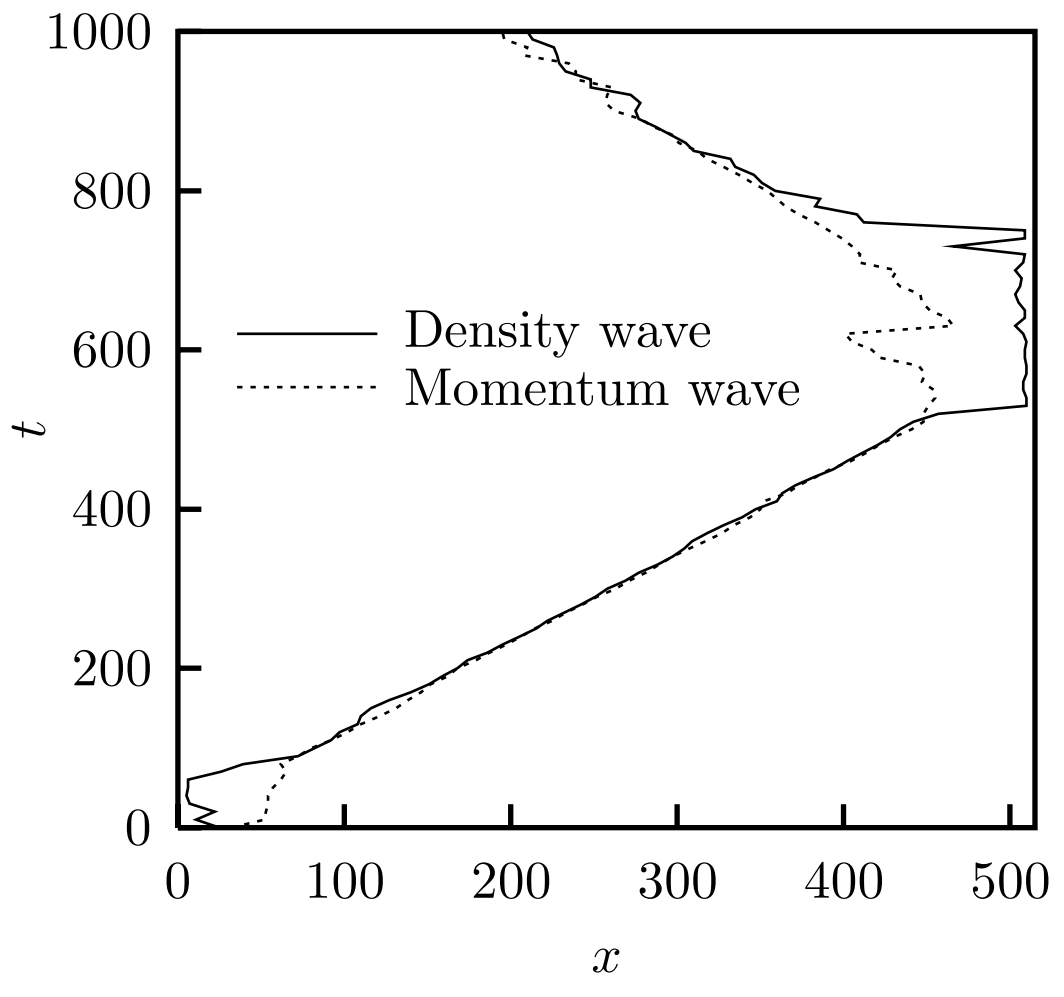

**Figure 8.14:** Path taken by the peaks of density wave and momentum wave in the tube for $n_1 = 0.11$, $n_2 = 0.04$, and STP set 1.

The above happens because the width of initially selected high density region at the left side of the container is non-zero (50 lattice sites). As a result, even after removing the partition between the left and the right chambers at $t = 0$, the maxima in density continues to occur towards the left wall for $\approx 60$ time steps. Since during this time interval the mean velocity of particles remains zero in this zone, the $x$-momentum only shows thermal fluctuates about zero. On the other hand, the particles at the interface of the chambers start moving towards the right wall immediately after the partition is removed.

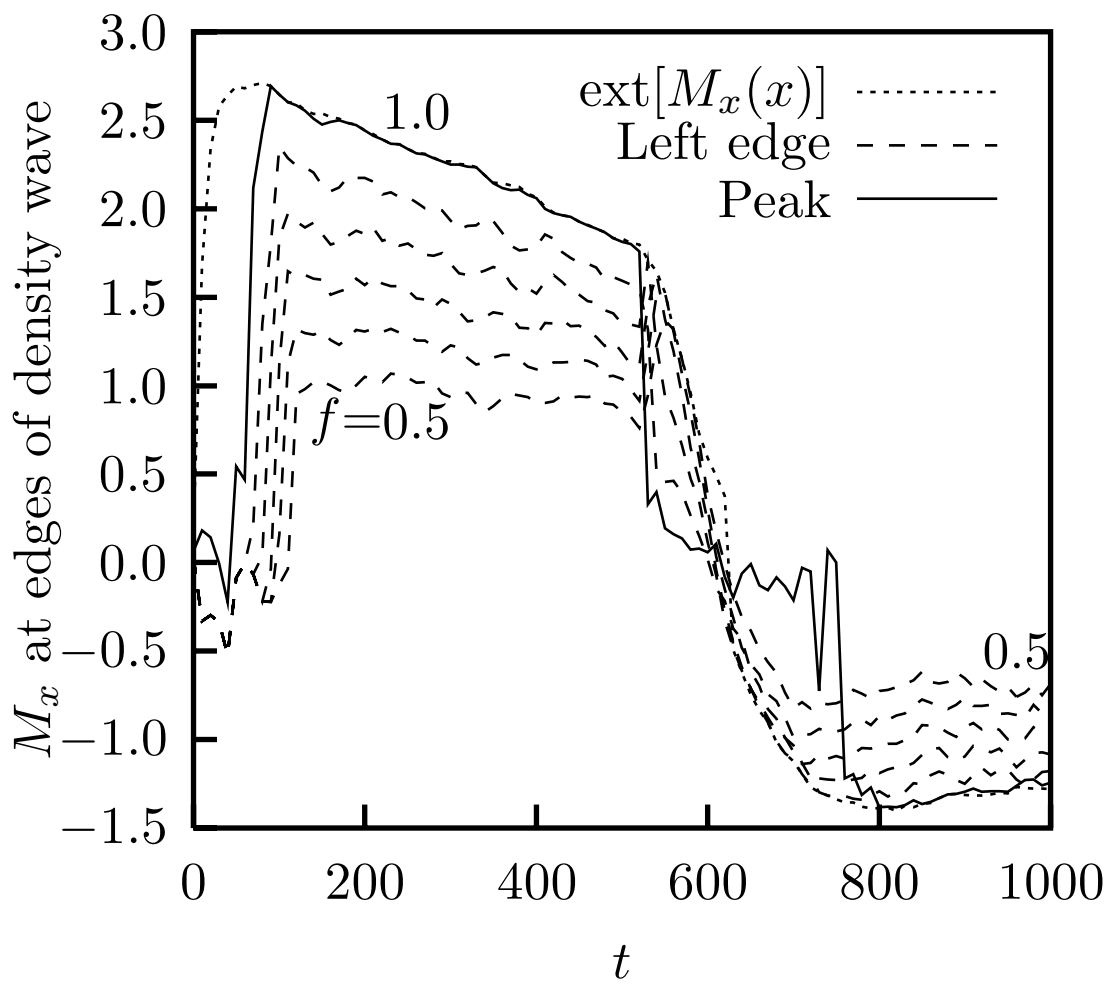

**Figure 8.15:** Variation of $M_x$ at the peak and left edges of density wave in the tube with time for $n_1 = 0.11$, $n_2 = 0.04$, and STP set 1.

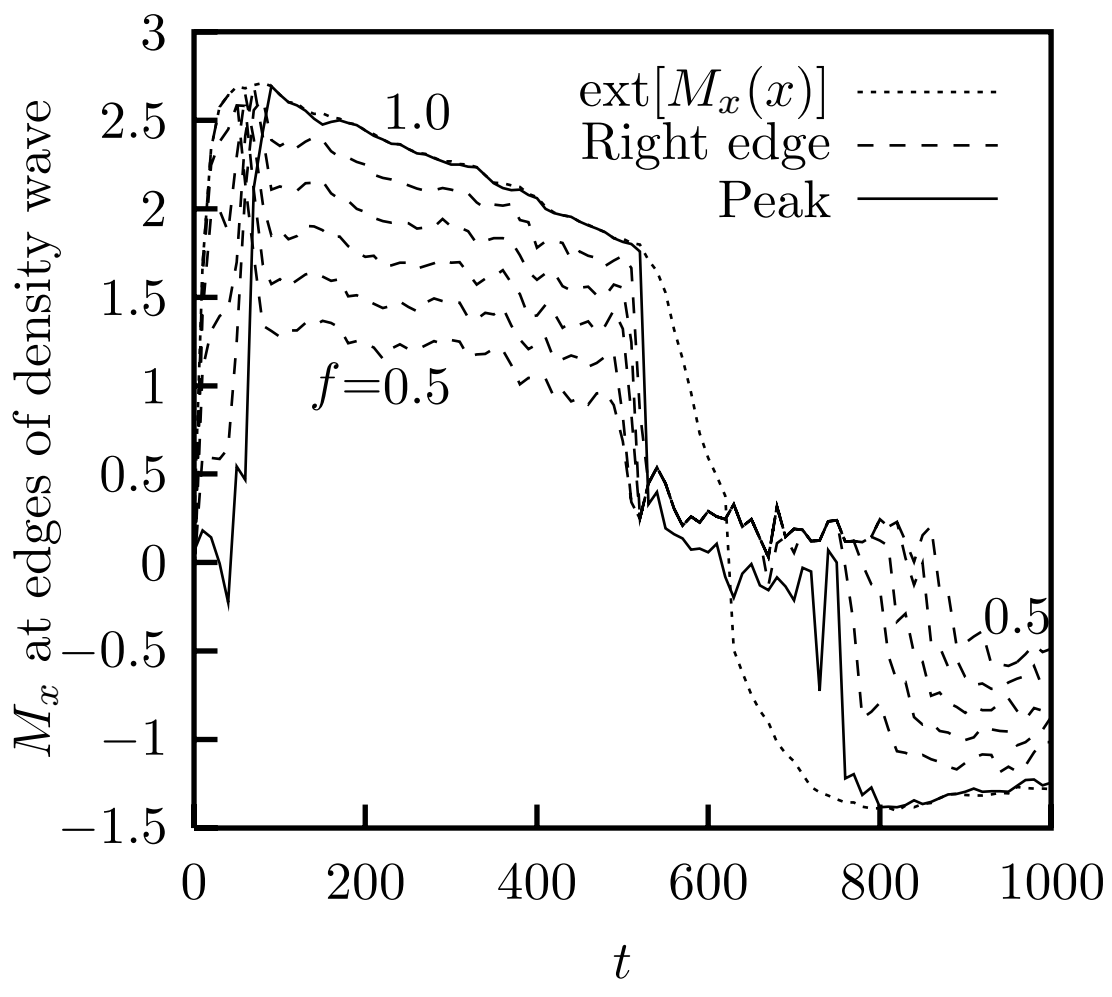

**Figure 8.16:** Variation of $M_x$ at the peak and right edges of density wave in the tube with time for $n_1 = 0.11$, $n_2 = 0.04$, and STP set 1.

As a result, the momentum wave grows rapidly at the interface and $x$-momentum attains a peak in $\approx 60$ time steps. By this time, the initial high density zone near the left wall becomes very thin, reflects from the left wall, and acquires net $x$-momentum towards the right. As this density front moves towards the right wall its peak decreases very rapidly accompanied by equally rapid increase in the $x$-momentum. From $t \approx 60$ till $t \approx 90$ the density front continuously approaches the momentum front and its peak merges with the peak of the momentum front at $t \approx 90$. Following this the density and momentum fronts move towards the right wall in phase. As the fronts approach the right wall they separate out once again, showing a behavior that is exactly opposite of what has been described above. The behavior of density and momentum waves and phase difference between them is clearly visible in Fig. 8.14 wherein the path taken by the peaks of these waves has been shown in the $x$-$t$–plane. This figure also shows that though the density wave actually hits the wall before reflecting, the momentum wave does not do so. The momentum wave never actually touches the wall but gets reflected from a distance only.

Analysis of the observed behavior of $x$-momentum outlined above is supplemented by Figs. 8.15 and 8.16 which show the variation of $x$-momentum at the peak and the left and right edges of density front along with the behavior of the peak of momentum front. These figures bring out an additional fact that the decay rate of $x$-momentum at the edges of the density wave decreases as one moves away from the peak either towards the leading or the trailing edge of the wave, *i.e.*, as $f$ increases. Furthermore, during the time interval of motion of the density wave from left to right in the tube, the decay rate at the trailing (left) edge of the density wave decreases faster compared to that at the leading (right) edge.

### 8.3.5 Variation of Velocity

In sections 8.3.3 and 8.3.4 all the references made to velocity were related to velocity of the density and momentum fronts. The velocities were computed from the path taken by these fronts in the $x$-$t$–plane. In this section, however, the behavior of velocity of particles occupying the peaks of these fronts as well as the behavior of maximum velocity of particles in the field has been analyzed. Discussion on spatio-temporal variation of $y$-velocity has been skipped because it does not give any information in addition to that already obtained from Fig. 8.7 which shows the variation of $y$-momentum.

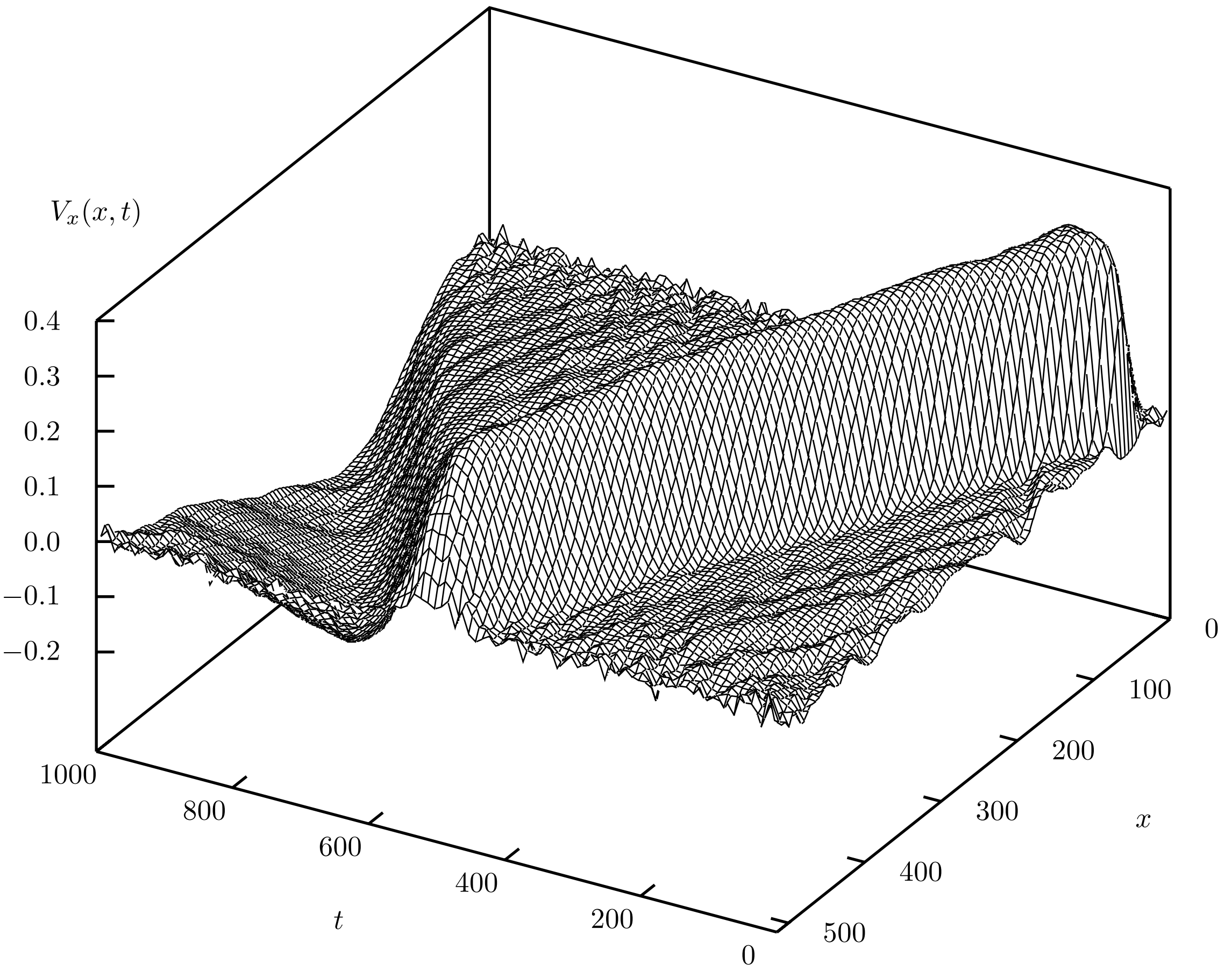

**Figure 8.17:** Spatio-temporal variation of $x$-velocity of particles in the tube for $n_1 = 0.11$, $n_2 = 0.04$, and STP set 1.

The computed variation of $x$-velocity of particles is shown in Fig. 8.17. Qualitatively, this figure appears almost similar to Fig. 8.9 with the difference that the platues in this figure are more rough and the peak of $x$-velocity front shows smaller decay rate compared to that shown by the $x$-momentum front in Fig. 8.9. These similarities and differences stand out when Fig. 8.12 is compared with Fig. 8.18 wherein the behavior of maxima, minima, and extrema of the $x$-velocity front in the tube is shown.

The time variation of $x$-velocity of particles at the peak of density front and at extrema of $x$-momentum front, along with extrema of $x$-velocity front, is shown in Fig. 8.19. In this figure, the extrema of $x$-velocity front and $x$-velocity of particles at the extrema of $x$-momentum front coincide. This is expected because in a fluid of identical particles the maximum/minimum particle velocity should occur at the point where the momentum of particles is maximum/minimum. This indicates that the velocity and momentum fronts are identical and indistinguishable at all times. This further implies that there are only two independent fronts (or, waves) that propagate in the tube: (i) the density front, and (ii) the momentum/velocity front. Furthermore, at this point it is worth recalling from Sec. 8.3.4 that these two fronts are independent only near the walls and merge together in free space.

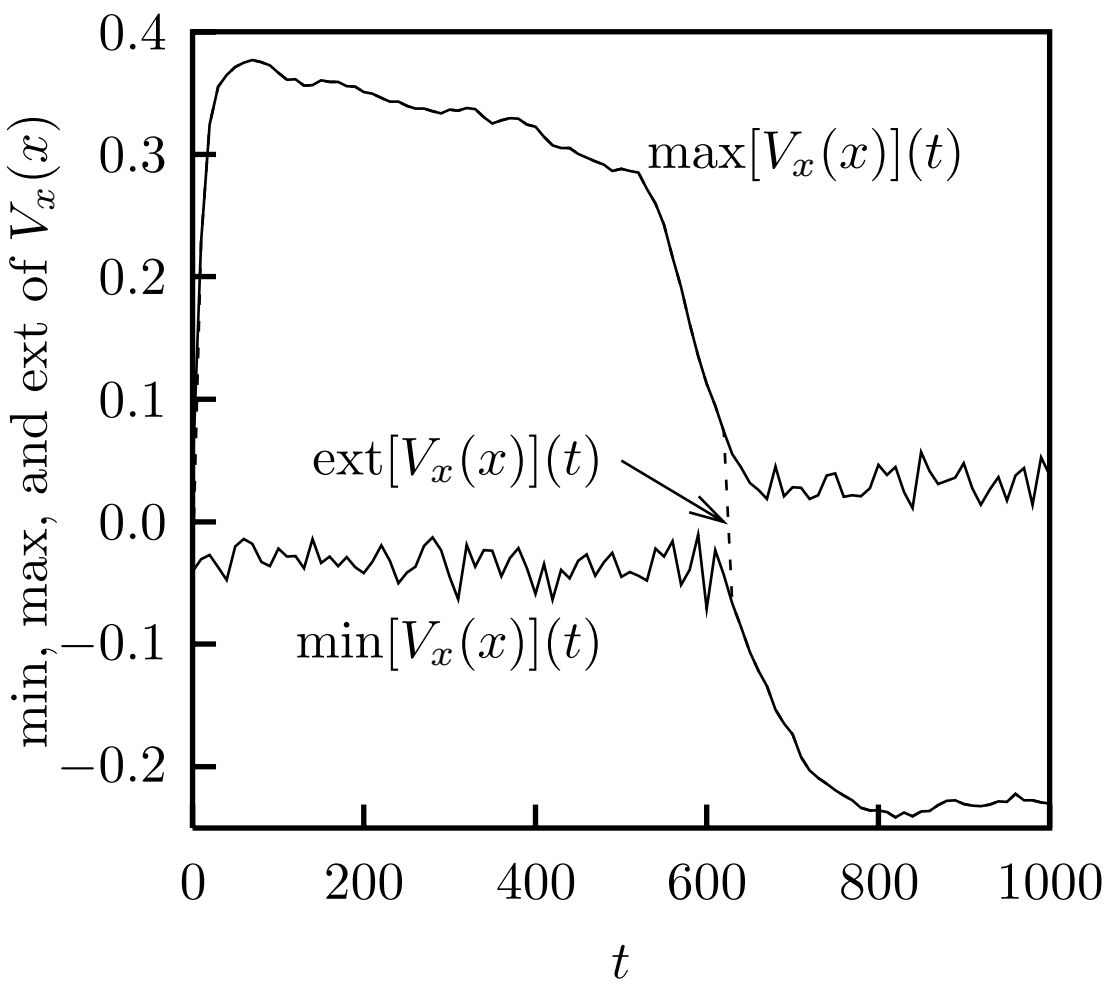

**Figure 8.18:** Variation of minima, maxima, and extrema of $V_x(x)$ in the tube with time for $n_1 = 0.11$, $n_2 = 0.04$, and STP set 1.

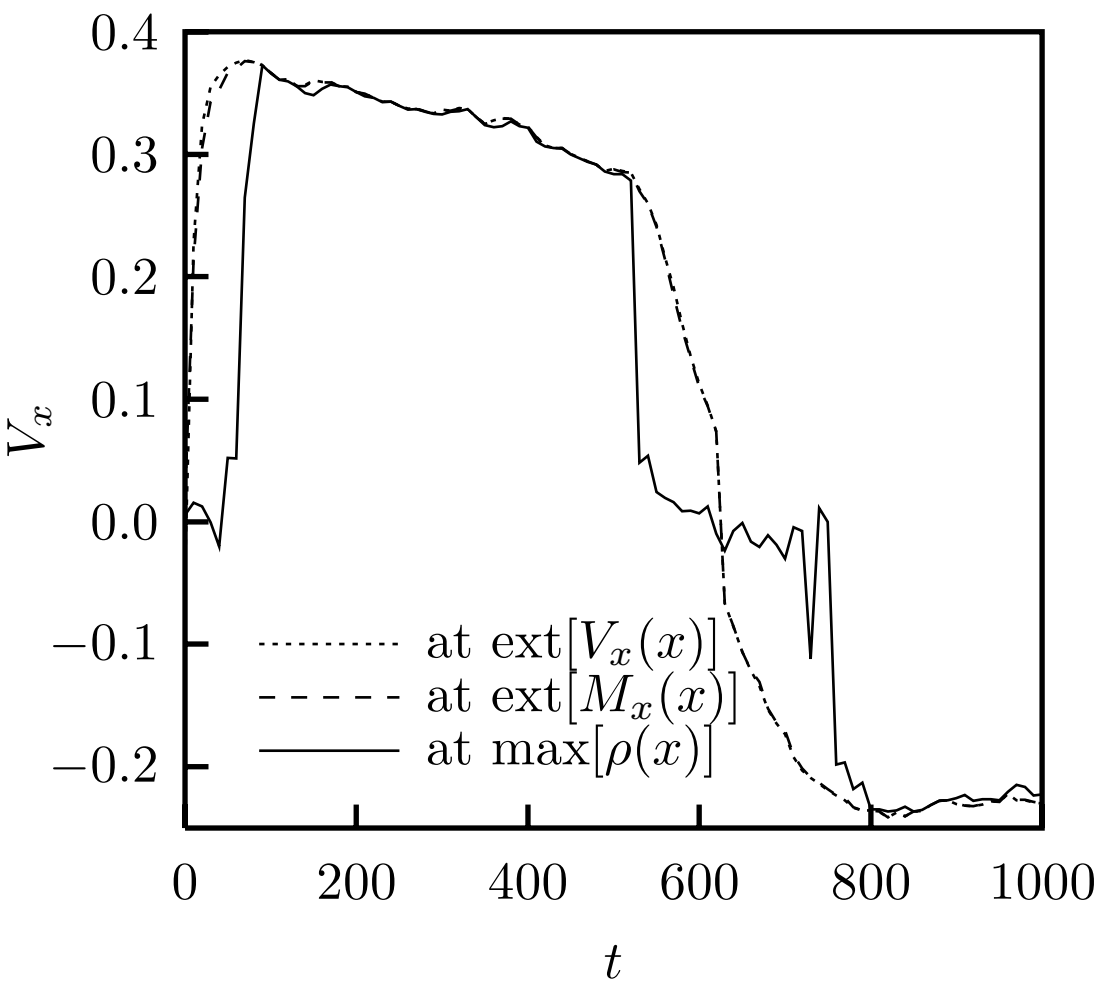

**Figure 8.19:** Variation of $x$-velocity of particles at the peak/extrema of density and momentum fronts and at the extrema $x$-velocity in the tube with time for $n_1 = 0.11$, $n_2 = 0.04$, and STP set 1.

Figs. 8.17–8.19 show that the peak $x$-velocity of particles in the tube is bounded below 0.4 sites per time step. This is considerably low compared to the velocity of density front (and momentum front) given in table 8.1. This implies that the motion of fronts is independent of the motion of particles. This is a classical observation that highlights the differences between the wave motion and mass motion.

## 8.4 Conclusions

This chapter relates to simulation of relaxation of strong density perturbation in a system of particles enclosed in a finite length tube using the SPLG-1 model. The results are new in the literature and no comparison has been possible. Besides several classical observations, some new observations regarding the propagation and interaction of waves with solid boundaries are also made in these simulations. All the phenomena observed in these simulations are physically explicable and consistent. The major conclusions that can be obtained from the simulation results presented in this chapter are as follows:

1) On removing partition separating high density and low density chambers a wave forms and propagates into the tube from high density side to the low density side. Near the walls, the mass and momentum components of the wave—in terms of maxima of density and maxima of momentum—separate out and go in different phases. Their development takes few mean free times following which they detach from the wall, come in phase with each other, merge, and propagate into the tube.

**2)** Relaxation/decay of density perturbation in tube shows linear behavior with time when the wave is away from the wall. In the vicinity of the wall, highly nonlinear relaxation pattern is observed.

**3)** The wave, when away from solid boundaries, propagates at a speed which is less than the maximum possible speed of particles. In general, the speed of the wave will never exceed (and can only become equal to) that of the maximum speed of particles when the wave is away from the walls (or, in free space). Thus, when the wave is away from the wall its Mach number will never exceed $\sqrt{2}$.

**4)** When the wave is approaching the wall and interacting with it, its speed increases vary rapidly and becomes many times more compared to the maximum speed of particles. In the simulations presented in this chapter, the peak of the density front is observed to travel at the speed of 5.3 lattice sites per time step during its approach towards the right wall. This corresponds to a Mach number of 7.49533.

**5)** In the process of approach and reflection from the wall, the wave—the maxima of density—stays with the wall for considerable time (few mean free times) before the wave reflects and starts moving away from the wall.

**6)** The motion of wave is independent of the motion of particles. This is in the sense that although the wave can propagate from one point in space to another, the particles comprising the medium may not and need not. This is a classical observation on wave propagation, *e.g.*, propagation of sounds waves and of waves on the surface of water, and is clearly visible in the simulations.

Several of the phenomena observed in the simulations presented in this chapter, in as much as the literature suggests, have not been observed earlier. This is probably because microscopic simulation of wave propagation at time scales much smaller compared the mean free time—like in the simulations presented in herein—have not been carried out earlier. As a result, comparative evaluation of the results presented in this chapter is not possible.

# Chapter 9

# Conclusions

That was long walk. Did you notice these roads do not end?
*Yes, sire.*
They bend and fork yet the path is straight.
*Yes, sire.*
What did you see—the path or its bends and forks?
*Both, sire.*
How was the walk?
*Fine, sire.*
What did you collect—pebbles or shells?
*Pearls, sire.*
Oh! The round white pebbles. Let me see.
. . .

The main objective of the present investigation has been to find a method for constructing molecular dynamics like models using cellular automata for simulation of fluid dynamic systems. To achieve this objective, which by its nature is highly interdisciplinary, many different areas have been analyzed from a single perspective and then correlated to form a single consistent whole. Consequently, the findings of this investigation are multifarious and relate to several different areas which have been investigated. Although the findings related to specific investigations have already been summarized at the end of respective chapters, all of them have been recollected, recollated, recategorized, and summarized under different and more appropriate section headings below.

## 9.1 Cellular Automata as Models of Physics and Physical Systems

The conclusions regarding applicability of cellular automata for modeling of physics and physical systems are as follows:

**1)** The formalism of cellular automata satisfies all the requirements of being an appropriate formalism for modeling of physics and physical systems.

**2)** Cellular automata models of fluid dynamic systems, being fully discrete, are not subject to problems encountered in simulation of physical systems on digital computers using integral and differential calculus based models existing in continuum space-time.

**3)** Since cellular automata are abstract mathematical systems their application for modeling and simulation of physics and physical systems is based on and requires appropriate interpretation of their basic elements.

**4)** Interpretation of the basic elements of cellular automata, not being internal to the formalism itself, has a certain context dependent flexibility which permits construction of many different types of cellular automata models of physical systems.

**5)** It is possible to develop molecular dynamics like models of fluid dynamic systems using the formalism of cellular automata by employing appropriate interpretation to the basic elements of cellular automata. A class of such models has been conceived and defined in the present investigation as *"single particle lattice gases"*.

## 9.2 Relationship of Microdynamics with Macrodynamics in Physical Systems

The objective of this investigation has necessitated studies on relationship of microdynamics and macrodynamics in physical systems, particularly fluid dynamic systems. The conclusions regarding this relationship arrived at in the present investigation are as follows:

**1)** The nonlinear terms of Navier-Stokes equation encode the overall dynamics of spatial momentum redistribution. The coefficient of nonlinear terms of Navier-Stokes equation arises from combined effect of spatial momentum redistribution during particle translation and that during interparticle interactions.

**2)** In physical systems interactions occur among particles occupying different points in space and the distance among particles never goes to zero during interactions. In physical systems the model of particles (in most general terms) is that of *point centers of forces with a hard core* (*i.e.*, non-zero range of interaction).

**3)** In physical systems spatial momentum redistribution occurs *via* two mechanism, *viz.*, (i) interparticle interactions, and (ii) particle translation. During interparticle interactions spatial momentum redistribution occurs instantaneously across distance of the order of range of interactions and smaller (but greater than zero). During particle translation spatial momentum redistribution occurs due to transfer of momentum associated with the particles from one point in space to another along with the particles.

**4)** Spatial momentum redistribution due to both particle translation as well as interparticle interactions is necessary for origin and propagation of compressible phenomena and compressibility effects in physical systems. Thus, the presence of both these mechanism is necessary in models of physical systems for correct simulation of compressible systems and phenomena.

## 9.3 Lattice Gases

General conclusions regarding lattice gases, *i.e.*, the cellular automata models of physical systems in which lattice site values are interpreted as representing the presence or absence of particles and their states at the lattice sites, are as follows:

1) Lattice gases whose collision rules have been developed using either different velocity sets or different interaction neighborhoods must be considered as different lattice gases. Different lattice gases having the same velocity set but different interaction neighborhoods will show different micro- and macrodynamical behavior.

2) Using different interpretations of the basic elements of cellular automata, different types/classes of lattice gases can be developed. At least two different classes of lattice gases—having different interpretations of basic elements of cellular automata—exist. One class, taken from literature and analyzed in the present investigation, is known as *"multiparticle lattice gases"*. The other class discovered and described in the present investigation is known as *"single particle lattice gases"*.

### 9.3.1 Multiparticle Lattice Gases

Multiparticle lattice gases cellular automata models of physical systems developed, analyzed, and used for simulation of many systems in the literature by several investigators. In these lattice gases the lattice sites values are interpreted as representing the presence or absence of one or many particles along with their states at the lattice sites. Alternatively, in these lattice gases more than one particles are allowed to occupy a lattice site at a time step. Multiparticle lattice gases, although appear similar to molecular dynamics, are not exact discrete counterparts of molecular dynamics. Major conclusions related to various aspects of multiparticle lattice gases are outlined below.

**Representation of Multiparticle Lattice Gases:** The conclusions about the nature and physical consistency of various possible representations of multiparticle lattice gases arrived at by analysis of the existing literature are as follows:

1) All multiparticle lattice gases can be represented using either multiple particle representation or partitioned spatial lattice representation. These representations, though different from each other, are equivalent and can be transformed into each other.

2) Partitioned spatial lattice representation shows number of peculiarities when the cells comprising the spatial lattice are treated as independent units. These peculiarities are: (i) the spatial location and velocity of particles become correlated, (ii) the structure of the phase space (of each particle; thus also of the entire system) becomes different from that in physical systems, and (iii) it becomes possible to define lattice gases whose evolution rules have non-zero impact parameter (these lattice gases cannot be transformed correctly to multiple particle representation because in multiple particle representation the impact parameter is necessarily zero). The peculiarities (i) and (ii) make this representation physically inconsistent.

3) The peculiarities of partitioned spatial lattice representation can be bypassed by treating the blocks, rather than the cells (or, lattice sites), comprising the spatial lattice as independent units whose location is identified with one coordinate point each. This treatment, in essence, transforms the partitioned spatial lattice representation into multiple particle representation which is a physically consistent representation.

4) In multiple particle representation the impact parameter of particles is always zero because collisions occur among particles occupying the same lattice site.

**Dynamics and Problems of Multiparticle Lattice Gases (Literature):** The conclusions about the simulation capabilities, dynamics, and problems of multiparticle lattice gases available and abstracted from the existing literature are as follows:

**1)** Multiparticle lattice gases are capable of simulating diffusive systems (*i.e.*, systems with zero mean mass motion velocity) and incompressible fluid dynamic systems (after appropriate rescaling). In strict sense, they are suitable for simulation of diffusive systems only. Utility of these lattice gases for simulation of fluid dynamic systems is restricted because of several problems of which two major problems are: (i) violation of Galilean invariance (or, non-Galilean invariance) and (ii) inability of correctly reproducing compressible phenomena (*i.e.*, incompressibility). Under incompressible flow conditions, the problem of violation of Galilean invariance can be overcome in systems consisting of identical particles by appropriate rescaling.

**2)** In the literature, violation of Galilean invariance in multiparticle lattice gases is thought to be because of (i) Fermi-Dirac nature of equilibrium distribution function, (ii) structure and discreteness of the underlying spatial lattice and discreteness and finiteness of magnitude and number of velocity vectors of particles, and (iii) breaking of translation invariance because the underlying discrete spatial lattice provides a preferred Galilean reference frame and constrains the motion of particles to be only along the links of the lattice. All these explanations are, in essence, identical.

**3)** In the literature, incompressibility of multiparticle lattice gases is thought to be a consequence of their non-Galilean invariance. From this interlink, usually, it is inferred that recovery of Galilean invariance would allow simulation of compressible systems. This inference, however, being subject to counter evidence arising from studies conducted on lattice Boltzmann automata, is incorrect. This implies that non-Galilean invariance is not the *only* cause of incompressibility of multiparticle lattice gases.

**Dynamics and Problems of Multiparticle Lattice Gases (Analysis of the Present Investigation):** The conclusions about the simulation capabilities, dynamics, and problems of multiparticle lattice gases obtained by rigorous analysis of multiparticle lattice gases in the present investigation are as follows:

**1)** In multiparticle lattice gases interaction occur among particles occupying the same lattice site (*i.e.*, the same point in space) and the distance among them is always zero during interactions. This dynamics is unphysical and differs from that in physical systems. The consequence of this dynamics is that the model of particles encoded in multiparticle lattice gases is that of *hard point particles* (*i.e.*, zero range of interaction).

**2)** In multiparticle lattice gases spatial momentum redistribution occurs only during particle translations and not during interparticle interactions. This is a consequence of the (unphysical) nature of interparticle interactions in multiparticle lattice gases.

**3)** Absence of spatial momentum redistribution during interparticle interactions leads to insufficient redistribution of momentum in space in multiparticle lattice gases which reflects in the form of appearance of the multiplicative Galilean invariance breaking parameter in the nonlinear terms of coarse grained momentum equation.

Thus, in multiparticle lattice gases violation of Galilean invariance occurs because there is no spatial momentum redistribution during interparticle interactions.

**4)** Multiparticle lattice gases are not able to correctly simulate compressible systems and phenomena because in them spatial momentum redistribution does not occur during interparticle interactions. Alternatively, the problem of incompressibility in multiparticle lattice gases is because it is not possible to invoke collective response from particles occupying different lattice sites in them.

**5)** Since dynamics of interparticle interactions is encoded into interaction rules, whether or not spatial momentum redistribution occurs during interparticle interactions depends upon the interaction rules. In view of this, points 3 and 4 above imply that non-Galilean invariance and incompressibility of multiparticle lattice gases is because of the structure and construction of their interaction (collision) rules.

**6)** Although, both non-Galilean invariance and incompressibility of multiparticle lattice gases originate from the same deficiency in the microdynamics of multiparticle lattice gases, they are neither interrelated nor consequences of each other in that it is not necessary that overcoming the problem of non-Galilean invariance is sufficient for overcoming the problem of incompressibility as well. This is because it is possible to (partially) overcome the problem of non-Galilean invariance in ways, *e.g.*, by addition of rest particles, which leave the problem of incompressibility unresolved.

**7)** For overcoming the problems in multiparticle lattice gases, it is necessary to restore spatial momentum redistribution during interparticle interactions which can be done only by allowing at most one or zero particle, irrespective of its state, to occupy a lattice site at a time step (the *"single particle exclusion principle"*. This, however, cannot be done in multiparticle lattice gases because in them two or more must be allowed to occupy a lattice site simultaneously in any dimension (by their definition). As a result, it is not possible to restore spatial momentum redistribution during interparticle interactions in multiparticle lattice gases in a physically consistent manner. Thus, it is not possible to overcome the problems of non-Galilean invariance and incompressibility of multiparticle lattice gases.

**8)** If the definition of multiparticle lattice gases is changed to incorporate single particle exclusion principle, the resulting lattice gases become radically different in all aspects from multiparticle lattice gases and are called as *"single particle lattice gases"*.

### 9.3.2 Single Particle Lattice Gases

Single particle lattice gases are cellular automata models of physical systems developed, analyzed, and used for simulation of two systems in the present investigation. In these lattice gases the lattice site values are interpreted as representing the presence or absence exactly one particle along with its state at the lattice sites (the *"single particle exclusion principle"*. Alternatively, in these lattice gases at most one particle is allowed to occupy a lattice site at a time step. Single particle lattice gases are discrete analogs of molecular dynamics based on the formalism of cellular automata (sought in the present investigation). Major conclusions related to various aspects of single particle lattice gases are as follows:

**Definition, Structure, Constructions, and Dynamics of Single Particle Lattice Gases:**

**1)** In single particle lattice gases spatial momentum redistribution occurs during interparticle interactions as well as during particle translation. Spatial momentum redistribution during interparticle interactions occurs because of *single particle exclusion principle* embedded into their definition. Thus, the microdynamics of these lattice gases has all the elements present in the microdynamics of physical systems. As a result, these lattice gases are physically consistent, Galilean invariant, and capable of correctly simulating compressible phenomena and systems.

**2)** In single particle lattice gases evolution of the system during one time step is decomposed into two sub-steps, *viz.*, (i) particle translation, and (ii) interparticle interactions. During translation step particles move from one lattice site to another as pointed out by their velocity vectors. During interaction step, interparticle interactions are processed iteratively and the rules are constructed for processing to be carried out in each iteration in a way that all the relevant conservation laws are satisfied. These rules are called as *"fractional interaction rules"*. At each time step particle translation follows processing of interparticle interactions. This amounts to decomposing the evolution operator $\mathcal{E}$ in terms of translation operator $\mathcal{T}$ and interaction operator $\mathcal{C}$ as

$$\mathcal{E} = \mathcal{T}\mathcal{C}$$

where $\mathcal{C}$ is further decomposed into fractional interaction operator $\mathcal{C}'$ as

$$\mathcal{C} = \underbrace{\mathcal{C}' \cdots \mathcal{C}'}_{N_{\text{iter}}}$$

where $N_{\text{iter}}$ is the number of iterations required for processing interparticle interactions. $N_{\text{iter}}$ cannot be estimated *a priori* because $\mathcal{C}'$ is, in general, probabilistic.

**3)** In single particle lattice gases particles evolve in a globally coupled manner during interparticle interaction. As a result, the exact dynamics of single particle lattice gases can be represented only in terms of evolution of $N$-particle joint probability distribution function. Use of fractional interaction rules for evolution, however, makes it possible to decompose the $N$-particle distribution function in terms smaller $r$-particle distribution functions which appear in the fractional interaction rules.

**4)** The method of construction of single particle lattice gases (or, their fractional evolution rules) differs drastically from that of multiparticle lattice gases. This is because for construction of single particle lattice gases several considerations, in addition to those in multiparticle lattice gases, are required. These additional considerations surface because of enforcement of the single particle exclusion principle.

**5)** The method of construction of single particle lattice gases, though quite involved, is fully algorithmic and can be programmed on conventional digital computers for constructing single particle lattice gases with desired parameters.

**6)** In single particle lattice gases the lattice parameters and time step can be selected as required and desired interaction potentials can be associated with particles. This makes single particle lattice gases fully discrete analogs of molecular dynamics based on the formalism of cellular automata. These lattice gases, as defined in the present investigation, exist in the limit of collision time being negligible compared to the time step. This constraint is not inherent to single particle lattice gases and can be removed with minor modifications, if required.

**Construction and Properties of an Example Single Particle Lattice Gas:** A single species single speed single particle lattice gas existing over square spatial lattice has been constructed and analyzed to elucidates the generalized method of construction of single particle lattice gases conceived and described in the present work. This single particle lattice gas is called as *"SPLG-1"* model. The findings about the SPLG-1 model and about single particle lattice gases in general can be summarized as follows:

**1)** The SPLG-1 model, as defined in the present investigation, exists in the limit of collision time being negligible compared to the mean free time.

**2)** Fractional interaction rules (*i.e.*, the BCC lookup tables) have to be constructed separately for processing interactions among fluid particles as well for processing interactions of fluid particles with stationary solid boundary particles. The interaction rules must be constructed in compressed form. In this form, in the SPLG-1 model, the rules for processing interactions among fluid particles contain only 5 BCCs. On the other hand, if the rules were constructed in uncompressed form using the reduced contact interaction neighborhood they would contain 3125 configurations and if the entire contact interaction neighborhood was used they would contain $1220703125 \approx 1.22 \times 10^9$ configurations. Thus, defining the interaction rules in compressed form drastically reduced the memory requirements for usage of the model.

**3)** In the SPLG-1 model desired interaction potentials can be incorporated by appropriately selecting the STPs.

**4)** Using the SPLG-1 model, systems with various types of boundary conditions, *viz.*, (i) free or open boundaries, (ii) toroidal or periodic boundaries, and (iii) solid boundaries, can be simulated easily.

**5)** The SPLG-1 model has several fixed states under specific boundary conditions which form a closed repeating sequence of states during processing of interparticle interactions. These states, however, cannot occur during simulations unless the starting state itself is a fixed states with appropriate boundary conditions. These states are not strange attractors for the system.

**6)** Iterative decomposition of the interaction operator in terms of fractional interaction operator makes detailed mathematical analysis of the SPLG-1 model leading to various coarse grained equations very cumbersome and possible infeasible. An appropriate method of analysis wherein iterations can be treated properly has not been found yet.

**7)** Analytical expressions for the probability of occurrence of BCCs, probability of collision of particles, mean free path, mean free time, and speed of sound in the SPLG-1 model have been derived. It is found that both the mean free part and mean free time become unity as particle density becomes unity (*i.e.*, when the lattice is fully occupied). In this model, the speed of sound is a constant $(1/\sqrt{2})$ independent of density and other model parameters.

8) Finally, the construction of SPLG-1 model conclusively shows that the space of single particle lattice gases existing in the limit of collision time being negligible compared to the time step (or, alternatively, in the limit of collision time being negligible to the mean free time) is not empty.

**First Simulation Using the SPLG-1 Model:** The major findings from simulation of a system of particles enclosed in a box at equilibrium using the SPLG-1 model are as follows:

1) The motion of particles in the SPLG-1 model forms a Gaussian Markov process for STP sets which satisfy all the conservation laws including the law of conservation of angular momentum. The time variation of mean and mean square displacements of particle is as expected physically. Since the diffusive behavior of this model is identical to that reproduced by the Langevin equation, its the diffusive properties can be obtained by solving the Langevin equation. This conclusively shows that the SPLG-1 model is physically consistent and suitable for simulation of diffusive systems.

2) The STP sets in which the law of conservation angular momentum is violated, introduce bias in the motion of particles and force them to move in preferred directions. This results in dynamical behavior of lesser dimensionality compared to two as indicated by the dynamical exponent being less than unity. The exact nature of bias can be determined by analyzing the STP sets.

3) The time correlation function for velocity decays as expected for two-dimensional systems at long times. The dynamical exponents being unity indicate that the dynamical behavior of the model is two-dimensional and the model is physically consistent as expected.

4) The theoretical estimation of mean free path in the SPLG-1 model is strikingly accurate despite many simplifications and approximations involved.

**Second Simulation Using the SPLG-1 Model:** Simulation of relaxation of strong density perturbation in a system of particles enclosed in a finite length tube using the SPLG-1 model bring several new phenomena regarding propagation of waves and their interaction with solid boundaries to light. All these phenomena are physically explicable and consistent. The results, being new in the literature, cannot be subjected to comparative evaluation at present. The major findings from these simulations are as follows:

1) On removing partition separating high density and low density chambers a wave forms and propagates into the tube from high density side to the low density side. Near the walls, the mass and momentum components of the wave—in terms of maxima of density and maxima of momentum—separate out and go in different phases. Their development takes few mean free times following which they detach from the wall, come in phase with each other, merge, and propagate into the tube.

2) Relaxation/decay of density perturbation in tube shows linear behavior with time when the wave is away from the wall. In the vicinity of the wall, highly nonlinear relaxation pattern is observed.

3) The wave, when away from solid boundaries, propagates at a speed which is less than the maximum possible speed of particles. In general, the speed of the wave, when its is away from the walls or in free space, can never exceed and can only become equal to that of the maximum speed of particles. Thus, in the SPLG-1 model, when the wave is away from the wall its Mach number will never exceed $\sqrt{2}$.

**4)** When the wave approaches the wall and interacts with it, its speed increases vary rapidly and becomes many times more compared to the maximum speed of particles. In the simulations presented in this chapter, the peak of the density front is observed to travel at the speed of 5.3 lattice sites per time step during its approach towards the right wall. This corresponds to a Mach number of 7.49533.

**5)** In the process of approach and reflection from the wall, the wave—the maxima of density—stays with the wall for considerable time (few mean free times) before the wave reflects and starts moving away from the wall.

**6)** The motion of wave is independent of the motion of particles. This is in the sense that although the wave can propagate from one point in space to another, the particles comprising the medium may not and need not. This is a classical observation on wave propagation, *e.g.*, propagation of sounds waves and of waves on the surface of water, and is clearly visible in the simulations.

# Chapter 10

# Scope and Directions for Further Work

Did you notice the travellers wait at the forks and take a turn almost magically?
*I observed, sire.*
They seem to have a sense of direction and an ability to see and anticipate.
*Yes, sire.*
Can you tell what lies ahead?
*Wouldn't that be incorrect, sire?*
Why so?
*Free identifies with value less.*
Always?
...

The investigations carried out in this thesis on construction of molecular dynamics like cellular automata models, though directed towards simulation of fluid dynamic systems, are applicable in the more general and wider context of simulation of systems of particles. This wider applicability is, in fact, evident in the objective of this investigation since molecular dynamics is applicable for simulation of all the systems of particles. As a result, the class of cellular automata models, *viz.*, the single particle lattice gases, developed in this work are capable of simulating any system of particles. The only constraint is that the single particle lattice gases described in this work exist in the limit of collision time being negligible compared to a time step. This constraint, however, is neither a rigid constraint nor inherent to single particle lattice gases and can be easily removed as described in Sec. 5.2.3.6. In fact, whether this constraint should be kept or removed depends upon simulation requirements and is largely a free choice of the investigator. It will be useful to know the conditions under which this constraint must be kept or removed.

The method of construction of single particle lattice gases described in this thesis, although very general, has been used for construction of an example lattice gas which does not contain all the elements possible in a typical single particle lattice gas. In fact, the SPLG-1 model constructed in this thesis is the simplest possible two-dimensional model and employed the smallest symmetrical discrete velocity set in two-dimensional space. It will be advantageous to construct a few more one-, two-, and three-dimensional single particle lattice gases with bigger discrete velocity sets and carry out a comparative study of their dynamics. I suggest that single particle lattice gases be constructed in one-dimension using the discrete velocity sets

$$\begin{aligned}\mathcal{V}_{1.1} &= \{\pm 1\}\\ \mathcal{V}_{1.2} &= \{0, \pm 1\}\end{aligned}$$

$$\mathcal{V}_{1.3} = \{\pm 1, \pm 2\}$$
$$\mathcal{V}_{1.4} = \{0, \pm 1, \pm 2\}$$

and in two-dimensions using the discrete velocity sets

$$\begin{aligned}
\mathcal{V}_{2.2} &= \{(0,0),(\pm 1,0),(0,\pm 1)\} \\
\mathcal{V}_{2.3} &= \{(\pm 1,0),(0,\pm 1),(\pm 1,\pm 1)\} \\
\mathcal{V}_{2.4} &= \{(0,0),(\pm 1,0),(0,\pm 1),(\pm 1,\pm 1)\} \\
\mathcal{V}_{2.5} &= \{(\pm 1,0),(0,\pm 1),(\pm 1,\pm 1),(\pm 2,0),(0,\pm 2)\} \\
\mathcal{V}_{2.6} &= \{(0,0),(\pm 1,0),(0,\pm 1),(\pm 1,\pm 1),(\pm 2,0),(0,\pm 2)\}
\end{aligned}$$

and a comparative simulation study of their dynamics be carried out. Similar studies should be carried out in three-dimensions also. Studies in three-dimensions, however, appear to be extremely difficult at present because of large memory and simulation time requirements.

It will be very useful to carry out general theoretical analysis of single particle lattice gases and arrive at coarse grained dynamical equations. This, however, appear to be very difficult at present because of iterative decomposition of the interaction operator into fractional interaction operator and also because it requires a general method of treatment of arbitrary BCC lookup tables. A theoretical method for recovering the overall interaction operator from fractional interaction operators is needed (possibly it needs to be developed) before coarse grained dynamical equations can be rigorously obtained for single particle lattice gases. In fact, unless this problem is resolved, analytical expressions for viscosity in single particle lattice gases can also not be obtained.

On simulation side, only two simulation studies are presented in the present investigation. These studies are primarily intended for demonstrating the usage usability of the single particle lattice gases. Many more studies can and should be carried out for ascertaining the dynamical behavior of single particle lattice gases. Simulation studies, however, relate to specific single particle lattice gases only. As a result, for studying dynamical behavior of different single particle lattice gases through simulations, the simulations must be carried out afresh. For initial studies, it is desirable to simulate only those systems for which experimental and/or theoretical results are available.

In the single particle lattice gases formalized in this investigation the particles, in essence, are fermions which necessarily have non-zero rest mass. This is a consequence of the single particle exclusion principle. The method of construction of single particle lattice gases outlined in the present investigation is applicable only to fermionic systems. For simulating arbitrary physical systems it is necessary that Bosons should also be incorporated into the models. Thus, it is desirable to find a method of symbolically representing Bosons, of treating interactions among them, and of treating their interactions with fermions.

# Appendix A

# Re-rearrangement of Momentum Equation in Wolfram's Paper

I have some objections on the validity of Wolfram's analysis given in Secs. 2.5 and 2.6 of [1] wherein an attempt has been made to show the closeness of coarse grained momentum equation of lattice gases with the standard Navier-Stokes equation. The objections are concerned with (i) the validity of the form of standard Navier-Stokes equation used in [1], and (ii) the reduction of the coarse grained momentum equation of lattice gases to this form. Thus, in Sec. A.1 the relevant sections of Wolfram's analysis have been reproduced from [1], in Sec. A.2 the objections on the validity of this analysis are presented, and in Sec. A.4 the analysis has been redone systematically to arrive at the correct form of equations.

The numbering of equations in the following sections is according to the following scheme: Equations reproduced from [1] are numbered in the form (XX.YY.ZZ), the numbers being the same as those in [1]. Equations not taken from [1], *i.e.*, the equations reproduced from other sources or derived herein, are numbered in the form (XX.YY).

## A.1 Wolfram's Analysis

In Wolfram's paper [1] the equation for conservation of momentum in generalized form has been written as

$$\partial_t(nu_i) + \partial_j \Pi_{ij} = 0 \tag{2.4.10}$$

where $\Pi_{ij}$ obtained after appropriate considerations is

$$\Pi_{ij} = \frac{n}{2}\delta_{ij} + \frac{n}{4}c^{(2)}\left[u_i u_j - \frac{1}{2}|\boldsymbol{u}|^2\delta_{ij}\right] + \frac{n}{4}c_{\triangledown}^{(2)}\left[\partial_i u_j - \frac{1}{2}\boldsymbol{\nabla}\cdot\boldsymbol{u}\right] \tag{2.5.10}$$

Substituting the expression for $\Pi_{ij}$ from Eq. (2.5.10) in Eq. (2.4.10), the momentum equation has been obtained as

$$\partial_t(nu_i) + \frac{1}{4}nc^{(2)}\left\{(\boldsymbol{u}\cdot\boldsymbol{\nabla})\boldsymbol{u} + \left[\boldsymbol{u}(\boldsymbol{\nabla}\cdot\boldsymbol{u}) - \frac{1}{2}\boldsymbol{\nabla}|\boldsymbol{u}|^2\right]\right\} = -\frac{1}{2}\boldsymbol{\nabla}n - \frac{1}{8}nc_{\triangledown}^{(2)}\boldsymbol{\nabla}^2\boldsymbol{u} - \frac{1}{4}\Xi \tag{2.5.11}$$

where

$$\Xi = \boldsymbol{u}(\boldsymbol{u}\cdot\boldsymbol{\nabla})(nc^{(2)}) - \frac{1}{2}|\boldsymbol{u}|^2\boldsymbol{\nabla}(nc^{(2)}) + (\boldsymbol{u}\cdot\boldsymbol{\nabla})(nc_{\triangledown}^{(2)}) - \frac{1}{2}(\boldsymbol{\nabla}\cdot\boldsymbol{u})\boldsymbol{\nabla}(nc_{\triangledown}^{(2)}) \tag{2.5.12}$$

In Sec. 2.6 of [1], standard Navier-Stokes equation for continuum fluids in $\mathcal{D}$-dimensions has been written as

$$\frac{\partial n\boldsymbol{u}}{\partial t} + \mu n(\boldsymbol{u}\cdot\boldsymbol{\nabla})\boldsymbol{u} = -\boldsymbol{\nabla}p + \eta\nabla^2\boldsymbol{u} + \left(\zeta + \frac{1}{\mathcal{D}}\eta\right)\boldsymbol{\nabla}(\boldsymbol{\nabla}\cdot\boldsymbol{u}) \tag{2.6.1}$$

where $p$ is pressure, and $\eta$ and $\zeta$ are, respectively, shear and bulk viscosities. The coefficient $\mu$ of the convective terms is usually constrained to have value 1 by Galilean invariance.

Eq. (2.5.11) has been compared with Eq. (2.6.1) and closeness of the two has been pointed out with the approximations that the terms proportional to $\boldsymbol{u}\boldsymbol{\nabla}n$ and $\boldsymbol{u}(\boldsymbol{\nabla}\cdot\boldsymbol{u})$ must be neglected and the term proportional to $\boldsymbol{\nabla}|\boldsymbol{u}|^2$ can be combined with the $\boldsymbol{\nabla}n$ term to yield an effective pressure term which includes fluid kinetic energy distributions. This gives the values of the constants in Eq. (2.6.1) in terms of those in Eq. (2.5.11) as

$$\zeta = 0 \tag{2.6.2}$$

$$\eta = n\nu = -\frac{1}{8}nc_{\nabla}^{(2)} \tag{2.6.3}$$

and,

$$\mu = \frac{1}{4}c^{(2)} \tag{2.6.4}$$

where $\nu$ in Eq. (2.6.3) is kinematic viscosity.

## A.2 Objections on Wolfram's Analysis

The objections, on Wolfram's analysis partially reproduced above, are the following:

**1)** Eq. (2.6.1) is not the correct form of the standard Navier-Stokes equation. This statement is irrespective of the coefficient $\mu$ of the convective term.

**2)** The last term in the right hand side of Eq. (2.5.10), *i.e.*, the term $-\frac{1}{2}\boldsymbol{\nabla}\cdot\boldsymbol{u}$, is incorrect because this term is a scalar. In the correct form of the equation, a tensor term should have been there. The correct form of this term is $-\frac{1}{2}(\boldsymbol{\nabla}\cdot\boldsymbol{u})\delta_{ij}$.

**3)** Third term in the right hand side of Eq. (2.5.12) is incorrect because this term is a scalar. In the correct form of the equation, a vector term should have been there.

**4)** Eq. (2.4.10), with the expression of $\Pi_{ij}$ given by Eq. (2.5.10), does assume the form of Eq. (2.5.11).

**5)** Under the approximations invoked by Wolfram, Eq. (2.5.11) does not assume any form close to that of Eq. (2.6.1). This is in addition to the points made above.

## A.3 Forms of Standard Navier-Stokes Equation

The standard Navier-Stokes equation, in the absence of body forces and for small temperature variations so that shear and bulk viscosities remain constant, can be written [2,3]

$$\frac{\partial n\boldsymbol{u}}{\partial t} + \boldsymbol{\nabla}\cdot(n\boldsymbol{u}\boldsymbol{u}) = -\boldsymbol{\nabla}p + \eta\boldsymbol{\nabla}^2\boldsymbol{u} + \left(\zeta + \frac{1}{\mathcal{D}}\eta\right)\boldsymbol{\nabla}(\boldsymbol{\nabla}\cdot\boldsymbol{u}) \tag{A.1}$$

where $n\boldsymbol{uu}$ is a tensor, and $n$ is density. This equation can be reduced to the form

$$n\frac{\partial \boldsymbol{u}}{\partial t} + n(\boldsymbol{u}\cdot\boldsymbol{\nabla})\boldsymbol{u} = -\boldsymbol{\nabla}p + \eta\boldsymbol{\nabla}^2\boldsymbol{u} + \left(\zeta + \frac{1}{\mathcal{D}}\eta\right)\boldsymbol{\nabla}(\boldsymbol{\nabla}\cdot\boldsymbol{u}) \tag{A.2}$$

using the continuity equation

$$\frac{\partial n}{\partial t} + \boldsymbol{\nabla}\cdot(n\boldsymbol{u}) = 0 \tag{A.3}$$

and the expression

$$\boldsymbol{\nabla}\cdot(n\boldsymbol{uu}) = n(\boldsymbol{u}\cdot\boldsymbol{\nabla})\boldsymbol{u} + \boldsymbol{u}(\boldsymbol{\nabla}\cdot n\boldsymbol{u})$$

Both these forms of the Navier-Stokes equation, *i.e.*, Eqs. (A.1) and (A.2), are Galilean invariant. If one wants to *intuitively* generalize these equations for systems violating Galilean invariance, one puts a multiplicative factor $\mu$ (in Wolfram's symbolism; $g$ in the symbolism used in the present investigation and also used by Frisch *et. al.* [4]) in the nonlinear terms. With this generalization, each one of Eqs. (A.1) and (A.2) result in two different possible equations, *viz.*, Eq. (A.1) gives the equations

$$\frac{\partial n\boldsymbol{u}}{\partial t} + \mu\boldsymbol{\nabla}\cdot(n\boldsymbol{uu}) = -\boldsymbol{\nabla}p + \eta\boldsymbol{\nabla}^2\boldsymbol{u} + \left(\zeta + \frac{1}{\mathcal{D}}\eta\right)\boldsymbol{\nabla}(\boldsymbol{\nabla}\cdot\boldsymbol{u}) \tag{A.4}$$

and

$$\frac{\partial n\boldsymbol{u}}{\partial t} + \boldsymbol{\nabla}\cdot(\mu n\boldsymbol{uu}) = -\boldsymbol{\nabla}p + \eta\boldsymbol{\nabla}^2\boldsymbol{u} + \left(\zeta + \frac{1}{\mathcal{D}}\eta\right)\boldsymbol{\nabla}(\boldsymbol{\nabla}\cdot\boldsymbol{u}) \tag{A.5}$$

and the Eq. (A.2) gives the equations

$$n\frac{\partial \boldsymbol{u}}{\partial t} + \mu n(\boldsymbol{u}\cdot\boldsymbol{\nabla})\boldsymbol{u} = -\boldsymbol{\nabla}p + \eta\boldsymbol{\nabla}^2\boldsymbol{u} + \left(\zeta + \frac{1}{\mathcal{D}}\eta\right)\boldsymbol{\nabla}(\boldsymbol{\nabla}\cdot\boldsymbol{u}) \tag{A.6}$$

and

$$n\frac{\partial \boldsymbol{u}}{\partial t} + n(\boldsymbol{u}\cdot\boldsymbol{\nabla})\mu\boldsymbol{u} = -\boldsymbol{\nabla}p + \eta\boldsymbol{\nabla}^2\boldsymbol{u} + \left(\zeta + \frac{1}{\mathcal{D}}\eta\right)\boldsymbol{\nabla}(\boldsymbol{\nabla}\cdot\boldsymbol{u}) \tag{A.7}$$

Note again that Eqs. (A.4), (A.5), (A.6), and (A.7) are purely intuitive generalizations of Eqs. (A.1) and (A.2). On substituting $\mu = 1$ in these equations, they unconditionally reduce to their parent equations. Note also that none of these equations are similar to that used by Wolfram, *viz.*, Eq. (2.6.1), in [1]. Thus, Wolfram's generalization of the Navier-Stokes equation for including non-Galilean invariance is incorrect. Which equation is the correct generalization of the Navier-Stokes equation for lattice gases (more precisely, multiparticle lattice gases) has been determined in the next section (Sec. A.4).

Physical validity of Eqs. (A.4), (A.5), (A.6), and (A.7) is not of concern here because they have been written down only to show that Wolfram's (apparently *intuitive*) generalization of the Navier-Stokes equation, *i.e.*, Eq. (2.6.1) in [1], is incorrect.

## A.4 Redoing Wolfram's Analysis

Using Eq. (2.5.10) with the correction pointed out in Sec. A.2, the second term of Eq. (2.4.10) can be evaluated as

$$\partial_j \Pi_{ij} \;=\; \partial_j \frac{n}{2}\delta_{ij} + \partial_j \frac{n}{4}c^{(2)}\left[u_i u_j - \frac{1}{2}|\boldsymbol{u}|^2\delta_{ij}\right] + \partial_j \frac{n}{4}c^{(2)}_{\triangledown}\left[\partial_i u_j - \frac{1}{2}(\boldsymbol{\nabla}\cdot\boldsymbol{u})\delta_{ij}\right] \tag{A.8}$$

$$= \partial_j \frac{1}{2} n\delta_{ij} + \partial_j \frac{1}{4} nc^{(2)} u_i u_j - \partial_j \frac{1}{8} nc^{(2)} |\boldsymbol{u}|^2 \delta_{ij} + \partial_j \frac{1}{4} nc_{\nabla}^{(2)} \partial_i u_j - \partial_j \frac{1}{8} nc_{\nabla}^{(2)} (\boldsymbol{\nabla} \cdot \boldsymbol{u}) \delta_{ij} \tag{A.9}$$

$$= \underbrace{\partial_j \frac{1}{2} n\delta_{ij}}_{\text{(A)}} + \underbrace{\partial_j n\mu u_i u_j}_{\text{(B)}} - \underbrace{\partial_j \frac{1}{2} n\mu |\boldsymbol{u}|^2 \delta_{ij}}_{\text{(C)}} - \underbrace{2\partial_j \eta \partial_i u_j}_{\text{(D)}} + \underbrace{\partial_j \eta (\boldsymbol{\nabla} \cdot \boldsymbol{u}) \delta_{ij}}_{\text{(E)}} \tag{A.10}$$

where Eq. (A.10) is obtained from Eq. (A.9) on substituting from Eqs. (2.6.3) and (2.6.4).

Various terms of Eq. (A.10) can be written in the conventional form as

$$\begin{aligned} \text{Term-(A)} &= \partial_j \frac{1}{2} n\delta_{ij} \\ &= \frac{1}{2} \boldsymbol{\nabla} n \end{aligned} \tag{A.11}$$

$$\begin{aligned} \text{Term-(B)} &= \partial_j n\mu u_i u_j \\ &= u_i u_j \partial_j n\mu + n\mu \partial_j u_i u_j \\ &= u_i u_j \partial_j n\mu + n\mu u_i \partial_j u_j + n\mu u_j \partial_j u_i \\ &= \boldsymbol{u}(\boldsymbol{u} \cdot \boldsymbol{\nabla})(n\mu) + n\mu \boldsymbol{u}(\boldsymbol{\nabla} \cdot \boldsymbol{u}) + n\mu(\boldsymbol{u} \cdot \boldsymbol{\nabla})\boldsymbol{u} \qquad \text{(A.12)} \\ &= \boldsymbol{u}(\boldsymbol{\nabla} \cdot \boldsymbol{u}) n\mu + n\mu \boldsymbol{\nabla} \cdot (\boldsymbol{u}\boldsymbol{u}) \\ &= \boldsymbol{\nabla} \cdot (n\mu \boldsymbol{u}\boldsymbol{u}) \qquad \text{(A.13)} \end{aligned}$$

$$\begin{aligned} \text{Term-(C)} &= \partial_j \frac{1}{2} n\mu |\boldsymbol{u}|^2 \delta_{ij} \\ &= \frac{1}{2} n\mu \partial_j |\boldsymbol{u}|^2 \delta_{ij} + |\boldsymbol{u}|^2 \partial_j \frac{1}{2} n\mu \delta_{ij} \\ &= \frac{1}{2} n\mu \boldsymbol{\nabla} |\boldsymbol{u}|^2 + |\boldsymbol{u}|^2 \boldsymbol{\nabla} \left( \frac{1}{2} n\mu \right) \qquad \text{(A.14)} \\ &= \boldsymbol{\nabla} \left( \frac{1}{2} n\mu |\boldsymbol{u}|^2 \right) \qquad \text{(A.15)} \end{aligned}$$

$$\begin{aligned} \text{Term-(D)} &= 2\partial_j \eta \partial_i u_j \\ &= 2\eta \partial_j \partial_i u_j + 2\partial_i u_j \partial_j \eta - 2u_j \partial_j \partial_i \eta \\ &= 2\eta \partial_i \partial_j u_j + 2\partial_i u_j \partial_j \eta - 2u_j \partial_j \partial_i \eta \\ &= 2\eta \boldsymbol{\nabla}(\boldsymbol{\nabla} \cdot \boldsymbol{u}) + 2\boldsymbol{\nabla}(\boldsymbol{u} \cdot \boldsymbol{\nabla})\eta - 2(\boldsymbol{u} \cdot \boldsymbol{\nabla})\boldsymbol{\nabla}\eta \end{aligned} \tag{A.16}$$

$$\begin{aligned} \text{Term-(E)} &= \partial_j \eta (\boldsymbol{\nabla} \cdot \boldsymbol{u}) \delta_{ij} \\ &= \eta \partial_j (\boldsymbol{\nabla} \cdot \boldsymbol{u}) \delta_{ij} + (\boldsymbol{\nabla} \cdot \boldsymbol{u}) \partial_j \eta \delta_{ij} \\ &= \eta \boldsymbol{\nabla}(\boldsymbol{\nabla} \cdot \boldsymbol{u}) + (\boldsymbol{\nabla} \cdot \boldsymbol{u}) \boldsymbol{\nabla}\eta \end{aligned} \tag{A.17}$$

Using Eq. (A.10) and substituting from Eqs. (A.11), (A.12), (A.14), (A.16), and (A.17), the equation for conservation of momentum, Eq. (2.4.10), becomes

$$\begin{aligned} &\partial_t(nu_i) + \frac{1}{2} \boldsymbol{\nabla} n + \boldsymbol{u}(\boldsymbol{u} \cdot \boldsymbol{\nabla})(n\mu) + n\mu \boldsymbol{u}(\boldsymbol{\nabla} \cdot \boldsymbol{u}) + n\mu(\boldsymbol{u} \cdot \boldsymbol{\nabla})\boldsymbol{u} \\ &\quad - \frac{1}{2} n\mu \boldsymbol{\nabla} |\boldsymbol{u}|^2 - \frac{1}{2} |\boldsymbol{u}|^2 \boldsymbol{\nabla}(n\mu) - \eta \boldsymbol{\nabla}(\boldsymbol{\nabla} \cdot \boldsymbol{u}) - 2\boldsymbol{\nabla}(\boldsymbol{u} \cdot \boldsymbol{\nabla})\eta \\ &\qquad + 2(\boldsymbol{u} \cdot \boldsymbol{\nabla})\boldsymbol{\nabla}\eta + (\boldsymbol{\nabla} \cdot \boldsymbol{u})\boldsymbol{\nabla}\eta = 0 \end{aligned} \tag{A.18}$$

For comparison of Eq. (2.5.11) with the above equation, Eq. (2.5.11) can be rewritten, by collating all the terms in left hand side and substituting from Eqs. (2.5.12), (2.6.3), and (2.6.4), as

$$\begin{aligned}\partial_t(nu_i) + \frac{1}{2}\boldsymbol{\nabla} n + \boldsymbol{u}(\boldsymbol{u}\cdot\boldsymbol{\nabla})(n\mu) + n\mu\boldsymbol{u}(\boldsymbol{\nabla}\cdot\boldsymbol{u}) + n\mu(\boldsymbol{u}\cdot\boldsymbol{\nabla})\boldsymbol{u}& \\ -\frac{1}{2}n\mu\boldsymbol{\nabla}|\boldsymbol{u}|^2 - \frac{1}{2}|\boldsymbol{u}|^2\boldsymbol{\nabla}(n\mu) - \eta\boldsymbol{\nabla}^2\boldsymbol{u} + 2(\boldsymbol{u}\cdot\boldsymbol{\nabla})\eta + (\boldsymbol{\nabla}\cdot\boldsymbol{u})\boldsymbol{\nabla}\eta &= 0 \qquad \text{(A.19)}\end{aligned}$$

Comparison of Eq. (A.18) with Eq. (A.19) shown that Eq. (2.4.10) with the expression of $\Pi_{ij}$ given by Eq. (2.5.10) does not assume the form of Eq. (2.5.11). In the same continuation, Eq. (2.5.11) does not assume any form closer to Eq. (2.6.1).

Further rearrangement of Eq. (A.18) gives

$$\begin{aligned}\partial_t(nu_i) + \boldsymbol{\nabla}\cdot(n\mu\boldsymbol{u}\boldsymbol{u}) &= -\boldsymbol{\nabla}\left[\frac{1}{2}n\left(1-\mu|\boldsymbol{u}|^2\right)\right] + \eta\boldsymbol{\nabla}(\boldsymbol{\nabla}\cdot\boldsymbol{u}) \\ &\quad +2\boldsymbol{\nabla}(\boldsymbol{u}\cdot\boldsymbol{\nabla})\eta - 2(\boldsymbol{u}\cdot\boldsymbol{\nabla})\boldsymbol{\nabla}\eta - (\boldsymbol{\nabla}\cdot\boldsymbol{u})\boldsymbol{\nabla}\eta \qquad \text{(A.20)}\end{aligned}$$

Note that in the above equation a term of the form $(\boldsymbol{\nabla}\cdot\boldsymbol{\nabla})\boldsymbol{u}$ should have appeared; which has not happened. For this term to appear it is necessary that a term of the form $\partial_j u_i$ be present in the expression for $\Pi_{ij}$; which is missing. In the coarse grained equation obtained by Frisch *et. al.* [4], these terms are present. Thus, possibly, the expression for $\Pi_{ij}$ obtained in [1] is also erroneous.

The above also shows that the correct form of the Navier-Stokes equation on *intuitively* generalizing it to include non-Galilean invariant systems is the one given by Eq. (A.5).

# Appendix B

# Distance of Closest Approach of Particles in Two-Dimensional Space

Let two particles A and B of mass $m_A$ and $m_B$ be moving with velocities $\boldsymbol{v}_A$ and $\boldsymbol{v}_B$ on a collision path in two-dimensional space. Let $\Phi(r)$ be their mutual interaction potential. The distance of closest approach $r_{c_{min}}$ of these particles is computed as follows:

Since the system is not subjected to external forces, the velocity of its center of mass does not change with time even though the velocities of the particles change with time. Let $\boldsymbol{v}_{cm}$ be the velocity of center of mass of the system, then the law of conservation of linear momentum gives

$$\boldsymbol{v}_{cm}(m_A + m_B) = m_A\boldsymbol{v}_A + m_B\boldsymbol{v}_B \tag{B.1}$$

Similarly, at the point of closest approach of particles, the law of conservation of linear momentum gives

$$\boldsymbol{v}_{cm}(m_A + m_B) = m_A\boldsymbol{u}_A + m_B\boldsymbol{u}_B \tag{B.2}$$

where $\boldsymbol{u}_A$ and $\boldsymbol{u}_B$ are the velocities of particles at the point of closest approach, *i.e.*, when their distance is $r_{c_{min}}$.

At the point of closest approach of particles, the law of conservation of energy gives

$$\Phi(r_{c_{min}}) + \frac{1}{2}m_A\boldsymbol{u}_A^2 + \frac{1}{2}m_B\boldsymbol{u}_B^2 = \frac{1}{2}m_A\boldsymbol{v}_A^2 + \frac{1}{2}m_B\boldsymbol{v}_B^2 \tag{B.3}$$

This equation can be solved for the value of $r_{c_{min}}$ if $\boldsymbol{u}_A$ and $\boldsymbol{u}_B$ are known.

Let $\hat{\boldsymbol{v}}_{cm}$, defined as

$$\hat{\boldsymbol{v}}_{cm} = \frac{\boldsymbol{v}_{cm}}{|\boldsymbol{v}_{cm}|} \tag{B.4}$$

be the unit vector in the direction of $\boldsymbol{v}_{cm}$.

At the point of closest approach of particles the velocities of particles must necessarily be parallel to each other as well as to the velocity of center of mass of the system. Let $\boldsymbol{v}_A$ and $\boldsymbol{v}_B$ be resolved as components parallel and perpendicular to $\hat{\boldsymbol{v}}_{cm}$. The components perpendicular to $\hat{\boldsymbol{v}}_{cm}$ take the particles close to each other whereas the components parallel to $\hat{\boldsymbol{v}}_{cm}$ remain unchanged. The components perpendicular to $\hat{\boldsymbol{v}}_{cm}$ become zero at the point of closest approach of particles. Thus,

$$\boldsymbol{u}_A = \hat{\boldsymbol{v}}_{cm}(\boldsymbol{v}_A \cdot \hat{\boldsymbol{v}}_{cm}) \tag{B.5}$$

$$\boldsymbol{u}_B = \hat{\boldsymbol{v}}_{cm}(\boldsymbol{v}_B \cdot \hat{\boldsymbol{v}}_{cm}) \tag{B.6}$$

These equations can be used to solve Eq. (B.3) for value of $r_{c_{min}}$.

If

$$\Phi(r) = ar^{-b}$$

the solution for $r_{c_{\min}}$ is

$$r_{c_{\min}} = \left[ \frac{2a}{m_A[\boldsymbol{v}_A^2 - (\boldsymbol{v}_A \cdot \hat{\boldsymbol{v}}_{cm})^2] + m_B[\boldsymbol{v}_B^2 - (\boldsymbol{v}_B \cdot \hat{\boldsymbol{v}}_{cm})^2]} \right]^{1/b} \tag{B.7}$$

It is noteworthy that $r_{c_{\min}}$ decreases with increase in velocity of particles.

The procedure of computation of $r_{c_{\min}}$ for collisions in three-dimensional space can be found in [1,2].

# List of Publications

The following papers have been published as yet from this work:

1) H. Agrawal and E. Rathakrishnan. Single particle cellular automata models for simulation of the master equation: Non-reacting single specie systems. In M. Capitelli, editor, *Molecular Physics and Hypersonic Flows*, volume 482 of *NATO ASI Series C: Mathematical and Physical Sciences*, pages 759–770. Kluwer Academic Publishers (Dorcrechet, The Netherlands), 1996. Proceedings of the NATO Advanced Study Institute, on Molecular Physics and Hypersonic Flows, Maratea, Italy, May 21–June 3, 1995.

2) H. Agrawal, K. Srinivasan, and E. Rathakrishnan. Simulation of nonlinear relaxation of density differences in an athermal gas using an extended lattice-gas-automata model. Presented at the 20th International Symposium on Space Technology and Science, Gifu, Japan, May 19–25, 1996.